\documentclass[12pt,a4paper,twoside,openright]{report} %include openright here in the documentclass for the arXiv
\usepackage{dissertation}
\makeglossaries
\makeindex

\logo{EC}{School of Sciences}{}
\logoB{EC}{School of Sciences}{}

\author{Rafael Wagner}

\titleA{Coherence and contextuality}
\titleB{as quantum resources}
\titleC{}

\phd{Doctoral Program in Physics}
\supervisor{Ernesto Fagundes Galvão}
\cosupervisor{Rui Soares Barbosa\\ Mikhail Vasilevskiy}

\bibpunct[,]{(}{)}{;}{a}{,}{,}

\begin{document}
\setlength{\parindent}{0em}

%-- Covers
\begin{titlepage}
\pagecolor{PANTONECoolGray7C}
\afterpage{\nopagecolor}
\color{white}
\thelogo
\leading{20.5pt}
{\Large
\theauthor
\\
\\
\textbf{\thetitleA}
\\
\textbf{\thetitleB}
\\
\textbf{\thetitleC}
}

\vspace*{\fill}
{\footnotesize \myear}
\end{titlepage}

\null
\thispagestyle{empty}
\pagecolor{PANTONE1807C}
\afterpage{\nopagecolor}
\newpage

\begin{titlepage}
\color{PANTONECoolGray7C}
\thelogoB
\leading{20.4pt}
{\Large
\theauthor
\\
\\
\textbf{\thetitleA}
\\
\textbf{\thetitleB}
\\
\textbf{\thetitleC}
}

\vspace{55.2mm}
\leading{16.8pt}
{\large
Doctoral Thesis
\\
\thephd
\\
\thearea
Work carried out under the supervision of
\\
\textbf{\thesupervisor}
\\
\thecosupervisor}

\vspace*{\fill}
{\footnotesize \myear
}
\end{titlepage}

\null
\thispagestyle{empty}
\pagecolor{PANTONE1807C} %Substitute with white
\afterpage{\nopagecolor}
\newpage

%-- Document setup
\newgeometry{right=25mm, left=25mm, top=25mm, bottom=25mm}
\pagenumbering{roman}

\setlength{\parskip}{0pt}
\setlength{\parindent}{0pt}

%-- Preamble
\chapter*{Copyright and Terms of Use for Third Party Work}
\noindent
This dissertation reports on academic work that can be used by third parties as long as the internationally accepted standards and good practices are respected concerning copyright and related rights.

\noindent
This work can thereafter be used under the terms established in the license below.

\noindent
Readers needing authorization conditions not provided for in the indicated licensing should contact the author through the RepositóriUM of the University of Minho.

\section*{License granted to users of this work:}

%\textit{[Caso o autor pretenda usar uma das licenças Creative Commons, deve escolher e deixar apenas um dos seguintes ícones e respetivo lettering e URL, eliminando o texto em itálico que se lhe segue. Contudo, é possível optar por outro tipo de licença, devendo, nesse caso, ser incluída a informação necessária adaptando devidamente esta minuta]}

\noindent
\includegraphics[width=0.12\textwidth]{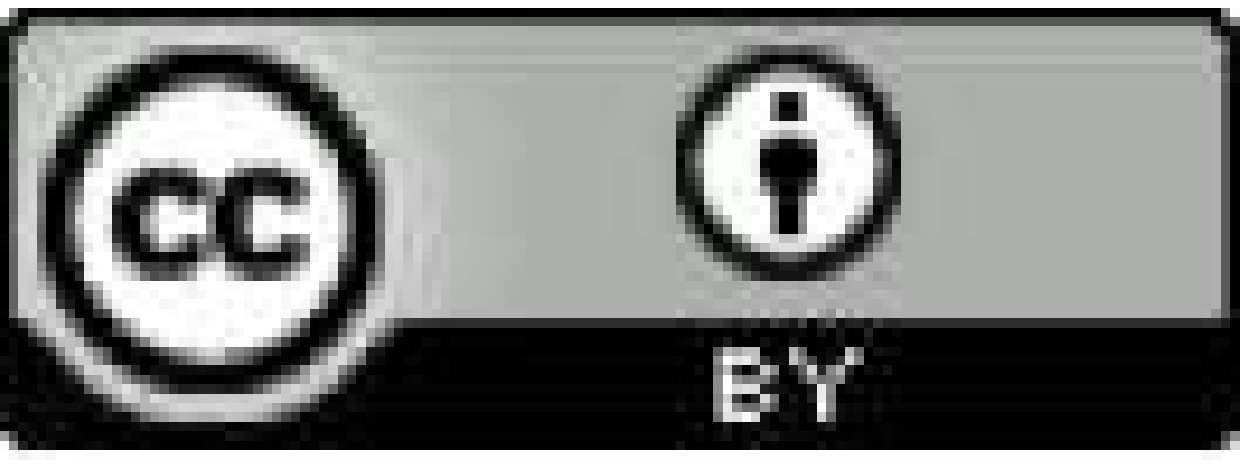}\newline 
\textbf{CC BY}\newline 
\url{https://creativecommons.org/licenses/by/4.0/}
\chapter*{Acknowledgements}

\myindent This PhD thesis was only possible thanks to the enormous help and support I received from my PhD advisors Ernesto Galvão, Rui Soares Barbosa, and Mikhail Vasilevskiy, as well as my Master’s advisor, Bárbara Amaral. Their kindness, patience, and critical feedback have been instrumental in my academic growth in ways that cannot be adequately expressed here. I am deeply grateful to all of them.

\myindent The original work presented in this thesis would not have been possible without the support of an extraordinary group of friends and collaborators. Doing my best to mention them all, I warmly thank Adán Cabello, Alexandra and Mafalda Ramoa, Alisson Tezzin, Amit Te’eni, Ana Neres, Luisa Madail, João Henriques, Anita Camillini, Antonio Ruiz-Molero, Carlos Fernandes, Carlos Tavares, Chiara Esposito, Costantino Budroni, Cristhiano Duarte, Danilo Cius, David Schmid, Eliahu Cohen, Emmanuel Zambrini Cruzeiro, Fabio Sciarrino, Felix Ahnefeld, Filipa Peres, Flavien Hirsch, Francesco Hoch, Giulio Camilo, Gonzalo Carvacho, Guilherme Zambon, Ilaria Andrei, Ismael L. Paiva, Jack Davis, John Selby, José Diogo Guimarães, Junseo Lee, Kyrylo Simonov, Laurens Walleghem, Leonardo Novo, Luis Santos, Marco Erba, Manisha Jain, Marcelo Terra Cunha, Marina Ancelli, Michael Oliveira, Naim Elias Comar, Nicoló Spagnolo, Osvaldo, Pedro Dieguez, Raman Choudhary, Roberto Baldijão, Lorenzo Catani, Som Kanjilal, Selma Memic, Rudi Pietsch, Roberto Osellame, Ron Cohen, Shuming Cheng, Taira Giordani, Yu Guo, Yìlè Yīng, Zohar Schwartzman-Nowik, and Pedro Azado. Special thanks to Sara Franco for proofreading early versions of this thesis.

\myindent I would also like to express my gratitude to professors Martin B. Plenio, Costantino Budroni, Luis Paulo Santos, Armando Nolasco Pinto, António Onofre, Mikhail Vasilevskiy, and João Viana Lopes for serving as members of the jury committee for my thesis defense and for their insightful comments and suggestions, which have contributed to improving this dissertation.

\myindent I am immensely thankful to my whole family---especially Mauro, Elaine, Dora, Anna, Mingo, Bianca, and Marta---and my friends Gustavo and Cauê, who supported me in countless ways. Above all, I owe my deepest gratitude to my wife, Elisabetta Girardi, for her unwavering support and love throughout this journey. 

\myindent I would like to acknowledge the financial support that made this PhD thesis possible. I am grateful to the Quantum Portugal Initiative and FCT—Fundação para a Ciência e a Tecnologia (Portugal) for funding my research through the PhD Grant SFRH/BD/151199/2021. I also acknowledge funding to disseminate the research outputs of this project from The Foundations of Quantum Computational Advantage - FoQaCia, Project GA no. 101070558; the Photonics Quantum Sampling Machine Phoqusing, Project FET- Future Emerging Technologies H2020 FETOPEN; the Quantum advantage via non-linear boson sampling QU-BOSS, Project  ERC Advanced Grant GA no. 884676; and the Bar-Ilan Institute of Nanotechnology and Advanced Materials (BINA). I also want to thank the International Iberian Nanotechnology Laboratory (INL) in Braga for hosting my research activities, and for Professor Costantino Budroni and the University of Pisa or hosting my research activities during a six months visiting period.

\myindent To conclude, I dedicate this thesis to the memory of my cousin,  Cézar Augusto Martinez.
\chapter*{Statement of Integrity}
\noindent
I hereby declare having conducted this academic work with integrity.

\noindent
I confirm that I have not used plagiarism or any form of undue use of information or falsification of results along the process leading to its elaboration.

\noindent
I further declare that I have fully acknowledged the Code of Ethical Conduct of the University of Minho.

\phantom{space}

\noindent
University of Minho, Braga, \myear

\vspace{25mm}
\noindent\theauthor
\chapter*{Resumo}

Neste trabalho, exploramos a interseção de dois subcampos fundamentais da teoria da informação quântica: a coerência quântica e a contextualidade. Apesar de suas diferenças aparentes, ambas as áreas abordam questões-chave relevantes para os fundamentos e as aplicações da teoria quântica. Por meio do desenvolvimento de uma abordagem que usa teoria de grafos, que estende um formalismo recentemente introduzido por Galvão e Brod~(Phys.~Rev.~A \textbf{101},~062110,~2020), estabelecemos uma conexão formal entre testemunhas de coerência quântica e desigualdades de não-contextualidade. Nossas principais contribuições incluem: o desenvolvimento do formalismo de grafos mencionado anteriormente para a geração de testemunhas de coerência e contextualidade; um mapeamento formal entre as desigualdades derivadas por Galvão e Brod e desigualdades de não contextualidade já existentes na literatura; a conceitualização de uma forma relacional de coerência quântica; uma demonstração de vantagem contextual na tarefa de interrogação quântica; e a descoberta de uma família infinita de testemunhas de coerência que também exigem estados quânticos definidos em espaços de Hilbert com dimensões específicas.

\paragraph{Palavras-chave} Coerência Quântica, Contextualidade, Informação Quântica, Informação Relacional, Testemunhas de Dimensão, Interrogação Quântica

\cleardoublepage

\chapter*{Abstract}

In this thesis, we explore the intersection of two fundamental subfields of quantum information theory: quantum coherence and contextuality. Despite their apparent differences, both areas address key issues relevant to the foundations and applications of quantum theory. By developing a novel graph-theoretic approach, extending a framework recently introduced by Galvão and Brod~(Phys.~Rev.~A~\textbf{101},~062110,~2020), we establish a formal connection between inequality-based witnesses of quantum coherence and noncontextuality inequalities. Our key contributions include: the development of a graph-theoretic framework for generating coherence and contextuality witnesses; a formal mapping between the inequalities that follows from the work by Galvão and Brod to existing noncontextuality inequalities; the conceptualization of a relational form of quantum coherence; a proof of contextual advantage for the task of quantum interrogation; and the discovery of an infinite family of coherence witnesses that also require quantum states in Hilbert spaces of specific dimensions.

\paragraph{Keywords} Quantum Coherence, Contextuality, Quantum Information, Relational Information, Dimension witnessing, Quantum Interrogation

\cleardoublepage

%Acronym entries
\newacronym{locc}{LOCC}{Local Operations and Classical Communication}
\newacronym{losr}{LOSR}{Local Operations assisted by a classical source of Shared Randomness}
\newacronym{gpt}{GPT}{General Probabilistic Theories (nothing to do with the chat)}
\newacronym{opt}{OPT}{Operational-probabilistic theory}
\newacronym{kcbs}{KCBS}{Klyachko, Can, Binicio\u{g}lu, and Shumovsky}
\newacronym{chsh}{CHSH}{Clauser, Horne, Shimony, and Holt}
\newacronym{csw}{CSW}{Cabello, Severini, and Winter}
\newacronym{ks}{KS}{Kochen and Specker}
\newacronym{pvm}{PVM}{Projection-valued measure }
\newacronym{povm}{POVM}{Positive operator-valued measure}
\newacronym{stab}{STAB}{Stable set polytope}
\newacronym{lsss}{LSSS}{Lostaglio, Senno, Schmid, and Spekkens}
\newacronym{pbr}{PBR}{Pusey, Barrett, and Rudolph}
\newacronym{epr}{EPR}{Einstein, Podolsky, and Rosen}
\newacronym{cglmp}{CGLMP}{Collins, Gisin,  Linden, Massar, and Popescu}
\newacronym{sdp}{SDP}{Semidefinite Programming}
\newacronym{npa}{NPA}{Navascués, Pironio, and Acín}
\newacronym{hom}{HOM}{Hong, Ou, and Mandel}
\newacronym{ifm}{IFM}{Interaction-free measurement}
\newacronym{mzi}{MZI}{Mach--Zehnder Interferometer}
\newacronym{bs}{BS}{Beam splitters}
\newacronym{ps}{PS}{Phase shifter}
\newacronym{pr}{PR}{Popescu and Rohrlich}
\newacronym{pu}{PU}{Projective (or projectively) unitary}
\newacronym{kd}{KD}{Kirkwood and Dirac}
\newacronym{rac}{RAC}{Random access codes}
\newacronym{si-c}{SI-C}{State-independent contextuality}
\newacronym{fab}{FAB}{Fine, Abramsky, and Brandenburger}
\newacronym{scs}{SCS}{Splitting Cone Solver}
\newacronym{cvxopt}{CVXOPT}{Python Software for Convex Optimization}
\newacronym{psd}{PSD}{Positive semidefinite}
\newacronym{nv}{NV}{Navascués and Vértesi}
\newacronym{obg}{OBG}{Oszmaniec, Brod, and Galvão}
%Oszmaniec--Brod--Galvão
%Navascués--Vértesi
% %Glossary entries
% \newglossaryentry{maths}
% {
%     name=mathematics,
%     description={Mathematics is what mathematicians do}
% }
% \newglossaryentry{latex}
% {
%     name=latex,
%     description={Is a markup language specially suited for 
% scientific documents}
% }
% \newglossaryentry{formula}
% {
%     name=formula,
%     description={A mathematical expression}
% }

% TiKZ style file generated by TikZiT. You may edit this file manually,
% but some things (e.g. comments) may be overwritten. To be readable in
% TikZiT, the only non-comment lines must be of the form:
% \tikzstyle{NAME}=[PROPERTY LIST]

% Node styles
\tikzstyle{copy}=[fill=white, draw=black, shape=circle]
%Default Styles
\tikzstyle{none}=[inner sep=0pt]
% Edge styles
\tikzstyle{new}=[-, thick, draw={rgb,255: red,81; green,41; blue,241}]
\tikzstyle{new_dashed}=[-, thick, dashed, draw={rgb,255: red,81; green,41; blue,241}]
\tikzstyle{alice}=[-, fill={rgb,255: red,124; green,231; blue,255}, draw={rgb,255: red,12; green,60; blue,216}]
\tikzstyle{bob}=[-, fill={rgb,255: red,255; green,123; blue,125}, draw={rgb,255: red,171; green,18; blue,21}]
\tikzstyle{alice_dashed}=[-, dashed, draw={rgb,255: red,74; green,201; blue,255}]
\tikzstyle{bob_dashed}=[-, dashed, draw={rgb,255: red,255; green,123; blue,125}]
% Edge styles
%Some declaring stuff
\pgfdeclarelayer{edgelayer}
\pgfdeclarelayer{nodelayer}
\pgfsetlayers{edgelayer,nodelayer,main}
\phantomsection
\tableofcontents

\cleardoublepage
\listoffigures

% List of tables
\renewcommand*{\listtablename}{List of Tables}
\listoftables
\clearpage

% Acronyms
\printglossary[type=\acronymtype,nonumberlist, title={Acronyms}]

% Glossary
\printglossary[title={Glossary}, nonumberlist]

\cleardoublepage
\pagenumbering{arabic}

%Introduction
\chapter{Introduction}\label{chapter: introduction}

\begin{quote}
``\textit{Some mathematicians are birds, others are frogs. Birds fly high in the air and survey broad vistas of mathematics out to the far horizon. They delight in concepts that unify our thinking and bring together diverse problems from different parts of the landscape. Frogs live in the mud below and see only the flowers that grow nearby. They delight in the details of particular objects, and they solve problems one at a time. (...) Mathematics needs both birds and frogs. (...)  It is stupid to claim that birds are better than frogs because they see farther, or that frogs are better than birds because they see deeper. The world of mathematics is both broad and deep, and we need birds and frogs working together to explore it.}'' \\
    (Freeman~\cite{dyson2009birds}, Birds and Frogs~\footnote{We would like to thank Professor Antonio Onofre for pointing to this work by Dyson during a MAP-fis annual conference, which we took as an inspiration for this introduction. })
\end{quote}

\section{Motivation}

\myindent We can replace `mathematics' with `quantum information science' in the quote above from Freeman Dyson's `Birds and Frogs'~\citep{dyson2009birds} with one key addition: quantum information science does not require only birds and frogs within a single field, say, physics or mathematics. Its significantly interdisciplinary nature demands the collaboration of \emph{several} distinct `species' of Birds and Frogs---mathematicians, computer scientists, physicists, both theoretical and experimental, philosophers, engineers, and chemists. This diversity is perhaps one of the field's greatest challenges, but also one of its greatest assets. Quantum information science is a relatively young and rapidly expanding field of research, growing exponentially in funding and personnel. Yet, it faces immense pressure to establish itself as a respected scientific discipline or risk being dismissed as another `hyped' idea in modern science. 

\myindent Such a dynamic and evolving environment faces a series of challenges. Firstly, as with any interdisciplinary field, there must be a concerted effort to bridge the gaps between subfields and disciplines. This includes overcoming superficial differences, such as variations in jargon and terminology, and breaking down silos that prevent meaningful interactions between subfields. Furthermore, it is crucial---and often overlooked---to carefully examine and reinforce the foundational pillars of the field. For quantum information science, its pillars are grounded on the field of \emph{quantum foundations}~\citep{hardy2010why,landsman2017foundations}. The `birds and frogs of quantum foundations' are constantly critically assessing which paths are worth pursuing, and which are not, serving as an---often ignored---foundational task force capable of recognizing challenges that transcend individual subfields, stemming instead from deeper, fundamental issues inherent to the discipline as a whole. Finally, we must develop tools that address the interdisciplinary nature of the field, resolving tensions between subfields in a meaningful and insightful way. Such tools should not only lead to experimental and technological advancements but also provide novel insights into the fundamental physics of the key underlying processes. 

\myindent These considerations form the primary motivation for the work presented in this thesis. We plan to keep fluctuating between a `bird' and a `frog' approach to research. With this broad perspective in mind, our research focuses on (i) bridging two distinct fields of study, (ii) motivating and developing overarching tools applicable to both fields (and potentially beyond), and (iii) investigating the novel perspectives these tools provide in terms of applications to existing problems across multiple subfields. This `big picture' goal serves as the guiding principle of our work and informs the structure and direction of this thesis. Having clarified our main motivation, we now proceed to specify the problems we address and the main questions that this thesis attempts to resolve, also motivating the relevance of the questions we pose. In summary, our main questions and findings fit into each of the three broad points (i)-(iii) made above.

\section{About this thesis}

\myindent Unsurprisingly, this thesis title gives away which fields of study we attempt to bring together: the subfield of quantum foundations interested in investigating the failure of \emph{noncontextuality}~\citep{budroni2021kochenspeckerreview}, a property held by ontological models~\citep{harrigan2010einstein} that attempt to explain operational-probabilistic descriptions of experimentally accessible data, and the subfield of quantum information science dedicated to investigating \emph{quantum coherence}~\citep{streltsov2017colloquium}, a property traditionally considered to be held by certain states of a quantum system. 

\myindent Research in noncontextuality theory has its roots in quantum foundations. It is linked to the idea that one could prove that quantum theory fails to provide a \emph{realistic} account of its predictions while, additionally, allowing for a form of \emph{classical explainability}. It is a field traditionally considered as `bird work', which is abstract, with little to no applicability. Contrastingly, coherence theory is a field that is traditionally considered `frog work', which is objective, grounded in the useful notion of a resource theoretic framework, which one can use to quantify a certain property of states of physical systems, and put such a resource to \emph{use} for solving information and computational tasks. At first, the two fields may seem to have entirely different concerns, and beyond that, even emphasize completely different operational aspects of an experiment: while noncontextuality has historically emphasized the relevance of \emph{measurements} and \emph{empirical statistics} resulting from laboratory operations, quantum coherence theory has traditionally emphasized the role that \emph{preparation procedures} selecting certain \emph{quantum states} play. To make matters worse, there are proofs of contextuality that work by performing measurements onto \emph{any state}, including maximally mixed states~\citep{kochen1975problem,mermin1993hidden,cabello2008experimentally,kirchmair2009state}, i.e., states that lack quantum coherence in a `maximal' manner, while known subtheories of quantum theory (e.g. the qudit stabilizer subtheory) allow for an explanation in terms of noncontextual models even though coherent states are present~\citep{spekkens2007evidence,catani2023whyinterference}. Beyond that, while quantum coherence is a notion that seems to require a quantum mechanical treatment, noncontextuality is a \emph{theory-independent} notion, which can be probed independent of the validity of quantum theory for describing processes in Nature.  

\myindent From this perspective, it seems hopeless to even attempt to compare the two notions. Moreover, even if one could find a meaningful way of comparing them, one would run the risk of simply `confusingly mixing' two notions that are better thought of as distinct. We need to be cautious when constructing bridges between different subfields, as the intention is to bring \emph{clarity} in our understanding of how two notions (or two subfields) relate, even (and especially) if this means formally drawing additional boundaries.  

\myindent With that in mind, we make a more concrete and focused description of why one might believe that it is possible (and well motivated) to propose broadly applicable scenarios where there is a clear-cut connection between coherence and contextuality. Our starting point is the work by Ernesto Galvão and Daniel Brod~\citep{galvao2020quantum} that proposed a family of inequalities on empirical probabilities (and its related quantum sets of correlations) that could witness a form of basis-independent quantum coherence. Albeit targeted to quantum mechanical properties of quantum states, their approach resonates with existing research on noncontextuality due to the methodology used  (which considered Boole's~\citep{boole1854investigation} and Pitowsky's~\citep{pitowsky2008newbell} rules for bounding joint probabilities of events). So, the first research question we pose is the following one:

\begin{question}\label{question: coherence witnesses and contextuality}
   Is there a formal mathematical mapping between the coherence witnesses considered by~\cite{galvao2020quantum} and witnesses of quantum contextuality?
\end{question}

\myindent This is not an entirely trivial question. While certain elements of their work bear a resemblance to treatments of contextuality and noncontextuality inequalities, several aspects of the proposal in~\cite{galvao2020quantum} appear to be largely unrelated to established research on noncontextuality theory. For example, their treatment is motivated by properties of sets of \emph{quantum states}, and not measurements, as in contextuality theory. Moreover, their treatment does not consider generic probability distributions but only that of two-state overlaps. On the other hand, some of the inequalities they have found can also be found in other works related to these notions, and as mentioned previously, the methodology considered suggests that \emph{one} possible answer to the question above is the simple one of `the inequality-based witnesses (re-)discovered by~\cite{galvao2020quantum} are noncontextuality inequalities'. This would be therefore one of the many examples of two communities, using different terminology, motivated by different applications, which concluded that the same tools are relevant to both, without noticing that the tools and concepts were the same. 

\myindent \cref{question: coherence witnesses and contextuality}~is a concrete and focused question, and can be interpreted as a step towards the more general goal of understanding the relationship between coherence and contextuality. Nevertheless, there is another interesting question that follows from the consideration by~\cite{galvao2020quantum} of basis-independent coherence witnesses based on bounds on two-state overlaps of sets of quantum states. Their methodology has graph-theoretic insights that naturally guide us to imagine that their witnesses can be viewed as special cases within a broader graph-theoretic framework. Graph theory is a standard tool for quantum information science, and has been applied to investigate and characterize Bell nonlocality, prepare-and-measure scenarios, noncontextuality, and entanglement theory. With this insight, we pose the following question:

\begin{question}\label{question: generalize galvao and broads results}
    Is it possible to generalize the framework considered by~\cite{galvao2020quantum}, providing a graph-theoretic framework capable of completely characterizing the necessary and sufficient conditions for witnessing coherence provided solely with information of tuples of two-state overlaps?
\end{question}

\myindent If we find such a framework, capable of completely characterizing the ability of two-state overlaps to `signal' the presence of basis-independent coherence of a set of states, we can then ask the \emph{broader} question,  generalizing~\cref{question: coherence witnesses and contextuality}, of whether the coherence witnesses in this more general framework can also be linked to quantum contextuality. This question is of independent interest, well motivated both foundationally, and also by the many applications pointed out by~\cite{galvao2020quantum}. However, it would be perhaps even more well-motivated if, given that we have found such a framework, we may find \emph{novel} coherence witnesses having \emph{novel} applications. 

\begin{question}\label{question: applications}
    If one could find a framework resolving~\cref{question: generalize galvao and broads results}, generalizing the framework initially proposed by~\cite{galvao2020quantum}, what novel applications can be found due to this extension?
\end{question}

\myindent We propose solutions to all the questions outlined above. In Chapter~\ref{chapter: event_graph_approach}, we develop a graph-theoretic framework to address~\cref{question: generalize galvao and broads results}. Chapter~\ref{chapter: relational coherence} introduces a novel perspective on coherence theory, which we term \emph{relational quantum coherence}—the type of coherence encapsulated by two-state overlaps and, more generally, by unitary invariants. While we highlight this idea as a potentially significant legacy of our findings, capable of inspiring future impactful research, it remains an ongoing topic with many open questions yet to be explored. Using this tool, we analyze connections between coherence and contextuality in Chapter~\ref{chapter: from overlaps to noncontextuality}, resolving not only~\cref{question: coherence witnesses and contextuality} but also its broader implications within our new framework. Finally, in Chapter~\ref{chapter: applications}, we propose novel applications leveraging the introduced framework, providing two case studies viewed as solutions to~\cref{question: applications}.

\myindent We proceed next to discuss the list of contributions that are present in this thesis, which attempts to resolve the main questions proposed earlier. We then conclude this introductory Chapter with a guide on how to read this thesis.

\section{List of publications}

\myindent The backbone of this thesis can be found in several articles, published or made available in open-source online repositories, which were written during its conception. They resolved the main questions posed in an early stage of our research project, and have also helped mature some of the original ideas and perspectives we endorse. The following papers are discussed in this thesis:

\begin{enumerate}    
    \item Rafael Wagner, Rui Soares Barbosa, and Ernesto F. Galvão, ``\emph{Inequalities witnessing coherence, nonlocality, and contextuality}'', Physical Review A 109 (3), 032220 (2024), \citep{wagner2024inequalities}.
    \item Rafael Wagner, Anita Camillini, and Ernesto F. Galvão, ``\emph{Coherence and contextuality in a Mach-Zehnder interferometer}'', Quantum 8, 1240 (2024), \citep{wagner2024coherence}.
    \item Taira Giordani, Rafael Wagner, Chiara Esposito, Anita Camillini, Francesco Hoch, Gonzalo Carvacho, Ciro Pentangelo, Francesco Ceccarelli, Simone Piacentini, Andrea Crespi, Nicolò Spagnolo, Roberto Osellame, Ernesto F. Galvão, and Fabio Sciarrino, ``\emph{Experimental certification of contextuality, coherence, and dimension in a programmable universal photonic processor}'', Science Advances 9 (44), eadj4249 (2023), \citep{giordani2023experimental}.
    \item Carlos Fernandes, Rafael Wagner, Leonardo Novo, and Ernesto F. Galvão, ``\emph{Unitary-invariant witnesses of quantum imaginarity}'', Physical Review Letters 133, 190201 (2024), \citep{fernandes2024unitary}.
    \item Rafael Wagner, Filipa C. R. Peres, Emmanuel Zambrini Cruzeiro, and Ernesto F. Galvão, ``\emph{Unitary-invariant method for witnessing nonstabilizerness in quantum processors}'', Journal of Physics A: Mathematical and Theoretical 58 285302 (2025), \citep{wagner2024certifying}.
\end{enumerate}

\myindent Some other works that are \emph{not} included in this thesis but that were completed during the time of this thesis project are the following ones:

\begin{enumerate}
    \item Rafael Wagner and Ernesto F. Galvão, ``\emph{Simple proof that anomalous weak values require coherence}'', Physical Review A 108 (4), L040202 (2023), \citep{wagner2023anomalous}.
    \item Rafael Wagner, Zohar Schwartzman-Nowik, Ismael L. Paiva, Amit Te’eni, Antonio Ruiz-Molero, Rui Soares Barbosa, Eliahu Cohen, and Ernesto F. Galvão, ``\emph{Quantum circuits for measuring weak values, Kirkwood–Dirac quasiprobability distributions, and state spectra}'', Quantum Science and Technology 9 (1), 015030 (2024), \citep{wagner2024quantumcircuits}.    
    \item Filipa C. R. Peres, Rafael Wagner, and Ernesto F. Galvão, ``\emph{Non-stabilizerness and entanglement from cat-state injection}'', New Journal of Physics 26 (1), 013051 (2024), \citep{peres2024nonstabilizerness}.
    \item David Schmid, Roberto D Baldijão, Yìlè Yīng, Rafael Wagner, and John H Selby, ``\textit{Kirkwood-Dirac representations beyond quantum states and their relation to noncontextuality}'', Physical Review A 110, 052206 (2024), \citep{schmid2024kirkwood}.
    \item Laurens Walleghem, Rafael Wagner, Yìlè Yīng, and David Schmid, ``\emph{Extended Wigner's friend paradoxes do not require nonlocal correlations}'', Physical Review A 112 (2), 022212 (2025), \citep{walleghem2024extended}.
    \item Naim E. Comar, Danilo Cius, Luis F. Santos, Rafael Wagner, and Bárbara Amaral, ``\emph{Contextuality in anomalous heat flow}'', PRX Quantum 6 (3), 030359 (2025), \citep{comar2024contextualityanomalousheatflow}.
    \item Roberto D. Baldijão, Rafael Wagner, Cristhiano Duarte, Bárbara Amaral, and Marcelo Terra Cunha, ``\emph{Emergence of noncontextuality under quantum Darwinism}'', PRX Quantum 2 (3), 030351 (2021), \citep{baldijao2021emergence}.
    \item Rafael Wagner, Roberto D. Baldijão, Alisson Tezzin, and Bárbara Amaral, ``\emph{Using a resource theoretic perspective to witness and engineer quantum generalized contextuality for prepare-and-measure scenarios}'', Journal of Physics A: Mathematical and Theoretical 56 (50), 505303 (2023), \citep{wagner2023using}.
    \item Tiago Santos, Rafael Wagner, and Bárbara Amaral, ``\emph{Convexity of noncontextual wirings and how they order the set of correlations}'', Physical Review A 110, 032217 (2024), \citep{tiago2024convexity}.
    \item Laurens Walleghem,  Yìlè Yīng, Rafael Wagner, and David Schmid, ``\emph{Connecting extended Wigner’s friend arguments and noncontextuality}'', Quantum 9, 1819 (2025), \citep{walleghem2024connectingextendedwignersfriend}.
    \item Kyrylo Simonov, Rafael Wagner, and Ernesto F. Galvão, ``\textit{Estimation of multivariate traces of states given partial classical information}'', arXiv preprint: arXiv 2505.20208,~\citep{simonov2025estimation}.
\end{enumerate}

\subsubsection{Author contributions}

\myindent Scientific research is a collaborative effort. This is the reason why we chose to write this thesis, and any other scientific article, in the first person of the plural: the role each `I' plays in science is insignificant compared to the role `We' play. However, as a PhD thesis also constitutes an individual evaluation, I momentarily switch to first-person writing to state in more detail my individual contribution to the research articles that form the basis of this dissertation. I am the first author or co-first author of all the research papers that constitute the backbone of this dissertation. In our research field, being (co)-first author is considered to signal some (subjective and imprecise, as always) `ordering' of author contribution. In effect, all authors mentioned in all works that appear in this thesis contributed crucially to the development of the research presented in the papers. For each result, I outline the specifics of each contribution, emphasizing my contributions to these mentioned works: 

\begin{itemize}
    \item The foundational work that is certainly the main research output of this thesis, and the research project that idealized it, is Ref.~\citep{wagner2024inequalities}, which was idealized by Ernesto Galvão and Rui Soares Barbosa as a follow up from Ref.~\citep{galvao2020quantum}. The main goal there was to formalize a broadly applicable framework and clarify its connections with existing Bell and Kochen--Specker noncontextuality inequalities in the literature. All other references we discuss~\citep{wagner2024coherence,giordani2023experimental,wagner2024certifying,fernandes2024unitary} showcase the intricate, and in our view interesting, avenues that were open by Ref.~\citep{wagner2024inequalities}. I have proved some of the main results in that paper, which were later better formalized by my supervisors Rui Soares Barbosa and Ernesto Galvão. 
    \item I have conceived the project of Ref.~\citep{wagner2024coherence}, proved some of the main theoretical results, and wrote the first draft, all with close (and equal) contributions from Anita Camillini. The manuscript's idea, conceptualization, and presentation were then significantly improved and refined by one of my supervisors, Ernesto Galvão. All authors contributed equally to the analysis of the numerical simulations performed, as well as the theoretical results. 
    \item Reference~\citep{giordani2023experimental} was made in collaboration with the experimental group led by Fabio Sciarrino to test the predictions from~\cite{wagner2024coherence}. I hereby comment only on the theoretical contributions of that work, leaving aside the experimental contributions. I have helped with the writing of the first draft of the theory background, theoretical motivation, conclusion, and theoretical supplemental information together with Ernesto Galvão and Anita Camillini, who contributed equally. Anita Camillini did some of the numerical investigations that we report here for completeness of the presentation (in Chapter~\ref{chapter: applications}, Table~\ref{tab:kn}). I have mathematical results that we report in this thesis that were present in the supplemental material of that work. 
    \item I have conceived a project to investigate properties of imaginary values of Bargmann invariants together with Ernesto Galvão. I made some preliminary plots of sets of third-order Bargmann invariants that inspired Ernesto Galvão, who has had the great insight to propose a project to investigate the geometrical properties of these sets, and their relevance to various physical phenomena, together with Carlos Fernandes (with whom I share a first co-author contribution) and Leonardo Novo, which eventually converged into Ref.~\citep{fernandes2024unitary}. I have proved one of the four main results presented therein, written the first draft of the manuscript, and showed some theoretical results that are presented in Chapter~\ref{chapter: relational coherence}. 
    \item To conclude, I have conceived a project and proved some preliminary theoretical results that led to Reference~\citep{wagner2024certifying}. The most important theoretical contributions of that work, together with some novel insights (that are not presented in detail in this thesis) were done in collaboration with Filipa C. R. Peres (with whom I share a first co-author contribution) with the supervision of Emmanuel Zambrini Cruzeiro and Ernesto Galvão. All authors contributed equally to the presentation, relevance, correctness, and motivation of the results. 
\end{itemize}

\section{How to read this thesis}

\myindent To conclude our introduction, we provide a guide on how to read this thesis and an overview of its structure. The thesis is divided into two main parts, excluding the Appendix: Part I introduces the foundational topics relevant to the thesis,  reviews the state-of-the-art, and systematically presents the supporting results in the existing literature that are used in Part II, where we discuss our main contributions. This division reflects a distinction in the nature of the contributions: Part I provides a review, while Part II introduces original results. We assume the reader is already familiar with the basics of quantum mechanics, linear algebra, and quantum information theory. Some familiarity with quantum foundations---especially in the context of investigating quantum correlations---is  beneficial for following Chapters~\ref{chapter: contextuality},~\ref{chapter: relational coherence} and~\ref{chapter: from overlaps to noncontextuality}.

\myindent Each chapter of Part I can be read independently, and the presentation is designed so that each chapter is self-contained. Chapter~\ref{chapter: quantum coherence} introduces the notion of quantum coherence, focusing on static resources and discussing some lesser known topics within this broad field, such as `quantum imaginarity' and basis-independent coherence. Chapter~\ref{chapter: contextuality} reviews the two leading notions of contextuality: Kochen--Specker contextuality and generalized contextuality. Chapter~\ref{chapter: information tasks} presents two information-theoretic tasks: dimension witnessing and quantum interrogation. Finally, in Chapter~\ref{chapter: Bargmann invariants}, we review the theory of unitary invariants, with a focus on Bargmann invariants: how they have been studied, the developments in the theory of measuring them, and the main results from~\cite{galvao2020quantum} that we aim to generalize and that are essential for addressing the central questions posed in this introduction, to be answered in Part~II.

\myindent Part II, in contrast, presents our original contributions, advancing the topics discussed in Part I with new insights and results. These contributions are rooted in the theoretical background provided earlier and aim to push the boundaries of the existing literature in quantum information science and foundations. We have already discussed our contributions above, where each Chapter is written with a focus on resolving the Questions~\ref{question: coherence witnesses and contextuality}--~\ref{question: applications}. 

\myindent For completeness, we include an appendix to introduce purely mathematical definitions and to review technical constructions used throughout the text. Appendix~\ref{app: basic algebra} covers basic results in operator and linear algebra, along with elementary notions such as relations, functions, equivalence relations, convexity, and related concepts. Appendix~\ref{sec: convex polytopes} reviews the mathematical theory of convex polytopes, and Appendix~\ref{sec: graph theory} presents the elements of graph theory used in Part~II.

%-- Dissertation 
\part{Preliminaries}
%\sloppy
\tolerance=2000          % Increase tolerance for uneven spacing (default 200)
\hbadness=10000          % Suppress underfull warnings for low-badness situations

\chapter{Quantum coherence}\label{chapter: quantum coherence}

\begin{quote}
    ``\textit{(...) in the Schrödinger’s cat experiment, having the technological ability to \emph{detect} that the cat was in a superposition state at all, implies having the ability to perform a unitary that revives a dead cat, an ability that one could call `quantum necromancy.' (...)  if reviving a dead cat is considered `hard'---for essentially any reasonable definition of `hard'---then distinguishing Schrödinger’s cat from a classical mixture is `hard' in that same sense, and the cat can be treated as effectively decohered.}''~\citep{aaronson2020hardnessdetectingmacroscopicsuperpositions}
\end{quote}

\myindent Though shocking, this quote captures fundamental reasons why investigating quantum coherence is crucial. To start, coherence lies at the core of foundational discussions in quantum theory, dating back to its inception---famously illustrated by~\cite{schrodinger1935gegenwartige} in his thought-provoking \emph{ Gedankenexperiment}. It also underpins recent technological advancements, including quantum computers, quantum sensors, and quantum communication devices. The quote above highlights one of the many reasons why coherence is relevant to computer science, emphasizing its implications for the analysis of computational complexity~\citep{arora2009computational}, which helped deepen our understanding of how quantum information processing~\citep{nielsen2002quantum,wilde2013quantum} fundamentally differs from its classical counterpart. 

\myindent This chapter provides an overview of the branch of quantum theory dedicated to quantifying, witnessing, classifying, characterizing, and applying \emph{quantum coherence}. Our overview is not comprehensive, as it focuses on key aspects relevant to our contribution to this body of literature, presented in Chapter~\ref{chapter: relational coherence}. We refer to the reviews by~\cite{streltsov2017colloquium},~\cite{chitambar2019quantum}, and the recent book by~\cite{gour2024resources} for a comprehensive presentation. These have considered the resource-theoretic perspective, and because of that, we refer to several of the constructions we mention as `resources' without going into great detail on the specifics of the resource-theoretic formalism. For basic introductions to the field of \emph{quantum} resource theories we refer to the references just mentioned. For a broader perspective on resource theories that are not necessarily based on quantum primitives (quantum states, measurements, channels, etc.) we refer to~\cite{coecke2016mathematical}, and~\cite{gonda2023monotones}. Finally, we refer to~\cite{abramsky2019comonadic},~\cite{wolfe2020quantifyingbell},~\cite{barbosa2023closingBell}, and~\cite{tiago2024convexity} to specific resource theories of \emph{behaviors}, a notion we  introduce in Chapter~\ref{chapter: contextuality}.

\myindent The most common introduction to coherence theory is via the phenomenon of \emph{quantum interference}, which is presented using the wave function formalism of quantum theory. In such initial encounters, it is often shown that the dynamical evolution of a wave function $\psi(x)$, which describes the state of a single photon or electron that has passed through two slits and is detected on a screen, must allow for superpositions to accurately reproduce the interference patterns observed experimentally. This serves as an example of the broader concept of \emph{quantum coherence}, specifically referring here to pure states $\vert \psi \rangle$ that are in superpositions with respect to the position basis $\vert x \rangle$. More generally, the concept of quantum coherence is encapsulated within the density matrix formalism of quantum theory~\citep{nielsen2002quantum,wilde2013quantum}, where quantum states of a physical system described by a Hilbert space $\mathcal{H}$ are associated with density matrices $\rho$. These may exhibit coherence---in the form of non-null off-diagonal elements---depending on the chosen basis for $\mathcal{H}$. In our presentation, we focus on \emph{finite-dimensional} Hilbert spaces $\mathcal{H} \simeq \mathbb{C}^d$, as this suffices for the purpose of this thesis.

\myindent As mentioned above, it is well established that quantum coherence is a necessary resource for quantum computation and quantum information processing in general. See a nontechnical early discussion of this fact by~\cite{deutsch1992quantumcomputation}, who himself had developed the first universal model of quantum computation~\citep{deutsch1985universal}. For more recent evidence for the relevance of coherence in quantum computing and information processing, we refer to ~\citep{ahnefeld2022coherence,naseri2022entanglement,wagner2024coherence,ahnefeld2025coherenceresourcephaseestimation}, and references therein. 

\myindent Having briefly highlighted the relevance of quantum coherence, it is worth mentioning that there are several distinct paths for its investigation, all contributing to a foundational and practical understanding of this phenomenon. One approach is to explore quantum coherence from a resource-theoretic standpoint~\citep{chitambar2019quantum,gour2024resources}, addressing its quantification and operational relevance. One may also search for formal guarantees that indeed quantum superposition enables higher computational (or information-processing) power---typically by investigating complexity-theoretic separations, such as those between \texttt{BQP}~\footnote{For a rigorous introduction to complexity-theoretic notions we refer to the book by~\cite{arora2009computational}. In simple terms, \texttt{BQP} is a \emph{set}, containing abstract elements representing \emph{problems}: all those that can be efficiently solved by a quantum computer. Informally, efficiency is a notion related to how much time it takes to solve the problem, as the problem instances `grow'.} and other complexity classes~\citep{aaronson2011bosonsampling,chen2022exponential,chen2023complexityofNISQ}. Building on this, a more pragmatic avenue focuses on proposing better algorithms on quantum hardware for solving existing problems~\citep{dalzell2023quantumalgorithms}. Alternatively, one may investigate \emph{why} quantum coherence is not commonly observed in Nature,  describing as accurately as possible which mechanisms are responsible for making quantum coherence such a fragile (i.e., sensitive to noise) phenomenon~\citep{zurek2003decoherence,schlosshauer2019quantumdecoherence}. Complementarily, another line of inquiry attempts at \emph{preventing} these mechanisms responsible for the loss of coherence taking place, so that we can process and control coherent states fault-tolerantly~\citep{lidar2013quantumerrorcorrection,terhal2015quantumerrorcorrection}. 

\myindent In this Chapter we focus on a different research direction: that of~\emph{certification of quantum coherence}. In the context of nonclassical resources such as quantum coherence, certification tasks are also known as \emph{detecting} or \emph{witnessing}~\hspace{0.03cm}tasks. The goal of these tasks is to provide a certificate of the presence of quantum coherence within a device, guaranteeing its presence based on the experimental statistics observed from it. We can illustrate this using our earlier example of quantum interference. By specifying all interference patterns that can be observed for incoherent states $\psi(x)$, statistical deviations thereof guarantee a `type' of pattern that requires quantum coherence. The nontriviality of such tests depends on prior information available about the device, or specific experimentally available tools. In practice, witnessing generic nonclassical features of quantum devices is challenging and aids in the characterization, benchmarking, and calibration of devices for their application in other tasks. Properly witnessing and characterizing nonclassical features of a device, such as coherence, is a relevant \emph{intermediate step} between device fabrication and its use in quantum technology. 
 
\myindent The structure of this chapter is as follows. We start by formally defining the standard notion of quantum coherence in Section~\ref{sec: basis-dependent coherence}, and motivate the problem of witnessing quantum coherence that is later discussed in~\cref{sec: coherence witnesses}. We then describe a particular instance of quantum coherence that has recently gained attention, known as \emph{quantum imaginarity}, in Section~\ref{sec: quantum imaginarity}. As the standard notion of coherence considers a fixed reference basis, we also review attempts at relaxing this constraint. In Section~\ref{sec: basis-independent coherence of a single state} we describe approaches to \emph{basis-independent coherence} focusing on a single state and optimizations over choices of basis and unitary-invariant functions of a state with itself, such as the purity. In Section~\ref{set: basis-independent coherence of a set of states} we extend the discussion to approaches that have instead considered coherence as a property of a \emph{set of states}. 

\section{Basis-dependent coherence}\label{sec: basis-dependent coherence}

\myindent The scenario in which one can properly investigate basis-dependent quantum coherence is the following: 

\begin{definition}[Basis-dependent coherence]\label{def: basis dependent coherence}
Let $\mathcal{H}$ be a fixed finite-dimensional~\footnote{For a discussion of quantum coherence in continuous-variable systems, hence in cases where one relaxes the assumption of finite-dimensionality of the Hilbert spaces investigated, we refer the reader to Section III-J of ~\cite{streltsov2017colloquium}.} Hilbert space, $d = \dim(\mathcal{H})$, and let $\mathbb{A} = \{\vert a_i\rangle \}_{i=1}^d$ be any orthonormal basis of $\mathcal{H}$. A state $\rho \in \mathcal{D}(\mathcal{H})$ is said to be \emph{incoherent} (also known as coherence-free) with respect to the basis $\mathbb{A}$ if 
\begin{equation}
    \forall i,j. \,\, i\neq j \implies  \langle a_i|\rho |a_j \rangle = 0.
\end{equation}
We denote the set of all such states as $\mathcal{I}(\mathcal{H},\mathbb{A})$. A state is \emph{coherent} if it is not incoherent, i.e., we say that a state $\rho$ is \emph{coherent with respect to $\mathbb{A}$} when $\langle a_i|\rho|a_j\rangle \neq 0$, for some $i\neq j$.
\end{definition}

\myindent Sometimes this definition of coherence is referred to as an example of a \emph{static quantum resource}~\citep{gour2024resources} as opposed to the possibility of investigating quantum coherence \emph{dynamically} as was done by~\cite{saxena2020dynamical}.~\footnote{Provided with a set $\mathcal{I}(\mathcal{H},\mathbb{A})$ one can define the completely dephasing channel $\mathcal{D}_{\mathbb{A}}$ acting as $\rho \mapsto \sum_{\ket a \in \mathbb{A}} \langle a|\rho|a\rangle \vert a \rangle \langle a \vert$. Investigating coherence \emph{dynamically} amounts to focusing on quantum channels $\mathcal{E}$, instead of quantum states $\rho$, and how such channels are affected when composed with the map $\mathcal{D}_{\mathbb{A}}$. The analogue in this case of an incoherent quantum state is not unique. One can consider channels $\mathcal{E}$ that do not create coherence, channels that do not detect coherence, or channels satisfying both properties, among other possibilities~\citep{saxena2020dynamical}. } This notion of coherence, introduced in Def.~\ref{def: basis dependent coherence}, is also sometimes referred to as \emph{speakable coherence}~\citep{marvian2016unspeakable}. A similar definition can be proposed that defines a broader class of situations where instead of considering only orthonormal bases $\mathbb{A}$ one can simply require \emph{linear independence}. This choice then translates into a different resource that has been considered by~\cite{theurer2017resource}.

\myindent Importantly, the above notion is then only well defined when one considers a specific choice of reference basis $\mathbb{A}$ for the Hilbert space of interest $\mathcal{H}$. The state $\vert +\rangle = \sfrac{1}{\sqrt{2}}(\vert 0\rangle + \vert 1 \rangle)$ is coherent with respect to $\{\vert 0\rangle, \vert 1 \rangle \}$ but incoherent with respect to $\{\vert +\rangle, \vert -\rangle\}$. As another example, useful to us for setting a basic choice of notation, the states $\sfrac{1}{\sqrt{2}}(\vert 0\rangle + \vert 1 \rangle)$ and $\sfrac{1}{\sqrt{2}}(\vert 0\rangle + \vert 2 \rangle)$ are two different states coherent with respect to the canonical (also known as standard) basis $\{\vert 0\rangle, \vert 1 \rangle, \dots, \vert d-1 \rangle\}$. We denote this basis as $\mathbb{D}_d$ or $\{\vert i \rangle\}_{i=0}^{d-1}$. Note therefore that $\mathbb{D}_2 = \{\vert 0\rangle, \vert 1 \rangle\}$.

\myindent A pure state $\rho = \vert \psi \rangle \langle \psi \vert$ is incoherent with respect to $\mathbb{A}$ if and only if $\vert \psi \rangle \in \mathbb{A}$. For any finite-dimensional Hilbert space there exists only one state that is incoherent with respect to \emph{every} choice of basis, and it is the maximally mixed state $ \mathbb{1}/d$. 

\myindent It may not be evident from the definition we have given for coherence, but there are different `types' or `classes' of quantum coherence. For a trivial example, take the two states
\begin{equation*}
    \vert +\rangle = \frac{1}{\sqrt{2}}(\vert 0\rangle + \vert 1 \rangle )
\end{equation*}
and 
\begin{equation*}
    \vert \Psi \rangle = \frac{1}{\sqrt{2}}(\vert 00\rangle + \vert 11 \rangle).
\end{equation*}
There are aspects of coherence that are present in one that are not present in the other. Ideally, we would want a framework that signals the differences between the two, in the sense that there are ways of guaranteeing that we have $\vert \Psi \rangle$ and not $\vert +\rangle$, which helps highlighting how they differ physically and mathematically. When distinguishing these two forms of coherence operationally, we might also be interested in situations that do not require the complete characterization of tomographic information of these states to observe their physical differences.

\myindent For this particular case, we consider the difference between the coherence of single systems and the coherence of multiple systems that is described by \emph{quantum entanglement} of pure states, and there \emph{is} a way of distinguishing this specific realization of coherence for multipartite states as opposed to any other kind of multipartite coherence that has no entanglement, and it is via the construction of Bell inequalities or entanglement witnesses. To the best of our knowledge, the corresponding program—that is, defining a comprehensive and meaningful set of diverse families of coherent states, along with proposing a methodology for characterizing coherence without requiring full state tomography—has received, arguably, relatively less attention in coherence theory than in entanglement theory.

\myindent Nevertheless, some `types' of basis-dependent coherence are well defined and can be systematically characterized. For example, it is intuitively clear that as we go to higher-dimensional systems we can have more `ways' in which a state $\rho$ can be coherent. Quantitatively, it could be the case that we also reveal more nuanced, nontrivial, and useful aspects of quantum coherence in higher-dimensional systems. For example, we can pose the question: 

\begin{tcolorbox}[
    colback=lightblue!10, % light blue background color with slight transparency
    colframe=lightblue,   % matching border-color
    width=\textwidth,     % adjust width to fit text width
    boxrule=0.5mm,        % border thickness
    sharp corners,
    title=Box 1: Statistics requiring both Hilbert space dimension and coherence,
    fonttitle=\bfseries,  % bold font for title
    title filled=true,    % filled title background
    coltitle=white        % white text color for title
]
\begin{question}\label{question: dimension and coherence?}Are there experimental statistics that signal the presence of quantum coherence for states $\rho \in \mathcal{D}(\mathcal{H})$ that \emph{could not} be observed when using coherent states of smaller dimension (i.e., $\sigma \in \mathcal{D}(\mathcal{K})$ with $\dim(\mathcal{K}) < \dim(\mathcal{H}))$?
\end{question}
\end{tcolorbox}

\myindent One simple approach to answer~\cref{question: dimension and coherence?} is via the notion of \emph{coherence rank}~\citep{killoran2016converting,chin2017coherencerankmixedstates}. Let $(\mathcal{H},\mathbb{A})$ be the scenario in which it is well defined to characterize the coherence of a certain state $\rho$. In such a case, the coherence rank $k$ of a given pure state $\rho = \vert \psi \rangle \langle \psi \vert $ is described by the number $k$ of non-zero coefficients when expanding $\vert \psi \rangle = \sum_{k=1}^d \alpha_k \vert a_{k} \rangle $ in the basis $\mathbb{A}$. This was first proposed by~\cite{levi2014quantitative}, where states with coherence rank $k$ were termed \emph{k-coherent delocalized states}. More recently the term \emph{multilevel coherent states} has been used by~\cite{ringbauer2018certification}. 

\begin{definition}[Coherence rank]
    Let $(\mathcal{H},\mathbb{A})$ be a pair of Hilbert space $\mathcal{H}$ and reference basis $\mathbb{A} = \{\vert a_i \rangle \}_{i=1}^d$. The (basis-dependent) coherence rank of a pure state $\vert \psi \rangle \in \mathcal{H}$ is given by 
    \begin{equation}
        R_C(\vert \psi \rangle ) =  \left |\{k \in [d]  \mid c_{k} \neq 0 \}  \right|
    \end{equation}
    where $\vert \psi \rangle = \sum_{k=1}^d c_{k} \vert a_{k} \rangle$ is the expansion of $\vert \psi \rangle$ in the basis $\mathbb{A}$ and $[d] \equiv \{1,2,\dots,d\}$.
\end{definition}

\myindent For example, if we consider $(\mathbb{C}^2, \mathbb{D}_2)$ the state $\vert +\rangle$ has coherence rank 2. If we consider $(\mathbb{C}^5,\mathbb{D}_5)$, the state 
\begin{equation}
    \frac{1}{\sqrt{3}}(\vert 0\rangle + \vert 2\rangle + \vert 4 \rangle )
\end{equation}
has coherence rank 3.~\footnote{One should not confuse the coherence rank with the matrix \emph{rank} of the states $\rho$, which in this case of pure states is always equal to $1$. } The coherence rank is 3 for the state just described, but the maximal coherence rank that a state can have is given by the dimension $d$ of the space $\mathbb{C}^d$, which in the case above would be $5$. 

\myindent Another simple example of how to approach~\cref{question: dimension and coherence?} is via the notion of \emph{coherence quantifiers} that are not normalized.  Coherence quantifiers are defined as follows~\citep{baumgratz2014quantifying}.

\begin{definition}[Coherence quantifiers]
    Let $\mathcal{H},\mathcal{H}'$ be quantum systems and $\mathbb{A},\mathbb{A}'$ be reference bases for each system, respectively. We say that a  function $C(\cdot | \mathbb{A}) : \mathcal{D}(\mathcal{H}) \to \mathbb{R}$ is a coherence quantifier if it satisfies the following:
    \begin{enumerate}
        \item For every element $\sigma \in \mathcal{I}(\mathcal{H},\mathbb{A})$ we have that  $C(\sigma | \mathbb{A}) = 0$.
        \item $C(\cdot|\mathbb{A})$ is a convex function.
        \item For every quantum channel $\mathcal{E}:\mathcal{D}(\mathcal{H}) \to \mathcal{D}(\mathcal{H}')$ such that $\mathcal{E}(\mathcal{I}(\mathcal{H},\mathbb{A})) \subseteq \mathcal{I}(\mathcal{H}',\mathbb{A}')$ then $C$ is monotonically decreasing with respect to the application of $\mathcal{E}$, i. e. $C(\mathcal{E}(\rho)|\mathbb{A}') \leq C(\rho|\mathbb{A})$ for every $\rho$.
    \end{enumerate}
\end{definition}

For example, take the $\ell_1$-coherence quantifier~\citep{baumgratz2014quantifying} that is defined by the mapping $C(\cdot \vert \mathbb{A}): \mathcal{D}(\mathcal{H}) \to \mathbb{R}$ where for every quantum state $\rho \in \mathcal{D}(\mathcal{H})$ we have 
\begin{equation}
    C(\rho|\mathbb{A}) := \sum_{i \neq j}|\langle a_i | \rho | a_j \rangle |.
\end{equation}

In this case, it is clear that there are quantum states $\rho_1 \in \mathcal{D}(\mathbb{C}^{d_1}),\rho_2 \in \mathcal{D}(\mathbb{C}^{d_2})$ such that, for example,  
\begin{equation}\label{eq:inequality_coherence_quantifiers}
    C(\rho_1|\mathbb{D}_{d_1}) \leq C(\rho_2|\mathbb{D}_{d_2})
\end{equation}
whenever $d_1 < d_2$. 

\begin{example}[Dimension gap between coherence quantifiers]
    Let us take $\rho_1 = \vert +\rangle \langle +\vert$, such that $d_1=2$ and $\rho_2 = \vert \psi \rangle \langle \psi \vert$ where $\vert \psi \rangle = \sfrac{1}{\sqrt{3}}(\vert 0 \rangle + \vert 1 \rangle + \vert 2 \rangle )$ such that $d_2 = 3$. Recall that $\mathbb{D}_2 = \{\vert 0 \rangle, \vert 1 \rangle\}$ and $\mathbb{D}_3 = \{\vert 0 \rangle, \vert 1 \rangle, \vert 2 \rangle\}$. Then, we have that Ineq.~\eqref{eq:inequality_coherence_quantifiers} holds since
    \begin{equation}
        C(\vert +\rangle \langle +\vert \mid\mathbb{D}_2) = \vert \langle 0\vert + \rangle \langle +\vert 1 \rangle \vert + \vert \langle 1\vert + \rangle \langle +\vert 0 \rangle \vert = \sfrac{1}{2}+\sfrac{1}{2}=1,
    \end{equation}
    which is the largest value of $C(\rho \mid \mathbb{D}_2)$ for all states $\rho \in \mathcal{D}(\mathbb{C}^2)$, and moreover
    \begin{align}
        C(\vert \psi \rangle \langle \psi \vert \mid \mathbb{D}_3) &= \vert \langle 0 \vert \psi \rangle \langle \psi \vert 1 \rangle \vert +\vert \langle 0 \vert \psi \rangle \langle \psi \vert 2 \rangle \vert +\vert \langle 1 \vert \psi \rangle \langle \psi \vert 2 \rangle \vert \\&+\vert \langle 1 \vert \psi \rangle \langle \psi \vert 0 \rangle \vert +\vert \langle 2 \vert \psi \rangle \langle \psi \vert 0 \rangle \vert +\vert \langle 2 \vert \psi \rangle \langle \psi \vert 1 \rangle \vert \\
        &= \sfrac{1}{3}+\sfrac{1}{3}+\sfrac{1}{3}+\sfrac{1}{3}+\sfrac{1}{3}+\sfrac{1}{3}=2.
    \end{align}
    
    In fact, it is easy to see that $C(\vert \psi_d \rangle \langle \psi_d \vert \mid \mathbb{D}_d) = d-1$ for states $\vert \psi_d\rangle = \sfrac{1}{\sqrt{d}}\,\sum_{i=0}^{d-1}\vert i \rangle $. 
\end{example}

\myindent Both the examples above (of coherence rank, and of coherence quantifiers) can capture aspects of when basis-dependent coherence \emph{requires} higher-dimensional systems, either to allow for a larger coherence rank or to allow for larger values of $C(\rho|\mathbb{A})$. Up until now, the two definitions described above capture coherence but it is unclear if they satisfactory answer Question~\ref{question: dimension and coherence?}. For instance, they rely on the complete characterization of the states via quantum tomography, which is well-known to be experimentally challenging as the Hilbert space dimension increases~\citep{nielsen2002quantum,flammia2012quantum,odonnel2015quantumspectrum,odonnell2016efficient,wright2016learn,yuen2023improved}. However, there is a way, in principle, of connecting the estimation of properties such as those just described to direct experimental observables, without relying on full quantum state tomography, via a technique that is known as \emph{witnessing}.

\subsection{Witnesses of coherence}\label{sec: coherence witnesses}

\myindent The idea of proposing nonclassicality witnesses was first considered in the context of witnessing quantum entanglement by~\cite{horodecki1996necessary} in the search of necessary and sufficient conditions for guaranteeing that quantum states are entangled~\citep{horodecki2001separability}. This is an idea that is extremely well motivated due to theoretical proofs showing that deciding whether a bipartite quantum state is entangled is, in general, a provably hard task~\citep{gurvits2004classical}. To the best of our knowledge, the person who coined the term `witness' was Barbara Terhal in her seminal work~\citep{terhal2000bell} showing that every Bell inequality can formally give rise to an entanglement witness.

\myindent In the standard framework of quantum and static resource witnesses~\cite[Sec. 9.4]{gour2024resources}, one defines a \emph{resource witness} to be a Hermitian operator, commonly denoted with the letter $W$, defined relative to some subset $\mathcal{F} \subseteq \mathcal{D}(\mathcal{H})$ of quantum states. More generally, we consider a definition of witness that does not necessarily rely on the fact that the set $\mathcal{F}$ is compact and convex, nor that it is a subset of the set of all density matrices of a certain system. 

\begin{definition}[Witness]\label{def: witness}
    Let $\mathcal{F}\subseteq \mathcal{O}$ be any subset of a set $\mathcal{O}$ and both are nonempty. We say that a function $f: \mathcal{O} \to \mathbb{R}$ is a \emph{witness} for an element $o \in \mathcal{O} \setminus \mathcal{F}$ if there exists some value $\alpha \in \mathbb{R}$ such that  for every element $o_f \in \mathcal{F}$ 
    \begin{equation}
        f(o_f) \leq \alpha 
    \end{equation}
    \emph{but}
    \begin{equation}
        f(o)> \alpha.
    \end{equation}
\end{definition}

\myindent The intuition behind this definition is simply that we say that $f$ is a witness for an element $o \in \mathcal{O}\setminus \mathcal{F}$, meaning that $f$ can certify the property ``$o$ is not an element of $\mathcal{F}$'', whenever the action of $f$ on all elements of $\mathcal{F}$ is bounded by some scalar  $\alpha$, while $f(o)$ surpasses this bound, i.e. $f(o)> \alpha$.

\myindent Above, it is irrelevant whether one defines witnesses with respect to the choice $f(o_f)\leq \alpha$ or with respect to the choice $f(o_f)\geq \alpha$ for all $o_f$, as long as one fixes a specific choice. For the case where the witnesses are convex-linear functionals $f$ acting on a finite-dimensional vector space, the bound $\alpha$ can always be taken to be equal to zero without loss of generality. In other words, if there exists a convex-linear witness for $o$ with respect to a certain $\alpha$ there must exist another convex-linear witness for $o$ with respect to $0$. Resource witnesses that are defined relative to some Hermitian operator are specific instances of~\cref{def: witness} where $f: \mathcal{D}(\mathcal{H}) \to \mathbb{R}$ is given by the linear functionals $f_W(\cdot):= \text{Tr}(W(\cdot))$, and by choosing $\alpha = 0$. We introduce this specific definition for witnesses via Hermitian operators for future reference. To distinguish them from the more general case, we  refer to such witnesses as \emph{operator witnesses}.

\begin{definition}[Operator witnesses]\label{def: quantum resource witness}
    Let $\mathcal{F} \subseteq \mathcal{D}(\mathcal{H})$ any nonempty subset of quantum states. We say that a Hermitian operator $W$ is an \emph{operator witness} for a state $\rho \in \mathcal{D}(\mathcal{H}) \setminus \mathcal{F}$ if, for every element $\sigma \in \mathcal{F}$ we have that 
    \begin{equation}
        \text{Tr}(W\sigma) \leq 0,
    \end{equation}
    while
    \begin{equation}
        \text{Tr}(W\rho) > 0.
    \end{equation}
\end{definition}

\myindent Following~\cref{def: witness}, every coherence quantifier (or measure) is, in particular, a witness of quantum coherence, yet not an operator witness as by~\cref{def: quantum resource witness}. For comprehensive reviews focused on quantifiers and measures of coherence, we refer to~\cite{streltsov2017colloquium} for a theoretical account and to~\cite{wu2021experimental} for experimental progress. Various witnesses of coherence exist and, in particular, great relevance has been put on the coherence witnesses that yield exactly the value of a certain coherence quantifier. It is important to note that finding witnesses yielding the value of a certain quantifier is not a simple task in general as it often requires the use of optimization techniques. Moreover, even when one \emph{does} find such a witness, it may be an extremely difficult observable $W$ to estimate in practice. Yet, finding coherence witnesses usually simplifies considerably the search for a more practical approach to certifying nontrivial properties of quantum resources in large quantum systems. Alternatively, it provides useful insights into how one might design methods that favor practical implementations.

\myindent Focusing on the question we have posed in Box 1 however, to the best of our knowledge, there have been only a few attempts to construct witnesses of \emph{both} basis-dependent coherence \emph{and} dimension, within the resource theoretic perspective. Early work by~\cite{chin2017coherencerankmixedstates,chin2017generalizedcoherence} considered coherence quantifiers that could capture the information of the coherence rank, but they had the usual issue of quantification strategies of being experimentally accessible only via exponentially many measurements. To the best of our knowledge, the first proposal for a family of witnesses of basis-dependent coherence \emph{and} dimension (hence multilevel coherence) was made by~\cite{ringbauer2018certification}. In their work, they proposed a family of witnesses of multilevel coherence capable of capturing basis-dependent coherence present in $d$-dimensional pure quantum states, and they also experimentally tested this prediction for systems of dimension  $d=2,3$ and $4$. 

\myindent As we will see later in Chapter~\ref{chapter: information tasks}, by relaxing the assumption that one already \emph{knows} what the Hilbert space dimension is, answering the question posed by Box 1 becomes more interesting, and it also becomes possible to reveal statistical results requiring both coherence and dimension of the prepared quantum states via the notion of \emph{prepare-and-measure scenarios}. 

\section{Quantum imaginarity: Quantum coherence due to complex numbers}~\label{sec: quantum imaginarity}

\myindent One `type' of coherence that has recently gained a great amount of attention is  \emph{quantum imaginarity}, first discussed in the context of quantum resource theories  by~\cite{hickey2018quantifying}.~\footnote{An early mention of this resource can be found in~\citep{gour2017quantumresource}. There, the author simply noted that a resource theory where real density matrices are taken to be free objects is an example of an affine static quantum resource theory that lacks a resource-destroying map~\citep{liu2017resourcedestroying}. To the best of our knowledge, this remains the only known example of such a static resource theory.} In this case, we again need to specify a pair $(\mathcal{H},\mathbb{A})$ of Hilbert space and basis of reference. Yet, we only focus on states that have coherence due to imaginary values. We make this precise in the following definition:

\begin{definition}[Basis-dependent imaginarity]
Let $\mathcal{H}$ and $\mathbb{A}$ be a fixed Hilbert space and reference basis of it. We say that a given state $\rho \in \mathcal{D}(\mathcal{H})$ has \emph{basis-dependent imaginarity} if there exist $i,j$ such that $\text{Im}[\langle a_i |\rho |a_j \rangle] \neq 0$. On the contrary, we say that a state is \emph{imaginarity-free}, or simply \emph{real-represented with respect to $\mathbb{A}$}, if $\langle a_i |\rho |a_j \rangle \in \mathbb{R}$ for every $i,j$. We define the subset of imaginarity-free states as $\mathcal{R}(\mathcal{H},\mathbb{A}) \subseteq \mathcal{D}(\mathcal{H})$. 
\end{definition}

\myindent Clearly from the definition we have that for every pair $(\mathcal{H},\mathbb{A})$ the following chain of inclusions holds: $$\mathcal{I}(\mathcal{H},\mathbb{A}) \subseteq \mathcal{R}(\mathcal{H},\mathbb{A}) \subseteq \mathcal{D}(\mathcal{H}).$$
As there is no review on the specific topic of quantum imaginarity, and since existing reviews on coherence have not described recent developments on this topic, we  provide a more thorough literature review of it in what follows.

\myindent We start by pointing that some early results in the literature suggested that imaginarity had \emph{no} relevant role, as, for instance, various real-only formulations of quantum mechanics were known~\citep{stueckelberg1960quantum,stueckelberg1961quantum,antoniya2013real,hardy2011limited,wootters2010entanglement,wootters2015optimal}. Moreover, quantum theory restricted to real amplitudes was shown to be universal for quantum computation by~\cite{rudolph2002rebit}, and by~\cite{aharonov2003simple}, and capable of reproducing the statistics of any Bell experiment by~\cite{McKague2009simulating}. Any generic complex-valued quantum computation can be evaluated using only real amplitudes by using an extra auxiliary qubit, enlarging the system from $\mathcal{H} \mapsto \mathbb{C}^2 \otimes \mathcal{H}$. In such a way, free states $\rho^{\mathbb{R}}$ and observables $H^{\mathbb{R}}$ defined in the larger system reproduce all predictions of the resourceful states $\rho$ and observables $H$ via $\text{Tr}(\rho H) = \text{Tr}(\rho^{\mathbb{R}} H^{\mathbb{R}})$. This construction is also known as a \emph{Stueckelberg mapping} or \emph{simulation}~\citep{ying2025quantumtheoryneedscomplex}. 

\myindent This is a crucial aspect of the notion of imaginarity just described, which we expand on now. Unless we focus on the pair $(\mathcal{H},\mathbb{A})$ as the defining description of our scenario, i.e., the setting in which it is meaningful to talk about imaginarity, the notion itself trivializes. Similarly to the case of coherence, for every Hilbert space, there \emph{always exist} some basis $\mathbb{A}_\rho$ such that a state $\rho$ is imaginarity free (in fact, diagonal if we take its spectral basis). Therefore, unless one fixes the reference basis, there is no notion of `imaginarity of a state'. Perhaps less trivially, and marking a clear departure from coherence here, if we do not fix the Hilbert space, then, as mentioned before, there is a \emph{dilation} such that any imaginary-represented statistics can be reproduced by some real-represented one, on a larger Hilbert space. Precisely, suppose that we fix, for each dimension $d$, a certain canonical basis $\mathbb{A}$ to characterize imaginarity. We want to show that every statistic that considers states (and observables) that have imaginarity can be entirely reproduced by states (and observables) that are imaginarity-free but on a higher dimensional space. If $\rho \in \mathcal{D}(\mathcal{H})$ has imaginarity, then the simple operation
\begin{equation*}
    \rho \mapsto \rho^{\mathbb{R}} = \frac{1}{2}\left(\begin{matrix}
        \text{Re}[\rho] & - \text{Im}[\rho]\\
        \text{Im}[\rho] & \text{Re}[\rho]
    \end{matrix}\right) \in \mathcal{R}(\mathbb{C}^2 \otimes \mathcal{H},\mathbb{A})
\end{equation*}
sends any state having imaginarity into a real-only one, in a higher dimensional Hilbert space. If we also do the same to the observables
\begin{equation*}
    H \mapsto H^{\mathbb{R}} = \left(\begin{matrix}
        \text{Re}[H] & - \text{Im}[H]\\
        \text{Im}[H] & \text{Re}[H]
    \end{matrix}\right) 
\end{equation*}
a simple calculation shows that, for every $\rho,H \in \mathcal{B}(\mathcal{H})$ such that $\rho$ is a quantum state and $H$ is a Hermitian operator, then  $\rho^{\mathbb{R}},H^{\mathbb{R}} \in \mathcal{B}(\mathbb{C}^2\otimes \mathcal{H})$ are a state and a Hermitian operator such that 
\begin{equation*}
    \text{Tr}(\rho H) = \text{Tr}(\rho^{\mathbb{R}} H^{\mathbb{R}}).
\end{equation*}

\myindent Despite these drawbacks, quantum imaginarity has found several applications. It has been shown that imaginarity can be made operationally meaningful in a precise sense by~\cite{wu2021operational,wu2021resource,wu2023resource}. Imaginarity of quantum theory was shown to have crucial effects in certain discrimination tasks~\citep{wu2021operational,herzog2002minimum}, hiding and masking~\citep{zhu2021hidingmasking}, machine learning~\citep{sajjan2023imaginary}, pseudorandomness~\citep{haug2023pseudorandom}, multiparameter metrology~\citep{carollo2018uhlmann,carollo2019quantumness,miyazaki2022imaginarityfree}, outcome statistics of linear-optical experiments~\citep{jones2023distinguishability,menssen2017distinguishability,shchesnovich2018collective}, Kirkwood-Dirac quasiprobability distributions~\citep{wagner2024quantumcircuits,budiyono2023operational,budiyono2023quantifying,budiyono2023quantum}, and weak-value theory~\citep{wagner2023anomalous,kedem2012usingtechnical,dixon2009ultrasensitive,hosten2008observation,brunner2010measuringsmall,hofmann2011uncertainty,kunjwal2019anomalous,hofmann2012complex,bollen2010direct}. Because of that it is also relevant to attempt to characterize (if possible) the quantum imaginarity of states and measurements of a physical system. Nevertheless, due to the dilation and basis-dependence aspects just described, it seems that such tests would be significantly challenging. 

\subsection{Witnesses of quantum imaginarity}

\myindent Surprisingly,~\cite{renou2021quantum} found an experimental scenario that cannot be exactly modeled using quantum theory with real-valued amplitudes only. These findings have been experimentally verified~\citep{li2022testing,chen2022ruling,wu2022experimental}, and a few other similar scenarios have since been proposed~\citep{bednorz2022optimal,yao2024proposals}. The requirements to be met for such experiments are demanding, since they are based on network nonlocality scenarios. For example, every source needs to be independent, and measurements 
 from different parties must be space-like separated. In simple terms, requiring this independence -- and therefore a certain tensor product structure to be preserved -- guarantees that the trick of extending $\mathcal{H} \mapsto \mathbb{C}^2 \otimes \mathcal{H}$ fails. A substantially more critical analysis of the work by~\cite{renou2021quantum} from the perspective of \emph{foil} theories can be found in~\citep{ying2025quantumtheoryneedscomplex}.

\myindent The protocol described by~\cite{renou2021quantum} can be interpreted as an imaginarity witness via our~\cref{def: witness}. In that reference, they showed that there are  functionals $f$ acting on statistics of Network scenarios.~\footnote{We refer to reference~\citep{tavakoli2022Bell} for a recent review on this topic. } They then show that, if respecting the constraints of the Network scenarios, and if standard quantum theory is taken to be correct, the functional value \emph{certifies} that imaginarity (of states or measurements) was used to generate the statistics. Operator witnesses of imaginarity, in the sense of~\cref{def: quantum resource witness}, have not been considered (we show an infinite family of witnesses in Chapter~\ref{chapter: relational coherence} for the imaginarity of product states $\rho_1 \otimes \rho_2 \otimes \dots \otimes \rho_n$, see for instance  Sec.~\ref{subsec: relational imaginarity}). However, several \emph{quantifiers} of quantum imaginarity have been investigated~\citep{hickey2018quantifying,wu2021operational}.

\myindent Network nonlocality witnesses of the type used to demonstrate the need for imaginarity are found via relaxations of semidefinite programming problems~\citep{tavakoli2023semidefinite}. They are basis-independent, dimension-independent, semi-device independent, and independent of extra assumptions such as the possibility of realizing local tomography. The requirement of only weak assumptions certainly makes the foundational argument more compelling, but it has a price. The possible witnesses and experimental scenarios are few and challenging, requiring precise clock control of independent---or \emph{partially} independent~\citep{weilenmann2025partial}---sources, space-like separation, and measurements performed by many parties. 

\myindent Therefore, in the current state of the art we have two distinct treatments of imaginarity. On one hand, we have a basis-dependent, device-dependent description, which is tied to the standard approach of \emph{static quantum resource theories}~\cite{chitambar2019quantum}. In such a case, the characterization and quantification of quantum imaginarity are possible and much is known, however, one requires significant prior information about the device and the specification of a pair $(\mathcal{H},\mathbb{A})$. On the other hand, we have the approach outlined by~\cite{renou2021quantum}, which do not require the assumption of such pairs $(\mathcal{H},\mathbb{A})$, nor complete prior information about the device.

\section{Basis-independent coherence of a single state}~\label{sec: basis-independent coherence of a single state}

\myindent One aspect of the notion of coherence as we have described is its basis dependence. This particular aspect is not physically well-motivated in general, and one needs to appeal to specific aspects of the physical scenario in consideration. Physical systems do not come equipped with a preferred notion of reference basis, and this is only well motivated if we know other aspects of the system such as its dynamics, or if we have complete control of the system such as in quantum computation, or also when it is possible to characterize how the system decoheres, in which case the environment `selects' a preferred reference basis. 

\myindent There are various other physical situations in which a reference basis is an entirely \emph{arbitrary choice} which do not need to have a specific profound physical meaning. Linear optics is one such example. For example, a mere rotation of the waveplates changes the reference basis for polarization.

\myindent In most situations however one would prefer to make statements (or to consider) physical properties only if they \emph{do not} depend on our representation of the system. Arguably, when certifying a certain nonclassical feature of a quantum device, it is unnatural to say that `my device cannot be reproduced, nor substituted, by a classical device only if you use this specific representation to describe it'. And if we consider the notion of quantum information or computational advantage, it becomes even clearer that we would like a certain information task, or a certain computation, to be in some sense `better' than any classical counterpart, in a way that does not specify a certain choice of basis. This is not to say that in various situations one does not have a reference basis and is interested in investigating properties \emph{relative to it} but there are other situations in which this choice becomes arbitrary, or worse, weakens the physical intuition we have about a certain phenomenology. 

\myindent To remedy this intrinsic basis dependence of the description of the standard notion of coherence, a first approach that has been considered in the literature has been given in terms of the following choice: We consider that the \emph{only} truly incoherent state is the maximally mixed state $\mathbb{1}/d$ for a given Hilbert space of dimension $d = \dim(\mathcal{H})$. Again, since the topic that we  refer to as \emph{basis-independent coherence of a single state} has not been the main focus, or even a part, of any existing review, we provide a more comprehensive description of the  state of the art.

\myindent To the best of our knowledge, the first attempt at introducing a notion of basis-independent quantum coherence was done by~\cite{yao2015quantum}. Their results connected such a notion of ``basis-independent coherence'' with the notion of \emph{quantum discord}~\citep{modi2012discord}. They considered an entropic coherence quantifier given by a function $C_{\mathrm{S}}(\cdot | \mathbb{A}) : \mathcal{D}(\mathcal{H}) \to \mathbb{R}$
\begin{equation*}
    C_{\mathrm{S}}(\rho|\mathbb{A}) = \min_{\sigma \in \mathcal{I}(\mathcal{H},\mathbb{A})}S(\rho \Vert \sigma),
\end{equation*}
applied specifically to the case of multipartite states, i.e., quantum states $\rho \in \mathcal{D}(\mathcal{H}^{\otimes N})$. They then proceed to do a minimization with respect to all possible \emph{local unitaries} for a certain multipartite quantum state, defining a function $\widetilde{C}_{\mathrm{S}}:\mathcal{D}(\mathcal{H}^{\otimes N}) \to \mathbb{R}$ via 
\begin{equation*}
    \widetilde{C}_{\mathrm{S}}(\rho) = \min_{\vec U}C_{\mathrm{S}}(\vec{U}\rho\vec{U}^\dagger|\vec U\mathbb{A}),
\end{equation*}
that they referred to as the `basis-free quantum coherence' of a multipartite quantum state $\rho$. Above we have written $\vec U \mathbb{A} \equiv \{\vec U \vert a_i\rangle \}_{i=1}^{d^N}$, representing a change of basis, $\mathbb{A}$ some separable basis of $\mathcal{H}^{\otimes N}$, and $\vec{U} \equiv U_1 \otimes \dots \otimes U_N $. Arguably, the focus on local unitaries is a restriction on the set of all possible reference choices, implying that to some extent one does not truly have a `basis-free' notion of coherence captured by this approach.

\myindent A similar yet more general avenue was taken by~\cite{yu2016total}. Let $\mathsf{d}$ denote a `distance' defined on $\mathcal{B}(\mathbb{C}^d)$. They have considered optimizations over \emph{all} possible unitaries via
\begin{equation*}
    C_{\mathsf{d}}(\rho) = \max_{U} \mathsf{d}\left(U \rho U^\dagger, \mathcal{D}_{\mathbb{A}}( U\rho U^\dagger )\right), 
\end{equation*}
where $\mathcal{D}_{\mathbb{A}}( \rho ) = \sum_i\langle a_i| \rho | a_i \rangle \vert a_i \rangle \langle a_i \vert $ and $\mathbb{A}$ is fixed to be equal to $\mathbb{D}_d$. For different choices of `distances', they show that various quantifiers of basis-independent coherence can be proposed:
\begin{equation*}
    C_{\Vert \cdot \Vert_2}(\rho) = \text{Tr}(\rho^2)-\frac{1}{n}, \,\, C_{\mathrm{S}}(\rho) = \log(n)-S(\rho),
\end{equation*}
if choosing, respectively,  $\Vert X-Y\Vert_2, S(X\Vert Y)$ being the $\ell_2$-norm and the relative entropy. 
We note that, indeed, the above quantifiers are related to unitary-invariant functions (see Chapter~\ref{chapter: Bargmann invariants} for a formal definition) of a state. Following this basis-independent interpretation, it became relevant to investigate what is the largest value a certain monotone can attain, depending on the choice of basis of reference, a problem that was later investigated by~\cite{yao2016maximal},~\cite{hu2017maximal}, and~\cite{streltsov2018maximal}.

\myindent Another related approach is to quantify coherence via a distance to the only state diagonal with respect to every basis, which is the maximally mixed state. This has been the approach considered by~\cite{yu2016total},~\cite{yao2016frobenius}, and~\cite{wang2017intrinsic}. These attempts usually culminated in quantifiers based on unitary-invariant functions such as the purity or other quantities as just described, which even led to the proposal of investigating resource theories of \emph{purity} alone~\citep{streltsov2018maximal}. On a related note, given that the maximally mixed state plays such an important role in such approaches, it was also the main focus of some resource-theoretic approaches, such as was considered by~\cite{ma2019operational} and ~\cite{radhakrishnan2019basisindependent}.  Ideas of considering \emph{purity} given by $\text{Tr}(\rho^2)$ as a basis-independent figure of merit of some form of coherence have been investigated much earlier~\citep{brukner2001conceptual,galvao2002foundations}, but without an emphasis on viewing this notion as a resource and only as a carrier of some form of unitary-invariant quantum information. 

\myindent It is worth pointing out that some of these contributions found relevance not just in a mathematical sense, but in more practical scenarios. For instance, this notion of basis-independent coherence has been shown to power magnetic sensing~\citep{le2020basis}, and to distinguish between `classical' and `quantum' randomness~\citep{ma2019operational}. It also has an appeal for quantifying unitary-information in linear-optical systems as motivated by the discussions in~\cite{brukner2001conceptual} and ~\cite{galvao2002foundations}.

\myindent One can argue that defining the basis-independent coherence of a state in terms of the presence of \emph{some} coherence with respect to \emph{some} state is somewhat too permissive. This notion trivializes the set of incoherent states to be only the singleton $\{\mathbb{1}/d\}$ and the set of possible quantifiers of coherence to be only those unitary-invariant functions of a state with itself. In fact, one example of the `triviality' of this approach to quantum coherence is the fact that any purity value $\text{Tr}(\rho^2) \neq \sfrac{1}{d} = \text{Tr}(\sfrac{\mathbb{1}}{d^2})$ signals the presence of this purported form of basis-independent coherence of a single state, making the task of witnessing trivial, even in situations where one does not need to assume anything about the system such as Hilbert space dimension, purity or reference basis. Despite being the first interesting take on this problem, it has perhaps a limited scope of applicability. In what follows, we  discuss a more promising recently proposed avenue that does not consider only the idea of basis-independent coherence of a single state, but of a \emph{set} of states.

\section{Set coherence: Basis-independent coherence of a set of states}~\label{set: basis-independent coherence of a set of states}

\myindent Within the framework of quantum resource theories, we see a novel perspective emerging in recent literature. Rather than focusing on resources defined relative to a single specific quantum object----such as a quantum state, measurement, instrument, or channel---as we have considered before for quantum coherence and imaginarity, and as can similarly be done for quantum nonstabilizerness and entanglement, this new approach considers \emph{sets} or \emph{collections} of quantum elements as the primary objects of study~\citep{uola2019quantifying,galvao2020quantum,ducuara2020multiobject,martins2020quantum,buscemi2020complete,designolle2021set,miyazaki2022imaginarityfree,salazar2022resource,selby2023contextuality,wagner2024inequalities,wagner2024certifying}. 

\myindent Defining phenomena based on properties of collections of objects, rather than on the properties of individual objects, is an approach that has long been established in the quantum foundations literature. The violation of Bell inequalities, the violation of noncontextuality inequalities, and the analysis of joint measurability are all examples of notions that can be probed only as a \emph{collective property}.  All these are examples of phenomena that can only be well defined if one considers a collection of objects, forming what is often called a \emph{scenario} or a \emph{fragment of a physical theory}.

\myindent Of particular interest to us are the notions of set-coherence~\citep{designolle2021set} and set-imaginarity~\citep{miyazaki2022imaginarityfree}. In this case, we fix a physical system of interest that, in quantum theory, is equivalently described by some Hilbert space. We take subsets $\{\rho_i\}_i \subseteq \mathcal{D}(\mathcal{H})$ to be the objects of interest and define the following.

\begin{definition}[Set coherence]\label{def: set coherence}
    Let $\mathcal{H}$ be a Hilbert space. The set of states $\{\rho_i\}_i$ is \emph{set incoherent} if there exists a unitary $U:\mathcal{H} \to \mathcal{H}$ such that $U \rho_i U^\dagger = \sigma_i$ for all $i$ where $\sigma_i$ are diagonal density matrices with respect to some basis $\mathbb{A}$. In other words if there exists \emph{some} basis $\mathbb{A}$ with respect to which $\{\rho_i\}_i \subseteq \mathcal{I}(\mathcal{H},\mathbb{A})$. If this does not hold, we say that $\{\rho_i\}_i$ is \emph{set coherent}.
\end{definition}

\myindent Equivalently, the set $\{\rho_i\}_i$ is incoherent if $[\rho_i,\rho_j] = 0$ for all $i,j$. In words, the above definition is surprisingly simple and intuitive, and it is also surprising that it had not been introduced earlier. We note that noncommutativity has always been a key element when considering \emph{measurements and observables}, and was fundamental for the analysis of early results in the theory of measurement incompatibility, and for the proof of the Kochen--Specker theorem~\citep{kochen1975problem}. However, it is \emph{operationally different} to consider non-commutativity in the preparation of physical systems even if \emph{mathematically} equivalent to incompatibility of observables~\footnote{Here taken to be dual to projection-valued measures, which in finite dimensions are equivalent to sets of orthogonal projectors $\{\Pi_{a}\}_a$. In this case, incompatibility and noncommutativity are mathematically equivalent notions.} since in both cases, noncommutativity is the source of the effect.

\myindent Considering sets provides a simple manner of analyzing coherence in a basis-independent way, without the criticisms we have pointed out earlier for the attempts of considering a single state. For example, if one considers the quantum states described by $\vert 0\rangle \langle 0 \vert $ and $\vert +\rangle \langle + \vert $, as a \emph{pair} of states they are set coherent. This implies that for all choices of reference bases for the Hilbert space, one sees coherence terms, i.e. off-diagonal non-null terms in their density matrix representation,  in either one the states, or even both. 

\myindent It is interesting that in some situations, other works have focused on the operational relevance of states as opposed to that of measurements. Famously, Spekkens' notion of contextuality~\citep{spekkens2005contextuality} generalized the notion of Kochen--Specker contextuality, only applicable to measurements, to apply to both quantum states (viewed as equivalence classes of preparation procedures) and quantum channels (viewed as equivalence classes of transformation procedures). We  present this notion in detail in Chapter~\ref{chapter: contextuality}. More recently, a series of works by Anubhav Chaturvedi, Debashis Saha, and colleagues~\citep{chaturvedi2020quantum,chaturvedi2021quantum,manna2024unbounded} have also investigated in depth nonclassical properties of sets of states that go beyond the  Spekkens’ formalism.

\myindent One of the most relevant distinctions between set incoherence of objects $\{\rho_i\}_i$ and incoherence of objects $\rho$ with respect to some basis $\mathbb{A}$ is related to \emph{convexity}. The standard convention for considering the linear combination of two sets is the following: Take $X, Y$ any pair of subsets of the same vector space, we have that the new set $\alpha X + \beta Y$ for $\alpha,\beta \in \mathbb{R}$ is defined as
\begin{equation}
    \alpha X + \beta Y := \{\alpha x + \beta y \mid x \in X, y \in Y\}.
\end{equation}
Let us denote as usual the power set of a set $X$, the set of all possible subsets of $X$, by $2^X$. Then, we denote the set of all set incoherent $\{\rho_i\}_i$ as $\mathcal{S}_{\mathrm{inc}} \subseteq 2^{\mathcal{D}(\mathcal{H})}$. It is simple to see that, while $2^{\mathcal{D}(\mathcal{H})}$ is closed under convex combinations, meaning that every convex combination of sets of states yields another set of states, the set incoherent portion $\mathcal{S}_{\mathrm{inc}}$ of the power set is \emph{not} closed under convex combinations. To see this, following an example from~\citep{designolle2021set},~\footnote{Formally, ~\cite{designolle2021set} considers the set-coherence of \emph{ordered sets}, in which case the convex combination is defined element-wise. The proof that the \emph{ordered} convex combination of incoherent tuples also yields a coherent tuple is similar. We provide an independent one based on the values of Bargmann invariants in~\cref{subsec: equality constraints}. } it is enough to take 
\begin{equation}\label{eq: set coherence nonconvex}
\mathcal{S}_1 = \{\vert 0\rangle \langle 0 \vert, \frac{1}{3}\vert 0\rangle \langle 0 \vert + \frac{2}{3}\vert 1\rangle \langle 1 \vert \}, \mathcal{S}_2 = \{\vert +\rangle \langle + \vert, \frac{1}{4}\vert +\rangle \langle + \vert + \frac{3}{4}\vert -\rangle \langle - \vert\} \in \mathcal{S}_{\mathrm{inc}}.
\end{equation}
In this case, $\sfrac{1}{2}\,\mathcal{S}_1 + \sfrac{1}{2}\,\mathcal{S}_2$ yields $\{\rho_a,\rho_b,\rho_c,\rho_d\}$ where 
\begin{align*}
    \rho_a &= \frac{1}{2}\vert 0\rangle \langle 0 \vert + \frac{1}{2}\vert +\rangle \langle + \vert\\
    \rho_b &= \frac{1}{2}\vert 0\rangle \langle 0 \vert + \frac{1}{8}\vert +\rangle \langle + \vert+ \frac{3}{8}\vert -\rangle \langle - \vert \\
    \rho_c &= \frac{1}{6}\vert 0\rangle \langle 0 \vert + \frac{2}{6}\vert 1\rangle \langle 1 \vert + \frac{1}{2}\vert +\rangle \langle + \vert \\
    \rho_d &= \frac{1}{6}\vert 0\rangle \langle 0 \vert + \frac{2}{6}\vert 1\rangle \langle 1 \vert +\frac{1}{8}\vert +\rangle \langle + \vert+ \frac{3}{8}\vert -\rangle \langle - \vert
\end{align*}
which is set coherent since $[\rho_a,\rho_d]\neq 0$, hence outside of $\mathcal{S}_{\mathrm{inc}}$. This is a distinctive feature with respect to considering the relevant objects to be the elements of $\mathcal{I}(\mathcal{H},\mathbb{A})$ which is a convex set. In resource theories, convexity is a much-wanted property as it allows us to easily use existing tools that apply only to convex sets. Arguably, this is a technical drawback of the notion of set incoherence as opposed to the basis dependent treatment of incoherent states. In Chapter~\ref{chapter: relational coherence} we discuss how we can circumvent this drawback by focusing on a system-agnostic notion of coherence.

\myindent This research approach is still in its infancy, yet some `types' of set coherence have already been investigated and analyzed in the literature, such as the notions of set imaginarity by~\cite{miyazaki2022imaginarityfree} and the notion of set magic by~\cite{wagner2024certifying}. Let us define set imaginarity in detail.

\begin{definition}[Set imaginarity]\label{def: set imaginarity}
    Let $\mathcal{H}$ be a Hilbert space. The set $\{\rho_i\}_i$ is \emph{free of imaginarity, imaginarity-free} or  \emph{real-valued representable} if the exists a unitary $U:\mathcal{H} \to \mathcal{H}$ such that $U \rho_i U^\dagger = \rho_i^{\mathbb{R}}$ for all $i$ where $\rho_i^{\mathbb{R}}$ are all density matrices having real terms only, with respect to some basis $\mathbb{A}$. In other words, if there exists \emph{some} basis $\mathbb{A}$ with respect to which $\{\rho_i\}_i \subseteq \mathcal{R}(\mathcal{H},\mathbb{A})$. If this is not possible we then say that the set $\{\rho_i\}_i$ has basis-independent imaginarity, or that $\{\rho_i\}_i$ has set imaginarity.
\end{definition}

\myindent As before, we denote the set of all sets of states that are imaginarity-free as a subset of the power set $\mathcal{S}^{\mathbb{R}} \subseteq 2^{\mathcal{D}(\mathcal{H})}$. Note that, similarly to the case of set incoherent states, a certain set $\{\rho_i\}_i$ may have imaginarity with respect to some basis $\mathbb{A}$ even if it is real-valued representable. For instance, the set $\{\vert +_i\rangle \langle +_i \vert, \vert -_i\rangle\langle -_i \vert \}$ has imaginarity with respect to the basis $\mathbb{D}_2 = \{\vert 0\rangle, \vert 1 \rangle\}$. However, as there is a unitary transformation from $\mathbb{Y} = \{\vert +_i \rangle, \vert -_i \rangle \}$ to $\mathbb{D}_2$, the set $\{\vert +_i\rangle \langle +_i \vert , \vert -_i\rangle \langle -_i \vert \}$ is real representable relative to $\mathbb{Y}$, and hence has no set imaginarity. On the contrary, the set $\{\vert 0\rangle \langle 0 \vert , \vert +_i\rangle \langle +_i \vert , \vert +\rangle \langle + \vert \}$ \emph{has} set imaginarity, since for every unitary $U:\mathbb{C}^2 \to \mathbb{C}^2$ acting on all elements in the set, at least one of the states must have complex-valued off-diagonal terms with non-null imaginary part.

\subsection{Witnesses of set coherence}

\myindent Naturally, many witnesses of set coherence were proposed before the introduction of this terminology by~\cite{designolle2021set}. For example, in most prepare-and-measure scenarios~\citep{gallego2010device,poderini2020criteria,degois2021general} (see Chapter~\ref{chapter: contextuality} for other scenarios), the experimental statistics observed can be used to witness forms of set coherence. Of particular importance to us are the prepare-and-measure scenarios that estimate \emph{two-state overlaps} or \emph{Bargmann invariants} described by~\cite{galvao2020quantum} and~\cite{oszmaniec2024measuring}, respectively.

\myindent \cite{galvao2020quantum} have developed families of two-state inequalities that are defined as
\begin{equation}\label{eq: overlap cycle inequalities}
    \sum_{i=1}^{n-1}\text{Tr}(\rho_i\rho_{i+1})-\text{Tr}(\rho_1\rho_n) \leq n-2,
\end{equation}
for any set of quantum states $\{\rho_i\}_{i=1}^n$ on any Hilbert space $\mathcal{H}$.  Above, permutations of the minus sign multiplying $\text{Tr}(\rho_1\rho_n)$, i.e. changes $\{1,n\} \leftrightarrow \{j,j+1\}$ for any $j$, yield other $n-1$ equally valid inequalities. They have shown moreover that violations of such overlap inequalities are witnesses of set coherence for the set $\{\rho_i\}_{i=1}^n$. The simplest such inequality is given by 
\begin{equation}
    \text{Tr}(\rho_1\rho_2)+\text{Tr}(\rho_1\rho_3)-\text{Tr}(\rho_2\rho_3) \leq 1.
\end{equation}
They have also completely characterized the set of all possible points $$(\text{Tr}(\rho_1\rho_2),\text{Tr}(\rho_1\rho_3),\text{Tr}(\rho_2\rho_3)) \in [0,1]^3$$ that are able to witness set coherence. We return to this characterization in Chapter~\ref{chapter: Bargmann invariants} when we  focus on existing literature concerning the theory of Bargmann invariants. %Moreover, some of our main results in this thesis generalize the ideas considered by~\cite{galvao2020quantum} that we  present in Chapter~\ref{chapter: event_graph_approach} as an entirely graph-theoretic mathematical framework and later in Chapter~\ref{chapter: relational coherence} when we apply the mathematical framework to the problem of witnessing coherence of sets of states. 

\myindent Related to the point above,~\cite{oszmaniec2024measuring} have also made a remark that the complex-valued multivariate traces of states $\text{Tr}(\rho_1\rho_2 \dots \rho_n)$, also known as Bargmann invariants, are witnesses of set imaginarity. This is simple to see as these are unitary-invariant functions. We shall go back to this point in Chapter~\ref{chapter: Bargmann invariants} when we  formally introduce the notion of Bargmann invariants. Part of the contributions in this thesis lie in further developing this simple yet interesting connection between higher order Bargmann invariants and quantum imaginarity. 

\myindent We conclude our discussion on coherence with a remark that coherence may not be the best-motivated notion of \emph{nonclassicality} even if it is indeed a well-motivated resource for quantum information and computation. Some simple arguments can be made to that effect. For example, it is questionable if focusing so much on a property of \emph{states} may not imply that we are unnecessarily committing ourselves to a specific `picture' of the quantum formalism~\citep{thomas2024rolecoherencequantumcomputational}. Moreover, since quantum theory may not be the best theory for representing any phenomena in Nature, and since we may never be able to find \emph{the} best theory (if it exists), it would be good practice to develop a \emph{theory-independent} notion of nonclassicality. One proposal for such a notion (and likely the most accepted one in quantum foundations as a good notion of nonclassicality) is the notion of \emph{contextuality}, which we proceed to discuss. 
\chapter{Contextuality}\label{chapter: contextuality}

\begin{quote}
    ``\textit{However that may be, long may Louis de Broglie continue to inspire those who suspect that what is proved by impossibility proofs is lack of imagination.}''~\footnote{Arguably, it may also be a proof of `lack of imagination' using the same quote in both the Master Dissertation and the PhD thesis.} \\ (John~\cite{bell1982impossible})
\end{quote}

\begin{quote}
    ``\textit{[O]ur present [quantum mechanical] formalism is not purely epistemological; it is a peculiar mixture describing in part realities of Nature, in part incomplete human information about Nature---all scrambled up by Heisenberg and Bohr into an omelette that nobody has seen how to unscramble. Yet we think that the unscrambling is a prerequisite for any further advance in basic physical theory. For, if we cannot separate the subjective and objective aspects of the formalism, we cannot know what we are talking about; it is just that simple.}''\\
    (E. T.~\cite{jaynes1990complexity}, quote taken from~\cite{deronde2016unscrambling})
\end{quote}

\myindent The history surrounding the development of the notion of noncontextuality is fascinating.~\footnote{For other brief accounts on the history of the Kochen--Specker theorem we refer to the online recordings~\citep{cabello2017whattolearn,cabello2023historical}.} A relevant starting point is Louis de Broglie's ``\textit{mécanique ondulatoire}''~\citep{deBroglie1927lamecanique}, later known as de Broglie--Bohm pilot wave theory,~\citep{deBroglie1927lamecanique,bohm1952suggested_one,bohm1952suggested_two} which aimed to resolve a series of outstanding issues of the old quantum theory~\footnote{The term used today for presentations and interpretations of the quantum formalism prior to~\cite{vonNeumann1932mathematische}.}. His model re-established a notion of determinism, positing that the indeterminism in quantum theory's predictions could be understood as a lack of knowledge about a particle's position. This approach resolved the apparent need for concepts like instantaneous collapse or observer-dependent descriptions of experiments. In this model, a particle’s position remains experimentally inaccessible, even though it is assumed to truly `exist' and is thus considered `hidden'. This proposal was presented and debated at the famous 1927 Solvay Conference~\citep{bacciagaluppi2009quantum}, and became a key topic in the famous Einstein--Bohr debate~\citep{landsman2006Whenchampionsmeed}. 

\myindent De Broglie's model and similar ones became known (for obvious reasons) as \emph{hidden-variable models}~\citep{bohm1952suggested_one,bohm1952suggested_two,kochen1975problem,mermin1993hidden}. A more modern jargon for such a class of models is that of \emph{ontological models}~\citep{harrigan2010einstein,leifer2014isthe}, where variables (denoted $\lambda$) are not necessarily treated as `hidden', but assumed to provide a complete description of the state of a given system (also known as a complete description of `state of affairs'~\citep{spekkens2019ontological} or also as an `ontic state'~\citep{spekkens2007evidence}). We will recall these terminologies later, as these are still of great relevance in quantum foundations and not commonly used by the broad readership of physicists. 

\myindent Research in this topic was largely ignored by the physics community in the last century. Two major factors that contributed to this lack of interest by the overwhelming majority of the community of researchers working with quantum mechanics were a no-go result by~\cite{vonNeumann1932mathematische}, and Bohr's response~\citep{bohr1935can} to the critique to the quantum formalism put forth by Einstein, Podolsky, and Rosen (\acrshort{epr})~\citep{einstein1935can}. It was only through the seminal works by Ernst~\cite{specker1960dielogik}, John~\cite{bell1964ontheEPRparadox,bell1966ontheproblem} and later~\cite{kochen1975problem} that hidden variable models were reconsidered. In their works, they revisited the ideas in von Neumann's impossibility result, specifically the assumptions he made.~\footnote{Indeed, some commentators sometimes oversimplify von Neumann's no-go theorem by claiming it was based on `silly assumptions'. This view was popularized by~\cite{mermin1993hidden}. When doing so, some aspects are generally overlooked: For example, as pointed out by~\cite{hermens2010quantummechanicsrealismintuitionism}, von Neumann's result \emph{predates} the mathematical axioms of probability theory as we know them today~\citep{kolmogorov1933grundbegriffe,kolmogorov1956foundations}. Notably, no notion of `definite values of observables' nor of probability space of hidden variables is used in von Neumann's proof. Moreover, assuming the possibility of valuations of incompatible observables may be viewed as a \emph{mathematically} natural assumption; it only becomes a clearly invalid \emph{physical assumption} once we establish, first, a notion of incompatibility among observables and, second, that these cannot be jointly observed. Both aspects were understood by only a few researchers at the time, as the mathematical structures we know today---including matrix and functional analysis, operator algebra, and probability and measure theory---were still in their infancy. Therefore, while one is free to believe von Neumann's assumption is indeed unjustified, it may be too simplistic to call it a `silly' argument considering the historical context. Perhaps, a more concerning point is that most researchers, with the notable exception of Grete~\cite{hermann1935dienaturphilosophischen}, accepted it uncritically.} In the work of Bell, he proposed an experimentally accessible scenario where a conjunction of assumptions made could be tested and ruled out. Their foundational work sparked a fruitful line of inquiry in physics: formulating experimentally meaningful no-go theorems to exclude certain assumptions in physical theories. Furthermore, the Bell and Kochen–Specker (\acrshort{ks}) no-go theorems have generated significant interest due to their practical relevance to quantum information advantage in information protocols~\citep{brunner2014bell,budroni2021kochenspeckerreview}.

\myindent This Chapter introduces the notion of Kochen--Specker noncontextuality. We comment on how it relates to Bell's notion of local causality~\citep{brunner2014bell}, and introduce a generalization of these two as proposed by~\cite{spekkens2005contextuality}. For simplicity, as was our choice in Chapter~\ref{chapter: quantum coherence}, our introduction and overview will be far from comprehensive, focusing mostly on the relevant assumptions, concepts, and definitions that we will use later in Chapter~\ref{chapter: from overlaps to noncontextuality}. Our discussion will be focused and technical and will emphasize simple conceptual aspects that are not common to the broad physics readership, but that are basic in the context of quantum foundations (e.g., we will motivate and discuss the relevance of concepts such as \emph{ontological models}, \emph{nonclassicality}, and \emph{scenarios}). For a comprehensive review of the topic, we refer to~\cite{budroni2021kochenspeckerreview}. For introductory texts on the graph approach to \acrshort{ks} noncontextuality, we refer to the book by~\cite{amaral2018graph}; for a formal algebraic introduction, we refer to~\cite[Chapter 6]{landsman2017foundations} and to~\cite[Chapter 2]{hermens2010quantummechanicsrealismintuitionism}. For a classic presentation, we refer to the book by~\cite{peres2006quantumtheory}. 

\myindent The structure of this chapter is as follows. We introduce the notion of Kochen--Specker noncontextuality in Sec.~\ref{sec: KS noncontextuality}. This notion is precise and well defined. But to write down the definition of it meaningfully we will need to first get used to jargon that is more common to the quantum foundations literature. We start by introducing the notion of compatibility scenarios and joint measurements in~\cref{sec: compatibility scenarios}. While we aim to keep our discussion concise and refer readers to more comprehensive resources where possible, we also present simple illustrative examples. These are intended to gently introduce concepts to readers who may be entirely unfamiliar with the field, helping them to ‘see through’ the challenging definitions and understand the motivations behind them. Readers familiar with the topic are invited to skip Examples~\ref{example: boxes},~\ref{example: ideal measurements} and~\ref{example: KS ontological models}. We then describe behaviors and the notion of realization of a behavior, emphasizing quantum realizations in Sec.~\ref{sec: behaviors and their realizations}. From there, we define Kochen--Specker noncontextuality in Section~\ref{sec: no disturbance and KS noncontextuality}, together with the notion of noncontextuality inequalities, some of which are presented in Sec.~\ref{sec: n-cycle KS noncontextuality inequalities}. We then comment on how graph theory is useful in this context in Sec.~\ref{sec: graph approaches}. We conclude this topic with a brief remark on the connection with Bell's notion of local causality in Sec.~\ref{sec: Bell and KS}. 

\myindent In Section~\ref{sec: Spekkens noncontextuality} of this Chapter, ~we then proceed to discuss a distinct notion of noncontextuality introduced by Robert Spekkens, known as \emph{generalized noncontextuality}~\citep{spekkens2005contextuality}. We briefly discuss the types of scenarios relevant to us, the so-called \emph{prepare-and-measure scenarios}, which are only a fairly small fraction of all the possible scenarios that can be used to study generalized noncontextuality. We start in Sec.~\ref{sec: operational theories and theory independent approaches} with a discussion on \emph{operational-probabilistic theories}, which bring a formal treatment of what we mean by `theory-independent' descriptions when talking about notions of nonclassicality. In the same section, we define prepare-and-measure scenarios, and the notion of \emph{operational equivalences} between procedures in such scenarios. Sec.~\ref{sec: LSSS scenarios} limits our discussion to the types of prepare-and-measure scenarios we need in this thesis and, finally, we conclude in Sec.~\ref{sec: Preparation noncontextuality} with the definition of preparation noncontextuality (the instance of generalized noncontextuality that will suffice for our purposes in this thesis) and with some final considerations.

\section{Kochen--Specker noncontextuality}\label{sec: KS noncontextuality}

\myindent In what follows, our presentation of Kochen--Specker (\acrshort{ks}) noncontextuality will be close to the one that can be found in~\cite{tiago2024convexity}. Intuitively, \acrshort{ks} contextuality can be viewed as the impossibility of modeling statistical results of measurements as revealing pre-existing objective properties of that system, which are independent of the actual set of possible joint measurements one chooses to make~\citep{budroni2021kochenspeckerreview,kochen1975problem}. As described, it is unclear how one can propose a test to exclude (i.e., falsify) such an abstract and generic class of models. A \emph{hard} aspect of understanding elementary aspects of noncontextuality theory lies in the difficulty of overcoming the cumbersome and unusual terminology that is used. Let us provide a simple illustrative example of the kind of experiments we are interested in, which shall clarify the terminology we employ. We encourage readers familiar with the framework to skip the examples in this Chapter.

\begin{example}[Joint measurements and (compatibility) scenarios]\label{example: boxes}
    Consider a preparation device in the production line of a factory that sends boxes containing shoes to two workers (see Fig.~\ref{fig: boxes_compatibility_scenario}). The workers are lined up in sequence. The boxes are sent to the workers according to some probability that the device has of preparing different shoes. There are four different measurements that the workers are assigned to make. They need to open the box and check: (1) whether there is a pair of shoes, (2) whether the shoes are black, (3) whether the shoes are for running, and finally (4) whether there are defects in the shoes. Each option represents a \emph{measurement} of the system, and each worker is assigned to make a single measurement, one after the other. Each measurement above is also \emph{dichotomic},~\footnote{A dichotomic measurement is also called a \emph{binary-outcome} measurement. Both are used  interchangeably.} meaning that each measurement has only two possible outcome labels: $x_1 = \{2,\neg 2\}$, $x_2 = \{\mathrm{black},\neg\mathrm{black}\}$, $x_3 = \{\mathrm{running},\neg\mathrm{running}\}$ and $x_4 = \{\mathrm{defects}, \neg\text{defects}\}$, where we have used $\neg$ to represent a negation.

    \begin{figure}
        \centering
        \includegraphics[width=0.94\linewidth]{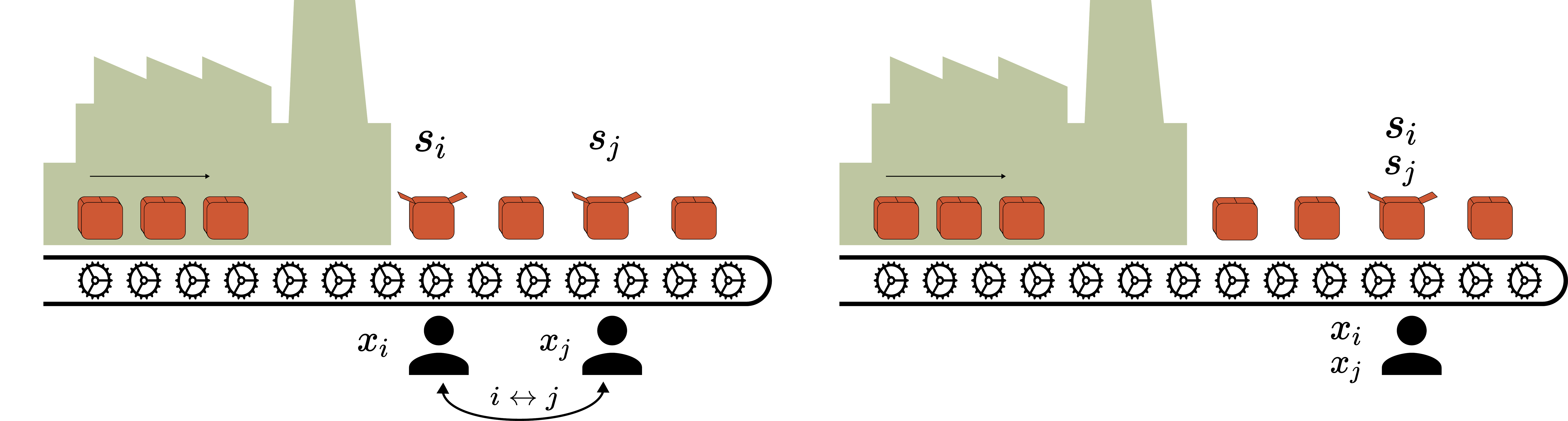}
        \caption{\textbf{Joint measurements, and the notion of a compatibility scenario.} A device prepares boxes sent to two workers, lined up in sequence one after the other. (left) Each worker is allowed to make one single dichotomic measurement, and for each box, they choose pairs of measurements $\{x_i,x_j\}$ for each of them to make. It does not matter the order in which the measurements are made. For each box, they mark their joint outcomes and at the end of the day, they end up with statistics $\{p(s_is_j|x_ix_j)\}$ related to their joint operations. The pairs $\{x_i,x_j\}$ are called contexts, and the operation is said to be a \emph{joint measurement} having joint outcomes $s_i$ and $s_j$, respectively. (right) Alternatively, a single worker can perform both measurements and report their joint statistics. In our~\cref{example: boxes}, it is simple to see that this is possible because, say, estimating $x_1$ (the number of shoes inside the box) and $x_3$ (whether the shoes are for running) can both be learned jointly when the worker opens the box and be used to gather the final statistics $\{p(s_is_j|x_ix_j)\}$. }
        \label{fig: boxes_compatibility_scenario}
    \end{figure}
    
    \myindent Any pair of measurements can be made \emph{jointly}. Suppose that in a given round they estimate $x_1$ and $x_3$. It is equivalent to letting the first measure $x_1$ and the second measure $x_3$, or vice-versa (left illustration from Fig.~\ref{fig: boxes_compatibility_scenario}). \emph{Alternatively}, one of the workers could step away (take a lunch break), and let a single worker measure both (right illustration from Fig.~\ref{fig: boxes_compatibility_scenario}). Measurements that can be jointly~\footnote{Some authors use the word `simultaneous' to mean `jointly'. In such cases, writing `compatible measurements can be performed simultaneously' is merely an abuse of terminology, and notions of simultaneity and time are not crucial here. Compatible measurements \emph{do not require} any form of clock synchronization and can be performed sequentially. Arguably, \emph{concurrently} is a better term than jointly or simultaneously. It means sequentially in any order, or simultaneously, or space-like separated (or rather, it means the equivalence class consisting of all those possible realizations).} made are called \emph{compatible}.~\footnote{Perhaps a technical remark is necessary here. Formally, our description accounts for joint \emph{ideal} measurements. We will later see what `ideal' means in this operational sense. This is relevant because for non-ideal measurements this notion of commutativity between measurements and being able to jointly perform them, as shown in Fig.~\ref{fig: boxes_compatibility_scenario} does not hold anymore. In quantum theory, this implies that there are noncommuting generalized (not projective) measurements that \emph{can} be jointly measured~\citep{heinosaari2016invitation}, and are therefore compatible. In summary, compatibility and commutativity are not equivalent notions, unless one takes measurements to be ideal.} At each round, only two of the four possible measurements are performed, and after varying the test with all possible pairs of measurements, the workers will have collected statistics of the form $p(s_is_j|x_ix_j)$ where $s_i$ is an outcome of $x_i$ and $s_j$ is an outcome of $x_j$. At the end of the day, their job consists of delivering the table $\{p(s_is_j|x_ix_j)\}_{i,j}$.  

    \myindent In this example, the measurements are always performed on the same physical state of the system, and in this case a \emph{mixed state} is characterized by a probabilistic ensemble of possible box types being prepared.

    \myindent This set-up is a specific realization of what we will later refer to as a \emph{scenario}, where the idea is to select specific operational aspects that can be followed by a broad class of different experimental situations. In the case just explained, the scenario is the prescription of preparing the same state of a physical system (in this case a mixed ensemble of possible boxes), later sent to be probed by pairs of joint measurements, with four dichotomic measurements in total. We can `label' this scenario as $(\mathcal{M},\mathcal{C},\mathcal{O}^{\mathcal{M}})$ where $\mathcal{M} = \{x_i\}_{i=1}^4$ denotes the measurements, elements in $\mathcal{C} \subseteq 2^{\mathcal{M}}$~\footnote{We denote $2^X$ the power set of a set $X$, which is the set of all possible subsets of $X$.} tell which measurements can be jointly performed, in this case $\{x_i,x_j\}$ for any $i,j$, and $\mathcal{O}^{\mathcal{M}} = \mathcal{O}^{x_1} \times \dots \times \mathcal{O}^{x_4}$ where $\mathcal{O}^{x_i}$ are the measurement outcomes of $x_i$. The idea of a scenario is more general than the idea of a specific implementation of it. It is easy to imagine infinitely many other experimental realizations of experiments that prepare a state and perform $4$ measurements, by collecting statistics of each pair of joint measurements allowed. 
\end{example}

\myindent As we progress we continue to refer back to~\cref{example: boxes}, as it is a simple example for illustrating some of the technical terms we use. In our treatment, \acrshort{ks} noncontextuality (or its failure) is a property of \emph{behaviors} defined with respect to \emph{compatibility scenarios} (a.~k.~a. measurement scenarios~\citep{barbosa2023closingBell}), both notions that we will shortly define. Before we start presenting the formal definitions for these terms, let us first sketch a `big picture' of our methods. While the mathematical details may be hard to follow, the following step-by-step methodology aims to make the main ideas accessible even to non-experts. 

\begin{enumerate}
    \item We first define a relevant class of scenarios of interest. This allows us to focus on a specific class of experiments and phenomena, clearly delimiting the scope of our test, and later the interpretations of our results.
    \item In a given scenario, we are interested in probabilistic data. This is what we call a \emph{behavior}. We later provide intuition for the notion of a \emph{realization} of a given behavior.
    \item We then bound all the possible behaviors in such a scenario, depending on some assumptions we make on the data, or on the kind of physical theory that could explain how this data is produced. This will induce a certain geometry on the possible set of behaviors that we review in Appendix~\ref{sec: convex polytopes}. In essence, this will lead to certain inequalities bounding the statistics depending on the assumptions we make on data arising in the scenario. 
    \item To conclude, given these inequalities, we try to investigate if different physical theories we can come up with (such as quantum theory) are capable of violating these inequalities or not. This is why the relevance of the notion of \emph{realizations of behaviors} is crucial, as different theories  `model' the behaviors differently, even if they may recover the same statistical predictions. Violations  imply that \emph{the conjunction of all assumptions} we have considered to build these inequalities cannot be valid. 
\end{enumerate}

\myindent Having an inequality-based methodology is not a crucial step but will be our focus in this thesis. For instance, the Kochen--Specker theorem was originally an inequality-free statement. This exact methodology (with or without inequality-based arguments) is employed in virtually all existing no-go results in quantum foundations~\citep{bell1964ontheEPRparadox,kochen1975problem,spekkens2005contextuality,pusey2012onthereality,colbeck2012freerandomness,frauchiger2018quantum,bong2020strong}. Having provided a high-level overview of the generic methodology, we now go back to our main presentation.

\subsection{Compatibility scenarios}\label{sec: compatibility scenarios}

\myindent In the \acrshort{ks} approach to noncontextuality we call \emph{contexts} the sets of measurements that can be jointly performed. This means that, if $\gamma \equiv \{x,y\}$ is a context containing the measurements $x,y$ that can yield outcomes $s_x,s_y$ there exists empirically accessible statistics described by $\{p(s_xs_y|xy)\}$.

\begin{definition}
    [Compatibility scenarios]\label{def: compatibility scenarios} A \emph{compatibility scenario} (also termed~\emph{measurement scenario}) is a triplet $\pmb{\Upsilon} := (\mathcal{M},\mathcal{C},\mathcal{O}^{\mathcal{M}})$, where $\mathcal{M}$ is a finite set of measurements, $\mathcal{C} \subseteq 2^{\mathcal{M}}$, and $\mathcal{O}^{\mathcal{M}} = \prod_{x \in \mathcal{M}}\mathcal{O}^x$, where $\mathcal{O}^x$ are the outcomes of $x \in \mathcal{M}$. We further impose that $\mathcal{C}$ is an \emph{antichain}, hence satisfying that for all $\gamma,\gamma' \in \mathcal{C}$, $\gamma \subseteq \gamma'$ implies $\gamma = \gamma'$.
\end{definition}

\myindent The elements of the antichain $\mathcal{C}$ are called \emph{maximal contexts}, where `maximal' here is with respect to the order defined by set inclusion.  The above definition and notation are common~\citep{amaral2018noncontextual,amaral2019resource, amaral2018graph}, but presentation and terminology for describing similar constructions can vary considerably~\citep{abramsky2019comonadic}. We will only consider \emph{finite scenarios} where each set $\mathcal{M},\mathcal{C},\mathcal{O}^{\mathcal{M}}$ is finite. For a treatment of scenarios allowing infinite sets of measurement outcomes $\mathcal{O}^x$, we refer to~\cite{barbosa2022continuous}. Each context $\gamma \in \mathcal{C}$ represents a set of measurements in $\mathcal{M}$ that can be jointly performed. We note that the notion of `measurement' expressed in~\cref{def: compatibility scenarios} is not necessarily the one described by the quantum formalism. It is a theory-independent formulation of any physical process yielding outcomes, which can be analyzed in what is now commonly referred to as an operational-probabilistic theory (\acrshort{opt})~\citep{dariano2017quantum,schmid2024structuretheorem}. We will also encounter this notion and provide an example of it later in Sec.~\ref{sec: Spekkens noncontextuality}. Similarly, the notion of compatibility (or joint measurability) expressed by the maximal contexts $\gamma \in \mathcal C$ can also be made entirely theory-independent, described solely by statistical results~\citep{amaral2018graph}. 

\myindent For each context $\gamma$, the set of all its possible joint outcomes is  $\mathcal{O}^\gamma = \prod_{x \in \gamma}\mathcal{O}^x$. When we jointly perform the measurements in $\gamma$, our output is encoded in a tuple $\pmb{s} \in \mathcal{O}^\gamma$.~\footnote{There is a one-to-one correspondence between a tuple $\pmb{a}=(a_1,\dots,a_n)\in O^n$ and a function $a: \{1,\dots,n\} \to O$ on some set $O$ defined by  $a(i)=a_i$. In this manuscript, we reserve bold symbols for tuples and plain symbols for their functional counterparts. Although these notations refer to the \emph{same} mathematical object, the distinction helps clarify whether we are emphasizing its interpretation as a tuple or as a function. }  Later we will consider scenarios where $\mathcal{O}^x = \mathcal{O}^{x'}$ for all $x,x' \in \mathcal{M}$, i.e. where all measurements have the same outcomes, and we will simply denote such sets of outcomes as $\mathcal{O}^x = \mathcal{O}^{x'} = \mathcal{O}$. Whenever this happens, we will also simply write $\pmb{\Upsilon} = (\mathcal{M},\mathcal{C},\mathcal{O})$ instead of $(\mathcal{M},\mathcal{C},\mathcal{O}^{\mathcal{M}})$ to simplify the notation.

\subsection{Behaviors and their realizations}\label{sec: behaviors and their realizations}

\myindent As mentioned before, \acrshort{ks} noncontextuality will be viewed as a property that may be satisfied (or not) by probabilistic data over compatibility scenarios. We now define what we mean by such probabilistic data:

\begin{definition}[Behaviors]
    Given a scenario $\pmb{\Upsilon} =(\mathcal{M},\mathcal{C},\mathcal{O}^{\mathcal{M}})$, a \emph{behavior} (a.~k.~a. an \emph{empirical model}) $B$ in this scenario is a family of probability distributions, one for each maximal context $\gamma \in \mathcal{C}$,

\begin{equation}\label{eq: behavior}
    B = \Bigg\{ p_\gamma : \mathcal{O}^\gamma \to [0,1]  \Bigg |  \sum_{\pmb{s} \in \mathcal{O}^\gamma} p_\gamma (\pmb{s}) = 1, \gamma \in \mathcal{C} \Bigg\}.
\end{equation}
\end{definition}

\myindent Experimentally, behaviors are the result of `running many times'~\footnote{This is true if one takes a \emph{frequentist} approach to the statistical data. This is the most common take when investigating experimental violations of Bell and Kochen--Specker noncontextuality inequalities in practice. Clearly, our formalism does not require one to specifically adhere to a frequentist or a Bayesian view.} a protocol that prepares a certain system and performs joint (or sequential, or concurrent) ideal measurements in $\gamma \in \mathcal{C}$ returning joint outcomes $\pmb{s} \in \mathcal{O}^\gamma$. Because of that, they are also called \emph{correlations}, \emph{probabilistic data-tables}, or simply \emph{data-tables}. In~\cref{example: boxes} we have described a scenario with four dichotomic measurements, with the contexts being all possible pairs of measurements. In this example, the behaviors are given by statistics $B$ that can be put in a data table (see Table~\ref{table: first example of a behavior}). Usually, this is how such behaviors are represented in small scenarios.

\begin{table}[h!]
\centering
\renewcommand{\arraystretch}{1.2} % Adjust row height if needed
\begin{tabular}{c|c|c|c|c}
\hline
 & 00 & 01 & 10 & 11 \\ 
\hline
\hline
$\{x_1,x_2\}$ & $1$ & $0$ & $0$ & $0$ \\ 
$\{x_1,x_3\}$ & $1$ & $0$ & $0$ & $0$ \\ 
$\{x_1,x_4\}$ & $1$ & $0$ & $0$ & $0$ \\ 
$\{x_2,x_3\}$ & $1$ & $0$ & $0$ & $0$ \\ 
$\{x_2,x_4\}$ & $1$ & $0$ & $0$ & $0$ \\ 
$\{x_3,x_4\}$ & $1$ & $0$ & $0$ & $0$ \\ 
\hline
\end{tabular}
\caption{\textbf{Table representation of behaviors.} This scenario has four measurements, all measurements are dichotomic and the maximal contexts are given by all sets $\{x_i,x_j\}$ of cardinality two. Each entry shows information  $p(s_is_j|x_ix_j)$ for a specific joint outcome $(s_i,s_j)$ of a specific joint measurement of a context $\{x_i,x_j\}$. If we suppose that this is a behavior from the test we have hypothesized in~\cref{example: boxes}, and if we assume that the $0$ label denotes the outcomes $1$ for $x_1$, black for $x_2$, running for $x_3$, and defects for $x_4$ the box above shows a situation that we call \emph{deterministic}, representing the fact that \emph{every box} that arrives has a single black running shoe that has some defect. }
\label{table: first example of a behavior}
\end{table}

\myindent Sometimes behaviors are also called \emph{boxes}, specifically if we take a theory-independent perspective and are not interested in which physical operations were responsible for generating the statistics of the behavior. Both terms are encountered in the literature. The intuition behind the term box is the following: Imagine the elements of $\mathcal{M}$ as buttons of the box, and, for each measurement $x$, we imagine the box having $|\mathcal{O}^x|$ output lights that inform us of the result of the measurements. The box has, therefore, certain rules as certain buttons cannot be jointly pressed (corresponding to certain measurements being incompatible). The information of allowed buttons to be jointly pressed is provided by maximal contexts $\gamma$.  In this thesis we will adhere to the following choice: we will reserve the term `box' to be used when referring to statistics that \emph{cannot} be obtained by quantum theory, such as in the case of the well-known boxes discovered by~\cite{popescu1994nonlocality}. To be more precise, we will only refer to `boxes' as the particular behaviors that \emph{satisfy} the no-disturbance condition (see Def.~\ref{def: no-disturbance}), but that have no quantum realization (see Def.~\ref{def: quantum realizations of behaviors in compatibility scenarios}). 

\myindent We have mentioned above the notion of \emph{ideal measurements} that is crucial for \acrshort{ks} noncontextuality, and that we now illustrate with an example.

\begin{example}[Ideal measurements]\label{example: ideal measurements}
An \emph{ideal measurement} is one that: (i)
yields the same outcome when it is repeated on the same
system and (ii) does not disturb other compatible measurements.~\footnote{For an example of a toy model reproducing the statistics of contextual behaviors by violating the assumption of ideal measurements, we refer to~\citep{tezzin2023violating}. Some authors have also considered more requirements for a measurement to be called `ideal'~\citep{xu2019necessary} that will not be of relevance to what follows. Note also that (ii) implies (i) because any measurement $x$ shares a context with itself. We are separating them as different assumptions for clarity of the presentation, as was done in~\cite[pg. 37]{budroni2021kochenspeckerreview}.} We provide later some intuition on the notion of `disturbance' relevant here. Let us go back to our description from~\cref{example: boxes} to illustrate this idea. To probe if their measurements are ideal, the workers can do the following. Both workers estimate the same measurement for some boxes and see if their outcomes coincide, i.e., the first worker makes a measurement $x_1$ that tests the number of shoes inside the box (two or not) and the second worker does the same, in sequence. They proceed in this way with all measurements. They may also like to make more than two repeated measurements in sequence, to increase their confidence that the outcomes indeed do not change when they make the same test over the same system over and over again. 

\myindent They then proceed to test if making measurements that are in the same context the results when probing one do not disturb the other. To test this, they can take compatible measurements (say $x_1$ and $x_2$ that test for the number of shoes and their color, respectively), perform them in different orders (say for $x_1$ then $x_2$, and vice-versa), and see if the outcomes of one influence or disturb the outcomes of the other. 

\myindent Classically, this is trivial and very intuitive. All these measurements performed by the workers will satisfy these properties. Yet, when performing experiments where systems and measurements can be described by quantum mechanics, the validity of the assumption of measurements being ideal becomes less obviously true and must be probed experimentally.
\end{example}

\begin{figure}
    \centering
    \includegraphics[width=1\linewidth]{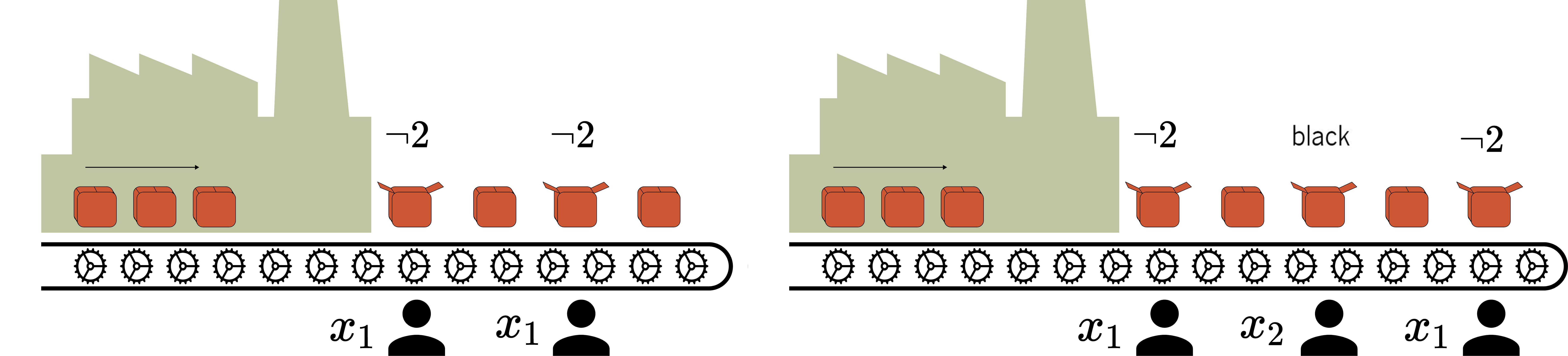}
    \caption{\textbf{Testing the assumption of ideal measurements.} (left) If the measurements performed are ideal, they always yield the same outcome when repeated over the same system. For instance, if the first worker performs the measurement $x_1$ and finds out that instead of a pair of shoes, there is some other number of shoes inside the box, say, one, when the second worker makes the measurement $x_1$ the same outcome must be observed. (right) If the measurements are ideal, they cannot disturb the outcomes of measurements in the same context. Hence if the first worker makes a measurement $x_1$ and obtains $\neg 2$, the second worker makes a measurement $x_2$, testing if the shoes are black or not, and obtains $\mathrm{black}$, because $\{x_1,x_2\}$ is a context if a third measurement of $x_1$ is made afterward, the outcome of it must always be $\neg 2$.}
    \label{fig: ideal measurements}
\end{figure}

\myindent The notion above is crucial as the assumption of \acrshort{ks} noncontextuality can only be tested if one probes the assumption that the measurements are ideal. Testing this property is a \emph{necessary requirement} for any experiment aiming to test \acrshort{ks} contextuality of quantum theory.~\footnote{Tests of \acrshort{ks} noncontextuality are extremely difficult to perform, and require various subtle and strict experimental requirements. Some experimental subtleties have been pointed out by~\cite{amselem2013comment}, and~\cite{budroni2021kochenspeckerreview}. For two recent attempted tests of loophole-free \acrshort{ks} contextuality experiments we refer to~\cite{wang2022significant} and~\cite{hu2023self}.} In quantum mechanics, ideal measurements are described by projection-valued measurements (\acrshort{pvm}). These are also sometimes called `sharp' measurements. 

\myindent Let us remark on the notion of \emph{realizations} of a given behavior. Take again~\cref{table: first example of a behavior} as an example of a behavior. As we have pointed out, if we implement the test imagined in~\cref{example: boxes} with the workers receiving shoes, after making all measurements, the deterministic behavior shown in~\cref{table: first example of a behavior} is a conceivable result. However, \emph{the same behavior} can also be the output of an entirely different experimental setup, which satisfies the constraints of the scenario. The systems that are prepared in some state may be any other physical system (cars, graphene, light from the cosmos, etc.) and the measurements performed can be any other set of measurements probing these systems (speed, position, color, frequency, temperature, etc.). The only requirement is that they satisfy the specific constraints of the scenario. As a single behavior is conceivable in any such test, we say that these are \emph{experimental realizations}, or simply \emph{realizations}, of the behavior $B$. There may be infinitely many, for each nondisturbing (see Def.~\ref{def: no-disturbance}) behavior, and \emph{none} as well (at least with our current understanding of physical theories), which is perhaps a less evident fact~\citep{popescu1994nonlocality}.

\myindent With some degree of abstraction, instead of considering a specific experimental realization we may consider any realization in a \emph{class}. For instance, given a probability table such as the one in~\cref{table: first example of a behavior}, with respect to a specific scenario, we may ask: Are there quantum states $\rho$ and (sharp) measurements $A_i = \{\Pi_{s_i|i}\}_{s_i}$, satisfying the compatibility requirements of the scenario, such that by jointly measuring $\rho$ with pairs of dichotomic projective quantum measurements $x,y \in \{A_1,A_2,A_3,A_4\}$ we can recover the same~\cref{table: first example of a behavior} using the Born rule? If this is possible, we will say the behavior is \emph{quantum realizable}. Note that in doing so we are not focusing on a specific experimental realization, but on the class of experimental realizations capable of preparing quantum states $\rho$ and of performing the dichotomic measurements mentioned (e.g. when we denote a state $\rho=\vert 0\rangle \langle 0 \vert $ as a density matrix this is just a mathematical `label' representing an equivalence class of distinct experimental implementations). This is an instance of a class of problems known as \emph{quantum realizability problems}~\citep{fraser2023estimationtheoreticapproachquantum}. In Chapter~\ref{chapter: relational coherence} we will see another instance of such problems.

\begin{definition}[Quantum realizations of behaviors in compatibility scenarios]\label{def: quantum realizations of behaviors in compatibility scenarios}
    Let $\pmb{\Upsilon} = (\mathcal{M},\mathcal{C},\mathcal{O}^{\mathcal{M}})$ be a compatibility scenario and $B = \{p_\gamma(\pmb{s})\}$ a behavior in that scenario. We say that $B$ is \emph{quantum realizable} in a Hilbert space $\mathcal{H}$ if there exists a quantum state $\rho \in \mathcal{D}(\mathcal{H})$, and a set of observables $\{A_i\}_i \in \mathcal{B}_{sa}(\mathcal{H})$, satisfying:
    \begin{enumerate}
        \item To each element $x_i \in \mathcal{M}$ we associate an observable $A_i$.
        \item For every context $\gamma \in \mathcal{C}$, if $x_i,x_j \in \gamma$ then $[A_i,A_j] = 0$.
        \item Let $A_i = \{\Pi_{s_i|i}\}_i$ its associated projection-valued measure (\acrshort{pvm}), then we have that, for every context $\gamma \in \mathcal{C}$ and every possible joint outcome $\pmb{s} \in \mathcal{O}^\gamma$, the statistics of the behavior follows from the Born rule
        \begin{equation*}
            p_\gamma(\pmb{s}) = \text{Tr}\left(\rho \prod_{x_i \in \gamma}\Pi_{s_i|i}\right),
        \end{equation*}
        where $s_i$ is our notation for the outcomes of the \acrshort{pvm} $\{\Pi_{s_i|i}\}_i$.
    \end{enumerate}
\end{definition}

\myindent Above, we follow the standard notation of representing measurement procedures by sets of projection operators. An observable (i.e. self-adjoint operators) $A_i$ is associated with a set of projectors $\{\Pi_{s_i|i}\}_{s_i}$ via the spectral theorem; these projectors formally define a \acrshort{pvm}. By \acrshort{pvm} we mean, as usually, a family of orthogonal projections that sum to the identity. This association is not one-to-one: the same \acrshort{pvm} can be associated with different observables by assigning different eigenvalues to the projectors. Nevertheless, each self-adjoint operator $A_i$ determines a unique \acrshort{pvm}, so the choice of \acrshort{pvm} for a given observable is unambiguous. 

\subsection{No-disturbance, global sections and KS noncontextuality}\label{sec: no disturbance and KS noncontextuality}

\myindent Behaviors may or may not satisfy what we call the \emph{no-disturbance condition}. Given two contexts
$\gamma$ and $\gamma '$, no-disturbance implies that the marginals for their intersection are well defined, and agree. If we have, for example, $\gamma = \{x, y\}$ and $\gamma' = \{y, z\}$, the no-disturbance condition implies:

\begin{equation*}
    \sum_{a} p_{\{x,y\}} (a,b) = \sum_{c} p_{\{y,z\}} (b,c).
\end{equation*}
\begin{definition}[No-disturbance set of behaviors]\label{def: no-disturbance}
 The \emph{no-disturbance set} $\mathrm{ND}(\pmb{\Upsilon})$ is the set of behaviors that satisfy the no-disturbance condition for any intersection of contexts in the scenario $\pmb{\Upsilon} = (\mathcal{M}, \mathcal{C}, \mathcal{O}^{\mathcal{M}})$. This is the set defined by all behaviors $B = \{p_\gamma(\pmb{s})\}$ in $\pmb{\Upsilon}$ such that for every pair $\gamma,\gamma' \in \mathcal{C}$ for which $\gamma \cap \gamma' \neq \emptyset$ we have that
\begin{equation}
    p_\gamma|_{\gamma \cap \gamma'}(\pmb{s}) = \sum_{\pmb{t} \in \mathcal{O}^\gamma: \pmb{t}|_{\gamma \cap \gamma'}=\pmb{s}}p_\gamma(\pmb{t})   = \sum_{\pmb{t} \in \mathcal{O}^{\gamma'}: \pmb{t}|_{\gamma \cap \gamma'}=\pmb{s}}p_{\gamma'}(\pmb{t}) = p_{\gamma'}|_{\gamma \cap \gamma'}(\pmb{s}).
\end{equation}
\end{definition}

\myindent Therefore, in more general scenarios it is not enough to look at the marginals of each measurement to guarantee that a behavior $B$ satisfy the no-disturbance condition.  One needs to look at the marginals of \emph{all} the intersections between different contexts in $\mathcal{C}$. 

\begin{example}[No-disturbance]\label{example: disturbance and no-disturbance}
For example, take Table~\ref{table: first example of a behavior}. In this case, we have a table that \emph{satisfies} the no-disturbance condition. If we consider, as an example, the two contexts $\{x_1,x_2\}$ and $\{x_2,x_4\}$ we have that
    \begin{align*}
        \sum_a p_{\{x_1,x_2\}}(ab) =
        \sum_c p_{\{x_2,x_4\}}(bc) 
    \end{align*}
     for any possible outcome $b \in \{0,1\}$ of $x_2$. If $b=0$ we have that $$p_{\{x_1,x_2\}}(00)+p_{\{x_1,x_2\}}(10) = 1 + 0 = 1 = 1 + 0 = 
        p_{\{x_2,x_4\}}(00)+p_{\{x_2,x_4\}}(01) $$ while if $b=1$ we have that $$p_{\{x_1,x_2\}}(01)+p_{\{x_1,x_2\}}(11) = 0 + 0 = 0 = 0 + 0 = 
        p_{\{x_2,x_4\}}(10)+p_{\{x_2,x_4\}}(11). $$ 
        Recognizing that this holds for all intersections between the contexts of this scenario, this proves that the behavior shown in Tab.~\ref{table: first example of a behavior} satisfies the no-disturbance condition. 
\end{example}
  
\myindent There is a simple way of defining \acrshort{ks} noncontextuality that uses only a statistical characterization, and avoids any discourse on hidden-variable models. We present this simple definition and then show later how it relates to the usual hidden-variable model construction. Intuitively, the idea is to acknowledge that every behavior has a canonical  \acrshort{ks} model. Specifically, if a behavior admits a \acrshort{ks} noncontextual hidden variable model then it must admit one where the hidden variable space (a.~k.~a.~the ontic state space) is $\mathcal{O}^{\mathcal{M}}$~\citep{abramsky2011sheaf}.  

\myindent Suppose that for a given behavior $B$ it is possible to assign a single probability distribution to the whole set $\mathcal{O}^{\mathcal{M}}$, which has marginals in each maximal context consistent with the behavior $B$~\citep{abramsky2011sheaf}. We call such probability distributions $p_{\mathcal{M}} : \mathcal{O}^{\mathcal{M}} \to [0,1]$ a \textit{global section} for the behavior $B$, which must satisfy
\begin{equation}
    p_{\mathcal{M}}\vert_{\gamma}(\pmb{s}) := \sum_{\pmb{t} \in \mathcal{O}^{\mathcal{M}}:\pmb{t}|_{\gamma} = \pmb{s}}p_{\mathcal{M}}(\pmb{t}) = p_\gamma(\pmb{s}) 
\end{equation}
for all contexts $\gamma \in \mathcal{C}$ of $\pmb{\Upsilon}$ and all $\pmb{s}\in \mathcal{O}^\gamma$. Note that $p_\gamma(\pmb{s})$ is the empirically accessible statistics, given by the behavior, while we may not know $p_{\mathcal{M}}$. We are merely interested in understanding if such distribution exists or not, and not in directly accessing it experimentally. If this is possible, we say that the behavior $B=\{\{p_\gamma(\pmb{s})\}_{\pmb{s}\in \mathcal{O}^\gamma}\}_{\gamma \in \mathcal{C}}$ is \acrshort{ks} noncontextual.

\begin{definition}[\acrshort{ks} noncontextual behaviors]
 Let $\pmb{\Upsilon}$ be any compatibility scenario. We say that a behavior $B$ in this scenario is \acrshort{ks} noncontextual if there exists a global section reproducing the behavior $B$ as marginals in the maximal contexts. The \emph{noncontextual} set $\mathrm{NC}(\pmb{\Upsilon})$ is the set of \acrshort{ks} noncontextual behaviors.
\end{definition}

\myindent Let us now discuss how this modern perspective relates to the usual description in terms of hidden-variable models. From the Fine-Abramsky-Brandenburger (\acrshort{fab}) Theorem~\citep{abramsky2011sheaf,fine1982hiddenvariablesPRL}, \acrshort{ks} noncontextual behaviors can be equivalently written as

\begin{equation}\label{eq: noncontextual factorizable}
    p_\gamma (\pmb{s}) = \sum_\lambda \mu(\lambda) \prod_{x \in \gamma} p_{x} (s_x|\lambda),
\end{equation}
where $\lambda \in \Lambda$ are any set of variables, and $\mu(\lambda)$ is a probability distribution over these variables, i.e., $\mu$ satisfies $\sum_\lambda \mu(\lambda)=1$ and $0 \leq \mu(\lambda) \leq 1, \forall \lambda$. Also, $p_{x}(s_x|\lambda)$ are so-called \emph{response functions}, satisfying that for any given $\lambda$ the mapping $p_{x}(\cdot|\lambda)$ yields a valid probability distribution over $\mathcal{O}^{x}$, for any $x\in \gamma$ and also any $\gamma \in \mathcal{C}$. 

\begin{theorem}[\acrshort{fab} theorem]\label{theorem:FAB theorem}
    Let $B$ be a behavior in a compatibility scenario $\pmb{\Upsilon}=(\mathcal{M},\mathcal{C},\mathcal{O}^{\mathcal{M}})$. Then, the following are equivalent:
    \begin{enumerate}
        \item There exists a global section $p_{\mathcal{M}}: \mathcal{O}^{\mathcal{M}} \to [0,1]$ reproducing $B$ as marginals in the maximal contexts of $\pmb{\Upsilon}$. 
        \item There exists a set $\Lambda$, a probability distribution $\mu$ on $\Lambda$, and response functions $p_x(s_x\vert \cdot):\Lambda \to [0,1]$ for every $x \in \mathcal{M}$ such that $B$ is given by Eq.~\eqref{eq: noncontextual factorizable}.
        \item There exists a set $\Lambda'$, a probability distribution $\mu'$ on $\Lambda'$, and \emph{deterministic} response functions $p'_x(s_x\vert \cdot):\Lambda' \to \{0,1\}$ for every $x \in \mathcal{M}$ such that $B$ is given by Eq.~\eqref{eq: noncontextual factorizable}.
    \end{enumerate}
\end{theorem}

The above description (given by Eq.~\eqref{eq: noncontextual factorizable}) is exactly what we have referred to before as a noncontextual hidden-variable model. More specifically, the above is the explicit description of the form of any  \emph{ \acrshort{ks} noncontextual hidden-variable model} for the behavior. When a behavior $B \equiv \{\{p_\gamma(\pmb{s})\}_{\pmb{s}}\}_\gamma$ has this precise form, it is said to be \emph{factorizable}~\citep{abramsky2011sheaf,barbosa2022continuous}. 

\myindent Note also that the difference between points 2 and 3 above are merely that in point 3 the response functions are assumed to be \emph{deterministic}. 

\myindent All noncontextual behaviors satisfy the no-disturbance condition. Different notions of noncontextuality have been proposed for behaviors that do not respect no-disturbance~\citep{kujala2015necessary,amaral2018necessary,amaral2019characterizing,tezzin2020contextualitybydefault}. Any such approach will have certain drawbacks, specified by the impossibility theorem proved by~\cite{tezzin2022impossibility}.

\begin{example}[An illustrative example of \acrshort{ks} noncontextual models]~\label{example: KS ontological models}
    Let us go back to our experimental situation from~\cref{example: boxes}. We will see how every possible data table from this experiment must have a \acrshort{ks} noncontextual model, in the usual view of hidden-variable models. This will also help to build an intuitive understanding of why it is natural to assume that such models can be called \emph{classical}. 

    \myindent If we recall that test, we've assumed that there is a machine delivering boxes to the workers. Let us denote by $\Lambda$ the set of all possible `values' of properties defining the `state' of the shoe boxes being prepared, and denote each element of this set as $\lambda \in \Lambda$. Note that $\lambda$ \emph{is a collection of such values} defining entirely all properties of the shoes inside the boxes that are to be probed, including those that will be measured by the workers. For instance, we have $\lambda = (\lambda_2,\lambda_{\mathrm{black}},\lambda_{\mathrm{running}},\lambda_{\mathrm{burned}},\dots)$ that are variables signaling the fact that the prepared box has two black shoes, for running, etc. This always happens for such a (classical) experiment. We can say in this description that the machine is preparing boxes that have a state completely characterized by some choice $\lambda \in \Lambda$. If one adheres to the view that these states represent \emph{aspects of the reality} of our system, these states are called \emph{ontic}. See Figure~\ref{fig: boxes_shoes_ontological_model}.

    \myindent Naturally, the machine produces boxes according to some probability distribution $\mu(\lambda)$, which is independent of the measurements and outcomes performed by the workers. The shoes are already there and have already been selected, before the workers have decided which measurements to perform and before they obtain their outcomes. 

    \myindent Once a box is prepared, each measurement performed by the workers merely reveal properties that have already been completely fixed by the specific preparation of the ontic state $\lambda$. Therefore, once the shoes are put in the box, which we are modeling as a specific choice of $\lambda \in \Lambda$, there is a deterministic probability distribution $p(s_x|x,\lambda) \equiv p_{x}(s_x|\lambda)$ of the worker measuring $x$ and obtaining $s_x$. For example, if the worker decides to make the test $x_1$ testing number of shoes, given that a state $\lambda = (\lambda_1,\dots)$ was chosen the probability becomes $p(1|x_1,\lambda) = 1$ and $p(2|x_1,\lambda) = 0$ since the variable $\lambda$ completely determines if there is one, two, or more shoes inside the box.

    \begin{figure}
        \centering
        \includegraphics[width=0.75\linewidth]{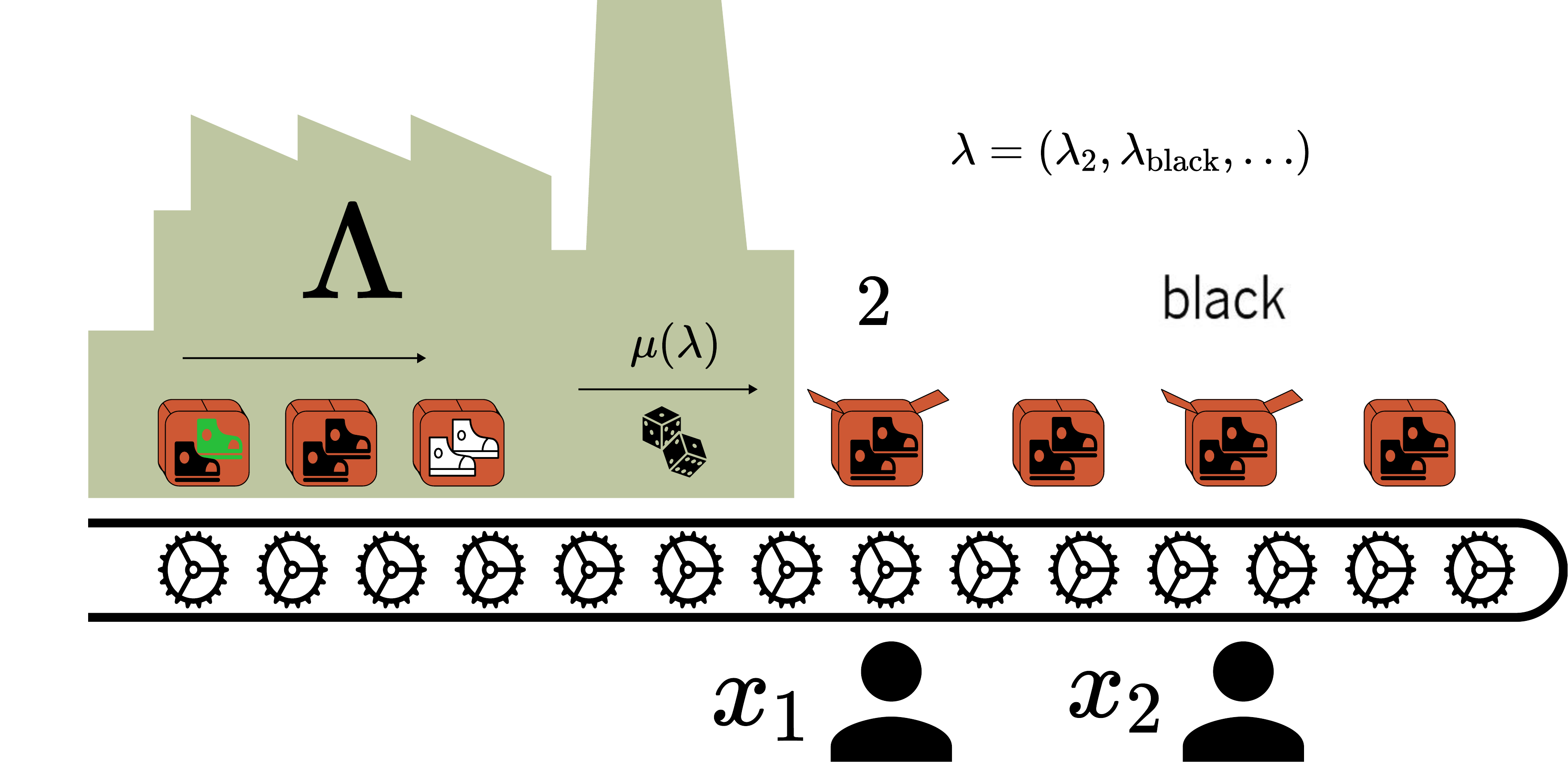}
        \caption{\textbf{Kochen--Specker noncontextual ontological model.} If there is a \acrshort{ks} noncontextual model for a given behavior $B$ in a scenario, the model  provides a `story' for it similarly as above. Some initial source of randomness $\mu(\lambda)$ prepares the system in some unknown state $\lambda$ that  completely characterizes the state of affairs of the physical system. Once this is done, measurements in the same context can be interpreted as merely revealing pre-determined properties that are deterministically fixed by $\lambda$ and by the chosen measurement. The probability of observing these outcomes is entirely independent of which context they are being performed in. In the figure, as soon as the shoes are put inside the box, measurements merely reveal the properties of the shoes, which are fixed. The case shows that there are two shoes and their color is black.}
        \label{fig: boxes_shoes_ontological_model}
    \end{figure}
    
    \myindent Note, moreover, that these outcomes are independent of which contexts one is measuring, i.e., suppose that we have the two contexts $\gamma^{(1)} = \{x_1,x_2\}$ and $\gamma^{(2)} = \{x_1,x_3\}$. Then, we have that the probability of observing an outcome from a measurement of $x_1$, given some $\lambda$, is independent of which other measurement is being performed, either $x_1$, or $x_2$, i.e., $$p(s_1|(x_1,\gamma^{(1)}),\lambda) = p(s_1|(x_1,\gamma^{(2)}),\lambda) \equiv p(s_1|x_1,\lambda).$$ The same is true for every possible context, such that $x_1$ is an element of. This is why the model is said to be \emph{noncontextual}. 

    \myindent Putting together all these aspects, we see that we can model any behavior $\{p(s_is_j|x_ix_j)\}$ in this scenario as 
    \begin{equation*}
        p(s_is_j|x_ix_j) = \sum_{\lambda\in \Lambda} \mu(\lambda) p(s_i|x_i,\lambda)p(s_j|x_j,\lambda)
    \end{equation*}
    where $p(s_i|x_i,\lambda) \in \{0,1\}$, $\sum_{s_i} p(s_i|x_i,\lambda) = 1$ for any measurement $x_i$ and any ontic state $\lambda$, and also $0 \leq \mu(\lambda) \leq 1$ with $\sum_\lambda \mu(\lambda) = 1$. 
    
    \myindent The probability distribution $\mu$ accounts for some source of (uncontrollable) randomness in the states of the systems being prepared. The variables $\lambda$ completely determine the state of the system. Measurements can be interpreted as simply revealing `pre-existing' properties of the state of the system (described by the variable $\lambda$). The statistical results of each measurement are independent of the context in which they are being performed. This simple scheme describes what is known as a Kochen--Specker noncontextual ontological model for the behavior, also known as a noncontextual hidden-variable model.   
\end{example}

\myindent The set of all behaviors in a compatibility scenario $\pmb{\Upsilon}$ that have a \acrshort{ks} noncontextual ontological model, which we have denoted as $\mathrm{NC}(\pmb{\Upsilon})$, is a  \emph{subset} of the set of all behaviors that satisfy the no-disturbance condition, i.e., 
\begin{equation}
    \mathrm{NC}(\pmb{\Upsilon}) \subseteq \mathrm{ND}(\pmb{\Upsilon}).
\end{equation}
Moreover, it is well known that both these sets are \emph{convex polytopes} (see Appendix~\ref{sec: convex polytopes}) and can hence be described both as the convex hull of a set of finitely many behaviors and as a set of finitely many intersections of closed half-spaces, described by facet-defining inequalities. The facet-defining inequalities of $\mathrm{NC}(\pmb{\Upsilon})$ are known as \acrshort{ks} \emph{noncontextuality inequalities}. 

\myindent Behaviors violating \acrshort{ks} noncontextuality inequalities cannot be described by any \acrshort{ks} noncontextual model. More interestingly, there are behaviors that both violate such inequalities \emph{and} are quantum realizable, implying that there are statistical results that can (in principle) be implemented in the laboratory that will falsify the possibility of Nature being described by one such model. If we denote the set of quantum realizable behaviors in a compatibility scenario as $\mathcal{Q}(\pmb{\Upsilon})$, the following sequence of set relations is known to hold:
\begin{equation}
    \mathrm{NC}(\pmb{\Upsilon}) \subseteq \mathcal{Q}(\pmb{\Upsilon}) \subseteq \mathrm{ND}(\pmb{\Upsilon}).
\end{equation}
Scenarios $\pmb{\Upsilon}$ where $\mathcal{Q}(\pmb{\Upsilon}) \neq \mathrm{NC}(\pmb{\Upsilon})$ are said to be \emph{contextuality-witnessing}, and have been completely characterized by~\cite{xu2019necessary}, using results from~\cite{vorobyev1962consistent}.~\footnote{Strictly speaking, this only holds for scenarios $\pmb{\Upsilon}$ that are captured by \emph{compatibility graphs}, which are the clique completions of their $2$-skeleton of \emph{compatibility hypergraphs}~\citep{amaral2018graph}, and suffice to study Kochen--Specker \emph{quantum} contextuality.}

\myindent For generic scenarios, completely characterizing $\mathrm{NC}(\pmb{\Upsilon})$ is a hard problem. Nevertheless, some noncontextuality inequalities are known, and we will now provide a brief overview of a family of known ones, and refer to~\citep{budroni2021kochenspeckerreview} for a comprehensive presentation of the state of the art. 

\subsection{The $n$-cycle noncontextuality inequalities}~\label{sec: n-cycle KS noncontextuality inequalities}

\myindent The only infinite family of compatibility scenarios (nonequivalent to some Bell scenarios) that have been completely characterized is the family of $n$-cycle scenarios, as shown  by~\cite{araujo2013all}, which we will denote as $\pmb{\Upsilon}_n$. We will soon see why these scenarios are called in such a way. In such scenarios, one has $n$ dichotomic measurements  $\mathcal{M} = \{x_i\}_{i=1}^n$ and all maximal contexts are given by $\{x_i,x_{i+1}\}$, where here summation is taken to be modulo $n$. All vertices and facet-defining inequalities of the polytope $\mathrm{ND}(\pmb{\Upsilon}_n)$ are known~\citep{araujo2013all}.~\footnote{More generally, it is easy to characterize the H-representation (i.e. the complete list of facet-defining inequalities) of $\mathrm{ND}(\pmb{\Upsilon})$ for any scenario $\pmb{\Upsilon}$. The difficulty lies in finding the V-representation. Contrastingly, it is easy to characterize the V-representation of $\mathrm{NC}(\pmb{\Upsilon})$ while, in this case, the difficulty lies in finding the H-representation. We refer to~\cref{sec: convex polytopes} for an introduction to terminology related to convex polytope theory.} Moreover, the set of all facet-defining inequalities of $\mathrm{NC}(\pmb{\Upsilon}_n)$ is also known, and given by 

\begin{equation}\label{eq: noncontextuality inequalities}
    I_{\pmb{a}}^{(n)}(\pmb{B}) = \sum^{n-1}_{i=0} a_i \langle x_i x_{i+1} \rangle \leq n-2,
\end{equation}
with each value $\pmb{a} = (a_0,\dots,a_{n-1}) \in \{-1,+1\}^n$ labeling a possible facet-defining inequality, being associated with a particular choice of values such that the number of terms $a_i = -1$ is odd. Above, we are mapping behaviors $B$ from $\pmb{\Upsilon}_n$ to two-point correlation functions via, letting $p_{\{x_i,x_j\}}(s_is_j) \equiv p(s_is_j|x_ix_j)$, 
\begin{align}\label{eq: two-point correlations}
\langle x_i x_{i+1}\rangle &= +p(00|x_ix_{i+1})+p(11|x_ix_{i+1})-p(10|x_ix_{i+1})-p(01|x_ix_{i+1})\\&=p(x_i=x_{i+1})-p(x_i\neq x_{i+1})\nonumber.
\end{align}
Each label $\pmb{a}$ is therefore associated with a different facet of $\mathrm{NC}(\pmb{\Upsilon}_n)$, for each fixed choice $n$. The sets of points $\left\{\langle x_i x_j\rangle \right\}_{i,j}$ are called the \emph{sets of correlations}.  For such scenarios, there is a one-to-one correspondence between the set of behaviors and the set of correlations~\citep{araujo2013all}.

\myindent Various properties of the polytopes $\mathrm{NC}(\pmb{\Upsilon}_n)$ are known. For every contextual behavior $B$ there is a unique $\pmb{a}$ for which $I_{\pmb{a}}^{(n)} (\pmb{B}) > n-2$.  \cite{choudhary2024lifting} investigated liftings of these inequalities to other compatibility scenarios. The values for the maximal quantum violations of the inequalities~\eqref{eq: noncontextuality inequalities} are known, and given by:

$$
    I_{Q}^{\mathrm{max}} = \begin{cases} \frac{3n \cos{(\frac{\pi}{n})} -n}{1 + \cos{(\frac{\pi}{n})}}, & \mbox{for odd } n, \\ n \cos{(\frac{\pi}{n})}, & \mbox{for even } n. \end{cases}
$$
The values for $n$ odd were shown by~\cite{araujo2013all}, and the case of $n$ even was shown by~\cite{wehner2006tsirelson} (see also~\cite{bharti2022graph}). Behaviors for which the value of $I_{\pmb{a}}^{(n)}(B)$ is larger than $I_{Q}^{\mathrm{max}}$ are known as \emph{post-quantum behaviors}, or as we have mentioned before, \emph{boxes}. 

\myindent The first instances of the above family of noncontextuality inequalities are of special relevance. For example, when $n=3$ we get the following inequality:
\begin{equation*}
    \langle x_0x_1\rangle + \langle x_1x_2 \rangle - \langle x_0x_2\rangle \leq 1, 
\end{equation*}
which \emph{does not have} any quantum violation~\citep{vorobyev1962consistent,xu2019necessary}. If we consider $n=4$, and denote $x_0 \equiv A_0, x_1 \equiv B_0, x_2 \equiv A_1, x_3 \equiv B_1$ we end up with the following inequality
\begin{equation*}
    \langle x_0x_1\rangle + \langle x_1 x_2 \rangle + \langle x_2 x_3 \rangle - \langle x_3 x_0 \rangle \leq 2 \iff
\end{equation*}
\begin{equation}\label{eq: chsh inequality}
    \langle A_0B_0\rangle + \langle B_0 A_1 \rangle + \langle A_1 B_1 \rangle - \langle A_0 B_1 \rangle \leq 2,
\end{equation}
which is the celebrated Bell inequality obtained by Clauser, Horn, Shimony, and Holt (\acrshort{chsh})~\citep{clauser1969proposed} that can also be re-written in the following form~\citep{collins2002bell}
\begin{equation}\label{eq: chsh collins et al}
    p^{A_0B_0}_=+p^{A_0B_1}_= +p^{A_1B_1}_=- p^{A_1B_0}_=  \leq 2,
\end{equation}
where $p_=^{A_iB_j} \equiv p(A_i = B_j) = p(00|A_iB_j)+p(11|A_iB_j)$ as we are considering dichotomic measurements. Finally, if we consider $n=5$ we obtain the celebrated Kochen--Specker noncontextuality inequality found by Klyachko, Can, Binicio\u{g}lu,
 and Shumovsky (\acrshort{kcbs})~\citep{klyachko2008simple} (see also~\citep{klyachko2002coherent}), 
\begin{equation}\label{eq: KCBS correlations}
    \langle x_0x_1\rangle + \langle x_1x_2 \rangle + \langle x_2x_3 \rangle + \langle x_3 x_4 \rangle + \langle x_0 x_4\rangle \geq -3.
\end{equation}

\subsection{The exclusivity graph approach}\label{sec: graph approaches}

\myindent In our description of \acrshort{ks}-noncontextuality, the key elements were the sets of measurements and the compatibility relations between such sets of measurements described by the contexts considered in a given compatibility scenario. A natural framework for representing such a structure is provided by graph theory~\citep{amaral2018graph}. For the reader unfamiliar with basic notions of graph theory, we provide introductory material to the topic in Appendix~\ref{sec: graph theory}. For example, consider a (trivial) compatibility scenario described by two measurements $\mathcal{M} = \{x,y\}$, one context $\gamma = \mathcal{M}$, and such that each measurement is dichotomic, i.e. $\mathcal{O}^x = \mathcal{O}^y = \{0,1\}$.  We can represent this information in a graph. For example, in the \emph{compatibility hypergraph}, we consider \emph{nodes} to represent measurements and \emph{edges} to represent contexts. This is why the inequality family described before is known as the $n-$cycle noncontextuality inequalities: they are facet-defining noncontextuality inequalities for a family of scenarios $\pmb{\Upsilon}_n$ that are represented in the compatibility hypergraph approach using \emph{n}-cycle graphs $C_n$.

\myindent There are many different graph approaches for representing such scenarios~\citep{amaral2018graph,leifer2020noncontextuality}, and it is by now a standard and well-established toolbox for analyzing properties of behaviors in such scenarios. In this thesis, part of our contribution is to develop yet another graph approach in Chapter~\ref{chapter: event_graph_approach}.  We will show our graph framework to be related to graph approaches to \acrshort{ks} noncontextuality in Chapter~\ref{chapter: from overlaps to noncontextuality}.

\myindent The relevance of such graph approaches stems from the fact that analyzing the polytopes $\mathrm{NC}(\pmb{\Upsilon})$ is a computationally nontrivial task, as the scenarios increase in complexity. Graph approaches enable the study of large scenarios by exploiting properties that are more effectively captured through their graph-theoretic relationships. Of relevance to us will be only the graph approach introduced by Cabello, Severini, and Winter (CSW)~\citep{cabello2014graph}.

\begin{figure}
    \centering    \includegraphics[width=0.4\linewidth]{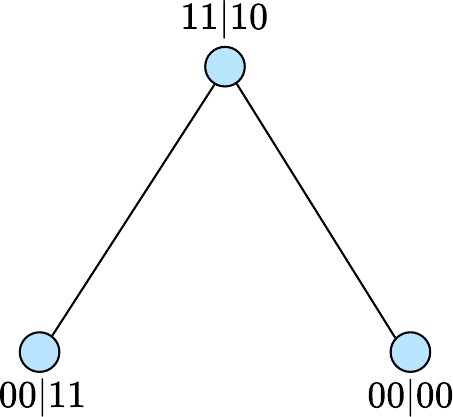}
    \caption{\textbf{Example of an exclusivity graph.} The nodes of the graph represent events that can be related to \emph{some} compatibility scenario. Not all events are necessarily present. There exists an edge between nodes if and only if the events are exclusive. For example $00|11$ and $11|10$ are exclusive because of the measurement labeled as $1$, as well as $11|10$ and $00|00$ because of the measurement labeled as $0$.}
    \label{fig: exclusivity_graph_v}
\end{figure}

\myindent In the CSW framework of~\cite{cabello2014graph}, contextuality scenarios are described by so-called \emph{exclusivity graphs}. Hence this formalism is also known as the \emph{exclusivity graph approach}~\citep{amaral2018graph,amaral2014phdthesis}. The vertices of an exclusivity graph $H$ represent `events' of joint measurements (a term used to refer to an ordered pair $\pmb{s}|\pmb{x}$ describing joint outcomes $\pmb{s}$ of a sequence of compatible measurement $\pmb{x}$), and its edges indicate
exclusivity between events, where two events are exclusive if they can be distinguished by a single measurement procedure. In other words, events $\pmb s|\pmb x$ and $\pmb{s}'|\pmb x$ are exclusive for any $\pmb s \neq \pmb s'$ since they are associated with the same measurement and only one of them can happen as a joint outcome. 

\begin{definition}[Exclusive events]
    Let $\pmb{x} = (x_1,\dots,x_n)$ be a finite sequence of compatible measurements, and let $s_i$ denote the possible outcomes $s_i \in \mathcal{O}^{x_i}$ of each measurement $x_i$. We refer to the pairs $\pmb{s}|\pmb{x} \equiv s_1s_2\cdots s_n|x_1x_2 \cdots x_n$ as \emph{measurement events}, or simply as \emph{events}. Two events $\pmb{s}|\pmb{x}, \pmb{s}'|\pmb{x}'$ are  \emph{exclusive} if and only if there exists $i,j \in \{1,2,\dots,n\}$ such that $x_i = x_j'$ while $s_i \neq s_j'$.
\end{definition}

\myindent Let us take a clarifying example of three events $$v_1 = 00|11,\,\,v_2 = 11|10,\,\,v_3 = 00|00$$ where the events are denoting now $s_1s_2|x_1x_2$. We have that $v_1$ and $v_2$ are exclusive because both these joint measurements implement measurement labeled as $1$ but have different outcomes $s_1=0$ for $v_1$ while $s_1=1$ for $v_2$. Similarly, $v_3$ and $v_2$ are exclusive due to events associated now with measurement labeled as $0$.  Finally, $v_1$ and $v_3$ are not exclusive. We show the associated exclusivity graph in~\cref{fig: exclusivity_graph_v}. Importantly, the `order' in this specific representation does not matter and only the measurement choices do. Both $v_2 = 11|10$ and $v_2' = 11|01$ are exclusive to $v_3=00|00$ because the same measurement labeled $0$ is implemented in both $v_2,v_2'$ having different outcomes from the one in the event $v_3$. We will soon see a similar example of exclusivity graph when we discuss the \acrshort{kcbs} inequality.

\myindent Even though the \acrshort{csw} framework is theory-independent, in the sense of being applicable to any \acrshort{opt} and not specifically to quantum theory viewed as an \acrshort{opt}, it is helpful for clarity of exposition to consider its instantiation in quantum theory, to better convey the underlying intuitions. In quantum theory, (sharp) measurement events are represented by projectors (\acrshort{pvm} elements) on a Hilbert space, or equivalently, by closed subspaces of the Hilbert space. Exclusivity is captured by orthogonality, which characterizes when two projectors may appear as elements of the same \acrshort{pvm}, \ie events from \emph{the same} measurement procedure. Given a set of projectors $\{\Pi_v\}_{v \in V}$ on a fixed Hilbert space, the corresponding contextuality scenario is thus described by its \emph{orthogonality graph}. This graph has a set of vertices $V$ and has an edge $\{u,v\}$ if and only if the projectors $\Pi_u$ and $\Pi_v$ are orthogonal to each other, i.e. when $\Pi_u\Pi_v=0$.

\myindent In what follows, we assume the reader is familiar with basic concepts and terminology of graph theory reviewed in Appendix~\ref{sec: graph theory}. In the \acrshort{csw} approach, a \emph{non-negative vertex weighting} $\vartheta:V(H) \to \mathbb{R}_{\geq 0}$ on the exclusivity graph $H$
determines a noncontextuality inequality on the probabilities $p(v)$ of measurement events $v \in V(H)$:
\begin{equation}\label{eq: CSW noncontextuality inequality}
\sum_{v \in V(H)} \vartheta(v) p(v)   \;\leq\; \alpha(H,\vartheta),
\end{equation}
where $\alpha(H,\vartheta)$ is the \emph{independence number} of the vertex-weighted graph (see Def.~\ref{def_app:independence_number}). In the quantum case, this yields a noncontextuality condition on the
statistics predicted by a given quantum state $\psi$:
$$\sum_{v \in V(H)} \vartheta(v) \langle \psi \vert \Pi_v \vert \psi \rangle   \;\leq\; \alpha(H,\vartheta) .$$

\myindent The discussion above introduces, therefore, the notion of a \emph{quantum realization of the probabilities $p(v)$} associated with some exclusivity graph $H$. There will be some quantum realization for $p(v)$ whenever there are projectors $\Pi_v$ for every $v \in V(H)$---which respect the exclusivity constraints of the graph, and thus satisfy $\Pi_u\Pi_v = 0$ for all $\{u,v\} \in E(H)$---and a quantum state $\vert \psi \rangle$ such that 
\begin{equation*}
    p(v) = \langle \psi \vert \Pi_v \vert \psi \rangle,
\end{equation*}
for all $v \in V(H)$. 

\myindent Such noncontextuality inequalities above determine the polytope of noncontextual behaviors for any exclusivity graph $H$.
This polytope, known as the stable set polytope (\acrshort{stab}) of $H$, denoted $\mathrm{STAB}(H)$,
is most readily defined by its V-representation, which we now present, following \cite[Chapter 3]{amaral2018graph}. For convenience, we recall~\cref{def: stable set} from Appendix~\ref{sec: graph theory}.

\begin{definition}[Stable set of a graph]
Let $H$ be a graph. A subset $S \subseteq V(H)$ of vertices is called a \emph{stable set} if no two vertices of $S$ are adjacent in $H$, i.e. for all $v, w \in S$, $\{v,w\} \not\in E(H)$. Write $\mathcal{S}(H)$ for the set of stable sets of $H$.
\end{definition}

\myindent To any subset of vertices $W\subseteq V(H)$ corresponds its characteristic map, the vertex $\{0,1\}$-labeling $\chi_W:V(H)\to \{0,1\}$ given by: 
\begin{equation}\label{eq: characteristic map V}
    \chi_W(v) := \begin{cases}
    1 & \text{ if $v \in W $,}\\
    0 & \text{ if $v \notin W$.}
    \end{cases}
\end{equation}
Through the inclusion $\{0,1\} \subseteq [0,1]$, one regards a vertex $\{0,1\}$-labeling (equivalently, a subset of vertices) as a point in $[0,1]^{V(H)} \subseteq \mathbb{R}^{V(H)}$.
Those arising from stable sets $S \in \mathcal{S}(H)$ correspond to the deterministic noncontextual models, which determine the whole convex set of noncontextual behaviors.

\begin{definition}[Stable set polytope of a graph]
\label{def: STAB}
The \emph{stable set polytope} of a graph $H$, denoted $\mathrm{STAB}(H)$, 
is the convex hull of the points $\chi_S \in [0,1]^{V(H)}$ with $S$ ranging over all stable sets of $H$,
\[
\mathrm{STAB}(H) := \mathrm{ConvHull}\left \{\chi_S | S \in \mathcal{S}(H)\right\}.
\]
\end{definition}

\myindent To get the intuition underlying this description, one may think of a vertex $\{0,1\}$-labeling ${\chi_W}:{V(H)}\to{\{0,1\}}$ as a deterministic assignment of truth values to all measurement events (vertices of the exclusivity graph). In this interpretation, the subset of vertices $W \subseteq V(H)$ is the set of measurement events that are assigned \textit{true}. The stability condition indicates that no two adjacent vertices of the exclusivity graph $H$ are labeled with $1$, that is, two exclusive measurement events cannot be jointly true. This captures the exclusivity condition at the deterministic level, thus yielding the deterministic noncontextual models.

\subsection{Relation between exclusivity graphs and measurement scenarios}

\myindent We have introduced the exclusivity graph approach practically as an independent framework for studying \acrshort{ks} noncontextuality. Having done that, we now have essentially two different descriptions of `scenarios' and also two different descriptions of `noncontextuality polytopes'. One is the notion of a scenario given by $\pmb{\Upsilon}$, with its corresponding noncontextuality polytope given by $\mathrm{NC}(\pmb{\Upsilon})$. The other is that of an exclusivity graph $H$ with its corresponding noncontextuality polytope $\mathrm{STAB}(H)$. The connection between the two can be found in the book by~\cite{amaral2018graph}, as well as in the original work by~\cite{cabello2014graph}. To put it simply, to every scenario $\pmb{\Upsilon}$ one can associate (many) exclusivity graphs $H_{\pmb{\Upsilon}}$ by selecting a few exclusive joint measurement events pertaining to $\pmb{\Upsilon}$. The probabilities $\{p(v)\}_{v \in H_{\pmb{\Upsilon}}}$ can be viewed as a behavior $B$ from $\pmb{\Upsilon}$ restricted to these selected joint measurement events. The inequalities  $\mathrm{STAB}(H_{\pmb{\Upsilon}})$ for any such constructed exclusivity graph constitute valid (often tight) \acrshort{ks} noncontextuality inequalities of $\mathrm{NC}(\pmb{\Upsilon})$. This implies that if a given collection of probabilities $\{p(v)\}_{v \in H_{\pmb{\Upsilon}}}$ violate an inequality from $\mathrm{STAB}(H_{\pmb{\Upsilon}})$ then any behavior $B$ for which one can construct $\{p(v)\}_{v \in H_{\pmb{\Upsilon}}}$ must be such that $B \notin \mathrm{NC}(\pmb{\Upsilon})$.

\myindent Some remarks are in order. In the exclusivity-graph approach, valid behaviors must preserve the exclusivity relations encoded in the graph $H$. In other words, every behavior $\{p(v)\}_{v \in V(H)}$ must satisfy, for every edge $\{u,v\} \in E(H)$,  
\begin{equation}
    p(u)+p(v)\leq 1.
\end{equation}
This can be viewed, to some extent, as an additional constraint on classicality. As discussed in~\cite[Sec. IV.A.5]{budroni2021kochenspeckerreview}, any inequality obtained from an exclusivity graph (which therefore assumes some form of exclusivity condition between edges) can be transformed into a valid noncontextuality inequality that does not assume such exclusivity relations. In this way, every inequality from the \acrshort{csw} approach can be converted into a generic \acrshort{ks}-type noncontextuality inequality, as described in terms of compatibility scenarios and the corresponding noncontextual polytopes. 

\myindent Since to every $\pmb{\Upsilon}$ we can associate some exclusivity graph $H_{\pmb{\Upsilon}}$, we can re-write some of the inequalities we have encountered before for the scenarios $\mathrm{NC}(\pmb{\Upsilon}_n)$ using the exclusivity graph approach. Let us consider the \acrshort{kcbs} inequality, which is a facet-defining inequality of $\mathrm{NC}(\pmb{\Upsilon}_5)$. Another way of viewing this inequality is as a facet-defining inequality of the $\mathrm{STAB}(H)$ polytope of a 5-cycle graph $H=C_5$. We see this in the following example:

\begin{figure}
    \centering    \includegraphics[width=0.4\linewidth]{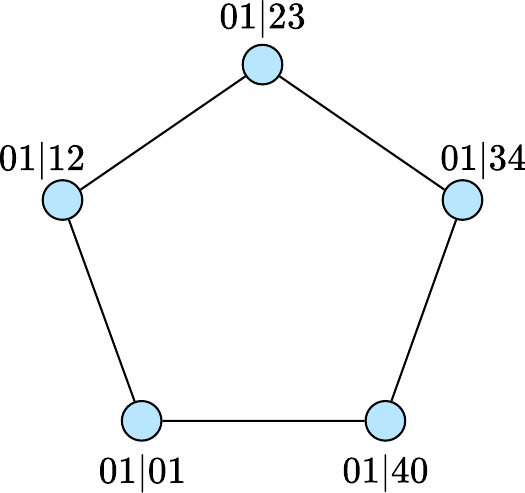}
    \caption{\textbf{Exclusivity graph for the \acrshort{kcbs} inequality.} Nodes of the graph represent events present in the \acrshort{kcbs}  inequality~\eqref{eq: exclusivity KCBS inequality}.  There exists an edge between nodes if, and only if (iff) the events are exclusive, i.e., the same measurement in both events yields different measurement outcomes. For example, $01|01$ and $01|40$ are exclusive because of the measurement labeled as $0$. The events $01|01$ and $01|12$ are exclusive because of the measurement labeled as $1$.}
    \label{fig: exclusivity_graph_kcbs}
    \end{figure}

\begin{example}
    First, recall the Inequality~\eqref{eq: KCBS correlations} that we have presented before. We have mentioned that this is a facet-defining noncontextuality inequality from $\mathrm{NC}(\pmb{\Upsilon}_n)$. Now, by a series of manipulations following the results of~\cite{badziag2010pentagrams} and ~\cite{cabello2014graph}, the same inequality can be re-written as
    \begin{equation}\label{eq: exclusivity KCBS inequality}
        p(01|01)+p(01|12)+p(01|23)+p(01|34)+p(01|04) \leq 2.
    \end{equation}
    If we consider the events appearing in this inequality, we can associate them to some exclusivity graph $H$ that, in this case, is isomorphic to the 5-cycle graph $C_5$. We show this graph in Fig.~\ref{fig: exclusivity_graph_kcbs}. 

    \myindent For this exclusivity graph, if one finds the \acrshort{stab} for this graph using the toolbox we describe in detail in Appendix~\ref{sec: convex polytopes}, the Ineq.~\ref{eq: exclusivity KCBS inequality} is found to be a facet-defining inequality of it. We also note that $2$ is equal to the independence number $\alpha(C_5,\vartheta)$ for this graph when the vertex weights are all equal to $+1$, i.e., $\vartheta(v) = +1, \forall v \in V(C_5)$ (see Appendix~\ref{sec: graph theory} for details on the independence number). Therefore, this inequality is also of the form of Eq.~\eqref{eq: CSW noncontextuality inequality}.     
\end{example}

\myindent Above, we note the usefulness of the graph-theoretic treatment, allowing us to obtain generic valid noncontextuality inequalities such as those described by~\cref{eq: CSW noncontextuality inequality}, using only information provided by the graph of exclusivity, hence avoiding the complete characterization of the polytopes $\mathrm{NC}(\pmb{\Upsilon})$. Importantly, in this approach, there are no inequalities `missing', in the sense that one may have behaviors $B$ violating an inequality from $\mathrm{NC}(\pmb{\Upsilon})$ while not violating any inequality $\mathrm{STAB}(H)$ for all possible exclusivity graphs $H$. If we consider any possible facet-defining noncontextuality inequality of $\mathrm{NC}(\pmb{\Upsilon})$ associated to any compatibility scenario $\pmb{\Upsilon}$, there exists some exclusivity graph $H$ for which this same inequality is also a valid inequality for $\mathrm{STAB}(H)$ (see~\cite{amaral2018graph} for a pertinent discussion on the connection between different approaches, as well as~\cite[Sec.~IV.A.5]{budroni2021kochenspeckerreview}).  

\subsection{Relation between KS noncontextuality and Bell's notion of local causality}~\label{sec: Bell and KS}

\myindent We will conclude this section by noticing that there exists a formal mathematical relationship between the notion of Kochen--Specker noncontextuality and  Bell's notion of local causality~\citep{abramsky2011sheaf}. In essence, every Bell scenario can be understood as a compatibility scenario $\pmb{\Upsilon}_{\mathrm{Bell}}$, in which case \acrshort{ks} noncontextuality inequalities are mathematically equivalent to Bell inequalities. The \emph{local polytope} in a Bell scenario is equal to the polytope $\mathrm{NC}(\pmb{\Upsilon}_{\mathrm{Bell}})$. Hence, any statement that is made capturing generic facet-defining noncontextuality inequalities will \emph{also} capture facet-defining Bell inequalities.

\myindent From the description of \acrshort{ks} noncontextual behaviors given above in Eq.~\eqref{eq: noncontextual factorizable} it is also possible to see that when considering Bell scenarios $\pmb{\Upsilon}_{\mathrm{Bell}}$ the \acrshort{ks} noncontextual ontological models become equivalent to what is known as a local hidden-variable model~\citep{brunner2014bell}. For simplicity, let us take the simple case of a bipartite Bell scenario. In this case, we have that Bell's notion of local causality implies that the ontological model reproducing the behavior in such a scenario must satisfy
\begin{equation*}
    p(ab|xy) = \sum_\lambda \mu(\lambda)p(a|x,\lambda)p(b|y,\lambda)
\end{equation*}
where $x$ and $y$ are now measurements performed in space-like separated laboratories, which guarantees their compatibility. By comparing this expression with that of Eq.~\eqref{eq: noncontextual factorizable} the equivalence becomes evident. 

\myindent However, and here lies a fundamental and subtle point, the two notions are \emph{not physically equivalent}. Bell's notion requires an extremely specific operational set-up while \acrshort{ks} noncontextuality allows for other operational set-ups. The experimental tests associated with probing the two notions are not equivalent, and neither are the assumptions associated with them. This is evident once one understands that the violation of a Bell inequality \emph{per se} does not mean that one has excluded the possibility of local hidden-variable models reproducing the observed statistics. One must also probe the \emph{assumptions underlying Bell's theorem} and \emph{operational aspects} of the Bell scenario. Even if mathematically the inequalities are the same, \acrshort{ks} noncontextuality inequalities do not require the same assumptions nor the same operational requirements. 

\myindent For example, it is possible to violate \acrshort{ks} noncontextuality inequalities that have \emph{the exact same form} of a \acrshort{chsh} inequality, as was done by~\cite{wang2022significant}, by performing sequential measurements on a single system. Clearly, even if the inequality is the same, local hidden-variable models are not falsified by the violation of this inequality in this experimental setup.

\section{Generalized noncontextuality}\label{sec: Spekkens noncontextuality}

\myindent So far, we have noticed that the notion of \acrshort{ks} noncontextuality has only considered \emph{various measurements} that can be performed on \emph{a single system}. Because of that, only the notion of  noncontextuality of measurements has been considered, leaving the possibility of investigating if some form of `noncontextuality of preparations of states' may be meaningfully introduced. Moreover, as we have discussed before, \emph{ideal measurements} were a crucial concept. In quantum theory, these are projective measurements, but projective measurements are not the most general notion of measurement. As is well known (see, for example~\cite{nielsen2002quantum}), a more general notion  is that of  positive operator-valued measures (\acrshort{povm}s). Arguably, these are points with respect to which \acrshort{ks} noncontextuality is silent. In what follows, we present an approach by Robert~\cite{spekkens2005contextuality} that has considered an extension of noncontextuality that resolved (among others) these points mentioned.

\myindent Spekkens's notion of \emph{generalized noncontextuality} is associated to operational probabilistic theories~\citep{dariano2017quantum,schmid2024structuretheorem,kunjwal2019beyondcabello}. In what follows, we will present the formalism in its traditional description, similar to the one considered originally by~\cite{spekkens2005contextuality}. In this description, we will focus on defining generalized noncontextuality as a property of ontological models reproducing the statistics of behaviors in prepare-and-measure scenarios. Similarly to the duality between \acrshort{ks} noncontextual models and global sections, a more modern perspective can be taken where, instead of talking about ontological models, one can talk about the geometry of general probabilistic theories~\citep{schmid2021characterization,selby2023accessible,schmid2024structuretheorem}, a topic that we will not discuss here.  

\myindent Before introducing the main concepts, we note that Spekkens' notion of generalized noncontextuality is significantly different, conceptually, from that of \acrshort{ks} noncontextuality. The two are related, as was shown by~\cite{spekkens2005contextuality},~\cite{leifer2013maximally}, and~\cite{kunjwal2016contextualitykochenspeckertheorem}. But, for what we will consider in this thesis, and for other practical reasons, it is best to think in the two approaches as entirely different definitions of noncontextuality. This will translate into the use of an entirely different notation, as well as different construction of scenarios, and a notion of behaviors.

\subsection{Operational probabilistic theories}~\label{sec: operational theories and theory independent approaches}

\myindent The description of an operational-probabilistic theory (\acrshort{opt}) starts with a set of basic (operational) physical processes: in the simplest scenarios, one considers preparations and measurements.
One focuses on experiments consisting of a preparation $P$ followed by a measurement $M$ that returns an outcome $k$.
A probability rule associates a probability $p(k | M, P)$ of obtaining outcome $k$ when performing measurement $M$ after the preparation $P$. More precisely, it associates a probability distribution over outcomes $k$ to each choice of preparation $P$ and measurement $M$. Formally, we can view such scenarios as the specification of a part of an \acrshort{opt} that we will focus on, which is accessible experimentally, and that is sometimes called a \emph{fragment}.

\myindent As prepare-and-measure scenarios are defined in other contexts beyond that of investigating generalized noncontextuality, we introduce a formal definition of such scenarios here.

\begin{definition}[Prepare-and-measure scenarios]\label{PM scenario}
    Let $I,J \in \mathbb{N}$ and for every $j \in [J]$ we also let $K_j \in \mathbb{N}$.~\footnote{Recall that we denote $[n] = \{1,2,\dots,n\}$ for any $n \in \mathbb{N}$ natural number, and we do not assume zero to be natural.} We say that these lists define a prepare-and-measure scenario with $I$ preparation inputs, $J$ measurement inputs, and $K_j$ outputs for each measurement input $j$. If $K_j = K$ for all $j$ we denote such a scenario as a triplet of natural numbers $(I,J,K)$. To each $i \in [I]$ we associate a preparation $P_i$ and to each $k_j|j$ we associate a measurement effect $k_j|M_j$. Behaviors in these scenarios are defined as the list of probability distributions $p(k_j|M_j,P_i)$ for every possible $k_j,j$, and $i$. 
\end{definition}

\myindent We remark that while above we have considered triplets of natural numbers for defining a scenario, we will not bound our notation to always necessarily representing preparations as $P_1,P_2,\dots,P_I$. The numbers $I,J$ and $K$ are truly the only numbers required to define the scenario, but we take the freedom of labeling the procedures differently if necessary. For example, we might have a scenario where $I=4$ in which case $P_1,P_2,P_3,P_4$  and $P_1,P_{1^\perp}, P_2, P_{2^\perp}$ are equally good labels for the preparations. We could consider $I, J$ and $K$ to be sets of labels, but this is not the standard choice in the literature dedicated in investigating prepare-and-measure scenarios. For a dichotomic measurement $M$, \ie one with only two possible outcomes $0$ and $1$, we simplify notation and write $p(M | P)$ for $p(1 | M, P)$. Similarly, as before, we have also the description of quantum realizable behaviors in a prepare-and-measure scenario.

\begin{definition}[Quantum realizable behavior in a prepare-and-measure scenario]\label{def: quantum realizable PM behavior}
    Let $(I,J,K)$ label a generic prepare-and-measure scenario. We say that a behavior $B = \{p(k|M_j,P_i)\}$ for this scenario is \emph{quantum realizable} in some Hilbert space $\mathcal{H}$ if there exists quantum states $\{\rho_i\}_{i \in I}$ and generalized quantum measurements $M_j = \{E_k^j\}_j$ such that 
    \begin{equation}
        p(k|M_j,P_i) = \text{Tr}(\rho_iE_{k}^j)
    \end{equation}
    for every $i,j,k$. 
\end{definition}

\myindent Above we have focused on scenarios $(I,J,K)$, considering that every measurement has the same number of outcomes, for simplicity of the presentation.~\footnote{One can also turn this simplification to a formal argument that only these scenarios need to be considered by noticing that we can choose the measurement $j$ with the largest number of outcomes $K_j$ and let all other measurements have the same number of outcomes, where the additional outcomes can be viewed as `dummy' ones that never happen.} A crucial---if sometimes overlooked---aspect is that the full set of procedures in an \acrshort{opt} includes also classical probabilistic mixtures (\ie convex combinations) of basic procedures, with the probability rule extended accordingly (\ie linearly).

\myindent Given an \acrshort{opt}, one defines an equivalence relation identifying indistinguishable procedures.
Following~\cite{spekkens2005contextuality}, we can introduce a definition of operationally equivalent preparations. 

\begin{definition}[Operational equivalence for preparation procedures]\label{def: preparation operational equivalences}
    Let $P$ and $P'$ be two preparation procedures in an operational-probabilistic theory. We say that $P$ and $P'$ are \emph{operationally equivalent}, and write $P \simeq P'$, iff
    \begin{equation*}
        p(k|M,P) = p(k|M,P'),
    \end{equation*}
    for all pairs $k|M$ in the theory.
\end{definition}

Similarly, we can introduce a notion of operational equivalence for measurement effects.

\begin{definition}[Operational equivalence for measurement effects]\label{def: effects operational equivalences}
    Let $k|M$ and $k'|M'$ be two measurement effects in an operational-probabilistic theory. We say that $k|M$ and $k'|M'$ are \emph{operationally equivalent}, and write $k|M \simeq k'|M'$, iff
    \begin{equation*}
        p(k|M,P) = p(k'|M',P),
    \end{equation*}
    for all preparations $P$ in the theory.
\end{definition}

\myindent We remark that the above are \emph{informal} `definitions' of an operational equivalence between two procedures in an  \acrshort{opt}. Any formal description of operational equivalences is out of the scope of this thesis, as it would involve category-theoretic concepts as discussed by~\cite{schmid2021unscrambling,schmid2024structuretheorem}, and by~\cite{chiribella2014dilation}. There is a whole literature devoted to critically understanding, analyzing, revisiting, and testing the underlying assumptions in this definition. We will not detail those here, but refer the interested reader to discussions present elsewhere~\citep{hermens2011theproblem,pusey2019contextuality,gitton2022solvablecriterion,gitton2022onthesystem,schmid2024addressing,schmid2024shadows,wagner2025structuretheoremcomplexvaluedquasiprobability}.

\myindent When one treats quantum theory as an operational theory, quantum states $\vert \phi \rangle \langle \phi \vert $ correspond to equivalence classes of operational procedures. For instance, a state $\ket{0}\langle 0 \vert$ may represent preparing a ground state of a nitrogen atom or preparing the horizontal polarization in photonic qubits. We relax this terminology and refer to `the preparation $P$ associated with a state $\ket{\phi}\langle \phi \vert $', even though strictly speaking $P$ is only an instance of an equivalence class of procedures. 
Such relaxation is safe for our purposes.
In effect, it corresponds to treating pure quantum states as the basic procedures.
The interesting operational equivalences relevant for preparation contextuality
go beyond these, holding between classical mixtures of basic procedures.
For example, in quantum theory, the preparation procedure corresponding to an equal mixture of pure qubit states $\ket{0}\langle 0 \vert$ and $\ket{1}\langle 1 \vert $
is operationally equivalent to that corresponding to an equal mixture of states $\ket{+}\langle + \vert $ and $\ket{-}\langle - \vert $.
Indeed, both these classical mixtures define the same qubit mixed state, the maximally mixed state.

\myindent In Spekkens' framework, the notion of a \emph{context} is then different than the one we have encountered before. Given two procedures $P \simeq P'$ that are operationally equivalent, we say that a context is the label (or descriptions) of the two procedures \emph{despite the fact} that they always yield the same statistics. 

\myindent A prepare-and-measure scenario~\citep{schmid2018all} in which one can investigate noncontextuality will be one where a set of preparations $\{P_i\}$, and measurement effects $\{k|M_j\}_{k,j}$ is considered, together with some list of operational equivalences~\citep{tavakoli2021bounding,chaturvedi2021quantum,wagner2023using}. 
\begin{definition}[Prepare and measure noncontextuality scenarios]
    A prepare-and-measure noncontextuality scenario is defined by a tuple $$\mathcal{PM} := (I,J,K,\mathcal{OE}_{P},\mathcal{OE}_{M}) = (I,J,K)\sqcup (\mathcal{OE}_{P},\mathcal{OE}_{M})$$ where the first three labels are numbers defining a scenario $(I,J,K)$ while  $(\mathcal{OE}_{P},\mathcal{OE}_{M})$ are sets used to define operational equivalences. The operational equivalences $\mathcal{OE}_{P}$ are completely characterized by 
    \begin{equation}\label{eq: operational equivalences P}
        \mathcal{OE}_{P} := \{S_{\iota},\{\alpha_{\iota}(i)\}_i\}_{{\iota}=1}^R,
    \end{equation}
    where $\{{S}_{\iota}\}_{\iota=1}^R$ is a partition of $[I]$ into $R$ disjoint sets. For each $i \in [I]$, $\alpha_{\iota}(i) \geq 0$. Moreover, $\sum_{i\in S_{\iota}} \alpha_{\iota}(i) = 1$. Similarly, for measurement effects we define sets of equivalences using 
    \begin{equation}\label{eq: operational equivalences M}
        \mathcal{OE}_{M} := \{V_\ell,\{\beta_\ell(k|j)\}\}_{\ell=1}^L
    \end{equation}
    where $\{V_\ell\}_{\ell=1}^L$ is a partition of $[K] \times [J]$ into $L$ disjoint sets. For each $\ell$, $\beta_\ell(k|j)\geq 0$ and $\sum_{k|j \in V_\ell} \beta_\ell(k|j)=1$.
\end{definition}

\myindent Usually, however, the description above is more relevant algorithmically, as inputs to some vertex-enumeration program, than it is theoretically. For clear presentation purposes one usually writes operational equivalences as $$\sum_{i \in S_\iota} \alpha_\iota (i) P_i \simeq \sum_{j \in S_{\iota'}} \alpha_{\iota'}(j) P_j,$$ where $\iota \neq \iota'$ are two different labels of partitions. The sets $\mathcal{OE}_P$ and $\mathcal{OE}_M$ can then be viewed as a way of representing sets of operational equivalences characterized by convex mixtures of procedures in a prepare and measure scenario $(I,J,K)$. The labels $\iota, \ell$ describe how many of these operational equivalences are being considered. To clarify the definition just given we consider the example of the simplest non-trivial prepare and measure noncontextuality scenario that can be constructed. 

\begin{example}[Simplest scenario]\label{example: simplest scenario}
    As an example, we consider the simplest non-trivial scenario that was introduced by~\cite{pusey2018simplest} and that has been completely characterized~\citep{schmid2018all,wagner2020resourcetheory} and robustly analysed~\citep{khoshbin2024alternative}. In this case $I=4$, and we have four preparation procedures $P_1,P_2$ and $P_{1^\perp}, P_{2^\perp}$ satisfying the operational equivalence
    \begin{equation*}
        \frac{1}{2}P_1 + \frac{1}{2}P_{1^\perp} \simeq \frac{1}{2}P_2 + \frac{1}{2}P_{2^\perp}.
    \end{equation*}
    This notation is used since we later view this  scenario as the first element within an infinite family of operational scenarios. There are two dichotomic measurement procedures $M_1$ and $M_2$. These are all the operational elements necessary to characterize the simplest prepare-and-measure scenario. The operational equivalence above means that, as we have described in Def.~\ref{def: preparation operational equivalences}, for every measurement outcome $k|M$ it must hold that
    \begin{equation*}
        \frac{1}{2}p(k|M,P_1) + \frac{1}{2}p(k|M,P_{1^\perp}) = \frac{1}{2}p(k|M,P_2)+\frac{1}{2}p(k|M,P_{2^\perp}).
    \end{equation*}
    In this manner, we have two partitions $S_1 = \{1,1^\perp\},S_2 = \{2,2^\perp\}$, of the set $\{1,1^\perp,2,2^\perp\}$. The associated convex weights are   $\alpha_1(1)=\alpha_1(1^\perp)=\sfrac{1}{2}$ and $\alpha_2(2)=\alpha_2(2^\perp)=\sfrac{1}{2}$. 
\end{example}

\myindent For our purposes, we can focus on very specific types of scenarios that we will refer to as the Lostaglio, Senno, Schmid, and Spekkens (\acrshort{lsss}) prepare-and-measure scenarios. These scenarios were introduced due to their relevance in quantum information advantage provided by contextuality~\citep{schmid2018discrimination,lostaglio2020contextualadvantage}, and their operational structure resembles other scenarios such as the one known as the simplest scenario~\citep{pusey2018simplest}.

\subsection{LSSS prepare-and-measure scenarios}~\label{sec: LSSS scenarios}

\myindent A description of prepare-and-measure scenarios starts with the prescription of a list of preparations and measurement effects. We then consider a set of preparation procedures $\{P_i\}_i \cup \{P_{i^\perp}\}_i$, in other words, for label $i$ the scenario has two procedures $P_i$ and $P_{i^\perp}$. For every preparation $P_i$ we assume that there exists a corresponding dichotomic `test measurement' $M_i$ satisfying that $p(M_i|P_i) = 1$. Moreover, $p(M_i|P_{i^\perp}) = p(M_j | P_{j^\perp}) = 0$. More generally, we may allow for some deviation of this property, in which case we write that $p(M_i|P_{i^\perp}) \geq \varepsilon_i$ and $p(M_i|P_{i}) \geq 1-\varepsilon_i$ for some small scalar $\varepsilon_i > 0$.

\myindent A second aspect of such scenarios is the prescription of operational equivalences that must be satisfied by the list of procedures we have considered above. The operational equivalences we will focus on take the following form:
\begin{equation*}
    \frac{1}{2}P_i + \frac{1}{2}P_{i^\perp} \simeq \frac{1}{2}P_j + \frac{1}{2}P_{j^\perp}.
\end{equation*}
for every pair $\{P_i,P_j\}$ of preparation procedures $P_i$ and $P_j$. Note that these constraints apply to preparation procedures. We do not assume any operational equivalences for measurement procedures. Therefore, the scenarios under consideration aim to probe preparation contextuality only.

\myindent The probabilities $p(M_i| P_j)$
are usually called the \stress{confusability}~\citep{lostaglio2020certifying,schmid2018discrimination}, because they may be interpreted as the probability of guessing \emph{incorrectly} that the preparation performed had been $P_i$ instead of $P_j$. These probabilities provide a theory-independent, operational treatment of two-state overlaps, which reduces to the familiar notion in the case of quantum theory viewed as an operational theory:
\[p(M_i | P_j)  \stackrel{\mathrm{QT}}{=} \Tr\left(\ket{\phi_i}\!\bra{\phi_i}\ket{\phi_j}\!\bra{\phi_j}\right) = \vert \langle \phi_i \vert \phi_j \rangle \vert^2 .\]

\myindent This aspect will have an enormous impact on our results to come in Chapter~\ref{chapter: applications}, where we present a proof of quantum information advantage for the task of quantum interrogation provided by contextuality. We will introduce this task in Chapter~\ref{chapter: information tasks}, and proceed now to define the notion of \emph{generalized noncontextuality} for preparation procedures that we have been anticipating.

\subsection{Preparation noncontextuality}~\label{sec: Preparation noncontextuality}

\myindent When faced with an operational theory, a natural question is whether it admits an explanation in terms of a noncontextual ontological model. We have already described such models in the \acrshort{ks} framework of the last section. Here, the notion of an ontological model is similar and follows a similar reasoning. However, the very notion of noncontextuality changes considerably. Let us start by providing how, in general, an ontological model attempts to explain the statistical predictions of a prepare-and-measure scenario (that can be viewed as a mere fragment of the entire operational theory).

\myindent In general, an ontological model consists of a measurable space $(\Lambda, \mathcal{F}_\Lambda)$~\footnote{A measurable space is an ordered pair $(\Lambda,\mathcal{F}_\Lambda)$ where $\Lambda$ is a set and $\mathcal{F}_\Lambda$ is a $\sigma$-algebra. Both are elementary definitions in measure and probability theory. We use this definition, however, as has been showed by~\cite{schmid2018all}, when considering finitely defined prepare-and-measure scenarios, without loss of generality one can take the ontic spaces to be \emph{finite}, easing the mathematical treatment of ontological models.} of \stress{ontic} states equipped with ontological interpretations for preparation and measurement procedures~\citep{harrigan2010einstein}:
preparation procedures $P$ determine probability measures $\mu_P$ on $\Lambda$,
whereas measurement procedures $M$ determine measurable functions $\xi_M$ mapping each ontic state $\lambda \in \Lambda$ to (a distribution on) outcomes. For example, if we recall our~\cref{example: KS ontological models}, the machine was performing a single preparation procedure $P$, selecting shoes and putting them in the boxes. We had modelled the state of affairs describing each box by variables $\lambda$, and we have associated to it some distribution $\mu_P$. In what follows we will consider situations where there might be more than one single preparation. In this case, each preparation $P_1, \ldots, P_I$ is  modeled as a  different distribution $\mu_{P_1},\ldots,\mu_{P_n}$ over $\Lambda$.      

\myindent Still in~\cref{example: KS ontological models}, the measurable functions $\xi_M$ where given by the probabilities $\xi_{M=x_i}(k=s_i|\lambda) \equiv p(s_i|x_i,\lambda)$. Recall that if a behavior $B \in \mathrm{NC}(\pmb{\Upsilon})$ then there exists a model for which these response functions are \emph{deterministic}. This is sometimes referred to as the \emph{outcome-determinism} assumption~\cite{spekkens2014statusofoutcome}, and it is something that follows from the \acrshort{fab} theorem we stated in~\cref{theorem:FAB theorem}. In the Spekkens' formalism, these response functions need not be deterministic. 

\myindent The probability $\mu_P$ plays a central role in quantum foundations. When $P$ corresponds to the preparation of some quantum state, say $\vert \psi \rangle \langle \psi \vert $, the associated distributions $\mu_{\psi}$ are referred to as the \textit{epistemic states} of the ontological model. In such a case, epistemic states carry only some knowledge about reality, itself described by the ontic states $\lambda$. The theories for which quantum states are interpreted as epistemic states are generically known as \textit{epistemic interpretations of quantum mechanics}. For an overview, we refer to~\cite{leifer2014isthe}. Some celebrated results in quantum foundations are presented in this language, such as Spekkens' noncontextual toy theory~\citep{spekkens2007evidence,catani2023whyinterference}, and the Pusey--Barrett--Rudolph (\acrshort{pbr}) no-go theorem~\citep{pusey2012onthereality}. 

\myindent Classical mixtures of preparation procedures must be determined \emph{linearly} from that of basic procedures, \eg $$\mu_{\frac{1}{2}P+\frac{1}{2}Q} = \frac{1}{2}\mu_P+\frac{1}{2}\mu_Q,$$ for every two preparation procedures $P, Q$. The composition of the interpretations of a preparation and a measurement (going via the ontic space $\Lambda$) is required to recover the empirical or operational predictions, \ie
\[p(\cdot | M, P) \;=\; \int_\Lambda \xi_M \, \mathrm{d} \mu_P ,\]
or with variables,
\begin{equation}\label{eq: empirical adequacy}
p( k | M, P) \;=\; \int_\Lambda \xi_M(k | \lambda) \, \mathrm{d} \mu_P(\lambda) .
\end{equation}
Everything that we have described so far merely describes a generic ontological model. Shortly then, an ontological model is a prescription of ontological counterparts to the \acrshort{opt} that accurately and consistently combine (according to~\cref{eq: empirical adequacy}) to produce the statistical predictions of the theory. Eq.~\eqref{eq: empirical adequacy} is also sometimes referred to as the condition of \emph{empirical adequacy}, forcing the ontological model not only to have ontological counterparts $P \to \mu_P$ and $k|M \to \xi_{k|M}$ but also to correctly reproduce the data.

\myindent Such a realization by an ontological model is said to be noncontextual if operationally equivalent procedures are given the same interpretation.
For preparations, the requirement is that two operationally equivalent preparation procedures determine the same probability measure on $\Lambda$.
We refrain from going into detail on the general definition, as the characterization that follows suffices.

\begin{definition}[Preparation noncontextuality]\label{def: preparation noncontextuality}
    Consider a prepare-and-measure noncontextuality scenario $\mathcal{PM}$ having preparations $\{P_i\}_i$, measurements effects $\{k|M_j\}_{k,j}$ and generic operational equivalences $\mathcal{OE}_P$ for preparation procedures given by
    \begin{equation}
        \sum_{i \in S_\iota}\alpha_\iota (i) P_i \simeq \sum_{j \in S_{\iota'}} \alpha_{\iota'}(j) P_j,
    \end{equation}
    where $(S_\iota,\{\alpha_\iota(i)\}_i),(S_{\iota'},\{\alpha_{\iota'}(j)\}_j) \in \mathcal{OE}_P$ for any pair $\iota,\iota'$. We say that an ontological model reproducing the prepare-and-measure statistics $\{p(k|M_j,P_i)\}_{i,j,k}$ via 
    \begin{equation}
        p(k|M_j,P_i) = \int_\Lambda \xi_{M_j}(k|\lambda)\mathrm{d}\mu_{P_i}(\lambda)
    \end{equation}
    is \emph{preparation noncontextual}, if the ontological counterparts $\mu_{P_i}$ associated to the preparation procedures $P_i$ satisfy that
    \begin{equation}
        \sum_{i \in S_\iota}\alpha_\iota (i) \mu_{P_i} = \sum_{j \in S_{\iota'}} \alpha_{\iota'}(j) \mu_{P_j},
    \end{equation}
    for every $\iota$.
\end{definition}

\myindent \cite{schmid2018discrimination}, and later~\cite{lostaglio2020contextualadvantage} have shown that the \acrshort{lsss} constraints imply that any preparation noncontextual model explaining preparation procedures $P_i$ as probability measures $\mu_i$ on $\Lambda$ must satisfy
\begin{equation}\label{equation: noncontextual overlaps}
    p(M_i| P_j) 
    = 1 - \|\mu_i - \mu_j\|_{_{\mathsf{TV}}} ,
\end{equation}
where $\|\cdot - \cdot\|_{_{\mathsf{TV}}}$ denotes the total variation distance between probability measures,
given for an arbitrary measurable space $(\Lambda, \mathcal{F}_\Lambda)$ by
\[\|\mu_i - \mu_j\|_{_{\mathsf{TV}}} = \sup_{E \in \mathcal{F}_\Lambda}|\mu_i(E) - \mu_j(E)|.\]
In the case when $\Lambda$ is discrete (which is effectively all we need),
this distance is related to the $l_1$ norm:~\footnote{In the continuous case, it is often rendered as $\|\mu_i - \mu_j\|_{_{\mathsf{TV}}} = \int_\Lambda \vert \mu_i(\lambda) - \mu_j(\lambda) \vert \,\mathrm{d}\lambda$ in terms of a reference measure such as the Lebesgue measure on the real line.}
\begin{align*}
    \|\mu_i - \mu_j\|_{_{\mathsf{TV}}} = \frac{1}{2}\|\mu_i - \mu_j\|_{_1} = \frac{1}{2}\sum_{\lambda \in \Lambda}|\mu_i(\lambda)-\mu_j(\lambda)|.
\end{align*}

From the above considerations, we recall the following result. 

\begin{theorem}[Adapted from~\cite{lostaglio2020contextualadvantage}]\label{theorem: lostagio_senno}
Consider any prepare-and-measure scenario having the \acrshort{lsss} constraints detailed before. Then, a preparation noncontextual ontological model explaining the statistics of $p(M_i|P_j)$ must satisfy the following inequality,
\begin{equation}
    \vert \Vert \mu_i - \mu_j \Vert_1 - 2(1-p(M_i \vert P_j)) \vert \leq 2\varepsilon_i 
\end{equation}
in the limit of ideal measurements $\varepsilon_i \to 0$ we have,
\begin{equation}
    p(M_i \vert P_j) = 1 - \frac{1}{2}\int_\Lambda \vert \mu_i(\lambda) - \mu_j(\lambda)\vert \mathrm{d}\lambda.
\end{equation}
\end{theorem}

\myindent The relevance of \acrshort{ks} noncontextuality and generalized noncontextuality stems from the fact that they provide stringent criteria for nonclassicality. From our discussion, it is clear that when there exists \emph{some} ontological model that reproduces the data of an experiment they give in return \emph{some} notion of classical explainability. In a sense, they function similarly to how statistical mechanics functions: they postulate some complete set of variables able to, in principle, completely explain the states of the theory (in that case, $\lambda= (p,q)$ position and momenta coordinates) and any source of randomness, uncertainty, or unpredictability happens solely because our best description of the states of the system is only provided by some form of `partial knowledge' of which are the true states (in this case, $\mu_P = \mu_P(p,q)$ some probability distribution over phase-space). Ontological models abstract away this idea, allowing for a more broad and generic notion of `phase space' than that restricted to the context of statistical mechanics, and allowing for a more broad and generic notion of `complete set of variables' than that of phase-space points in classical mechanics. 

\myindent Interestingly, it has been shown that there are situations in which such models similar to classical statistical mechanics---or better saying `toy models' that hold on to Leibniz's principle of indiscernibles~\citep{spekkens2019ontological}---can reproduce the phenomenology captured by some quantum phenomena traditionally viewed as puzzling or intriguing. When such models explain the data, even if certain quantum resources are present (such as quantum coherence, quantum entanglement, non-zero discord, and so on) one cannot have any form of quantum-over-classical advantage in information, computational, or thermodynamic tasks because essentially they can `simulate' the same tasks classically. Finding bounds for when such models can be constructed (even in principle) is therefore of profound foundational and technological relevance. Many such bounds exist for quantum computation and information tasks. One can say that these bounds \emph{define} what it means for an information task to be truly advantageous with respect to \emph{every possible} classical model, where classical has the precise meaning of `allowing for an explanation in terms of a noncontextual ontological model'.

\myindent Provided with this formal understanding of what means to have a quantum information advantage, which was made possible due to the quantum foundational findings we have presented, in Chapter~\ref{chapter: information tasks} we proceed to investigate two technologically relevant information tasks that we will be interested in this thesis: that of dimension witnessing~\citep{gallego2010device}, which was one of the motivating tasks for analyzing nonclassical aspects in prepare-and-measure scenarios, and that of quantum interrogation~\citep{elitzur1993quantum}, which has been a source of interpretational debate~\citep{catani2023whyinterference}, and showcases one of the many puzzling features of quantum theory by the ability to detect the presence of an object without ever `interacting' with it, due to counterfactual reasoning.

\chapter{Two quantum information tasks}~\label{chapter: information tasks}

\begin{quote}
    ``\textit{A theorist is invited to a lab. The experimentalists, not entirely happy with the nuisance, decide to submit the visitor to the ordeal `Guess what we are measuring.' Hardly distinguishing lasers from vacuum chambers, the
theorist cannot hope to identify the system under study, and
asks for a black-box description of the experiment in order
to disentangle at least the physics from the cables. (...) the experimentalists
show the data: the frequencies $p(ab|xy)$ of occurrence of a given pair of outcomes for each pair of measurements. The theorist makes some calculations and
delivers a verdict. (...) `You are using systems of
dimension at least $d$'.} ''
\\~\citep{brunner2008testing}
\end{quote}

\begin{quote}
    ``\textit{In one world the photon is scattered by the object, and in two others it does not. Since all worlds take place in the physical universe, we cannot say that nothing has `touched' the object'. We get information about the object without touching it in one world, but we `pay' the price of interacting with the object in the other world.}''\\~\citep{elitzur1993quantum} 
\end{quote}

Physicists are traditionally motivated by a desire to understand the fundamental properties of Nature, often without regard for practical applications. Conversely, applying this understanding to improve people's everyday lives through technological development has traditionally been the domain of engineering. Occasionally, however, this distinction `blurs': physicists may consider practical challenges to deepen their understanding of Nature, while engineers may explore foundational principles to gain better intuition about what is (or is not) possible to engineer. This interplay has proven to be remarkably successful, as seen during the development of thermodynamics and, more recently, in the development of the field of quantum information.

\myindent In this Chapter we discuss two types of quantum information tasks that teach intriguing aspects of Nature, while simultaneously providing insights for novel technologies: the task of Hilbert space dimension witnessing and the task of quantum interrogation. The reason we focus on these tasks is entirely incidental, and it is correlated to the results presented in Chapter~\ref{chapter: applications}. We describe them as instances of the fruitful dialogue between foundations and applications, which is a theme of this thesis. Our focus here can then be viewed as setting the stage for the case study in Chapter~\ref{chapter: applications}, applying the formalism we build in Chapters~\ref{chapter: event_graph_approach} and~\ref{chapter: relational coherence}.

\myindent The structure of this chapter is as follows. We discuss each protocol separately, starting with dimension witnessing in Sec.~\ref{sec: dimension witnessing} and later discussing quantum interrogation in Sec.~\ref{sec: quantum interrogation}. We discuss dimension witnesses by means of violations of Bell inequalities (in Sec.~\ref{sec: dimension witnesses Bell scenarios}), Kochen--Specker noncontextuality inequalities (in Sec.~\ref{sec: dimension witnesses KS scenarios}) and prepare-and-measure inequalities (in Sec.~\ref{sec: dimension witnesses PM scenarios}). We conclude Sec.~\ref{sec: dimension witnessing} discussing stringent requirements beyond just Hilbert space dimensionality in Sec.~\ref{sec: stringent dimension}. Then, to introduce the standard scheme for quantum interrogation we introduce the basic experimental setup of a Mach--Zehnder interferometer in Sec.~\ref{subsec: MZI} and then introduce the quantum interrogation via its original description in Sec.~\ref{sec: bomb testing}. We conclude by describing a figure of merit for the success of a quantum interrogation task in Sec.~\ref{sec: bomb test figure of merit}. 

\section{Dimension witnessing}~\label{sec: dimension witnessing}

\myindent Various definitions of dimension are mathematically related, yet physically unrelated. For example, we refer to spatial dimensions as those associated with the three spatial coordinates $(x,y,z)$. Each point in space is associated with such a triplet of coordinates, telling us how to move around in $3$-dimensional space relative to some fixed choice of origin $(0,0,0)$. We also have a notion of the dimension of the \emph{phase space}. This is the space of all relevant variables characterizing, if we consider for example classical mechanics, the description of all possible states that a certain physical system can have. For one classical point particle moving in a three-dimensional physical space, it suffices to list all the generalized coordinates of momentum $p_x,p_y,p_z$ and position $q_x,q_y,q_z$ to completely characterize its state. This space having points $(p_x,p_y,p_z,q_x,q_y,q_x)$ has six dimensions. Itself, the set of all possible \emph{states} that the physical system can have is described by yet another space, called the \emph{state space}. The three concepts are mathematically linked via the notion of the dimension of a vector space but are physically describing different ideas.  Our focus on dimension witnessing is related to the dimension of the \emph{state space} of a given system as modeled by quantum theory. 

\myindent Having clarified the `type' of dimensionality we are interested in, let us also mention some basic aspects of it. Dimensionality is linked to how we express the relevant properties of a system (physical or not) of interest using a list. For example, if we are interested in describing the statistical properties of a coin flip, we can characterize it by points in a line segment $(p,1-p)$ connecting $(1,0)$ and $(0,1)$ where $0 \leq p \leq 1$. The set of all such points is called the \emph{state space} of the system that we are referring to as a generic coin, where each pair $(p,1-p)$ describes the state of the system. A physicist's way of thinking about the dimension relevant to a problem may be the number of degrees of freedom characterizing that problem. The state of a coin is completely characterized by the value $p$ alone (1 degree of freedom), and therefore the state-space is one dimensional.

\myindent The notion of dimension is made formal using \emph{linear algebra}. We  say that an abstract vector space has a certain dimension $d$ if there are exactly $d< \infty$ vectors that both (i) span the entire space and that are (ii) linearly independent.~\footnote{In Chapter~\ref{chapter: event_graph_approach} we also encounter a related notion of dimension, that of \emph{polytope dimension} given by its affine dimension, which we define in Appendix~\ref{sec: convex polytopes}.} If we extend our consideration to infinite-dimensional spaces, where these lists can become infinitely large or even uncountable, the situation becomes significantly more nuanced. Nevertheless, it is still possible to discuss dimensionality meaningfully, provided certain additional mathematical tools are employed. 

\myindent In the example of the coin just given above, we have the two dimensions of the ambient space $\mathbb{R}^2 $ of the state space $\{(p,1-p) \in \mathbb{R}^2 \mid 0 \leq p \leq 1\}$. This is a two dimensional space since this is the cardinality of the canonical basis (that we have already encountered in Chapter~\ref{chapter: quantum coherence}) given by $\vert 0\rangle \equiv (1,0)$ and $\vert 1\rangle \equiv  (0,1)$. The state space of the coin is one-dimensional, and given by the (affine extension of the) line segment connecting these two points. 

\myindent It is important to recall that the dimension of a vector space \emph{depends on the underlying scalar field} associated with the vector space. If we consider a generic Hilbert space $\mathcal{H}$ it can have a certain dimension $d_{\mathbb{C}}$ associated with the field of complex numbers $\mathbb{C}$, as well as a certain dimension $d_{\mathbb{R}}$, associated to the field of real numbers $\mathbb{R}$. Generally, $d_{\mathbb{R}} \neq d_{\mathbb{C}}$.

\myindent When describing abstract state spaces and their dimension, it becomes evident that \emph{different} physical systems can share the same number of degrees of freedom and, consequently, be represented as tuples of points within the same abstract vector space. For instance,  coins of different denominations (25 or 50 cents) can have their statistical properties characterized by points $(p_1,1-p_1)$ or $(p_2,1-p_2)$ in the same abstract vector space $\mathbb{R}^2$, with respect to the same dimension, in this case, two. The two coins are different physical systems, but their behavior regarding being heads or tails once we flip them is described within the same abstract space.

\myindent Another interesting point to make is that, when expressing the relevant ambient abstract space we are considering, we always make some `choice'. This choice corresponds to our hypothesis on what is the state space of the system. When describing a coin, instead of describing it in terms of $(p,1-p)$ we could instead have considered $(p_1, p_2, 1-p_1-p_2)$ where $p_1$ is the probability that the coin lands heads, $p_2$ is the probability that the coin lands tails, and $1-p_1-p_2$ is the probability that the coin lands `up' (i.e., on its side). Every state described by $(p,1-p)$ is also described by a state $(p_1, p_2, 1-p_1-p_2)$ simply making $p_1+p_2=1$. Of course, for all practical purposes, the probability that the coin lands up is extremely small, yet conceivably different than zero depending on the properties of the coin, which implies that effectively a more precise description of the ambient dimension of this abstract vector space characterizing the statistical properties of the coin could have been a space of dimension $3$ instead of $2$. 

\myindent Therefore, we can easily perform our first dimension witnessing experiment. We flip a coin. We start and obtain heads, hence so far it seems that we can describe perfectly well the state of our coin using just a zero-dimensional (since it is completely described by a single point) state space characterizing the `heads' outcome. Now, we flip it again, and as soon we obtain tails we have witnessed that the dimension of the abstract state space describing the statistics of our coin is one. Maybe if we are extremely lucky, we can even see the coin land up and witness that in fact we need two dimensions to describe the coin's state precisely. 

\myindent If this would be it, dimension witnessing would certainly not be an entire section of this thesis. It is trivial to `witness' the dimension of a system given that we have complete knowledge of the relevant degrees of freedom that must govern the state (or dynamics) of a certain physical system. Another important thing is that this example showcases the fact that `upper bounding' the dimension is non-trivial (and perhaps~\footnote{Mathematically speaking, the task as described \emph{is} impossible since on can reproduce any statistics from a lower dimensional space using a higher dimensional one, by just remaining within a lower dimensional subspace of it. However, this does not exclude the possibility of elaborating \emph{physically} meaningful arguments, or statistical tests, in favor of upper bounding the dimension of state space of a physical system.} impossible) as there may always be some additional irrelevant degrees of freedom that play no role.

\myindent To make things more interesting, we can make the following changes in the protocol where we assume significantly less prior information of the functioning of the device (in the case above, the coin flipping procedure). Suppose that instead of knowing the physical system, we only have access to the statistics (described by the behaviors) of the scenario. The only thing we know is a certain probability $p(b|y,x)$, and that they were obtained in a prepare-and-measure setting, where $y$ labels different measurements and $x$ labels different preparations. For some data tables, we cannot say anything about the dimension. Suppose that a colleague handles us the data table $\{p(b|y,x)\}_{b,y,x}$, given by
\begin{equation*}
    \left\{p(0|0,1),p(1|0,1);p(0|0,2),p(1|0,2);p(0|0,3),p(1|0,3)\right\} = \left\{\frac{1}{2},\frac{1}{2};\frac{1}{2},\frac{1}{2};\frac{1}{2},\frac{1}{2}\right\},
\end{equation*}
can we infer the dimension from the behavior only? Clearly, in this case, we cannot know the dimension of the state space. It may be that the state space of the system has any dimension, and we cannot say anything better than the trivial lower bound of a $0$-dimensional system where the state space of the object is completely characterized by a single point. Therefore, if we do not know which measurements are performed and which states are prepared, or any other detail of the system, it gets less clear that we can still witness the dimension of the state space. However,  upon receiving the statistics
\begin{equation*}
    \left\{p(0|0,1),p(1|0,1),p(0|0,2),p(1|0,2),p(0|0,3),p(1|0,3)\right\} = \left\{1,0;0,1;\frac{1}{2},\frac{1}{2}\right\},
\end{equation*}
we may conclude that the system must at least have dimension \emph{two}. This is because, for the exact same measurement (labeled as `$0$') there were two preparations yielding exactly \emph{opposite} outcomes, hence distinguishable preparations. We can then relate these preparations with two distinguishable states of our system. Therefore, while for generic data we may not be able to witness the dimension of the state space,~\footnote{Of course, modulo trivial conclusions regarding the dimension of the state space as discussed before.} in some situations this is clearly possible, even classically.

\myindent If we consider \emph{quantum theory} a physical system $\mathrm{A}$ is assumed to have an associated Hilbert space $\mathcal{H}_A$. However, it is not assumed that this Hilbert space is the \emph{unique} Hilbert space in which the system $\mathrm{A}$ can be represented. Moreover, it is not assumed that different physical systems $\mathrm{A},\mathrm{B}$ must be associated to different Hilbert spaces $\mathcal{H}_A,\mathcal{H}_B$. In effect, quite the contrary! Normally, a single system $\mathrm{A}$ may be equivalently described by infinitely many Hilbert spaces, and infinitely many different physical systems may have their behavior and statistical predictions associated with the exact same Hilbert space. A finite-dimensional Hilbert space is one whose underlying vector space is also finite-dimensional, and we have already described how we talk about dimensions of vector spaces. Since physical systems are modeled by Hilbert spaces, their dimension is associated with the system's degrees of freedom. The task of dimension witnessing can then be described as follows:

\begin{tcolorbox}[
    colback=lightblue!10, % light blue background color with slight transparency
    colframe=lightblue,   % matching border color
    width=\textwidth,     % adjust width to fit text width
    boxrule=0.5mm,        % border thickness
    sharp corners,
    title=Box 3: Hilbert space dimension witnessing task,
    fonttitle=\bfseries,  % bold font for title
    title filled=true,    % filled title background
    coltitle=white        % white text color for title
]
\begin{task}\label{task: dimension witnessing} Provided solely with the behavior in a given fixed experimental scenario, and under the assumption that the behavior was generated via preparing and measuring quantum states with respect to some Hilbert space $\mathcal{H}_A$ of a physical system $\mathrm{A}$, provide the greatest  lower bound $d_{\mathrm{min}}$ such that $\dim(\mathcal{H}_A) \geq d_{\mathrm{min}}$. 
\end{task}
\end{tcolorbox}

\myindent We say that a lower bound $d_{\mathrm{min}}$ is trivial when $d_{\mathrm{min}}=1$. The lower bound $d_{\mathrm{min}}$ plays the role of a `minimal' dimension that the state space $\mathcal{H}_A$ of the system must have. We start with the elementary results for proposed witnesses that \emph{do not} distinguish between classical and quantum information resources. 

\subsection{Dimension witnesses based on Bell inequalities}\label{sec: dimension witnesses Bell scenarios}

\myindent \cite{brunner2008testing} introduced the idea of dimension witnessing, presenting it in the general context we have discussed before for behaviors in experimental scenarios, but focusing on \emph{behaviors in Bell scenarios}, described by specific correlations of the form
\begin{equation*}
    p(s_1s_2|x_1x_2) = \text{Tr}(\rho E_{s_1}^{x_1} \otimes E_{s_2}^{x_2})
\end{equation*}
where $\rho \in \mathbb{C}^d \otimes \mathbb{C}^d$ is some bipartite quantum state for some $d$ we want to lower bound and $M_{x_i} = \{E_{s_i}^{x_i}\}_{s_i}$ are local generalized (i.e. \acrshort{povm}) measurements. Therefore, in this setting we are granted the premise that the statistics generated are that of a Bell experiment, and therefore that our physical system $\mathrm{A}$ is associated with a Hilbert space of the form $\mathcal{H}_A = \mathcal{H}^{\otimes 2} = \mathbb{C}^d \otimes \mathbb{C}^d$ with total dimension $d_T = \dim(\mathcal{H}_A) = d^2$. The goal in this case is to therefore witness a lower bound $d \geq d_{\min}$, if possible, and use it to then infer a lower bound on $d_T$. 

\myindent The essence of the task is then to provide tools for which we can show that certain behaviors $B = \{p(s_1s_2|x_1x_2)\}_{s_1,s_2,x_2,x_2}$ require a Hilbert space with minimum dimension $d_{\mathrm{min}}$ to be generated. We establish this description as an instance of a quantum realization problem that we have already discussed before in Chapters~\ref{chapter: quantum coherence} and~\ref{chapter: contextuality}. 

\begin{definition}[Dimension dependent quantum realizability in compatibility scenarios]\label{def: dimension dependent compatibility scenario}
    Let $\pmb{\Upsilon}$ be any compatibility scenario. A behavior $B$ is said to be $d$-dimensional realizable if it is quantum realizable (according to Def.~\ref{def: quantum realizations of behaviors in compatibility scenarios}) in a Hilbert space of dimension $d$. The set of all such behaviors is denoted as $\mathcal{Q}^{(d)}(\pmb{\Upsilon})$. 
\end{definition}

\myindent Therefore, a generic dimension witnessing protocol for solving~\cref{task: dimension witnessing} is to provide witnesses (see Def.~\ref{def: witness}) capable of detecting gaps
\begin{equation}
    \mathcal{Q}^{(d_1)}(\pmb{\Upsilon}) \subsetneq \mathcal{Q}^{(d_2)}(\pmb{\Upsilon}),
\end{equation}
when $d_1 < d_2$. We note that while it is well known that for every compatibility scenario $\pmb{\Upsilon}$ the set $\mathcal{Q}(\pmb{\Upsilon})$ is convex, it is also known that, fixing a specific dimension, the set $\mathcal{Q}^{(d)}(\pmb{\Upsilon})$ is, in general, nonconvex~\citep{donohue2015identifying,sikora2016minimum,yu2024characterizing}.  Naturally, for Bell scenarios $\pmb{\Upsilon}_{\mathrm{Bell}}$, it is straightforward to consider witnesses based on optimizations over Bell inequalities. Given known Bell inequalities $I(\pmb{B}) \leq b$ bounding $\mathrm{NC}(\pmb{\Upsilon}_{\mathrm{Bell}})$, for some Bell scenario $\pmb{\Upsilon}_{\mathrm{Bell}}$, one may search for gaps by finding instances where
\begin{equation}
    I(\pmb{B}) \stackrel{L}{\leq} b \stackrel{d_1}{\leq} b_{d_1} \stackrel{d_2}{<} b_{d_2} \stackrel{}{\leq} b_{\max}^Q 
\end{equation}
where $d_1 < d_2$, $b$ is the so-called local bound making $I(\pmb{B}) \leq b$ a facet-defining inequality, and $b_{\max}^Q$ is the largest value that the functional $I(\pmb{B})$ can achieve for any $B \in \mathcal{Q}(\pmb{\Upsilon}_{\mathrm{Bell}})$. Similarly, $b_{d}$ is the largest value that $I(\pmb{B})$ can achieve for any $B \in \mathcal{Q}^{(d)}(\pmb{\Upsilon}_{\mathrm{Bell}})$. To present such witnesses,~\cite{brunner2008testing} used the family of Bell inequalities introduced by Collins, Gisin, Linden, Massar, and 
Popescu (\acrshort{cglmp}), where they have found qutrit witnesses of $\mathbb{C}^3 \otimes \mathbb{C}^3$. Independently, Vértesi and Pál (see~\citep[Table V]{pal2008efficiency} and~\citep{vertesi2008generalized})  also found various similar gaps between behaviors achievable using two-qubit and two-qutrit quantum systems in bipartite Bell scenarios.~\footnote{Interestingly,~\cite{pal2008efficiency} have shown numerically and theoretically that there was no gap between optimal violations of bipartite Bell inequalities using complex Hilbert spaces or real Hilbert spaces, a result later generalized by~\cite{McKague2009simulating} for any Bell scenario as we have mentioned in Chapter~\ref{chapter: quantum coherence}.} 

\myindent These results have impulsed a new field of research devoted both to investigating the sets of quantum behaviors $\mathcal{Q}^{(d)}(\pmb{\Upsilon}_{\mathrm{Bell}})$ as well as proposing linear and nonlinear functionals that can be interpreted as dimension witnesses in the context of Bell scenarios. One well-known systematic approach to such a problem is by using semidefinite programming (\acrshort{sdp}) techniques~\citep{tavakoli2023semidefinite} that allow for a dimension restriction~\citep{moroder2013device,mironowicz2014properties,navascues2015bounding,navascues2015characterizing}. Dimension witnesses based on Bell inequalities were experimentally tested and have since became a standard benchmark when characterizing quantum hardware~\citep{lo2016experimental}. Such gaps have also been connected to randomness certification by~\cite{li2013relationship}.

\myindent On a similar note, as noted by~\cite{acin2006grothendieck} and~\cite{brunner2008testing}, there exists a relationship between local hidden variable models of Werner states~\citep{werner1989quantum} and \emph{Grothendieck’s constant}. This has led to improvements in local hidden variable models for quantum realizable behaviors in high-dimensional Hilbert spaces. The connection between Hilbert space dimension witnesses via Bell inequalities, local hidden variable models (specifically those for Werner states and projective measurements) and the Grothendieck’s constant have been used by~\cite{briet2011generalized} as a tool for investigating behaviors $B \in \pmb{\Upsilon}_{\mathrm{Bell}}$ in Bell scenarios requiring \emph{both} large Hilbert space dimension and entanglement. Since then, bounding Grothendieck’s constant by improving bounds on local hidden variable models in Bell scenarios with many measurement settings has become an interesting subfield of research~\citep{pitowsky2008newbell,hirsch2017betterlocalhidden,diviansky2017qutrit,designolle2024betterboundsgrothendieckconstants,designolle2024improved} at the intersection of quantum foundations and pure mathematics. 

\myindent Much more can be said about dimension witnessing via the violation of Bell inequalities~\citep{perez2008unbounded,eltschka2013negativity,navascues2014characterization}. To conclude this brief overview, we point out simple aspects of the tests described so far. First, it is clear that in order to violate a Bell inequality the states and measurements considered need to have some form of coherence, as it is well known that entanglement is necessary for the violation of Bell inequalities. Therefore, the gaps considered witness \emph{both} Hilbert space dimension, and that models satisfying Bell's notion of local causality cannot reproduce the observed statistics. We will see later that the same is not immediately true for witnesses based on inequalities in prepare-and-measure scenarios. Second, one could also point out that \emph{any} violation of a bipartite Bell inequality can witness that $d_T = \dim(\mathcal{H} \otimes \mathcal{H}) \geq 2^2 = 4$ since one needs at least two entangled qubits for violating any Bell inequality. Finally, and motivated by this last remark, one can ask if it is possible to witness $d_T \geq 3$ if instead of considering Bell scenarios $\pmb{\Upsilon}_{\mathrm{Bell}}$ we consider more general compatibility scenarios. As we will presently see, \acrshort{ks} noncontextuality inequalities can also be used for witnessing Hilbert space dimension. 

\subsection{Dimension witnesses based on Kochen--Specker scenarios}\label{sec: dimension witnesses KS scenarios}

\myindent As has been proven by Kochen and Specker~\citep{kochen1975problem} there exists a \acrshort{ks} noncontextual ontological model for $\mathcal{H} \simeq \mathbb{C}^2$, implying that for any compatibility scenario $\pmb{\Upsilon}$ we have that $\mathcal{Q}^{(2)}(\pmb{\Upsilon}) \subseteq \mathrm{NC}(\pmb{\Upsilon})$. For a modern account and more pedagogical presentation of this result we refer the interest reader to~\cite{wright2023contextualityin}, which generalizes the \acrshort{ks} noncontextual ontological model first introduced by~\cite{kochen1975problem} to the fragment of quantum theory consisting of generic $n$-qubit unentangled measurements and $n$-qubit fully separable (product) states.

\myindent Using this fact,~\cite{guhne2014bounding} showed that \acrshort{ks} noncontextuality inequalities could also be used as dimension witnesses, even when one relaxes the constraint that the behaviors must satisfy the compatibility conditions of their associated scenario. Using the \acrshort{csw} graph-theoretic approach, and the insights that exclusive outcomes are able to provide some information on the dimension of the Hilbert space (as we have discussed previously),~\cite{ray2021graphdimension}  introduced a more systematic analysis of how \acrshort{ks} noncontextuality inequalities and graph approaches can yield dimension witnesses. There the authors provide explicit examples of dimension witnesses able to encounter lower bounds up to Hilbert space dimension $d_{\min} = 7$, and explicitly construct an infinite family of exclusivity graphs $G_k$---dubbed $k$-Qite graphs---yielding inequalities given by 
\begin{equation}\label{eq: Qite dimension witnesses}
    \sum_{v \in V(G_k)} p(v) \leq k
\end{equation}
for which the violation of these inequalities is possible if and only if the underlying Hilbert space dimension is at least $d_{\min} = k+1$. 

\myindent Although we introduce, in Chapters~\ref{chapter: event_graph_approach} and~\ref{chapter: applications}, dimension witnesses via a graph-theoretic inequality-based approach, these are not immediately dimension witnesses due to their relation to \acrshort{ks} noncontextuality inequalities. This is because our dimension witnesses require significantly fewer constraints than those imposed for such tests (as we discuss in Chapter~\ref{chapter: applications}). In fact, as opposed to \acrshort{ks} noncontextuality inequalities the dimension-witnessing inequalities we construct have, in general, violations by states and measurements on systems $\mathcal{H}\simeq \mathbb{C}^2$. The dimension witnesses we introduce later in Chapter~\ref{chapter: applications} are not related to Bell inequalities either, but related to specific behaviors in prepare-and-measure scenarios. We now proceed to review what is known about dimension witnesses in such scenarios.

\subsection{Dimension witnesses based on prepare-and-measure inequalities}\label{sec: dimension witnesses PM scenarios}

\myindent We have already encountered prepare-and-measure scenarios before in Chapter~\ref{chapter: contextuality}, when considering Spekkens's notion of generalized noncontextuality. Other notions of nonclassicality can be investigated in such scenarios (see for instance~\citep{degois2021general}), one of which being the notion of set coherence as we show in Chapter~\ref{chapter: relational coherence}. However, and perhaps most relevantly, these scenarios have been largely investigated in connection with their usefulness in communication games, such as parity oblivious games~\citep{spekkens2009preparation}, quantum random access codes~\citep{ambainis1998densequantumcodinglower,ambainis2002dense}, and randomness certification~\citep{bowles2014certifying,vicente2017shared}. Witnessing dimension is, in fact, important in such scenarios as in most cases the underlying dimension of the message is a \emph{key} element in making the scenario (and tasks based on them) useful.

\myindent We start defining the set of behaviors in a prepare-and-measure scenario that are quantum (or classical) realizable by state spaces of a certain dimension. 

\begin{definition}[Dimension dependent  realizability in prepare-and-measure scenarios]\label{def: dimension dependent PM scenario}
    Let the triplet of positive integers $(I, J, K)$ define a prepare-and-measure scenario as by Def.~\ref{PM scenario}, with $I$ preparations, $J$ measurements having $K$ outcomes each. We say that a behavior $B = \{p(k|M_j,P_i)\}$ in $(I,J,K)$ is 
    \begin{enumerate}
    \item[(i)] $d$-dimensional quantum realizable if it is quantum realizable (according to Def.~\ref{def: quantum realizable PM behavior}) in a Hilbert space of dimension $d$, i.e., $\mathcal{H} \simeq \mathbb{C}^d$. We denote the set of all such behaviors as $\mathcal{Q}^{(d)}(I, J, K)$.
    \item[(ii)] $d$-dimensional incoherent realizable (for states) if it is $d$-dimensional quantum realizable but where $\{\rho_i\}_i \subseteq \mathcal{D}(\mathbb{C}^d)$ is set incoherent. We denote this set as $\mathbb{P}_{I,J,K}^{d}$.
    \item[(iii)] $d$-dimensional incoherent realizable powered by some shared randomness if it is in the convex hull of $\mathbb{P}_{I,J,K}^{d}$. We denote this set as $\mathfrak{P}_{I,J,K}^{d} = \mathrm{ConvHull}(\mathbb{P}_{I,J,K}^{d})$.
    \end{enumerate}
\end{definition}

\myindent We have denoted $\mathfrak{P}_{I,J,K}^{d}$ in this manner as this is our notation for convex polytopes. We also note that, as we have already pointed out before, dimension restrictions usually make the relevant set of behaviors in a given scenario \emph{nonconvex}. A family of nonconvex dimension witnesses has been introduced by~\cite{bowles2014certifying}, related to scenarios $\mathbb{P}_{I,J,K}^d$. To the best of our knowledge, the first to introduce and investigate the polytopes $\mathfrak{P}_{I,J,K}^{d}$ were~\cite{gallego2010device} and the first work to have considered dimension witnesses in generic prepare-and-measure scenarios was~\cite{wehner2008lowerbound}. 

\myindent In their work,~\cite{gallego2010device} have characterized facet-defining inequalities for the convex polytopes $\mathfrak{P}_{3,2,2}^{2}$ and $\mathfrak{P}_{3,2,2}^{3}$, and have provided an infinite family of valid inequalities (that are dimension witnesses) for the polytopes $\mathfrak{P}_{N,N-1,2}^{d}$ given by 
\begin{equation}\label{eq: gallego PM witnesses}
    I_{N,d}(\pmb{B}) := \sum_{j=1}^N \Delta p(M_j|P_1) + \sum_{i=2}^{N}\sum_{j=1}^{N+1-i}\alpha_{ij}\Delta p(M_j|P_i) \leq \frac{N(N-3)}{2}+2d-1
\end{equation}
where $\Delta p(M_j|P_i) := p(0|M_j,P_i)-p(1|M_j,P_i)$ and $\alpha_{ij} = 1$ if $i+j \leq N$ and $-1$ otherwise. In particular, they have also showed that $I_{N,N-1}(B) \leq N(N+1)/2-1$ are not only valid but also facet-defining for the polytopes $\mathfrak{P}_{N,N-1,2}^{N-1}$ for all $N \leq 5$. These witnesses were then experimentally verified by~\cite{hendrych2012experimental}.

\myindent One important aspect of prepare-and-measure scenarios is that they are only interesting---in the sense of having a gap between what is achievable by quantum realizations or by set incoherent realizations---if one imposes some restrictions on the messages (preparation procedures) that are implemented. If this is not done, then any correlation is merely equivalent to a classical correlation. We have already encountered one such possible restriction in the form of the operational equivalences for preparation procedures that we have introduced in Chapter~\ref{chapter: contextuality}, with the \acrshort{lsss} constraints being our most relevant example.~\cite{gallego2010device} has considered a different restriction, more commonly imposed in the context of (quantum) random access codes, which is of assuming an \emph{upper bound} on the dimension of the state space of the message. Specifically, they have assumed that the Hilbert space dimension $d$ of the systems $\rho_x$ is smaller than $I$, the number of preparations. In fact, various other restrictions are possible such as energy-based or information-based restrictions, each of which defines a different polytope of behaviors for the associated prepare-and-measure scenarios~\citep{tavakoli2022informationally,pauwels2024informationcapacityquantumcommunication}. One major output of this thesis is the formalization of the notion of an \emph{overlap restriction} for such prepare-and-measure scenarios, which differs from what has already been considered (for example, by~\cite{shi2019semideviceindependent}). In our case, we assume that the behavior is described by tuples of two-state overlaps, which leads to the convex polytopes we shall study in Chapter~\ref{chapter: event_graph_approach}. 

\myindent As a matter of fact, one example of a dimension witness based solely on overlaps has been considered by~\cite{galvao2020quantum}, and tested by~\cite{giordani2021witnesses}. These were dimension witnesses given by the $c_3(r) \leq 1$ inequality we have presented in Eq.~\ref{eq: overlap cycle inequalities}. There, the authors have shown that prepare-and-measure scenarios, together with the assumption that we restrict our attention to behaviors of the form $p(k|j,i) = \text{Tr}(\rho_i\rho_{k|j})$, allow for dimension witnesses as there is a gap between overlap triplets $(\text{Tr}(\rho_1\rho_2),\text{Tr}(\rho_1\rho_3),\text{Tr}(\rho_2\rho_3))$ that can be achieved using two or three-dimensional Hilbert spaces. We discuss more on this topic in Chapters~\ref{chapter: Bargmann invariants} and~\ref{chapter: applications}. 

\myindent An important aspect of the witnesses considered by~\cite{gallego2010device} is that, for each dimension $d$, there is a gap between what is possible using set incoherent and set coherent realizations (vide Def.~\ref{def: quantum realizable PM behavior}), but these gaps are hard to find numerically. This issue was resolved by~\cite{brunner2013dimension}, who proposed a family of dimension witnesses for any dimension $d$ having distinct classical and quantum bounds found analytically. They have considered prepare-and-measure scenarios $(I,J,K)$ where $I = N, J = N(N-1)/2$ and $K=2$. They have also noted that the witnessing task is only nontrivial if the Hilbert space dimension satisfy $d < N$, as we have mentioned previously.

\myindent Their witnesses are based on the following quantity:
\begin{equation}\label{eq: PM witnesses navascues}
    W_N(\pmb{B}) := \sum_{i>i'}|p(M_{i,i'}|P_i)-p(M_{i,i'}|P_{i'})|^2 
\end{equation}
where all $N(N-1)/2$ dichotomic measurements are labeled using $(i,i')$ with $i,i'=1,\dots,N$ and $i > i'$. Hence,  for example, for $N=3$ they have the $3(3-1)/2=3$ measurements $M_{(3,2)},M_{(3,1)},M_{(2,1)}$. Recall that we denote $p(M_j|P_i) \equiv p(1|M_j,P_i)$.  We now note that the witness described above is a convex functional, implying that optimal values can be taken to be described by pure states. This is easy to see from the following calculations:

\begin{align*}
    &W_N(\alpha \pmb{B} + (1-\alpha)\tilde{\pmb{B}}) =\\
    &=\sum_{i>i'}|\left(\alpha p(M_{i,i'}|P_i)+(1-\alpha)\tilde{p}(M_{i,i'}|P_i)\right)-\left(\alpha p(M_{i,i'}|P_{i'})+(1-\alpha)\tilde{p}(M_{i,i'}|P_{i'})\right)|^2 \\
    &=\sum_{i>i'}|\alpha\left( p(M_{i,i'}|P_i)-p(M_{i,i'}|P_{i'})\right)+(1-\alpha)\left(\tilde{p}(M_{i,i'}|P_i)-\tilde{p}(M_{i,i'}|P_{i'})\right)|^2 \\
    &\leq \alpha \sum_{i>i'}|p(M_{i,i'}|P_i)-p(M_{i,i'}|P_{i'})|^2+(1-\alpha)\sum_{i<i'}|\tilde{p}(M_{i,i'}|P_i)-\tilde{p}(M_{i,i'}|P_{i'})|^2\\
    &=\alpha W_N(\pmb{B}) + (1-\alpha)W_N(\tilde{\pmb{B}}),
\end{align*}
where we have used that $|sa+(1-s)b|^2 \leq |sa|^2+|(1-s)b|^2 = s^2|a|^2+(1-s)^2|b|^2 \leq s|a|^2+(1-s)|b|^2$ for any $0 \leq s \leq 1$ and $a,b \in \mathbb{R}$.

\myindent  Then,~\cite{brunner2013dimension} proceed to show that, whenever the Hilbert space dimension of the message states $d = \dim(\mathcal{H})$ satisfies $d \leq N$ it follows that 
\begin{equation*}
    W_N(\pmb{B}) \stackrel{\mathrm{inc}}{\leq} C_d \leq Q_d
\end{equation*}
where $C_d$ is a bound on $d$-dimensional quantum realizations by set incoherent states given by  $$C_d = \frac{N(N-1)}{2} - \left\lfloor \frac{N}{d}\right\rfloor \left(N - \frac{d}{2} \left( \left\lfloor \frac{N}{d}\right\rfloor +1\right) \right)$$ and $Q_d$ is a generic bound for $d$-dimensional quantum realizations by set coherent states $$Q_d = \frac{N^2}{2}\left(1-\frac{1}{d}\right).$$
These predictions were later experimentally verified by~\cite{sun2016experimental}.~\footnote{They have also, interestingly, witnessed set imaginarity.}

\begin{figure}[t]
    \centering
    \includegraphics[width=0.85\linewidth]{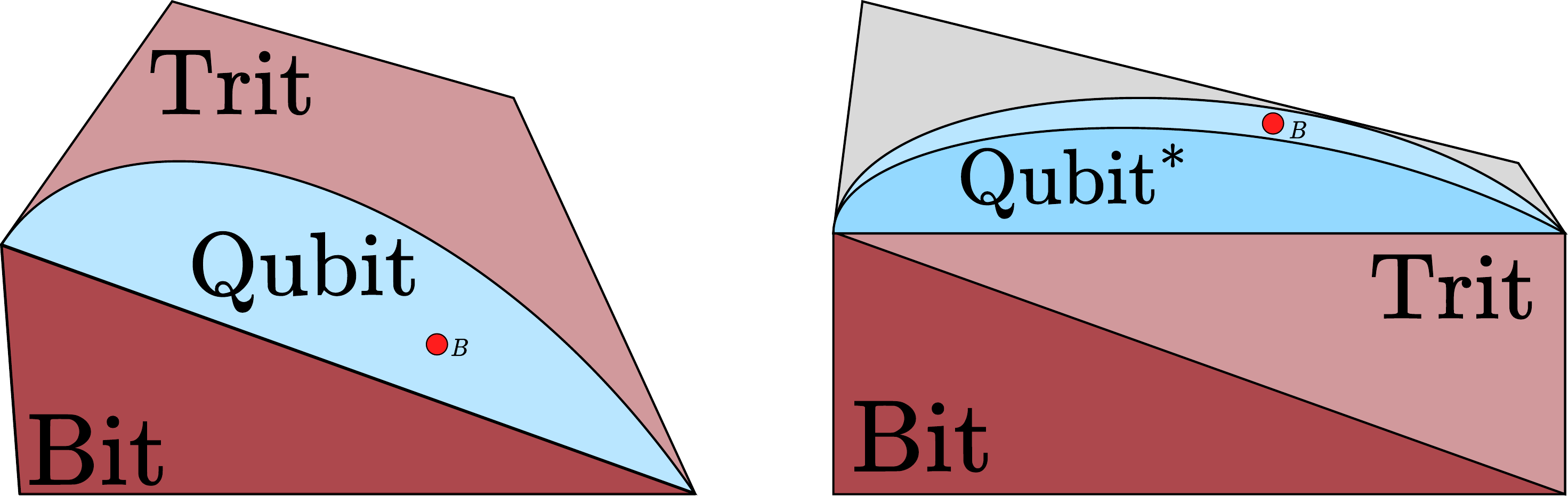}
    \caption{\textbf{Distinguishing between dimension witnesses that can test for coherence or not.} On the left, we have a situation where it is possible to witness Hilbert space dimension. A behavior $B$  is shown in red. In this situation, every behavior $B$ is quantum realizable. Therefore, we can certify that the behavior shown is not described by a bit (set incoherent pair $\mathbb{D}_2 = \{\vert 0\rangle, \vert 1\rangle\}$ or convex combinations thereof), but it could, in principle, be described either by a trit (set incoherent triplet $\mathbb{D}_3$) \emph{or} by a qubit. We cannot distinguish between incoherent realizability and coherent realizability. On the right, there is a distinction between coherent and incoherent realizability. We show the set of $2$-dimensional coherent realizable behaviors as `Qubits$^*$' since we are showing the convex hull of the set. The behavior $B$ shown (red circle) is guaranteed to be quantum realizable by coherent states spanning a Hilbert space of dimension at least $d_{\min} = 3$. }
    \label{fig: coherence and dimension part I}
\end{figure}

\subsection{Stringent tests on dimension}\label{sec: stringent dimension}

\myindent It is important to note that the witnesses just described in Eqs.~\eqref{eq: gallego PM witnesses} and~\eqref{eq: PM witnesses navascues},  \emph{cannot} witness the difference between dimension arising due to nontrivial coherent states of Hilbert spaces, or simply from the fact that we have more dimensions available, such as access to $\mathbb{D}_d$ for some $d=2,3,\dots$ etc. Similarly, the witnesses proposed by~\citep{galvao2020quantum} cannot distinguish between witnessing the state space of a classical trit and a single qubit, although in their case they have witnesses of nonclassicality that are independent of Hilbert space dimension (as we discuss more on Chapter~\ref{chapter: relational coherence}) implying that they can distinguish between a qubit and a classical $d$it for all possible choices of $d \in \mathbb{N}$. 

\myindent This is illustrated in Figure~\ref{fig: coherence and dimension part I}. In this figure, we see that unless we assume the message has a certain dimension, say $d$, the value of the witness alone cannot be used to distinguish if the behavior was realized by set coherent quantum states or by set incoherent quantum states of higher dimension. This is because these existing witnesses are associated with spaces such as those shown in Fig.~\ref{fig: coherence and dimension part I} on the left. 

\myindent This raises the question of whether we are not only interested in witnessing the Hilbert space dimensionality of a system, but also in having guarantees that we are \emph{nontrivially processing information} in these additional degrees of freedom, beyond what is possible using only incoherent states. This leads to the idea of stringent dimension witnesses that certify \emph{both} the Hilbert space dimension \emph{and} some additional nontrivial aspects of the system's dimension. One such possibility is shown in Fig.~\ref{fig: coherence and dimension part I} on the right, where we have a prepare-and-measure scenario, and we find that there are witnesses whose values \emph{cannot} be achieved by any set incoherent realization.

\begin{tcolorbox}[
    colback=lightblue!10, % light blue background color with slight transparency
    colframe=lightblue,   % matching border color
    width=\textwidth,     % adjust width to fit text width
    boxrule=0.5mm,        % border thickness
    sharp corners,
    title=Box 4: Witnessing coherence and dimension,
    fonttitle=\bfseries,  % bold font for title
    title filled=true,    % filled title background
    coltitle=white        % white text color for title
]
\begin{question}\label{question: dimension and coherence witnessing} Is it possible to define a prepare-and-measure scenario, and some restriction on the behaviors of these scenarios (e.g., upper bound on the dimension or information of the message), such that there exists a witness $W$ satisfying $$W(\pmb{B}) \leq C \leq Q_{d_1} < Q_{d_2},$$
where $C$ bounds \emph{every} $d$-dimensional incoherent realizable behavior $B$ (see Def.~\ref{def: dimension dependent PM scenario}), and $Q_{d_i}$ bounds $d_i$-dimensional quantum realizations, with $Q_{d_1} \neq Q_{d_2}$ and $d_1 < d_2$?
\end{question}
\end{tcolorbox}

\myindent To the best of our knowledge~\cref{question: dimension and coherence witnessing} has not yet been answered in the context of prepare-and-measure scenarios. We provide an affirmative answer to this question in Chapter~\ref{chapter: applications} using the methods we introduce in Chapters~\ref{chapter: event_graph_approach} and~\ref{chapter: relational coherence}. 

\myindent We have already encountered one notion of \emph{basis-dependent} coherence and dimension witnesses in Chapter~\ref{chapter: quantum coherence}. Since here we are focusing on behaviors, it is natural to focus on resources that are basis-independent, however, we recall that the results from~\cite{ringbauer2018certification} have introduced basis-dependent witnesses of coherence and dimension, using the notion of coherence rank. 

\myindent Certifying both coherence and dimension, as we have described in~\cref{question: dimension and coherence witnessing}, is our first proposal for a stringent test of Hilbert space dimension.~\footnote{In future work we intend to consider other stringent tests of Hilbert space dimension. } However, we note that there are other proposals in the literature that have also been considered in the context of prepare-and-measure scenarios. The first one is the notion of \emph{irreducible dimension}, first introduced by~\cite{cong2017witnessing} in the context of Bell scenarios and dimension witnesses based on behaviors in Bell scenarios $\pmb{\Upsilon}_{\mathrm{Bell}}$, which was later considered (and experimentally implemented) in prepare-and-measure scenarios by~\cite{aguilar2018certifying}. Describing the idea shortly and in an oversimplified manner, one would like to distinguish between processing information with a four-dimensional quantum system from using two distinguishable instances of the same two-dimensional system in sequence. As showed by~\cite{cong2017witnessing}, the dimension witnesses introduced considered in~\cite{brunner2008testing} based on the \acrshort{cglmp} family of inequalities that we have discussed earlier are \emph{not} witnesses of this form of irreducible dimension. Recently, another proposal for a stringent test on dimension based on simulability has been considered by~\cite{bernal2024absolutedimension}, which they have termed the \emph{absolute dimension} of a set of states.

\section{Quantum interrogation}~\label{sec: quantum interrogation}

\myindent In this Section we describe a task known as \emph{quantum interrogation} that was first introduced by~\cite{elitzur1993quantum}. For this thesis, we are focused only on this first original description of the task, which we refer to as the \emph{standard} quantum interrogation scheme, also known in the literature as the \emph{standard} interaction-free measurement (\acrshort{ifm}). The test introduced by~\cite{elitzur1993quantum} is now known as the  \emph{standard} \acrshort{ifm} test because ever since this introduction various related schemes for \acrshort{ifm} have been developed. For example, in the standard test the optimal efficiency of the protocol is $\eta = \sfrac{1}{2}$. This has been experimentally verified and improved by~\cite{kwiat1995interactionfree,kwiat1999high} by showing how a slight modification of the scheme (using the quantum Zeno effect) can lead, in principle, to efficiencies that tend to perfection, i.e., $\eta \to 1$. This same scheme was then robustly analyzed by~\cite{rudolph2000better}. 

\myindent The seminal proposal by~\cite{elitzur1993quantum} led to various breakthroughs in quantum imaging, quantum communication, and quantum computation. \emph{Quantum imaging} refers to the field of research exploring how quantum states of light can be used to enhance the resolution (or other aspects) of imaging protocols through nonclassical correlations, such as those arising from quantum coherence and quantum entanglement. For example, since the experiment introduced  by~\cite{elitzur1993quantum} can detect a bit of information indicating the presence or absence of an object in an interferometer it can serve as the basis for performing what is now known as \emph{interaction free imaging}~\citep{white1998interactionfree}. Such schemes have been applied, for example, in the context of performing interaction-free versions of ghost imaging~\citep{pittman1995optical} by~\cite{zhang2019interactionfree}. One notable, albeit speculative, opportunity lies in performing imaging with biological tissue, as it could theoretically reduce the light dose absorbed by the object~\citep{kwiat1996quantumseeing,cohen2025lossresilient}.

\myindent A related scheme for imaging was developed by~\cite{lemos2014quantum}(see also~\citep{lahiri2015theory}), which enables quantum imaging in which the photons interacting with the object are never detected, and the signal is constructed solely from photons that have never interacted with the object.~\footnote{In this context, and in our following discussion, the term `interaction' refers to a 'particle-like' interaction, such as the photon interpreted as a particle either passes through an object or is detected by it, rather than a quantum mechanical wave-function interference between the quantum state of a photon and the quantum state of the object, which is the more accurate description. We use the term `interaction' in this sense as it is traditional in this literature. } This technique has also been implemented with what is known as `classical light'~\citep{cardoso2019classical}, which is jargon of the field of quantum optics for light beams that are outputs of a laser. More broadly speaking, in quantum optics one refers to classical light or a classical state of light as one that has a positive quasiprobability Glauber--Sudarshan $P$ representation~\citep{glauber1963coherent,sudarshan1963equivalence}. Interestingly, the technique is sufficiently powerful to allow for complete state tomography~\citep{fuenzalida2024quantum}. We refer to~\citep{lemos2022quantum} for a recent review and tutorial on how to perform such tests. 

\myindent As we have already mentioned in the end of last Chapter, the test considered by~\cite{elitzur1993quantum} works by following a type of reasoning known in philosophy and logic as a \emph{counterfactual}~\citep{lewis1979counterfactuals}. In the experiment we apply the counterfactual `if the interferometer had no object inside, detector two would never click'. As another example, in Italy, one has the famous sarcastic counterfactual `If my grandmother had wheels, she would be a bicycle'.~\footnote{``\textit{Se mia nonna avesse le ruote sarebbe una carriola.}''} Albeit simple, the idea that quantum coherence and interferometric experiments allow, using counterfactual reasoning, to unambiguously learn one bit of information (in our case, indicating whether an object is present inside the interferometer) led to the development of protocols that are now known as \emph{counterfactual computation}~\citep{mitchison2001counterfactual,hosten2006counterfactual} and \emph{counterfactual communication}~\citep{salih2013protocol}. We refer the interested reader to the thesis by~\cite{hance2023interplay} and references therein.

\myindent The setup necessary for understanding the standard interrogation protocol uses the Mach--Zehnder interferometer (\acrshort{mzi}), which we now introduce.

\subsection{Mach--Zehnder Interferometer}\label{subsec: MZI}

\begin{figure}[t]
    \centering
    \includegraphics[width=0.6\textwidth]{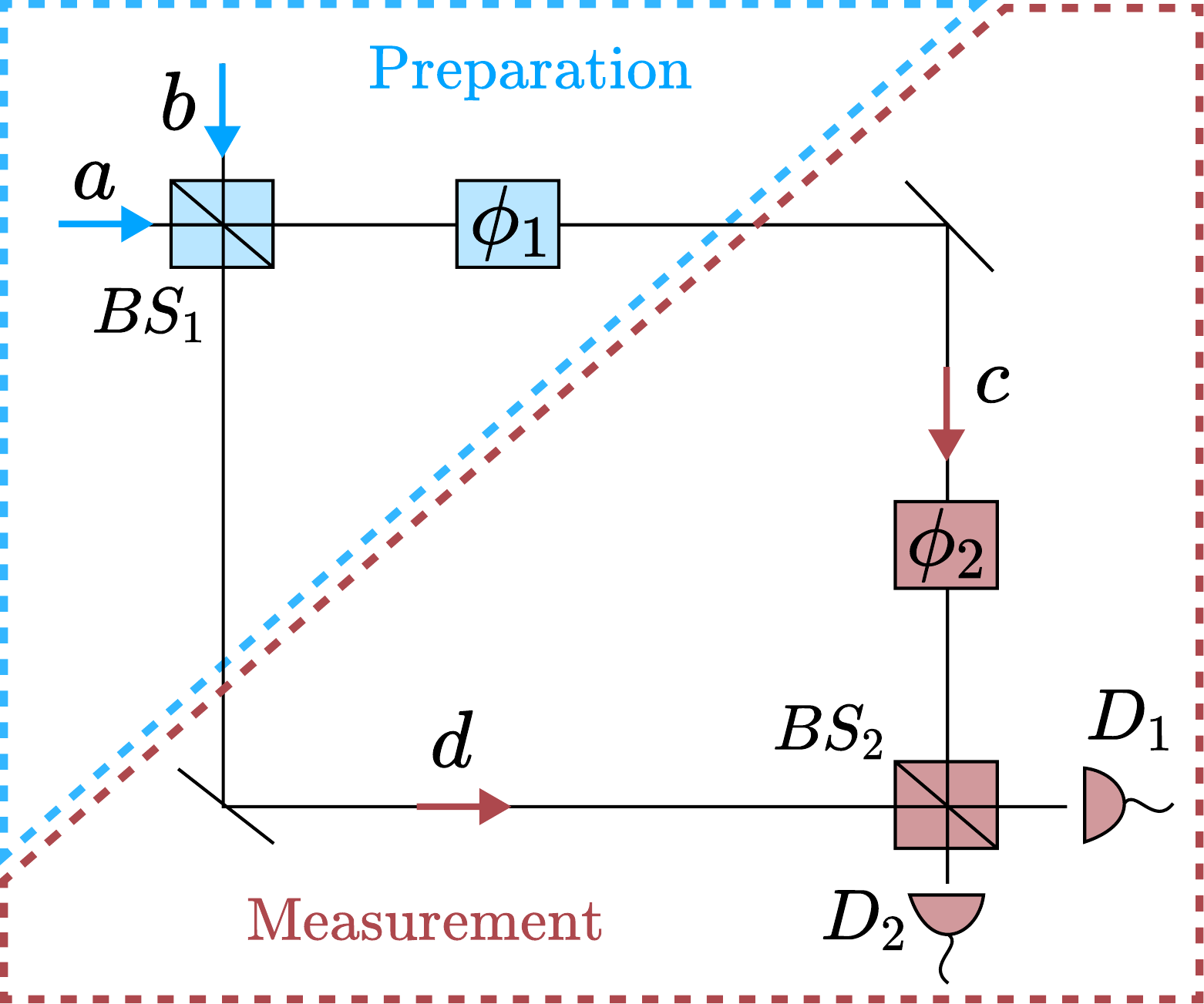}
    \caption{\textbf{Mach-Zehnder Interferometer as a prepare-and-measure experiment.} In the preparation stage the interferometer is fed with a single photon. The first beam-splitter, $BS_1$ generates superposition between the spatial modes, according to some angle $\theta_1$. Qubit path encoding is chosen as $\ket{0}$ and $\ket{1}$ for modes \textit{a} and \textit{b}, respectively. A phase-shifter $\phi_1$, is placed after $BS_1$. The measurement stage is made of a phase-shifter $\phi_2$, and a second beam-splitter $BS_2$ parametrized by some angle $\theta_2$, with photo-detectors placed after each output mode. Note that it suffices in practice to use only a single phase-shifter to adequately describe the statistics of a \acrshort{mzi}. Here, we have separated into two since we are interpreting the \acrshort{mzi} as a prepare-and-measure experiment.}
    \label{fig:MZI}
\end{figure}

\myindent A Mach--Zehnder interferometer (\acrshort{mzi}) is a particularly simple device capable of demonstrating the wave-like behavior of photons~\citep{loudon2000quantum,zetie2000does}. We show this device in Fig.~\ref{fig:MZI}. In its standard configuration, a \acrshort{mzi} is made of two beam-splitters (\acrshort{bs}s) and two mirrors, with optical paths of equal length. Quantum information is encoded on the photons' path, and single photon interference is captured by the setting of a phase-shifter (\acrshort{ps}) tuned inside one of the arms. 

\myindent To build a connection with prepare-and-measure contextuality scenarios later in Chapter~\ref{chapter: applications}, we interpret the \acrshort{mzi} as a prepare-and-measure device. The two \acrshort{bs}s are defined with a fixed $\sfrac{\pi}{2}$ phase-shift between the optical modes and tunable transmissivity in terms of parameters $\theta_k$,
\begin{equation}\label{eq: BS}
    U_{\theta_k} := \left(\begin{matrix}\cos\theta_k&i\sin\theta_k\\i\sin\theta_k&\cos\theta_k\end{matrix}\right),
\end{equation}
where $k=1,2$ labels which \acrshort{bs} is considered.  In this configuration, the \emph{preparation stage} plays the role of a universal one-qubit state generator. Given the orthonormal basis $\{\ket{0},\ket{1}\}$, corresponding to the upper and the lower optical paths respectively, we can encode any pure state using a single photon that enters $BS_1$, for instance in mode $a$, using:
\begin{align}\label{eq: state inside the interferometer}
    \ket{\psi(\theta_1,\phi_1)} = e^{i\phi_1}\cos\theta_1 \ket{0}+i\sin\theta_1\ket{1}.
\end{align}
In optics, this encoding of a qubit into the presence or absence of a single photon in two waveguides is known as a \emph{dual-rail encoding}. Here, we have used that each \acrshort{ps} modeled unitarily as $U_{\phi_i}$ is represented by the matrix~\footnote{We note our choice is different than what is usually considered, where the phase is applied to state $\vert 1\rangle$, so the phase appears in the lower diagonal element instead.} 
\begin{equation}\label{eq: PS}
    U_{\phi_k} = \left(\begin{matrix}e^{i\phi_k}& 0\\ 0& 1\end{matrix}\right).
\end{equation}

\myindent Similarly, the \emph{measurement stage} is made of a \acrshort{ps} defined by a unitary $U_{\phi_2}$ and a \acrshort{bs} with photo-detectors at both outputs, a configuration that allows projection onto any qubit state. For instance, if we choose values $\theta_2,\phi_2$ such that  $U_{\phi_1}U_{\phi_2}=\mathbb{1}$ and $U_{\theta_1}U_{\theta_2}=\mathbb{1}$, the measurement stage performs exactly the conjugated operation of the preparation stage, and thus a measurement that exactly corresponds to a projector onto state $\ket{\psi(\theta_1,\phi_1)}\bra{\psi(\theta_1,\phi_1)}$. In the ideal case, detector $D_1$ will click with probability $1 = \vert \langle \psi(\theta_1,\phi_1)\ket{\psi(\theta_1,\phi_1)} \vert^2$, while detector $D_2$ will click with probability $0$. In such a case, the measurement stage performs a dichotomic measurement $M=\{\ket{\psi(\theta_1,\phi_1)}\bra{\psi(\theta_1,\phi_1)}, \mathbb{1}- \ket{\psi(\theta_1,\phi_1)}\bra{\psi(\theta_1,\phi_1)}\}$. 

\myindent It is worth noting that this same measurement $M$ can also perfectly distinguish between the case with an input photon in mode ${a}$, and the one in mode ${b}$. If we input the photon in mode ${b}$, detector $D_1$ will never click, as state $\ket{\psi(\theta_1,\phi_1)^\perp}\bra{\psi(\theta_1,\phi_1)^\perp}$---orthogonal to $\ket{\psi(\theta_1,\phi_1)}\bra{\psi(\theta_1,\phi_1)}$---is prepared inside the interferometer instead.

\myindent It is interesting to look at the case where the two \acrshort{bs}s are characterized by \textit{different} parameters $\theta$ and $\phi$. By interpreting the measurement stage as a time-reversed one-qubit state generator for a pure state $\ket{\psi(\theta_2,\phi_2)}\bra{\psi(\theta_2,\phi_2)}$, the overall action of the interferometer is to project the state prepared in the first stage onto the state prepared in the second
\begin{align}
    |\bra{0} U_{BS_{2}}U_{\phi_2}U_{\phi_1}U_{BS_1}\ket{0}|^2 = |\langle \psi(\theta_2,\phi_2)\vert \psi(\theta_1,\phi_1) \rangle|^2.\label{eq: PM MZI as overlaps}
\end{align}
This perspective allows to interpret the \acrshort{mzi} as a natural device for estimating quantum two-state overlaps from the frequency of clicks in the detectors $D_1, D_2$, given various choices of \acrshort{ps}s and \acrshort{bs}s. This is instrumental for our results in Chapter~\ref{chapter: applications}, and was used by~\cite{giordani2021witnesses,giordani2023experimental} to certify coherence inside interferometers using measurements of two-state overlaps. 

\myindent One aspect that we do not discuss in this thesis, but that was used extensively by~\cite{giordani2023experimental}, is that the \acrshort{mzi} serves as the building block for describing \emph{any} multimode interferometer~\citep{reck1994experimental,clements2016optimal}. In such cases, we can also have the same interpretation and view such multimode interferometers as prepare-and-measure devices, which can also be used to estimate two-state overlaps, and witness the superpositions of photonic states in Hilbert spaces of higher dimension (see our results in Chapter~\ref{chapter: applications}, Section~\ref{sec: dimension and coherence witnessing section}). 

\subsection{Bomb-testing gedanken experiment}\label{sec: bomb testing}

\begin{figure}[t]
    \centering
    \includegraphics[width=1\linewidth]{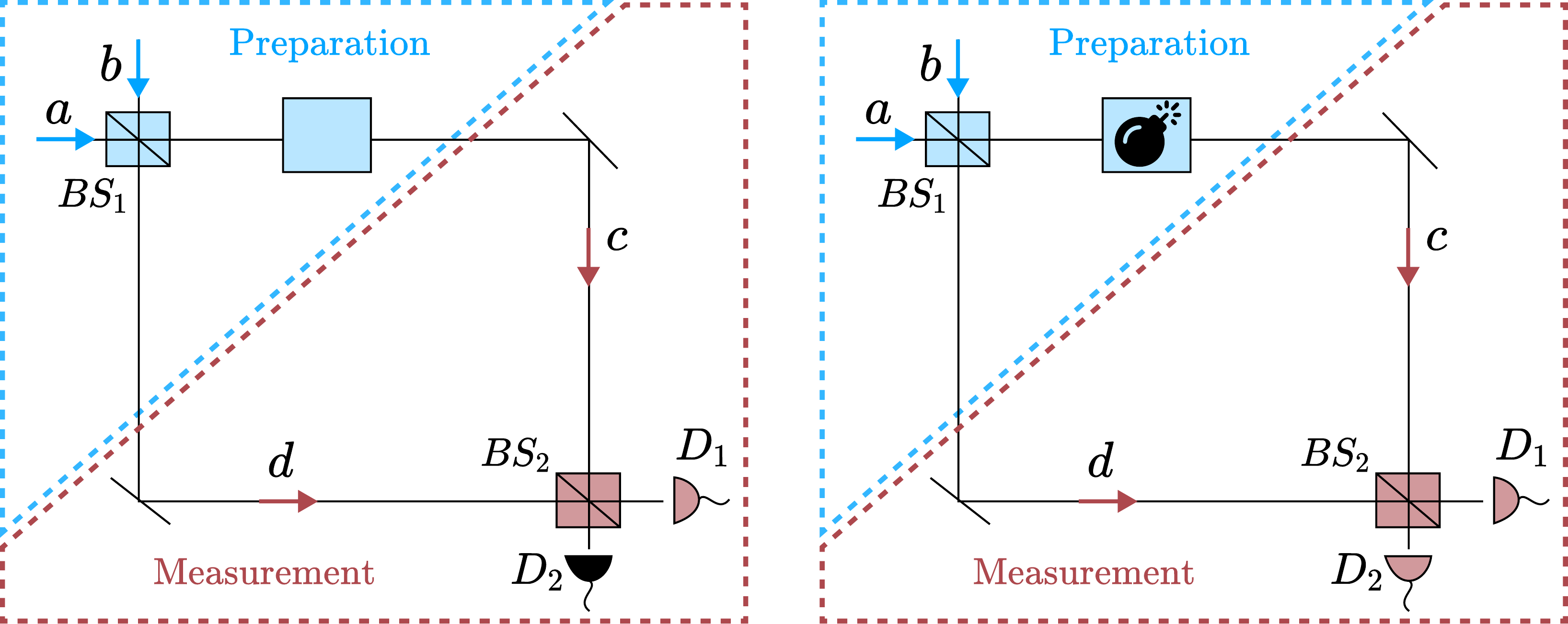}
    \caption{\textbf{Quantum interrogation with an \acrshort{mzi}.} The goal of the quantum interrogation task is to determine that there is an object that can be in one of the two paths of the interferometer, without interacting with it (in the `particle-like' sense). The object is assumed to be extremely sensitive, and when it absorbs a photon we fail the task. To detect the object we calibrate the \acrshort{mzi} so that: (Left) When there is no object inside the interferometer, whenever we input a photon in mode $a$ the only detector that ever clicks is the detector $D_1$ while $D_2$ never clicks, illustrated in this case as a dark detector. To do so, we just need to make sure that for any $U_{\theta_1}$ of the first \acrshort{bs} is reversed by the second \acrshort{bs} via $U_{\theta_2} = U_{\theta_1}^\dagger$. (Right) In the presence of an object, we now have three different possibilities: (1) the object absorbs the photon and we fail the task, (2) the photon is detected on detector $D_1$ and we need to re-run the experiment as this click is also consistent with lack of object in the interferometer, and finally, (3) the photon is detected on detector $D_2$ and we unambiguously conclude that there is an object inside the interferometer. We have assumed ideal detectors that do not have dark counts. }
    \label{fig: Bomb_MZI}
\end{figure}

\myindent One of the several important applications for such a simple linear optics device consists of performing the standard quantum interrogation task~\citep{elitzur1993quantum,vaidman1996interaction}. For this task, a totally opaque object~\footnote{The results that follow do not consider semitransparent objects. Transparent or semitransparent objects require a separate treatment~\citep{white1998interactionfree}. } is placed on one of the arms of the \acrshort{mzi}, typically chosen to be a bomb for historical reasons~\citep{elitzur1993quantum}. The bomb is assumed to explode if and only if it interacts with a single photon. We choose $U_{\theta_1}=U_{\theta_2}^\dagger, |\theta_1|=|\theta_2|=\theta$ and $\phi_1=\phi_2=0$, so that the single photon superposition inside the interferometer is controlled only by the \acrshort{bs} $U_\theta$. Thus, 
\begin{align*}
    \vert \psi(\theta_1,\phi_1)\rangle &\equiv \vert \psi(\theta)\rangle = U_\theta\vert 0\rangle =  \cos(\theta)\vert 0 \rangle + i\sin(\theta)\vert 1 \rangle, 
\end{align*}
and, moreover, one of the detectors is always dark since after the second \acrshort{bs} described by unitary $U_{\theta}^\dagger$ the state of the photon returns back to $|0\rangle$ since the \acrshort{mzi} is set-up to satisfy  $U_{\theta_2}^\dagger U_{\theta_1} = U_{\theta}^\dagger U_\theta =\mathbb{1}$. In practice, this implies that only detector $D_1$ clicks, for any choice of angle $\theta$. Note that $U_\theta^\dagger = U_{-\theta}$, where 
\begin{equation}
    U_{-\theta}|0\rangle = \cos(\theta) \vert 0\rangle -i\sin(\theta)\vert 1 \rangle. 
\end{equation}
We show this configuration of an \acrshort{mzi} in Figure~\ref{fig: Bomb_MZI}. 

\myindent In this scenario, we input the photon in mode \textit{a} (Fig.~\ref{fig: Bomb_MZI}). The experiment thus consists of testing between two possibilities: 

\begin{enumerate}
    \item \textit{Hypothesis 1)} There is a dud in the left arm of the \acrshort{mzi}, hence it never interacts with the photon, detector $D_1$ will always click, and detector $D_2$ (usually referred to as the dark detector) will never click. 
    \item \textit{Hypothesis 2)} There is an active bomb in the left arm of the \acrshort{mzi}, and therefore the photon will hit the bomb with probability $\cos^2(\theta)$, detonating it, while it will choose the different path with probability $\sin^2(\theta)$. 
\end{enumerate}

\myindent The latter case in which there is an object on the left arm of the \acrshort{mzi}, yet its presence collapses the photon's state inside the interferometer to be $\ket{1}\bra{1}$,  gives us a chance to detect the object/bomb without exploding it. The idea is to exploit the bomb as a complete path-information measurement device.  In fact, after the photon in state $\vert 1 \rangle \langle 1 \vert $ passes the second \acrshort{bs}, detector $D_1$ will click with probability $\sin^2(\theta)$, and detector $D_2$ will click with probability $\cos^2(\theta)$. Note, $D_2$ clicks \textit{only} in the case of an unexploded active bomb, thus detecting the presence of the object. With this protocol in mind, we define the following information task:

\begin{tcolorbox}[
    colback=lightblue!10, % light blue background color with slight transparency
    colframe=lightblue,   % matching border color
    width=\textwidth,     % adjust width to fit text width
    boxrule=0.5mm,        % border thickness
    sharp corners,
    title=Box 3: Quantum interrogation task,
    fonttitle=\bfseries,  % bold font for title
    title filled=true,    % filled title background
    coltitle=white        % white text color for title
]
\begin{task}\label{task: quantum interrogation} Using as many photons as needed, detect the presence of an active bomb \emph{without} exploding it, with the highest possible probability.
\end{task}
\end{tcolorbox}

\subsection{Figure of merit for quantum interrogation}\label{sec: bomb test figure of merit}

\myindent If the \acrshort{bs}s are symmetric, detector $D_2$ clicks with probability $\sfrac{1}{4}$. Therefore we may need to run the experiment many times to see any $D_2$ event, having the drawback that, by doing so, the bomb may explode during new rounds. Let us call $p_{\mathrm{succ}}$ the probability of success, which is equal to the probability that we see a click event in detector $D_2$ and the probability $p_{\mathrm{inconclusive}}$ that detector $D_1$ clicks.~\footnote{In our discussion we are considering the idealized situation where we have perfect detectors, and that there are no photon losses during the process.} We hence run the experiment until either we fail (when the bomb explodes) or succeed ($D_2$  clicks) using as many photons as necessary. The probability we will be able to \emph{eventually} succeed in our task is given by~\citep{rudolph2000better},
\begin{equation*}
    \eta = \sum_{k=0}^\infty p_{\mathrm{succ}}p_{\mathrm{inconclusive}}^k.
\end{equation*}
In other words, the above is the correct probability of eventually succeeding because we perform a second test only if we have not failed but also not succeeded in the first round of the test, and we perform a third test only if we have not failed and neither succeeded the two previous ones, and so on, leading to the sequence
\begin{figure}[t]
    \centering
    \includegraphics[width=0.65\textwidth]{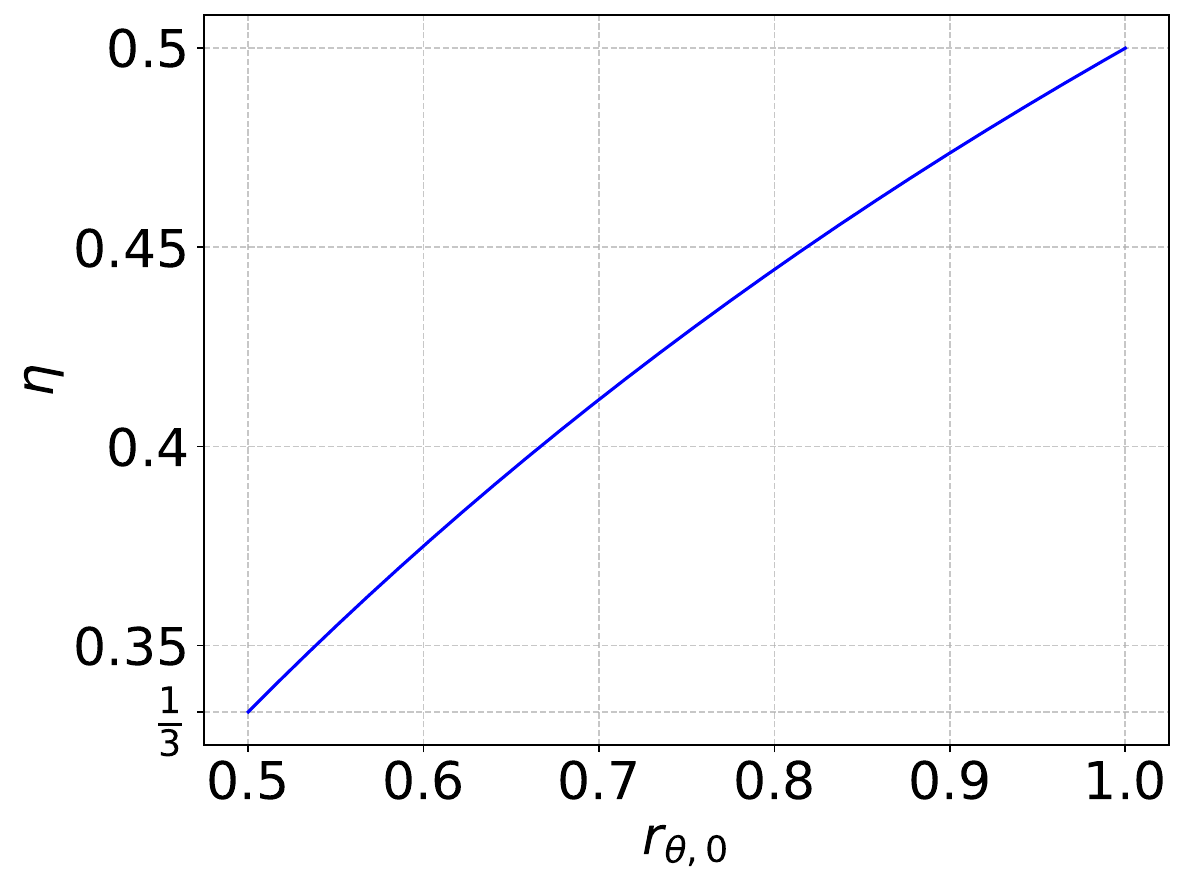}
    \caption{\textbf{Efficiency of the standard quantum interrogation task.} Varying the parameter $\theta$ of the \acrshort{bs}s we vary $r_{\theta,0} \equiv r_{\theta 0} = |\langle 0|\psi(\theta) \rangle|^2$, which is a two-state overlap completely characterizing $\eta$ given by $\eta(r) = r/(1+r)$.}
    \label{fig: efficiency quantum interrogation preliminaries}
\end{figure}

\begin{equation*}
    \eta = \underbrace{p_{\mathrm{succ}}}_{\text{Detected in the first run}} + \underbrace{p_{\mathrm{succ}}p_{\mathrm{inconclusive}}}_{\text{Detected in the second run}}+\underbrace{p_{\mathrm{succ}}p_{\mathrm{inconclusive}}^2}_{\text{Detected in the third run}}+\dots
\end{equation*}
Once we notice that this is an infinite sum of a geometric progression with initial value $p_{\mathrm{succ}}$ and common ratio $p_{\mathrm{inconclusive}}$, we conclude that
\begin{equation*}
    \eta = \frac{p_{\mathrm{succ}}}{1-p_{\mathrm{inconclusive}}}.
\end{equation*}
Noticing that $p_{\mathrm{inconclusive}} = 1-p_{\mathrm{succ}}-p_{\skull}$,  given by the probability that the dark detector clicks plus the probability that the bomb explodes $p_{\skull}$, we end up with Eq.~\eqref{eq: efficiency}. The figure of merit for the efficiency of this task $\eta$ is therefore defined operationally as~\citep{elitzur1993quantum,kwiat1995interactionfree}
\begin{equation}\label{eq: efficiency}
    \eta =   \frac{p_{\mathrm{succ}}}{p_{\mathrm{succ}} + p_{\skull}}.
\end{equation}
In case of symmetric \acrshort{bs}s, as pointed out by~\cite{elitzur1993quantum}, we have $$\eta = \frac{1/4}{1/4+1/2} = \frac{1}{3}.$$ The same efficiency can be achieved by the noncontextual model of~\cite{catani2023whyinterference}. In more generality, if we allow $U_{\theta}$ to be any as given by Eq.~\eqref{eq: BS} we have that 
\begin{equation}
    \eta^Q = \frac{\cos^2(\theta)}{\cos^2(\theta)+1}
\end{equation}
where we have used that $$p_{\skull} = |\langle 1|\psi(\theta_1) \rangle |^2 = \sin^2(\theta)$$ and $$p_{\mathrm{succ}} = |\langle 0|\psi(\theta_1) \rangle |^2 |\langle 1|\psi(\theta_2) \rangle |^2 = \cos^2(\theta)\sin^2(\theta)$$ as well as $|\theta_1| = |\theta_2| = \theta$. We plot $\eta^Q$ as a function of the overlap $|\langle 0\vert  \psi(\theta) \rangle |^2 = r_{\theta,0}$ in Fig.~\ref{fig: efficiency quantum interrogation preliminaries}.

\myindent For our purposes, these are all the relevant aspects regarding the \acrshort{mzi} and the quantum interrogation task. As we have seen, the quantum efficiency $\eta^Q$ of the standard quantum interrogation protocol can be expressed in terms of two-state overlaps  $r_{\theta,0} = |\langle 0|\psi(\theta) \rangle|^2 = \cos^2(\theta)$. Notice that such overlaps can also be written as traces of the corresponding density operators, $r_{\theta,0} = \text{Tr}(\rho_0\rho_\theta)$, with $\rho_0 = |0\rangle \langle 0 \vert$ and $\rho_\theta = \vert \psi(\theta) \rangle \langle \psi(\theta) \vert $. In this sense, our discussion so far has naturally led to the relevance of multivariate traces of states, which we now proceed to review in depth in the next Chapter.

\chapter{Bargmann invariants}\label{chapter: Bargmann invariants}

\begin{quote}
    ``\textit{Any significant statement in quantum theory is therefore a statement about unit rays.}''\\(Valentine~\cite{bargmann1964note})
\end{quote}

\begin{quote}
    ``\textit{Invariant theory has already been pronounced dead several times, and like the phoenix it has been again and again rising from its ashes.}''\\~\citep{dieudonne1970invariant}
\end{quote}

\myindent Invariant theory has long been a cornerstone of physics. It is widely accepted that any mature physical theory must produce predictions independent of arbitrary choices of gauge, reference frames, or coordinate systems. These foundational principles are deeply rooted in the mathematics of \emph{invariant} and \emph{group theory}, which have led to some of the most profound and universally applicable physical frameworks. For instance, Einstein's theory of general relativity is grounded in the invariance of physical laws under \emph{changes of reference frame}, such as the constancy of the speed of light in the vacuum. Similarly, quantum field theory---our most rigorously tested physical theory to date---relies on concepts of \emph{gauge invariance}, \emph{symmetry} (and symmetry breaking), and hallmarks of invariant theory like Noether's theorem~\citep{noether1918invariante} and Wigner's theorem~\citep{wigner1931gruppen}.

\myindent Wigner's theorem, later generalized by~\cite{uhlhorn1963representation},~\cite{molnar2002orthogonality}, and~\cite{semrl2003generalized},~\footnote{An interesting aspect is that Wigner's theorem \emph{does not hold} for quaternionic quantum mechanics for Hilbert spaces of dimension $2$, as showed by~\cite{bargmann1964note}. Such a difference could lead to experimental tests capable of falsifying quaternionic (or, more unexpectedly, complex) quantum theory similar to proposals by~\cite{renou2021quantum}. } served as the primary motivation for the introduction of the main topic of this chapter: \emph{Bargmann invariants}. These were first introduced by Valentine Bargmann in 1964~\citep{bargmann1964note} as an example of an expression that is gauge invariant. In his seminal paper, Bargmann openly stated that his work was ``expository and contains no new results''~\citep{bargmann1964note}, and the expressions that we now term Bargmann invariants do not even make a relevant part of his main result. Wigner's theorem asserts that linear transformations acting on quantum states $\vert \psi \rangle$ and satisfying certain constraints, such as preservation of inner products, must be either unitary or anti-unitary. While Bargmann did not introduce fundamentally new results, as he acknowledged, he pointed out an insightful construction, formulated as a product of inner products, or as in our more general description, in terms of multivariate traces of states. This construction, perhaps well-known in the mathematical literature of invariant theory of that period, was novel to the physics community. 

\myindent We have already encountered Bargmann invariants in Chapter~\ref{chapter: quantum coherence}, and in this Chapter we  provide them with a formal definition (see~\cref{definition: Bargmann invariants}). We denote them as
\begin{equation*}
    \Delta_{n}(\rho_1,\dots,\rho_n) = \text{Tr}(\rho_1 \cdots \rho_n),
\end{equation*}
with the same letter $\Delta$ as used by~\cite{bargmann1964note}. The label $n$ denotes the \emph{order} of the invariant. 

\myindent The study of multivariate traces of matrices to explore quantum mechanical properties dates back to the early days of quantum theory, notably in investigations of \emph{$n$-point correlation functions} in quantum statistical mechanics~\citep{fano1957description}. These values are termed invariants because they remain unchanged under the transformation 
\begin{equation}\label{eq: simultaneous conjugation}
    \pmb{X} \equiv (X_1, \dots, X_n) \mapsto  S \pmb{X}S^{-1} \equiv (S X_1 S^{-1}, \dots, S X_n S^{-1}) 
\end{equation}
for any invertible  $S$ and any tuple of square matrices $\pmb{X} \equiv (X_1,\dots,X_n)$. The most important foundational result in invariant theory was discovered by~\cite{hilbert1890ueber}, who showed that the ring of all polynomials on the symbols $\pmb{X}$ invariant under any transformation of the form of Eq.~\eqref{eq: simultaneous conjugation} is actually \emph{finitely generated}. Much research has been devoted to finding optimal solutions for such a finite set, and the form of these generators. It is now known that multivariate traces on the symbols $\pmb{X}$ constitute finite families of generators~\citep{donkin1992invariants,wigderson2019mathematics}, and for various case studies---e.g., restricting $S$ to be unitary and the number of symbols $n$ to be $1$---, all the generators (or at least their number) are known~\citep{pearcy1962complete}. In this Chapter, we briefly discuss one such construction that was found by~\cite{chien2016characterization}, and by~\cite{oszmaniec2024measuring}. For interesting recent research concerning multivariate traces that are \emph{not focused on} Bargmann invariants, but on generic symbols $\pmb{X}$, we refer to recent work by~\cite{klep2021positive,klep2021optimization} investigating optimization problems for polynomials of such quantities.~\footnote{These results build upon seminal work by  Navascués, Pironio, and Acín (\acrshort{npa})~\citep{navascues2014characterization} and semidefinite programming techniques, as well as on the implications of the recently proved $\mathrm{MIP}^*=\mathrm{RE}$ to Banach space theory~\citep{ji2021mipstar}. }

\myindent In the case of pure states, Bargmann invariants take the form
$$\Delta_{n}(\pmb{\Psi}) = \langle \psi_1|\psi_2 \rangle \langle \psi_2|\psi_3 \rangle \cdots \langle \psi_{n-1}|\psi_n \rangle \langle \psi_n|\psi_1 \rangle,$$
for $\pmb{\Psi} \equiv (\vert \psi_1\rangle, \dots, \vert \psi_n\rangle) \in \mathcal{H}^n$.~\footnote{In what follows, we denote $\mathcal{H}^n = \underbrace{\mathcal{H} \times \cdots \times \mathcal{H}}_{n\text{ times}}$.} Simon,  Mukunda, and collaborators significantly developed the theory related to these constructions~\citep{simon1993Bargmann,mukunda2001Bargmann,mukunda2003Bargmann,mukunda2003Wigner,akhilesh2020geometric}, motivated by their connection to \emph{geometric phases}~\citep{berry1984quantal}. These invariants have multiple applications, motivating the development of efficient ways for measuring them. There is a burgeoning field of research devoted to investigating applications and resources connected to Bargmann invariants and their measurements. In photonics, the  phases of Bargmann invariants are known as \emph{collective photonic phases}~\citep{shchesnovich2015partial,shchesnovich2018collective}, and have been measured for invariants of up to fourth order~\citep{bong2018strong,jones2020multiparticle,pont2022quantifying,jones2023distinguishability}. They are essential for the complete characterization of partial multiphoton indistinguishability~\citep{rodari2024experimentalobservationcounterintuitivefeatures,rodari2024semideviceindependentcharacterizationmultiphoton}, as required by photonic quantum computation protocols with single photons~\citep{shchesnovich2015partial,oszmaniec2024measuring}. We devote this Chapter to the presentation of the relevant known information on Bargmann invariants that shall be used in the second part of this thesis. 

\myindent The structure of this Chapter is as follows. We start by defining the notions of gauge invariance, unitary invariance and projective unitary invariance in~\cref{sec: gauge and unitary invariance}, where we introduce the frame graph framework and the role played by \emph{Gram matrices} to resolve the so-called unitary invariance problem.~\cref{sec: measuring Bargmann invariants} briefly discusses how to measure Bargmann invariants and~\cref{sec: applications of Bargmann invariants} motivates the study of Bargmann invariants reviewing several applications. Section~\ref{subsec: applications of two-state overlaps} focuses on the applications of two-state overlaps, and, in particular, on the work by~\cite{galvao2020quantum}, relevant to the statements of some of the main problems this thesis aims to resolve (such as Question~\ref{question: coherence witnesses and contextuality}). Section~\ref{subsec: applications of higher-order Bargmann invariants} concludes with a brief overview of applications of higher-order invariants and with a few concluding remarks for Part I of this thesis.  

\section{Gauge and unitary invariance}~\label{sec: gauge and unitary invariance}

\myindent We begin by introducing two notions of invariance: \emph{gauge} invariance and \emph{unitary} invariance, with an emphasis on their role in the context of \emph{tuples of states}. Gauge transformations, and the associated study of gauge invariants, are more commonly encountered in what is often referred to as `pure-state quantum mechanics,' where one investigates the kinematics and dynamics of vector states $\vert \psi \rangle \in \mathcal{H}$. 

\begin{definition}[Abelian gauge-transformation and gauge-invariance]
    Let $\mathcal{H}$ be a Hilbert space. An \emph{Abelian gauge-transformation} $M_\theta$ is an element of the one-parameter group of transformations defined by the mapping $$M_\theta(\vert \psi \rangle )= e^{i\theta}\vert \psi \rangle,$$ $\theta \in [0,2\pi)$, for every $\vert \psi \rangle \in \mathcal{H}$. A function $f:\mathcal{H}^n \to \mathbb{C}$ is \emph{gauge-invariant} if it is invariant under gauge transformations, i.e., if $$f(M_{\pmb{\theta}}(\pmb{\Psi})) \equiv f(M_{\theta_1}\vert \psi_1\rangle, M_{\theta_2}\vert \psi_2 \rangle, \dots, M_{\theta_n}\vert \psi_n\rangle) = f(\vert \psi_1\rangle, \vert \psi_2\rangle, \dots, \vert \psi_n\rangle) \equiv f(\pmb{\Psi})$$ for every $\pmb{\theta} = (\theta_1,\theta_2,\dots,\theta_n) \in [0,2\pi)^n$.
\end{definition}

\myindent Since we will only consider the notion of Abelian gauge-transformations just described, we will simply refer to any such transformation as a \emph{gauge transformation} for brevity. Above, we have introduced the notation $$M_{\theta_1} \times \cdots \times M_{\theta_n} \equiv M_{\pmb{\theta}}:\mathcal{H}^n \to \mathcal{H}^n$$ for the map that acts  as $(\vert \psi_1\rangle,\dots, \vert \psi_n\rangle) \mapsto (e^{i\theta_1}\vert \psi_1\rangle, \dots, e^{i\theta_n}\vert \psi_n\rangle)$. It is well known that properties of quantum systems should be considered only in terms of the equivalence classes $[\vert \psi \rangle ] : = \{e^{i\theta}\vert \phi \rangle \mid \theta \in \mathbb{R}\}.$  Such equivalence classes are termed \emph{rays}, and the space of these rays is known as the \emph{projective Hilbert space}. 

\myindent From a more modern perspective, when considering `density matrix quantum mechanics', equivalence classes are described by rank-1 projections $\psi = \vert \psi \rangle \langle \psi \vert $ and generic quantum states $\rho$ become elements within the (closed) convex hull of all such projections. We have been denoting this set as $\mathcal{D}(\mathcal{H})$. By construction, density matrices are already gauge-invariant and, correspondingly, any function $f$ such that  $f(\vert \psi_1 \rangle, \dots, \vert \psi_n \rangle ) = g({\psi_1},\dots,{\psi_n})$ for some other function $g$ is also gauge-invariant.

\myindent Another notion of invariance that we consider is that of \emph{unitary invariance}. 

\begin{definition}[Unitary transformations and unitarily-invariant functions of states]\label{def: global unitary transformations}
    Given a unitary $U:\mathcal{H} \to \mathcal{H}$ we define \textit{unitary transformations} on tuples $\pmb{\Psi} \in \mathcal{H}^n$ by the mapping \[\pmb{\Psi} \mapsto U\pmb{\Psi} = (U\ket{\psi_1},\ldots,U\ket{\psi_n}).\]
    
    A function $f:\mathcal{H}^n \to \mathbb{C}$ is \emph{unitary-invariant} if it is invariant under unitary transformations, i.e., $$f(U \pmb{\Psi}) = f(\pmb{\Psi})$$ for every unitary $U$ and every tuple $\pmb{\Psi} \in \mathcal{H}^n$.
\end{definition}

\myindent In words, the definition above is stating that the action of the unitary group on $\mathcal{H}$ lifts to unitary actions on tuples. We can also translate the definition above to the case of tuples of bounded operators $\pmb{X} \in \mathcal{B}(\mathcal{H})^n$, and unitaries $\mathcal{U}:\mathcal{B}(\mathcal{H}) \to \mathcal{B}(\mathcal{H})$, where we have instead the mapping
$$\pmb{X} \mapsto \mathcal{U}(\pmb{X}) \equiv U\pmb{X} U^\dagger = (UX_1 U^\dagger, \dots, U X_n U^\dagger).$$
When a certain function $f$ is both unitary-invariant \emph{and} gauge-invariant we say that it is \emph{projectively unitary (\acrshort{pu}) invariant}. Note that the operation $\pmb{\rho} \mapsto U\pmb{\rho}U^\dagger$, for some unitary $U$, defines an equivalence relation $\pmb{\rho} \sim_u \pmb{\sigma} \iff \exists U: \pmb{\rho} = U\pmb{\sigma}U^\dagger$ over $\mathcal{D}(\mathcal{H})^n$. Similarly for the case of gauge-transformations and projectively-unitary transformations defined over $\mathcal{H}^n$. 

\begin{definition}[Gauge, isometric, unitary, and projectively-unitary equivalent tuples]\label{def: equivalences_of_tuples} Let $\mathcal{H}, \mathcal{K}$ be two Hilbert spaces, and fix $n \in \mathbb{N}$. We say that two tuples $\pmb{\Psi} \in \mathcal{H}^n$ and $\pmb{\Phi} \in \mathcal{K}^n$ are:
\begin{enumerate}
\item[(i)] \emph{Gauge equivalent}, written $\pmb{\Psi} \sim_g \pmb{\Phi}$, if $\mathcal{H}=\mathcal{K}$ and there exists some $\pmb{\theta} \in [0,2\pi)^n$ such that $M_{\pmb{\theta}}(\pmb{\Psi}) = \pmb{\Phi}$. 
\item[(ii)] \emph{Isometrically equivalent}, written $\pmb{\Psi} \sim_{iso} \pmb{\Phi}$, if there exists a Hilbert space isometry $V: \mathrm{span}(\pmb{\Psi}) \to \mathrm{span}(\pmb{\Phi})$ such that $V(\pmb{\Psi}) = \pmb{\Phi}$. 
\item[(iii)] \emph{Unitarily equivalent}, written $\pmb{\Psi} \sim_u \pmb{\Phi}$, if there exists a surjective isometry $U$ such that $U(\pmb{\Psi}) = \pmb{\Phi}$. 
\item[(iv)] \emph{Projectively-unitary (\acrshort{pu}) equivalent}, written $\pmb{\Psi} \sim_{pu} \pmb{\Phi}$, if they are \emph{both} gauge and unitarily equivalent. 
\end{enumerate}
We write $[\pmb{\Psi}]_g$, $[\pmb{\Psi}]_{iso}$, $[\pmb{\Psi}]_u$, $[\pmb{\Psi}]_{pu}$ for the equivalence class of the tuple $\pmb{\Psi}$ under $\sim_g$, $\sim_{iso}$, $\sim_u$, and $\sim_{pu}$, respectively.
\end{definition}

\myindent If the ambient Hilbert spaces match, i.e., $\mathcal{H}=\mathcal{K}$, two tuples are isometrically equivalent iff they are also unitarily equivalent. Also, if $n=1$ gauge equivalence implies unitary equivalence; yet for $n \neq 2$ this is no longer the case since gauge equivalence acts differently in each entry of the tuples. All relations in Def.~\ref{def: equivalences_of_tuples} can be straightforwardly translated to the case when the Hilbert space is $\mathcal{B}(\mathcal{H})$. For example, we say that two tuples of mixed states $\pmb{\rho},\pmb{\sigma}$ are \emph{unitarily equivalent}, written $\pmb{\rho}\sim_u \pmb{\sigma}$, if and only if there exists some unitary $U$ such that the two tuples are related by a unitary transformation $\mathcal{U}(\pmb{\rho}) = U\pmb{\rho}U^\dagger = \pmb{\sigma}$. When considering tuples of states $\pmb{\rho}$, two tuples are gauge equivalent iff they are equal, and they are \acrshort{pu}-equivalent iff they are unitarily equivalent. 

\myindent All relations in Def.~\ref{def: equivalences_of_tuples} are equivalence relations since they are all group actions of a group in a set.\footnote{We discuss this mathematical aspect in detail in Appendix~\ref{app: basic algebra}.}  For example, take tuples of operators $\pmb{\rho},\pmb{\sigma}\in\mathcal{D}(\mathcal{H})^n$ and $\sim_u$. Clearly $\pmb{\rho} \sim_u \pmb{\rho}$ choosing $U=\mathbb{1}$; $\pmb{\rho} \sim_u \pmb{\sigma}$ implies that $\pmb{\sigma} \sim_u \pmb{\rho}$ from the fact that $U^{-1} = U^\dagger$ is also a unitary whenever $U$ is; and finally, $\pmb{\rho} \sim_u \pmb{\sigma}$, $\pmb{\sigma} \sim_u \pmb{\eta} $ implies that $\pmb{\rho} \sim_u \pmb{\eta}$ from the fact that unitaries are closed under composition.

\myindent A fundamental problem of interest is whether two given tuples of states are unitarily equivalent. We call this the \emph{unitary equivalence problem}, or for vector states, the projective unitary \emph{(\acrshort{pu}) equivalence problem}. We state these as a decision problems:

\begin{tcolorbox}[
    colback=lightblue!10, % light blue background color with slight transparency
    colframe=lightblue,   % matching border-color
    width=\textwidth,     % adjust width to fit text width
    boxrule=0.5mm,        % border thickness
    sharp corners,
    title=Box 4: Equivalence problems,
    fonttitle=\bfseries,  % bold font for title
    title filled=true,    % filled title background
    coltitle=white        % white text color for title
]
\begin{task}\label{task: unitary invariance problem} Let $\mathcal{H}$ be a Hilbert space,  $\mathcal{H}_1 := \{\vert \psi \rangle \in \mathcal{H} \mid \langle \psi|\psi\rangle = 1\}$,  and $n \in \mathbb{N}$. Then, 
\begin{enumerate}
\item The \textit{unitary equivalence problem} on $\mathcal{H}_1^n$ is defined by the task of deciding if $\pmb{\Psi} \sim_{u} \pmb{\Phi}$ for any given pair $\pmb{\Psi},  \pmb{\Phi} \in \mathcal{H}_1^n$.
\item The projective unitary equivalence problem on $\mathcal{H}_1^n$, or simply the \textit{\acrshort{pu} problem}, is defined by the task of deciding if $\pmb{\Psi} \sim_{pu} \pmb{\Phi}$ for any given pair $\pmb{\Psi},  \pmb{\Phi} \in \mathcal{H}_1^n$.
\item The unitary equivalence problem on $\mathcal{D}(\mathcal{H})^n$ is defined by the task of deciding if $\pmb{\rho} \sim_u \pmb{\sigma}$ for any given pair $\pmb{\rho},\pmb{\sigma} \in \mathcal{D}(\mathcal{H})^n.$
\end{enumerate}
\end{task}
\end{tcolorbox}

\myindent We will soon see that Bargmann invariants are useful for resolving  equivalence problems. For future reference we recall the constructions of Bargmann invariants in a definition:

\begin{definition}[Bargmann invariants]\label{definition: Bargmann invariants} Let $\mathcal{H}$ be a fixed Hilbert space. 

\begin{enumerate}
    \item[(i)] We define a function $\Delta_n: \mathcal{H}_1^n \to \mathbb{C}$ which to any $n$-tuple of pure normalized quantum state vectors $\pmb{\Psi} \equiv (\vert \psi_1\rangle, \dots, \vert \psi_n \rangle) \in \mathcal{H}_1^n$, where $\psi_i \equiv \vert \psi_i\rangle \langle \psi_i \vert$, associates its \emph{pure-state Bargmann invariant}   $$\Delta_n(\pmb{\Psi}) = \text{Tr}\left(\prod_{k=1}^n \psi_k\right).$$ 
    \item[(ii)]  We define a function $\Delta_n: \mathcal{D}(\mathcal{H})^n \to \mathbb{C}$ which to any $n$-tuple of quantum states $\pmb{\rho} \equiv (\rho_1,\dots,\rho_n) \in \mathcal{D}(\mathcal{H})^n$ associates its \emph{mixed state Bargmann invariant}  $$\Delta_n(\pmb{\rho}) = \text{Tr}\left(\prod_{k=1}^n \rho_k\right).$$ 
\end{enumerate}

We say that $n$ is the \emph{order} of the Bargmann invariant.     
\end{definition}

\myindent It is elementary to see that pure state Bargmann invariants are \acrshort{pu} invariants, and that mixed state Bargmann invariants are unitary invariants, from the fact that the trace is cyclic. We now provide some simple examples of Bargmann invariants that are of relevance for the discussions presented later on in Chapter~\ref{chapter: relational coherence}.

\begin{example}[Purely complex Bargmann invariants]\label{example: purely imaginary invariant} 
    We start with the example constructed by~\cite{bargmann1964note} of a third-order Bargmann invariant that has zero real part. Let $\vert \psi_1\rangle = \vert 0\rangle, \vert \psi_2\rangle = \vert -\rangle $ and $\vert \psi_3\rangle = \sfrac{1}{\sqrt{3}}(\vert 0\rangle + (1-i)\vert 1\rangle)$. In this case, we have that all these are normalized vector states (note that $\langle \psi_3|\psi_3\rangle = \sfrac{1}{3}(1+(1+i)(1-i)) = \sfrac{1}{3}(1+1-i^2)=1$). Moreover, we have that
    \begin{equation*}
        \Delta_3(\pmb{\Psi}) = \text{Tr}(\psi_1\psi_2\psi_3) = \langle 0|-\rangle \langle-|\psi_3\rangle \langle \psi_3|0\rangle = \frac{1}{\sqrt{2}}\,\frac{1-(1-i)}{\sqrt{6}} \,\frac{1}{\sqrt{3}}=\frac{i}{6}.
    \end{equation*}  
\end{example}

\myindent From this example we learn that in general $\Delta_3(\pmb{\Psi}) \notin \mathbb{R}$, and moreover that there are invariants with zero real part. Bargmann invariants can also be purely real-valued, as the following example shows.  

\begin{example}[Purely real Bargmann invariants]
    We now proceed to give an example of a third-order invariant satisfying $\text{Im}[\Delta_3(\pmb{\Psi})] = 0$. Let  $\vert \psi_1\rangle = \vert 0\rangle, \vert \psi_2\rangle = \vert +\rangle$, and $ \vert \psi_3\rangle = \sfrac{1}{\sqrt{3}}\vert 0\rangle + \sqrt{\sfrac{2}{3}}\vert 1\rangle$. In this case, we have that  
    \begin{equation*}
        \Delta_3(\pmb{\Psi}) = \frac{1}{\sqrt{2}}\left(\frac{1}{\sqrt{6}} + \frac{\sqrt{2}}{\sqrt{6}}\right)\frac{1}{\sqrt{3}} = \frac{1+\sqrt{2}}{6}.
    \end{equation*}
\end{example}

\myindent More generally, for tuples of single-qubit vector states, any triplet of states lying along a great circle of the Bloch sphere has a purely real-valued Bargmann invariant. 

\begin{example}[Purely real-valued negative Bargmann invariants]
   We now proceed to give an example of a third-order invariant satisfying $\text{Re}[\Delta_3(\pmb{\Psi})] < 0$ \emph{and} $\text{Im}[\Delta_3(\pmb{\Psi})] = 0$. Take for example~\footnote{See reference~\cite[Pg. 21]{wagner2024quantumcircuits} or also the Appendix of reference~\citep{galvao2020quantum}.} the case $\vert \psi_1\rangle = \vert 0\rangle, \vert \psi_2\rangle = \sfrac{1}{2}(\vert 0\rangle + \sqrt{3}\vert 1\rangle ), \vert \psi_3\rangle = \sfrac{1}{2}(\vert 0\rangle - \sqrt{3}\vert 1\rangle).$ In this case   $\Delta_3(\pmb{\Psi}) = -\sfrac{1}{8}$.
\end{example}

\myindent We will prove in Chapter~\ref{chapter: relational coherence} that this is the smallest possible value that $\text{Re}[\Delta_3(\pmb{\rho})]$ can take for every triple of normalized quantum states $\pmb{\rho}$ (see also Appendix~\ref{app: basic algebra}). To the best of our knowledge, the earliest identification of this value as the minimal real part of a third-order Bargmann invariant is due to~\cite{allahverdyan2014nonequilibrium}. 

\begin{example}[Bargmann invariants that \emph{cannot} be complex-valued]
    We have seen that, in general, Bargmann invariants are complex numbers. However, some Bargmann invariants \emph{cannot} be complex. The simplest and most relevant example to us is the class of second-order Bargmann invariants $\text{Tr}(\rho_1\rho_2)$. It is simple to see that these are always real-valued from the fact that they can be interpreted as statistics (given by the Born rule) of experimental results following a prepare-and-measure experiment. However, other classes of Bargmann invariants are also real-valued only. For example, $\text{Tr}(\rho\sigma\rho\sigma)$ or $\text{Tr}(\rho\sigma\rho)$ are also always real-valued higher-order Bargmann invariants. These types of Bargmann invariants are relevant for the direct estimation of \emph{norm distances}, as shown by~\cite{quek2024multivariatetrace}. We come back to this connection between distances of states and Bargmann invariants in Chapter~\ref{chapter: relational coherence},~\cref{example: distance between two states}.
\end{example}

\subsection{A warning regarding the notation}

\myindent It is useful to introduce several equivalent notations for denoting Bargmann invariants. To start, the one above is one of the most compact notations, but it is too agnostic to the tuple. In many situations, it is useful to write $$\Delta_n(\pmb{\Psi}) = \Delta_{1234\dots n}(\pmb{\Psi}) \equiv \Delta_{1,2,3,4,\dots,n}(\pmb{\Psi}),$$
using the comma in the (rare) occasions in which $n \geq 10$. While in the current context the ordering of states is explicit due to our use of tuples, we will later focus on scenarios where only the unordered set of states $\Psi = \{\vert \psi_i\rangle\}_{i=1}^n$ is given. In these cases, we are interested in extracting information from the set alone, and several conventions may be used to map the set into an ordered tuple for evaluation. To streamline notation, we adopt a labeling convention where $\Psi$ denotes sets, bold symbols $\pmb{\Psi}$ denotes tuples, and we write expressions such as $\Delta_{123}(\pmb{\Psi}), \Delta_{132}(\pmb{\Psi}),$ etc., to represent $\text{Tr}(\psi_1\psi_2\psi_3), \text{Tr}(\psi_1\psi_3\psi_2)$, and so forth. Since the underlying set remains fixed, this notation allows us to concisely represent \emph{multiple} Bargmann invariants without ambiguity of the underlying states considered.

\myindent Also, one commonly wants to consider \emph{repeated entries}, such as tuples where $\psi_i = \psi_j$ for some pair of labels $i,j$ in which case it is also useful to consider the notation
$$\Delta_n(\pmb{\Psi}) = \Delta_{\psi_1,\psi_2,\dots,\psi_n}.$$

\myindent In most of this work,  whenever we consider Bargmann invariants that are only \emph{real-valued}, we use the letter $r$ instead of $\Delta$ to represent these values. This is especially crucial for the case of two-state overlaps, where we will write $r_{i,j}(\pmb{\Psi}) \equiv r_{\psi_i\psi_j} = |\langle \psi_i|\psi_j\rangle|^2$. 

\myindent We conclude this small detour to discuss notation with a remark that is relevant to us for our results in Part II of this thesis, when investigating the geometry of tuples of Bargmann invariants. There, we shall consider tuples for which different Bargmann invariants have different labels, in which case we have $$\pmb{\Delta}(\pmb{\rho}) = (\Delta_w(\pmb{\rho}))_w$$ where each $w$ is a family of numbers (sometimes called `\emph{words}' or `\emph{finite sequences}') such as $(1,2,3)$ or $(1,3)$, etc. In such cases we have, for example, $\Delta_{1212}(\pmb{\rho}) = \text{Tr}(\rho_1\rho_2\rho_1\rho_2)$. No matter what, whenever we consider Bargmann invariants in this thesis, we always label them making reference to the tuple of (pure or mixed) quantum states. When we  write, in later Chapters, $\Delta_{123}$ or $\Delta_{12}$, this does \emph{not} necessarily denote a Bargmann invariant, but merely denote a labeled complex-valued coordinate. We come back to this later as this is a subtle (yet important) choice of notation.

\subsection{Gram matrices and their relevance to unitary equivalence}

\myindent It is possible to relate the problem of deciding whether $\pmb{\Psi} \sim_{u} \pmb{\Phi}$ to the problem of deciding whether two Gram matrices are equal. A \emph{Gram matrix}, also known as a \emph{Gramian} matrix or simply as a \emph{Gramian}, is a matrix defined from a tuple of vectors (not necessarily quantum states) $\pmb{v} \equiv (v_1,\dots,v_n) \in V^n$ in an  inner-product space $(V,\langle \cdot, \cdot \rangle)$. We review in detail the basic properties of Gram matrices in  Appendix~\ref{app: basic algebra}. Given $\pmb{v}$ its Gram matrix is the $n\times n$ complex matrix $G_{\pmb{v}}$ given by $(G_{\pmb{v}})_{ij} = \langle v_i|v_j\rangle$. Every Gram matrix is a \textit{positive semidefinite matrix} (\acrshort{psd}), and every positive semidefinite matrix is the Gram matrix of some set of vectors. When the vectors are pure normalized quantum vector states over a finite-dimensional Hilbert space we have that $(G_{\pmb{\Psi}})_{ii} = 1$ for all $i$. \acrshort{psd} matrices with leading diagonal equal to $\pmb{1} = (1,1,\dots,1)$ are also known as \emph{correlation matrices}. 

\begin{theorem}
    Let $\pmb{\Psi}\in \mathcal{K}^n,\pmb{\Phi} \in \mathcal{H}^n$. Then, $\pmb{\Psi} \sim_{iso} \pmb{\Phi}$ if and only if $G_{\pmb{\Psi}} = G_{\pmb{\Phi}}$.
\end{theorem}

\begin{proof}
    Clearly, $\pmb{\Psi} \sim_{iso} \pmb{\Phi} \implies (G_{\pmb{\Psi}})_{ij} = \langle \psi_i|\psi_j\rangle = \langle \phi_i|V^\dagger V |\phi_j \rangle = \langle \phi_i|\phi_j\rangle = (G_{\pmb{\Phi}})_{ij}$. Now, assuming that $G_{\pmb{\Psi}} = G_{\pmb{\Phi}}$, we can define the map $V: \mathrm{span}(\pmb{\Psi}) \to \mathrm{span}(\pmb{\Phi})$ via $$V\left(\sum_i\alpha_i\vert \psi_i \rangle \right) := \sum_i \alpha_i\vert \phi_i\rangle.$$ This map $V$ is well-defined since different linear combinations of the same vector $\vert v \rangle \in \mathrm{span}(\pmb{\Psi})$ are all mapped to the same element $V\vert v \rangle \in \mathrm{span}(\pmb{\Phi})$, i.e., let $\vert v \rangle = \sum_i \alpha_i \vert \psi_i\rangle = \sum_j \beta_j \vert \psi_j\rangle$ be two different linear decompositions of the same vector $\vert v \rangle \in \mathrm{span}(\pmb{\Psi})$, then
\[
\sum_i \alpha_i \vert \phi_i\rangle = \sum_j \beta_j \vert \phi_j\rangle,
\]
because $\sum_i \alpha_i \vert \psi_i\rangle = \sum_j \beta_j \vert \psi_j\rangle$ implies $\sum_k (\alpha_k - \beta_k) \vert \psi_k \rangle = 0$, and since $G_{\pmb{\Psi}} = G_{\pmb{\Phi}}$, we have (writing $\gamma_k \equiv \alpha_k-\beta_k$)
\begin{align*} 
0&=\left \Vert \sum_k \gamma_k  \ket{\psi_k} \right \Vert^2 = 
\left\langle \sum_k \gamma_k \psi_k \middle| \sum_l \gamma_l \psi_l \right\rangle = \sum_{k,l}\gamma_k^*\gamma_l \langle \psi_k|\psi_l\rangle = \sum_{k,l}\gamma_k^*\gamma_l \langle \phi_k|\phi_l\rangle \\
&= \left \Vert \sum_k \gamma_k \vert \phi_k\rangle  \right\Vert^2  \quad \Rightarrow \quad \sum_k (\alpha_k - \beta_k) \vert \phi_k \rangle =  0.
\end{align*}
So the image under $V$ is independent of the choice of decomposition. Hence, $V$ is well-defined. Moreover, it is an isometry by construction  due to the equality of Gram matrices, i.e. for any $\vert v\rangle = \sum_i \alpha_i|\psi_i\rangle, \vert w \rangle = \sum_j \beta_j \vert \psi_j\rangle  \in \mathrm{span}(\pmb{\Psi})$, 
    $$\langle v|w\rangle = \sum_{i,j}\alpha_i^* \beta_j \langle \psi_i|\psi_j\rangle = \sum_{i,j}\alpha_i^* \beta_j \langle \phi_i|\phi_j\rangle = \langle V(v)|V(w)\rangle.$$
\end{proof}

\myindent This is a common knowledge fact that was first stated (without proof) by~\cite{halperin1962onthegrammatrix}, and later used by~\cite{chien2016characterization} to advance the \acrshort{pu} equivalence problem, which we will discuss shortly. Related variations were also developed by~\cite{chefles2004physicaltransformations} and~\cite{jozsa2000distinguishability}. When we have tuples of quantum states over the same Hilbert space $\mathcal{H}= \mathcal{K}$ we have the corollary:

\begin{corollary}\label{proposition: unitary equivalence equality Gram matrices}
    Let $\pmb{\Psi},\pmb{\Phi} \in \mathcal{H}^n$. Then, $\pmb{\Psi} \sim_u \pmb{\Phi}$ if and only if $G_{\pmb{\Psi}} = G_{\pmb{\Phi}}$. 
\end{corollary}

\myindent This provides a simple scheme for deciding if two sets of vectors are unitarily equivalent, simply by checking if their Gram matrices match. While this resolves the task for unitary equivalence \emph{it does not} resolve the task of \acrshort{pu} equivalence. This is because it is possible that $\pmb{\Psi} \sim_{pu} \pmb{\Phi}$ while $G_{\pmb{\Psi}} \neq G_{\pmb{\Phi}}$. This happens since, as we have mentioned before, $\pmb{\Psi} \sim_{pu} \pmb{\Phi}$ iff there exists a unitary $U$ \emph{and} a list $\pmb{\theta} = (\theta_1,\dots,\theta_n)$ such that $\vert \psi_i\rangle = e^{i\theta_i}U\vert \phi_i\rangle$. In this case, the Gram matrices relate via
\begin{equation*}
    (G_{\pmb{\Psi}})_{ij} = \langle \psi_i|\psi_j\rangle = e^{i(\theta_j-\theta_i)}\langle \phi_i|\phi_j\rangle = e^{i(\theta_j-\theta_i)}(G_{\pmb{\Phi}})_{ij} = e^{-i\theta_i}(G_{\pmb{\Phi}})_{ij}e^{i\theta_j} ,
\end{equation*}
and so whenever $\theta_i \neq \theta_j$ we have $G_{\pmb{\Psi}} \neq G_{\pmb{\Phi}}$. Hence, for the case of \acrshort{pu} equivalence, we must have instead: 
\begin{proposition}\label{proposition: gram matrices of pu equivalence}
    Let $\pmb{\Psi},\pmb{\Phi} \in \mathcal{H}^n$. Then, $\pmb{\Psi} \sim_{pu} \pmb{\Phi}$ if and only if there exists a \emph{diagonal} unitary matrix $U_{\pmb{\theta}} = \mathrm{diag}(e^{i\theta_1},\dots,e^{i\theta_n})$ such that $G_{\pmb{\Psi}} = U_{\pmb{\theta}}^\dagger G_{\pmb{\Phi}} U_{\pmb{\theta}}$. 
\end{proposition}

\myindent Note, however, that for the \acrshort{pu} problem, instead of investigating the equality of a Gram matrix we need to consider equality \emph{up to} some diagonal unitary matrix. In other words, $G_{\pmb{\Psi}} = G_{\pmb{\Phi}}$ is not sufficient to guarantee \acrshort{pu} equivalence: unitary equivalence implies $G_{\pmb \Psi} = G_{\pmb\Phi}$ and thus \acrshort{pu} equivalence from~\cref{proposition: gram matrices of pu equivalence} (letting $U_{\pmb \theta} = U_{\pmb 0} = \mathbb{1}$) but the converse is not necessary. Since, however, this difference is only due to a gauge difference it is interesting to investigate a framework where it is possible to construct some matrix that solves the projective unitary equivalence problem, similarly to how the Gram matrices solve the unitary equivalence problem. One such construction is possible using the theory of frame graphs introduced by~\cite{chien2016characterization} and further developed by~\cite{oszmaniec2024measuring}.

\subsection{Frame graphs and their relevance to \acrshort{pu} equivalence}

\myindent If we recall our hints from the idea that going from an analysis based on vector states $\vert \psi \rangle$ to an analysis based on density matrices $\rho$ allows us to get rid of any gauge-dependence, it is natural to try to investigate if a similar simple trick holds for the Gram matrices, where we simply try to `rewrite' a standard Gram matrix as some function of quantities that are not only invariant to global unitary transformations but also gauge-invariant. In fact, it is possible to relate a Gram matrix $G_{\pmb{\Psi}}$ with another matrix $G_{\pmb{\psi}}$ (where we note the difference that while $\pmb{\Psi} \in \mathcal{H}^n$ is an $n$-tuple of vectors we have that $\pmb{\psi} \in \mathcal{D}(\mathcal{H})^n$ is an $n$-tuple of pure state density matrices $\psi \equiv \vert \psi\rangle \langle \psi\vert$) that is merely a function of pure state Bargmann invariants $\Delta_n(\pmb{\Psi})$ of order up to $n$. 

\myindent We start presenting the \emph{frame graph framework} by defining what is a \emph{frame graph}.

\begin{definition}[Frame graphs]~\label{def: frame graphs}
    Let $\mathcal{H}$ be a fixed Hilbert space and  $\pmb{\Psi} = (\vert \psi_i \rangle )_{i=1}^n \in \mathcal{H}^n$. A \emph{frame graph} $F_{\pmb{\Psi}} = (V(F_{\pmb{\Psi}}), E(F_{\pmb{\Psi}}))$ is a vertex-labeled graph with vertex set  $V(F_{\pmb{\Psi}}): = [n] \equiv \{1,\dots,n\}$, vertex labeling $[n] \ni i \mapsto \vert \psi_i \rangle \in \pmb{\Psi}$, edge set  $E(F_{\pmb{\Psi}}):= \{ \{u,v\} \mid  u \neq v,\langle \psi_u | \psi_v \rangle \neq 0\}.$ 
\end{definition}

\myindent In words, a frame graph $F_{\pmb{\Psi}}$ is a graph where the vertices are essentially the elements of the tuple $\pmb{\Psi}$ and there is an edge in the graph iff the corresponding vectors in $\pmb{\Psi}$ are non-orthogonal. A simple example is shown in Fig.~\ref{fig: 3-cycle frame graph} for the tuple $\pmb{\Psi} = (\vert 0 \rangle, \vert + \rangle, \vert +_i \rangle)$. 

\myindent Frame graphs, as defined, are \emph{simple graphs} since there are no loops ($e = \{u,u\} \notin F_{\pmb{\Psi}}$), no multi-edges ($e_1,e_2 \in F_{\pmb{\Psi}}$, and $u\neq v$ with $ u,v \in e_1 \cap e_2 \Rightarrow e_1=e_2$), edges are not directed ($e=\{u,v\}$), and neither edges nor vertices are weighted (see Appendix~\ref{sec: graph theory}). Note, importantly, that for each tuple $\pmb{\Psi}$ there is a corresponding frame graph $F_{\pmb{\Psi}}$, implying that the tuples are the primitive notion. For any pair of tuples $\pmb{\Psi},\pmb{\Phi} \in \mathcal{H}^n$ we have that $\pmb{\Psi} \sim_u \pmb{\Phi}$ or $\pmb{\Psi} \sim_{pu} \pmb{\Phi}$ imply that $F_{\pmb{\Psi}} = F_{\pmb{\Phi}}$. 

\begin{lemma}\label{lemma: equivalent tuples imply equal frame graph}
    Let $\mathcal{H}$ be a Hilbert space and $n \in \mathbb{N}$. For any pair  $\pmb{\Psi},\pmb{\Phi} \in \mathcal{H}^n$ and any relation $\sim$ from~\cref{def: equivalences_of_tuples} we have that $\pmb{\Psi}\sim\pmb{\Phi} \implies F_{\pmb{\Psi}} = F_{\pmb{\Phi}}$.
\end{lemma}

\myindent In what follows, we consider \emph{connected} (see~\cref{sec: graph theory}) frame graphs $F_{\pmb{\Psi}}$. For a detailed analysis of the more general case, we refer to~\cite{chien2016characterization}. 

\myindent Given the frame graph $F_{\pmb{\Psi}}$ of a certain tuple $\pmb{\Psi}$ we now consider a \emph{spanning tree} (see~\cref{sec: graph theory}), denoted as $\tau$, of the graph  $F_{\pmb{\Psi}}$. Since we consider only connected frame graphs, there is always at least one, and often more than one, spanning tree $\tau$ for $F_{\pmb{\Psi}}$. We now follow a procedure described by~\cite{oszmaniec2024measuring} for relating a certain Gram matrix $G_{\pmb{\Psi}}$ with another  $G^\tau_{\pmb{\psi}}$, which depends on the specific choice of spanning tree $\tau$, and that is a function solely of Bargmann invariants. Specifically, if we recall~\cref{proposition: gram matrices of pu equivalence}, we can provide a construction where, for every $\pmb{\Psi}$ and $\tau$ we find $U_{\pmb{\theta},\tau}$ such that $G_{\pmb{\Psi}} = U_{\pmb{\theta},\tau}^\dagger G_{\pmb{\psi}}^\tau U_{\pmb{\theta},\tau}$, where each element of $G_{\pmb{\psi}}^\tau$ is a function of Bargmann invariants.

\myindent To make it clear \emph{why} we are interested in this construction, let us assume that we have successfully done this re-writing. Our goal is simple: we want to make an argument now that this construction allows us to make a statement such as the one of~\cref{proposition: unitary equivalence equality Gram matrices}, but as an equality of the matrices $G^\tau_{\pmb{\psi}}$ given solely by Bargmann invariants. A good feature of this construction is that, when doing this, we do not need to explicitly find the matrices $U_{\pmb{\theta},\tau}$ to provide a complete characterization of \acrshort{pu} equivalence. Given a frame graph and a fixed choice of spanning tree, we are able to shift focus to \emph{a single Gram matrix}, instead of an infinite family of Gram matrices, parametrized by $\pmb{\theta}$. As a matter of fact, one \emph{can} use this construction to find all possible (infinitely many) Gram matrices completely characterizing \acrshort{pu} equivalence, as was done by~\cite{chien2016characterization}. They have considered the description of $G_{\pmb{\psi}}^\tau$ as a way of finding the family of all possible matrices of the form $U_{\pmb{\theta},\tau} G_{\Psi} U_{\pmb{\theta},\tau}^\dagger$ parametrized by the gauge parameters responsible for defining $U_{\pmb{\theta},\tau}$. 

\myindent Having briefly motivated it, let us now proceed to this construction. For any fixed tuple  $\pmb{\Psi} \in \mathcal{H}^n$, and associated frame graph $F_{\pmb{\Psi}}$, take $\tau$ to be an arbitrary choice of spanning tree. In this case, we have that there is a privileged set of inner products defined through the spanning tree. Note that, since $\tau$ is a spanning tree, and the number of vertices of $F_{\pmb{\Psi}}$ is $n$, the number of edges in $\tau$ is then $n-1$. We define the unitary operation  $U_{\pmb{\theta},\tau} = \text{diag}(e^{i\theta_1},\dots,e^{i\theta_n})$ as the operation satisfying that for every $\{i,j\} \in \tau$ we have
\begin{equation}\label{eq: definition of Utheta and tau}
\langle M_{\theta_i}\psi_i|M_{\theta_j}\psi_j\rangle = |\langle \psi_i|\psi_j\rangle|,
\end{equation}
i.e. the operation for which the mapping 
$$G_{\pmb{\Psi}} \mapsto U_{\pmb{\theta},\tau}^\dagger G_{\pmb{\Psi}}U_{\pmb{\theta},\tau} = G^\tau_{M_{\pmb{\theta}}(\pmb{\Psi})}$$
is such that the inner products  $(G^\tau_{M_{\pmb{\theta}}(\pmb{\Psi})})_{ij}$ associated to edges in $\{i,j\} \in E(\tau)$ are real-valued and positive. Note that from~\cref{lemma: equivalent tuples imply equal frame graph} $F_{\pmb{\Psi}} = F_{M_{\pmb{\theta}}(\pmb{\Psi})}$, for any $\pmb{\theta}$, which implies that this operation preserves the frame graph we focus on.  Our goal is to show that $(G^\tau_{M_{\pmb{\theta}}(\pmb{\Psi})})_{ij} = (G^\tau_{\pmb{\psi}})_{ij},$ meaning that all its elements are functions of Bargmann invariants. 

\myindent Effectively, the action described above is a gauge-transformation $\pmb{\Psi} \mapsto M_{\pmb{\theta}}({\pmb{\Psi}})$ for which, for all $\{i,j\} \in \tau$ we have that 
\begin{equation}\label{eq: overlaps of the final gram matrix}
(G^\tau_{M_{\pmb{\theta}}(\pmb{\Psi})})_{ij} = \Delta_{\psi_i,\psi_j}^{\frac{1}{2}} = \sqrt{\text{Tr}(\psi_i\psi_j)} =  \sqrt{|\langle \psi_i|\psi_j\rangle |^2},
\end{equation}
given solely in terms of the square roots of two-state overlaps.~\footnote{Here is a good point to recall that some authors refer to the overlap between two states as the inner-products $\langle \psi_i|\psi_j\rangle$. In our notation and terminology, two-state overlaps are $\text{Tr}(\rho\sigma)$ of two states $\rho$ and $\sigma$, pure states or not.}

\myindent It remains to be seen that after $\pmb{\Psi} \mapsto M_{\pmb{\theta}}(\pmb{\Psi})$, given by $U_{\pmb{\theta},\tau}$, all the remaining phases of the inner-products $(G^\tau_{M_{\pmb{\theta}}(\pmb{\Psi})})_{ij}$ with $\{i,j\} \in E(F_{\pmb{\Psi}})\setminus  E(\tau)$ can \emph{also} be inferable from Bargmann invariants. To see this, we start by introducing some notation. First, note that since $\tau$ is a spanning tree, for any edge $e' = \{i',j'\} \notin E(\tau)$ there exists a \emph{cycle path} $$\mathsf{c}_{e'}: i' \to \tau_1 \to \tau_2 \to \dots \to \tau_m \to j' \to i'$$ such that $\{i',\tau_1\},\{\tau_1,\tau_2\},\dots,\{\tau_m,j'\} \in E(\tau)$. To this path, we associate the Bargmann invariant
\begin{equation*}
    \Delta_{\mathsf{c}_{e'}}(\pmb{\Psi}) = \langle \psi_{i'}|\psi_{\tau_1}\rangle \langle \psi_{\tau_1}|\psi_{\tau_2}\rangle \dots \langle \psi_{\tau_m}|\psi_{j'}\rangle \langle \psi_{j'}|\psi_{i'}\rangle = \Delta_{\mathsf{c}_{e'}}(M_{\pmb{\theta}}(\pmb{\Psi})).
\end{equation*}
To every $e' \notin E(\tau)$ we can associate a cycle path $\mathsf{c}_{e'}$ just as the one above, and therefore its associated Bargmann invariant $ \Delta_{\mathsf{c}_{e'}}(\pmb{\Psi})$. The phases of $\langle M_{\theta_{j'}}\psi_{j'}|M_{\theta_{i'}}\psi_{i'}\rangle$ are completely determined by the phases of $\Delta_{\mathsf{c}_{e'}}(\pmb{\Psi})$ via 
\begin{align*}
    (G^\tau_{M_{\pmb{\theta}}(\pmb{\Psi})})_{i'j'} &= \langle M_{\theta_{j'}}\psi_{j'}|M_{\theta_{i'}}\psi_{i'}\rangle \\
    &= \frac{\Delta_{\mathsf{c}_{e'}}(\pmb{\Psi})}{|\langle \psi_{i'}|\psi_{\tau_1}\rangle||  \langle \psi_{\tau_1}|\psi_{\tau_2}\rangle| \dots |\langle \psi_{\tau_m}|\psi_{j'}\rangle|} \\&= \frac{\Delta_{\mathsf{c}_{e'}}(\pmb{\Psi})}{\sqrt{\Delta_{i',\tau_1}(\pmb{\Psi})}\sqrt{\Delta_{\tau_1,\tau_2}(\pmb{\Psi})} \dots \sqrt{\Delta_{\tau_m,j'}(\pmb{\Psi})}}.
\end{align*}
Using this equation, we can then see that it is possible to write, for any $\{i',j'\} \in E(F_{\pmb{\Psi}}) \setminus  E(\tau)$ the associated inner products in the frame graph as

\begin{equation}\label{eq: invariants of the final Gram matrix}
    (G^\tau_{M_{\pmb{\theta}}(\pmb{\Psi})})_{i'j'} = \sqrt{\Delta_{i',j'}(\pmb{\Psi})}e^{i\phi_{\mathsf{c}_{e'}}}, \,\,e^{i\phi_{\mathsf{c}_{e'}}} = \frac{\Delta_{\mathsf{c}_{e'}}(\pmb{\Psi})}{|\Delta_{\mathsf{c}_{e'}}(\pmb{\Psi})|}.
\end{equation}

In this way, using both Eqs.~\eqref{eq: overlaps of the final gram matrix} and~\eqref{eq: invariants of the final Gram matrix}, we conclude that it is possible to write, for any given $\pmb{\Psi}$, the equation $G_{\pmb{\Psi}} = U_{\pmb{\theta},\tau} G_{\pmb{\psi}}^\tau U_{\pmb{\theta},\tau}^\dagger$ where now $G_{\pmb{\psi}}^\tau = G_{M_{\pmb{\theta}}(\pmb{\Psi})}^\tau$ is defined with respect to only \acrshort{pu} invariants. 

\myindent Since this procedure is generic, the same holds for any other $\pmb{\Phi}$ that is \acrshort{pu} equivalent to $\pmb{\Psi}$, with the only difference that $U_{\pmb{\theta},\tau}$ would change to some other diagonal unitary $U_{\pmb{\theta}',\tau}$, but we are guaranteed that $U_{\pmb{\theta}',\tau}$ exists just from the fact that these two tuples are \acrshort{pu} equivalent. In summary, from this discussion,  whenever $G_{\pmb{\psi}}^\tau = G_{\pmb{\phi}}^\tau$ we have that

$$G_{\pmb{\Psi}} = U_{\pmb{\theta},\tau} G_{\pmb{\psi}}^\tau U_{\pmb{\theta},\tau}^\dagger = U_{\pmb{\theta},\tau} G_{\pmb{\phi}}^\tau U_{\pmb{\theta},\tau}^\dagger = U_{\pmb{\theta},\tau} {U}_{\pmb{\theta}',\tau}^\dagger G_{\pmb{\Phi}} {U}_{\pmb{\theta}',\tau} U_{\pmb{\theta},\tau}^\dagger$$
from which we conclude, using~\cref{proposition: gram matrices of pu equivalence}, that $\pmb{\Psi} \sim_{pu} \pmb{\Phi}$. From that, we have the following theorem.

\begin{theorem}\label{theorem: equivalence depending on the tree}
    Let $\pmb{\Psi},\pmb{\Phi} \in \mathcal{H}^n$ and $F_{\pmb{\Psi}} \simeq F_{\pmb{\Phi}}$. Let $\tau$ be any spanning tree for $F_{\pmb{\Psi}}$ (and  $F_{\pmb{\Phi}}$) yielding a construction of matrices  $G_{\pmb{\psi}}^\tau$ and $G_{\pmb{\phi}}^\tau$ via Eqs.~\eqref{eq: overlaps of the final gram matrix} and~\eqref{eq: invariants of the final Gram matrix}. Then, $\pmb{\Psi} \sim_{pu} \pmb{\Phi}$ if and only if $G_{\pmb{\psi}}^\tau = G_{\pmb{\phi}}^\tau$.
\end{theorem}

\myindent Given that the construction just discussed above is mathematically cumbersome to present, we now enrich our presentation with a few examples that help clarify the algorithm for a few simple cases.

\begin{example}[Frame graph given by the  $C_3$ cycle graph]
    Let us consider $\pmb{\Psi} \in \mathcal{H}_1^3$ with $\mathcal{H} \simeq \mathbb{C}^2$ given by $$\pmb{\Psi} = (\vert 0\rangle, \vert +\rangle, \vert +_i\rangle).$$
    In this case, we have that $\langle \psi_i|\psi_j\rangle \neq 0$ for all $i\neq j$ implying that $F_{\pmb{\Psi}}$ is the $3$-cycle graph. The associated frame graph is shown in Fig.~\ref{fig: 3-cycle frame graph}.
    \begin{figure}
        \centering
        \includegraphics[width=0.35\linewidth]{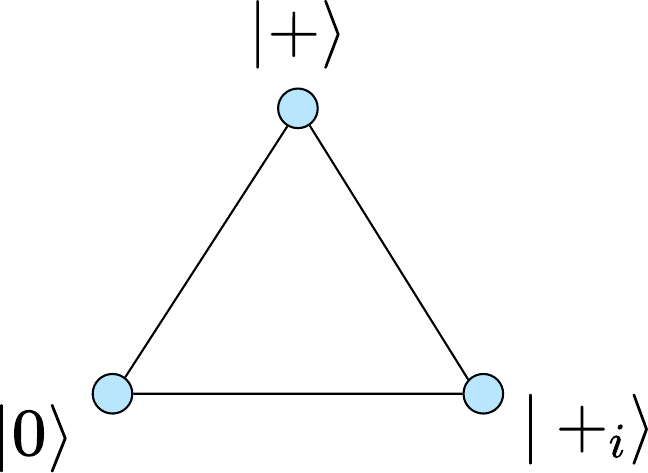}
        \caption{\textbf{Frame graph isomorphic to a 3-cycle graph $C_3$.} In this case $\pmb{\Psi} = (\vert 0\rangle, \vert +\rangle, \vert +_i\rangle)$ where $\vert +\rangle = \sfrac{1}{\sqrt{2}}(\vert 0\rangle + \vert 1\rangle )$ and $\vert +_i\rangle = \sfrac{1}{\sqrt{2}}(\vert 0\rangle + i\vert 1\rangle )$. All three inner products have non-zero values. $F_{\pmb{\Psi}} = F_{(\vert 0\rangle, \vert +\rangle, \vert +_i\rangle )} \simeq C_3$ is shown, where we have labeled the vertices of the graph with the vector states $\vert \psi_1\rangle, \vert \psi_2\rangle$, and $ \vert \psi_3\rangle$. }
        \label{fig: 3-cycle frame graph}
    \end{figure}
    
    \myindent Let us construct $G_{\psi}^\tau$. Take $\tau$ to be the spanning tree given by $e_1 = \{1,2\}$ and $e_2 = \{2,3\}$. In this case we have that $(G_{\pmb{\psi}}^\tau)_{12} = \sqrt{|\langle \psi_1|\psi_2\rangle |^2}$ and $(G_{\pmb{\psi}}^\tau)_{23} = \sqrt{|\langle \psi_2|\psi_3\rangle |^2}$. Note that $(G_{\pmb{\psi}}^\tau)_{11} = (G_{\pmb{\psi}}^\tau)_{22} = (G_{\pmb{\psi}}^\tau)_{33} = 1$ since we have that $\pmb{\Psi}$ is defined with respect to normalized quantum states. Finally, we have that the only edge not in $\tau$ is the one given by $\{1,3\}$. In this case, we have that 
    \begin{equation*}
        (G_{\pmb{\psi}}^\tau)_{1,3} = \sqrt{|\langle \psi_1|\psi_3\rangle|^2} e^{i\phi}
    \end{equation*}
    where $$e^{i\phi} = \frac{\langle \psi_1|\psi_3\rangle \langle \psi_3|\psi_2\rangle \langle \psi_2|\psi_1\rangle}{|\langle \psi_1|\psi_3\rangle \langle \psi_3|\psi_2\rangle \langle \psi_2|\psi_1\rangle|} = \frac{\text{Tr}(\psi_1\psi_3\psi_2)}{|\text{Tr}(\psi_1\psi_3\psi_2)|}.$$ This yields the following matrix:
    \begin{equation*}
        G_{\pmb{\psi}}^\tau = \left(\begin{matrix}
            1 & \sqrt{\text{Tr}(\psi_1\psi_2)} & \sqrt{\text{Tr}(\psi_1\psi_3)}e^{i\phi}\\
            \sqrt{\text{Tr}(\psi_1\psi_2)} & 1 & \sqrt{\text{Tr}(\psi_2\psi_3)}\\
            \sqrt{\text{Tr}(\psi_1\psi_3)}e^{-i\phi} & \sqrt{\text{Tr}(\psi_2\psi_3)} & 1
        \end{matrix}\right).
    \end{equation*}
    \myindent A different choice of spanning tree leads to a different matrix that is a function of Bargmann invariants. Take now $\tilde{\tau}$ to be the spanning tree defined with respect to $e_1 = \{1,2\}$ and $e_{3} = \{1,3\}$ instead. In this case, we have that 
    \begin{equation}\label{eq: Gram matrix of invariants phase}
        G_{\pmb{\psi}}^{\tilde{\tau}} = \left(\begin{matrix}
            1 & \sqrt{\text{Tr}(\psi_1\psi_2)} & \sqrt{\text{Tr}(\psi_1\psi_3)}\\
            \sqrt{\text{Tr}(\psi_1\psi_2)} & 1 & \sqrt{\text{Tr}(\psi_2\psi_3)}e^{i\tilde{\phi}}\\
            \sqrt{\text{Tr}(\psi_1\psi_3)} & \sqrt{\text{Tr}(\psi_2\psi_3)}e^{-i\tilde{\phi}} & 1
        \end{matrix}\right),
    \end{equation}
    where we have used the fact that now $(G_{\pmb{\psi}}^{\tilde{\tau}})_{13} = \sqrt{|\langle \psi_1|\psi_3\rangle |^2}$ since $\{1,3\} \in E(\tilde \tau)$ and, in this case, the phase $\tilde{\phi}$ becomes
    \begin{equation*}
        e^{i\tilde{\phi}} = \frac{\text{Tr}(\psi_1\psi_2\psi_3)}{|\text{Tr}(\psi_1\psi_2\psi_3)|}.
    \end{equation*}
    
    \myindent With the explicit tuple $\pmb{\Psi} = (\vert 0 \rangle, \vert + \rangle, \vert +_i \rangle)$ in this example we have that $\mathrm{Tr}(\psi_i\psi_j) = \sfrac{1}{2}$ for all $i,j$ and $e^{i\phi} = (1-i)/|1-i| =  (1-i)/\sqrt{2} = e^{-i\pi /4}$. This leads to the following matrix: 
    \begin{equation*}
        G_{\pmb{\psi}}^\tau = \left(\begin{matrix}
            1 & \sqrt{\frac{1}{2}} & \sqrt{\frac{1}{2}}\,e^{-i\pi/4}\\
            \sqrt{\frac{1}{2}} & 1 & \sqrt{\frac{1}{2}}\\
            \sqrt{\frac{1}{2}}\,e^{i\pi/4} & \sqrt{\frac{1}{2}} & 1
        \end{matrix}\right).
    \end{equation*}
    And for the case of $\tilde\tau$ the overlaps are all the same, but we have instead  $e^{i\tilde\phi} = e^{i\pi/4} = e^{-i\phi}$.
\end{example}

\begin{example}[Frame graph given by the $K_4$ complete graph of four nodes]
    Let us consider now the tuple $\pmb{\Psi} \in \mathcal{H}^4$ given by $\pmb{\Psi} = (\vert 0\rangle, \vert +\rangle, \vert +_i\rangle, \vert 0\rangle)$. We show its frame graph in Fig.~\ref{fig: K4 frame graph}. 
    \begin{figure}
        \centering
        \includegraphics[width=0.35\linewidth]{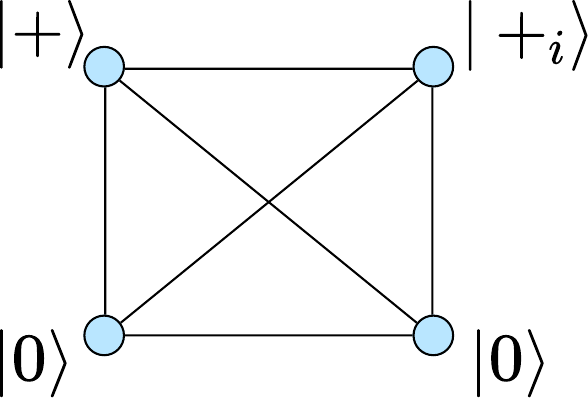}
        \caption{\textbf{Frame graph isomorphic to a complete graph $K_4$.} In this case we consider $\pmb{\Psi} = (\vert 0\rangle, \vert +\rangle, \vert +_i\rangle, \vert 0\rangle)$. All inner products have non-zero values. $F_{\pmb{\Psi}} \simeq K_4$ is shown, where we have labeled the vertices $1,2,3,4$ of the graph with the vector states $\vert \psi_1\rangle,\vert \psi_2\rangle, \vert \psi_3\rangle, \vert \psi_4\rangle $ of the tuple $\pmb{\Psi}$.}
        \label{fig: K4 frame graph}
    \end{figure}

    \myindent Following a similar procedure, we can construct $G_{\pmb{\psi}}^\tau$ for a given spanning tree of $K_4$. One may consider the spanning tree of frame graphs isomorphic to complete graphs $K_n$ as the star subgraph defined by edges $\{1, i\}$. We denote this choice of spanning tree as $\tau \equiv  \star$. In this case, we have that $G_{\pmb{\psi}}^\star$ is given by 
    \begin{equation}\label{eq: Gram matrix K4}
        G_{\pmb{\psi}}^\star = \left(\begin{matrix}
            1 & \sqrt{\text{Tr}(\psi_1\psi_2)} & \sqrt{\text{Tr}(\psi_1\psi_3)} & \sqrt{\text{Tr}(\psi_1\psi_4)}\\
            \sqrt{\text{Tr}(\psi_1\psi_2)} & 1 & \sqrt{\text{Tr}(\psi_2\psi_3)}e^{i\phi_{123}} & \sqrt{\text{Tr}(\psi_2\psi_4)}e^{i\phi_{124}}\\
            \sqrt{\text{Tr}(\psi_1\psi_3)} & \sqrt{\text{Tr}(\psi_2\psi_3)}e^{-i\phi_{123}} & 1 & \sqrt{\text{Tr}(\psi_3\psi_4)}e^{i\phi_{134}}\\
            \sqrt{\text{Tr}(\psi_1\psi_4)} & \sqrt{\text{Tr}(\psi_2\psi_4)}e^{-i\phi_{124}} & \sqrt{\text{Tr}(\psi_3\psi_4)}e^{-i\phi_{134}} & 1
        \end{matrix}\right),
    \end{equation}
    where we have added labels to the phases $\phi$ to facilitate understanding the specific cycle considered, and hence, the associated Bargmann invariant. In those cases we have
    \begin{equation*}
        e^{i\phi_{ijk}} = \frac{\text{Tr}(\psi_i\psi_j\psi_k)}{|\text{Tr}(\psi_i\psi_j\psi_k)|}.
    \end{equation*}
    Relative to the specific choice of $\pmb{\Psi} = (\vert 0 \rangle, \vert + \rangle, \vert +_i \rangle, \vert 0 \rangle)$ we have then that all overlaps are equal to $\sfrac{1}{2}$ except $\mathrm{Tr}(\psi_1\psi_4) = 1$. For the phases, we have that $e^{i\phi_{123}} = e^{i\pi/4}$ similarly from the last example, and $e^{i\phi_{1k4}} = \mathrm{Tr}(\psi_1\psi_k\psi_4)/|\mathrm{Tr}(\psi_1\psi_k\psi_4)| = \mathrm{Tr}(\psi_1\psi_k)/\mathrm{Tr}(\psi_1\psi_k)=1$ for $k \in \{2,3\}$ since $\psi_1 = \psi_4$.  This yields the following matrix: 
    \begin{equation}
        G_{\pmb{\psi}}^\star = \left(\begin{matrix}
            1 & \sqrt{\frac{1}{2}} & \sqrt{\frac{1}{2}} & 1\\
            \sqrt{\frac{1}{2}} & 1 & \sqrt{\frac{1}{2}}e^{i\pi/4} & \sqrt{\frac{1}{2}}\\
            \sqrt{\frac{1}{2}} & \sqrt{\frac{1}{2}}e^{-i\pi/4} & 1 & \sqrt{\frac{1}{2}}\\
            1 & \sqrt{\frac{1}{2}} & \sqrt{\frac{1}{2}} & 1
        \end{matrix}\right).
    \end{equation}
    
\end{example}

\myindent As a final remark, we want to point out that there is a relationship between the order of the Bargmann invariants that are required to write $G_{\pmb{\psi}}^\tau$ and how sparse the frame graph $F_{\pmb{\Psi}}$ is. In the two examples that we have just considered, the phases were all written in terms of third-order pure-state Bargmann invariants. However, if the frame graph is not complete as in these two examples, there are situations in which higher-order invariants are \emph{necessary}, in the sense that for all possible spanning trees, at least one of the phases is given in terms of fourth (or higher) order Bargmann invariants (the simplest example being the case where $F_{\pmb{\Psi}} \simeq C_n$ with $n\geq 4$). This specific fact is not  of particular relevance in this thesis, but it is certainly overlooked in various treatments~\citep{avdoshkin2023extrinsic}.

\myindent From our discussion so far, we have considered that the frame graph is not the primitive notion, but the tuples of vectors $\pmb{\Psi}$ are. Later in Chapter~\ref{chapter: relational coherence} we consider a different situation, in which the \emph{graphs are the primitive notion}, defining the class of all possible tuples of vectors for which the graph can be viewed as a frame graph. This simple change in perspective allow us to learn interesting properties about the \emph{geometry} of the sets of all possible Bargmann invariants. 

\myindent We have seen how Bargmann invariants completely characterize the problem of projective unitary equivalence for tuples of vector states, and the problem of unitary equivalence for tuples of pure rank-1 projective quantum states. To conclude, we also mention that similar results hold for more general mixed states, in which case, however, the situation becomes much more complicated as invariants of the type $\text{Tr}(\rho^2\sigma^3\eta \sigma \eta)$ start to be relevant. For a discussion of the more general problem of determining the necessary multivariate traces to decide unitary equivalence of tuples of mixed states, we refer to the recent book by~\cite{wigderson2019mathematics}, specifically the subsection on simultaneous conjugation.

\section{Measuring Bargmann invariants}\label{sec: measuring Bargmann invariants}

\myindent Note that the results just discussed are, to a certain extent, profound in the sense that they provide a complete set of quantities that completely characterize relational properties of sets of quantum states, where relational properties here are understood as properties characterized by \acrshort{pu} invariant functions over tuples of quantum states (or vector states). Given that we have already seen a relevant problem that Bargmann invariants solve and completely characterize, we now briefly comment on different proposals for \emph{estimating} them. 

\begin{figure}[t]
    \centering
    \includegraphics[width=0.5\linewidth]{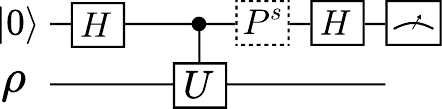}
    \caption{\textbf{Hadamard test.} The (pair of) quantum circuit(s) estimating $\langle U \rangle_\rho$. We input a product state $\vert +\rangle \langle + \vert \otimes \rho$. A controlled unitary between the auxiliary qubit and the two states is performed. If $s=0$, we apply a Hadamard gate and measure the auxiliary qubit in the computational $\{0,1\}$ basis. In this case the probability of observing $0$ is given by $p(0) = (1+\text{Re}[\text{Tr}(U\rho)])/2$. If $s=1$ we apply a phase gate $P=\mathrm{diag}(1,i)$ followed by a Hadamard gate and measure the auxiliary qubit in the computational $\{0,1\}$ basis. In this case the probability of observing $0$ is given by $p(0) = (1+\text{Im}[\text{Tr}(U\rho)])/2$.}
    \label{fig: Hadamard test}
\end{figure}

\myindent The most prominent examples of quantum circuits used to estimate Bargmann invariants are instances of the well-known \emph{Hadamard test}.~\footnote{To the best of our knowledge, it is unknown which was the first work to have introduced the Hadamard test. For textbook presentations we refer to the lecture notes by~\cite{lin2022lecturenotesquantumalgorithms}.} Before explaining each of these instances separately, we briefly describe the general structure of the Hadamard test. A quantum circuit implementing this test is shown in~\cref{fig: Hadamard test}. Hadamard tests are key subroutines in many quantum algorithms where one needs to estimate the expectation value of a unitary operator $U$ with respect to a quantum state $\rho$, i.e. $\langle U \rangle_\rho = \text{Tr}(U\rho)$. 

\myindent Since unitary operators are generally not Hermitian, this quantity is typically complex-valued. The Hadamard test therefore involves a pair of quantum circuits---labelled $s=0,1$ depending on whether a phase gate is applied---designed to separately estimate $\text{Re}[\text{Tr}(U\rho)]$ and $\text{Im}[\text{Tr}(U\rho)]$. 

\myindent The test proceeds as follows. We initialize an auxiliary qubit $\mathbb{C}^2$ in the coherent state $\vert +\rangle = \sfrac{1}{\sqrt{2}}(\vert 0\rangle + \vert 1\rangle) $, and prepare in parallel the system $\mathcal{H}$ in a quantum state $\rho \in \mathcal{D}(\mathcal{H})$. A controlled-unitary operation is then applied, given by $\vert 0\rangle \langle 0 \vert \otimes \mathbb{1}+\vert 1\rangle \langle 1 \vert \otimes U$, where $U$ is the unitary we want to find the average of. It is only after this controlled unitary that we distinguish between the two circuits implementing the test: To estimate $\text{Re}[\text{Tr}(U\rho)]$ we apply a Hadamard gate $H$ 
to the auxiliary qubit and measure it in the computational basis. In this case, $s=0$ and the probability of measuring outcome $0$ is  $$p(0) = \frac{1}{2}(1+\text{Re}[\text{Tr}(U\rho)]).$$ To instead estimate $\text{Im}[\text{Tr}(U\rho)]$ we apply a phase gate $P=\text{diag}(1,i)$, followed by a Hadamard $H$, and then measure it in the computational basis. In this case, $s=1$ and the probability of observing $0$ becomes $$p(0) = \frac{1}{2}(1+\text{Im}[\text{Tr}(U\rho)]).$$

\myindent In what follows, we build upon this idea to review quantum circuits capable of estimating multivariate traces of quantum states.

\subsection{Estimating two-state overlaps}

\myindent Starting with protocols for measuring $\text{Tr}(\rho \sigma)$, the first we consider is the SWAP test~\citep{buhrman2001quantumfingerprinting}. Let us denote $U_{\mathtt{SWAP}}: \mathcal{H} \otimes \mathcal{H} \to \mathcal{H} \otimes \mathcal{H}$ as the unitary operation such that $U_{\mathtt{SWAP}}(\vert u\rangle  \otimes \vert v\rangle ) = \vert v\rangle  \otimes \vert u\rangle $. In terms of a basis $\{\vert i\rangle\}_i$ for $\mathcal{H}$ we can write its action as~\citep{bandyopadhyay2023efficient} 
\begin{equation}\label{eq: SWAP unitary basis dec}
U_{\mathtt{SWAP}} = \sum_{i,j} \vert i\rangle \langle j\vert \otimes \vert j \rangle \langle i \vert,
\end{equation}
from which it is elementary to show that $U_{\mathtt{SWAP}}(\vert \psi \rangle \otimes \vert \phi \rangle )= \vert \phi \rangle \otimes \vert \psi \rangle $ and that for generic states $\rho_1,\rho_2$ we have that 
\begin{equation}\label{eq: overlap as average of SWAP on rho1rho2}
    \text{Tr}(\rho_1\rho_2) =  \text{Tr}(U_{\mathtt{SWAP}}(\rho_1 \otimes \rho_2)).
\end{equation}
This follows from the calculation:
\begin{align*}
    \mathrm{Tr}(U_{\mathtt{SWAP}}(\rho_1\otimes \rho_2)) &\stackrel{\eqref{eq: SWAP unitary basis dec}}{=} \mathrm{Tr}\left(\left(\sum_{i,j}\vert i\rangle \langle j\vert \otimes \vert j\rangle \langle i\vert\right)\, \rho_1 \otimes \rho_2 \right)\\
    &=\sum_{i,j}\mathrm{Tr}\left(\vert i\rangle \langle j \vert \rho_1\right)\mathrm{Tr}(\vert j\rangle \langle i \vert \rho_2)\\ 
    &=\sum_{i,j}\langle j \vert \rho_1 \vert i \rangle \langle i \vert \rho_2 \vert j\rangle \\
    &= \sum_j \langle j \vert \rho_1\rho_2 \vert j\rangle = \text{Tr}(\rho_1\rho_2).
\end{align*}

\begin{figure}[t]
    \centering
    \includegraphics[width=0.35\linewidth]{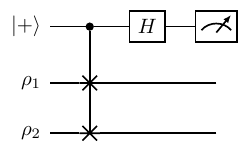}
    \caption{\textbf{SWAP test.} The quantum circuit estimating the two-state overlaps $\text{Tr}(\rho_1\rho_2)$. We input the product quantum state $\vert +\rangle \langle + \vert \otimes \rho_1 \otimes \rho_2$. A controlled SWAP (also known as a \emph{Fredkin gate}) between the auxiliary qubit and the two states is performed. We then measure the auxiliary qubit in the basis $\{\vert +\rangle, \vert -\rangle\}$, or equivalently, apply a Hadamard and measure with respect to the computational $\{0,1\}$ basis. The two-state overlap is recovered via the relation $p(0) = (1+\text{Tr}(\rho_1\rho_2))/2$.}
    \label{fig: SWAP test}
\end{figure}

\myindent It is then possible to use this relation to propose a quantum circuit for estimating the two-state overlap known as the SWAP test~\citep{buhrman2001quantumfingerprinting}, which performs a {Hadamard test} where the unitary $U = U_{\mathtt{SWAP}}$. When necessary, we write $U_{\mathtt{SWAP}_{a,b}}$ where $a,b$ are labels for the two systems being swapped. The SWAP test was recently re-discovered using machine learning techniques~\citep{schiansky2023demonstration,larocca2022group} and it is shown in Fig.~\ref{fig: SWAP test}. For a generalization of this test to generic \emph{qudit} systems $\mathcal{H} = \mathbb{C}^d$ or even infinite dimensional systems we refer to~\cite{fujii2003exchange} and~\cite{foulds2024generalising}.~\cite{foulds2021controlledSWAP} have also investigated applications for entanglement witnessing. 

\begin{figure}[t]
    \centering
    \includegraphics[width=0.35\linewidth]{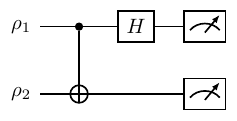}
    \caption{\textbf{Destructive SWAP test.} We input a product state $\rho_1 \otimes \rho_2$ of two single-qubit states $\rho_1,\rho_2 \in \mathcal{D}(\mathbb{C}^2)$. We then apply a $\mathrm{CNOT}$ gate, followed by a Hadamard on the first qubit, and perform local measurements in the computational basis on each qubit. We probability of observing $1$ in both qubits is $p(11) = {(1 - \text{Tr}(\rho_1\rho_2))}/{2}$. The same test can be generalized to the case where $\rho_1,\rho_2 \in \mathcal{D}((\mathbb{C}^2)^{\otimes n}).$ For the generic case with multiqubit systems we refer to~\citep{bandyopadhyay2023efficient}.}
    \label{fig: destructive SWAP}
\end{figure}

\myindent We note from Fig.~\ref{fig: SWAP test} that, to estimate the two-state overlap, we only perform measurements in the auxiliary qubit used to implement the Hadamard test. The product state $\rho_1 \otimes \rho_2$ is then projected onto another quantum state, but it is not destroyed in the process, i.e., other quantum gates could be applied to these systems. Contrastingly, the  \emph{destructive} SWAP \emph{test}~\citep{garcia2013swap,bandyopadhyay2023efficient,galvao2020quantum},  uses the fact that the SWAP operator can be expended in the Bell basis, and effectively destroys the two quantum states. The quantum circuit implementation is shown in Fig.~\ref{fig: destructive SWAP}. The destructive SWAP test uses one CNOT and one Hadamard gate. No non-Clifford gate is required. However, it is challenging to perform a similar test for systems $\mathcal{H} = (\mathbb{C}^d)^{\otimes n}$ where $d > 2$~\citep{garcia2013swap}. 


\myindent Note that for both tests shown in Figs.~\ref{fig: SWAP test} and~\ref{fig: destructive SWAP} we can imagine that a form of delegated quantum computation is happening, where we do not have complete information about the quantum states that are sent to us by another party. We can perform the quantum computation for this party, and return to them the overlap information without ever knowing the actual states that were prepared. 

\myindent Another simple possibility for estimating generic two state overlaps $\text{Tr}(\rho\sigma)$ is to use a prepare-and-measure scenario. In this situation, we consider a set-up where a state preparator prepares states $\rho$ and sends them to be measured by a binary-outcome \acrshort{povm} given by $\{\sigma,\mathbb{1}-\sigma\}$. In this case, the Born rule statistics for this prepare and measure scenario are given by $\{\text{Tr}(\rho\sigma),1-\text{Tr}(\rho\sigma)\}$. 

\subsection{Estimating  higher-order invariants}

\begin{figure}[t]
    \centering
    \includegraphics[width=0.55\textwidth]{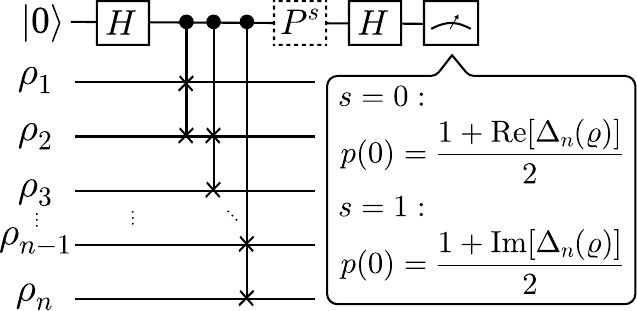}
    \caption{\textbf{Cycle test.} The quantum circuit estimating higher-order Bargmann invariants using the cycle test, inputs quantum states $\rho_1 \otimes \dots \otimes \rho_n$ and an auxiliary qubit system in the state $\vert +\rangle \langle + \vert$. A sequence of controlled Fredkin gates is applied. We then measure the auxiliary qubit in the $\{\vert + \rangle, \vert - \rangle \}$ basis, as in a standard Hadamard test to infer the average of $U_{\mathtt{CYC},n}$ over the input product states, hence estimating the values of Bargmann invariants due to Eq.~\eqref{eq: cycle operator relation}.}
    \label{fig: cycle test}
\end{figure}

\myindent We can generalize the argument made for the SWAP test to estimate generic higher-order Bargmann invariants with different implementations of a Hadamard test~\citep{oszmaniec2024measuring,quek2024multivariatetrace}. We start noticing that, for every $n$-tuple of quantum states $\pmb{\rho} = (\rho_i)_{i=1}^n \in \mathcal{D}(\mathcal{H})^n$, we can write 
\begin{equation}\label{eq: cycle operator relation}
    \Delta_n(\pmb{\rho}) = \text{Tr}(U_{\mathtt{CYC},n} \rho_1  \otimes \dots \otimes \rho_n)
\end{equation}
where $U_{\mathtt{CYC},n}$ is the unitary operator corresponding to the cyclic permutation $\pi_n$ of $n$ elements 
\begin{equation}\label{eq: cycle permutation}
    (a_1,a_2,a_3\dots,a_{n-1},a_n) \stackrel{\pi_{n}}{\mapsto} (a_2,a_3,\dots,a_{n-1},a_n,a_1)
\end{equation}
under the natural (faithful) representation of the symmetric group $S_n$ into the unitary group  $U(\mathcal{H}^{\otimes n})$, acting as $\pi \mapsto U_\pi$, with 
\begin{equation}
    U_\pi(\vert \psi_1\rangle \otimes \cdots \otimes \vert \psi_n\rangle ) = \vert \psi_{\pi^{-1}(1)}\rangle \otimes \cdots \otimes \vert \psi_{\pi^{-1}(n)}\rangle.
\end{equation}
When $n=2$ we have that $U_{\mathtt{CYC},2} \equiv U_{\mathtt{SWAP}}$. 
\myindent To perform the Hadamard test, we simply need to find decompositions of the cycle test into different elementary gates. The most natural choice is the  sequence of SWAPs,
\begin{equation}
    U_{\mathtt{CYC},n} = U_{\mathtt{SWAP}_{1,2}}U_{\mathtt{SWAP}_{2,3}}\cdots\, U_{\mathtt{SWAP}_{n-1,n}}.
\end{equation}
This leads to the Hadamard test shown in Fig.~\ref{fig: cycle test}, known as the \emph{cycle test}~\citep{oszmaniec2024measuring}. Note that when $n=2$ this reduces to the SWAP test from Fig.~\ref{fig: SWAP test}.

\myindent Different quantum circuits for estimating Bargmann invariants arise depending on the decomposition of the $n$-cycle unitary operator $U_{\mathtt{CYC},n}$ into SWAP gates, as well as on the specific gate set available for implementing these unitaries. One way of improving the depth of the quantum circuit just described is by using entangled states~\citep{oszmaniec2024measuring,quek2024multivariatetrace} as auxiliary systems, instead of a single-qubit control system as in Fig.~\ref{fig: cycle test}. In this case, different decompositions of Fredkin gates that use circuits of smaller depth are also possible~\citep{oszmaniec2024measuring,quek2024multivariatetrace}. We conclude this section with a disclaimer, noticing that \emph{much more} can be said about the estimation of multivariate traces using quantum circuits~\citep{faehrmann2025intheshadowofhadamard,simonov2025estimation}, or even using higher-order quantum operations which go beyond the quantum circuit paradigm~\citep{azado2025measuringunitaryinvariantsquantum}, but since estimation is not the main focus of Part II of this thesis, we refer the interested reader to the references pointed out above. 

\section{Applications of Bargmann invariants}\label{sec: applications of Bargmann invariants}

\myindent Next, we make a brief review of how Bargmann invariants have been useful for applications in quantum information science. We separate this section into two: applications of second-order Bargmann invariants and of higher-order Bargmann invariants. Our main focus on the applications of two-state overlaps are those related to the work by~\cite{galvao2020quantum}.

\subsection{Applications of two-state overlaps}\label{subsec: applications of two-state overlaps}

\myindent Two-state overlaps are in some sense ubiquitous in quantum mechanics. In particular, we can write the expectation value $\langle A \rangle_\rho$ of a Hermitian operator $A$ solely in terms of two-state overlaps. Therefore, to some extent, it is perhaps meaningless to consider `applications' of such a fundamental quantity. Unsurprisingly, we could list an enormous number of relevant uses for the estimation and characterization of two-state overlaps, ranging from applications in benchmarking quantum devices, quantum machine learning, quantum metrology for the estimation of the quantum Fisher information, etc.  Because of that, instead of stating the obvious relevance of two-state overlaps for the quantum formalism, we consider specific tests that  motivate our considerations in Part II of this thesis. More specifically, we review here the applications that were relevant to \emph{linear optics}, which motivated~\cite{galvao2020quantum} to introduce the framework and ideas that we generalize in Part II.

\subsubsection{Characterizing multiphoton indistinguishability}

\myindent We start pointing out that two-state overlaps---and more generally, \emph{Gram matrices}---play a major role in the theory of photonic indistinguishability. The most relevant starting point for an analysis of this fact is the \emph{Hong--Ou--Mandel (\acrshort{hom}) effect}~\citep{hong1987measurement}. In this case, we have that two single photons with internal degrees of freedom characterized by states $\vert \phi\rangle$ and $\vert \psi \rangle$ enter a beam-splitter. The output statistics depend solely on the two-state overlap $|\langle \psi|\phi\rangle|^2$ and the effect is perfectly observed if and only if these two single photon states are indistinguishable in the sense that $|\langle \phi|\psi\rangle|^2 = 1$. See~\citep{randles2025interferenceinterferenceeffects} for a recent gentle introduction to the topic, and~\citep{brod2021Bosons} for a gentle introduction to its relevance to computational complexity. This effect showcases the now better understood role played by such overlaps of states $\vert \phi_1\rangle, \dots, \vert \phi_n\rangle$ of $n$-photons entering a generic multimode interferometer (of which the beam splitter is an example having two modes) in characterizing multiphoton indistinguishability: for perfectly indistinguishable photons, all two-state overlaps need to be equal to $1$. 

\myindent  Various works have therefore proposed generic tools for characterizing and analyzing bosonic indistinguishability from the values of two-state overlaps~\citep{jones2020multiparticle,jones2023distinguishability,giordani2020experimental,giordani2021witnesses}, and investigating properties of bosonic indistinguishability (and its relevance to computational complexity and quantum advantage) is a growing research field. 

\subsubsection{Bounding unknown overlap values with values of known ones}

\myindent One interesting idea that motivated~\cite{galvao2020quantum} to introduce a formalism that we generalize in Chapter~\ref{chapter: event_graph_approach}, is to consider solutions to the following problem: Given an $n$-tuple of normalized vector states $\pmb{\Psi} \in \mathcal{H}_1^n$, is it possible to estimate fewer than all pairwise overlaps and \emph{bound} the remaining ones? As the number of input single-photon states grows, the number of two-state overlaps necessary to characterize indistinguishability also grows, and this characterization becomes ever more challenging. Their idea was then to understand if it was possible to estimate just some two-state overlaps and \emph{bound} the remaining ones that are \emph{never} measured, easing the experimental cost of  characterizing photonic indistinguishability. 

\myindent To do so, they propose to start with the following foundational question: Given any triple of normalized pure vector states $\Psi \in \mathcal{H}_1^3$, in any Hilbert space $\mathcal{H}$, what are all the possible \emph{values} that the two-state overlap tuple
\begin{equation*}
    \pmb{\Delta}(\pmb{\Psi}) = \left(\Delta_{1,2}(\pmb{\Psi}),\Delta_{1,3}(\pmb{\Psi}),\Delta_{2,3}(\pmb{\Psi})\right)
\end{equation*}
can take, where $\pmb{\Psi} = (\vert \psi_1\rangle, \vert \psi_2\rangle, \vert \psi_3\rangle)$ and $\Delta_{i,j}(\pmb{\Psi}) = |\langle \psi_i|\psi_j\rangle|^2 \equiv r_{i,j}(\pmb{\Psi})$. 

\myindent For triples  $\pmb{\Psi} \in \mathcal{H}_1^3$, ~\cite{galvao2020quantum} have proved the following theorem.

\begin{theorem}[Bounding unknown two-state overlaps]\label{theorem: bounding unknown two-state overlaps}
    Let $\pmb{\Psi} \in \mathcal{H}_1^3$, for any finite-dimensional Hilbert space $\mathcal{H}$. Then, if we let $r_2({\psi_i,\psi_j}) := \mathrm{Tr}(\psi_i\psi_j)$ denote two-state overlaps, and assuming we know the values of $r_2({\psi_1,\psi_2}), r_2({\psi_1,\psi_3})$, we have then the following upper bounds $$r_2({\psi_2,\psi_3}) = \left \{ \begin{matrix}
        r_- & \mathrm{if} \,\,\,r_2({\psi_1,\psi_2})+r_2({\psi_1,\psi_3}) > 1\\ 0,&\mathrm{otherwise}
    \end{matrix} \right. $$
    and lower bounds $$r_2({\psi_2,\psi_3}) \leq r_+,$$ where we have used $$r_{\pm} := \left(\sqrt{r_2({\psi_1,\psi_2})r_2({\psi_1,\psi_3})} \pm \sqrt{(1-r_2({\psi_1,\psi_2}))(1-r_2({\psi_1,\psi_3}))}\right)^2.$$
\end{theorem}

\myindent Theorem~\ref{theorem: bounding unknown two-state overlaps} can be used for the task of bounding unknown two-state overlaps from known ones. Suppose that we know $r_2({\psi_1,\psi_2})$ and $r_2({\psi_1,\psi_3})$. In this case, knowing that the possible values of $r_2({\psi_2,\psi_3})$ are not independent of these two due to Theorem~\ref{theorem: bounding unknown two-state overlaps} it is possible to provide non-trivial bounds on the values of $r_2({\psi_2,\psi_3})$ \emph{from the values} of $r_2({\psi_1,\psi_2})$ and $r_2({\psi_1,\psi_3})$. This certainly eases the experimental cost of estimating all overlaps to characterize, for example, photonic indistinguishability.

\myindent Let us provide a concrete example. Suppose that $r_2({\psi_1,\psi_2}) = r_2({\psi_1,\psi_3}) = 0.98$. Using Theorem~\ref{theorem: bounding unknown two-state overlaps} we can show that $r_2({\psi_2,\psi_3}) \geq 0.9216$ without ever measuring it.  As we have shown, overlaps alone do not provide the complete characterization of \acrshort{pu} equivalence, and hence cannot completely characterize multiphoton indistinguishability. Nevertheless, they are a necessary and useful benchmark. 

\myindent Theorem~\ref{theorem: bounding unknown two-state overlaps} yields the following corollary. 

\begin{corollary}\label{corollary: boundary of triplets of overlaps}
    Let $\pmb{\Psi} \in \mathcal{H}_1^3$ for any finite-dimensional Hilbert space $\mathcal{H}$. Then, the tuple $$\pmb{r}(\pmb{\Psi}) = \left(r_2({\psi_1,\psi_2}),r_2({\psi_1,\psi_3}),r_2({\psi_2,\psi_3})\right)$$ of two-state overlaps $r_2({\psi_i,\psi_j})=\mathrm{Tr}(\psi_i\psi_j)$ must satisfy the trivial bounds $0 \leq r_2({\psi_i,\psi_j}) \leq 1$ for all $i,j$ and the non-trivial bound
    \begin{equation}\label{eq: boundary of Q(C_3)}        r_2({\psi_1,\psi_2})+r_2({\psi_1,\psi_3})+r_2({\psi_2,\psi_3})-2\sqrt{r_2({\psi_1,\psi_2})r_2({\psi_1,\psi_3})r_2({\psi_2,\psi_3})} \leq 1.
    \end{equation}
\end{corollary}

With that,~\citep{galvao2020quantum} has also provided a solution to the foundational question of what are all the possible tuples of two-state overlaps of three vector states. However, they have also noticed that not all two-state overlaps \emph{require} coherent states. If we imagine for instance three maximally mixed states, their two-state overlaps are equivalent to the states that one would associate with three states of an ideal coin-flipping experiment. They proceed to bound what they have referred to as two-state overlaps following from \emph{coherence-free} states, later termed \emph{set incoherent} by~\cite{designolle2021set} as we have discussed in Chapter~\ref{chapter: quantum coherence}.

\subsubsection{Basis-independent coherence witnesses}

\myindent Consider the following situation, and here we follow closely~\cite{galvao2020quantum}, where we have three (possibly mixed) states that are \emph{set incoherent}. Hence we take quantum states  $\pmb{\rho} = (\rho_1,\rho_2,\rho_3) \in \mathcal{D}(\mathcal{H})^3$, of any finite-dimensional Hilbert space $\mathcal{H}$. \emph{If} these are incoherent, with respect to \emph{some} basis of reference $\mathbb{A}$ for $\mathcal{H}$ we have that each of these states must satisfy $$\rho_i = \sum_{\vert a\rangle  \in \mathbb{A} } \langle a|\rho_i|a\rangle \vert a \rangle \langle a \vert. $$
Let us denote the probability distribution $p_a^{(i)} := \langle a|\rho_i|a\rangle $, which according to the Born rule can be interpreted as the probability that upon performing the measurement $\{\vert a\rangle \langle a \vert \}_{\vert a\rangle \in \mathbb{A}}$ on the quantum state $\rho_i$ we observe $a$. In this case, all the states $\rho_i$ are diagonal with respect to this same basis of reference, and hence pairwise commute. Their overlaps then read:

\begin{equation*}
    r_{i,j}(\pmb{\rho}) = \text{Tr}(\rho_i\rho_j) = \sum_{\vert a\rangle \in \mathbb{A}}\sum_{\vert a'\rangle \in \mathbb{A}} \langle a|\rho_i|a\rangle \langle a'|\rho_j|a'\rangle |\langle a|a'\rangle |^2 = \sum_{\vert a \rangle \in \mathbb{A}}p_a^{(i)}p_{a}^{(j)}.
\end{equation*}

\myindent Therefore, overlaps of pairwise commuting states (hence satisfying the equation above) can be interpreted as the probability that when independently measuring $\rho_i$ and $\rho_j$ with \acrshort{pvm} $\{\vert a \rangle \langle a \vert \}_{\vert a \rangle \in \mathbb{A}}$, one obtains the same outcome. Using just this fact,~\cite{galvao2020quantum} proceed to prove the following theorem.

\begin{theorem}[Adapted from~\citep{galvao2020quantum}]\label{theorem: n-cyclic inequalities}
    Let $\pmb{\rho} \subseteq \mathcal{I}(\mathcal{H},\mathbb{A})^3$ for some Hilbert space $\mathcal{H}$ and some basis $\mathbb{A}$. Then, the two-state overlaps $r_2({\rho_{i},\rho_{i+1}}) = \mathrm{Tr}(\rho_i\rho_{i+1})$ must satisfy the $n$-cycle inequalities~\eqref{eq: overlap cycle inequalities}. Where the summation of the labels is taken modulo $n$. 
\end{theorem}

\myindent Then, in particular, it holds that set incoherent two-state overlaps satisfy the set of inequalities 
    \begin{align}
        +r_2({\rho_1,\rho_2}) + r_2({\rho_1,\rho_3})-r_2({\rho_2,\rho_3}) \leq 1,\label{ineq: overlap triplets 1} \\
        +r_2({\rho_1,\rho_2}) - r_2({\rho_1,\rho_3})+r_2({\rho_2,\rho_3}) \leq 1, \label{ineq: overlap triplets 2}\\
        -r_2({\rho_1,\rho_2}) + r_2({\rho_1,\rho_3})+r_2({\rho_2,\rho_3}) \leq 1. \label{ineq: overlap triplets 3}
    \end{align}
that we have already presented in Chapter~\ref{chapter: quantum coherence}, Inequalities~\eqref{eq: overlap cycle inequalities}. We note that the conjunction of all these inequalities defines a convex polytope in $[0,1]^3$. We review the main theory of convex polytopes in Appendix~\ref{sec: convex polytopes}. This implies that sets of quantum states violating these inequalities must be set coherent. Of relevance to us is the technique that Galvão and Brod used to prove~\cref{theorem: n-cyclic inequalities}: they used the \emph{transitivity of equality}. 

\myindent Their argument starts by contemplating the possibility that we interpret each quantum state $\rho$ of the $n$-tuple $\pmb{\rho}$ as a vertex of a cycle graph. Then, they notice that, from the assumption that these are all pairwise commuting density matrices, it must hold that $$r_2({\rho_i,\rho_{i+1}}) = \sum_ap_a^{(i)}p_a^{(i+1)} = p(A_i = A_{i+1}),$$ where, in this last part, we have re-written the overlap using $A_i$ and $A_{i+1}$ to be independent random variables taking values on the set of labels $a$, satisfying that $ p(A_i = a) = p_a^i$. The core of their argument lies then on Boole's rule~\citep{boole1854investigation} that for a given pair of logical propositions $\mathfrak{p},\mathfrak{q}$ it holds that 
$$p(\mathfrak{p}\wedge \mathfrak{q}) = p(\mathfrak{p}) + p(\mathfrak{q}) - p(\mathfrak{p}\vee \mathfrak{q}) \geq p(\mathfrak{p}) + p(\mathfrak{q})-1,$$
where $p(\mathfrak{p})$ is the probability that proposition $\mathfrak{p}$ is true. Considering $A_i = A_{i+1}$ as propositions, the inequalities from Theorem~\ref{theorem: n-cyclic inequalities} follow. For brevity, and because we recover these results in Chapters~\ref{chapter: event_graph_approach} and~\ref{chapter: quantum coherence} we do not present their argument in full detail.

\myindent Let us comment on two important points for us. Since their work is at the heart of most of our main questions it is useful to make some simple and immediate comments on their findings. To start, similar inequalities have appeared before and have been investigated by different communities. For example, the same inequalities for the case of $n=3$ were found by~\cite{caves2002conditions} when investigating \emph{antidistinguishability}. In the context of bounding ontological models,~\cite{suppes1981when} have also shown that these inequalities bound Bell local models when substituting $r_{i,j}(\pmb{\rho})$ with the two-point correlation functions $\langle A_iA_j\rangle$. When re-writing these inequalities in terms of such two-point correlation functions they become mathematically equivalent to those found by~\cite{araujo2013all}, and to the `original' Bell inequalities~\citep{larsson2014loopholes}. Moreover, after the appearence of reference~\citep{galvao2020quantum} the same characterization of the set of quantum correlations provided by~\cref{corollary: boundary of triplets of overlaps} was found by~\cite{le2023quantum} when characterizing the set of quantum correlations in the simplest Bell scenario. All these suggest that there might be a simple connection (or mapping) between these inequalities and existing Bell and noncontextuality inequalities, as we have pointed out in the introduction.

\myindent Another aspect is that~\cite{galvao2020quantum} have had the insight of considering a graph representation for organizing and presenting their results. Above, the states $\rho_i,\rho_{i+1}$ are viewed as labeling vertices of the $n$-cycle graph, and using the transitivity of equality together with Boole's rule they find their family of coherence witnesses. This suggests a simple generalization of their approach to any graph. 
\subsubsection{Classical dimension witnesses}

\myindent To conclude our review of applications on two-state overlaps, we comment on the work of~\cite{galvao2020quantum} and~\cite{giordani2021witnesses} that considered the task of dimension witnessing as a possible application. In their case, they have shown that there are regions inside the polytope characterized by the inequalities~\eqref{ineq: overlap triplets 1}-\eqref{ineq: overlap triplets 3} for which one can guarantee that triples of two-state overlaps must come from quantum states in Hilbert spaces of dimension larger than two. However, as already mentioned in Chapter~\ref{chapter: information tasks}, since these regions lie \emph{inside} the polytope just mentioned, one cannot guarantee that the states in the tuple $\pmb{\rho}$ are set coherent. In Chapter~\ref{chapter: applications} we show two-state overlap inequalities for which this separation is also possible, but \emph{outside} the polytope formed by considering tuples of two-state overlaps $\pmb{r}(\pmb{\rho})$ for which the set of states in the tuple $\pmb{\rho}$ are set incoherent.

\subsection{Applications of higher-order Bargmann invariants}\label{subsec: applications of higher-order Bargmann invariants}

\myindent To conclude this chapter, we briefly highlight some notable applications of Bargmann invariants $\text{Tr}(\rho_1 \cdots \rho_n)$ with $n\geq 3$. While these are not the central focus of this thesis, we believe they represent a fertile area of research that is likely to see significant developments shortly.

\myindent One of the most impactful applications  lies, again, in characterizing multiphoton indistinguishability. A pioneer in discussing the relevance of Bargmann invariants to multiphoton indistinguishability has been~\cite{shchesnovich2015partial} (see also~\cite{shchesnovich2018collective}). His findings were later experimentally investigated by~\cite{menssen2017distinguishability} and~\cite{jones2020multiparticle}. More recently, Bargmann invariants have been used to explore nuanced aspects of multiphoton indistinguishability, as demonstrated in works by~\cite{jones2023distinguishability} and~\cite{pont2022quantifying}. Given the importance of indistinguishability in the computational task of boson sampling~\citep{aaronson2011bosonsampling,brod2021Bosons} foundational studies have investigated how \emph{partial} indistinguishability affects phenomena like boson bunching~\citep{seron2023boson,rodari2024experimentalobservationcounterintuitivefeatures,rodari2024semideviceindependentcharacterizationmultiphoton}.

\myindent Beyond quantum optics, Bargmann invariants have relevant implications for condensed matter physics. As we have mentioned in the introduction, the notion of geometric (also known as Berry) phases~\citep{berry1984quantal} and that of Bargmann invariants have been connected by~\cite{simon1993Bargmann} (see also~\citep{akhilesh2020geometric}). However, the relevance of Bargmann invariants to condensed matter physics is still an active topic of research. Recent studies by~\cite{avdoshkin2023extrinsic} used Bargmann invariants to characterize the \emph{extrinsic} geometry of manifolds parameterized by vector states.~\cite{reascos2023quantum} used the fact that there is a relationship between \emph{spin chirality} and Bargmann invariants, proposing a quantum circuit scheme using the Hadamard test to estimate this property.

\myindent Another emerging frontier is the relationship between higher-order Bargmann invariants and quasiprobability distributions, such as those introduced by~\citep{kirkwood1933quantum} and~\cite{dirac1945analogy} (commonly referred to as \acrshort{kd} distributions~\citep{arvidsson_Shukur2024properties}). While \acrshort{kd} distributions have traditionally been applied to quantum states, recent extensions by~\cite{schmid2024kirkwood} have expanded these representations to encompass generic quantum processes such as measurements and transformations. The first work to point out that these quasiprobability distributions of states are simply given in terms of Bargmann invariants was~\citep{wagner2024quantumcircuits}. This brought together the community interested in Bargmann invariants---traditionally interested in condensed matter physics due to its connection with Berry phase, or in multiphoton indistinguishability as mentioned above--with a community interested in a wide range of topics such as quantum thermodynamics~\citep{lostaglio2022kirkwood,gherardini2024quasiprobabilities}, scrambling of quantum information~\citep{halpern2018quasiprobability,gonzalez2019otoc}, quantum computation~\citep{thio2025kirkwooddiracnonpositivitynecessaryresource}, weak value theory~\citep{wagner2023anomalous,thio2024contextualityverifiednoncontextualexperiments}, and quantum metrology~\citep{arvidssonShukur2020quantum}. 

\myindent In our view, these diverse applications firmly establish the importance of Bargmann invariants and underscore the need for a unified and broadly applicable framework to explore their foundational implications. This thesis builds upon the groundwork laid by~\cite{shchesnovich2015partial} and~\cite{galvao2020quantum}, particularly for two-state overlaps, and extends these insights to higher-order Bargmann invariants. The forthcoming chapters present the main findings of this thesis, which aim to contribute scientifically rigorous results to this vibrant subfield of quantum information science.

\part{Results}

\chapter{Event graph framework}\label{chapter: event_graph_approach}

\begin{quote}
    ``\emph{The origins of graph theory are humble, even frivolous. Whereas many branches of mathematics were motivated by fundamental problems of calculation, motion, and measurement, the problems which led to the development of graph theory were often little more than puzzles, designed to test the ingenuity rather than to stimulate the imagination. But despite the apparent triviality of such puzzles, they captured the interest of mathematicians, with the result that graph theory has become a subject rich in theoretical results of a surprising variety and depth.}''\\
    ~\citep{biggs1986graph}
\end{quote}

\myindent Graph theory is a well-established branch of mathematics with applications spanning several fields. Its role in physics, however, has been somewhat more specialized. While not traditionally central to the physicist's toolkit, graph theory has found its relevance in multiple specialized corners of quantum information science. For example, as we have mentioned in Chapter~\ref{chapter: contextuality}, various \emph{graph approaches} provide natural frameworks for organizing and investigating key quantum mechanical concepts such as contextuality, Bell nonlocality, and measurement incompatibility. Beyond foundational studies, graph-theoretic methods also appear in practical contexts, including optimization, quantum search algorithms, and causal inference.

\myindent Arguably, this wide range of applications motivates one to learn basic graph-theoretic constructions, particularly for researchers exploring the intersection of mathematics and quantum science. However, it also raises an important question: Should we develop \emph{yet another} graph approach? 

\myindent Before we start our first Chapter of this thesis that considers our research outputs, let us provide some arguments motivating the relevance of our framework, answering the question above. Of course, sufficient motivation would be the fact that it is used to resolve Questions~\ref{question: coherence witnesses and contextuality} and~\ref{question: generalize galvao and broads results} that are part of the main focus of this thesis. We have already argued in favor of these back in Chapter~\ref{chapter: introduction}, yet now we take the opportunity to motivate the framework \emph{beyond} the scientific purpose of answering the main questions of this thesis. In other words, we believe that arguments can be made for the relevance of our framework even if one \emph{does not find} the main questions we have outlined in Chapter~\ref{chapter: introduction} interesting. 

\begin{enumerate}
    \item (Usefulness.) Of course, our most relevant argument is that it is \emph{useful} for some purposes. For example, it allows us to uncover new applications for quantum foundations and quantum information, not previously considered for coherence (and contextuality) as we show later on in Chapter~\ref{chapter: applications}. 
    \item (Simplicity.) To some extent, the necessary concepts and ideas we consider are not (and do not require) profound results from graph theory. In essence, we need only some familiarity with basic types of graphs and graph-theoretic definitions to be able to prove interesting aspects of the framework. Arguably, our appeal here is that one, in fact, \emph{does not need to commit} to a deep understanding of graph theory to start contributing, nor to understand the statements of most of the results we present.  
    \item (Encompassing.) As we show in Chapter~\ref{chapter: from overlaps to noncontextuality}, our framework can be translated to other frameworks (namely the \acrshort{csw} graph-approach) from which we argue that our framework can serve as a useful starting point for those interested in graph theory and its connection to \acrshort{ks} contextuality.  
    \item (Coherence.) A particularly novel---and perhaps hardest to motivate---aspect of our work is the application of graph-theoretic techniques to quantum coherence, an area not typically associated with graph-based methods (unlike contextuality). We address this motivation more thoroughly in Chapter~\ref{chapter: relational coherence}, where we show, among other things, that our framework leads to a \emph{comprehensive classification} of basis-independent quantum coherence witnessed by two-state overlaps.
    \item (Open problems.) Any proposal for a topic of research is as good as the questions it raises. While we propose and advance the graph-theoretic framework considered here (that we term the \emph{event graph approach}) we leave various open questions regarding its structure, which we discuss later at the end of this Chapter. We also leave some open problems on the connection between this graph approach and its associated quantum realizability problems in Chapter~\ref{chapter: relational coherence}. 
    \item (Unification.) As we also stress later in Chapter~\ref{chapter: relational coherence}, by investigating a specific type of quantum realizability problem related to this graph approach it is possible to investigate \emph{within a single framework} more than one aspect of quantum theory, including coherence (see Chapter~\ref{chapter: relational coherence}), the failure of Bell's notion of local causality and Kochen and Specker's notion of contextuality (see Chapter~\ref{chapter: from overlaps to noncontextuality}), generalized contextuality (see, again, Chapter~\ref{chapter: from overlaps to noncontextuality}), and bosonic indistinguishability (see Chapters~\ref{chapter: Bargmann invariants} where we have reviewed advances in this direction). \emph{This} is, in our view, a powerful motivation for investigating this approach when compared to other existing graph approaches. 
\end{enumerate}

\myindent We now return to the focused questions we attempt to answer in this thesis. In this Chapter, we extend the formalism that has been introduced by~\cite{galvao2020quantum}, as considered and developed by~\cite{wagner2024inequalities,wagner2024coherence,wagner2024certifying}. This Chapter is our proposal for resolving~\cref{question: generalize galvao and broads results}, one of the main questions we tackle. We discuss here the mathematical aspects of constructing what we call the \emph{event graph polytopes}, characterized by facet-defining \emph{event graph inequalities}, and their geometrical properties. For brevity of the presentation, we assume in this Chapter some familiarity of the reader with basic concepts in graph theory and convex geometry that are reviewed in detail in Appendices~\ref{sec: graph theory} and~\ref{sec: convex polytopes}, respectively. Since we commonly make use of pictorial representations and illustrative examples we believe that readers unfamiliar with graph theory could follow the discussion, at least intuitively and in broad strokes. However, we advise that to have a good understanding of the technical contributions herein, some familiarity with basic graph-theoretic constructions is necessary. 

\myindent The structure of this Chapter is as follows. We start in Sec.~\ref{sec: event graph approach} by presenting the framework, defining the class of graphs $G$ we are interested in, which we dub event graphs, and how these graphs relate to a subclass of convex 0/1-polytopes, dubbed the event graph polytopes, denoted as $\mathfrak{C}(G)$. We also show that every element $r \in \mathfrak{C}(G)$ can also be viewed as `realized' by a set of jointly distributed random variables, in a very specific sense. Provided with the rules for defining polytopes $\mathfrak{C}(G)$ from event graphs $G$ we classify and investigate the properties of various families of event graph polytopes (organized by families of graphs) in~\cref{sec: notable event graphs}. We then investigate in~\cref{sec: subpolytopes} what happens if we consider further restrictions on random variables realizing elements $r \in \mathfrak{C}(G)$, inducing sub polytopes of $\mathfrak{C}(G)$.
 We make a few final remarks in~\cref{sec: discussions event graph}.  

\section{Event graphs and event graph polytopes}\label{sec: event graph approach}

\myindent Let $G = (V(G), E(G))$ be a simple graph. Since these graphs are so relevant to our framework, we refer to $G$ as an \emph{event graph}.  We consider edge weightings~\footnote{See~\cref{def: vertex labeling and colouring} in Appendix~\ref{sec: graph theory}.} $r: E(G) \to [0,1]$, which assign a weight $r(e) \equiv r_e = r_{ij}$ to each edge $e=\{i,j\}$. Recall the usual bijective correspondence between a function $r: E(G) \to [0,1]$ and the associated tuple $\pmb{r} = (r_e)_{e \in E(G)} \in [0,1]^{|E(G)|}$. We use various equivalent notations for the elements $r(e)$ with $e=\{i,j\}$ of a tuple $\pmb{r}$, such as $r_e \equiv r_{i,j} \equiv r_{ij} $. When the domain of an edge weighting is $\{0,1\}$ instead of $[0,1]$ we refer to it as a \emph{deterministic edge weighting}, or as a \emph{0/1-labeling}. 

\myindent To define the sets of tuples $\pmb{r}$ we are interested in, namely, the event graph polytopes, we start by defining an equivalence relation on event graphs. Given a deterministic edge weighting $\alpha: E(G) \to \{0,1\}$, define a relation $\sim_\alpha$ on the set of vertices of an event graph $G$ whereby
$v \sim_\alpha w$ if and only if there is a path~\footnote{Including the path of length zero. See~\cref{def: walks and paths of a graph} in Appendix~\ref{sec: graph theory}.} from $v$ to $w$ through edges labeled by $1$.

\begin{definition}[Equivalence relation for deterministic edge weightings]\label{def: equivalence relation in graphs from paths}
    Let $G$ be an event graph and $\alpha: E(G) \to \{0,1\}$ be any deterministic edge weighting for $G$. Let us define the relation $R_\alpha  \subseteq V(G) \times V(G)$ as follows: 
    \begin{enumerate}
    \item[(a)] If $v \in V(G)$ then $(v,v) \in R_\alpha$. 
    \item[(b)] If $v,w \in V(G)$ with $v \neq w$, then $(v,w) \in R_\alpha$ if and only if there is a finite sequence $e_1, \ldots, e_n \in E(G)$ such that $v \in e_1$, $w \in e_n$, $e_i \cap e_{i+1} \neq \emptyset$, and $\alpha(e_i)=1$ for every $i$. 
    \end{enumerate}
    We use the common notation to write $(v,w) \in R_\alpha$ as $v \sim_\alpha w$. 
\end{definition}

\myindent Note that the above is a well-defined relation, which we now show it to be an \emph{equivalence relation}. 

\begin{lemma}
    Let $G$ be any event graph and $\alpha$ be any deterministic edge weighting. The relation  $\sim_\alpha$ from Def.~\ref{def: equivalence relation in graphs from paths} is an equivalence relation. 
\end{lemma}

\begin{proof}
    By construction $(v,v) \in R_\alpha$. Now suppose that $v \sim_\alpha w$. This implies that there is a path $e_1,\dots,e_n$ from $v$ to $w$ for which $\alpha(e_i) = 1$ for every $i$ (note that $v$ and $w$ need not be adjacent vertices). From the backwards path $e_n,e_{n-1},\dots,e_1$ we conclude that $w \sim_\alpha v$. To conclude, suppose that we have $v,u,w \in V(G)$ such that $v \sim_\alpha u$ and $u\sim_\alpha w$. Let $e_1,\dots,e_n$ be the path from $v$ to $u$ and $s_1,\dots,s_n$ be the path from $u$ to $w$ we have that the sequence $e_{e_1},\dots,e_{e_n},e_{s_1},\dots,e_{s_n}$ for which $e_{x} = x$ (e.g. $e_{e_1} = e_1$) satisfy that  $\alpha(e_x) = 1$ for all labels $x$ and that $v \in e_{e_1}, w \in e_{s_n}$ from which we conclude that $v \sim_\alpha w$.
\end{proof}

\begin{figure}
    \centering
    \includegraphics[width=0.5\linewidth]{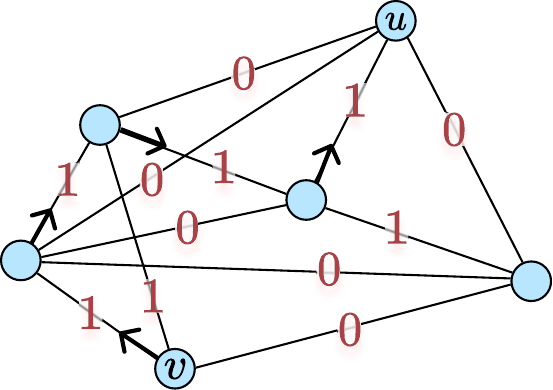}
    \caption{\textbf{An example of an event graph, a choice of $\{0,1\}$-labeling, and two nodes pertaining to the same equivalence class.} The labeling $\alpha: E(G) \to \{0,1\}$ is shown. Note that $u \sim_\alpha v$ due to the path described by the arrows. Note also that this is not the only path that can be used to show that $u \sim_\alpha v$. This is an example of labeling that is \emph{not} a point of the event graph polytope $\mathfrak{C}(G)$ as it is easy to find two nodes $i\sim_\alpha j$ for which $\alpha(\{i,j\}) = 0$.}
    \label{fig:equivalence_graph_relation}
\end{figure}

\myindent Figure~\ref{fig:equivalence_graph_relation} provides an illustration of an example of an event graph $G$, a deterministic edge weighting $\alpha$ for $G$, and two vertices $u,v$ for which $u \sim_\alpha v$ for a specific choice of $\alpha$. Having this notion of  equivalence relation at our disposal we can now define the following polytope:

\begin{definition}[Event graph convex polytope]\label{def: event graph polytope}
    Let $G$ be an event graph. Let $\alpha: E(G) \to \{0,1\}$ be a deterministic edge weighting such that for every edge $\{i,j\} \in E(G)$, we have that  $$i \sim_\alpha j \implies \alpha(\{i,j\}) = 1.$$ Denoting the set of all such $\alpha$ as $\mathcal{V}_G$ we define $\mathfrak{C}(G):= \text{ConvHull}(\mathcal{V}_G)$ as the convex hull of all such deterministic edge weightings.  
\end{definition}

\myindent The set $\mathfrak{C}(G)$ constructed above is a convex polytope for every possible choice of event graph $G$ because it is the convex hull of a subset of $[0,1]^{E(G)}$ of finitely many points (always fewer than $2^{|E(G)|}$). Suppose that $\alpha$ is a deterministic edge weighting for which $u \sim_\alpha w$ \emph{but} $\alpha(\{u,w\}) = 0$. Since $G$ is an event graph, it is a simple graph, and hence $\{u,w\} \in E(G)$ implies that $u \neq w$. Moreover, the condition $u \sim_\alpha w$ together with $\alpha(\{u,w\}) = 0$ holds if and only if there exists a closed path $e_1, \dots, e_n, \{u,w\}$ in which all edges except $\{u,w\}$ are assigned the value $1$ by $\alpha$, while $\{u,w\}$ is assigned $0$. This leads to an alternative characterization of the set $\mathcal{V}_G$.

\begin{definition}[Equivalent definition of $\mathcal{V}_G$]\label{def:kernel_loops}
    Let $\mathcal{C}(G)$ be the set of all subgraphs of $G$ isomorphic to some cycle graph $C_n$, with $n\geq 3$. Then $$\mathcal{V}_G = \{\alpha \in \{0,1\}^{E(G)}\mid |\text{Ker}(\alpha|_{\ell})|\neq 1, \forall \ell \in \mathcal{C}(G)\},$$ where $\text{Ker}(\alpha) = \{e \in E(G) \mid \alpha(e) = 0\}$ and $\alpha|_{\ell}$ denotes the restriction of $\alpha$ to the edges $E(\ell) \subseteq E(G)$ of $\ell \in \mathcal{C}(G)$.
\end{definition}

\myindent Various simple facts can be immediately shown to hold. For example, note that since $\mathfrak{C}(G) \subseteq [0,1]^{E(G)}$ we have trivially that $0 \leq r_e \leq 1$ for every edge weighting $r \in \mathfrak{C}(G)$. We refer to these as \emph{trivial inequalities}.

\begin{example}[$\mathcal{V}_{C_3}$]
    \begin{figure}[t]
        \centering
        \includegraphics[width=0.65\linewidth]{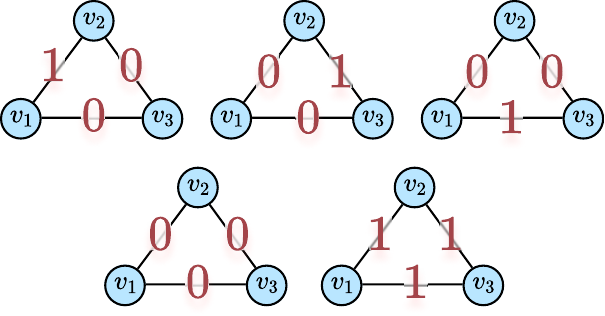}
        \caption{\textbf{Extreme points of $\mathfrak{C}(C_3)$}. The $3$-cycle graph $C_3$ is showed with $V(C_3) = \{v_1,v_2,v_3\}$. The five different graphs are labeled by the five possible deterministic edge weightings in  $\mathcal{V}_{C_3}$.}
        \label{fig: extremal_of_C_3_verticse}
    \end{figure}
    Let $C_3$ be the $3$-cycle graph. The possible 0/1-labelings $\alpha: E(C_3) \to \{0,1\}$ for this graph are 
    \begin{equation*}
        \left\{ (0,0,0), (1,0,0), (0,1,0), (0,0,1), (1,1,0), (0,1,1), (1,0,1), (1,1,1) \right \},
    \end{equation*}
    where we have considered the (arbitrarily chosen) ordering
    $$\pmb{\alpha} = (\alpha(\{v_1,v_2\}),\alpha(\{v_1,v_3\}),\alpha(\{v_2,v_3\})) \equiv (\alpha_e)_{e \in E(C_3)}.$$
    This is the list corresponding to the $8$ vertices of the cube $[0,1]^3$. From that set, we note that $\pmb{\beta} = (1,0,1) \notin \mathcal{V}_{C_3}$ since $v_1 \sim_{\beta} v_3$ but $\beta(\{v_1,v_3\}) = 0$. Similarly, $(1,1,0),(0,1,1) \notin \mathcal{V}_{C_3}$. All the other deterministic edge weightings satisfy the definition of $\mathcal{V}_{C_3}$ described by Def.~\ref{def: event graph polytope}. Equivalently, Fig.~\ref{fig: extremal_of_C_3_verticse} shows that according to Def.~\ref{def:kernel_loops} all the tuples not in $\mathcal{V}_{C_3}$ have a cycle subgraph (in this case, the only possibility being the graph itself) with a single 0 assignment on its edges.
\end{example}

\myindent There are some notable edge weightings that we can immediately show to be in \emph{every} event graph polytope. 

\begin{proposition}\label{prop:simplex_inside_event_polytopes}
    Let $\mathfrak{S}^{n}$ denote the standard $n$-simplex.~\footnote{Usually the $n$-dimensional simplex is denoted as $\Delta_n$. We avoid this notation because we use $\Delta_n(\pmb{\rho})$ to denote Bargmann invariants.} For every event graph $G$ we have that $\mathfrak{S}^{|E(G)|} \subseteq \mathfrak{C}(G)$.
\end{proposition}

\begin{proof}
    The extreme points of $\mathfrak{S}^{|E(G)|}$ are isomorphic to edge $\{0,1\}$-labelings such that for a fixed $e^\star \in E(G)$ one has $\alpha_{e^\star} = 1$ while for all $e \neq e^\star$ one has $\alpha_{e}= 0$. Therefore, by construction, any such deterministic edge weighting is an element of $\mathfrak{C}(G)$, and since $\mathfrak{C}(G)$ is convex the convex hull of all such points is also contained in $\mathfrak{C}(G)$. 
\end{proof}

\myindent Note that the simplex $\mathfrak{S}^{|E(G)|}$ is always a \emph{proper} subset of $\mathfrak{C}(G)$, implying that any element $r \in \mathfrak{C}(G)$ can be described by more than one convex decomposition in terms of elements in $\mathcal{V}_G$. The fact that the $|E(G)|$-simplex is always a proper subset is guaranteed by the following lemma. 

\begin{lemma}\label{lemma: constants are classical}
    For any $G$ the constant edge weighting, i.e., $r(e) = s \in [0,1]$ for all $e \in E(G)$, is an element of $\mathfrak{C}(G)$.
\end{lemma}

\begin{proof}
    Clearly the constant deterministic edge weightings assigning $r_0(e)=0, \forall e$ and $r_1(e)=1,\forall e$ are in $\mathfrak{C}(G)$. We can then write any constant edge weighting as $r = s r_1 + (1-s)r_0$, for any $s \in [0,1]$, and since $\mathfrak{C}(G)$ is convex we get that $r \in \mathfrak{C}(G)$.
\end{proof}

\myindent As shown by~\cite{wagner2024inequalities}, the problem of deciding if a certain deterministic edge weighting is an extreme point of the polytope (i.e. an element in $\text{ext}(\mathfrak{C}(G))$) is computationally easy, but the problem of finding \emph{all} extreme points is computationally hard. 

\begin{corollary}
    The set $\mathfrak{C}(G)$ is compact.
\end{corollary}

\myindent The above is a corollary from a well-known result in convex polytope theory, which states that any convex polytope is compact (see~\cref{theorem: polytopes are compact} in Appendix~\ref{sec: convex polytopes}). From the Heine--Borel theorem, $\mathfrak{C}(G)$ is also closed and bounded. From Minkowski's (or in this case, equivalently, the Krein--Milman) theorem we get that $\mathfrak{C}(G) = \text{ConvHull}(\text{ext}(\mathfrak{C}(G)))$, i.e., it is the convex hull of extreme points. All these supporting theorems are reviewed in Appendix~\ref{sec: convex polytopes}. This implies that $\mathcal{V}_G \supseteq \text{ext}(\mathfrak{C}(G))$. It could be that $\mathcal{V}_G$ has more points than $\text{ext}(\mathfrak{C}(G))$ (since this constitutes the so-called \emph{minimal} V-representation of a convex polytope), in case some elements of $\mathcal{V}_G$ can be written as convex combinations of the others. Their convex hull would then define the \emph{same} convex polytope. However, this is not the case. We can see this by noting that every element of $\mathcal{V}_G$ is also a vertex of the hypercube $[0,1]^{E(G)}$. Therefore, these are elements in $\text{ext}([0,1]^{E(G)})$. Because of that, we conclude that  $\mathcal{V}_G = \text{ext}(\mathfrak{C}(G))$.

\myindent To some extent, the discussion above has presented the mathematical backbone of the framework. Of course, above we have merely a mathematical construction devoid of interpretation, but this is the basic idea of the framework: to every event graph $G$ we can associate convex polytopes that we term event graph polytopes. The facet-defining inequalities characterizing the minimal H-representation for such polytopes are called the \emph{event graph inequalities}. In~\cref{sec: notable event graphs} we compute several of these inequalities. The relevance and importance of these constructions are further investigated in Chapter~\ref{chapter: relational coherence}. We now show how such polytopes are related to jointly distributed random variables.

\subsection{Edge weightings from jointly distributed random variables}

\myindent We want to consider an equivalent characterization of the event graph polytope. The fact that this is an equivalent characterization will only become clear later (in fact, we will state this equivalence as a theorem), while for now, we momentarily treat it as an entirely different construction of a convex polytope. To define the \emph{event-graph polytope associated to jointly distributed random variables}, which for now we denote as $\mathfrak{C}_{\mathbb{P}}(G) \subseteq [0,1]^{E(G)}$, take each vertex $i \in V(G)$ to `represent' a random variable $A_i$ with values belonging to some alphabet $\Lambda$, and suppose these are jointly distributed.  This determines an edge weighting $r$ where each weight $r_{ij}$ is the probability that the random variables corresponding to vertices $i$ and $j$ output equal values, i.e. $$r_{ij} = p(A_i = A_j) .$$
An edge weighting $r$ is in $\mathfrak{C}_{\mathbb{P}}(G)$ if it arises in this fashion from jointly distributed random variables $(A_i)_{i \in V(G)}$. Each weight $r_{ij}$ is then a measure of the correlation between the output values of $A_i$ and of $A_j$.
In the case of a dichotomic alphabet  $\Lambda = \{ +1, -1\}$, this quantity is related to the expectation value of the product by  $\left\langle A_i A_j\right\rangle = 2r_{ij}-1$. Note that we do not assume a fixed finite outcome set $\Lambda$ or a bound on its size. The set $\mathfrak{C}_{\mathbb{P}}(G)$ consists of the edge weightings that arise from jointly distributed random variables with outcomes in \emph{some} set $\Lambda$. We frame this in terms of the notion of \emph{realizability by jointly distributed random variables}. 

\begin{definition}[Event graph polytope from jointly distributed random variables]\label{def:realizable_joint_dist}
    Let $G=(V(G),E(G))$ be an event graph. Let the mapping $v \mapsto A_v$ associate to every vertex of the graph a random variable $A_v$ taking values from an alphabet  $\Lambda$, such that $(A_v)_{v \in V(G)}$ is a jointly distributed set of random variables. We say that an edge weighting  $r \in \mathfrak{C}_{\mathbb{P}}(G) \subseteq [0,1]^{E(G)}$ if and only if for every $\{i,j\}\in E(G)$ we have  $$r_{ij} =  p(A_i = A_j),$$ where $p$ is the joint probability distribution for $(A_v)_{v \in V(G)}$. If this holds, we say that $r$ is \emph{realizable} by a set of jointly distributed random variables.   
\end{definition}

\myindent From the definition above $\mathfrak{C}_{\mathbb{P}}(G)$ is a convex set since, for any two edge weightings $r^{(1)}, r^{(2)} \in \mathfrak{C}(G)$ and $a \in (0,1),$ we have that for all edges $e \in E(G)$ it holds that  
\begin{align*}
    \tilde{r}_e &:= a r_e^{(1)} + (1-a) r_e^{(2)} \\
    &= a \,p(A_i^{(1)}= A_j^{(1)}) + (1-a) p(A_i^{(2)}= A_j^{(2)})\\
    &= p(A_i=A_j),
\end{align*}
where we have considered an extended joint distribution for a new set of random variables $(A_i)_{i \in V(G)}$ now taking values in $\Lambda \sqcup \Lambda'$. Let us consider an ordering of the finite set of labels $|V(G)|$ for the tuples of random variables $(A_i)_{i \in V(G)} = (A_1, A_2,\dots, A_{|V(G)|})$. In such a new set we have that these random variables are jointly distributed via the following construction: If $\tilde{\lambda} = (1, \lambda), \lambda \in \Lambda$ then, for every $A_i$,  $$p_{A_1,A_2,\dots,A_{|V(G)|}}(\tilde \lambda_1,\tilde{\lambda}_2, \dots, \tilde{\lambda}_{|V(G)|}) = a \,  p_{A_1^{(1)},A_2^{(1)},\dots,A_{|V(G)|}^{(1)}}(\lambda_1,\lambda_2,\dots,\lambda_{|V(G)|})$$ and if $\tilde{\lambda} = (2,\lambda'), \lambda' \in \Lambda'$ then 
$$p_{A_1,A_2,\dots,A_{|V(G)|}}(\tilde \lambda_1,\tilde{\lambda}_2, \dots, \tilde{\lambda}_{|V(G)|}) = (1-a) \,  p_{A_1^{(2)},A_2^{(2)},\dots,A_{|V(G)|}^{(2)}}(\lambda_1',\lambda_2',\dots,\lambda_{|V(G)|}').$$
For all the remaining assignments, constituting the cases where  $(\tilde{\lambda}_1,\dots,\tilde{\lambda}_{|V(G)|})$ has some elements $\tilde{\lambda} = (1,\lambda)$ and others $\tilde{\lambda}' = (2,\lambda')$, we set   $p_{A_1,\dots,A_{|V(G)|}}(\tilde{\lambda}_1,\dots,\tilde{\lambda}_{|V(G)|}) = 0$. By definition, this implies that $\tilde{r} \in \mathfrak{C}(G)$. Explicitly, given $e=\{i,j\}$ for simplicity let us arrange $(\lambda_1,\lambda_2,\dots,\lambda_{|V(G)|},\lambda_i,\lambda_j) \equiv (\pmb{\lambda},\lambda_i,\lambda_j)$. we have then
\begin{align*}
    \tilde{r}_{i,j} &= p(A_i=A_j) = \sum_{\tilde{\lambda} \in \Lambda \sqcup \Lambda'}\sum_{\widetilde{\pmb{\lambda}} \in (\Lambda \sqcup \Lambda')^{V(G)\setminus \{i,j\}}}p_{A_1,\dots,A_{|V(G)|}}(\widetilde{\pmb{\lambda}},\tilde{\lambda},\tilde{\lambda})\\
    &= a \sum_{\lambda \in \Lambda}\sum_{\pmb{\lambda} \in \Lambda^{V(G)\setminus \{i,j\}}} p_{A_1^{(1)},\dots,A_{|V(G)|}^{(1)}}(\pmb{\lambda},\lambda,\lambda) \\
    &\,\,\,\,\,\,+ (1-a)\sum_{\lambda' \in \Lambda'}\sum_{\pmb{\lambda'} \in (\Lambda')^{V(G)\setminus \{i,j\}}} p_{A_1^{(2)},\dots,A_{|V(G)|}^{(2)}}(\pmb{\lambda'},\lambda',\lambda')\\
    &=a \sum_\lambda p_{A_i^{(1)},A_j^{(1)}}(\lambda,\lambda)+(1-a)\sum_{\lambda' \in \Lambda'}p_{A_i^{(2)},A_j^{(2)}}(\lambda',\lambda')\\
    &=a p(A_i^{(1)}=A_j^{(1)})+(1-a)p(A_i^{(2)} = A_j^{(2)}) = a r^{(1)}_{i,j}+(1-a)r^{(2)}_{i,j}.
\end{align*}

\myindent This shows the following lemma:

\begin{lemma}\label{lemma: C_P is convex}
    The set $\mathfrak{C}_{\mathbb{P}}(G)$ is a convex subset of $[0,1]^{E(G)}$.
\end{lemma}

\myindent Since edge-weightings $r$ are realizable by jointly distributed random variables, whenever they assign weight $1$ to an edge $\{i,j\}$ one has that $A_i = A_j$. This implies that deterministic edge weightings $\alpha$ arising from such distributions cannot be such that $i \sim_\alpha j$ with $\alpha_{i,j}= 0$ because otherwise, we would be violating \emph{the transitivity of equality}. This hints at the possibility that both sets $\mathfrak{C}_{\mathbb{P}}(G)$ and $\mathfrak{C}(G)$ are the same, which is what we now prove.

\begin{theorem}\label{theorem: C_G = C_P_G}
    Let $G$ be any event graph. Then $\mathfrak{C}(G) = \mathfrak{C}_{\mathbb{P}}(G)$.
\end{theorem}

\begin{proof}  $\left(\mathfrak{C}(G) \subseteq \mathfrak{C}_{\mathbb{P}}(G)\right)$. Suppose that $r \in \mathfrak{C}(G)$, which implies that $r = \sum_\alpha p_\alpha \alpha$ where $\alpha \in \text{ext}(\mathfrak{C}(G))$, and therefore $\alpha:E(G) \to \{0,1\}$ is an edge 0/1-labeling for which no cycle from $G$ has a single zero edge label (by Def.~\ref{def:kernel_loops}). We now consider, for each such $\alpha$, the construction of a family of random variables $(A_i^\alpha)_{i \in V(G)}$ to be any such that $A_i^\alpha = A_j^\alpha$ if $\alpha_{ij} = 1$ and, when $\alpha_{ij} = 0$ we let $A_i^\alpha$ and $A_j^\alpha$ be random variables taking values on disjoint sets $\Lambda_i, \Lambda_j \subseteq \Lambda$, i.e., $\Lambda_i \cap \Lambda_j = \emptyset$ for all edges $\{i,j\} \in E(G)$ such that $\alpha_{i,j} = 0$. Therefore, whenever $\alpha_{ij} = 0$ we have that $p(A_i = A_j) = 0$ by construction. Note that this construction is only possible because $\alpha \in \text{ext}(\mathfrak{C}(G))$. Again, by construction, these are all independent random variables, hence jointly distributed by the product of each distribution of each random variable, and $p(A_i = A_j) = \alpha_{ij}$ implying that  $\alpha \in \mathfrak{C}_{\mathbb{P}}(G)$ for all $\alpha \in \text{ext}(\mathfrak{C}(G))$. Since, from Lemma~\ref{lemma: C_P is convex},  $\mathfrak{C}_{\mathbb{P}}(G)$ is a convex set and $p_\alpha$ are convex weights it implies that $r \in \mathfrak{C}_{\mathbb{P}}(G)$. 
    
    \myindent $\left(\mathfrak{C}(G) \supseteq \mathfrak{C}_{\mathbb{P}}(G)\right)$. We will show that if $r$ is an edge weighting then $r \notin \mathfrak{C}(G) \implies r \notin \mathfrak{C}_{\mathbb{P}}(G)$. In this case, for every convex decomposition of $r = \sum_\alpha p_\alpha \alpha$ there will always be at least one deterministic edge weighting  $\alpha^\star$ satisfying that $\{u,v\} \in E(G)$ with $u \sim_{\alpha^\star} v$ and $\alpha^\star_{u,v} = 0$. In all such cases, $\alpha^\star \notin \mathfrak{C}_{\mathbb{P}}(G)$, due to a violation of transitivity of equality. Moreover, since any such $\alpha^\star$ is an extreme point, it cannot be described as the convex combination of other points in $[0,1]^{E(G)}$, and in particular, cannot be described by convex combinations of other points in the convex subset $\mathfrak{C}_{\mathbb{P}}(G)$. Therefore, any edge weighting $r \notin \mathfrak{C}(G)$ of this form must be described by some convex combination of points, where at least one point $\alpha^\star$ cannot be an element of $\mathfrak{C}_{\mathbb{P}}(G)$. Hence $r \notin \mathfrak{C}_{\mathbb{P}}(G)$. 
\end{proof}

\myindent Because of this theorem, from now on we simply write $\mathfrak{C}(G)$ as the event graph polytope associated with an event graph $G$, and interpret Def.~\ref{def:realizable_joint_dist} as an equivalent definition of $\mathfrak{C}(G)$. Both descriptions of this polytope are useful.

\myindent If $\mathfrak{C}(G) \neq \mathfrak{C}(G')$ are two polytopes (having the same full dimensionality, hence differing in terms of their minimal H-representation) then their event graphs must also be different (i.e., there can be no graph isomorphism between $G$ and $G'$), which trivially follows from the fact that $\mathfrak{C}$ is a function from event graphs $G$ to $0/1$-polytopes. 

\begin{proposition}\label{proposition: different graphs with the same event polytope}
    There exist non-isomorphic event graphs $G$ and $G'$ for which $\mathfrak{C}(G) = \mathfrak{C}(G')$. 
\end{proposition}

\begin{proof}
    Let $G$ and $G'$ be two graphs with the same number of nodes and edges, but we let $G$ to be a star graph while $G'$ to be a path graph. For both of these graphs, $\mathfrak{C}(G) = \mathfrak{C}(G')$, but $G$ and $G'$ are not isomorphic. In fact, in this case, $\mathfrak{C}(G) = \mathfrak{C}(G') = [0,1]^{|E(G)|}$, which we refer to as the `trivial' event graph polytope (see Theorem~\ref{theorem: tree theorem for event graphs}). 
\end{proof}

\myindent We now showcase a bit of the usefulness of relating graphs with convex polytopes: we select and organize classes of convex polytopes depending on the properties of the associated graphs. This possibility was already somewhat hinted at by~\cref{proposition: different graphs with the same event polytope} from what we can infer that any graph $G$ that has no cycles (hence, a tree) but the same number of edges and nodes is related to the same event graph (trivial) polytope $\mathfrak{C}(G) = [0,1]^{|E(G)|}$ (see Theorem~\ref{theorem: tree theorem for event graphs}). 

\subsection{Event graph polytopes from graph decompositions}

\myindent In this section, we prove some general facts that relate to the classical polytopes of different graphs. In particular, we show that some methods of combining graphs to build larger graphs do not give rise to new non-trivial classes of facet-defining inequalities. Or, seen analytically rather than synthetically, that some graphs $G$ can be decomposed into smaller component graphs in a way that reduces the question of characterizing $\mathfrak{C}(G)$ to that of characterizing the polytopes of these components. These observations help trim down the class of graphs that is worth analyzing in the search for new inequalities, which is physically motivated in Chapters~\ref{chapter: relational coherence} and~\ref{chapter: contextuality}.

\begin{proposition}\label{prop:disjointunion}
Let $G_1$ and $G_2$ be event graphs, and write $G_1 + G_2$ for their disjoint union. Then
\[\mathfrak{C}(G_1 + G_2) = \mathfrak{C}({G_1}) \times \mathfrak{C}({G_2}) = \{(r_1,r_2)\mid r_1 \in \mathfrak{C}(G_1), r_2 \in \mathfrak{C}(G_2)\}.\]
\end{proposition}
\begin{proof}
Two finite-dimensional convex polytopes are equal if and only if they have equal extreme points. We have that $\alpha$ is a deterministic edge weighting for $G = (V(G_1)\sqcup V(G_2), E(G_1)\sqcup E(G_2))$ if and only if $\alpha: E(G_1) \sqcup E(G_2) \to \{0,1\}$ such that for all $\{u,v\}$ we have that $u \sim_\alpha v$ implies $\alpha_{uv}=1$. Since the disjoint union of a graph describes a \emph{disconnected graph}, every edge in this graph comes only from either one graph or the other, and therefore $u \sim_\alpha v$ implies both are vertices of the same graph. Because of that,  $\alpha \in \text{ext}(\mathfrak{C}(G_1+G_2))$ if, and only if, there exists $\alpha \equiv [\alpha_1,\alpha_2]: E(G_1) \sqcup E(G_2) \to \{0,1\}$ such that 
\begin{equation}
    \alpha(e) = \left \{ \begin{matrix}
        \alpha_1(e), & \text{if }e \in E(G_1) \\
        \alpha_2(e), & \text{if }e \in E(G_2)
    \end{matrix}\right.
\end{equation}
where $\alpha_1 \in \text{ext}(\mathfrak{C}(G_1))$ and any $\alpha_2 \in \text{ext}(\mathfrak{C}(G_2))$. This concludes the proof.
\end{proof}

\myindent Since both $\mathfrak{C}(G_1),\mathfrak{C}(G_2)$ are convex polytopes, in such a situation the H-representation of the product is simply given in terms of the collection of both H-representations of each polytope individually~\citep{wagner2023using}. This result implies that every non-trivial class of facet-defining inequalities is simply given by the inequalities from each polytope individually. We can then use~\cref{prop:disjointunion} to justify our focus solely on \emph{connected} simple graphs. From now on, unless stated otherwise, event graphs are taken to be \emph{connected} simple graphs. 

\myindent If instead of considering the binary operation yielding the disjoint union of two event graphs, we consider binary operations that `glue' the two, in some circumstances, it is still possible to infer aspects of the final polytope from the two polytopes considered \emph{before} the `gluing' happened. See Appendix~\ref{sec: graph theory} for the formal description of graph compositions we consider, which go under the name of `gluing'. 

\begin{figure}[t]
    \centering
    \includegraphics[width=0.8\linewidth]{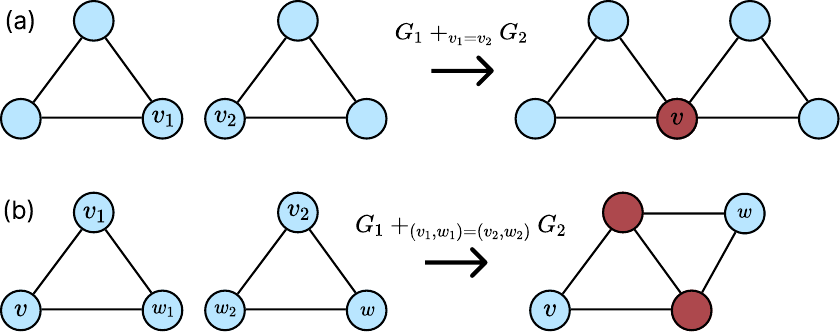}
    \caption{\textbf{Illustration of gluing along a vertex and gluing along an edge.} (a) We let two event graphs $G_1\simeq G_2 \simeq C_3$ and we either glue them along a vertex, so that $v_1 \in V(G_1)$ and $v_2 \in V(G_2)$ become $v \in V(G_1 +_{v_1=v_2} G_2)$. (b) We now consider gluing along an edge for the edges labeled by $\{v_1,w_1\} \in E(G_1)$ and $\{v_2,w_2\} \in E(G_2)$. From~\cref{prop:glue_vertex} and~\cref{prop:glu_edge} we learn that the H-representation of the graphs on the left can be found to be either trivial inequalities $0 \leq r_{ij} \leq 1$ or relabelings of the non-trivial facet-defining inequalities of $\mathfrak{C}(C_3)$ related to the new labelings for the nodes of the constructed graphs. }
    \label{fig: gluings example}
\end{figure}

\myindent Let us start considering gluing any two graphs along a single vertex (see Fig.~\ref{fig: gluings example}-(a) for an illustration). In this simple case, we can show that the event graph polytope resulting from such a gluing is identical to that obtained from the disjoint union.

\begin{lemma}[Gluing along a vertex does not generate new non-trivial facets]\label{prop:glue_vertex}
Let $G_1$ and $G_2$ be graphs, $v_1 \in V(G_1)$ and $v_2 \in V(G_2)$, then $\mathfrak{C}({G_1 +_{v_1 = v_2} G_2}) = \mathfrak{C}({G_1}) \times \mathfrak{C}({G_2})$.
\end{lemma}

\begin{proof}
    We start noting that every deterministic assignment $\alpha$ can be written as $\alpha \equiv [\alpha_1,\alpha_2]$. This lemma will follow again from the fact that  $\alpha_1 \in \text{ext}(\mathfrak{C}(G_1)), \alpha_2 \in \text{ext}(\mathfrak{C}(G_2))$ if, and only if, $\alpha \equiv [\alpha_1,\alpha_2]  \in \text{ext}(\mathfrak{C}(G_1 +_{v_1=v_2} G_2))$. To see this, suppose that there is some deterministic $\alpha$ that is not an extreme point of $\mathfrak{C}(G_1 +_{v_1=v_2} G_2)$. This happens iff an edge $\{u,v\}$ exists in this graph such that $u \sim_\alpha v$ but $\alpha_{uv}=0$. Since, by definition, this graph is the gluing along a vertex of $G_1$ and $G_2$ we have that there exists $i \in \{1,2\}$ for which $\{u,v\} \in E(G_i)$ such that  $\alpha_i = \alpha|_{E(G_i)}$ is not a deterministic edge weighting from $\text{ext}(\mathfrak{C}(G_i))$. Note that, crucially, we have used that any path establishing $u \sim_\alpha v$ must be entirely contained in one of the two graphs since there is only one vertex connecting them. 
\end{proof}

\myindent Read analytically, if $G$ is a graph with a \emph{cut vertex} $v$, \ie a vertex whose removal disconnects the graph into two components with vertex sets $V_1$ and $V_2$, then its polytope can be characterized in terms of the polytopes of the induced subgraph on $V_1 \cup \{v\}$ and $V_2 \cup \{v\}$.
In particular, the facet-defining inequalities of $\mathfrak{C}(G)$ are those of each of these two components.

\begin{corollary}
    Let $(G_i)_{i=1}^n$ be a finite sequence of event graphs. Let $v_i \in V(G_i)$ for any $i\in \{1,\dots,n\}$.  If we denote $\tilde{G}$ the sequential gluing along the vertices, i.e., 
    $$\tilde{G} = G_1 +_{v_1 = v_2} G_2 +_{v_2=v_3} G_3 +_{v_3=v_4} \dots +_{v_{n-1}=v_n} G_n, $$
    then we have that $$\mathfrak{C}(\tilde{G}) = \prod_{i=1}^n \mathfrak{C}(G_i),$$
    where above we have used that, given the fixed sequence of vertices, gluing along a sequence of vertices is associative. 
\end{corollary}

\myindent As an aside, this result is the missing ingredient for fully characterizing the event graphs that cannot display any nonclassicality, i.e. for which all edge weightings $E(G) \to [0,1]$ are classical.
This could be seen as an analog of Vorob{\textquotesingle}ev's  theorem~\citep{vorobyev1962consistent} in our framework.
\begin{theorem}\label{theorem: tree theorem for event graphs}
A graph $G$ is such that $\mathfrak{C}(G) = [0,1]^{E(G)}$ if and only if it is a tree.
\end{theorem}
\begin{proof}
For `only if' part, if $G$ has a cycle then any edge labeling $E(G) \to \{0,1\}$ that restricts to $(1,\ldots,1,0)$ on said cycle is not in $\mathfrak{C}(G)$.
For the `if' part,
apply \cref{prop:glue_vertex} multiple times, following the construction of a tree as a sequence of gluings along a vertex of copies of $K_2$, whose event graph polytope is $[0,1]$.
\end{proof}

\myindent If we glue graphs along an \emph{edge} instead of along a vertex (as showed in Fig.~\ref{fig: gluings example}-(b)) we do not have that one remains with the same notion of a combination of convex polytopes as the two cases above. Nevertheless, one does not generate new non-trivial facet-defining inequalities. 

\begin{proposition}[Gluing along an edge does not generate new non-trivial facets]\label{prop:glu_edge}
    Let $\,G_1$ and $G_2$ be graphs, $v_1,w_1 \in V(G_1)$ and $v_2,w_2 \in V(G_2)$ such that $e_i := \{v_i,w_i\} \in E(G_i)$. Writing
\[G := G_1 +_{(v_1,w_1)=(v_2,w_2)} G_2,\]
we have
\begin{align*}
    \mathfrak{C}({G}) =& \setdef{r \in [0,1]^{E(G)}}{r|_{E(G_1)} \in \mathfrak{C}({G_1}), r|_{E(G_2)} \in \mathfrak{C}({G_2})}
    \\ \cong& \setdef{(r,s)}{r \in \mathfrak{C}({G_1}), s \in \mathfrak{C}({G_2}), r_{e_1} = s_{e_2}} 
    \\ \cong& (\mathfrak{C}({G_1}) \times [0,1]^{E(G_2)\setminus\enset{e_2}}) \cap ([0,1]^{E(G_2)\setminus\enset{e_1}} \times \mathfrak{C}({G_2})) ,
\end{align*}
where for the last line we assume that $\mathfrak{C}({G_1})$ is written with $e_1$ as its last coordinate and $\mathfrak{C}({G_2})$ with $e_2$ as its first coordinate.
\end{proposition}

\myindent The proof of this result can be found in the appendix of reference~\citep{wagner2024inequalities}. Note therefore from the above that it is simple to infer the H-representation of $\mathfrak{C}(G)$ when it is the gluing along an edge of two other graphs, but that this is not equivalent to the H-representation given by the operation of gluing along a vertex. Also, the above shows that nontrivial facet-defining inequalities can only appear when it is impossible to consider a graph as a composition of graphs glued along an edge or a vertex. As we will see later, this implies that various notable graph-theoretic constructions have the same event graph polytopes up to the relabeling of edge weightings. 

\begin{corollary}\label{corollary:sequential_gluing}
    Let $(G_i)_{i=1}^n$ be a finite sequence of event graphs, and let $e_i \equiv \{v_i,w_i\} \in E(G_i)$ for every $i \in \{1,\dots,n\}$. Then, let $\tilde{G}$ be the event graph that is the sum of all such graphs along the same edge, i.e.,
    $$\tilde{G} := G_1 +_{e_1=e_2} G_2 +_{e_1=e_3} G_3 +_{e_1=e_4} \dots +_{e_{1}=e_n} G_n,$$
    then we have that 
    $$\mathfrak{C}(\tilde{G}) = \left \{(r_i)_{i=1}^n \mid \forall i, r_i \in \mathfrak{C}(G_i), r_1(e_1)=r_2(e_2)=\dots = r_n(e_n) \right\},$$
    where we have used that gluing along the same edge is associative.
\end{corollary}

\myindent Gluings, in general, can be made along more edges. Generalizing the corollary above to more general types of gluings is not a trivial task. This is because gluings, in general, are capable of generating nontrivial facet-defining inequalities. We will see an example of this later on, but one will be able to see this since gluing two graphs $C_3$ along an edge and then gluing another graph $C_3$ on the resulting graph along \emph{two} edges can generate a graph $K_4$ that has nontrivial facet-defining inequalities that are \emph{not} present in $C_3$ graphs.  

\myindent One can notice that a graph  introduce more constraints on the possible deterministic edge weightings than any of its subgraphs since having fewer edges implies fewer cycle subgraphs. Intuitively, this suggests that the event graph polytope of a graph should be viewed as a sub polytope of any event graph polytope constructed from its subgraphs. The following theorem makes this intuition precise. 

\begin{theorem}[Event graph polytopes of any subgraph ]
    Let $G'$ be a subgraph of $G$, such that $V(G) \subseteq V(G')$ and $E(G')\subseteq E(G)$. Then $\mathfrak{C}(G)$ is a subpolytope of $\mathfrak{C}(G') \times [0,1]^{E(G) \setminus E(G')}$
\end{theorem}

\begin{proof}
    We want to show that $\text{ext}(\mathfrak{C}(G)) \subseteq \text{ext}\left(\mathfrak{C}(G') \times [0,1]^{E(G) \setminus E(G')}\right)$. If $\alpha \in \text{ext}(\mathfrak{C}(G))$, for every $e = \{u,v\} \in E(G')$, $u \sim_\alpha v$ implies $\alpha_e = 1$ since $\{u,v\} \in E(G)$. Moreover, for every $e \notin E(G')$, we have $\alpha_e \in \{0,1\}$. Because of that, \begin{align*}\alpha \in \text{ext}(\mathfrak{C}(G')) \times \{0,1\}^{E(G)\setminus E(G')} &= \text{ext}(\mathfrak{C}(G')) \times \text{ext}\left([0,1]^{E(G)\setminus E(G')}\right) \\&= \text{ext}\left(\mathfrak{C}(G') \times [0,1]^{E(G) \setminus E(G')}\right),\end{align*}
    as we wanted.
\end{proof}

\myindent This theorem is crucial as it justifies investigating more carefully a specific family of graphs, the densest possible graphs of $n$ vertices, which are the complete graphs $K_n$.

\begin{corollary}[Event graph polytopes of complete graphs]\label{corollary: event graph polytopes of complete graphs}
    $\mathfrak{C}(K_n)$ is a subpolytope of $\mathfrak{C}(G) \times [0,1]^{|E(K_n)\setminus E(G)|}$ any event graph having $n$ vertices.
\end{corollary}

\section{Notable event graph polytopes and their H-representations}\label{sec: notable event graphs}

\myindent In what follows we use the construction discussed in the previous section to present some interesting families of event graph polytopes. Algorithmically, the H-representation is found from the V-representation provided by the characterization of $\mathfrak{C}(G)$ from deterministic edge weightings in $\text{ext}(\mathfrak{C}(G))$ we have introduced before. All event graph inequalities we discuss here can be found in~\cite{wagner2022github}.

\subsection{Cycle graphs}

\begin{figure}[t]
    \centering
    \includegraphics[width=0.7\linewidth]{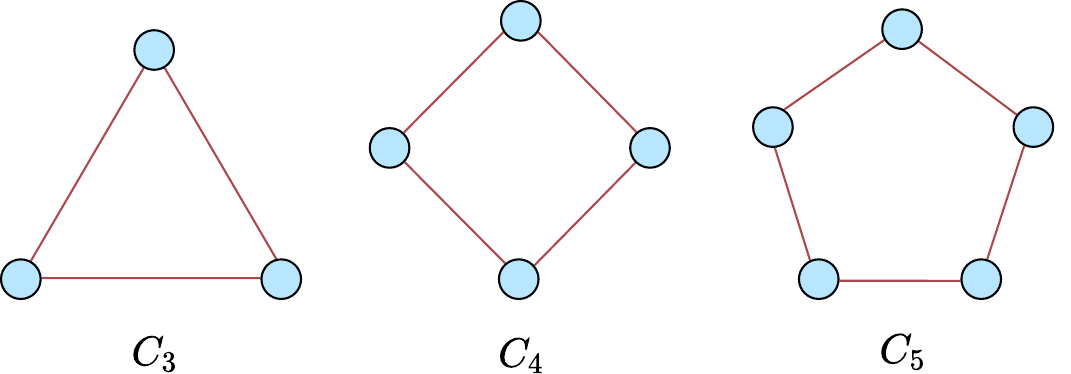}
    \caption{\textbf{Cycle graphs $C_n$.} We show the first three elements of the family of cycle graphs $C_n$ for $n=3,4$ and $5$ yielding non-trivial facet-defining inequalities of $\mathfrak{C}(C_n)$. The complete list of facet-defining inequalities for $\mathfrak{C}(C_n)$ for all $n$ is given by trivial inequalities of the type $0 \leq r_e \leq 1$ and by the family of inequalities found by~\cite{galvao2020quantum}. }
    \label{fig: cycle_graphs_figure}
\end{figure}

\myindent We now show how, in our formalism, we recover the inequalities found by~\cite{galvao2020quantum}. In our framework, and at this point, these are not interpreted as overlap inequalities. For now these are facet-defining inequalities for the convex polytope $\mathfrak{C}(G)$ \emph{or}, interpreting each $r_{ij}$ as a `measure' of the correlation between two jointly distributed and independent random variables $A_i$ and $A_j$, inequality bounds on such quantities. The connection between these two views arises due to Theorem~\ref{theorem: C_G = C_P_G}. Every inequality found by~\cite{galvao2020quantum} is a facet-defining inequality of $\mathfrak{C}(C_n)$ for some $n$. The first graphs of the family of cycle graphs are shown in~\cref{fig: cycle_graphs_figure}. Since we get these inequalities as a particular case, of a more comprehensive framework valid for any graph and not only for cycle graphs, we argue this is a solution of~\cref{question: generalize galvao and broads results} posed in Chapter~\ref{chapter: introduction} as one of the main goals of this thesis. 

\myindent The family of convex polytopes $\mathfrak{C}(C_n)$ has an H-representation described for all $n \geq 1$ by the trivial inequalities $0 \leq r_e \leq 1$ together with the nontrivial inequalities
\begin{equation}\label{eq:cycle_inequalities}
    c_n(\pmb{r}) := -r_e + \sum_{e'\neq e \in E(C_n)}r_{e'} \leq n-2
\end{equation}
for every $e \in E(C_n)$. Therefore this H-representation recovers the inequalities found by~\cite{galvao2020quantum} and shows, moreover, that they form a complete set of facet-defining inequalities for an infinite family of event graph polytopes. For both the singleton graph, and the graph $C_2$, one has only the trivial edge weighting inequalities $0 \leq r_e \leq 1$. The first nontrivial cycle inequality appears in the H-representation of $\mathfrak{C}(C_3)$ that is given by $0 \leq r_{ij} \leq 1$ together with 
\begin{align}
    &-r_{12}+r_{13}+r_{23} \leq 1, \\
    &+r_{12}-r_{13}+r_{23} \leq 1, \\
    &+r_{12}+r_{13}-r_{23} \leq 1.
\end{align}

\myindent The fact that~\cref{eq:cycle_inequalities} describes facet-defining inequalities is clear from the fact that it is simple to find a set of affinely independent points $F_n \subseteq \mathfrak{C}(C_n)$ saturating these inequalities (i.e., satisfying $c_n(\pmb{r}) = n-2$), and having as many points as $n$, the dimension of the polytope $\mathfrak{C}(C_n)$. For $C_3$ and $-r_{12}+r_{13}+r_{23} \leq 1$ one can simply use the set  $$F_3 = \{(0,0,1),(0,1,0),(1,1,1)\}.$$ For $C_4$ one has, e.g., the inequality
\begin{equation}\label{eq: c4 event graph inequality}
    -r_{12}+r_{23}+r_{34}+r_{14} \leq 2
\end{equation}
where one can use the set $$F_4=\{(0,1,1,0),(0,0,1,1),(0,1,0,1),(1,1,1,1)\},$$ and so on.

\subsection{Complete graphs}

\begin{figure}[t]
    \centering
    \includegraphics[width=0.7\linewidth]{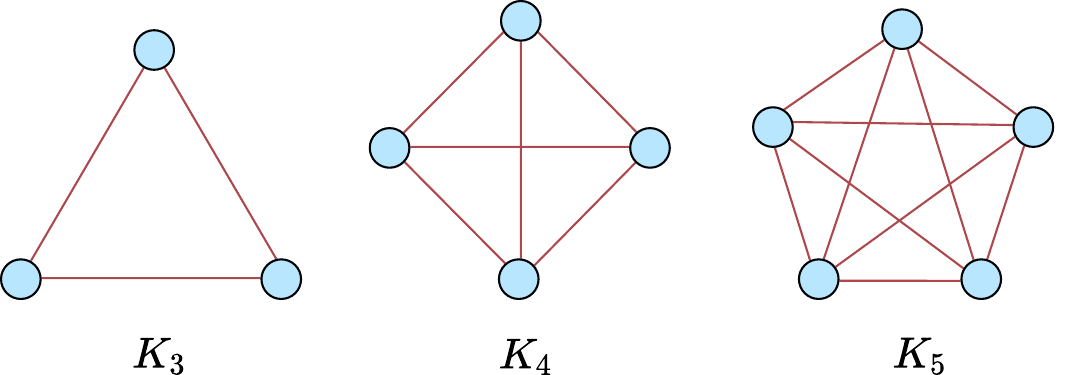}
    \caption{\textbf{Complete graphs $K_n$.} We show the first three elements of the family of complete graphs $K_n$ for $n=3,4$ and $5$ yielding non-trivial facet-defining inequalities of $\mathfrak{C}(K_n)$. Note that $K_3 = C_3$. The complete list of facet-defining inequalities for these three graphs can be found in~\cite{wagner2022github}. }
    \label{fig: complete_graphs_figure}
\end{figure}

\myindent The first nontrivial noncycle inequality, i.e. an inequality different than $0 \leq r_{ij} \leq 1$ and those from~\cref{eq:cycle_inequalities}, appears for the event graph polytope $\mathfrak{C}(K_4)$, where $K_4$ is the complete graph of four nodes.~\cref{fig: complete_graphs_figure} shows some complete graphs relevant to us, including this graph.  Let us denote by $r_{ij}$ the edge weightings associated to edges $\{i,j\}$ in this graph, where $V(K_4) = \{1,2,3,4\}$. The H-representation for this polytope has all possible trivial and $3$-cycle inequalities, but it also has the following inequalities
\begin{equation}\label{eq:k4_inequalities}
    r_{12}+r_{13}+r_{14}-r_{23}-r_{24}-r_{34} \leq 1,
\end{equation}
and all other elements in this class are found by permutations of the labels in the vertices. This class of inequalities is the second of an infinite family of facet-defining inequalities for the event graph polytopes $\mathfrak{C}(K_n)$, with $n \geq 3$, as we show in~\cref{theorem:hn_is_facet_of_Kn}. We have researched the literature to see where similar inequalities could have been found, and we have found no analog (even an indirect one).

\myindent Fix a natural number $n \geq 2$.
Write $V_n = \enset{1, \ldots, n}$ for the vertices of $K_n$, and let
$E_n$ denote the set of edges of $K_n$, \ie all two-element subsets of $V_n$.
Consider a partition of $E_n$ into the subsets $G_n, R_n \subseteq E_n$ given as  
\begin{align*}
G_n &\defeq \setdef{\enset{1,i}}{i=2,\ldots,n}
\\
R_n &\defeq E_{n} \setminus G_n.
\end{align*}
The edges in $R_n$ determine a complete subgraph of $K_n$ with one fewer vertex, \ie a subgraph isomorphic to $K_{n-1}$.
In turn, the edges in $G_n$ form a subgraph isomorphic to $K_{1,n-1}$, a star graph with $n$ vertices.  
We use this specific partition of $E_n$ to define a generalized version of the inequality from \cref{eq:k4_inequalities}:  
\begin{equation}\label{eq:new_K4_generalized}
    h_n(\pmb{r}) \defeq \sum_{e\in G_n}r_e - \sum_{e\in R_n} r_e \leq 1.
\end{equation}
It is simple to see that this inequality recovers the one from~\cref{eq:k4_inequalities} for the case $n=4$. We can also note that $h_3(\pmb r)$ is a cycle inequality from $\mathfrak{C}(C_3) \equiv \mathfrak{C}(K_3)$. As a remark, we note that we can also write this family recursively via the equation
\begin{equation}\label{eq:hn_recursively}
    h_n(\pmb{r}) = h_{n-1}(\pmb{r})+r_{1,n}-\sum_{i=2}^{n-1} r_{i,n} \leq 1
\end{equation}
starting from $n=1$ letting $h_1(\pmb{r})=0$.~\footnote{Recall that we write $r_{i,j} \equiv r_{ij}$ whenever necessary.} The first three  inequalities are then $h_1(\pmb{r})=0 \leq 1, h_2(\pmb{r})=r_{12} \leq 1$, and $ h_3(\pmb{r})=r_{12}+r_{13}-r_{23} \leq 1$.

\myindent With the construction from above we can write the following theorem.

\begin{theorem}\label{theorem:hn_is_facet_of_Kn}
    Let $K_n$ be the complete graph with $n$ vertices and $\mathfrak{C}(K_n)$ its associated event graph polytope. Then $h_n(\pmb{r}) \leq 1$ defined by~\cref{eq:new_K4_generalized} is a facet-defining inequality for $\mathfrak{C}(K_n)$, for all $n\geq 3$.
\end{theorem}

\begin{proof}
We establish this result by finding the set of vertices (i.e. extreme points) of the polytope $\mathfrak{C}(K_n)$ that belongs to---and therefore determines---this facet.
It suffices to find a set of points $F$ in the space (of edge weightings) such that:
\begin{enumerate}
\item[(i)] all the points in $F$ belong to the polytope $\mathfrak{C}(K_n)$,
\item[(ii)] all the points in $F$ saturate the inequality, \ie belong to the hyperplane determined by it,
\item[(iii)]
the set $F$ is affinely independent,
and 
\item[(iv)] $F$ contains as many points as the dimension $D$ of the polytope so that it generates an affine subspace of dimension $D-1$.
\end{enumerate}
In our proof, the chosen points are moreover vertices of the polytope, as they are edge $\enset{0,1}$-labelings.

\myindent We construct a set $F$ of polytope vertices.
This consists of two kinds of edge labelings:
those that assign $1$ to exactly one edge of $G_n$ (and $0$ to all other edges of $E_n$) and those that assign $1$ precisely to a triangle consisting of two edges from $G_n$ and another from $R_n$.
More formally, we define a family of edge $\enset{0,1}$-labelings indexed by subsets of size $1$ or $2$
of the vertex set $\enset{2,\ldots,n}$, as follows:
for each $i = 2, \ldots, n$,
define the edge $\enset{0,1}$-labeling $r^{(i)}$
with
$r^{(i)}_{1i} = 1$
and $r^{(i)}_e=0$ for all other edges $e$;
for each pair
$i,j = 2, \ldots, n$ with $i \neq j$
define the edge $\enset{0,1}$-labeling $r^{(i,j)}$
with $r^{(i,j)}_{1i} = r^{(i,j)}_{1j} = r^{(i,j)}_{ij} = 1$
and $r^{(i,j)}_e=0$ for all other edges $e$.
The set $F$ is then given by
\[
F \defeq \setdef{r^{(i)}}{i = 2, \ldots, n} 
\cup \setdef{r^{(i,j)}}{i,j = 2, \ldots ., n, i \neq j}.
\]
\Cref{fig:principle_argument_newk4_infinite} depicts the construction of the set $F$ for the case of $n=5$.
We now check conditions (i)--(iv) to establish the desired result.

\begin{figure}[t]
    \centering
    \includegraphics[width=0.8\textwidth]{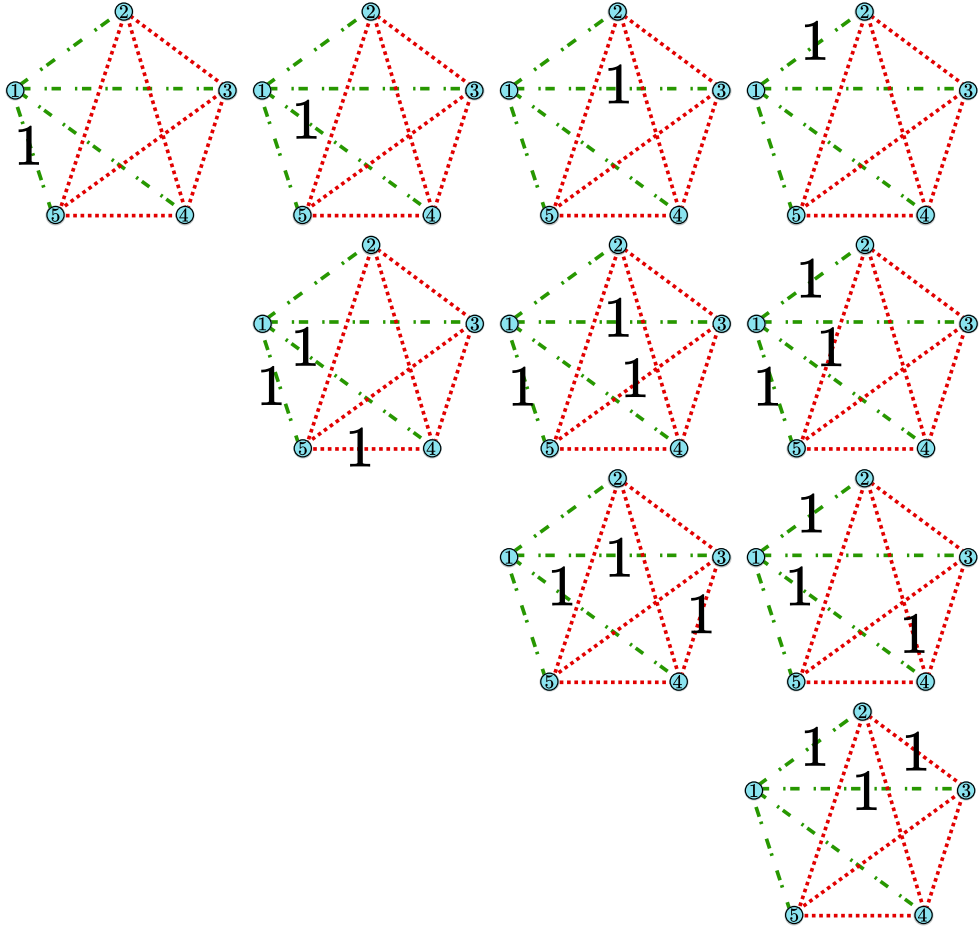}
    \caption{\textbf{The construction of the set $F$ for the $K_5$ graph.} Each edge is labelled $1$ where explicitly noted, otherwise it is labelled $0$ (to keep the figures easy to read). The first row shows the four labelings of the form $r^{(i)}$ with only one edge labeled $1$ from $G_5$.
    The remaining rows show the labelings of the form $r^{(i,j)}$, which assign label $1$ to exactly one triangle consisting of two edges from $G_5$ and the connecting edge from $R_5$.}
    \label{fig:principle_argument_newk4_infinite}
\end{figure}

\myindent For condition (i), we use \cref{def:kernel_loops} to show that all the edge labelings in the set $F$ are vertices of the polytope $\mathfrak{C}({K_n})$. Indeed, no cycle can have exactly one edge with the label $0$. In the case of the labelings of the form $r^{(i)}$, this is immediate as there is only one edge not labeled $0$.
For the labelings of the form $r^{(i,j)}$, no triangle (\ie subgraph isomorphic to $C_3$) has exactly one edge labeled $0$: if one chooses two edges labeled $1$ then the remaining edge that completes the 3-cycle also has label $1$. Moreover, any larger cycle can have at most two edges labeled $1$.

\myindent Condition (ii) is directly checked: for each $i = 2, \ldots, n$ we have
\[
\sum_{e\in G_n}r^{(i)}_e - \sum_{e\in R_n} r^{(i)}_e = r^{(i)}_{1i} - 0 = 1 - 0 = 1 ,
\]
and for each pair $i, j = 2, \ldots, n$ with $i \neq j$,
\[
\sum_{e\in G_n}r^{(i,j)}_e - \sum_{e\in R_n} r^{(i,j)}_e = r^{(i,j)}_{1i} + r^{(i,j)}_{1j} - r^{(i,j)}_{ij} = 2 - 1 = 1. 
\]

\myindent For condition (iii), 
affine independence can be verified by inspecting the matrix whose columns are the vectors corresponding to the edge-labelings in $F$. Ordering the components of each vector (corresponding to the edges of $K_n$) in lexicographic order and listing $r^{(i)}$ followed by $r^{(i,j)}$ also in that order, the matrix is arranged to be triangular with diagonal entries all equal to 1, hence its determinant is equal to $1$, implying linear independence of the vectors.

\myindent We can give as an example the deterministic edge weightings for $\mathfrak{C}(K_5)$ shown in Fig.~\ref{fig:principle_argument_newk4_infinite}. In this case we have $$\pmb{r} = (r_{12},r_{13},r_{14},r_{15},r_{23},r_{24},r_{25},r_{34},r_{35},r_{45})$$ and  
\begin{equation}
    \left (\begin{matrix}
          \pmb{r}^{(2)}\\
          \pmb{r}^{(3)}\\
          \pmb{r}^{(4)}\\
          \pmb{r}^{(5)}\\
          \pmb{r}^{(2,3)}\\
          \pmb{r}^{(2,4)}\\
          \pmb{r}^{(2,5)}\\
          \pmb{r}^{(3,4)}\\
          \pmb{r}^{(3,5)}\\
          \pmb{r}^{(4,5)}
    \end{matrix}\right)
    =
    \left(\begin{matrix}
        1 & 0 & 0 & 0 & 0 & 0 & 0 & 0 & 0 & 0\\ 
        0 & 1 & 0 & 0 & 0 & 0 & 0 & 0 & 0 & 0\\ 
        0 & 0 & 1 & 0 & 0 & 0 & 0 & 0 & 0 & 0\\ 
        0 & 0 & 0 & 1 & 0 & 0 & 0 & 0 & 0 & 0\\ 
        1 & 1 & 0 & 0 & 1 & 0 & 0 & 0 & 0 & 0\\ 
        1 & 0 & 1 & 0 & 0 & 1 & 0 & 0 & 0 & 0\\ 
        1 & 0 & 0 & 1 & 0 & 0 & 1 & 0 & 0 & 0\\ 
        0 & 1 & 1 & 0 & 0 & 0 & 0 & 1 & 0 & 0\\ 
        0 & 1 & 0 & 1 & 0 & 0 & 0 & 0 & 1 & 0\\ 
        0 & 0 & 1 & 1 & 0 & 0 & 0 & 0 & 0 & 1
    \end{matrix}\right)
\end{equation}
which is triangular with diagonal entries all equal to $1$ as claimed. 

\myindent Finally, for condition (iv), 
as all these labelings are distinct, one can count the number of elements of $S$ from the way they were constructed:
\[|F| = \binom{n-1}{1} + \binom{n-1}{2} = \binom{n}{2} = \frac{n(n-1)}{2}.\]
We conclude that it is the same as the dimension of the ambient space (of edge labelings) where the polytope lives, and thus also of the polytope itself.
\end{proof}

\myindent We see numerically when finding the complete list of all facet-defining inequalities of $\mathfrak{C}(K_4)$ that both $h_3(\pmb{r}) \leq 1$ and $h_4(\pmb{r})\leq 1$ are facet-defining inequalities. Similarly for $\mathfrak{C}(K_5)$ and $\mathfrak{C}(K_6)$. Because of that, we conjecture the following (which would follow from the last theorem in case we would have a good understanding of how \emph{lifting} works in the context of event graph polytopes).

\begin{conjecture}\label{conjecture: hn inequalities are always facet-defining}
Consider any event graph polytope $\mathfrak{C}(K_n)$, for $n\geq 3$. Then $\{h_i(\pmb{r})\leq 1\}_{i=3}^n$ are facet-defining inequalities of $\mathfrak{C}(K_n)$.
\end{conjecture}

\myindent It is interesting to note that for the complete event graph $K_n$ of $n$ vertices the number of deterministic edge labelings realizable by jointly distributed random variables, \ie vertices of the polytope $\mathfrak{C}({K_n})$, is given by a well-known sequence, the \emph{Bell} or \emph{exponential number sequence}~\citep{OEISBell,bell1934exponential}. We show these numbers in~\cref{tab:Bell_sequence}. The $n$-th Bell number is the number of partitions, or equivalence relations, of a set of size $n$.
It is clear that edge $\enset{0,1}$-labelings of $K_n$ are in one-to-one correspondence with symmetric reflexive relations on the set of vertices $\enset{1, \ldots, n}$, where the label of an edge $\enset{v,w}$ determines whether the pairs $(v,w)$ and $(w,v)$ are in the relation. Among these, the edge labelings in $\text{ext}(\mathfrak{C}(K_n))$ correspond to the equivalence relations (which additionally satisfy transitivity), with the underlying vertex labeling determining a partition of the vertices.
For a general graph $G$, it is still true that such edge labelings arise from partitions, or equivalence relations, on the set of vertices, determined by the underlying vertex labeling. The difference is that an edge labeling does not carry enough information to characterize a relation fully. So, in particular, different vertex partitions may give rise to the same edge labeling. This fact was used by~\cite{wagner2024inequalities} to propose an algorithm for finding elements of $\text{ext}(\mathfrak{C}(G))$ for any given event graph $G$. One remark about such an algorithm is that, since different vertex partitions may give rise to the same edge labelings apart from graphs $G=K_n$ the algorithm outputs repeated vertices, which for large sparse graphs makes it significantly memory inefficient. 

\begin{table}[t]
    \centering
    \begin{tabular}{c|c}
      $n$   &  $|\text{ext}(\mathfrak{C}(K_n))|$\\
      \hline \hline
      3 & 5\\
      4 & 15\\
      5 & 52\\
      6 & 203\\
      7 & 877\\
      8 & 4140\\
      9 & 21147\\
      10& 115975\\
      11& 678570\\
      12& 4213597\\
      13& 27644437\\
      14& 190899322\\
    \end{tabular}
    \caption{\textbf{Number of vertices of the event graph polytopes $\mathfrak{C}(K_n)$.} The number of vertices is given by the Bell sequence~\cite{OEISBell,bell1934exponential}. In~\cite{wagner2022github} the complete list of vertices is presented for $3 \leq n \leq 11$. The complete H-representation of $\mathfrak{C}(K_n)$ is known for $3 \leq n \leq 6$ and partial lists of facet-defining inequalities are known for $n=7$, and they can be found in~\cite{wagner2022github}.}
    \label{tab:Bell_sequence}
\end{table}

\myindent We have found numerically the obvious fact that~\cref{eq:new_K4_generalized} are not the only nontrivial facet-defining inequalities in such polytopes. The complete list of classes of facet-defining inequalities---hence up to symmetries of the polytope---for $\mathfrak{C}(K_5)$ is the following:
\begin{align}
    &c_3(\pmb r) := -r_{12}+r_{15}+r_{25}\leq 1 \label{eq:K5_class_1}\\
    &h_4(\pmb r) := +r_{15}+r_{25}+r_{35}-(r_{12}+r_{13}+r_{23})\leq 1\label{eq:K5_class_2}\\
    &h_5(\pmb{r}) := +r_{12}+r_{13}+r_{14}+r_{15}-(r_{23}+r_{24}+r_{25}+r_{34}+r_{35}+r_{45})\leq 1\label{eq:K5_class_3}\\
    &I_{(K_5,4)}(\pmb r) :=  +r_{12}+r_{14}+r_{15}+r_{23}+r_{34}+r_{35}-(r_{13}+r_{24}+r_{25}+r_{45})\leq 2\label{eq:K5_class_4}\\
    &I_{(K_5,5)}(\pmb{r}):=+r_{12}+r_{15}+r_{23}+r_{34}+r_{45}-(r_{13}+r_{14}+r_{24}+r_{25}+r_{35})\leq 2\label{eq:K5_class_5}\\
    & +2r_{12}+2r_{23}+2r_{24}+2r_{25}-(r_{13}+r_{14}+r_{15}+r_{34}+r_{35}+r_{45})\leq 3 \label{eq:K5_class_6}\\
    &I_{(K_5,6)}(\pmb r):=+r_{13}+r_{14}+2r_{24}+r_{34}+2r_{45}-(2r_{12}+2r_{25}+2r_{35})\leq 3\label{eq:K5_class_7}\\
    &I_{(K_5,7)}(\pmb{r}) := +2r_{12}+2r_{14}+2r_{15}+r_{23}+r_{35}-(2r_{13}+2r_{24}+r_{25}+2r_{45})\leq 3\label{eq:K5_class_8}\\
    &I_{(K_5,8)}(\pmb r):= +2r_{13}+2r_{14}+2r_{23}+2r_{24}+3r_{35}+3r_{45}-(2r_{12}+4r_{15}+4r_{25}+r_{34})\leq 5.\label{eq:K5_class_9}
\end{align}
The first three inequalities are given by the functionals $h_3,h_4$ and $h_5$ defined inductively via~\cref{eq:hn_recursively}. All the remaining inequalities are not equivalent to such inequalities and are not $n$-cycle inequalities. We have included a few labels for these inequalities for referencing later on Chapter~\ref{chapter: relational coherence}. One of these has no label because we  prefer to label another element of the same inequality class, shown in~\cref{eq: h52 as a new class of inequalities}. 

\myindent We have observed a notable property of the inequalities from $\mathfrak{C}(K_n)$ numerically, although we have not yet provided a proof. It seems that \emph{all} the facet-defining inequalities of the event graph polytopes $\mathfrak{C}(K_n)$ are elements of \emph{some} non-trivial recursively defined infinite family of facet-defining inequalities. To give a concrete example, consider the inequality shown in~\cref{eq:K5_class_6}. Relabeling the vertices we can write an element of this class as follows
\begin{equation}\label{eq: h52 as a new class of inequalities}
    h_5^{(2)}(\pmb r) =+2r_{12}+2r_{13}+2r_{14}+2r_{15}-r_{23}-r_{24}-r_{25}-r_{34}-r_{35}-r_{45}\leq  3.
\end{equation}
We find in the H-representation of $\mathfrak{C}(K_6)$ the following facet-defining inequality
\begin{equation}
    h_6^{(2)}(\pmb r) =+2r_{12}+2r_{13}+2r_{14}+2r_{15}+2r_{16}-r_{23}-r_{24}-r_{25}-r_{26}-r_{34}-r_{35}-r_{36}-r_{45}-r_{46}-r_{56}\leq 3
\end{equation}
suggesting the following conjecture.
\begin{conjecture}\label{conjecture: another infinite family of inequalities}
    Let $G=K_n$ be a complete graph and $G_n,R_n$ defined as in inequality~\eqref{eq:new_K4_generalized}. Then, for any $n\geq 5$ the inequalities
    \begin{equation}\label{eq: hn2 for all n}
        h_n^{(2)}(\pmb{r}):= 2\sum_{e \in G_n}r_e-\sum_{e \in R_n}r_e \leq 3
    \end{equation}
    are facet-defining inequalities of $\mathfrak{C}(K_n)$.
\end{conjecture}
\myindent We have introduced the notation for functional $h_n^{(2)}(\pmb{r})$ since we have that $h_n^{(1)}(\pmb{r}) \equiv h_n(\pmb r)$, suggesting a straightforward generalization of these functionals for every $m$ via $h_n^{(m)}(\pmb r)$. In fact, a similar conjecture could be drawn for every inequality in Eqs.~\eqref{eq:K5_class_4}-~\eqref{eq:K5_class_8}! But we do not need to stop at $K_5$. For example, $\mathfrak{C}(K_6)$ has the following facet-defining inequality
\begin{align}
    &h_6^{(3)}(\pmb{r}) := +3r_{12}+3r_{13}+3r_{14}+3r_{15}+3r_{16}\nonumber \\&-r_{23}-r_{24}-r_{25}-r_{26}-r_{34}-r_{35}-r_{36}-r_{45}-r_{46}-r_{56}\leq 6 \label{eq: h63 for the specific case}
\end{align}
from which one can start inferring patterns of families of facet-defining inequalities to conjecture, such as $h_n^{(m)}(\pmb{r}) \leq m(m+1)/2$ for integers $1 \leq m \leq n-2$ and $n \geq 3$. The complete set of facet-defining inequalities of $\mathfrak{C}(K_6)$ can be found in reference~\citep{wagner2022github}. In total $\mathfrak{C}(K_6)$ has an H-representation with 50652 inequalities. From analyzing them, in comparison with the H-representation of $\mathfrak{C}(K_5)$ we can write the following conjecture.

\begin{conjecture}\label{conjecture: C(Kn) has facets of any other C(Kn) for larger n}
    Every facet-defining inequality for $\mathfrak{C}(K_n)$ is also a facet-defining inequality for $\mathfrak{C}(K_{n'})$ for all $n' \geq n$.
\end{conjecture}

\myindent Note that from~\cref{corollary: event graph polytopes of complete graphs} every facet-defining inequality for $\mathfrak{C}(K_n)$ is at least a \emph{valid} inequality for $\mathfrak{C}(K_{n'})$ for all $n' \geq n$. We can see that the above holds by inspection of the facet-defining inequalities found for $n$ up to $6$. One way of showing this conjecture to be true would be to investigate the profound properties of liftings in the event graph approach. 

\myindent The largest complete H-representation found for a complete graph was $\mathfrak{C}(K_6)$ but some partial sets of facet-defining inequalities can be found for other event graph polytopes $\mathfrak{C}(K_n)$ since we have found the complete V-representation for $n$ up to $11$. One such interesting inequality of $\mathfrak{C}(K_7)$ is the following:
\begin{equation}\label{eq:kappa_ineq}
    \kappa(\pmb{r}) := -2r_{12} + r_{14} + r_{16} - 2r_{23} + 2r_{27} + 2r_{34} - 2r_{35} -2r_{36} + 2r_{37} + r_{45} + r_{46} + r_{47} + r_{57} \leq 6.
\end{equation}
\myindent The reason why this inequality may be considered interesting is because it can be considered a candidate of an inequality able to robustly witnessing nonstabilizerness, as we comment on in Chapter~\ref{chapter: relational coherence}. However, mathematically we can already say that this inequality satisfies a specific (and uncommon) feature that also the inequality $c_3(\pmb r)$ satisfies, which is the following one. Let us write $\kappa(\pmb{r}) = \pmb{\gamma} \cdot \pmb{r} $ where $\pmb{\gamma} = (-2,0,1,0,\dots,1,0)$ is the vector of coefficients defining the functional $\kappa$ while $\pmb{r} = (r_{12},r_{13},r_{14},\dots,r_{57},r_{67})$. Then, we can re-write $$\kappa(\pmb{r}) =  \sum_{e \in P_\kappa(G)}p_er_e+\sum_{e \in N_\kappa(G)}n_er_e$$ where $p_e \in \mathbb{Z}$ are the positive coefficients of the functional (e.g. $p_{14}=1$ and $p_{16}=1$), while $n_e \in \mathbb{Z}$ are the negative coefficients (e.g. $n_{12}=-2$ and $n_{23} = -2$). Also, we have denoted the subgraphs $P_\kappa(G),N_\kappa(G)$ having positive and negative coefficients defined by $\kappa$, respectively. Then, it holds that $\kappa$ cannot be violated by edge-weightings where $r_e = 1/2$ for all $e \in P_\kappa(G)$ and $r_e=0$ for all $e \in N_\kappa(G)$. For this case we have that $\kappa(\pmb{r}) = 6$. This specific feature makes such inequalities \emph{good candidates} for inequalities where it is easy to show that they are bounding a specific `type' of coherence known as \emph{nonstabilizerness} (or as \emph{magic})~\citep{wagner2024certifying}.

\subsection{Triangular book graphs}

\begin{figure}[t]
    \centering
    \includegraphics[width=0.7\linewidth]{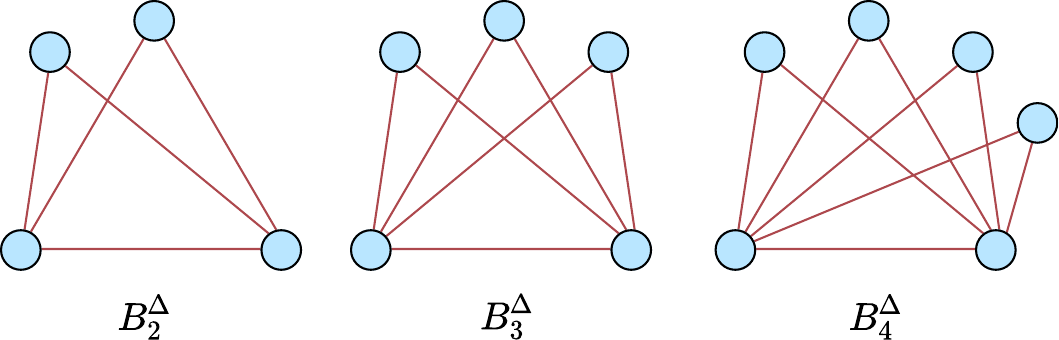}
    \caption{\textbf{Triangular book graphs $B_n^\Delta$.} We show the first triangular book graphs different from the $3$-cycle graph (since $B_1^\Delta \equiv C_3$), i.e., $B_2^\Delta, B_3^\Delta$ and $B_4^\Delta$. The facet-defining inequalities of $\mathfrak{C}(B_n^\Delta)$ are given by all the possible $3$-cycle inequalities related to labelings of the vertices of these graphs, due to~\cref{corollary:sequential_gluing}. }
    \label{fig: triangular_book_graphs_figure}
\end{figure}

\myindent We start now discussing some relevant event graphs that are not complete nor cycle graphs. The first such graph we discuss is what is known as a \emph{triangular book graph}, which is the sequential gluing of $C_3$ graphs along the same edge. We denote these as $B_n^\Delta$ where $n$ is the number of $C_3$ graphs that are glued. These are only an instance of the family of graphs known as \emph{book graphs} that are constructed by such gluings along the same edge of, most commonly, cycle graphs. We denote book graphs as $B_n(C_m)$, where $C_m$ is the type of cycle graph being glued and $n$ is the number of glued graphs. In our notation $B_n(C_3) \equiv B_n^\Delta$. Triangular book graphs are shown in~\cref{fig: triangular_book_graphs_figure}.  From~\cref{corollary:sequential_gluing} we learn that every inequality from this graph is either a trivial inequality or a $\mathfrak{C}(C_3)$ facet inequality. Albeit seemingly trivial, this is relevant since this is the structure of event graphs considered by~\cite{wagner2023anomalous}. There, the authors have considered relational properties related to the possible values that a \emph{weak value}~\citep{dressel2014colloquium} can take, in which case one is interested in relational properties of two states, together with a set of basis states. The relevant graph in this case becomes a triangular book graph, where the number of triangular books depends on the number of dimensions. We have also numerically checked the validity of~\cref{corollary:sequential_gluing} to hold for triangular book graphs up to $11$ vertices ($9$ $C_3$ subgraphs glued along the same edge) to benchmark our code.

\myindent As we discuss more in Chapter~\ref{chapter: relational coherence}, because of~~\cref{corollary:sequential_gluing} all the relational information that overlaps can capture regarding coherence for such a graph is just that described by 3-cycle inequalities given by~\cref{eq:cycle_inequalities}. To obtain \emph{further} relational information (specifically, regarding set coherence) one \emph{must} consider invariants beyond two-state overlaps, which was one of the main points raised by~\cite{wagner2023anomalous}. 

\subsection{Bipartite graphs}

\begin{figure}[t]
    \centering
    \includegraphics[width=0.7\linewidth]{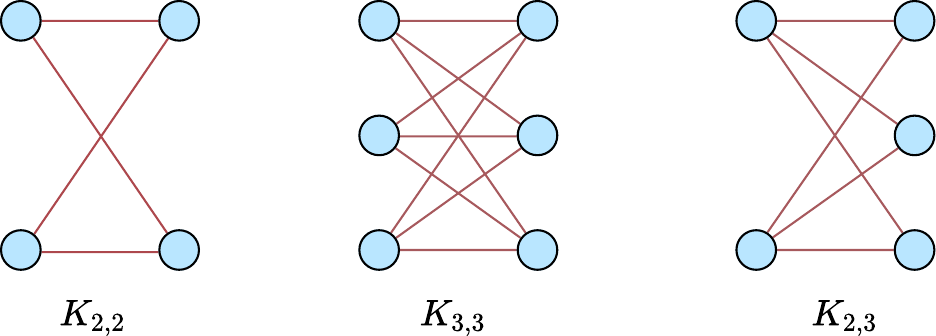}
    \caption{\textbf{Complete bipartite graphs $K_{n,m}$.} We show three different complete bipartite graphs, $K_{2,2}$, $K_{3,3}$ and $K_{2,3}$. Note that $K_{2,2} \equiv C_4$ hence $\mathfrak{C}(K_{2,2}) = \mathfrak{C}(C_4)$. }
    \label{fig: compelte_bipartite_graphs}
\end{figure}

\myindent In general, complete bipartite graphs (some of which are shown in~\cref{fig: compelte_bipartite_graphs}) have facet-defining inequalities that are not equivalent to those we have encountered before. Let us investigate the first noncycle and noncomplete graph inequality in a complete bipartite graph. We start by showing that we need to go for complete bipartite graphs beyond $K_{2,n}$.

\begin{conjecture}
    Let $G=K_{2,n}$ be an event graph, for any $n\geq 3$. Every nontrivial facet-defining inequality of $\mathfrak{C}(K_{2,n})$ is given by nontrivial facet-defining inequalities of $\mathfrak{C}(C_4)$.
\end{conjecture}

\myindent We have numerically shown the above to hold for $n\in \{3,4,5,6,7\}$. The first candidate of complete bipartite graph that can have some novel nontrivial inequalities is then $K_{3,3}$. Indeed, this graph has two novel classes of inequalities that are not equivalent to $4$-cycle inequalities and are given by
\begin{align}
    &I_{(K_{3,3},1)}(\pmb r) := + r_{1,\tilde{1}}+ r_{1,\tilde{2}}+ r_{1,\tilde{3}}+ r_{2,\tilde{1}}    - r_{2,\tilde{3}}- r_{3,\tilde{1}}+ r_{3,\tilde{2}}- r_{3,\tilde{3}} \leq  3,\label{eq: first K33 inequality}\\
    &I_{(K_{3,3},2)}(\pmb r):= +3r_{1,\tilde{1}}+2r_{1,\tilde{2}}+ r_{1,\tilde{3}}+2r_{2,\tilde{1}}-2r_{2,\tilde{2}}-2r_{2,\tilde{3}}+ r_{3,\tilde{1}}-2r_{3,\tilde{2}}+ r_{3,\tilde{3}} \leq  6,\label{eq: second K33 inequality}
\end{align}
where we have written the vertex set $V(K_{3,3}) = \{1,2,3,\tilde{1},\tilde{2},\tilde{3}\}$. 

\myindent As such graphs are relevant to bipartite communication problems, we have investigated some instances that arise in such cases. Specifically, the class of event graphs of interest to quantum random access codes is the class of complete bipartite graphs $K_{n,2^n}$. We have found the complete V-representation for $\text{ext}(\mathfrak{C}(K_{3,8}))$ and a partial V-representation for $\text{ext}(\mathfrak{C}(K_{4,16}))$. We have found that $|\text{ext}(\mathfrak{C}(K_{3,8}))| \leq 84706$ and, also, we have numerically found a list of $977793$ vertices of $\text{ext}(\mathfrak{C}(K_{4,16}))$.~\footnote{The code, however, produces many repeated vertices, implying that the actual number of vertices found may be much smaller. } 

\subsection{Wheel graphs}

\begin{figure}[t]
    \centering
    \includegraphics[width=0.7\linewidth]{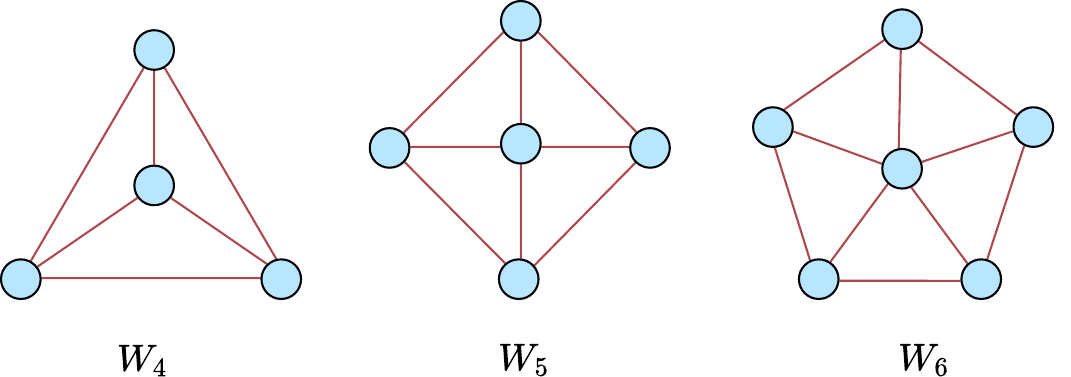}
    \caption{\textbf{Wheel graphs $W_{n}$.} The family of wheel graphs $W_n$ is shown. This family is structurally similar to the family of cycle graphs, with the difference that they have one more node (hence $|V(W_n)| = |V(C_{n+1})|$) connected to all other nodes, hence inducing a star subgraph. The family itself is not necessarily relevant to us by there are two relevant points here: first, this construction of having an additional node can be viewed as a formal graph-theoretic operation that is called the \emph{suspension}. In this case, wheel graphs are the suspension graph of a cycle graph. Moreover, the graph $W_6$ is particularly relevant to us as we show, in Chapter~\ref{chapter: from overlaps to noncontextuality}, how to translate facet-defining inequalities of $\mathfrak{C}(W_6)$ into known noncontextuality inequalities (namely \acrshort{kcbs} inequalities).}
    \label{fig: wheel_graphs_figure}
\end{figure}

\myindent Let us now consider a family of graphs that is not complete, $n$-cycle, or bipartite, but that can be constructed from gluings of $C_3$ graphs but yielding nontrivial novel event graph polytope inequalities. This family is the family of wheel graphs $W_n$. These are shown in~\cref{fig: wheel_graphs_figure}. It can be shown that these are the suspension graphs of cycle graphs (see Appendix~\ref{sec: graph theory} for the definition of a suspension graph). Most notably, the event graph polytope $\mathfrak{C}(W_6)$ has the following facet-defining inequality
\begin{equation}\label{eq:kcbs_event_graph_inequality}
    h_{\mathrm{KCBS}}(\pmb r):= -r_{12}-r_{23}-r_{34}-r_{45}-r_{15}+r_{16}+r_{26}+r_{36}+r_{46}+r_{56} \leq 2.
\end{equation}
We have found the complete H-representation of $\mathfrak{C}(W_n)$ for $n \in \{3,4,5,6,7\}$, which can be found in~\cite{wagner2022github}.

\myindent To conclude this catalog of different event graphs and their related event graph polytopes, there are various families of graphs---Mongolian tent graphs, the Wagner graph, the Hershel graph, hypercube graphs, cubic graphs, the Frankin graph, ladder graphs, (stacked) book graphs, hypercube graphs, suspended bipartite graphs, etc.---for which we have characterized V-representations, H-representations, or both, for the event graph polytopes. Some of these choices were arbitrary, either to prove numerically some of the propositions found regarding equivalences between event graph polytopes via gluings or to test the algorithms we have developed. These can all be found at~\cite{wagner2022github}.

\section{Subpolytopes of event graphs by restricting deterministic edge weightings}\label{sec: subpolytopes}

\myindent For a given event graph $G$ there are two polytope operations that we now investigate. The first is to analyze sub polytopes of $\mathfrak{C}(G)$ constructed via the convex hull of subsets of elements in $\text{ext}(\mathfrak{C}(G))$. The second is to investigate \emph{cross-sections} of $\mathfrak{C}(G)$. Recall that, traditionally, one takes the cross-section of an $n$-dimensional polytope to be the nonempty intersection of that polytope with an $(n-1)$-dimensional hyperplane, that yields another convex polytope. We consider the cross-section of $\mathfrak{C}(G)$, embedded in the $|E(G)|$ space, as the nonempty intersection with generic $|E(G')|$-dimensional subspaces defined by equations $\bigwedge_{e \in E(G')}r_e = 0$ (this considers edge-weightings for which all edges of the subgraph $G'$ of $G$ are equal to zero) or $\bigwedge_{e \in E(G')}r_e = 1$ (this considers edge-weightings for which all edges of the subgraph $G'$ of $G$ are equal to one) where $G'$ is a subgraph of $G$.

\myindent At first, these two seem to be purely motivated by improving the geometric understanding of event graph polytopes. However, and as will be clear in later chapters, they are linked to physical interpretations. Investigating sub polytopes of $\mathfrak{C}(G)$ by discarding elements in $\text{ext}(\mathfrak{C}(G))$ is connected to imposing \emph{restrictions} on realizability in terms of jointly distributed random variables, allowing us to construct families of polytopes $\mathfrak{C}_{\mathbb{P},d}(G)$ depending on the cardinality of $|\Lambda|$, the sample space of such random variables. Investigating the cross sections characterized by $\bigwedge_{e \in E(G')}r_e = 0$ will reveal an equivalence between \acrshort{ks} noncontextuality inequalities in the \acrshort{csw} framework and some event graph facet inequalities. Investigating cross sections $\bigwedge_{e \in E(G')}r_e = 1$ will be relevant for the problem of quantum realizations of edge weightings.

\begin{definition}[Cross sections of event graph polytopes]
    Let $G$ be an event graph and let $G'$ be any subgraph of $G$. We define the cross sections $\mathfrak{C}^i_{G'|G}$, with $i\in \{0,1\}$ as
    \begin{equation}
        \mathfrak{C}^i_{G'|G} := \left \{r \in \mathfrak{C}(G) \mid \forall e \in E(G'), r_e=i\right\}.
    \end{equation}
\end{definition}
When $G$ is clear from the context we simply write $\mathfrak{C}_{G'}^i$. If $G' \simeq C_2$ with $E(G') = \{e\}$ we simply write $\mathfrak{C}^i_e$.

\subsection{Cross section in $n$-cycle event graphs}

\myindent To start we investigate the simplest possible cross section in $n$-cycle event graphs. We want to show that, for a given fixed $\tilde{e}$ the cross section of $\mathfrak{C}(C_n)$ along $r_{\tilde{e}} = 1$ is isomorphic to the event graph polytope $\mathfrak{C}(C_{n-1})$.

\begin{lemma}\label{lemma: equal_no_violation}
    Let $G=C_n$ be the $n$-cycle graph, with $n\geq 4$ and fix some edge $\tilde{e} \in E(C_n)$. Then, we have that $\mathfrak{C}_{\tilde{e}}^1 = \{1\} \times  \mathfrak{C}(C_{n-1})$  while $\mathfrak{C}_{\tilde{e}}^0 \neq \{0\} \times \mathfrak{C}(C_{n-1})$. 
\end{lemma}

This lemma shows that if we define the cross-section $\mathfrak{C}_{\tilde{e}}^1$ of the polytope $\mathfrak{C}(C_n)$ along the direction $r_{\tilde{e}}=1$, the resulting polytope is isomorphic to $\mathfrak{C}(C_{n-1})$. This implies that given any facet-defining inequality of $\mathfrak{C}(C_n)$, if we force an edge to be equal to one, the resulting inequality is a facet-defining inequality of $\mathfrak{C}(C_{n-1})$.

\begin{proof}[Proof of Lemma~\ref{lemma: equal_no_violation}]
    Because $\mathfrak{C}(C_n)$ is a convex polytope, the same is true for the cross-section $\mathfrak{C}_{\tilde{e}}^1$. Let us assume, without loss of generality, an ordering $$\pmb{r}=(r_e)_{e \in E(C_n)} = (r_{\tilde{e}},r_{e_1},\dots,r_{e_{n-1}}).$$ We want to show that 
    $$\text{ext}(\mathfrak{C}_{\tilde{e}}^1) = \{(1,\pmb{s}): \pmb{s} \in \text{ext}(\mathfrak{C}(C_{n-1}))\}.$$
    
    ($\subseteq$) Let $\tilde{\pmb{r}} \in \text{ext}(\mathfrak{C}_{\tilde{e}}^1)$. By construction, we must have that $\tilde{\pmb{r}} \equiv (1,\tilde{\pmb{s}}) \in \mathfrak{C}(C_n)$. Since $\tilde{\pmb{r}}$ is assumed to be an extreme point $\tilde{\pmb{s}}$ is a deterministic edge weighting. Moreover, $\tilde{s}:E(C_{n-1}) \to \{0,1\}$ satisfy  $|\text{Ker}(\tilde{s})| \neq 1$ since $\text{Ker}(\tilde{r}) = \text{Ker}(\tilde{s})$. Therefore, from Def.~\ref{def:kernel_loops}, $\tilde{\pmb{s}} \in \text{ext}(\mathfrak{C}(C_{n-1}))$.
    
    ($\supseteq$) This direction follows trivially.
    Hence, we have that $\mathfrak{C}_{\tilde{e}}^1 = \{1\} \times \mathfrak{C}(C_{n-1}),$ where $\{1\}$ is the singleton convex polytope, as we wanted. 
    
    \myindent To conclude, note that the constant deterministic edge weighting $s_1:E(C_{n-1}) \to \{0,1\}$ $s_1(e)=1$ is in $\mathfrak{C}(C_{n-1})$ due to~\cref{lemma: constants are classical}, from which even though  $(0,\pmb{s}_1) \in \{0\} \times \mathfrak{C}(C_{n-1})$ we conclude that  $(0,\pmb{s}_1) \notin \mathfrak{C}(C_n)$ and since, by definition, $\mathfrak{C}_{\tilde{e}}^0 \subseteq \mathfrak{C}(C_n)$ we have also that $(0,\pmb{s}_1) \notin \mathfrak{C}_{\tilde{e}}^0$. Therefore $\mathfrak{C}_{\tilde{e}}^0 \neq \{0\}\times \mathfrak{C}(C_{n-1})$. 
\end{proof}

\myindent We now discuss the case $n=3$ that was left out of Lemma~\ref{lemma: equal_no_violation}.

\begin{example}[The cross section $\mathfrak{C}_{\{1,2\}\vert C_3}^1$]\label{example:C_3_cross_section}
    Let us consider $G=C_3$, where $E(C_3) = \{\{1,2\},\{1,3\},\{2,3\}\}$. We want to find $\mathfrak{C}_{\{1,2\}|C_3}^1 \equiv \mathfrak{C}_{\{1,2\}}^1$. If we consider the ordering $\pmb{r} = (r_{12},r_{13},r_{23})$, these points are described by 
    \begin{equation}
        \mathfrak{C}_{\{1,2\}}^1 \ni \pmb {r} \iff \pmb{r} \in  \{(1,a,b)\in \mathfrak{C}(C_3)\}.  
    \end{equation}
    Because of that, we have that $\pmb{r} = \sum_\alpha p_\alpha (1,a_\alpha,b_\alpha)$ of deterministic edge weightings $(1,a_\alpha,b_\alpha) \in \text{ext}(\mathfrak{C}(C_3))$. However, we now note that for every $\alpha$ we must have that $a_\alpha = b_\alpha$, to satisfy the assumption that these are extreme points of $\mathfrak{C}(C_3)$ implying that every point of $\mathfrak{C}_e^1$ is of the form $p(1,0,0)+(1-p)(1,1,1)$. 
\end{example}

\myindent The example above suggests that the cross sections $\mathfrak{C}_{G'|G}^1$ must satisfy `additional' constraints on the possible values of edge weightings to preserve the set-theoretic relation   $\mathfrak{C}_{G'|G}^1 \subseteq \mathfrak{C}(G)$. This intuitive idea is well captured by the notion of \emph{edge contraction} when we consider more general graphs. After an edge contraction, loops and multiple edges must have the same edge weightings, for them to be valid edge weightings of event graph polytopes. 

\subsection{Restricting event graph polytopes by restricting random variables}

\myindent If we recall~\cref{def:realizable_joint_dist} we  note that we have allowed the random variables $(A_v)_{v \in V(G)}$ to be jointly distributed with respect to \emph{some} set $\Lambda$ in order to realize values $r \in \mathfrak{C}(G)$. We now restrict the sets $\Lambda$ by imposing \emph{upper bounds} on their cardinality. We start with a simple example to provide some intuition on \emph{why} we may want to do this, which resonates with some of our comments on Chapter~\ref{chapter: information tasks} regarding dimension witnessing.

\begin{example}[Restricted event graph set $\mathfrak{C}^{(2)}(C_3)$]\label{example:restricted_C3_event_graph_polytope} Let us consider the simplest nontrivial event graph $G=C_3$. Suppose that $r \in \mathfrak{C}(C_3)$, but further impose that any jointly distributed family of random variables $(A_i)_{i=1}^3$ realizing $r$ take values from an alphabet $\Lambda = \{\lambda,\lambda^\perp\}$. Denote such a subset of $\mathfrak{C}(C_3)$ by $\mathfrak{C}^{(2)}(C_3)$. 

\myindent Let us start by noting that the constant deterministic edge weighting 
$r_0(e) = 0$ for all $e \in E(C_3)$ is such that $r_0 \notin \mathfrak{C}^{(2)}(C_3)$. If this would be possible, we would need to satisfy the conjunction of the following equations:

\begin{align}
    &r_0(\{1,2\}) = p(A_1=A_2) = p(A_1=\lambda,A_2=\lambda)+p(A_1=\lambda^\perp,A_2=\lambda^\perp) =  0,\\
    &r_0(\{1,3\}) = p(A_1=A_3) = p(A_1=\lambda,A_3=\lambda)+p(A_1=\lambda^\perp,A_3=\lambda^\perp)=  0,\\
    &r_0(\{2,3\}) = p(A_2=A_3) = p(A_2=\lambda,A_3=\lambda)+p(A_2=\lambda^\perp,A_3=\lambda^\perp) =  0,
\end{align}
    which is \emph{impossible} as a solution to the problem above is dual to finding a $\{0,1\}$-coloring for $C_3$, something that we make formal briefly (using~\cref{def: lambda realizable edge weightings}) but that follows from any exhaustive attempt of resolving the above forcing vertices to either satisfy $A_i=\lambda$ always or $A_i=\lambda^\perp$ always. We recall that in our notation we have that 
    \begin{equation*}
        p(A_i = A_j) = \sum_\lambda p(A_i = \lambda,A_j = \lambda) = \sum_\lambda p_{A_i,A_j}(\lambda,\lambda),
    \end{equation*}
    and if, in addition to being jointly distributed, the random variables are also \emph{independent}, they satisfy moreover that $$p_{A_1, A_2,\dots, A_{|E(G)|}}(\lambda_1,\lambda_2,\dots,\lambda_{|E(G)|}) = p_{A_1}(\lambda_1)p_{A_2}(\lambda_2)\dots p_{A_{|E(G)|}}(\lambda_{|E(G)|}).$$ The calculations above imply that restricting the cardinality of the alphabet $\Lambda$ restricts the realizability via random variables of points in $\mathfrak{C}(G)$.
\end{example}

\myindent We now proceed to investigate properties of the sets $\mathfrak{C}^{(d)}(G)$. We start by defining them.

\begin{definition}[Restricted realizability for edge weightings]
    Let $G$ be an event graph, $d \in \mathbb{N}_{\geq 2}$ and $(A_i)_i$ denote jointly distributed random variables taking values on $\Lambda$. We define
    \begin{align}
        \mathfrak{C}^{(d)}(G) := \Bigr\{r  \mid \exists \Lambda=\{\lambda_k\}_{k=1}^d, \exists (A_i)_{i} \text{ s.t. }  \forall \{i,j\} \in E(G), r_{i,j} = \sum_{k=1}^d p(A_i=\lambda_k,A_j=\lambda_k) \Bigr\}
    \end{align}
    as the set of all edge weightings realizable by random variables restricted to draw from an alphabet $|\Lambda|\leq d$.
\end{definition}

\myindent We call an element $r \in \mathfrak{C}^{(d)}(G)$ as an edge weighting realized (or realizable) by restricted random variables. The following simple lemma holds:
\begin{lemma}
    For every event graph $G$ and every $d \in \mathbb{N}_{\geq 2}$, we have that $\mathfrak{C}^{(d)}(G) \subseteq \mathfrak{C}(G)$.
\end{lemma}

\myindent It is also simple to see from the definition that to every event graph $G$ it holds that 
\begin{equation}
    \mathfrak{C}^{(2)}(G) \subseteq \mathfrak{C}^{(3)}(G) \subseteq \mathfrak{C}^{(4)}(G) \subseteq \dots \subseteq \mathfrak{C}^{(|V(G)|)}(G) = \mathfrak{C}(G).
\end{equation}

\myindent We have already mentioned in~\cref{example:restricted_C3_event_graph_polytope} that there exists some relation between colorability of $G$ and realizability of edge weightings $r$ by restricted random variables. We can make this relation formal, but for that, we need to introduce some definitions. We start by defining the notion of a \emph{$\Lambda$-realization} of deterministic edge weightings.

\begin{definition}[$\Lambda$-realizable edge weightings]\label{def: lambda realizable edge weightings}
    Let $\Lambda$ be any alphabet and let $\alpha: E(G) \to \{0,1\}$ be a deterministic edge weighting for an event graph $G$. We say that $\alpha$ is \emph{$\Lambda$-realizable} if there exists a vertex $\Lambda$-labeling $\lambda: V(G) \to \Lambda$ such that $\alpha_{u,v} = \delta_{\lambda(v),\lambda(u)}$ for every $\{u,v\} \in E(G)$ where
    $$\delta_{\lambda(v),\lambda(u)} = \begin{cases} 1 & \text{if $\lambda(v)=\lambda(u)$,} \\ 0 & \text{if $\lambda(v) \neq \lambda(u)$.}\end{cases}$$
\end{definition}

\myindent We can use the definition above to note that a deterministic edge weighting $\alpha \in \text{ext}(\mathfrak{C}(G))$ if and only if it is $\Lambda$-realizable by some $\Lambda$, implying that we have yet another equivalent characterization of the set $\mathfrak{C}(G)$ (in fact, this characterization is just a rewriting of the fact that for deterministic edge weightings $\Lambda$-realizations are dual to jointly distributed random variables forced to reproduce the deterministic constraints of realizing the edge weightings via $r_{ij} = p(A_i=A_j)$). However, this rewriting is also useful as it allows us to note that deterministic edge weightings $\alpha \in \mathfrak{C}^{(d)}(G)$ are those that are $\Lambda$-realizable where $\vert \Lambda \vert = d$. 

\begin{proposition}\label{proposition: extremal_from_realizability}
    For every event graph $G$, $\alpha \in \mathrm{ext}(\mathfrak{C}(G))$ if, and only if, $\alpha$ is $\Lambda$-realizable for some $\Lambda$. Moreover, if $\alpha$ is $\Lambda$-realizable then it is with $\Lambda$ of a size at most $|V(G)|$. 
\end{proposition}

\begin{proof}
    Suppose that $\alpha$ is $\Lambda$-realizable. Then, take any $\{u,v\} \in E(G)$. If $u \sim_\alpha v$ it implies that $\alpha$ assigns $1$ to every edge in the path from $u$ to $v$. Without loss of generality denote the vertices in such a path by the finite sequence $(u,s_1,s_2,\dots,s_k,v)$ for some $0 \leq k < |V(G)|$. Since $\alpha$ is $\Lambda$ realizable we must have that $\alpha_{u,s_1} = \dots = \alpha_{s_k,v} = 1 \implies \lambda(u) = \lambda(s_1)=\dots = \lambda(v)$, and therefore $\alpha_{u,v} = \delta_{\lambda(u),\lambda(v)}=1$ which implies that $\alpha \in \text{ext}(\mathfrak{C}(G))$ (by Def.~\ref{def: event graph polytope}). 
    
    \myindent On the other hand, if $\alpha \in \text{ext}(\mathfrak{C}(G))$ we can algorithmically construct a vertex labeling $\lambda: V(G) \to \Lambda$ for which $\alpha$ is $\Lambda$-realizable. For each equivalence class $[v] := \{v' \in V(G) \mid v' \sim_\alpha v\}$ we assign a different vertex label, i.e.,  $\forall v \in [v], \lambda(v) = \lambda_v$. Hence $\lambda(v) \neq \lambda(w)$ if and only if $[v] \neq [w]$. By construction, this is a vertex $\Lambda$-labeling $\lambda: V(G) \to \Lambda$ and $\Lambda$ has the cardinality given by the number of equivalence classes $[v]$. Moreover, $\alpha_{u,v} = 1$ if and only if $u,v \in [v]$, and therefore if and only if $\lambda(u)= \lambda(v)$. Therefore we conclude that, indeed $\alpha$ is $\Lambda$ realizable. This concludes the proof.
\end{proof}

\begin{corollary}
    For every event graph $G$, a deterministic edge weighting $\alpha:E(G) \to \{0,1\}$ is an element of $\mathfrak{C}^{(d)}$ if, and only if, $\alpha$ is $\Lambda$-realizable for some $\Lambda$ of cardinality at most $d$.
\end{corollary}

\myindent From the results above we have that $\mathfrak{C}^{(|V(G)|)}(G) = \mathfrak{C}^{(d')}(G)$ for every $d' \geq |V(G)|$ and every event graph $G$. This is because we have that $\alpha \in \mathfrak{C}(G)$ is always $\Lambda$-realizable by a $\Lambda$ which has the same cardinality as the number of equivalence classes $[v]$ defined through $\sim_\alpha$, and in the worst-case scenario, these are equal to $V(G)$, which happens when $\alpha = \alpha_0$.

\myindent Note also that from above we were very close to constructing a coloring by considering the equivalence classes induced by the deterministic edge weighting. It is not a formal coloring because many adjacent vertices (namely those for which the deterministic edge weighting equals $1$) have the same label. Hence, it is intuitive that we could `color' the graph where vertices are equivalence classes. To relate the $\Lambda$ realizability with graph colorings we make the following definition of a \emph{possibly loopy graph induced by an edge $\{0,1\}$-labeling}, which makes the above intuition formal.

\begin{definition}[Possibly loopy graph induced by a deterministic edge weighting]\label{def: possibly loopy graph}
    Let $G$ be an event graph and $\alpha: E(G) \to \{0,1\}$ any deterministic edge weighting. We define a new graph (not necessarily an event graph) $G/\alpha$ by defining $V(G/\alpha) = \{[v]\}_{v \in V(G)}$, i.e., vertices are given by the equivalence classes $[v]:= \{v'\in V(G) \mid v \sim_\alpha v' \}$ and edges $E(G/\alpha)$ such that $\{[v],[w]\} \in E(G/\alpha)$ if there exists $v' \in [v]$ and $w' \in [w]$ such that $\{v',w'\} \in E(G)$ with $\alpha_{v',w'} = 0$. We call $G/\alpha$ the \emph{possibly loopy graph induced by the deterministic edge weighting $\alpha$}, or simply the \emph{quotient graph} with respect to $\sim_\alpha$.
\end{definition}

\myindent We state and prove~\cref{theorem: colorings}  below assuming familiarity with the notion of what a coloring of a graph is. In case the reader is not familiar, we review and define this notion in Appendix~\ref{sec: graph theory}. Briefly, colorings are vertex labelings of a graph, so that adjacent vertices are always labeled differently.  With these technical ingredients, we can now prove the following theorem.

\begin{theorem}~\label{theorem: colorings}
Let $\fdec{\alpha}{E(G)}{\enset{0,1}}$ and $\Lambda$ be any set.
There is a one-to-one correspondence between $\Lambda$-realizations of $\alpha$ and $\Lambda$-colourings of $\Galpha$.
\end{theorem}
\begin{proof}
Let $\fdec{\lambda}{V(G)}{\Lambda}$ such that $\alpha = \delta_{\lambda(u),\lambda(v)}$ for every $\{u,v\} \in E(G)$.
If $v \sim_\alpha w$, then $\lambda(v)=\lambda(w)$, by propagating equality along the path labelled by $1$.
Hence, the map $\fdec{\kappa}{V(\Galpha)}{\Lambda}$ given by $\kappa([v])\defeq\lambda(v)$ is well defined.
Now, an edge $e \in E(\Galpha)$
is of the form $e=\enset{[v],[w]}$ for some
$v, w \in V(G)$ such that $\alpha(\enset{v,w}) = 0$.
Since $\alpha_{v,w} = \delta_{\lambda(v),\lambda(w)}$, this means that $\lambda(v) \neq \lambda(w)$, hence $\kappa([v]) \neq \kappa([w])$. Thus, $\kappa$ is a $\Lambda$-colouring of $G/\alpha$.

\myindent Conversely, given a colouring $\fdec{\kappa}{V(\Galpha)}{\Lambda}$, set
$\lambda(v) \defeq \kappa([v])$. Let $e = \enset{v, w} \in E(G)$. 
If $\alpha(e)=1$, then $[v]=[w]$, hence $\lambda(v)=\lambda(w)$ because $\kappa$ is a colouring.
If $\alpha(e)=0$, then $\enset{[v],[w]}\in E(\Galpha)$, hence $\lambda(v)\neq\lambda(w)$. In either case, $\alpha(e) = \delta_{\lambda(v),\lambda(w)}$ from which we conclude that $\alpha$ is $\Lambda$-realizable.

\myindent The two processes just described are inverses of one another.
\end{proof}

\begin{corollary}\label{cor:realizablecolourable}
An edge $\enset{0,1}$-labeling is $\Lambda$-realizable if and only if the possibly loopy graph $G/\alpha$ induced by $\alpha$ is $\Lambda$-colourable.
\end{corollary}

\begin{corollary}\label{corollary: dimensionally restricted from colorability}
    Let $d \in \mathbb{N}_{\geq 2}$. Then, a deterministic edge weighting $\alpha \in \mathfrak{C}^{(d)}(G)$ for a given event graph $G$ if and only if $G/\alpha$ is $d$-colorable.
\end{corollary}

\begin{proof}
    This follows from the same proof of the theorem above, by restricting the cardinality of $\Lambda$ to be $d$.
\end{proof}

\begin{corollary}
    A deterministic edge weighting $\alpha \in \mathfrak{C}^{(2)}(G)$ if, and only if, the possibly loopy graph $G/\alpha$ induced by $\alpha$ is bipartite. 
\end{corollary}

\begin{proof}
    This follows from the fact that a graph is bipartite if and only if it is 2-colorable.
\end{proof}

\myindent Note that from the definition of the quotient graphs $G/\alpha$, whenever we have that $\alpha$ assigns zero to every edge, the quotient graph is isomorphic to the graph itself.  

\begin{lemma}\label{lemma: G/r0 is equal to G}
    Let $G$ be an event graph and $r_0$ the deterministic edge weighting defined by $r_0(e)=0$ for every edge $e$ of $G$. Then, $G/r_0 \simeq G$.
\end{lemma}

\myindent This lemma, together with~\cref{corollary: dimensionally restricted from colorability}, allow us to conclude the following:

\begin{proposition}
    Let $n \in \mathbb{N}$. The deterministic edge weighting $r_0 $ is an element of $ \mathfrak{C}^{(d)}(K_n)$ if and only if $d \geq n$. 
\end{proposition}

\begin{proof}
    From~\cref{lemma: G/r0 is equal to G} we have that $K_n/r_0 \simeq K_n$, which is $d$-colorable iff $d\geq n$. The conclusion from the proposition then follows from~\cref{corollary: dimensionally restricted from colorability}.
\end{proof}

\myindent We have seen that $\mathfrak{C}(G) = \text{ConvHull}(\text{ext}(\mathfrak{C}(G)))$, which is given by deterministic edge weightings that satisfy `transitivity of equality'.  However, it is \emph{not} clear that the same holds for $\mathfrak{C}^{(d)}(G)$. It is not clear that such a set is convex as, for example, in our proof that $\mathfrak{C}_{\mathbb{P}}(G)$ is convex we have used the fact that we could increase the cardinality of $\Lambda$ at will. It could still be, nevertheless, that $\mathfrak{C}^{(d)}(G)$ is a nonconvex polytope. From a more practical side, one can in principle work with the convex hull of this set, or of the set of extreme points of $\alpha \in \text{ext}(\mathfrak{C}(G))$ making $G/\alpha$ $d$-colorable. It is, however, often the case that when we impose a \emph{restriction} to the dimensionality of realizability by jointly distributed random variables one \emph{loses} convexity. Because of that, we conjecture that the sets $\mathfrak{C}^{(d)}(G)$ are \textit{not} convex, neither are the convex hull of the extreme points just described. 

\begin{conjecture}\label{conjecture: inside the convex hull}
    Let $\mathcal{V}_{G,d} \subseteq \mathrm{ext}(\mathfrak{C}(G))$ denote the set of all extreme points $\alpha$ for which $G/\alpha$ is $d$-colorable, for any event graph $G$. There exists some event graph $G$ and some $d \in \mathbb{N}_{\geq 2}$ such that $\mathfrak{C}^{(d)}(G) \subsetneq \mathrm{ConvHull}(\mathcal{V}_{G,d})$. 
\end{conjecture}

\begin{conjecture}\label{conjecture: nonconvexity of C_Pd}
    There is $d<|V(G)|$ and an event graph $G$ such that $\mathfrak{C}^{(d)}(G)$ is not convex.
\end{conjecture}

\myindent We would be interested to prove or provide a counterexample to these two conjectures.

\section{Discussion}\label{sec: discussions event graph}

\myindent In this section we take the opportunity to briefly describe our perspective for what the framework itself can be used to, from a purely mathematical point of view, as well as other theoretical opportunities. We leave comments related to quantum physics for Chapter~\ref{chapter: relational coherence}. 

\subsection{Perspective}

\myindent The framework itself so far treated solely the realization of certain edge weightings on graphs via random variables. It is therefore important to highlight what are the geometrical insights that such a framework allows that other polytopes constructed via jointly distributed random variables in the literature lack. For example, various of the inequalities we have described here have been previously discussed, as we have mentioned in Chapter~\ref{chapter: contextuality}. We will see later that our event graph polytope shares deep connections with other polytopes in the literature associated with contextuality. Perhaps it is relevant to pause and ponder on our findings, and think about how our framework can be distinguished from others from this purely geometrical point of view. 

\myindent Conceivably the best motivations for our framework lie in some of the simple facts one finds about our geometrical constructions. For instance, the symmetries associated with complete graphs, and the many patterns we have encountered when analyzing the polytopes $\mathfrak{C}(G)$ give hope that such polytopes are significantly simpler to investigate geometrically (than, say, $\mathrm{NC}(\pmb{\Upsilon})$), and could be one day \emph{completely characterized}. This is partly expressed by our conjectures~\ref{conjecture: another infinite family of inequalities} and~\ref{conjecture: C(Kn) has facets of any other C(Kn) for larger n}. Since, as we see in the next Chapters, there are formal dualities between these and some \acrshort{ks} (and Bell) polytopes, this would sort out a complete characterization of a relevant class of polytopes for quantum foundations.

\myindent We can also note that, given that the next Chapter  provides physical motivation for studying these structures, and later on we will see even `technological motivation' from the point of view of quantum information advantage and benchmarking nonclassical properties of quantum devices (cf.~\cref{chapter: applications}). Investigating these polytopes can be associated to bounding the ability of jointly distributed random variables to reproduce the prediction of tasks that can be framed in terms of edge weightings. This is, in our view, one of the main motivations for investigating the mathematical and geometrical properties of the structures we have introduced. Because of that, we make now an effort to highlight the most relevant mathematical problems in the framework that follow from our contributions. 

\subsection{Future directions}

\myindent Let us now comment on some open opportunities to investigate the framework we have just introduced, purely from a mathematical perspective. We discuss conceptual, experimental, and applied opportunities in later chapters. 

\myindent Perhaps the most promising line of research that could be developed for the framework we have introduced, from a mathematical point of view, is to investigate \emph{liftings} of facet-defining inequalities for $\mathfrak{C}(G)$. Will there be different `kinds' of liftings, such as measurement and outcome liftings in Kochen--Specker~\citep{choudhary2024lifting} or Bell~\citep{pironio2005lifting,pironio2014allCHSHpolytopes} scenarios? As a simple example of a result, one could show that every non-trivial event graph polytope, \ie $\mathfrak{C}(G) \neq [0,1]^{E(G)}$, has all possible $3$-cycle inequalities. Moreover, we could understand better why $n$-cycle inequalities, for $n\geq 4$, cannot be lifted to $\mathfrak{C}(K_n)$ as we see numerically that these are never facet-defining inequalities of complete graphs. Another simple example of a conjecture that has numerical evidence is the fact that every $\mathfrak{C}(K_n)$ has all the $h_n(\pmb r)\leq 1$ inequalities up to $n$ as facet-defining inequalities, a conjecture that we have made before (see~\cref{conjecture: hn inequalities are always facet-defining}), which would easily follow from an understanding of lifting in such event graph polytopes. Lifting results are important as they point out which facet-defining inequalities are \emph{not} truly arising due to non-trivial new connections generated by more complex graphs.

\myindent Another aspect is that many general results could be found regarding the investigation of cross sections of event graphs for more general graphs than those we have considered, namely, the very simple cases of $\mathfrak{C}_e^i$ for $i=0,1$. It would be useful to find out if generic statements (or statements for other families of graphs) can be made for the cross sections $\mathfrak{C}_{G'|G}^i$ for more generic subgraphs $G'$ of $G$, or even other cross sections entirely. As we have mentioned previously, we believe that in such cases the idea of edge contraction may become relevant, for the cases of deterministic edge weightings and cross sections considering a certain subset of edges equal to $1$. The idea behind this `clue' is that in such cases, one should be `forced' to interpret the two random variables realizing the deterministic edge weighting as equal. We leave this technical investigation for future work.

\myindent For any event graph $G$, the suspended graph $\nabla G$ is again an event graph. This construction becomes relevant in Chapter~\ref{chapter: from overlaps to noncontextuality}, and it appeared indirectly in the present chapter when we noted that every wheel graph is the suspension of a cycle graph (see also Appendix~\ref{sec: graph theory}). This observation naturally raises the following question: is there a general structural relationship between $\mathfrak{C}(G)$and the polytope associated with the suspension  $\mathfrak{C}(\nabla G)$? We believe this could be an interesting direction for future work.

\myindent In the context of `restricted polytopes', it is of pressing relevance to understand if these are convex polytopes or not (~\cref{conjecture: inside the convex hull} and~\cref{conjecture: nonconvexity of C_Pd}). This is because in such a case there could exist scenarios in which one bounds the cardinality of random variables (for instance, due to the rules of some communication game) for which a convex combination of strategies are \emph{not} possible, which is usually taken to be a problematic feature. 

\chapter{Quantum coherence from the event graph approach}\label{chapter: relational coherence}

\begin{quote}
    ``\textit{Hilbert space does not come equipped with a preferred basis; we should be able to deduce that the position basis (or some other one) is useful to our analysis of the system, rather then assuming we have been given it from the start.}''\\(Sean~\cite{carroll2022reality})
\end{quote}

\myindent In this chapter we introduce, motivate, and develop a new paradigm in quantum coherence theory that we term \emph{relational quantum coherence}. The term \emph{relational} emphasizes that we do not consider coherence as a property of single objects alone, as is traditionally considered, but as a property only well-defined \emph{relative to} (a collection of) other objects. This is similar to how set coherence is defined and motivated (see Chapter~\ref{chapter: quantum coherence}), but there is one nontrivial distinctive aspect of our approach: ours is a \emph{system-agnostic} property characterized in terms of experimentally accessible statistics. In other words, relational coherence as we define is \emph{not}  a property of a fixed set of quantum states with respect to a fixed Hilbert space. It is a property of tuples of numbers described by statistics realizable by Bargmann invariants. These tuples signal that \emph{every} possible set of quantum states, with respect to any quantum system capable of realizing their values, must fail to be set incoherent. 

\myindent This is not a profoundly new idea. It clearly borrows understanding from the field of quantum foundations devoted to understand the Bell and Kochen--Specker theorems. When these topics are presented, most of the emphasis is on the fact that  Bell's theorem (or the \acrshort{ks} theorem) allows us to falsify classes of hidden-variable models satisfying certain constraints. However, there are some more nuanced features that we learn from these no-go results. Let us discuss some underappreciated lessons from Bell's  theorems, which serve as a preliminary motivation for our proposed notion of coherence.

\myindent It is well established that a violation of a Bell inequality \emph{does not} reveal a property of states alone (such as entanglement) nor a property of measurements alone  (such as measurement incompatibility) but it reveals a property held by the \emph{conjunction} of a given state and the various measurements performed on it, revealed by experimental statistics. Even though one can use Bell inequalities as, e.g.,  entanglement witnesses~\citep{terhal2000bell}, the fact that from the violation of a Bell inequality we see the failure of Bell's notion of local causality is a property of the experimental setup as a whole. From this perspective, what is commonly termed `Bell nonlocality' is actually a relational property of one state and many measurement effects, satisfying specific causal relations.

\myindent It is also well known that the violation of a Bell inequality implies that not only a given collection of states and measurements, satisfying the scenario constraints, fails to be described by ontological models satisfying Bell's notion of local causality, but that \emph{every possible} collection, satisfying the same scenario constraints, and yielding the same inequality violation, fails to have such an explanation. In other words, if a behavior $B$ of a Bell scenario $\pmb{\Upsilon}_{\mathrm{Bell}}$ that is realized operationally by a certain state and a set of measurements is such that $B \notin \mathrm{NC}(\pmb{\Upsilon}_{\mathrm{Bell}})$, not only this specific realization fails to have a Bell local explanation, but the same must hold for \emph{every possible realization} of $B$, quantum theoretic or not. In case $B \in \mathcal{Q}(\pmb{\Upsilon}_{\mathrm{Bell}})$, this conclusion holds regardless of the Hilbert space of the system considered, and of the reference basis used to describe quantum operators.   

\myindent Also, there is a sense in which the `zoo' of different Bell inequalities formally \emph{organize and classify} all the possible ways in which quantum theory can be made distinct from such models considered by Bell. This may look as a trivial point to make, but this is only so \emph{because we know} that there is a notion of a `complete list of Bell inequalities' for a given Bell scenario. These lists formally characterize (and answer) the informal question `what are all the possible ways in which the statistical results of a Bell scenario can fail to be represented by local hidden variable models?' The fact that different statistical results, for the same scenario, violate different inequalities evidences the non-trivial fact that there are formally different ways in which a data-table can fail to be locally representable. 

\myindent Our intention here in the present Chapter is therefore to propose a similar framework for coherence satisfying the features described above, familiar to the theory of Bell nonlocality~\citep{brunner2014bell}: (i) We want to define relational coherence as a basis-independent and system-independent property witnessed by experimental statistics;~\footnote{This comparison would be made even stronger if it would also possible to motivate and justify relational coherence as a \emph{theory-independent} notion using generalized probabilistic theories. There is a well-defined way in which one can talk about coherence beyond quantum theory using such a landscape of physical theories. We have, however, failed to find meaningful ways of doing so using the event graph approach described in Chapter~\ref{chapter: event_graph_approach}. We leave the possibility of doing so as an open problem. } (ii) and, moreover, we also want to propose a systematic classification of all the possible ways in which these statistics, which in our case are described by tuples of Bargmann invariants, can fail to be relationally \emph{in}coherent, meaning, having a quantum realization by pairwise commuting sets of states. 

\myindent In this Chapter, we show how quantum realizability of edge weightings introduced in Chapter~\ref{chapter: event_graph_approach} can be viewed as a partial answer to the two points above, and as one of the applications of this framework that we believe can be interpreted as \emph{a} solution to~\cref{question: applications}. We show that the event graph inequalities found in Chapter~\ref{chapter: event_graph_approach}, and more broadly the event graph approach, provide a comprehensive (yet incomplete) classification and organization of linear functionals over tuples of states that serve as witnesses of relational coherence. By construction, these witnesses can \emph{also} be interpreted as witnesses of coherence in the standard treatment previously introduced in Chapter~\ref{chapter: quantum coherence}.  

\myindent Before reviewing the structure and results of this Chapter, we also take the opportunity to mention that the term `relational' has been used for different (but possibly related) reasons within quantum information and quantum foundations. For example, the idea that coherence is a relational property was key to the  `coherence as a myth' debate initiated by the work of~\cite{molmer1997optical}. In summary, the debate stems from two different views on how a single quantum state $\rho$ (and its coherence) ought to be described. See the review by~\cite{smolin2004continuous} for an introduction to this debate. Here we consider the interpretation of it from~\cite{bartlett2006dialogue}, focusing on coherent states of light. One party (the `factists') interprets states $\rho$ as described relative to a basis $\mathbb{S}_1$ that is an external reference frame,  independent of $\rho$, with respect to which we define what it means for $\rho$ to be coherent (this is the treatment that goes according to what was presented in Chapter~\ref{chapter: quantum coherence}). The other party (the `fictionists') views both $\rho$ and $\mathbb{S}_1$ as characterizing internal degrees of freedom of a quantum state, and here $\mathbb{S}_1$ is an internal reference frame while \emph{a second} different reference frame $\mathbb{S}_2$ is viewed as the correct reference frame for representing $\rho$ and its coherence. While the `factists' conclude from their analysis that $\rho$ must be a coherent state, the `fictionists' conclude the \emph{exact opposite}, arguing that a careful analysis of experimental tests show that $\rho$ can always be taken to be incoherent, hence proposing that coherence has no true element of reality, even if a convenient `fiction' to make calculations.

\myindent Part of the resolution to this debate came from acknowledging that proponents of the `factist' party  considered coherence as a relational property between $\rho$ and $\mathbb{S}_1$, while proponents of the `fictionist' party considered coherence as a relational property between $\rho$ and \emph{the different frame} $\mathbb{S}_2$~\citep{bartlett2006dialogue}. Viewed in this way, it is perhaps less surprising that in one case the state $\rho$ is coherent while in the other the state can be incoherent. Similarly, it is also less surprising that both views reproduce the same statistics. Our framework uses the term `relational' in a similar way, but instead of focusing on the relation between a state $\rho$ and a reference  $\mathbb{A}$ we consider the arguably weaker relation between a state $\rho$ and \emph{any other set of states}  $\{\sigma_i\}_i$ regardless of whether they make (or do not make) a basis of a Hilbert space. 

\myindent Another use of the word `relational' can be found in Carlo Rovelli's interpretation of quantum mechanics dubbed the \emph{relational interpretation of quantum mechanics}~\citep{rovelli1996relational}. In that case, the attempt is to resolve the measurement problem~\citep{schlosshauer2005decoherence} by noting that there is no notion of a `measurement'. Every measurement one makes is merely an interaction between different systems, and the distinction between measuring system and measured system is merely relational. We do not comment on how our framework relates to (or differs from)  this interpretation. We simply consider it a different context and leave possible connections to be investigated by future research. 

\myindent Yet a third use of the word relational appears in the literature devoted to the Page--Wooters formalism~\citep{page1983evolution}. In this formalism one has a Hilbert space (that can be viewed as the `Hilbert space of the universe') $\mathcal{H}_T$ that can be partitioned as  $\mathcal{H}_T = \mathcal{H}_1 \otimes \mathcal{H}_2 \otimes \dots \otimes \mathcal{H}_n$. Investigating the ways in which one can subdivide a Hilbert space $\mathcal{H}_T$ into different tensor product components is known as \emph{quantum mereology}~\citep{carroll2021mereology}. The term `relational' in the Page--Wooters formalism comes from the view that we should describe the dynamical evolution of a system $\mathcal{H}_1$ as viewed by some other system $\mathcal{H}_2$ \emph{differently} than the dynamical evolution of $\mathcal{H}_1$ as viewed by a third party, say, $\mathcal{H}_3$. In this case, the time evolution predicted by the Schrödinger equation is a  relational property dependent on which pair of systems are being considered. For the purposes of this thesis, we do not consider this use of the word relational, and we draw no connection between the Page--Wooters formalism and ours.

\subsubsection{Outline}

\begin{figure}[t]
    \centering
    \includegraphics[width=1\textwidth]{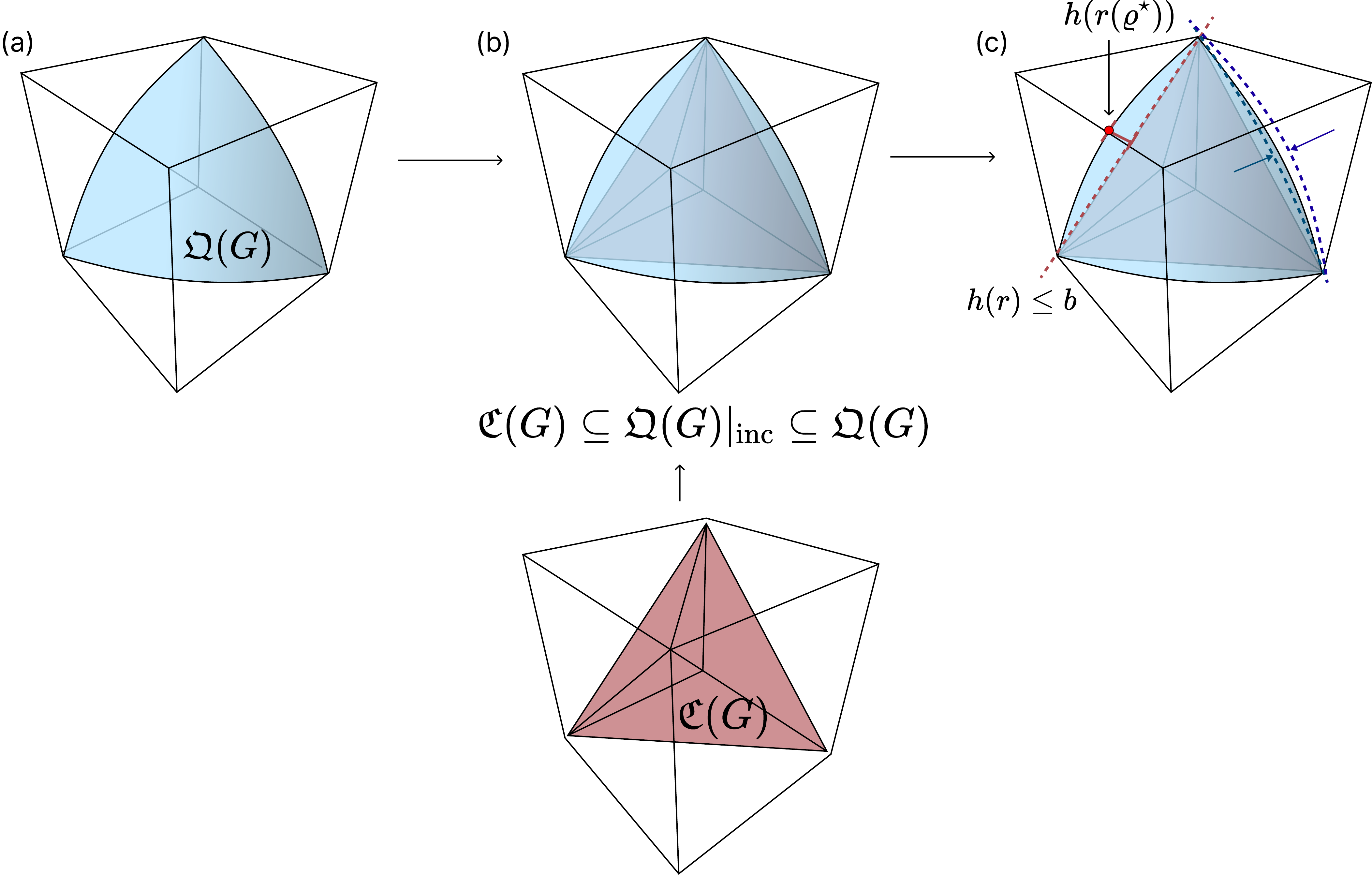}
    \caption{\textbf{Illustration of the structure of this Chapter.} (a) Section~\ref{sec: sets of Bargmann invariants tuples} focuses on sets of Bargmann invariants $\mathfrak{Q}(G)$, and its geometry. (b) Sections~\ref{sec: relational coherence} and~\ref{sec: prepare and measure overlap} motivate and conceptualize the notion of relational coherence, using our results from Chapter~\ref{chapter: event_graph_approach}. (c) Lastly, we conclude the Chapter with Sections~\ref{sec: relational coherence from the event graph approach} and~\ref{sec: relational coherence higher order Bargmanns} where we study numerical optimization tools for investigating gaps between the convex polytopes $\mathfrak{C}(G)$ and the quantum correlation sets we introduce here. 
    }
    \label{fig:chapter outline}
\end{figure}

\myindent This Chapter can be thought of as having three main parts (structured in a manner illustrated in Fig.~\ref{fig:chapter outline}). The first one focuses on defining and investigating geometrical properties of sets of Bargmann invariants, which can be found in  Sec.~\ref{sec: sets of Bargmann invariants tuples}. The second part focuses on \emph{motivating, operationalizing, and conceptualizing} the notion of \emph{relational coherence} captured by points in such sets of correlations and can be found in Sec.~\ref{sec: relational coherence} and Sec.~\ref{sec: prepare and measure overlap}. Points that cannot be realized by any set of incoherent states, in any Hilbert space, are dubbed \emph{relationally coherent}. The last part focuses on providing tools for \emph{characterizing and optimizing} the search for relational coherence using the event graph approach, and beyond. 

\myindent In Section~\ref{sec: sets of Bargmann invariants tuples}, we define \emph{sets of tuples of Bargmann invariants} from quantum realizability of edge weightings, as well as more general weightings which relate to Bargmann invariants of order higher than two. We significantly advance the theory for investigating and characterizing this new set of quantum correlations, which we believe has implications for quantum foundations beyond the specific context of relational coherence, which is the focus of this Chapter.

\myindent Section~\ref{sec: relational coherence} and provides detailed motivation and conceptualization to the notion of relational coherence. In there, we argue that a relational view of coherence encompasses previous notions of coherence; that relational quantum coherence is both a physically and conceptually well motivated notion; and that it organizes rigorously different types of coherence previously considered, such as quantum imaginarity. We also discuss, in Section~\ref{sec: prepare and measure overlap}, the relevance of quantum-realizable edge-weightings to the specific setting of prepare-and-measure scenarios, which has been considered to be a paradigmatic example of semi-device independent tests of nonclassicality. We propose that one can view quantum-realizable edge weightings as correlations arising in prepare-and-measure scenarios. 

\myindent In Sections~\ref{sec: relational coherence from the event graph approach} and~\ref{sec: relational coherence higher order Bargmanns} we focus on the specific ways in which Bargmann invariants can have {relational coherence}. We show how to use \acrshort{sdp} techniques for optimally violating event-graph inequalities, providing several lower and upper bounds for quantum violations of the inequalities introduced in Chapter~\ref{chapter: event_graph_approach}. We also investigate the role played by the purity of the states for obtaining optimal violations. Moreover, we discuss the relevance of higher-order Bargmann invariants to the characterization of relational coherence, such as relational imaginarity. We conclude this Chapter in~\cref{sec: discussion Bargmann relational chapter} commenting on theoretical opportunities left open for future research. 

\subsubsection{Structure of contributions reported in this Chapter}

\myindent In this Chapter we report the contribution of various of the works mentioned in Chapter~\ref{chapter: introduction} forming the backbone of this thesis, and not on contributions of a single article. A small number of the results presented here have not been published, and are original contributions from this thesis to the body of literature. To improve reading, our presentation is not divided by the published works but by fluidity of the discourse. To simplify the reviewing process of this thesis, we now give a detailed description of where the published material reported here can be found. Complementary, formal contributions in the form of Propositions, Lemmas, Definitions and Theorems that have been considered in published material  have a reference to the specific work in which they first appeared. When a reference is not made, the contribution is, to the best of our knowledge,  novel to the body of literature. 

\myindent Starting with the investigation of the geometry of quantum realizable edge-weigthings, the content of Sec.~\ref{sec: quantum realizations} associated to event graphs is present in references~\citep{wagner2024inequalities,wagner2024coherence,wagner2024certifying}. The definitions and analysis that go beyond the event graphs, hence considering higher-order Bargmann invariants, are original contributions that have not been published elsewhere (such as~\cref{proposition: cyclic is equivalent} and~\ref{proposition: word invariants and real realizability}). In Section~\ref{subsec: sets of tuples of Bargmann invariants}, the definitions of the sets $\mathfrak{Q}(G)$ and $\mathfrak{B}_n$ were adapted from references~\citep{wagner2024inequalities},~\citep{wagner2024certifying} and~\citep{fernandes2024unitary}.~\cref{lemma: projectors and event graph correlations},~\cref{theorem:convexity_tuples_of_order_m} as well as all its Corollaries are original contributions that had not previously appeared elsewhere. To conclude, the results presented in~\cref{subsec: candidates of Gram matrices} and~\cref{subsec: different sets of Bargmann invariants} are adapted from reference~\citep{fernandes2024unitary}.~\cref{theorem: galvao states reaching the boundary},~\cref{proposition: hadamard product}, and~\cref{proposition: unit disc asymptotic} are simple original contributions that had not previously been discussed in the literature. 

\myindent Moving to the middle of this Chapter we have Sections~\ref{sec: relational coherence} and~\ref{sec: prepare and measure overlap} that focus on motivating and conceptualizing the notion of relational coherence, and its relevance. Most of this discussion is adapted from results that appear in references~\citep{wagner2024inequalities},~\citep{giordani2023experimental},  and~\citep{wagner2024coherence}, or also in some of our work that has not been the main focus of this thesis, such as~\citep{wagner2024quantumcircuits} and~\citep{wagner2023anomalous}. 

\myindent The final sections of this Chapter are devoted to investigating properties of relational coherence including optimal violations, specific symmetric sets of single-qubit states that are set-coherent and achieve large violations of event graph inequalities, as well as the notion of relational imaginarity. In Section~\ref{sec: relational coherence from the event graph approach} we present results adapted (and sometimes improved, such as the results from  Table~\ref{tab: seesaw bound} that improve the bounds reported in~\cite{wagner2024inequalities}) from references~\citep{wagner2024certifying},~\citep{fernandes2024unitary},~\citep{giordani2023experimental},   and~\citep{wagner2024inequalities}. Section~\ref{sec: relational coherence higher order Bargmanns} presents original results, especially the brief discussion on equality constraints from Sec.~\ref{subsec: equality constraints}, but has discussions adapted and influenced by reference~\citep{fernandes2024unitary}.

\section{Sets of Bargmann invariants}\label{sec: sets of Bargmann invariants tuples}

\myindent Our notion of relational coherence is tied to the notion of \emph{quantum realization}. We have already encountered in previous chapters some instances of quantum realizability when we have discussed prepare and measure scenarios (see~\cref{def: quantum realizable PM behavior} and~\cref{def: dimension dependent compatibility scenario}), and when we have discussed compatibility scenarios (see~\cref{def: quantum realizations of behaviors in compatibility scenarios}). In what follows, we introduce both the notion of quantum realization of edge weightings of a given event graph and a broader notion of quantum realization that is not described by a graph-theoretic formalism.

\myindent Our notion of realization is naturally tied to sets of tuples of Bargmann invariants. These are either sets of all possible quantum realizations of edge weightings in terms of two-state overlaps or sets of all possible higher-order Bargmann invariants. As we have mentioned in the beginning of this Chapter, there are various parallels between the notions we introduce (and the research directions we motivate) and that of Bell nonlocality theory. In Bell nonlocality, there is a great effort for characterizing the sets of quantum correlations, and understanding its geometric properties, which in that context corresponds to the characterization of the sets $\mathcal{Q}(\pmb{\Upsilon}_{\mathrm{Bell}})$. Here, similarly, we introduce sets of tuples of Bargmann invariants and stress that characterizing their geometry is a well-motivated topic in quantum foundations. That is, we introduce yet another set of quantum correlations worthy of investigation.

\subsection{Quantum realizations}\label{sec: quantum realizations}

\myindent Recall that for a given event graph $G$ we call any mapping $r: E(G) \to [0,1]$ as an edge weighting for $G$. For any $e = \{u,v\}$ we have $r(e) = r_{u,v} \in [0,1]$ is a real number between $0$ and $1$. The notion of a quantum realization for $r$ is related to the natural question of when these values $r_{u,v}$ can be interpreted as the two-state overlap $r_{u,v}= \text{Tr}(\rho_u\rho_v)$ of two quantum states $\rho_u,\rho_v \in \mathcal{D}(\mathcal{H})$, for some Hilbert space $\mathcal{H}$. We make this intuition formal by the following definition.

\begin{definition}[Quantum realizations of edge weightings, adapted from~\cite{wagner2024inequalities}] \label{def:Bargmann_event_graph_realizations}
    Let $G$ be any event graph. We say that an edge weighting $r:E(G) \to [0,1]$ is \emph{quantum realizable} in a Hilbert space $\mathcal{H}$ if there exists some vertex labeling $\varrho: V(G) \to \mathcal{D}(\mathcal{H})$ such that $r_{ij} = \mathrm{Tr}(\rho_i\rho_j)$ for every edge $\{i,j\} \in E(G)$. In such a case, we write $r=r(\varrho)$.
\end{definition}

\myindent Note that quantum realizability is a notion tied to the functions $\varrho: V(G) \to \mathcal{D}(\mathcal{H})$. Because of that, we often use the terminology `$r$ is realized by $\varrho$', or equivalently in terms of the tuples, `$\pmb{r}$ is realized by $\pmb{\rho}$'. This means that for such a choice $\varrho$, it holds that $r = r(\varrho)$. Moreover,  just as there is a correspondence between $r$ (viewed as a function) and $\pmb{r} \equiv (r_e)_{e \in E(G)}$ (its related tuple), there is also a correspondence between $\varrho$ (viewed as a function) and $\pmb{\rho} \equiv (\rho_v)_{v \in V(G)}$ (its related tuple). Accordingly, as established in our notation so far, we write   $r(\varrho)$ for the function $r(\varrho):E(G)\to \mathcal{D}(\mathcal{H})$ mapping $\{i,j\} \mapsto \mathrm{Tr}(\rho_i\rho_j)$, and $\pmb{r}(\pmb{\rho}) = (\mathrm{Tr}(\rho_i\rho_j))_{\{i,j\}\in E(G)}$ for the tuple of quantum-realizable two-state overlaps, highlighting the specific tuple of states that realizes it. For a given vertex labeling $\varrho$, let us denote its image by $\{\rho_v\}_{v \in V(G)}$. For instance, it is possible to consider that the image $\varrho(V(G)) = \{\rho\}$, the singleton set having one state. This means that the vertex labeling assigns to every vertex \emph{the same state}, and that every overlap $r_{ij}(\varrho)$ equals the purity $\text{Tr}(\rho^2)$. To build some intuition on the notation and terminology of this notion of quantum realizability, we now provide some simple examples.

\begin{example}[Quantum realizations for $r_0$ and $r_1$]
    Let us consider the event graph $G=C_3$. The constant edge weighting $r_0(e)=0$ cannot be realized by vertex labelings $\varrho: V(C_3) \to \mathcal{D}(\mathbb{C}^2)$ since $r_0(e)=0$ for all $e \in E(C_3)$ implies that such quantum realizations must span a Hilbert space of dimension at least $3$.
    
    \myindent The constant edge weighting $r_1(e)=1$ can be realized by  a vertex labeling whose image is $\{\vert 0\rangle \langle 0 \vert \}$. It can assign every vertex to the state $\vert 0\rangle\langle 0 \vert $, i.e. $\varrho_{\vert 0\rangle }(i) = \vert 0\rangle \langle 0 \vert$. Hence, we have that $r_1(\varrho_{\vert 0\rangle}) =  r_1$. Trivially, any quantum realizable edge weighting has \emph{infinitely many} quantum realizations due to unitary-invariance. 
\end{example}

\myindent It is possible that the same edge weighting $r$ is quantum realizable in a Hilbert space $\mathcal{H}$ by two different vertex labelings $\varrho_1 : V(G) \to \mathcal{D}(\mathcal{H})$, and $\varrho_2 : V(G) \to \mathcal{D}(\mathcal{H})$ where the image of one realization $\mathcal{S}_1 = \varrho_1(V(G))$ is set coherent while the image of the other $\mathcal{S}_2 = \varrho_2(V(G))$ is set \emph{in}coherent. 

\begin{example}[Some weightings can be realized by both set coherent and set incoherent vertex labelings]\label{example: realization coherent and incoherent}
    Consider the event graph $G = C_3$ again, whose vertices we write as $V(G)=\{1,2,3\}$. We now consider the two sets $\mathcal{S}_1 = \{\vert 0\rangle\langle 0 \vert , \vert +\rangle \langle + \vert ,\vert +_i\rangle \langle +_i \vert \}$, and $\mathcal{S}_2 = \{\mathbb{1}/2\}$. The first is set coherent, while the second is set incoherent. The edge weighting $$\pmb{r} = (r_{12},r_{13},r_{23}) = (\sfrac{1}{2},\sfrac{1}{2},\sfrac{1}{2}) \in \mathfrak{C}(C_3)$$ is realizable by both  $\varrho^{\mathrm{diag}}(i) = \mathbb{1}/2, i \in V(G)$ and by $$\varrho^{\mathrm{cohe}}(1)=\vert 0\rangle\langle 0 \vert, \varrho^{\mathrm{cohe}}(2)=\vert +\rangle\langle + \vert,\varrho^{\mathrm{cohe}}(3)=\vert +_i\rangle\langle +_i \vert.$$
    Succinctly, we have that $r = r(\varrho^{\mathrm{diag}})= r(\varrho^{\mathrm{cohe}})$. 
\end{example}

\myindent We also consider another more generic situation that goes beyond the event graph approach. We have mentioned previously in Chapter~\ref{chapter: event_graph_approach} that we could extend the approach to encompass more than edge weightings. While we do not have yet a fully developed formalism to discuss more generic weightings extending the constructions from Chapter~\ref{chapter: event_graph_approach}, we can very easily introduce a notion of quantum realizability that extends the one above.

\myindent We denote by $\mathcal{W}$ a chosen subset of finite sequences of elements of a set $V$. The elements of $\mathcal{W}$ are called \emph{words}. For example, take $V = \{1,2,3\}$, one possible set is given by $$\mathcal{W}' = \{(1,2,3,2),(1,3,2,1),(2,3,2,1)\}. $$
The construction that follows does not rely on event graphs, as the sets $\mathcal{W}$ play the central role. To some extent, graphs become unnecessary in the extended notion of quantum realization we define shortly.

\myindent With respect to $\mathcal{W}'$ chosen above, we can define the tuple of Bargmann invariants $(\Delta_{\pmb{w}}(\varrho))_{\pmb{w} \in \mathcal{W}'}$ as all the possible tuples of numbers equal to 
$$
    \pmb{\Delta}(\pmb{\rho}) = (\text{Tr}(\rho_1\rho_2\rho_3\rho_2), \text{Tr}(\rho_1\rho_3\rho_2\rho_1), \text{Tr}(\rho_2\rho_2\rho_2\rho_1)),
$$
for some choice of quantum states $\rho_1,\rho_2,\rho_2,\rho_3 \in \mathcal{D}(\mathcal{H})$ with respect to some choice of $\mathcal{H}$. 

\myindent The above implies that, if we are given some set of finite sequences $\mathcal{W}$ taking values on the labels $V$, we can also define the following notion of quantum realization.

\begin{definition}[Quantum realizability beyond edge weightings]\label{def: Bargmann invariant quantum realizability}
    Let $\mathcal{W} \subseteq \mathbb{N}^*$ be a finite subset of finite sequences (or words) over the natural numbers.~\footnote{In other words, $\mathcal{W}$ is a finite subset of the \emph{free monoid} on the set of natural numbers  $\mathbb{N}$, written  $\mathbb{N}^*$, which is the \emph{monoid} whose elements are  finite sequences.} Define $V := \bigcup_{\pmb{w} \in \mathcal{W}} \{w \mid w \in \pmb{w}\}$ the set of all natural numbers appearing in the sequences of $\mathcal{W}$. We say that $\Delta: \mathcal{W} \to \mathbb{C}$ is \emph{quantum realizable} in a Hilbert space $\mathcal{H}$ if there exists a function $\varrho: V \to \mathcal{D}(\mathcal{H})$ such that 
    \begin{equation}
        \Delta_{\pmb{w}} = \text{Tr}\left( \varrho(w_1)\cdots \varrho(w_m)\right)
    \end{equation}
    for every $\pmb{w} = (w_1,\dots,w_m) \in \mathcal{W}$. In such a case, we write $\Delta = \Delta(\varrho)$. 
\end{definition}

\myindent As for edge weightings $r$, we write $\Delta = \Delta(\varrho)$ representing quantum realizability when considering $\Delta$ and $\varrho$ as functions, and $\pmb{\Delta} = \pmb{\Delta}(\pmb{\rho})$ when representing quantum realizability for the equivalent representation given by tuples. In what follows, we will denote $\mathbb{N}^*$ the set of all finite words over the natural numbers $\mathbb{N}.$

\myindent Considering~\cref{def: Bargmann invariant quantum realizability}, it is clear that if we choose \(\mathcal{W}_G = \{(i,j) \mid \{i,j\} \in E(G)\}\) for some event graph \(G\), quantum realizations for \(\mathcal{W}_G\) corresponds to the standard notion of quantum realization for edge weightings defined over \(G\) from~\cref{def:Bargmann_event_graph_realizations}. For example, taking $G=C_3$ we have $$\mathcal{W}_{C_3} = \{(1,2),(1,3),(2,3)\}$$
and thus a quantum realizable $\Delta: \mathcal{W}_{C_3}\to \mathbb{C}$ takes the form $$(\Delta_{\pmb w}(\varrho))_{\pmb{w} \in \mathcal{W}_{C_3}} = \left(\mathrm{Tr}(\varrho(1)\varrho(2)),\mathrm{Tr}(\varrho(1)\varrho(3)),\mathrm{Tr}(\varrho(2)\varrho(3))\right ),$$
which equals our definition of $\pmb{r}(\varrho)$. Whenever we want to bring the connection with event graphs we will make this choice, otherwise, for this extended description in terms of sets of words we will not need to mention event graphs anymore.

\myindent It is simple to see, from properties of Bargmann invariants, that since the trace is cyclic, quantum realizations as described above yield equal results to every word that is equivalent under cyclic permutations. We say that two words $\pmb{w}_1,\pmb{w}_2$ are cyclic conjugate of each other, and write $\pmb{w}_1 \sim_{\mathrm{cyc}}\pmb{w}_2$, if it is possible to reach one word from another using some number of cyclic permutations. For example, the words $(1,2,3)$ and $(3,1,2)$ are equivalent up to cyclic permutations since we have that 
\begin{equation*}
    (1,2,3) \mapsto (2,3,1) \mapsto (3,1,2).
\end{equation*}
Since this defines an equivalence relation for the words of a set $\mathcal{W}$ we use the notation that $(1,2,3) \sim_{\mathrm{cyc}} (3,1,2)$. With that, we can write the trivial proposition.

\begin{proposition}\label{proposition: cyclic is equivalent}
    Let $\mathcal{W}$ be a finite subset of $\mathbb{N}^*$. Suppose that two words $\pmb{w}_1, \pmb{w}_2$ are cyclic conjugate of each other, i.e. $\pmb{w}_1 \sim_{\mathrm{cyc}} \pmb{w}_2$. Then, $\Delta_{\pmb{w}_1}(\varrho) = \Delta_{\pmb{w}_2}(\varrho)$ for all possible $\varrho$.
\end{proposition}

\myindent This implies that it is not really $\mathcal{W}$ that is relevant for quantum realization in terms of Bargmann invariants, but $\mathcal{W}/~\sim_{\mathrm{cyc}}$, the quotient of this set under the  relation $\sim_{\mathrm{cyc}}$. In what follows, we therefore restrict to choices of sets $\mathcal{W}$ where all words $\pmb{w}$ are non-equivalent under $\sim_{\mathrm{cyc}}$.

\myindent It is also interesting to see that there is another relevant relation for elements in $\mathcal{W}$, which we denote as $*$. This is an \emph{involution} $(\cdot)^*:\mathcal{W} \to \mathcal{W}$ of words defined as follows. Let $\pmb{w} = (v_1,v_2,v_3,\dots,v_{n-2},v_{n-1},v_n) \in \mathcal{W}$ where $v_i \in \mathbb{N}^*$ for every $i$. Then, the operation $*$ acts such that $\pmb{w} \mapsto \pmb{w}^* = (v_n,v_{n-1},v_{n-2},\dots,v_3,v_2,v_1).$ With this, we can write the following lemma.

\begin{proposition}
    Let $\mathcal{W}$ be a finite subset of $\mathbb{N}^*$. Then, $\Delta_{\pmb{w}^*}(\varrho) = \Delta_{\pmb{w}}(\varrho)^*$ for every $\varrho$ and every $\pmb{w} \in \mathcal{W}$.
\end{proposition}

\begin{proof}
    The proof is trivial. Let $\pmb{w} = (v_1,v_2,v_3,\dots,v_{n-2},v_{n-1},v_n)$ be any. Then, if $\Delta_{\pmb{w}} = \Delta_{\pmb{w}}(\varrho)$ we have that 
    \begin{align*}
        \Delta_{\pmb{w}}^*(\varrho) &= \left(\text{Tr}(\rho_{v_1}\rho_{v_2}\rho_{v_3}\cdots\rho_{v_{n-2}}\rho_{v_{n-1}}\rho_{v_n})\right)^* = \text{Tr}((\rho_{v_1}\rho_{v_2}\rho_{v_3}\cdots\rho_{v_{n-2}}\rho_{v_{n-1}}\rho_{v_n})^\dagger)\\
        &= \text{Tr}(\rho_{v_n}\rho_{v_{n-1}}\rho_{v_{n-1}}\cdots \rho_{v_2}\rho_{v_2}\rho_{v_1}) = \Delta_{\pmb{w}^*}(\varrho)
    \end{align*}
    where we have used the notation $\varrho(v_i) = \rho_{v_i}$, and that multivariate traces satisfy $\text{Tr}(Z_1Z_2\cdots Z_n) = \text{Tr}((Z_1Z_2\cdots Z_n)^T)$. Above, we have also denoted complex conjugation of a complex number as $\Delta^*$. 
\end{proof}

\myindent This implies that whenever $\pmb{w}=\pmb{w}^*$,  $\Delta_{\pmb{w}}(\varrho) \in \mathbb{R}$. 

\begin{corollary}\label{proposition: word invariants and real realizability}
    Let $\mathcal{W}$ be a finite subset of $\mathbb{N}^*$. Then, if $\pmb{w} \in \mathcal{W}$ is such that $\pmb{w} = \pmb{w}^*$ we have that $\Delta_{\pmb{w}}(\varrho) \in \mathbb{R}$ for every $\varrho$.
\end{corollary}

\myindent The proposition above implies that $\text{Tr}(\rho \sigma \rho^2), \text{Tr}(\rho^2\sigma^2), \text{Tr}(\rho\sigma\rho\sigma\rho\sigma), \text{Tr}(\rho\sigma\rho\sigma^3)$  are all examples of  higher-order real-valued \emph{only} Bargmann invariants, for all possible choices of $\rho$ and $\sigma$. 

\subsection{Sets of tuples of Bargmann invariants}\label{subsec: sets of tuples of Bargmann invariants}

\myindent Given that we have introduced a notion of quantum realization $r(\varrho)$, of a given edge weighting $r$ of an event graph $G$ we can define the set of all such quantum realizable weightings. We are interested in investigating its geometrical properties, and its relation to the convex polytope $\mathfrak{C}(G)$ of an event graph $G$. 

\begin{definition}[Sets of tuples of two-state overlaps, adapted from~\cite{wagner2024inequalities}]
    Let $G$ be an event graph and let $r: E(G) \to [0,1]$ be a quantum realizable edge weighting. We denote $\mathfrak{Q}(G)$ the set of all such weightings. In other words, 
    \begin{equation}
        \mathfrak{Q}(G) := \{r \in [0,1]^{E(G)} \mid \exists \varrho \text{ such that } r = r(\varrho)\}
    \end{equation}
    where $\varrho: V(G) \to \mathcal{D}(\mathcal{H})$ for some Hilbert space $\mathcal{H}$.
\end{definition}

\myindent To the best of our knowledge,~\cite{galvao2020quantum} were the first to investigate the geometry of such sets, focusing on the geometry of $\mathfrak{Q}(C_3)$. Interestingly, the set $\mathfrak{Q}(C_3)$ has parallels to a known set (the \emph{elliptope}, whose connections with the quantum set of correlations in Bell scenarios were independently discovered by~\cite{tsirelson1987quantumanalogues},~\cite{landau1988empirical}, and~\cite{masanes2005extremalquantumcorrelationsn}) that was recently considered by~\cite{le2023quantum}, as it also characterizes the set of quantum realizable Bell correlations in the so-called minimal Bell scenario. Its boundary is given by Eq.~\eqref{eq: boundary of Q(C_3)} presented in Chapter~\ref{chapter: Bargmann invariants}. 

\myindent We introduce some notation to organize the various different subsets of $\mathfrak{Q}(G)$ we consider. One such example is considering the restriction to pure state quantum realizations, which we denote by
\begin{equation}
    \mathfrak{Q}(G)|_{\mathrm{pure}} = \left\{r\in \mathfrak{Q}(G) \mid \exists \psi:V(G) \to \text{ext}(\mathcal{D}(\mathcal{H})) \text{ such that  } r = r(\psi)\right\},
\end{equation}
where we have denoted the set of pure states as the extremal elements of $\mathcal{D}(\mathcal{H})$ and changed our notation from $\varrho$ to $\psi$ to highlight the fact that we are considering pure states only.~\footnote{Note that for the case of mixed states we have introduced a different notation for the states $\rho$ and for the vertex labelings $\varrho$. For the case of pure states, for simplicity, we use the same notation for both the states and the vertex labelings.} As in Chapter~\ref{chapter: Bargmann invariants}, we denote by $\pmb{\psi} = (\psi(v))_{v \in V(G)} = (\vert \psi_v\rangle \langle \psi_v \vert )_{v \in V(G)}$ the associated tuple of quantum states defined by the vertex labeling $\psi$. This imply that $$(\pmb{r}(\pmb{\psi}))_{ij} = \text{Tr}(\psi_i\psi_j) = \text{Tr}(\vert \psi_i\rangle \langle \psi_i \vert \vert \psi_j \rangle \langle \psi_j \vert ) =  |\langle \psi_i |\psi_j\rangle |^2.$$ We denote by $\mathfrak{Q}^{(d)}(G)$ the restriction to quantum realizations of a specific Hilbert space dimension, i.e. quantum realizations $r = r(\varrho)$ such that $\varrho: V(G) \to \mathcal{D}(\mathbb{C}^d)$. 

\myindent We define $\mathfrak{Q}(G)|_{\mathrm{inc}}$ to be the subset of $\mathfrak{Q}(G)$ of all edge-weightings that have a realization with respect to set incoherent quantum states. In other words, this set denotes the following subset: 

\begin{equation}\label{eq: edge-weightings realizable with set incoherent states}
\mathfrak{Q}(G)|_{\mathrm{inc}} := \{r \in \mathfrak{Q}(G) \mid \exists \varrho: V(G) \to \mathcal{I}(\mathcal{H},\mathbb{A}) \text{ s.~t. }r = r(\varrho)\text{ for some }\mathcal{H}\text{ and some }\mathbb{A}\}.
\end{equation}
Above, $\mathbb{A}$ is an orthonormal basis for $\mathcal{H}$ and $\mathcal{I}(\mathcal{H},\mathbb{A})$ is as defined in~\cref{chapter: quantum coherence} (cf.~\cref{def: basis dependent coherence}).

\myindent The final restriction we consider is related to imaginarity-free realizations. Recall that we say that a certain set of states $\mathcal{S} \subseteq \mathcal{D}(\mathcal{H})$ is imaginarity free (or `set real') if there exists some basis $\mathbb{A}$ such that $\mathcal{S} \subseteq \mathcal{R}(\mathcal{H},\mathbb{A})$, meaning that with respect to the basis $\mathbb{A}$ we have that $\forall \rho \in \mathcal{S}$ it holds that $\langle a|\rho|a'\rangle \in \mathbb{R}$ for all $\vert a\rangle, \vert a'\rangle \in \mathbb{A}$ (~cf.~\cref{def: set imaginarity}). This yields the following restriction

\begin{equation}\label{eq: edge-weightings realizable with set real states}
\mathfrak{Q}(G)|_{\mathrm{real}} := \{r \in \mathfrak{Q}(G) \mid \exists \varrho: V(G) \to \mathcal{R}(\mathcal{H},\mathbb{A}) \text{ s. t. }r = r(\varrho)\text{ for some }\mathcal{H}\text{ and some }\mathbb{A}\}.
\end{equation}
These notions of set incoherence and basis-independent imaginarity of a set of states have been discussed in Chapter~\ref{chapter: quantum coherence},~\cref{set: basis-independent coherence of a set of states} in Definitions~\ref{def: set coherence} and~\ref{def: set imaginarity}. From the definitions just provided we see that for every event graph $G$ one has the following set of inclusions 
\begin{equation}
    \mathfrak{Q}(G)|_{\mathrm{inc}} \subseteq \mathfrak{Q}(G)|_{\mathrm{real}} \subseteq \mathfrak{Q}(G).
\end{equation}

\myindent Let us now define more generally sets of tuples of Bargmann invariants, i.e., going beyond two-state overlaps, by considering realizations of $\Delta$ instead of $r$. 

\begin{definition}[Sets of tuples of Bargmann invariants]
    Let $\mathcal{W}$ be any finite subset of $\mathbb{N}^*$. Define $V := \bigcup_{\pmb{w} \in \mathcal{W}} \{w \mid w \in \pmb{w}\}$ the set of all natural numbers appearing in the sequences of $\mathcal{W}$. We denote $\mathfrak{Q}(\mathcal{W})$ the set of all quantum realizable $\Delta: \mathcal{W} \to \mathbb{C}$ as by Def.~\ref{def: Bargmann invariant quantum realizability}. In other words, 
    \begin{equation}
        \mathfrak{Q}(\mathcal{W}) := \{\Delta \in \mathbb{C}^{\mathcal{W}} \mid \exists \mathcal{H},\,\exists \varrho:V\to\mathcal{D}(\mathcal{H}) \text{ such that }\Delta = \Delta(\varrho)\}
    \end{equation}
    where $\varrho$ is defined in~\cref{def: Bargmann invariant quantum realizability}.  If   $\mathcal{W}= \{(1,2,\dots,n)\}$ we write $\mathfrak{Q}({\mathcal{W}})\equiv \mathfrak{B}_n$ and refer to it as the set of $n$-order Bargmann invariants. 
\end{definition}

\myindent Note from the definition above that if we consider $\mathcal{W}_G := \{(i,j) \mid \{i,j\} \in E(G)\}/\sim_{\mathrm{cyc}}$ we have that $\mathfrak{Q}(G) = \mathfrak{Q}(\mathcal{W}_G)$, hence showing that the definition above generalizes the definition of $\mathfrak{Q}(G)$. The sets $\mathfrak{B}_n$  can be written as
\begin{equation}\label{def: sets of all Bargmann invariants m}
    \mathfrak{B}_n := \{\Delta \in \mathbb{C} \mid \exists \pmb{\rho} \in \mathcal{D}(\mathcal{H})^n \text{ such that }\Delta = \Delta_n(\pmb{\rho})\text{ for some }\mathcal{H}\},
\end{equation}
where $\Delta_n(\pmb{\rho}) = \mathrm{Tr}(\rho_1 \cdots \rho_n)$ (cf.~\cref{definition: Bargmann invariants}).  These sets are of particular relevance to us since they are simple enough to be analytically characterized, and also to witness relational coherence and relational imaginarity. To gain some intuition and familiarity with the notation, we now consider two simple examples. 

\begin{example}[Tuples of Bargmann invariants relevant to spectrum estimation]
    Take $\mathcal{W}_d := \{(1),(1,1),(1,1,1),\dots,(\underbrace{1,1,\dots,1}_{d\text{ times}})\}.$ In this case we have that $\pmb{\Delta} \in \mathfrak{Q}({\mathcal{W}_d}) \subseteq [0,1]^d$ if, and only if, there exists some quantum state $\rho$ with respect to some Hilbert space $\mathcal{H}$ for which $$\pmb{\Delta} = (\text{Tr}(\rho^k))_{k=1}^d.$$
    These tuples of Bargmann invariants can be used to estimate the spectrum of a quantum state $\rho$ (See~\citep{wagner2024quantumcircuits} and references therein). 
\end{example}

\begin{example}[Tuples of Bargmann invariants encoding information of a distance between two states]\label{example: distance between two states}
    Take $\mathcal{W} := \{(1,1),(1,2),(2,2)\}$.  We have then that $\pmb{\Delta} \in \mathfrak{Q}(\mathcal{W}) \subseteq [0,1]^3$ iff $$\pmb{\Delta} = (\text{Tr}(\rho^2),\text{Tr}(\rho\sigma),\text{Tr}(\sigma^2)),$$ for some pair of quantum states $\rho,\sigma \in \mathcal{D}(\mathcal{H})$. Any such tuple provides complete information about the \emph{distance} $d(\rho,\sigma) := \Vert \rho - \sigma\Vert_2$ between two states since we have that the functional $d_2: [0,1]^3 \to \mathbb{R}$ given by 
    \begin{equation}
        d_2(\pmb{\Delta}) = \Delta_{1,1}+\Delta_{2,2}-2\Delta_{1,2}
    \end{equation}
    is such that, for any quantum realization $\pmb{\Delta} = \pmb{\Delta}(\rho,\sigma)$ we have 
    \begin{equation*}
        d_2(\pmb{\Delta}(\rho,\sigma)) = \text{Tr}(\rho^2)+\text{Tr}(\sigma^2)-2\text{Tr}(\rho\sigma) = \Vert \rho-\sigma \Vert_2^2
    \end{equation*}
    where $\Vert \cdot \Vert_2$ is the Schatten-2 norm~\citep{quek2024multivariatetrace}.
\end{example}

\subsubsection{On the relevance of event graphs}

\myindent Here we pause and question if it is relevant to consider event graphs in general, or if it is sufficient to always consider complete graphs $G=K_n$, and simply specify projections $\pi$ onto some finite set of coordinates of interest.~\footnote{We discuss projections in~\cref{sec: convex polytopes}. To give a simple example, if we define $\pi_{12,13}$ as the projection $(r_{12},r_{13},r_{23}) \mapsto (r_{12},r_{13})$ we have that $\pi_{12,13}(\mathfrak{Q}(C_3))$ correspond to all the points $(r_{12},r_{13})$ such that there is \emph{some} point $(r_{12},r_{13},r) \in \mathfrak{Q}(C_3)$, for some value of $r$.} This is because, essentially, for a given quantum realization $\varrho: V(G) \to \mathcal{D}(\mathcal{H})$ there are well-defined two-state overlaps between any two nodes, \emph{even} for those that are not adjacent in the event graph. Hence the question: do we really need event graphs? 

\myindent From our current understanding, it does not seem that the event graphs are \emph{essential}, but we can argue in favor of the fact that having a generic description of such graphs, and the associated resources, is useful. For instance, we can note the following: 

\begin{lemma}\label{lemma: projectors and event graph correlations}
    Let $G$ be any event graph. Then $\mathfrak{Q}(G) = \pi_G(\mathfrak{Q}(K_n))$, for any $n\geq |V(G)|$, for some projection $\pi_G$ onto coordinates $r_e$ such that $e \in E(G)$. The same applies to $\mathfrak{C}(G) = \pi_G(\mathfrak{C}(K_n))$.
\end{lemma}

\begin{proof}
    This lemma follows trivially from the definition of projections. In this case, by definition we have that   
    \begin{equation}
        \pi_G(\mathfrak{Q}(K_n)) := \left\{ (r_e)_{e \in E(G)} \mid \exists (\tilde{r}_e)_{e \in E(K_n)\setminus E(G)}, \text{ such that } (r,\tilde{r}) \in \mathfrak{Q}(K_n) \right\},
    \end{equation}
    from which we can  conclude that the lemma holds. 
\end{proof}

\myindent Therefore, the event graphs are a meaningful way of avoiding the \emph{top-down} approach of always finding a characterization of $\mathfrak{C}(K_n)$---or of $\mathfrak{Q}(K_n)$---to only then find the characterization of $\mathfrak{C}(G)$---or $\mathfrak{Q}(G)$---by means of a specific choice of projection. Using event graphs we can bypass this and directly analyze the sets $\mathfrak{C}(G)$---or $\mathfrak{Q}(G)$. This has algorithmic, experimental, and theoretical relevance since in most cases the sets related to the complete graph are significantly more convoluted to characterize.

\myindent Another point can be made by appealing to the fact that event graphs help define a concrete \emph{scenario of study}. One can interpret the tuples $\pmb{r}$ as those accessible or relevant for investigation within a given graph $G$—that is, they correspond to the types of invariants that one has access to (or chooses to focus on) in that particular setting. In this case, arguably, it is wasteful to analyse and completely characterize $\mathfrak{C}(K_n),\mathfrak{Q}(K_n)$, only so that later we trace out all the information that we are not interested in, or that is irrelevant to the purposes of the analysis we pursue.

\subsection{Convexity of sets of tuples of Bargmann invariants}

\myindent We now proceed to investigate a simple property of the geometry of the set of all possible tuples of Bargmann invariants: \emph{convexity}. Just from their definition, we see that the sets of Bargmann invariants consider lists of multilinear functionals,  
\begin{align*}
\Delta_{n}\left(\sum_{i_1}\alpha_{i_1}\rho_{i_1},\ldots,\sum_{i_n}\alpha_{i_n}\rho_{i_n}\right) &= \text{Tr}\left(\sum_{i_1}\alpha_{i_1}\rho_{i_1}\cdots\sum_{i_n}\alpha_{i_n}\rho_{i_n}\right) \\&= \sum_{i_1,\ldots,i_n}\alpha_{i_1}\cdots\alpha_{i_n}\text{Tr}(\rho_{i_1}\cdots \rho_{i_n}) \\
&= \sum_{i_1,\ldots,i_n}\alpha_{i_1}\cdots\alpha_{i_n}\Delta_n(\rho_{i_1},\ldots,\rho_{i_n}),
\end{align*}
and that tuples $\pmb{\Delta}(\pmb{\rho})$ may have the same state appearing to different lists. For example, in the case of $G=C_3$ we have that quantum realizable edge weightings $r: E(C_3) \to [0,1]$ are of the form 
\begin{equation*}
    \pmb{r}(\pmb{\rho}) = (\text{Tr}(\rho_1\rho_2),\text{Tr}(\rho_1\rho_3),\text{Tr}(\rho_2\rho_3)).
\end{equation*}
Above, the same state $\rho_i$ appears in at least two entries but never in all of them. Therefore, from the multilinearity of each entry one immediately has that given two triples of states $\pmb{\rho},\pmb{\sigma}$, and $p \in (0,1)$, one generally finds  that (by applying the rule of multilinarity of the trace to each component of these vectors) $$\pmb{r}(p\pmb{\rho}+(1-p)\pmb{\sigma}) \neq p\pmb{r}(\pmb{\rho})+(1-p)\pmb{r}(\pmb{\sigma}).$$ This suggests that, in general, the sets $\mathfrak{Q}(G)$, are \emph{non-convex} sets. Moreover, we have seen in Chapter~\ref{chapter: quantum coherence} that convex combinations of sets of states that are set incoherent can become set coherent, which also suggests that this lack of convexity could, in principle, translate to the non-convexity of $\mathfrak{Q}(G)|_{\mathrm{inc}}$, and ultimately of $\mathfrak{Q}(G)$ as well. However, at least for some choices of word sets $\mathcal{W}$, it turns out that this intuition is \emph{misleading} and the sets $\mathfrak{Q}(\mathcal{W})$ are indeed convex.

\begin{theorem}[Convexity of sets of tuples of Bargmann invariants]\label{theorem:convexity_tuples_of_order_m}
    Let $\mathcal{W}$ be a finite subset of $\mathbb{N}^*$ satisfying that: 
    \begin{enumerate}
    \item[(i)] Every word $\pmb{w} \in \mathcal{W}$ has at least two different labels, i.e., there exists $w$ and $w'$ elements of $\pmb{w}$ such that $w \neq w'$. 
    \item[(ii)] Every word $\pmb{w} \in \mathcal{W}$ has the same cardinality $m \in \mathbb{N}$.
    \end{enumerate}
    Then, the set $\mathfrak{Q}(\mathcal{W})$ is convex.
\end{theorem}

\begin{proof}
    For any $p \in (0,1)$, let $a = p^{1/m}$, $b = (1-p)^{1/m}$, and $c = 1-a-b$. Let $\pmb{\Delta}^{(1)},\pmb{\Delta}^{(2)} \in \mathfrak{Q}(\mathcal{W})$ be any two quantum realizable tuples with respect to the vertex labelings $\varrho^{(k)}:V \to \mathcal{D}(\mathcal{H}^{(k)})$ denoted $\varrho^{(k)}(i) \equiv \rho_i^{(k)}$, i.e.,  $\pmb{\Delta}^{(k)} = \pmb{\Delta}^{(k)}(\pmb{\rho}^{(k)})$ for $k=1,2$. As described in~\cref{def: Bargmann invariant quantum realizability}, we take $V := \bigcup_{\pmb{w} \in \mathcal W} \{w \mid w \in \pmb{w}\}$ the set of all natural numbers appearing in the sequences of $\mathcal{W}$.
    
    \myindent We construct a new function $\varrho : V \to \mathcal{D}(\mathcal{H})$ denoted  $\varrho(i) = \rho_i$, where $\mathcal{H}=\mathcal{H}^{(1)} \oplus \mathcal{H}^{(2)}\oplus \mathbb{C}^{|V|}$, from the functions $\varrho^{(1)} : V \to \mathcal{D}(\mathcal{H}^{(1)})$ and $\varrho^{(2)}: V \to \mathcal{D}(\mathcal{H}^{(2)})$ via 
    \begin{equation*}
        \varrho(i) := a\, \varrho^{(1)}(i)\oplus b\,\varrho^{(2)}(i) \oplus c\, \vert i\rangle \langle i \vert. 
    \end{equation*}
    Above, we have that for each  $i \in V$ we denote $\vert i\rangle \in \mathbb{D}_{|V|}$ an element of the canonical basis for $\mathbb{C}^{|V|}$. This formally defines a new labeling where quantum states are associated to some higher-dimensional Hilbert space. Note that by construction $\varrho(i)$ are positive semidefinite matrices for all $i$ because $\varrho^{(k)}(i)$ are, and the direct sum preserves positivity. Moreover, we have that $$\text{Tr}(\varrho(i)) = a\text{Tr}(\varrho^{(1)}(i))+b\text{Tr}(\varrho^{(2)}(i))+c\langle i|i\rangle = a+b+c=1,$$ which implies, by definition, that $\varrho(i)$ is a valid density matrix for all $i \in V$. 
    
    \myindent Now we note that  for every word $\pmb{w} = (w_1,w_2,\dots,w_m) \in \mathcal{W}$ we have that 
    \begin{align*}
        \Delta_{\pmb{w}} &= \text{Tr}\left(\prod_{i=1}^m \varrho(w_i)\right) \\
        &= \text{Tr}\left(\prod_{i=1}^m (a\, \rho^{(1)}_{w_i} \oplus b\, \rho^{(2)}_{w_i} \oplus c\, \vert {w_i}\rangle \langle {w_i} \vert ) \right)  \\ &=\text{Tr}\left(a^m\,\prod_{i=1}^m \rho_{w_i}^{(1)} \oplus b^m\, \prod_{i=1}^m \rho_{w_i}^{(2)} \oplus c^m\, \prod_{i=1}^m \vert {w_i}\rangle \langle {w_i} \vert \right)\\
        &=a^m \text{Tr}\left(\prod_{i=1}^m \rho_{w_i}^{(1)}\right)+b^m\text{Tr}\left(\prod_{i=1}^m \rho_{w_i}^{(2)}\right)+c^m \cdot 0 \\
        &= p \Delta^{(1)}_{\pmb{w}} + (1-p)\Delta^{(2)}_{\pmb{w}}.
    \end{align*}
    
    \myindent Above we have used that, by construction, $\text{Tr}\left(\prod_{i=1}^m \vert w_i\rangle \langle w_i| \right) = 0$, since, by assumption, there exists at least two vertices $w_i \neq w_j$ in $\pmb{w}$, implying that there exists one inner-product in this trace for which $\langle w_i|w_j\rangle = 0$ since $\vert w_i\rangle$ and $\vert w_j\rangle$ are two different states of the canonical basis $\mathbb{D}_{|V|}$. We have also used the known relations $$\text{Tr}(a M_1 \oplus b M_2) = \text{Tr}(a M_1)+\text{Tr}(b M_2) = a \text{Tr}(M_1)+ b \text{Tr}(M_2)$$ for any $a,b \in \mathbb{R}$, and that $$(M_1\oplus M_2 \oplus M_3)(M_4\oplus M_5 \oplus M_6) = M_1M_4 \oplus M_2M_5 \oplus M_3M_6.$$  This shows that $p \pmb{\Delta}^{(1)} + (1-p)\pmb{\Delta}^{(2)} = \pmb{\Delta}(\pmb{\rho}) \in \mathfrak{Q}(\mathcal{W}),$ and we conclude the proof. 
\end{proof}

\myindent The above theorem is somewhat surprising, and is one of the main outputs of this thesis. It has some important corollaries:

\begin{corollary}\label{corollary:Bm_is_convex}
    For every $n\geq 1$ the set $\mathfrak{B}_n$ is a convex subset of $\mathbb{C}$.
\end{corollary}

\myindent This corollary shows that the set of all $n$-order Bargmann invariants is a convex set. Also, and more directly, we have that the sets $\mathfrak{Q}(G)$ are convex, for all the possible event graphs $G$.

\begin{corollary}\label{corollary: quantum realizations of edge weightings are convex}
    Let $G$ be any event graph. Then, the set $\mathfrak{Q}(G)$ is a convex subset of $[0,1]^{|E(G)|}$. 
\end{corollary}

\myindent Another important corollary follows from the fact that to prove~\cref{theorem:convexity_tuples_of_order_m}, if the two vertex labelings $\varrho^{(1)}: V(G) \to \mathcal{D}(\mathcal{H}^{(1)})$ and $\varrho^{(2)}: V(G) \to \mathcal{D}(\mathcal{H}^{(1)})$ are such that both $\varrho^{(k)}(V(G))$ are set-imaginarity free, for $k = 1, 2$, the image of the constructed vertex labeling $\varrho(V(G))$ is \emph{also} set-imaginarity free. 

\begin{corollary}\label{corollary:quantum_realizable_incoherent_is_closed}
    Let $\mathcal{W}$ be a set satisfying the same constraints as those for Theorem~\ref{theorem:convexity_tuples_of_order_m}. Then, $\mathfrak{Q}(\mathcal{W})|_{\mathrm{real}}$ is a convex subset of $\mathfrak{Q}(\mathcal{W})$.
\end{corollary}

\subsection{Candidate Gram matrices and their relation to sets of Bargmann invariants}\label{subsec: candidates of Gram matrices}

\myindent Recall in Chapter~\ref{chapter: Bargmann invariants} that we have studied Gram matrices $G_{\pmb{\Psi}}$ of a given tuple of vector states $\pmb{\Psi}$, and their relation with unitary-invariance problems. There, we have also described how one could use frame graphs to construct Gram matrices  $G_{\pmb{\psi}}^{\tau}$, defined only in terms of Bargmann invariants. Clearly, since we are now viewing Bargmann invariants from the perspective of quantum realizability there is a natural sense in which we can talk about such matrices as quantum realizable. 

\myindent We can start with a simple clarifying example. Take any function $f(\pmb{r})$ defined over edge-weightings $r \in [0,1]^{E(G)}$. Given that we have a notion of quantum realizability for such weightings, we \emph{also} have a notion of quantum realizability of the function $f$ through the values of $\pmb{r}$. For example, if we take $f:[0,1]^{E(C_3)} \to \mathbb{R}$ defined as 
\begin{equation*}
    f(\pmb{r}) = r_{12}^2+r_{13}^2+r_{23}^2,
\end{equation*}
we can define a notion of `quantum realizability for $f$' by investigating the values that $f$ takes on the set $\mathfrak{Q}(C_3)$. Or equivalently, the above equation  proposes to investigate the question: what are all the possible values that $f$ can take if there exists some tuple of quantum states $\pmb{\rho} = (\rho_1,\rho_2,\rho_3)$ for which 
\begin{equation*}
    f(\pmb{r}(\pmb{\rho})) = \text{Tr}(\rho_1\rho_2)^2+\text{Tr}(\rho_1\rho_3)^2+\text{Tr}(\rho_2\rho_3)^2.
\end{equation*}
Clearly, our choice of $f$ was just an example. Note also that we might be interested in defining functions $g: \mathcal{D}(\mathcal{H})^{V(G)} \to \mathbb{R}$, acting on the states, from the relation $g(\pmb{\rho}) = f(\pmb{r}(\pmb{\rho}))$. We make this distinction whenever we want to emphasize that the two functions are \emph{different}. In~\cref{sec: relational coherence} we consider notions of quantum realizability for linear functionals $f:[0,1]^{E(G)} \to \mathbb{R}$  defining event-graph inequalities discussed in Chapter~\ref{chapter: event_graph_approach}, such as $h_n:[0,1]^{E(K_n)} \to \mathbb{R}$ from Eq.~\eqref{eq:hn_recursively} or $c_n: [0,1]^{E(C_n)} \to \mathbb{R}$ from Eq.~\eqref{eq:cycle_inequalities}.

\myindent We now consider a situation that is more relevant to us, but that is slightly more intricate. Let $G = C_3$ and $V(C_3) := \{1,2,3\}$. If we choose 
\begin{equation}\label{eq: W_123}
\mathcal{W}_{123} := \{(1,2),(1,3),(2,3),(1,2,3)\}
\end{equation}
we can denote the following function $H: [0,1]^3 \times \mathbb{C}\to \mathrm{Mat}_3(\mathbb{C})$,~\footnote{This is our notation for the set of $3 \times 3$ complex matrices. See also~\cref{app: basic algebra}.} defined by 

\begin{equation}\label{eq: candidate matrix}
    H(r_{12},r_{13},r_{23},\Delta_{123}) := \left( \begin{matrix}
        1 & \sqrt{r_{12}} & \sqrt{r_{13}} \\
        \sqrt{r_{12}} & 1 & \sqrt{r_{23}}e^{i\phi_{123}}\\
        \sqrt{r_{13}} & \sqrt{r_{23}}e^{-i\phi_{123}} & 1
    \end{matrix}\right), 
\end{equation}
where we have that $e^{i\phi_{123}} = \Delta_{123}/|\Delta_{123}|$ if $\Delta_{123} \neq 0$ and $e^{i\phi_{123}} = 1$ if $\Delta_{123} = 0$.  We  refer to $H(\pmb{\Delta})$ as a \emph{candidate Gram matrix}, or simply as a \emph{candidate}. This is because the matrix above is \emph{not} a Gram matrix, i.e., positive semidefinite, for all possible values of $\pmb{\Delta}$. For example, $\text{det}[H(1,1,0,0)] = -1$. Using the definition of $\mathcal{W}_{123}$ as from Eq.~\eqref{eq: W_123} we can show the following lemma. 

\begin{lemma}[Adapted from~\cite{fernandes2024unitary}]\label{lemma: nonzero overlaps candidates determine pure state realizability}
    Let $H$ be as defined in Eq.~\eqref{eq: candidate matrix}. A tuple $\pmb{\Delta} = (r_{12},r_{13},r_{23},\Delta_{123}) \in \mathbb{C}^{\mathcal{W}_{123}}$ where all $r_{ij}$ coordinates are non-zero is quantum realizable by pure states if, and only if, the candidate matrix $H(\pmb{\Delta})$ is positive semidefinite. In other words, if $r_{12},r_{13},r_{23} \neq 0$, then $\pmb{\Delta} \in \mathfrak{Q}(\mathcal{W}_{123})|_{\mathrm{pure}}$ iff $H(\pmb{\Delta})$ is positive semidefinite.
\end{lemma}

\begin{proof}
Note that if $\pmb{\Delta}$ is quantum realizable we have that $\Delta_{123} \neq 0 \iff r_{ij}\neq 0\,\, \forall i,j$ and that moreover  \begin{equation}|\Delta_{123}(\pmb{\psi})| = |\langle \psi_1|\psi_2\rangle | |\langle \psi_2|\psi_3\rangle | |\langle \psi_3|\psi_1\rangle | =  \sqrt{r_{12}(\pmb{\psi})r_{13}(\pmb{\psi})r_{23}(\pmb{\psi})}.\end{equation} Therefore, the matrix  $H(\pmb{\Delta})$ is positive semidefinite by construction since  $H(\pmb{\Delta}) = G_{\pmb{\psi}}^\tau$ given by Eq.~\eqref{eq: Gram matrix of invariants phase} which is a Gram matrix as discussed in Chapter~\ref{chapter: Bargmann invariants}, and every Gram matrix is positive semidefinite. 
    
\myindent Suppose now that $H(\pmb{\Delta})$ is positive semidefinite. The Cholesky decomposition of $H(\pmb{\Delta})$ yields a Gram matrix representation, so that  $H(\pmb{\Delta}) = G_{\pmb{\Psi}}$ for some tuple of vectors $\pmb{\Psi}$. We note that $\pmb{\Psi} \in \mathcal{H}_1^3$---recall that we denote $\mathcal{H}_1 = \{\vert \psi \rangle \in \mathcal{H} \mid \langle \psi|\psi \rangle = 1\}$ the subset of normalized vector states---since $(H(\pmb{\Delta}))_{ii}=1$ for $i=1,2,3$. Moreover, we know that $(G_{\pmb{\Psi}})_{ij} = \langle \psi_i|\psi_j\rangle = (H(\pmb{\Delta}))_{ij} \neq 0$ for all $(i,j) \neq 0$. Hence, the frame graph $F_{\pmb{\Psi}} \simeq C_3$, and therefore, there exists some diagonal unitary $U$ such that $G_{\pmb{\Psi}} = U G_{\pmb{\psi}}^{\tau} U^\dagger $ where $G_{\pmb{\psi}}^{\tau}$ is given by Eq.~\eqref{eq: Gram matrix of invariants phase}, due to the construction we have detailed in Sec.~\ref{sec: gauge and unitary invariance}. This implies that $$H(\pmb{\Delta}) = G_{\pmb{\Psi}} = U G_{\pmb{\psi}}^{\tau} U^\dagger,$$ for some tuple of states $\pmb{\psi}$ and $U = \text{diag}(e^{i\theta_1},e^{i\theta_2},e^{i\theta_3})$  some diagonal unitary, with $\theta_1,\theta_2,\theta_3 \in [0,2\pi)$. We now show that $U$ is, up to a phase, the identity. To see this, we simply use the relation we have found  $H(\pmb{\Delta}) = U G_{\pmb{\psi}}^{\tau} U^\dagger$, implying that
    \begin{align*}
        H(\pmb{\Delta}) &= \left(\begin{matrix}
        e^{i\theta_1} & 0 & 0\\
        0 & e^{i\theta_2} & 0 \\
        0 & 0 & e^{i\theta_3}
        \end{matrix}\right)\left(\begin{matrix}
        1& a & b\\
        a & 1 & ce^{i\phi}\\
        b & ce^{-i\phi} & 1
        \end{matrix}\right)\left(\begin{matrix}
        e^{-i\theta_1} & 0 & 0\\
        0 & e^{-i\theta_2} & 0 \\
        0 & 0 & e^{-i\theta_3}
        \end{matrix}\right)\\
        &= \left(\begin{matrix}
        e^{i\theta_1}& ae^{i\theta_1} & be^{i\theta_1}\\
        ae^{i\theta_2} & e^{i\theta_2} & ce^{i\phi}e^{i\theta_2}\\
        be^{i\theta_3} & ce^{-i\phi}e^{i\theta_3} & e^{i\theta_3}
        \end{matrix}\right)\left(\begin{matrix}
        e^{-i\theta_1} & 0 & 0\\
        0 & e^{-i\theta_2} & 0 \\
        0 & 0 & e^{-i\theta_3}
        \end{matrix}\right) \\
        &= \left(\begin{matrix}
        1& ae^{i(\theta_1-\theta_2)} & be^{i(\theta_1-\theta_3)}\\
        ae^{i(\theta_2-\theta_1)} & 1 & ce^{i\phi}e^{i(\theta_2-\theta_3)}\\
        be^{i(\theta_3-\theta_1)} & ce^{-i\phi}e^{i(\theta_3-\theta_2)} & 1
        \end{matrix}\right),
    \end{align*}
    where $$a=\text{Tr}(\psi_1\psi_2),b=\text{Tr}(\psi_1\psi_3),c=\text{Tr}(\psi_2\psi_3) \in \mathbb{R}_{>0}$$ and $e^{i\phi} = \text{Tr}(\psi_1\psi_2\psi_3)/|\text{Tr}(\psi_1\psi_2\psi_3)|$, all guaranteed to exist by $G_{\pmb{\psi}}^\tau$. But since by construction we have that $(H(\pmb{\Delta}))_{12},(H(\pmb{\Delta}))_{13} > 0$, then $ae^{i(\theta_1-\theta_2)}, be^{i(\theta_1-\theta_3)} > 0$, implying that we must have that $e^{i(\theta_1-\theta_2)} = 1$, and hence $\theta_1 =\theta_{2}(\text{mod } 2\pi)$. Similarly,  $\theta_2 = \theta_3(\text{mod } 2\pi)$. Therefore, the only solution for $\theta_1,\theta_2,\theta_3 \in [0,2\pi)$ is to choose $\theta_1=\theta_2=\theta_3$. From which we conclude that $U=e^{i\theta}\mathbb{1}$, and that $\pmb{\Delta} = \pmb{\Delta}(\pmb{\psi})$ is quantum realizable by pure states (given for example by the Cholesky decomposition) since $H(\pmb{\Delta}) = G_{\pmb{\psi}}^\tau$.
\end{proof}

\myindent We now remark that restricting to $r_{12},r_{13},r_{23} \neq 0$ was important. This is because the mere fact that the candidate $H(\pmb{\Delta})$ is positive semidefinite for some choice $\pmb{\Delta}$ \emph{does not} imply that $\pmb{\Delta}$ is quantum realizable by pure states. For instance, the tuple $\pmb{\Delta} = (0,\sfrac{1}{4},\sfrac{1}{4},1)$ is \emph{not} quantum realizable by pure states since it is impossible that $|\langle \psi_1|\psi_2\rangle |^2 = 0$ while $\Delta_{123} = 1$. Nevertheless, with $e^{i\phi_{123}} = 1/|1| = 1$, then  $$H(0,\sfrac{1}{4},\sfrac{1}{4},1) = \left(\begin{matrix}
    1 & 0 & \sqrt{\frac{1}{4}}\\ 0 & 1 & \sqrt{\frac{1}{4}}\\
    \sqrt{\frac{1}{4}} & \sqrt{\frac{1}{4}} & 1
\end{matrix}\right)$$ is positive semidefinite, with $\mathrm{det}(H(\pmb{\Delta})) = 1/2$ and spectrum $\{\frac{2+\sqrt{2}}{2},1,\frac{2-\sqrt{2}}{2}\}$. 

\myindent The above is not surprising since in order to write the candidate matrix $H(\pmb{\Delta})$ we have assumed a form given by $G_{\pmb{\psi}}^\tau$ that was constructed \emph{assuming the validity of a specific frame graph} isomorphic to $C_3$ and a certain spanning tree $\tau$ of $C_3$. For the case when there are some overlaps equal to zero, the frame graph considered does not hold anymore (since one has a different frame graph non-isomorphic to $C_3$). Nevertheless, there is a simple way to consider the same candidate matrix and reach conclusions regarding quantum realizability by pure states for all possible tuples $\pmb{\Delta}$. 

\begin{theorem}[Adapted from~\cite{fernandes2024unitary}]\label{theorem: candidates determine pure state realizability}
    Let $H$ be as defined in Eq.~\eqref{eq: candidate matrix}. Then, $\pmb{\Delta} \in \mathfrak{Q}(\mathcal{W}_{123})|_{\mathrm{pure}}$ if, and only if, $|\Delta_{123}| = \sqrt{r_{12}r_{13}r_{23}}$ and $H(\pmb{\Delta})$ is positive semidefinite.
\end{theorem}

\begin{proof}
        The case for when $r_{12},r_{13},r_{23} \neq 0$ follows from Lemma~\ref{lemma: nonzero overlaps candidates determine pure state realizability}. We then assume  one can have some of the values $r_{i,j} = 0$. Suppose that $\pmb{\Delta} = \pmb{\Delta}(\pmb{\psi})$ is quantum realizable, hence $\langle \psi_1|\psi_2\rangle \langle \psi_2|\psi_3\rangle \langle \psi_3|\psi_1\rangle = 0$. From Sylvester’s criterion (cf.~\cref{theorem: Sylvesters criteria} in~\cref{app: basic algebra}), we have that the only non-trivial condition for positive semidefiniteness is the condition
    \begin{align*}
        \text{det}(H(\pmb{\Delta})) &= 1-r_{12}-r_{13}-r_{23} +2\sqrt{r_{12}r_{13}r_{23}} \geq 0.
    \end{align*}
    Since by~\cref{corollary: boundary of triplets of overlaps} this inequality is always satisfied by pure quantum states, we conclude this direction of the proof. 
    Let us now assume that there exists at least one $r_e = 0$, while  $H(\pmb{\Delta})$ is a positive semidefinite matrix where $e^{i\phi_{123}} = 1$ since $|\Delta_{123}| = \sqrt{r_{12}r_{13}r_{23}} = 0$. Again, the Cholesky decomposition implies that there exists a tuple $\pmb{\Psi}$ such that $H(\pmb{\Delta}) = G_{\pmb{\Psi}}$, and this equality constraint implies that, for any $e \in E(C_3)$, $$\langle \psi_i|\psi_j\rangle = \sqrt{r_e} = |\langle \psi_i|\psi_j\rangle |,$$ yielding a quantum realization  $\pmb{\Delta} = \pmb{\Delta}(\pmb{\psi})$.  
      
\end{proof}

\myindent We note that~\cite{eggeling2001separability} presented a similar result to this in the form of $\mathrm{det}(H(\pmb{\Delta})) \geq 0$ as above. This theorem leads naturally to the following corollary:

\begin{corollary}[Adapted from~\cite{fernandes2024unitary}]
    Suppose that $H(\pmb{\Delta})$ is positive semidefinite, with $|\Delta_{123}| = \sqrt{r_{12}r_{13}r_{23}}$. Then, if we denote by $\pmb{\Psi}$ the vectors of its Cholesky decomposition, such that $H(\pmb{\Delta}) = G_{\pmb{\Psi}}$, we have that $\pmb{\Delta} = \pmb{\Delta}(\pmb{\psi})$. 
\end{corollary}

\myindent In other words, a Cholesky decomposition (see~\cref{theorem: Cholesky decomposition} in~\cref{app: basic algebra}) of $H(\pmb{\Delta})$ gives a quantum realization for $\pmb{\Delta}$. This is an \emph{incredibly useful} result since we can investigate quantum realizations of $\pmb{\Delta}$ from the Cholesky decompositions of the candidate matrix $H(\pmb{\Delta})$. For example, it is now trivial to see that if $H(\pmb{\Delta})$ is a \emph{real} positive semidefinite matrix, then $\pmb{\Delta} \in \mathfrak{Q}(C_3)|_{\mathrm{real}}$. 

\subsection{On the relation between different sets of Bargmann invariants}\label{subsec: different sets of Bargmann invariants}

\myindent It is natural to imagine that the geometry of the sets $\mathfrak{Q}(G)$ and all its specific subsets that we consider (e.g. $\mathfrak{Q}(G)|_{\mathrm{real}}$ or $\mathfrak{Q}(G)|_{\mathrm{inc}}$, etc.) is hard to characterize and compare. Still, there are some facts that we can readily show.

\begin{proposition}[Adapted from~\cite{fernandes2024unitary}]\label{proposition: pure and real are the same}
    Every edge weighting $r:E(C_3)\to [0,1]$ realizable by pure quantum states is also realizable by imaginarity-free quantum states. Succinctly, $\mathfrak{Q}(C_3)|_{\mathrm{pure}} \subseteq  \mathfrak{Q}(C_3)|_{\mathrm{real}}$. 
\end{proposition}

\begin{proof}
We first show this to hold in case that there is some $r_{ij} =0$, and then we consider the case where all $r_{ij} \neq 0$.

\myindent Denote $\pmb{\psi} = (\vert \psi_i\rangle \langle \psi_i \vert)_{i \in V(G)}$ a quantum realization for which $\pmb{r} = \pmb{r}(\pmb{\psi})$. Then, if there exists one $r_{ij}(\pmb{\psi}) = 0$ this implies that 
\begin{equation*}
    r_{ij}(\pmb{\psi}) = |\langle \psi_i|\psi_j\rangle |^2 =  0.
\end{equation*}
Suppose, without loss of generality, that we have $r_{23} = 0$. Then, we consider any basis $\mathbb{A}$ of $\mathcal{H}$ for which $\vert \psi_2\rangle, \vert \psi_3\rangle \in \mathbb{A}$, and write $\vert \psi_1\rangle$ with respect to that basis. In this case, we write $$\vert \psi_1\rangle = \alpha \vert \psi_2\rangle + \beta \vert \psi_3\rangle + \sum_i \gamma_i \vert \phi_i\rangle.$$ Note therefore that $|\alpha|^2+|\beta|^2 \leq 1$. This implies that
\begin{equation*}
    r_{12}(\pmb{\psi}) = |\alpha|^2, r_{13}(\pmb{\psi}) = |\beta|^2, r_{23}(\pmb{\psi}) = 0,
\end{equation*}
Now, choose the \emph{different}  quantum realization $\pmb{\psi}^{\mathbb{R}} := (\vert \psi_1^{\mathbb{R}}\rangle \langle \psi_1^{\mathbb{R}} \vert, \vert \psi_2^{\mathbb{R}}\rangle \langle \psi_2^{\mathbb{R}} \vert, \vert \psi_3^{\mathbb{R}}\rangle \langle \psi_3^{\mathbb{R}} \vert )$ where $$\vert \psi_2^{\mathbb{R}}\rangle = \vert 0\rangle, \,\,\vert \psi_3^{\mathbb{R}}\rangle = \vert 1\rangle, \,\,\vert \psi_1^{\mathbb{R}}\rangle = |\alpha|\vert 0\rangle + |\beta|\vert 1\rangle + \sqrt{(1-|\alpha|^2-|\beta|^2)}\vert 2\rangle. $$ Then, we have that $\pmb{r}(\pmb{\psi}) = \pmb{r}(\pmb{\psi}^{\mathbb{R}})$, showing that $\pmb{r} \in \mathfrak{Q}(C_3)|_{\mathrm{real}} \cap \mathfrak{Q}(C_3)|_{\mathrm{pure}}$.

\myindent To conclude the proof, we use Lemma~\ref{lemma: nonzero overlaps candidates determine pure state realizability}. Suppose that $\pmb{r} \in \mathfrak{Q}(C_3)|_{\mathrm{pure}}$ with $r_{12},r_{13},r_{23} \neq 0$. Consider the candidate matrix $H(\pmb{r},1)$, where $e^{i\phi_{123}} = 1$. From  Lemma~\ref{lemma: nonzero overlaps candidates determine pure state realizability}, if  $H(\pmb{r},1)$ is positive semidefinite. This implies that there exists some $\pmb{\psi}^{\mathbb{R}}$ such that $\pmb{r} = \pmb{r}(\pmb{\psi}^{\mathbb{R}})$ given by a Cholesky decomposition of $H(\pmb{r},1)$. All the states $\pmb{\psi}^{\mathbb{R}}$ are real with respect to some basis since $(H(\pmb{r},1))_{ij}$ is a real positive semidefinite matrix. 

\myindent That $H(\pmb{r},1)$ is positive semidefinite for any $\pmb{r}\in \mathfrak{Q}(C_3)|_{\mathrm{pure}}$ follows from the Sylvester's criterion and from~\cref{corollary: boundary of triplets of overlaps}. Hence, we conclude that $\pmb{r} \in \mathfrak{Q}(C_3)|_{\mathrm{real}}$. This concludes the proof. 
\end{proof}

\myindent We now make some conjectures regarding other geometrical properties of the sets of quantum correlations we have been considering. Let $X \subseteq \mathbb{C}^n$ denote any convex and compact set, we recall that we denote by $\partial X$ its \emph{boundary} and by $\text{ext}(X)$ its set of \emph{extremal points}. Our first conjecture is that the boundary of the set of quantum realizable edge-weightings, for any event graph $G$, is given by edge-weightings realizable by pure states.

\begin{conjecture}
     Let $G$ be any event graph. Every point in the boundary of $\mathfrak{Q}(G)$ is quantum realizable by pure states only, i.e., $\partial \mathfrak{Q}(G) \subseteq \mathfrak{Q}(G)|_{\mathrm{pure}}$.  
\end{conjecture}

\myindent Even if the conjecture above holds, it would be interesting to find out an example of $\pmb{r} \in \mathfrak{Q}(G)$ such that $\pmb{r} = \pmb{r}(\pmb{\rho})$ and at least one state $\rho_i$ of $\pmb{\rho}$ is \emph{not} pure. We also conjecture that the following holds.

\begin{conjecture}
    Let $G$ be any event graph. Then, $\mathfrak{Q}(G)|_{\mathrm{pure}}$ is a convex set. 
\end{conjecture}

\myindent A closely related conjecture to the one above is the following.

\begin{conjecture}
    Let $G$ be any event graph. Then, $\mathfrak{Q}(G) = \mathrm{ConvHull}(\mathfrak{Q}(G)|_{\mathrm{pure}})$, in other words, all quantum realizable edge-weightings are given by the convex hull of pure state realizable edge weightings. 
\end{conjecture}

\myindent Ultimately, all these conjectures are made since we believe that, for any $G$, it holds that $\mathfrak{Q}(G)|_{\mathrm{pure}} = \mathfrak{Q}(G)$ due to a few partial results obtained in~\cite{wagner2024quantumcircuits} and from several important results for the set of third-order Bargmann invariants $\mathfrak{B}_3$ (recall the definition of these sets from Eq.~\eqref{def: sets of all Bargmann invariants m}). For this case  $\mathfrak{B}_3|_{\mathrm{pure}} = \mathfrak{B}_3 = \text{ConvHull}(\mathfrak{B}_3^{(2)})$ as we show below. First, we mention the following theorem.

\begin{theorem}[Adapted from~\cite{fernandes2024unitary}.]\label{theorem: all third-order Bargmann invariants}
    Any complex number $\Delta = |\Delta|e^{i\phi_\Delta}$ is such that $\Delta \in \mathfrak{B}_3|_{\mathrm{pure}}$ if, and only if 
    \begin{equation}\label{eq: boundary of B_3}
        1-3|\Delta|^{\frac{2}{3}}+2|\Delta|\cos(\phi_\Delta) \geq 0.
    \end{equation}
\end{theorem}

\myindent We do not provide a proof of this Theorem. A proof of it can be found in~\cite{fernandes2024unitary}. The essence of the proof is to use Theorem~\ref{theorem: candidates determine pure state realizability}, and optimize over the non-trivial condition $\text{det}(H(\pmb{\Delta})) \geq 0$ that follows from Sylvester's criterion for positive semidefiniteness.  We have that $\Delta \in \partial \mathfrak{B}_3|_{\mathrm{pure}}$ iff the inequality from Eq.~\eqref{eq: boundary of B_3} is saturated. 

\begin{theorem}[Adapted from~\cite{fernandes2024unitary}]\label{theorem: convexity of B_3}
    The set $\mathfrak{B}_3|_{\mathrm{pure}}$ is convex. 
\end{theorem}

\begin{proof}
    From~\cref{theorem: all third-order Bargmann invariants}, if we define  $$f(x,y) := 1+2x-3(x^2+y^2)^{\frac{1}{3}},$$ we have that $\Delta \in \mathfrak{B}_3|_{\mathrm{pure}}$ iff $\Delta = x + iy$ satisfies that $|\Delta|^2 \leq 1$ and $f(x,y) \geq 0$. The curve defined by $f(x,y) = 0$ can be re-written as $$(27/8)y^2 = (x-1)^2(x+1/8),$$ where we are only interested in solutions for which $x \leq 1$ since we require that $|\Delta|^2 \leq 1$. This curve forms a loop defined in the interval $-1/8 \leq x \leq 1$ by the two equations $$y_{\pm}(x) = \pm \sqrt{\frac{8}{27}(x-1)^2(x+1/8)}.$$ Our region of interest, characterized by $f(x,y) \geq 0$ for $-1/8 \leq x\leq 1$, is then the intersection between the epigraph of $y_-$, $$\text{epi}(y_-) := \{(x,y) \in [-1/8,1]\times \mathbb{R} \,\, |\,\, y \geq y_-(x)\}$$ and the hypograph of $y_+$, $$\text{hypo}(y_+):= \{(x,y) \in [-1/8,1]\times \mathbb{R} \,\, |\,\, y \leq y_+(x)\}$$ for all $-1/8 \leq x \leq 1$. In this interval the function  $y_-$ is convex and therefore the function $y_+ = -y_-$ is concave. This implies that both $\text{epi}(y_-)$ and $\text{hypo}(y_+)$ are convex sets~\footnote{Here we have used the fact that a function $f$ is convex iff its epigraph $\text{epi}(f)$ is a convex set, and that a function $f$ is concave iff its hypograph $\text{hypo}(f)$ is a convex set.}. Since the intersection of finitely many convex sets is also a convex set, we get that the region where $f(x,y) \geq 0$ and $|x+iy|^2 \leq 1$ defines a convex region in $\mathbb{R}^2$, from which we conclude that $\mathfrak{B}_3|_{\mathrm{pure}}$ is a convex set. 
\end{proof}

\myindent The important (and immediate) corollary that follows from the theorem above is the following. 

\begin{corollary}[Adapted from~\cite{fernandes2024unitary}]\label{corollary: all Bargmanns are pure Bargmanns}
    $\mathfrak{B}_3 = \mathfrak{B}_3|_{\mathrm{pure}}$.
\end{corollary}

\begin{proof}
    Clearly, $\mathfrak{B}_3|_{\mathrm{pure}} \subseteq \mathfrak{B}_3$. However, take $\text{Tr}(\rho_1\rho_2\rho_3) \in \mathfrak{B}_3$. We know that every quantum state can be written as some convex combination of pure states. From that, and multilinearity of the trace, $$\text{Tr}(\rho_1\rho_2\rho_3) = \sum_{\lambda_1,\lambda_2,\lambda_3} q_{\lambda_1}q_{\lambda_2}q_{\lambda_3} \text{Tr}(\psi_1^{\lambda_1}\psi_2^{\lambda_2}\psi_3^{\lambda_3}),$$ where $q_\lambda$ are convex weights, $\sum_{\lambda_1,\lambda_2,\lambda_3} q_{\lambda_1}q_{\lambda_2}q_{\lambda_3}= 1$ and $$\pmb{\psi}^{(\lambda_1,\lambda_2,\lambda_3)} = (\vert \psi_1^{\lambda_1}\rangle \langle \psi_1^{\lambda_1}|,\vert \psi_2^{\lambda_2}\rangle \langle \psi_2^{\lambda_2}|,\vert \psi_3^{\lambda_3}\rangle \langle \psi_3^{\lambda_3}|) \in \text{ext}(\mathcal{D}(\mathcal{H}))^3$$ for all $(\lambda_1,\lambda_2,\lambda_3)$. This shows that $\mathfrak{B}_3 \subseteq \text{ConvHull}(\mathfrak{B}_3|_{\mathrm{pure}})$. But, since from~\cref{theorem: convexity of B_3} we have that $\mathfrak{B}_3|_{\mathrm{pure}}$ is convex, we conclude that $\mathfrak{B}_3 \subseteq \text{ConvHull}(\mathfrak{B}_3|_{\mathrm{pure}}) = \mathfrak{B}_3|_{\mathrm{pure}}$. This concludes the proof. 
\end{proof}

\begin{figure}[t]
    \centering
    \includegraphics[width=0.7\textwidth]{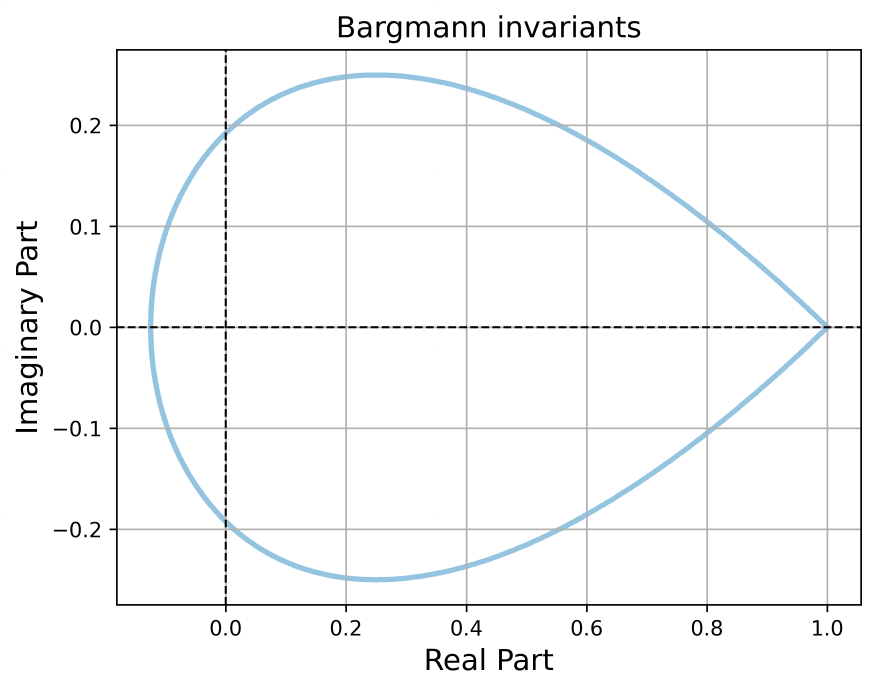}
    \caption{\textbf{Set of all third-order Bargmann invariants}. We show the curve defining the boundary $\partial \mathfrak{B}_3$, of the set $\mathfrak{B}_3$ of all the possible third-order Bargmann invariants, as showed by Theorem~\ref{theorem: all third-order Bargmann invariants} and~\cref{corollary: all Bargmanns are pure Bargmanns}. Each value $\Delta = x+iy$ is showed, where the $x$-axis shows $x = \text{Re}[\Delta]$, while the $y$-axis shows $y = \text{Im}[\Delta]$. We have that $-1/8 \leq x \leq 1$, both tight bounds, and $-1/4 \leq y \leq 1/4$. If we consider purely imaginary-valued Bargmann invariants, i.e. those for which $\text{Re}[\Delta] = 0$, the largest possible imaginary value is equal to $\text{Im}[\Delta] = 0.19245$, larger than the value $1/6$ from the example provided by~\cite{bargmann1964note} (see~\cref{example: purely imaginary invariant}).}
    \label{fig: third_order_invariants_B3}
\end{figure}

\myindent We show $\mathfrak{B}_3$ in Fig.~\ref{fig: third_order_invariants_B3}. We can make an interesting historical remark. The curve $f(x,y)=0$ defining the boundary of the set of all third-order invariants is known as a \emph{Tschirnhausen curve}~\citep{lawrence1972catalog} (discovered in 1690), a.~k.~a.~\emph{Catalan's trisectrix} (who provided a new description in 1832) or even \emph{l'Hospital's cubic} (who rediscovered the curve in 1696)~\citep{loria1902spezielle,wieleitner1908spezielle,weaver1942Tschirnhausen}. This is one in a class of curves defined by equations of the form $27ay^2 = x^2(x+9a)$, with $a \in \mathbb{R}$. Many other expressions exist for defining such a family of curves. In our case, we have  let $a=1/8$ and translate the curve from $x=0$ to $x=1$, to obtain our specific boundary given by the equations $(27/8)y^2 = (x-1)^2(x+1/8)$ and $x \leq 1$.

\myindent We conclude by investigating further properties of the set $\mathfrak{B}_n$, for other values of $n \geq 3$. For example, it is simple to see that $\mathfrak{B}_1 = \{1\}$ since $\text{Tr}(\rho) = 1$ for every quantum state by construction, and $\mathfrak{B}_2 = [0,1]$ that follows simply from the fact that $\text{Tr}(\rho_1\rho_2)$ is a particular instance of the Born rule. In fact, $$\mathfrak{B}_2 = \mathfrak{B}_2|_{\mathrm{real}} = \mathfrak{B}_{2}|_{\mathrm{inc}} = \mathfrak{B}_2|_{\mathrm{pure}}.$$ It follows from our results from above that $$\mathfrak{B}_1 \subsetneq \mathfrak{B}_2 \subsetneq \mathfrak{B}_3,$$
where the notation $\mathcal{S}_1\subsetneq \mathcal{S}_2$ indicates that $\forall s_1 \in \mathcal{S}_1 \implies s_1 \in \mathcal{S}_2$, but the converse does not hold since there exists at least one $s_2 \in \mathcal{S}_2 \setminus \mathcal{S}_1$. We conjecture that the same holds for all possible $n$, i.e., that $\mathfrak{B}_n \subsetneq \mathfrak{B}_{n+1}$. 

\begin{conjecture}\label{eq: difference between Bargmann sets}
    For all $n \geq 1$, $\mathfrak{B}_{n} \subsetneq \mathfrak{B}_{n+1}$.
\end{conjecture}

It is simple to show, however, that for pure states we have the following chain. 

\begin{proposition}\label{proposition: separation between sets B}
    For all integers $n \geq 1$, $\mathfrak{B}_n|_{\mathrm{pure}} \subseteq \mathfrak{B}_{n+1}|_{\mathrm{pure}}$.
\end{proposition}

\begin{proof}
    Given any $\Delta_n(\pmb{\rho}) = \text{Tr}(\psi_1\psi_2\dots \psi_n) \in \mathfrak{B}_n$, we have that $\Delta_n(\pmb{\rho}) = \text{Tr}(\psi_1\psi_2\dots \psi_n\psi_n) \in \mathfrak{B}_{n+1}$ where we have used that $\psi_n\psi_n = \vert \psi_n\rangle \langle \psi_n \vert  \vert \psi_n\rangle \langle \psi_n \vert = \vert \psi_n\rangle \langle \psi_n \vert = \psi_n$ since, by assumption, $\langle \psi_n|\psi_n\rangle = 1$.
\end{proof}

\begin{figure}[!htbp]
    \centering
    \includegraphics[width=0.8\linewidth]{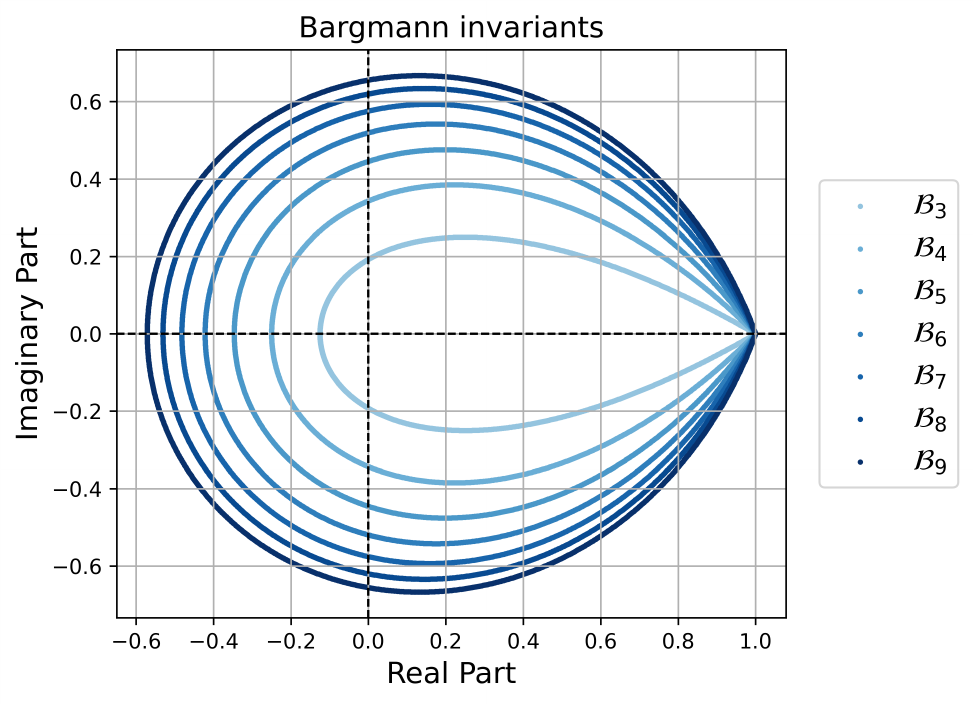}
    \caption{\textbf{Sets $\mathfrak{B}_n$ for $n=3,4,\dots,9$.} We show the lower bounds for the complex values $\Delta$ in the boundary of the sets $\mathfrak{B}_n$ found using seesaw optimization tools (see Eq.~\eqref{eq: seesaw boundary Bargmann invariants}). }
    \label{fig: sets of Bargmanns}
\end{figure}

\myindent We show in Fig.~\ref{fig: sets of Bargmanns} the sets $\mathfrak{B}_n$, numerically corroborating~\cref{proposition: separation between sets B}. Interestingly, if we restrict the dimensionality of $\mathfrak{B}_3$ to consider only single-qubit states, there exists a parametrized family of single-qubit states that characterizes the boundary $\partial \mathfrak{B}_3$. 

\begin{definition}[Oszmaniec--Brod--Galvão's (\acrshort{obg}) states, adapted from~\cite{oszmaniec2024measuring}]\label{def:OBG_states}
    Let $\{\vert \psi_k(\theta)\rangle \langle \psi_k(\theta) \vert  \}_{k=1}^n \subseteq \mathcal{D}(\mathbb{C}^2)$, for every integer $n \geq 3$, defined by  
\begin{equation}\label{eq: galvao states}
\ket{\psi_k (\theta)}=\cos(\theta/2)\ket{0}+\sin(\theta/2)e^{ 2\pi ik/n}\ket{1},
\end{equation}
for  $k \in \{0,1,2,\dots,(n-1)\}$, and arbitrary choice of $\theta \in [0,2\pi)$. We call this family of $n$-tuples of quantum states $\pmb{\psi}_\theta$ parametrized by $\theta$ the \emph{Oszmaniec--Brod--Galvão's tuples of states}, or simply \emph{Oszmaniec--Brod--Galvão's states}.
\end{definition}

\myindent With this definition, we can now state the following theorem.

\begin{theorem}\label{theorem: galvao states reaching the boundary}
    Let $\pmb{\psi}_\theta$ denote the tuple of states constructed from Eq.~\eqref{eq: galvao states}. For every $\theta \in [0,2\pi)$ we have that $\Delta_3(\pmb{\psi}_\theta) \in \partial \mathfrak{B}_3$. Moreover, if $\Delta \in \partial \mathfrak{B}_3$ then there exists some $\theta$ for which $\Delta = \Delta_3(\pmb{\psi}_\theta)$. 
\end{theorem}

\begin{proof}
    For any tuple  $\pmb{\psi}_\theta \in \mathcal{D}(\mathbb{C}^2)^3$ it holds that
\begin{equation}
\Delta_{3}(\pmb{\psi}_\theta)=[1+\sin^2(\theta/2) (e^{2\pi i/3}-1) ]^3. \label{eq:deltan}
\end{equation}
If we write $\Delta = |\Delta|e^{i\phi_\Delta}$ we have that $|\Delta|\cos(\phi_\Delta) = \text{Re}[\Delta]$. Let us show that for all $\theta$ we have that Ineq.~\eqref{eq: boundary of B_3} is saturated by $\Delta_3(\pmb{\psi}_\theta)$. From Eq.~\eqref{eq:deltan} we have that $\Delta_3(\pmb{\psi}_\theta) = a^3$ where $a=1+\sin^2(\theta/2) (e^{2\pi i/3}-1) $, and one can show that $$\text{Re}[a] = \frac{1}{4}(3\cos(\theta)+1), \,\, \text{Im}[a] = -\frac{\sqrt{3}}{4}(\cos(\theta)-1).$$
Therefore, from $a = \text{Re}[a]+i\text{Im}[a]$, 
\begin{equation*}
    a^3 = \text{Re}[a]^3-3\text{Re}[a]\text{Im}[a]^2 + i(3\text{Re}[a]^2\text{Im}[a]-\text{Im}[a]^3),
\end{equation*}
from where we get that $\text{Re}[\Delta] = \text{Re}[a^3] = \text{Re}[a]^3-3\text{Re}[a]\text{Im}[a]^2$, and since $|a^3|^{2/3} = |a|^2 = \text{Re}[a]^2+\text{Im}[a]^2$ we can now calculate
\begin{align*}
    &1-3|\Delta|^{2/3}+2\text{Re}[\Delta]= 1-3(\text{Re}[a]^2+\text{Im}[a]^2)+2(\text{Re}[a]^3-3\text{Re}[a]\text{Im}[a]^2) = \\
    &= 1-3\left(\frac{1}{4^2}(3\cos(\theta)+1)^2+\frac{{3}}{4^2}(\cos(\theta)-1)^2\right)\\
    &+2\left(\frac{1}{4^3}(3\cos(\theta)+1)^3-\frac{9}{4^3}(3\cos(\theta)+1)(\cos(\theta)-1)^2\right)\\
    &=1-\frac{3}{4}\left(1+3\cos^2(\theta)\right)+\frac{1}{4}\left(9\cos^2(\theta)-1\right) = 0,
\end{align*}
for all possible fixed choices of $\theta$. Note also that the converse holds, i.e., $\Delta \in \partial \mathfrak{B}_3$ implies that there exists some $\theta$ for which $\Delta = \Delta_3(\pmb{\psi}_\theta)$ since we can view the above as a parametrization of the curve $1-3|\Delta|^{2/3}+2\text{Re}[\Delta] = 0$ via $\Delta = (x(\theta)+iy(\theta))^3$, where $x(\theta) = \text{Re}[a] = (3\cos(\theta)+1)/4$ and $y(\theta) = \text{Im}[a] = -\sqrt{3}(\cos(\theta)-1)/4$. 
\end{proof}

\myindent The theorem we have just proved show that the boundary of $\mathfrak{B}_3$ is attained by invariants realized by tuples of single-qubit pure states, which implies that \emph{every} third-order Bargmann invariant can be obtained by constructing third-order Bargmann invariants of single-qubit pure states, and convex combinations thereof.

\begin{corollary}\label{corollary: B_3 is the convex hull of pure qubits}
    $\mathfrak{B}_3 = \mathrm{ConvHull}(\mathfrak{B}_3^{(2)}|_{\mathrm{pure}})$.
\end{corollary}

\myindent This is somewhat surprising as unitary-invariance merely suggests that every third-order Bargmann invariant $\Delta \in \mathfrak{B}_3|_{\mathrm{pure}}$ can be realized by pure states of dimension at most $d=3$ (as also discussed in~\citep{galvao2020quantum}) implying that it was known that $\mathfrak{B}_3|_{\mathrm{pure}} = \mathfrak{B}_3^{(3)}|_{\mathrm{pure}}$. 

\myindent Perhaps another surprising aspect is that \acrshort{obg} tuples of states from~\cref{def:OBG_states} satisfy an even more constraining property. Their Gram matrix $G_{\pmb{\Psi}_\theta}$ is a \emph{circulant} matrix (see Appendix~\ref{app: basic algebra}). This is also highlighted by the fact that when this happens, the Bargmann invariants  are equal to the \emph{power} of some complex number, to its order, i.e. $\Delta_m(\pmb{\psi}) = a^m$ for some complex number $a$. Naturally, one is led to believe that certainly $a$ must be the value of some other Bargmann invariant. Numerically, however, it is easy to see that this is not the case, since taking the root is not an operation that preserves quantum realizability (to see this, one can take $\Delta = -\sfrac{1}{8} \in \mathfrak{B}_3$ and note that $\Delta^\frac{1}{2} = \sfrac{i}{2\sqrt{2}} \notin \mathfrak{B}_3$). The following trivial proposition shows that, however, \emph{multiplying} (tuples of) Bargmann invariants \emph{does} preserve quantum realizability.  

\begin{proposition}\label{proposition: hadamard product}
    Let $\mathcal{W}$ be any finite subset of $\mathbb{N}^*$. Then, $\pmb{\Delta}_1,\pmb{\Delta}_2 \in \mathfrak{Q}(\mathcal{W})$ implies that $\pmb{\Delta}_1 \odot \pmb{\Delta}_2 \in \mathfrak{Q}(\mathcal{W})$ where $\odot$ is the Hadamard product. 
\end{proposition}

\begin{proof}
    Follows trivially from the fact that $\text{Tr}((A \otimes B)(C \otimes D)) = \text{Tr}(AC\otimes BD)= \text{Tr}(AC)\text{Tr}(BD)$.
\end{proof}

\myindent If we define the subset of $\mathfrak{B}_3|_{\mathrm{circ}} \subseteq \mathfrak{B}_3|_{\mathrm{pure}}$ of all Bargmann invariants realizable by triplets of pure quantum states such that their associated Gram matrix is also a circulant matrix, we can then refine~\cref{corollary: B_3 is the convex hull of pure qubits}.

\begin{corollary}
    $\mathfrak{B}_3 = \mathrm{ConvHull}(\mathfrak{B}_3^{(2)}|_{\mathrm{circ}})$.
\end{corollary}

\myindent From the discussions above we make the following conjecture.

\begin{conjecture}\label{conjecture: Bn and circ Bn}
    Consider any integer $n \geq 3$ and denote the sets $\mathfrak{B}_n$ of all possible Bargmann invariants. Then $\mathfrak{B}_n = \mathfrak{B}_n^{(2)}$ and, moreover, $\mathfrak{B}_n = \mathrm{ConvHull}(\mathfrak{B}_n^{(2)}|_{\mathrm{circ}})$.
\end{conjecture}

\myindent Finally, we conclude this analysis by showing that we can also use \acrshort{obg} tuples of states to show the following proposition.

\begin{proposition}\label{proposition: unit disc asymptotic}
    As $n \to \infty$ we have that $\mathfrak{B}_n \to \{\Delta \in \mathbb{C} \mid |\Delta| \leq 1\}$.
\end{proposition}

\begin{proof}    
Using the generalization of Eq.~\eqref{eq:deltan} to every $n$, we have that
\begin{equation}
    \Delta_n(\pmb{\psi}_\theta) = [1+\sin^2(\theta/2) (e^{2\pi i/n}-1) ]^n.
\end{equation}
Now, using the known limit relation $\lim_{n \to \infty}(1+\sfrac{a}{n})^n = e^a$ and the fact that we can expand $e^{2 \pi i/n} = 1 + \frac{2 \pi i}{n}+O\left(\frac{1}{n^2}\right)$ we have 
\begin{equation}
    \Delta_n(\pmb{\psi}_\theta) = \left[1+\sin^2(\theta/2)\left(\frac{2 \pi i}{n} +  O\left(\frac{1}{n^2}\right)\right)\right]^n,
\end{equation}
and therefore
\begin{equation}
    \lim_{n\to \infty}\Delta_n(\pmb{\psi}_\theta) = e^{2\sin^2(\theta/2)\pi i}.
\end{equation}
This means that by an appropriate choice of $\theta$, and as $n \to \infty$, $\Delta_n(\pmb{\psi}_\theta)$ approaches any complex number $e^{i\theta'}$, for any $ \theta' \in (0,2\pi]$.
\end{proof}

\myindent In this section, we have investigated in detail the geometrical properties of various types of quantum realizations for edge weightings---and even more general weightings---on a given event graph. In the following sections, we explore the physical motivations for studying these sets and focus on a specific class of quantum realizations that deserves particular attention: those pertaining to $\mathfrak{Q}(G)|_{\mathrm{inc}}$. 

\section{Relational coherence: motivation and conceptualization}\label{sec: relational coherence}

\myindent We now proceed to introduce and motivate our notion of `relational coherence', which for us is a property related to the notion of quantum realization of edge-weightings (or more generally, of the tuples $\pmb{\Delta}$ considered in Def.~\ref{def: Bargmann invariant quantum realizability}). Put it in simpler terms, relational coherence is a property that tuples of Bargmann invariants either have, or do not have. We want to distinguish it from the notion of \emph{set coherence} that was introduced back in Chapter~\ref{chapter: quantum coherence}. The distinction here is: 

\begin{enumerate}
    \item Set coherence is a property of a set of states $\{\rho_i\}_i$, for a fixed choice of Hilbert space $\mathcal{H}$.
    \item Relational coherence is a property of $\Delta: \mathcal{W} \to \mathbb{C}$, for a fixed choice of $\mathcal{W}$.
\end{enumerate}

\myindent The terminology above is  similar to when one says that a certain behavior $B$ from a compatibility scenario $\pmb{\Upsilon}$ is `contextual or not' depending on whether there exists a noncontextual model capable of explaining the statistics of the behavior $B$, while respecting the operational constraints imposed by $\pmb{\Upsilon}$. Similarly, we say that a certain tuple $\pmb{r}$, or a tuple $\pmb{\Delta}$, is `relationally coherent or not' depending on whether there exists a quantum realization in terms of set-coherent tuples of states, while respecting the structural relations determined by the labels of the event graph $G$, and expressed in the coordinate labels of $\pmb{r}$. 

\myindent Consistency among the coordinates is important. Clearly, every tuple of numbers $\pmb{r} \in [0,1]^{|E(G)|}$ can be reproduced by some set-incoherent (or set coherent) choice of two-state overlaps if they are allowed to not consistently respect the structure of an event graph, via some vertex-labeling. Concretely, take $G=C_3$ and the tuple of numbers $(1,0,1)$. There are, of course, two-state overlaps capable of realizing these values, e.g., $(|\langle 0|0\rangle|^2, |\langle 1|0\rangle |^2, |\langle 2|2\rangle|^2)$, but these realizations are \emph{not} considered as they do not respect the structure imposed by $G$ described by the labels $\pmb{r} = (r_{12},r_{13},r_{23})$, i.e., the realization by two-state overlaps just described cannot be viewed in terms of some vertex labeling for $C_3$. If one does require consistency with this labelings, it turns out that $(1,0,1)$ is not a quantum realizable tuple of edge-weightings for the event graph $C_3$. Having clarified the terminology we adopt, we now proceed to define and motivate the relevance of relational coherence.

\subsubsection{Relational  coherence}

\myindent We introduce relational coherence as follows.

\begin{definition}[Relational coherence]\label{def: quantum relational coherence}
    Let $G$ be any event graph. We say that $r$ has   \emph{relational coherence} if $r \in \mathfrak{Q}(G) \setminus \mathfrak{Q}(G)|_{\mathrm{inc}}$. More generally, for any finite set $\mathcal{W}\subseteq \mathbb{N}^*$, we say that $\Delta$ has \emph{relational coherence} if $\Delta \in \mathfrak{Q}(\mathcal{W}) \setminus \mathfrak{Q}(\mathcal{W})|_{\mathrm{inc}}$.
\end{definition}

\myindent In other words, the relational coherence of an edge weighting $r$ corresponds to the impossibility of reproducing the statistics $r$ using any set-incoherent quantum realization $r = r(\varrho)$, for any assignment $\varrho: V(G) \to \mathcal{D}(\mathcal{H})$ and any Hilbert space $\mathcal{H}$. 

\myindent We will later see in~\cref{lemma: incoherent_inside_C(G)} that $\mathfrak{Q}(G)|_{\mathrm{inc}} \subseteq \mathfrak{C}(G)$, implying that every facet-defining inequality of $\mathfrak{C}(G)$ can be interpreted as a (relational) coherence witness. We see in Chapter~\ref{chapter: applications} that our framework can be used to benchmark this form of coherence (and, consequently, set coherence and the usual basis-dependent description of quantum coherence) inside interferometers (and, more generally, any platform for quantum computation and quantum communication). 

\myindent The inclusion $\mathfrak{Q}(G)|_{\mathrm{inc}} \subseteq \mathfrak{C}(G)$ is also one of the motivations we have for distinguishing our dichotomy (relational coherence against incoherence), with that of \emph{set} coherence (opposed to set incoherence) that was introduced in Chapter~\ref{chapter: quantum coherence} by Def.~\ref{def: set coherence}. While it is unclear if $\mathfrak{Q}(G)|_{\mathrm{inc}}$ is a convex set or not, it is simple to motivate $\mathfrak{C}(G)$ as its meaningful `convexification' since the convex hull of $\mathfrak{Q}(G)|_{\mathrm{inc}}$ equals $\mathfrak{C}(G)$.~\footnote{This follows trivially from~\cref{lemma: incoherent_inside_C(G)} and from the fact that $\text{ext}(\mathfrak{C}(G)) \subseteq \mathfrak{Q}(G)|_{\mathrm{inc}}$.} The same is not clear for the notion of set incoherence. In this case, the counterpart of $\mathfrak{Q}(G)|_{\mathrm{inc}}$ to the case of set incoherence is a set that we denoted as $\mathcal{S}_{\mathrm{inc}}$. This set is known to be a nonconvex set as we have shown in Chapter~\ref{chapter: quantum coherence}, and its meaningful `convexification', which would be to consider $\text{ConvHull}(\mathcal{S}_{\mathrm{inc}})$, is quite an intricate set to consider, as it is a fairly nontrivial subset of the power set $2^{\mathcal{D}(\mathcal{H})}$ for some fixed Hilbert space of interest $\mathcal{H}$.~\footnote{Analysing set coherence as a resource from the perspective of $\text{ConvHull}(\mathcal{S}_{\mathrm{inc}})$ was left as an open problem by~\cite{designolle2021set}. We also comment on the fact that~\cite{designolle2021set} has considered the \emph{ordered} sets $\{ \pmb{\rho} \in \mathcal{D}(\mathbb{C}^d)^n \mid \pmb{\rho} \text{ is set-incoherent}  \}$ which implies that their convexification is, more formally, the convex hull of this subset of $\mathcal{D}(\mathbb{C}^d)^n$.} 

\myindent In what follows, we provide several  motivations for why we believe this notion of coherence is meaningful, and why it merits a detailed investigation. 

\subsubsection{Recalling motivations from previous chapters}

\myindent We have the burden of providing evidence that the notion we consider, i.e., that of relational coherence, is worth pursuing. In the beginning of Chapter~\ref{chapter: event_graph_approach} we have already provided some reasons for why we believe relational coherence is worth investigating. One of these reasons is now clear due to the fact that $\mathfrak{Q}(G)|_{\mathrm{inc}} \subseteq \mathfrak{C}(G)$ for any event graph $G$: relational coherence of $r$ implies coherence of any set of states $\pmb{\rho}$ for which $\pmb r=\pmb r(\pmb{\rho})$. Therefore, all the motivations for certifying and witnessing quantum coherence (see Chapter~\ref{chapter: quantum coherence}) using traditional schemes translate to those for considering quantum realizations of edge weightings. 

\myindent In Chapter~\ref{chapter: Bargmann invariants} we have talked about the \acrshort{hom} effect. This is a concrete  experimentally meaningful situation in which one can investigate properties of superposition that are \emph{agnostic} of the Hilbert space in question. Moreover, in Chapter~\ref{chapter: quantum coherence} we have mentioned the fact that there are different `types' of coherence. However, this notion of `type' of coherence was introduced only informally, based on several existing definitions in the literature that describe different ways in which quantum states can exhibit coherence, and how these distinctions give rise to different applications and uses of coherence as a quantum resource.

\myindent These preliminary discussions were all instances of the motivation we provide for investigating relational coherence, and the relevance of system-independence for our notion of coherence. We now structure our motivation in more than one way, complementing various of the arguments we have already presented in previous Chapters. 

\subsubsection{Relational coherence encompasses the usual notion of coherence and set coherence}

\myindent We begin with the obvious (yet important) point that we do not aim to replace the usual notion of coherence by that of relational coherence. Instead, our goal is to develop tools for  characterizing the ways in which quantum states (or sets of quantum states) can manifest their coherence from the values of unitary invariants. 

\myindent Let us start by recalling the following simple fact: there is no meaning to the phrase `the state $\rho$ is coherent' in the standard treatment of coherence~\citep{streltsov2017colloquium,baumgratz2014quantifying}, as we have seen in Chapter~\ref{chapter: quantum coherence}. One can view the pair $(\mathcal{H},\mathbb{A})$ of system and reference basis as the characterization of a `scenario', and it is only  meaningful to say instead that `the state $\rho \subseteq \mathcal{D}(\mathcal{H})$ is coherent with respect to $\mathbb{A}$', where $\mathbb{A} := \{\vert a \rangle\}$ characterizes this notion of scenario. But, obviously, such a definition of coherence is nothing more than the set coherence of the set $\{\rho\} \cup \{\vert a\rangle \langle a \vert \}_{\vert a \rangle \in \mathbb{A}}$. This is because a state $\rho$ is coherent with respect to a basis $\mathbb{A}$ if and only if $\{\rho\} \cup \{\vert a\rangle \langle a \vert \}_{\vert a \rangle \in \mathbb{A}}$ is set coherent.

\begin{proposition}
    Let $(\mathcal{H},\mathbb{A})$ be a pair of a quantum system and a choice of reference orthonormal basis for that system. Then, $\rho \in \mathcal{D}(\mathcal{H})$ is coherent with respect to the basis $\mathbb{A}$ iff $\{\rho\} \cup \{\vert a\rangle \langle a \vert \}_{\vert a \rangle\in \mathbb{A}}$ is set coherent.
\end{proposition}

\begin{proof}
    If $\{\rho\} \cup \{\vert a\rangle \langle a \vert \}_{\vert a \rangle\in \mathbb{A}}$ is set coherent, it is, in particular, set coherent if we choose $\mathbb{A}$ as the choice of basis. For the converse, if $\rho$ is coherent with respect to $\mathbb{A}$ this implies that there is at least one rank-1 state constructed from $\mathbb{A}$ that does not commute with $\rho$, which implies set coherence for $\{\rho\} \cup \{\vert a\rangle \langle a \vert \}_{\vert a \rangle\in \mathbb{A}}$.
\end{proof}

\myindent Therefore, in this sense we can note that the usual notion of basis-dependent coherence is just a very specific `type' of set coherence, one in which it is always possible to unitarily send all states in the set to be part of a reference basis \emph{but one}. Set coherence generalizes this idea by providing an operational framework for extending a notion traditionally associated with observables—namely, noncommutativity—to one concerning states, namely, set coherence. Hence, in our view, just as set coherence can be viewed as a simple generalization of the usual notion of coherence, ours can be viewed as a simple generalization of set coherence. 

\myindent As an example of use of this proposed notion of relational coherence captured by Bargmann invariants where one considers realizations of the form $\pmb{\Delta}(\varrho)$ where $\varrho: V \to \{\rho\} \cup \{\vert a\rangle \langle a \vert \}_{\vert a \rangle\in \mathbb{A}}$ we can mention our work~\citep{wagner2023anomalous}. There we provide a simple connection between relational coherence with \emph{anomalous weak values}. 

\subsubsection{We do not demand a paradigmatic shift}

\myindent We would also like to emphasize that our view does not require a profound  `paradigmatic' shift in the study of coherence. As we have pointed out in Chapter~\ref{chapter: information tasks}, a notion of nonclassicality based on set coherence realizability has already been explored using prepare-and-measure scenarios. Our contribution is, from this perspective, simply to recognize that this notion merits broader attention, beyond the specific restrictions previously considered.

\myindent For instance, recalling our discussion of quantum realization of prepare-and-measure scenarios from Chapter~\ref{chapter: information tasks}, or all the relevant uses of Bargmann invariants reviewed in Chapter~\ref{chapter: Bargmann invariants}, we see that prior work has already considered set-coherent realizations \emph{inferred} from statistical results---without relying on assumptions about a particular reference basis. From this perspective, we are not introducing a radically new concept but rather \emph{formalizing} an existing idea, and moreover, proposing a systematic toolbox for its investigation. 

\myindent We are also \emph{not} using new or extraneous tools for investigating these properties. Bargmann invariants have been proposed in the 1960's, and they appear in a whole variety of quantum related applications, as reviewed in Chapter~\ref{chapter: Bargmann invariants}. In line with our discussion in Chapter~\ref{chapter: introduction},---and to be precise about our motivations---our framework connects different research communities that, despite focusing on distinct topics, share common underlying tools. Though these communities have employed a broad variety of different approaches, their work can be seen as characterizing how nonclassical phenomena can be inferred from the values of Bargmann invariants. We continue this line of inquiry.

\subsubsection{Physically well motivated}

\myindent As we have pointed out when discussing basis-independent coherence in Chapter~\ref{chapter: quantum coherence}, there are reasons to favor properties of physical systems that are independent of a reference basis. Another physical motivation is that of considering the relevance of Bargmann invariants (and quantum superposition) for multi-photon interference. In these cases, one remains to some extent agnostic of what is the actual Hilbert space related to the internal degrees of freedom of the incoming photons. The values of Bargmann invariants of the different states completely characterize the outcomes of single-photon detectors in the output of such interferometers~\citep{shchesnovich2015partial,jones2020multiparticle}. This grants an \emph{operational} relevance to these values, and from it, the ways in which these values can be distinct from what one would obtain using just `classical resources', i.e., sets of pairwise commuting quantum states.

\subsubsection{Conceptual clarification}

\myindent As we have mentioned previously in the beginning of this Chapter, there has been a \emph{debate} regarding the existence of certain realizations of coherent states through what is known as `classical' optics. The debate surrounded precisely the fact that one could allegedly have the \emph{same} experimental results from two seemingly different perspectives, one that considers quantum states as incoherent mixtures of vector states, and one that considers coherent quantum states of light. 

\myindent While not equivalent to our notion of relational coherence, part of the debate was settled by the fact that it is better to view coherence of a state as a relational property. In that case, it was useful to note that the two sides of the debate were considering different reference bases, from which we note that it is \emph{obvious} that for one community a state $\rho$ could be coherent with respect to basis $\mathbb{S}$ while for the other side of the debate the same state could be incoherent, but now with respect to a \emph{different} basis $\mathbb{S}'$. This is natural, and unsurprising, but only once we acknowledge that coherence can be viewed from such a relational perspective. It is also useful to see this because part of the debate was that, essentially, both views were leading to the same experimental results. 

\myindent Using our framework, the same can happen. Take $\pmb{r} = (\text{Tr}(\rho_1\rho_2),\text{Tr}(\rho_1\rho_3),\text{Tr}(\rho_2\rho_3)) = \pmb{r}(\pmb{\rho})$.If $\pmb{r}$ violates an inequality from $\mathfrak{C}(C_3)$, we can conclude that $\pmb{r}$ is relationally coherent. Suppose now that two different parties find the same tuple $\pmb{r}$ from two different quantum realizations, $\pmb{r}(\pmb{\rho})$ and $\pmb{r}(\pmb{\sigma})$. According to party one, the state labelled as $1$, i.e., $\varrho(1) = \rho_1$ is coherent with the basis chosen by party one, while the state according to party two, having the same label, $\varsigma(1) = \sigma_1$ is \emph{incoherent} with respect to the basis chosen by party one. Both parties can conclude that their quantum states related to different quantum realizations indicate quantum coherence \emph{of their sets} even if they do not agree on which should be the basis of reference, and therefore on which states in the set are incoherent. 

\myindent This is a point that is relevant when we consider the relationship between coherence and contextuality in Chapter~\ref{chapter: from overlaps to noncontextuality}. There, we will see that even though there are behaviors $B$ yielding proofs of \acrshort{ks} contextuality by preparing states $\mathbb{1}/2 \otimes \mathbb{1}/2 \otimes \mathbb{1}/2$, incoherent with respect to \emph{every possible choice of basis}, there is a sense in which relational coherence is still required for $B$ to violate a \acrshort{ks} noncontextuality inequality. The necessary requirement there will be the relational coherence of the tuple $\pmb{r}$ `dual' to $B$ that also considers all the vector states in a \emph{state-independent contextuality set} (\acrshort{si-c}). Hence, even if basis-dependent quantum coherence of the prepared states is not necessary for quantum proofs of \acrshort{ks} contextuality, such a form of relational coherence \emph{is} necessary. Moreover, the relationship between the stabilizer subtheory and the Spekkens' toy model~\citep{spekkens2007evidence} illustrates that relational coherence, while necessary for contextuality in the generalized sense, is \emph{not} sufficient to guarantee it.

\myindent These short remarks from above showcase the conceptual clarity brought by viewing coherence as a relational property.

\subsubsection{Organizing types of quantum coherence via values of Bargmann invariants}

\myindent Recall that different sets of quantum states can fail to be unitarily equivalent to some set of incoherent states in significantly different ways. We have mentioned this feature of coherence back in Chapter~\ref{chapter: quantum coherence}. Since then, we have encountered various different examples of `types' of quantum coherence (e.g.,  imaginarity is here viewed, intuitively, as a `type' of coherence, while coherence that is necessarily encoded in qutrits is another `type'). Currently, there exists no precise way (to the best of our knowledge) to even propose a  classification, or an  organization, of the many ways in which a set of states can fail to be diagonalizable (hence set incoherent) with respect to some reference basis.  We now show that our framework is clearly moving towards a refined understanding of coherence in precisely this way, i.e., a possible framework where different types of coherence can be viewed as `organized and characterized' via the many ways in which Bargmann invariants can fail to be in the set $\mathfrak{Q}(G)|_{\mathrm{inc}}$ (or more generally in the sets $\mathfrak{Q}(\mathcal{W})|_{\mathrm{inc}}$). In what follows, to corroborate this view, we show that for every event graph $G$, we have that $\mathfrak{Q}(G)|_{\mathrm{inc}} \subseteq \mathfrak{C}(G).$ 

\myindent We start by noting that any set of states $\{\rho_i\}_{i} \subseteq \mathcal{D}(\mathcal{H})$ is set incoherent if and only if there exists some basis $\{\vert \lambda \rangle \}_{\lambda }$ such that $$\rho_i = \sum_{\lambda} \langle \lambda | \rho_i \vert \lambda \rangle P_\lambda$$ where $P_\lambda \equiv \vert \lambda \rangle \langle \lambda \vert $ are rank-1 projectors, for all $i$. This formally implies that, for any event graph $G$, any vertex labeling $\varrho:V(G) \to \{\rho_i\}_i$ yields a set incoherent quantum realization of $r$, where  
$$r_{ij}(\varrho) = \text{Tr}\left(\varrho(i)\varrho(j)\right) = \sum_{\lambda}p(\lambda|\varrho(i))p(\lambda|\varrho(j)).$$
Above we have written $p(\lambda|\varrho(i)) = \langle \lambda \vert \rho_i \vert \lambda \rangle $, which is the probability that upon measuring $\varrho(i)$ with $\{P_\lambda\}_\lambda$ we obtain $\lambda$. We conclude then that, by construction, there exists a mapping between a set incoherent realization to a realization in terms of a collection of jointly distributed \emph{and} independent random variables, from which we conclude that every $r$ that is quantum realizable by a set of incoherent states is such that $r \in \mathfrak{C}(G)$. 

\begin{lemma}\label{lemma: incoherent_inside_C(G)}
    Consider any event graph $G$ and any quantum realizable edge weighting $r = r(\varrho)$ such that $\varrho(V(G))$ is set incoherent. Then, $r \in \mathfrak{C}(G)$. 
\end{lemma}

\myindent Let us consider the following conjecture.

\begin{conjecture}\label{conjecture: convexity incoherent realizations}
    Let $G$ be any event graph. Then, the set $\mathfrak{Q}(G)|_{\mathrm{inc}}$ is a convex set. 
\end{conjecture}

\myindent If the conjecture above holds, we can show the following theorem.

\begin{theorem}\label{theorem:classical_overlaps_iff_incoherent_overlaps}
    Let $G$ be any event graph, and assume that~\cref{conjecture: convexity incoherent realizations} holds. An edge weighting $r\in \mathfrak{C}(G)$ if, and only if, $r$ is quantum realizable by some set of incoherent states. Succinctly, $\mathfrak{C}(G) = \mathfrak{Q}(G)|_{\mathrm{inc}}$.
\end{theorem}

\begin{proof}
    $(\Leftarrow)$ From~\cref{lemma: incoherent_inside_C(G)} we have that if $r$ is quantum realizable by incoherent states then $r \in \mathfrak{C}(G)$, for any $G$. $(\Rightarrow)$ Take $r \in \mathfrak{C}(G)$. It is simple to show that every element of $\text{ext}(\mathfrak{C}(G))$ is quantum realizable by sets of incoherent pure states. Because $\mathfrak{C}(G)$ is a convex polytope, $r$ is a convex combination $r = \sum_\omega p_\omega r_\omega$ where $r_\omega$ are quantum realizable by set incoherent states for all $\omega$. From~\cref{conjecture: convexity incoherent realizations} we have that $r \in \mathfrak{Q}(G)|_{\mathrm{inc}}$. 
\end{proof}

\myindent If true,~\cref{conjecture: convexity incoherent realizations} implies the theorem above, which would provide an immediate physical interpretation for all the framework described in Chapter~\ref{chapter: event_graph_approach}. Every facet-defining inequality of $\mathfrak{C}(G)$ can now be viewed as defining \emph{set coherence witnesses}, and providing a complete characterization of all the possible ways in which two-state overlaps can witness set coherence, with respect to any possible finite dimensional Hilbert space.~\footnote{Note that, this does not imply that the facet-defining inequalities are the best way to completely characterize all the set coherence that can be captured by two-state overlaps. In~\cref{subsec: relational imaginarity from overlaps} we show that \emph{non-linear} functionals, defined by the determinant of candidate matrices, may be useful as well. } Despite our motivation here being the conceptualization of the notion of relational coherence, this fact is instrumental for resolving one of the main question of this thesis (Question~\ref{question: applications}). 

\myindent From another perspective, if~\cref{conjecture: convexity incoherent realizations} holds then it would lead us to the following surprising corollary from~\cref{theorem:classical_overlaps_iff_incoherent_overlaps}.

\begin{corollary}
    Every $r \in \mathfrak{C}(G)$ can be realized by jointly distributed random variables that are, moreover, independent random variables.
\end{corollary}

\begin{proof}
    $r \in \mathfrak{C}(G)$ if and only if $r$ is quantum realizable by set incoherent states (via~\cref{theorem:classical_overlaps_iff_incoherent_overlaps}), which holds if and only if $r$ is realizable by some set of independent jointly distributed random variables. 
\end{proof}

\myindent Some remarks are needed. First, we note that $\mathfrak{C}(G) = \mathfrak{Q}(G)|_{\mathrm{inc}}$ does not imply that $r \in \mathfrak{C}(G)$ cannot have set coherent realizations. We have seen an example of this in~\cref{example: realization coherent and incoherent}. In Chapter~\ref{chapter: contextuality} we show that there exists a correspondence between $\mathfrak{C}^0_{G|\nabla G}$, where $\nabla G$ is the suspension graph of a graph $G$ (see Appendix~\ref{sec: graph theory}), and noncontextual polytopes characterized by the \acrshort{stab} polytope. From this perspective, a similar remark applies, that we \emph{cannot} conclude from the above that incompatible measurements cannot have \acrshort{ks} noncontextual models. 

\myindent Second, it is clear that the role played by convexity of the sets $\mathfrak{Q}(G)|_{\mathrm{inc}}$ and $\mathfrak{C}(G)$ was crucial. \emph{If}~\cref{conjecture: convexity incoherent realizations} turns out to be incorrect, the above would not hold. Finally, convexity and the arguments above rely heavily on the fact that we are \emph{not} restricting the dimensionality of the Hilbert space of the quantum realizations (nor the related cardinality of the sets $\Lambda$ on which the jointly distributed random variables are valued). Because of that, we do not expect that a similar result holds for $\mathfrak{Q}(G)|_{\mathrm{inc}}^{(d)}$. In other words, there could exist an event graph $G$ and some integer $d \geq 2$ such that $\mathfrak{Q}(G)|_{\mathrm{inc}}^{(d)} \neq \mathfrak{C}(G)^{(d)}$. 

\myindent From these results it follows a preliminary possible organization of types of coherence (at least for the case of two-state overlaps). We can organize these types depending on the different facet-defining inequality violations of $\mathfrak{C}(G)$, for different event graphs $G$. \emph{If} we would also conclude that, when considering higher order Bargmann invariants, the sets  $\mathfrak{Q}(\mathcal{W})|_{\mathrm{inc}}$ would also be given by convex polytopes, the same classification would follow. We leave this to future work.

\section{Two-state overlaps in prepare-and-measure scenarios: device-dependent or semi device-independent?}\label{sec: prepare and measure overlap}

\myindent We have mentioned in Chapters~\ref{chapter: contextuality} and~\ref{chapter: information tasks} (see Def.~\ref{PM scenario}) a specific class of scenarios known as prepare-and-measure scenarios. One motivation for their introduction was to certify forms of nonclassicality in quantum devices without the demanding requirements of device-independent tests based on Bell inequality violations. In effect, these scenarios trade off full device independence (hence the coined term \emph{semi}-device independent protocols) and the need for a Bell experiment in favor of a prepare-and-measure setup provided some specific promises about the device.

\myindent The first such promise considered was that of a fixed  dimensionality of the underlying Hilbert space of the system being prepared by the sender. As noted in Chapter~\ref{chapter: information tasks}, several other proposals have since been adopted, given that upper bounds on the dimension have been subject to criticism (cf.~\cite{vanHimbeeck2017semidevice}). In such scenarios, these restrictions are fundamental in order to have a non-trivial distinction between the statistics that one can obtain using quantum realizations when compared to classical realization (where, recall from Def.~\ref{def: dimension dependent PM scenario}, `classical' is used as equivalent to set incoherent realization of the states). 

\myindent For example, as illustrated in Fig.~\ref{fig: coherence and dimension part I} (right), every correlation of the scenario can be realized using set-incoherent resources (with a trit), but if we know that the system is two-dimensional we can see a gap between what can be realized by set-incoherent realizations with bits, and set-coherent realizations with qubits. In doing so, one is led to the several prepare-and-measure polytopes discussed previously in Chapter~\ref{chapter: information tasks}. Depending on the specific restrictions, constructing the associated polytope of set-incoherent realizable correlations can be more or less challenging.

\myindent Here, we take the opportunity to introduce the idea of a \emph{two-state overlap} promise, within the prepare-and-measure paradigm. In this case, one considers all the possible prepare-and-measure scenarios defined by the triplets $(I,J,2)$, where $I=J$ (see  Def.~\ref{def: quantum realizable PM behavior} for this notation). The premise here is that the observed data is that of (obviously) two-state overlaps
\begin{equation*}
    p(0|y,x) = \text{Tr}(\rho_x\rho_{0|y}),
\end{equation*}
where $p(1|y,x) = 1-p(0|y,x)$.

\myindent We now list some obvious positive aspects of this proposal. First, one \emph{does not need} to commit to an upper bound on the Hilbert space dimension. Second, these statistics are related to the notion of  \emph{confusability}, useful for probing the notion of quantum generalized contextuality (see~\cref{sec: Spekkens noncontextuality}). Third, there is now (as we show in~\cref{sec: relational coherence from the event graph approach} below by finding quantum violations of event graph inequalities, but also as we have already commented on when reviewing the results by~\cite{galvao2020quantum} in Chapter~\ref{chapter: Bargmann invariants}) a separation between set-coherent realizability and set-incoherent realizability in \emph{any} finite-dimensional Hilbert space. In other words, it is \emph{not} true that every correlation in such a prepare-and-measure scenario can be realized by set-incoherent realizations in a sufficiently large Hilbert space dimension, contrasting it with the situation showed in Fig.~\ref{fig: coherence and dimension part I} (right). Fourth, the polytope bounding set-incoherent realizability is simple to characterize. These are the polytopes $\mathfrak{C}(G)$ for an event graph $G$ that we have introduced in Chapter~\ref{chapter: event_graph_approach}. 

\myindent We conclude with the drawbacks. The first, and perhaps the most significant, is that one may argue that such a proposal cannot be considered semi-device independent as we have a promise on the power of the \emph{receiver} and not on the \emph{sender}.  Nevertheless, if one is convinced that semi-device independent certification is an interesting methodology to pursue, especially for certification of nonclassical properties in quantum devices, then one \emph{can} make our proposal into a semi-device independent scheme. In order to do so, one can propose a test that is semi-device independent by assuming a dimension upper bound and then use the tools from~\cite{miklin2021universalscheme} to certify that the statistics obtained is that of two-state overlaps. From this, one can then proceed to use the overlap statistics to violate the inequalities we have introduced in Chapter~\ref{chapter: event_graph_approach} that we  have shown to witness coherence (and dimension, as we show in more detail in Chapter~\ref{chapter: applications}). This makes the test semi-device independent (we are not even assuming anymore a restriction based on two-state overlaps), at the cost of assuming an upper bound in the dimension. 

\myindent In summary, we can take two perspectives:

\begin{enumerate}
    \item (Device dependent) One considers a prepare-and-measure scenario, where the receiver makes only dichotomic measurements. One assumes that the statistical results obtained are close to that of two-state overlaps. No upper bound on the dimension is required. No assumption of purity of states is needed. It is possible to witness relational coherence, or Hilbert space dimension (as we show in Chapter~\ref{chapter: applications}) from the violation of facet-defining inequalities characterizing the event graph polytope. 
    \item (Semi-device independent) One considers, again, a prepare-and-measure scenario, where the receiver makes only dichotomic measurements. One assumes an upper bound on the underlying Hilbert space dimension of the message of the sender. One uses the results by~\cite{miklin2021universalscheme} to certify that the statistics is close to that of two-state overlaps. It is possible to witness the same properties as above, again, from the violation of facet-defining inequalities characterizing the event graph polytope. 
\end{enumerate}

\myindent In our view, the device-dependent perspective of simply assuming that the statistics obtained is (close to being that of) two-state overlaps is more interesting for experimentalists benchmarking nonclassical behavior of quantum devices, but it is worth pointing out that both perspectives are valid and accounted for by the framework we introduce. 

\section{Relational coherence from the event graph approach}\label{sec: relational coherence from the event graph approach}

\myindent In this section we investigate \emph{when} we can detect that edge weightings have relational coherence. In other words, provided that relationally coherent edge weightings are those that are both quantum realizable \emph{and} violate some facet-defining inequalities of $\mathfrak{C}(G)$, we are interested in investigating properties of such violations. Additionally, we investigate maximal violations of event graph inequalities, the role of purity of the states, and if there is a systematic protocol for finding optimal violations via semidefinite programming techniques.  

\subsection{Pure states maximally violate facet-defining inequalities of event graph polytopes}\label{subsec: pure states violating}

\myindent We begin with the following simple lemma, which shows that any convex-linear functional $h: [0,1]^{E(G)} \to \mathbb{R}$ attains its maximum over quantum realizations on some realization by pure quantum states.

\begin{lemma}[Adapted from~\cite{wagner2024certifying}]\label{lemma: pure_state_realizations_larger}
    Let $h:[0,1]^{E(G)}\to \mathbb{R}$ be any convex-linear functional, acting over edge weightings $r \in [0,1]^{E(G)}$, for any event graph ${G}$. Then, for any quantum realization $r=r(\varrho)$ with respect to some vertex labeling $\varrho:V(G) \to \{\rho_i\}_{i}$, there exists a pure state quantum realization $r=r(\psi)$ with respect to a vertex labeling $\psi:V(G) \to \{\vert \psi_i \rangle \langle \psi_i \vert \}_{i}$, such that \begin{equation*}
        h(\pmb r(\pmb \rho)) \leq h(\pmb r(\pmb \psi)).
    \end{equation*}
\end{lemma}

\begin{proof}
    Each $\varrho(i) = \rho_i$ is a convex combination of pure states $\{\psi^{(i)}_{\omega_i}\}_{\omega_i \in \Omega_i}$ for some set of pure states $\Omega_i = \{\vert \omega_i \rangle \langle \omega _i\vert\}_{\omega_i}$. Let us choose an ordering so that $\pmb{\rho} = (\varrho(1),\varrho(2),\dots,\varrho(m))$, with $m = |V(G)|$. Noticing that $h(r)$ is, by assumption, a convex-linear functional,
    \begin{align*}
        &\forall i,\rho_i = \sum_\omega \lambda_\omega^{(i)}\vert \psi_\omega^{(i)} \rangle \langle \psi_\omega^{(i)} \vert \implies \pmb{\rho} = \sum_{\omega_1,\dots,\omega_m} \lambda_{\omega_1}^{(1)}\cdots \lambda_{\omega_m}^{(m)} (\psi_{\omega_1}^{(1)},\psi_{\omega_2}^{(2)},\dots,\psi_{\omega_m}^{(m)})\\ &\Longrightarrow h(\pmb r(\pmb{\rho})) = \sum_{\omega_1,\dots,\omega_m} \lambda_{\omega_1}^{(1)}\cdots \lambda_{\omega_m}^{(m)}h(\pmb r(\psi_{\omega_1}^{(1)},\psi_{\omega_2}^{(2)},\dots,\psi_{\omega_m}^{(m)})) = \sum_{\pmb{\omega}}\lambda_{\pmb{\omega}}h(\pmb r(\pmb{\psi}_{\pmb{\omega}})). 
    \end{align*}
    where, for every $\pmb{\omega} = (\omega_1,\omega_2,\dots,\omega_m)$, we have that $\pmb{\psi}_{\pmb{\omega}} = (\psi_{\omega_1}^{(1)},\psi_{\omega_2}^{(2)},\dots,\psi_{\omega_m}^{(m)})$   defines a vertex labeling of $V(G)$ with pure states. To conclude the above, one needs to introduce some redundant values of $1 = \sum_{\omega_i}\lambda_{\omega_i}^{(i)}$. The equation then follows from linearity with respect to $r$, and hence multilinearity with respect to the states due to the trace operation. 
    
    \myindent It is simple to see that $\{\lambda_{\pmb{\omega}}\}_{\pmb{\omega}}$ is also a set of convex weights.  Choosing now a particular $\pmb{\omega}^\star$ such that $\forall \pmb{\omega}, h(\pmb r(\pmb{\psi}_{\pmb{\omega}}) \geq h(\pmb r(\pmb{\psi}_{\pmb{\omega}^\star}))$ we see that, by construction, $$h(\pmb r(\pmb{\rho})) = \sum_{\pmb{\omega}}\lambda_{\pmb{\omega}} h(\pmb r(\pmb{\psi}_{\pmb{\omega}}))\leq \sum_{\pmb{\omega}}\lambda_{\pmb{\omega}} h(\pmb r(\pmb{\psi}_{\pmb{\omega}^\star})) = h(\pmb r(\pmb{\psi}_{\pmb{\omega}^\star})),$$
    since $\sum_{\pmb{\omega}} \lambda_{\pmb{\omega}}=1$. 
\end{proof}

\myindent This lemma shows that if we consider optimal violations of facet-defining inequalities $h(\pmb{r}) \leq b$ of event graph polytopes it suffices to consider just an optimization over pure states. In other words, the maximum that any functional $g(\pmb\rho) := h(\pmb r(\pmb\rho))$ can take, where vertex labelings $\varrho$ take values over a convex subset $\mathcal{S} \subseteq \mathcal{D}(\mathcal{H})$ is given by the extremal points $\text{ext}(\mathcal{S})$ of $\mathcal{S}$. 

\begin{corollary}[Adapted from~\cite{wagner2024certifying}]
    Let $\mathcal{S} \subseteq \mathcal{D}(\mathcal{H})$ be a compact convex subset of $\mathcal{D}(\mathcal{H})$, and $G = (V(G),E(G))$ any event graph. For every event graph inequality $h(\pmb{r}) \leq b$, if we denote by $\mathcal{S}^{V(G)}$ the set of all possible vertex $\mathcal{S}$-labelings $\varrho: V(G) \to \mathcal{S}$, the maximum  $$h(r(\varrho^\star)) = \max_{\varrho \in \mathcal{S}^{V(G)}}h(r(\varrho)),$$ is attained when $\varrho^\star(V(G)) \in \mathrm{ext}(\mathcal{S})$. 
\end{corollary}

\myindent Above, the optimization is one where we have considered $\mathcal{S}$ to have a fixed Hilbert space dimension. Having the results above we can now prove the following theorem.

\begin{theorem}[Adapted from~\cite{wagner2024certifying}]\label{theorem: pure states maximally violate event graph inequalities}
    The maximal quantum violation of any facet-defining inequality of $\mathfrak{C}(G)$, for any event graph $G$, is attained by pure states.
\end{theorem}

\begin{proof}
    Since $\mathfrak{C}(G)$ is a convex polytope, any facet-defining inequality from $\mathfrak{C}(G)$ is described by linear functionals $h(\pmb r)$, together with some $b \in \mathbb{R}$ satisfying $h(\pmb r) \leq b$. From~\cref{lemma: pure_state_realizations_larger} it is trivial now to conclude that maximal violations, which are associated to maximum values attained by $h(\pmb r)$ are given by pure states.
\end{proof}

\myindent Recall also that, from the results of~\cite{galvao2020quantum}, we further know that the optimal violations of facet defining inequalities of $\mathfrak{C}(G)$ must also be such that $\varrho^\star \in \mathcal{D}(\mathbb{C}^{|V(G)|})$.

\subsection{Lower bounds on optimal violations from seesaw iterations}\label{subsec: seesaw for optimal lower bounds}

\myindent In this subsection, and the following one, we delve into the investigation of optimal violations of facet-defining inequalities of the event graph polytope. We assume familiarity of the reader with semidefinite programming (\acrshort{sdp}). For more information we refer to~\cite{tavakoli2023semidefinite}. Every facet-defining inequality of $\mathfrak{C}(G)$ can be written as $h(\pmb r) \leq b$ where $h(\pmb r)$ is a linear functional. In full generality, these functionals take the form
\begin{equation}
    h(\pmb r) = \sum_{e \in E(G)}\gamma_e r_e \leq b,
\end{equation}
where $\pmb{\gamma} = (\gamma_e)_e \in \mathbb{Z}^{E(G)}$. For quantum realizable edge-weightings, each such functional becomes 
\begin{equation}
    h(\pmb r(\pmb \rho)) = \sum_{\{i,j\} \in E(G)}\gamma_{i,j} \text{Tr}(\rho_i\rho_j).
\end{equation}
It is simple to see that, denoting $h(\pmb{r}(\pmb{\rho})) = g(\pmb{\rho})$ with $g: \mathcal{B}(\mathcal{H})^{V(G)} \to \mathbb{R}$, if we pick a coordinate $\rho_i$ of $\pmb{\rho}$ and fix all the other quantum states, we can use a semidefinite program to optimize $g$ with respect to $\rho_i$ using
\begin{align}\label{eq: see saw Bargmanns}
    &\max_{X_i \in \matdCC} g(\rho_1,\rho_2,\dots,X_i, \dots \rho_{|V(G)|-1},\rho_{|V(G)|}) \\
    &\text{subject to } X_i \geq  0\nonumber \\
    &\text{Tr}(X_i) = 1 \nonumber 
\end{align}
where $X_i \geq 0$ denotes that $X_i$ is positive semidefinite (hence Hermitian, see~\cref{app: basic algebra}). Since we have picked an arbitrary coordinate $\rho_i$, it is clear that we can use a seesaw methodology for providing valid \emph{lower bounds} for the optimal quantum realizations $r(\varrho)$ for each Hilbert space dimension $d$. From the fact that $g(\pmb{\rho})$ is a function that is invariant under global unitary transformations (see Def.~\ref{def: global unitary transformations}), we have that the largest Hilbert space dimension that needs to be checked is $d= |V(G)|$ (as we have discussed in~\cref{chapter: Bargmann invariants}). Also, to start the seesaw, we need to input certain Ansatz matrices $\rho_2,\dots,\rho_{|V(G)|}$ that we can choose at random. Commonly, the performance of the seesaw technique depends on these initial random seeds to the seesaw optimization and on the number of iterations. One needs to choose a number of iterations that is sufficient to see a numerical convergence of the problem. 

\myindent The seesaw technique is significantly better to find lower bounds than, for instance, using Monte Carlo methods. Let us provide some simple (non-rigorous) numerical comparison to give intuition on how much better the seesaw technique is, at least compared to naive random sampling technique. Consider an example of a run of search for optimal values via random sampling, considering a naive \texttt{Python} implementation using \texttt{QuTiP}~\citep{johansson2012qutip,lambert2024qutip5}. Sampling over 5-tuples of Haar random pure quantum states $\pmb{\psi} = (\psi_1,\psi_2,\dots,\psi_5)$~\footnote{One possible simplification is to consider optimizations with respect to the subset $\mathfrak{Q}(G)|_{\mathrm{real}}$. } in dimension $d=4$ led to a violation of $h_5(\pmb r(\pmb{\psi})) = 1.243  > 1$ after about $511$ seconds of runtime. This was the best trade-off between runtime and size of violation (on another run we have obtained $h_5(r(\pmb{\psi})) = 1.1102$ after $\sim 7227$ seconds!). For the same inequality, 40 iterations of the seesaw~\footnote{We see numerically that about 8 iterations are enough for convergence, always finding for this functional, the same optimal value as with 40 iterations.} finds $h_5(r(\pmb{\psi})) = 1.3750$ in 
 about $8$ seconds.~\footnote{Of course, seconds are not a rigorous and universally valid way of quantifying efficiency since it is dependent on the actual implementation and on how good the computer used is. Here we simply want to provide intuition on the usefulness of such seesaw protocols compared to the simplest and most naive implementation.} 

\myindent Another important aspect is that such seesaw optimizations yield \emph{valid} quantum realizations since they bound the quantum set from inner approximations. We show  later in~\cref{tab: seesaw bound} various values obtained through seesaw-based quantum optimization, some of which exhibit interesting violations arising from symmetric single-qubit vertex labelings. However, the limitation is that we have no guarantee that these seesaw optimization techniques actually find the optimal violations of the inequalities. In order to do so we need procedures that provide  \emph{upper bound} estimates for the values of these functionals. In this case, these are found using hierarchies of relaxations of \acrshort{sdp} optimizations. When they converge, their bounds are \emph{not} guaranteed to be valid solutions to the problem. It is only when the upper and lower bounds match that we know that we have found a \emph{tight} optimal largest value for the functional. We now recall how one can use existing tools for the specific case of two-state overlaps. 

\subsection{Upper bounds from \acrshort{sdp} relaxation hierarchies}\label{subsec: upper bounds from sdp hierarchies}

\myindent In order to upper bound the possible values that $h(r(\varrho))$ can have, for all possible quantum realizations $r(\varrho) \in \mathfrak{Q}(G)$ for some event graph $G$ we can use existing techniques for prepare-and-measure scenarios~\citep{wagner2024certifying}. As we have already discussed above in Sec.~\ref{sec: prepare and measure overlap}, the statistics $\pmb{r}(\pmb{\rho}) = (\text{Tr}(\rho_i\rho_j))_{\{i,j\} \in E(G)}$ can be understood as part of a specific implementation of prepare-and-measure scenarios. Because of that, we can use existing \acrshort{sdp} techniques for bounding optimal values that have been considered in this framework~\citep{tavakoli2023semidefinite}. 

\myindent For example, without restricting the Hilbert space dimension, we can consider upper bounds of $h(r(\varrho))$ using the moment matrix of \emph{tracial moments}. In this case we choose the initial set of operators for defining the hierarchy of moment matrices as $\mathcal{L} := \{\mathbb{1},\pmb{\rho}\}$, with $\pmb{\rho} = (\rho_1,\dots,\rho_{|V(G)|})$, since these are the matrices relevant for the prepare-and-measure statistics of the two-state overlap scenario. Thus, we choose 
\begin{equation}
    \Gamma(S_i,S_j) = \text{Tr}(S_i^\dagger S_j)
\end{equation} 
where $S_i,S_j \in \mathcal{S}^{(k)}$ for any level $k$ of the hierarchy. The moment matrix is then a $|\mathcal{S}^{(k)}| \times |\mathcal{S}^{(k)}|$ matrix, for which we impose constraints of normalization such as $\Gamma(\rho_i,\mathbb{1}) = \Gamma(\mathbb{1},\rho_i) = 1$. 

\myindent From Theorem~\ref{theorem: pure states maximally violate event graph inequalities} we can also include as restrictions to the \acrshort{sdp} symmetries of the moment matrix such that $\rho_i^2 = \rho_i$ (since optimal values are attained by pure states). Moreover, from unitary-invariance, we know that the dimension of the solution problem must be at most $d=|V(G)|$. This can be seen as an `effective' dimension bound, which implies that we can use a specific dimensionally-restricted hierarchy of \acrshort{sdp}s introduced by~\cite{navascues2014characterization}. As another symmetry, we note that the moment matrix has a section giving the two-state overlaps $\text{Tr}(\rho_i\rho_j)$ since, by definition,  $$\Gamma(\rho_i,\rho_j) = \text{Tr}(\rho_i^\dagger \rho_j) = \text{Tr}(\rho_i\rho_j).$$ For higher levels of the hierarchy, the fact that states are pure and the fact that the trace is cyclic force various symmetries on the moment matrix. For instance, while the first level has moment matrices given by $$\mathcal{S}^{(1)} = \{\mathbb{1},\rho_1,\rho_2,\dots,\rho_{|V(G)|}\},$$ the second level of the hierarchy has $$\mathcal{S}^{(2)} = \mathcal{S}^{(1)} \cup \{\rho_1^2,\rho_2^2,\dots,\rho_1\rho_2,\dots\}$$ from which we have that $$\Gamma(\rho_1^2,\rho_2\rho_3) = \text{Tr}((\rho_1^2)^\dagger \rho_2\rho_3) = \text{Tr}(\rho_1^2\rho_2\rho_3) = \text{Tr}(\rho_1\rho_2\rho_3) = \Gamma(\rho_1,\rho_2\rho_3) = \Gamma(\rho_2\rho_1,\rho_3),$$
as an example of a symmetry to be imposed.~\footnote{Another related and more general approach, which we have not considered, is to view these problems within the class of trace polynomial optimization problems~\citep{klep2021optimization}. We leave this for future research.} 

\myindent For prepare-and-measure scenarios the difficult task is to propose such hierarchies that work under the restriction that one has an upper bound on the Hilbert space dimension, i.e, in our case instead of providing upper bounds with respect to the set $\mathfrak{Q}(G)$, providing upper bounds with respect to the sets $\mathfrak{Q}(G)^{(d)}$, for different fixed  choices of dimension $d$. This is because, so far, as we have described above, the moment matrix formalism is agnostic to the specific Hilbert space dimension (as in the \acrshort{npa} hierarchy). To impose a bound on the dimension, we can use the methodology proposed by~\cite{navascues2015bounding}, which we now discuss.

\myindent Instead of proceeding just as in the case of the usual \acrshort{npa} hierarchy as done by~\cite{navascues2014characterization}, which does not assume a specific Hilbert space dimension since it is based only on the optimization of a generic positive semidefinite moment matrix, one can do the following: For  each level $k$ of the hierarchy we randomly sample pure quantum states, forming a specific operator pool given by $L_1 = \{\mathbb{1},\pmb{\rho}\}$ and perform the matrix operations for finding $\mathcal{S}^{(k)}_1$ for this case. This leads to a `moment matrix' $(\Gamma_1(S_i,S_j))_{S_i,S_k \in \mathcal{S}^{(k)}_1}$ that satisfies all the constraints wanted, including positive semidefiniteness, by construction.~\footnote{Strictly speaking, these are not moment matrices. In standard moment-matrix constructions, one typically fixes an initial state $\psi$ and defines entries of the form $\Gamma_{ij} = \langle \psi | A_i^\dagger A_j | \psi \rangle$. In contrast, the present definition $\Gamma_{ij} = \mathrm{Tr}[\rho_i \rho_j]$ follows a different approach: the resulting matrix is still positive semidefinite, but it does not represent a true moment matrix, as its entries are not moments.}  We do the same process for obtaining several moment matrices $\{\Gamma_1,\Gamma_2,\dots,\Gamma_M\}$ and we truncate this procedure when we are certain that these matrices span a space of moment matrices compatible with the dimension $d$. To conclude, the \acrshort{sdp} is then defined as the problem
\begin{align*}
    &\max_{\pmb{c} \in \mathbb{R}^M} g(\Gamma)\\
    &\text{subject to }\Gamma = \sum_{j=1}^M c_j \Gamma_j,\\
    &\Gamma \geq 0,\\
    &\sum_{j}c_j = 1.
\end{align*}
Above, we have written as $g$ the functional acting on the moment matrices and yielding $h(r(\varrho))$ that we want to optimize. While the above relaxation is unknown to theoretically converge as $k \to \infty$, in some cases this relaxation matches the seesaw values. 

\subsection{Quantum violations of event graph inequalities}\label{subsec: optimal quantum violations of event graphs}

\myindent We now provide in detail optimal violations that were found using the \acrshort{sdp} techniques discussed above. We start with the simplest class of inequalities given by the $n$-cycle inequalities $c_n(\pmb r) \leq n-2$ from Eq.~\eqref{eq:cycle_inequalities}.

\subsubsection{Quantum violations of cycle inequalities}

\myindent The optimal solution of the maximization problem for $c_n(\pmb r)$ was found using \acrshort{sdp} techniques by~\cite{wagner2024certifying}. Using both seesaw optimization for the lower bounds and the hierarchy of \acrshort{sdp}s for upper bounds, we could infer the following family of optimal solutions. Take the set of states $\{|\psi_x\rangle\}_{x=1}^n$ of the form
\begin{equation}\label{eq: states max cycles}
    |\psi_x\rangle = \cos (\theta_x^{(n)}) |0\rangle + \sin (\theta_x^{(n)})|1\rangle\,.
\end{equation}
where 
\begin{equation*}
  \theta_x^{(n)} =
    \begin{cases}
      \frac{\pi}{2} - \frac{(x-1)\pi}{2n} & \text{if $n$ is odd,}\\
      \frac{\pi}{2} + \frac{(x-1)\pi}{2n} & \text{if $n$ is even.}
    \end{cases}       
\end{equation*}

\myindent This family of states $\pmb{\psi}$ was found explicitly as the solution to $$c(\pmb{r}(\pmb{\psi})) = \max_{\varrho \in \mathcal{D}(\mathbb{C}^{n})^{V(G)}}c_n(\pmb{r}(\pmb{\rho}))$$ with the  \acrshort{sdp} seesaw optimization technique. Using the Navascués--Vértesi (\acrshort{nv})~\citep{navascues2015bounding} hierarchy, we could then show that the quantum bound is \emph{tight} (up to $10^{-5}$ precision) for this family of states up to $n = 8$, since lower and upper bounds agreed. Due to numerical limitations, we have implemented only the seesaw technique for values smaller than $8$. For larger values~$8 \leq n \leq 20$, we have checked the lower bounds match the analytical formula (provided below in Eq.~\eqref{eq: analytical formula cn ineq}) up to the same precision. Because of these numerical results, we have the following conjecture:

\begin{conjecture}[Adapted from~\cite{wagner2024certifying}]\label{conjecture: optimal values tuples for cn inequalities}
    The tuples $\pmb{\psi} = (\psi_x)_{x=1}^n$ constructed using Eq.~\eqref{eq: states max cycles} provide the optimal quantum violations of the inequalities $c_n(\pmb{r}) \leq n-2$ for all $n$. 
\end{conjecture}

\myindent For the case $n=3$, the optimal values match those found by~\cite{galvao2020quantum}.  If~\cref{conjecture: optimal values tuples for cn inequalities} holds, the maximum quantum violation is then given by
\begin{equation}\label{eq: analytical formula cn ineq}
    c_n(\pmb r(\pmb\psi)) = (n-1)\cos^2\left(\frac{\pi}{2n}\right)-\cos^2\left[\left(1-\frac{1}{n}\right)\frac{\pi}{2}\right].
\end{equation}
On the other hand, the optimal incoherent quantum realization is $c_n(\pmb r(\pmb\psi^{inc})) = n-2$. The fraction between these two quantities goes to $1$ as $n$ grows.
\begin{proposition}[Adapted from~\cite{wagner2024certifying}]
    The limit of the ratio $c_n(\pmb r(\pmb\psi^{inc}))/c_n(\pmb r(\pmb\psi))$ when $n\to \infty$ is 1.
\end{proposition}

\begin{proof}
To see this, write
\begin{equation*}
    \frac{c_n(\pmb r(\pmb\psi^{inc}))}{c_n(\pmb r(\pmb\psi))} = \frac{n-2}{(n-1)\cos^2(\frac{\pi}{2n})-\cos^2[\frac{1}{2}(1-\frac{1}{n})\pi]}.
\end{equation*}

The term $\cos^2(\frac{1}{2}(1-\frac{1}{n})\pi)$ is bounded, therefore the limit reduces to
\begin{align*}
    \lim_{n\to\infty} \frac{c_n(\pmb r(\pmb\psi^{inc}))}{c_n(\pmb r(\pmb\psi))} &= \lim_{n\to\infty} \frac{n-2}{(n-1)\cos^2(\frac{\pi}{2n})} = \lim_{n\to\infty} \frac{n-2}{n-1} = 1.
\end{align*}
\end{proof}

\myindent To conclude this discussion on the violations of cycle inequalities, we want to mention that they satisfy the following property: If we let all negative weights on the inequality functional to be equal to zero, and all positive weights to be equal to $\sfrac{1}{2}$ the inequality is \emph{not} violated, i.e., 
\begin{equation*}
    c_n(\pmb{r}) = -r_{12}+r_{23}+\dots+r_{1n} = -0+\sum_{i=1}^{n-1}\frac{1}{2} = \frac{n-1}{2} \leq n-2,
\end{equation*}
for all $n \geq 3$. This is not a particularly common feature of event graph inequalities. For example, if we do the same with $h_4(\pmb{r}) \leq 1$ (see Eq.~\eqref{eq:k4_inequalities}) we have that $h_4(\pmb{r}) = 3/2 > 1$. 

\myindent While this is a seemingly trivial thing, it signals a specific type (or class) of inequalities that we refer to as \emph{magic witness candidates}. This is because there is a specific `type' of coherence, crucial for quantum computation, which is commonly referred to as \emph{magic} or also as \emph{nonstabilizerness}~\citep{bravyi2005universal}. States that lack magic are called \emph{stabilizer states} and they satisfy the specific constraint that pure multi-qubit states $\psi \in \mathcal{D}(\underbrace{\mathbb{C}^2 \otimes \dots \otimes \mathbb{C}^2}_{m\text{ times}})$ have discretized two-state overlaps, i.e., two pure stabilizer states must have overlaps such that $$\text{Tr}(\psi_i\psi_j) \in \{0,1,1/2,1/2^2,1/2^3,\dots,1/2^m\}.$$
Using this fact, it is simple to see that violations of cycle inequalities actually \emph{require} the states to be magic, as was shown by~\cite{wagner2024certifying}. 

\myindent While finding inequalities that can be considered candidates of magic witnesses is simple, proving that these are indeed magic witnesses with our current methods is non-trivial. The only \emph{other} inequality that we have found, satisfying the property described above, were the ones given by Eq.~\eqref{eq:kappa_ineq}. We have numerical evidence that those inequalities \emph{are} also magic witnesses, but have so-far failed to provide a proof. 

\subsubsection{Quantum violations of various facet-defining inequalities}\label{sec: quantum_violations_C(K_5)}

\myindent Having finished our focus on $n$-cycle graphs, we now investigate the violations of families of inequalities bounding $\mathfrak{C}(K_5)$.~\footnote{The results of this short subsection are adapted from an Appendix of reference~\citep{wagner2024inequalities}.} We  show that there are specific intriguing violations that have some interesting symmetry of states in the Bloch sphere. The optimal violations found using seesaw techniques are shown in Table~\ref{tab: seesaw bound}.  

\begin{figure}[t]
    \centering
    \includegraphics[width=0.8\linewidth]{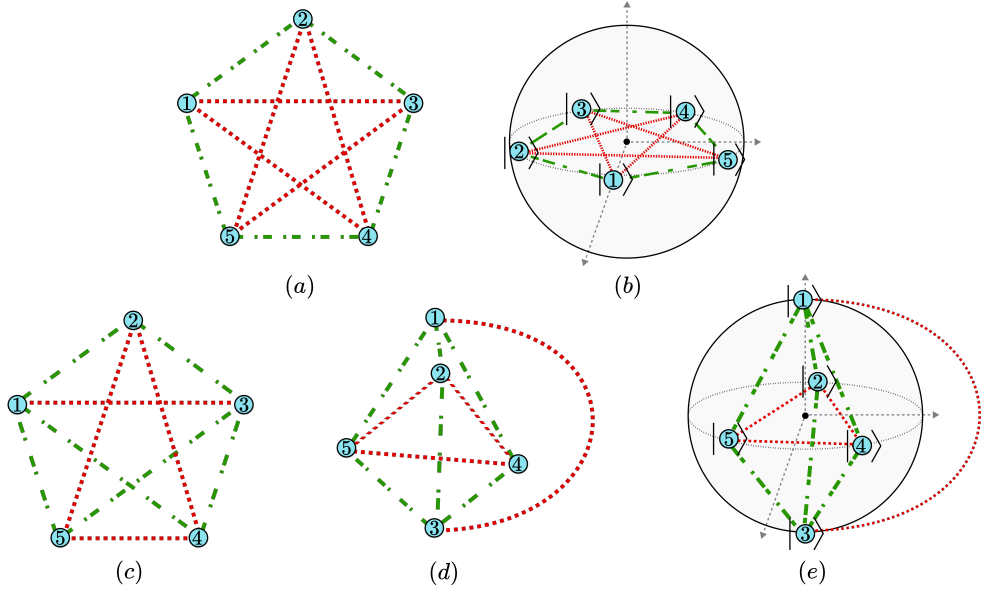}
    \caption{\textbf{Qubit states violating event-graph inequalities.} (a) depicts the inequality $r_{12}+r_{23}+r_{34}+r_{45}+r_{15}-r_{13}-r_{14}-r_{24}-r_{25}-r_{35}\leq 2$ with edges corresponding to positive terms in green (dash-dotted lines) and to negative terms in red (dashed-only lines). (b) shows a set of five pure states equally spaced over a great circle of the Bloch sphere, which violates this inequality attaining a value of $\sfrac{5\sqrt{5}}{4} > 2$. (c) depicts the  inequality $r_{12}+r_{14}+r_{15}+r_{23}+r_{34}+r_{35}-r_{13}-r_{24}-r_{25}-r_{45} \leq 2$ as in (a).
     (d) depicts the same inequality with the graph displayed in a different geometric configuration, mirroring that of a set of states in the Bloch sphere that largely violates it.
     (e) represents that set of five pure states in the Bloch sphere: three states equally spaced around the equator plus the two eigenstates of the Pauli matrix $Z$; this set of states attains a value of $\sfrac{9}{4} > 2$ for the inequality.}
    \label{fig: qubit_violations}
\end{figure}

\myindent Consider the inequality $I_{(K_5,5)}(\pmb r) \leq 2$ from Eq.~\eqref{eq:K5_class_5}, which we recall, is given by
$$I_{(K_5,5)}(\pmb r) := +r_{12}+r_{15}+r_{23}+r_{34}+r_{45}-(r_{13}+r_{14}+r_{24}+r_{25}+r_{35})\leq 2$$
It can be violated using a quantum realization provided by labeling the vertices with the vector states 
\begin{align}
    \vert \psi_k \rangle = \frac{1}{\sqrt{2}}\left(\vert 0 \rangle + e^{2\pi i k/5}\vert 1 \rangle\right)
\end{align}
with $k=0,\dots,4$.
This quantum realization attains a value of $\sfrac{5\sqrt{5}}{4}$ and hence a violation of $\sfrac{5\sqrt{5}}{4}-2\approx 0.79508$. As we see from Table~\ref{tab: seesaw bound}, these are the optimal lower bounds for this inequality. 

\myindent We have also found another interesting  violation with single-qubits symmetrically arranged in the Bloch sphere for the inequality~\eqref{eq:K5_class_4} given by 
$$I_{(K_5,4)}(\pmb r) = +r_{12}+r_{14}+r_{15}+r_{23}+r_{34}+r_{35}-(r_{13}+r_{24}+r_{25}+r_{45})\leq 2.$$
The quantum violation is found using single-qubit states depicted in~\cref{fig: qubit_violations}: choosing $\vert \psi_2 \rangle, \vert \psi_4 \rangle, \vert \psi_5 \rangle $ equally distributed on the equator of the Bloch sphere, i.e. separated by angles of $\sfrac{2\pi}{3}$, implying that $r_{24}=r_{25}=r_{45}=\sfrac{1}{4}$, and choosing $\vert \psi_1 \rangle = \vert 0 \rangle ,\vert \psi_3 \rangle = \vert 1 \rangle$, implying that $r_{13}=0$ and all remaining overlaps are equal to $\sfrac{1}{2}$.
This set of vectors attains the value $\sfrac{6}{2}-\sfrac{3}{4}=\sfrac{9}{4}$ and hence a violation of $\sfrac{9}{4}-2=\sfrac{1}{4}$. These symmetrically arranged qubit states are also the states used in the construction of the elegant joint measurement introduced by~\cite{gisin2019entanglement}. However, we could find a higher violation of the same inequality using qutrits as shown in Table~\ref{tab: seesaw bound}.

\myindent The maximal quantum violation found for the inequality $h_4(\pmb r) \leq 1$ used the following four qutrit vector states:
\begin{align*}
    \ket{\psi_1} &= \ket{0} \\
    \ket{\psi_2} &= \sqrt{\frac{5}{9}}\ket{0} + \sqrt{\frac{4}{9}}\ket{1} \\
    \ket{\psi_3} &= \sqrt{\frac{5}{9}}\ket{0} - \sqrt{\frac{1}{9}}\ket{1} + i\sqrt{\frac{1}{3}}\ket{2} \\
    \ket{\psi_4} &= \sqrt{\frac{5}{9}}\ket{0} - \sqrt{\frac{1}{9}}\ket{1} - i\sqrt{\frac{1}{3}}\ket{2}\\
\end{align*}
This set of states attains a value of $\sfrac{4}{3}$ and hence violation of $\sfrac{4}{3}-1=\sfrac{1}{3}$.

\begin{table}[t]
    \centering
    \begin{tabular}{cccccccc}
    \hline\hline
       Inequality &  & $d=2$  &  $d=3$ & $d=4$ & $d=5$ & $d=6$ \\
    \hline
        Eq.~\eqref{eq:cycle_inequalities} &$c_3(\pmb r) \leq 1$  & \textbf{1.2500} & 1.2500 & 1.2500 & 1.2500 & 1.2500 \\
        Eq.~\eqref{eq:hn_recursively} &$h_4(\pmb r) \leq 1$ & 1.0000 & \textbf{1.3333} & 1.3333 & 1.3333 & 1.3333 \\
        Eq.~\eqref{eq:hn_recursively}& $h_5(\pmb r) \leq 1$& 0.2500 & 1.0000 & \textbf{1.3750} & 1.3750 & 1.3750 \\
        Eq.~\eqref{eq:K5_class_4} &$I_{(K_5,4)}(\pmb r) \leq 2$ & \textbf{2.2500} & 2.3333 & 2.3333 & 2.3333 & 2.3333 \\
        Eq.~\eqref{eq:K5_class_5} &$I_{(K_5,5)}(\pmb r) \leq 2$ & \textbf{2.7951} & 2.7951 & 2.7951 & 2.7951 & 2.7951 \\
        Eq.~\eqref{eq: hn2 for all n}& $h_5^{(2)}(\pmb r) \leq 3$ & 3.0000 & \textbf{3.3333} & 3.5000 & 3.5000 & 3.5000 \\        
        Eq.~\eqref{eq:K5_class_7} &$I_{(K_5,6)}(\pmb r) \leq 3$ & \textbf{3.6592} & 3.8095 & 3.8095 & 3.8095 & 3.8095 \\
        
        Eq.~\eqref{eq:K5_class_8}& $I_{(K_5,7)}(\pmb r) \leq 3$ & \textbf{3.1250} & 3.5000 & 3.5000 & 3.5000 & 3.5000 \\
        
        Eq.~\eqref{eq:K5_class_9} &$I_{(K_5,8)}(\pmb r) \leq 5 $ & \textbf{5.6666} & 6.0000 & 6.0000 & 6.0000 & 6.0000 \\
        
        Eq.~\eqref{eq: first K33 inequality} &$I_{(K_{3,3},1)}(\pmb r) \leq 3$ & \textbf{3.5000} & 3.6893 & 3.6893 & 3.6893 & 3.6893 \\
        
        Eq.~\eqref{eq: second K33 inequality}& $I_{(K_{3,3},2)}(\pmb r) \leq 6$  & \textbf{7.1845} & 7.1871 & 7.1871 & 7.1871 & 7.1871 \\
        
        Eq.~\eqref{eq:kcbs_event_graph_inequality} &$h_{\mathrm{KCBS}}(\pmb r) \leq 2$ & \textbf{3.2500} & 3.2500 & 3.2500 & 3.2500 & 3.2500 \\
        Eq.~\eqref{eq: hn2 for all n}& $h_6^{(2)}(\pmb r) \leq 3$ & 2.2500 & 3.0000 & \textbf{3.3750} & 3.6000 & 3.6000 \\
        Eq.~\eqref{eq: h63 for the specific case} &$h_6^{(3)}(\pmb r) \leq 6$ & 6.0000 & \textbf{6.3333} & 6.5000 & 6.6000 & 6.6000 \\
    \hline\hline
    \end{tabular}
    \caption{\textbf{Seesaw bounds for the violation of event graph inequalities.} The seesaw iteration number for all shown values is 40. In bold we show the first Hilbert space dimension in which one has a violation of the event graph inequality. We also have numerical confirmation of the fact that optimal inequality violations for any event graph $G$ need to be tested solely for $2 \leq d \leq |V(G)|$ due to unitary invariance. We see that $h_4$, $h_5$, $h_5^{(2)}$, $h_6^{(2)}$, and $h_{6}^{(3)}$ are all (candidates of) witnesses of coherence and dimension (cf. Chapter~\ref{chapter: applications}). }
    \label{tab: seesaw bound}
\end{table}

\myindent We remark once more that the above violations are \stress{not} necessarily optimal. They were found using techniques of semidefinite programming over the quantum set. In future work, we aim to investigate---using hierarchy of \acrshort{sdp} relaxations---if these inequality violations are indeed optimal. We also note that Table~\ref{tab: seesaw bound} largely improves over the bounds found initially by~\cite{wagner2024inequalities} using Monte Carlo simulations. 

\subsubsection{Quantum violations of the infinite family of inequalities $h_n(\pmb{r}) \leq 1$}

\myindent In this subsection, we investigate the optimal quantum violations of the family of inequalities $h_n(\pmb{r}) \leq 1$ described in Eq.~\eqref{eq:hn_recursively}. Our main goal is to show that it is possible to re-write the problem of optimizing the quantum realizations of $h_n(\pmb r)$ as a \emph{quadratic} \acrshort{sdp}. We numerically implement such a family of \acrshort{sdp}s in Chapter~\ref{chapter: applications}, where we conclude that these can be used to certify numerically that  $h_n(\pmb r) \leq 1$, for $n \geq 4$, are witnesses of coherence \emph{and} Hilbert space dimension. 

\myindent We start with two preliminary lemmas. 

\begin{lemma}[Adapted from~\cite{giordani2023experimental}]\label{lemma: hn+=trX2}
Let $h_n^+(\pmb r)$  be the \textit{sum} of all edge weightings for an event graph $K_n$. Let $\psi:V(K_n) \to \mathrm{ext}(\mathcal{D}(\mathcal{H}))$ be any pure-state quantum realization, with $\psi(V(G)) = \{\vert \psi_i \rangle \langle \psi_i \vert \}_{i=1}^n$. Then, we have that
\begin{equation}
    h_n^+(\pmb r) = \frac{n^2}{2}\mathrm{Tr}(X^2)- \frac{n}{2}
\end{equation}
where $X = \frac{1}{n}\sum_{i=1}^n \vert \psi_i \rangle \langle \psi_i \vert$. 
\end{lemma}

We would like to point out that the lemma above is an elementary lemma, which has also been shown to hold (using a different language) by~\cite{brunner2013dimension}. 

\begin{proof}
    Noting that $X^2$ is given by
\begin{equation*}
    X^2 = \frac{1}{n^2} \sum_{i,j=1}^n \vert \psi_i \rangle \langle \psi_i \vert \psi_j \rangle \langle \psi_j \vert  = \frac{1}{n^2}\sum_{i\neq j}\vert \psi_i \rangle \langle \psi_i \vert \psi_j \rangle \langle \psi_j \vert  + \frac{1}{n^2}\sum_{i=j}\vert \psi_i \rangle \langle \psi_i \vert 
\end{equation*}
we find that 
\begin{equation*}
    \text{Tr}(X^2)  = \frac{1}{n^2}\sum_{i\neq j}r_{i,j}(\pmb{\psi}) + \frac{1}{n} = \frac{2}{n^2}\sum_{i<j}r_{i,j}+\frac{1}{n},
\end{equation*}
where we have used the fact that $r_{i,j} = r_{j,i}$. Now, since $h_n^+(\pmb r) = \sum_{i < j}r_{i,j}$ we conclude the following:
\begin{equation*}
    \text{Tr}(X^2) = \frac{2}{n^2}h_n^+(\pmb r) + \frac{1}{n} \implies h_n^+(\pmb r) = \frac{n^2}{2}\left(\text{Tr}(X^2) - \frac{1}{n}\right) = \frac{n^2}{2}\text{Tr}(X^2)-\frac{n}{2},
\end{equation*}
as we wanted to show.
\end{proof}

\myindent We may also note that, due to the recurrence description of $h_n(\pmb r)$ it is possible to use only $h_n^+(\pmb r)$ to describe the inequality.

\begin{lemma}[Adapted from~\cite{giordani2023experimental}]\label{lemma: hn = hn+-2hn+}
 The inequality functional $h_n(\pmb r)$ given by Eq.~\eqref{eq:hn_recursively} is equal to $h_n(\pmb r) = h_n^+(\pmb r)-2h_{n-1}^+(\pmb r)$.
\end{lemma}

\begin{proof}
As we have that 
\begin{equation*}
    \sum_{i=1}^n r_{i,n} = h_n^+(\pmb r)-h_{n-1}^+(\pmb r),
\end{equation*}
trivially we see then that 
\begin{equation*}
    h_n(\pmb r) = \sum_{i=1}^nr_{i,n} - h_{n-1}^+(\pmb r) = h_n^+(\pmb r)-h_{n-1}^+(\pmb r)- h_{n-1}^+(\pmb r)
\end{equation*}
as we wanted to show. 
\end{proof}

\myindent For the sake of clarity, let us consider an example of the above description for the lemma. Recalling that $h_3^+(\pmb r) = r_{1,2}+r_{1,3}+r_{2,3}$ then, 
\begin{align*}
    h_4^+(\pmb r)-2h_3^+(\pmb r) &= r_{1,2}+r_{1,3}+r_{1,4}+r_{2,3}+r_{2,4}+r_{3,4}-2(r_{1,2}+r_{1,3}+r_{2,3})\\
    &=r_{1,4}+r_{2,4}+r_{3,4}-r_{1,2}-r_{1,3}-r_{2,3} = h_4(\pmb r).
\end{align*}

This is equivalent to $h_4(\pmb r)$ up to a relabeling of vertices $1$ and $4$. We can use the lemmas above to write $h_n(\pmb r)$ only in terms of $X$, which eventually leads to a semidefinite programming description of the problem of maximizing $h_n(\pmb r)$ with quantum realizations of pure states.

\begin{theorem}[Adapted from~\cite{giordani2023experimental}]\label{theorem: h_n_SDP}
    The quantum realizations of the inequality functional $h_n(\pmb r)$ given by~\eqref{eq:hn_recursively} can be expressed as a quadratic semidefinite program optimized over  quantum states $X$. 
\end{theorem}

\begin{proof}
    From~\cref{lemma: hn = hn+-2hn+} and~\cref{lemma: hn+=trX2} together we can write $h_n(\pmb r)$ as,
    \begin{equation}
        h_n(\pmb r) = \left(\frac{n^2}{2}\text{Tr}(X^2)-\frac{n}{2}\right) - 2\left(\frac{(n-1)^2}{2}\text{Tr}(X_\star^2)-\frac{n-1}{2}\right)
    \end{equation}
where we have that $X_\star = \frac{1}{n-1}\sum_{i=1}^{n-1}\vert \psi_i \rangle \langle \psi_i \vert$. In other words, the second term of $h_n$ lacks $\vert \psi_n\rangle \langle \psi_n \vert$. It is clear that we may write $X$ in terms of $X_\star$ provided that we fix $\vert \psi_n \rangle$. As $h_n(\pmb r(\pmb{\rho}))$ is \acrshort{pu} invariant we can unitarily transform any set of states such that $\vert \psi_n\rangle = \vert 0 \rangle \in \mathbb{C}^d$ is the reference $d$-dimensional canonical basis vector of $\mathbb{C}^d$. In this case, we have that
\begin{align*}
    X = \frac{1}{n}\sum_{i=1}^n\vert \psi_i \rangle \langle \psi_i \vert = \frac{1}{n}\left(\sum_{i=1}^{n-1}\vert \psi_i\rangle\langle\psi_i \vert +\vert \psi_n \rangle \langle \psi_n \vert \right) = \frac{1}{n}((n-1)X_\star+\vert 0\rangle\langle 0 \vert )
\end{align*}
which implies that,
\begin{align*}
    X^2 &= \frac{1}{n^2}((n-1)X_\star+\vert 0\rangle\langle 0 \vert )((n-1)X_\star+\vert 0\rangle\langle 0 \vert ) \\
    &= \frac{1}{n^2}\left( (n-1)^2X_\star^2 + (n-1)X_\star \vert 0\rangle\langle 0\vert + (n-1)\vert 0\rangle\langle 0 \vert X_\star + \vert 0\rangle\langle 0\vert   \right)
\end{align*}
from which we infer that,
\begin{equation*}
    \text{Tr}(X^2) = \frac{1}{n^2}\left((n-1)^2\text{Tr}(X_\star^2) + 2(n-1)\langle 0 \vert X_\star \vert 0\rangle + 1\right).
\end{equation*}
This last expression allows us to write $h_n(\pmb r)$ in terms of $X_\star$ and $\vert 0\rangle \langle 0 \vert$ only.

\begin{align*}
    h_n(\pmb r) &= \frac{n^2}{2}\left( \frac{1}{n^2}\left((n-1)^2\text{Tr}(X_\star^2) + 2(n-1)\langle 0 \vert X_\star \vert 0\rangle + 1\right) \right) \\&- \frac{n}{2}-2\left(\frac{(n-1)^2}{2}\text{Tr}(X_\star^2)-\frac{n-1}{2}\right)\\
    &=\frac{(n-1)^2}{2}\text{Tr}(X_\star^2)+(n-1)\text{Tr}(\vert 0\rangle\langle 0\vert X_\star)+\frac{1}{2}-\frac{n}{2}-(n-1)^2\text{Tr}(X_\star^2)+n-1\\
    &=-\frac{(n-1)^2}{2}\text{Tr}(X_\star^2)+(n-1)\text{Tr}(\vert 0\rangle\langle 0\vert X_\star)+\frac{n-1}{2}
\end{align*}

If we fix $C := \vert 0\rangle\langle 0 \vert$ for $\vert 0\rangle \in \mathbb{C}^d$ the above form of $h_n$ is quadratic in $X_\star$. We can then write the following quadratic \acrshort{sdp} problem for any $n\in \mathbb{N}$,
\begin{align}
    \max_{X_\star \in \mathbb{C}_{d,d}} & A_n \langle X_\star,X_\star \rangle + B_n \langle C,X_\star \rangle + C_n \\
    \text{subject to } & X_\star \geq 0\\
    &\text{Tr}(X_\star)=1
\end{align}
for any $2 \leq d \leq n-1 $. Above we simply set, $A_n := -(n-1)^2/2$, $B_n := (n-1),$ and $C_n=(n-1)/2$. We also use the common notation for the trace inner product $\text{Tr}(A^\dagger B) = \langle A,B\rangle$.  
\end{proof}

\myindent The resulting optimized values of such quadratic \acrshort{sdp}, if they converge, provide \emph{upper bounds} for the values of $h_n(\pmb r)$ that can be reached with quantum realization using sets of $d$-dimensional pure quantum states. This is because the final optimization \emph{does not} take into consideration that $X_\star$ has the specific form described in terms of $\frac{1}{n}\sum_{i=1}^n \vert \psi_i\rangle \langle \psi_i \vert $, which implies that, at least in principle, the obtained values could be higher than what is actually achievable by matrices restricted to have the form $\frac{1}{n}\sum_{i=1}^n \vert \psi_i\rangle \langle \psi_i \vert $. 

\myindent We implement this specific \acrshort{sdp} later when we discuss applications of such inequalities in Chapter~\ref{chapter: applications}, since the optimal values of this family of inequalities leads to dimension witnesses. Note also that, once a few straightforward changes are made, the same technique applies for the family of inequalities that was conjectured to be facet-defining as well, given by \begin{equation}h_n^{(m)}(\pmb r) \leq m(m+1)/2\end{equation} for $1 \leq m \leq n-2$. These can also be solved using a similar quadratic \acrshort{sdp}.

\subsection{Relational imaginarity from two-state overlaps}\label{subsec: relational imaginarity from overlaps}

\myindent We have introduced the notion of relational coherence in~\cref{def: quantum relational coherence}, but one of the motivations for doing so was to have a characterization of different `types' of coherence. One important example being that of \emph{relational imaginarity}, that we now define.~\footnote{Recall that we denote by $\mathbb{N}^*$ the set of all finite sequences of natural numbers.} 

\begin{definition}[Relational imaginarity]\label{definition: relational imaginarity}
    Let $\mathcal{W}$ be any finite subset of $\mathbb{N}^*$. We say that $\Delta$ has relational imaginarity if $\Delta \in \mathfrak{Q}(\mathcal{W})\setminus \mathfrak{Q}(\mathcal{W})|_{\mathrm{real}}$. 
\end{definition}

\myindent In particular, we say that edge weightings  $r$ have relational imaginarity if they have relational coherence but $r \notin \mathfrak{Q}(G)|_{\mathrm{real}}$, i.e., they cannot be realized by imaginarity-free tuples of states. 

\myindent From the results we have presented so far, it appears that edge-weightings $r$ \textit{cannot} have relational imaginarity. For example, we show in~\cref{proposition: pure and real are the same} that there is no tuple $\pmb{r} \in \mathfrak{Q}(C_3)|_{\mathrm{pure}}$ having relational imaginarity. To show this, we have used the relationship between the candidate matrix $H(\pmb{\Delta})$ defined by Eq.~\eqref{eq: candidate matrix}, its positive semidefiniteness, and the quantum realizability of $\pmb{\Delta}$. We can now investigate quantum realizability relative to a different candidate matrix. If we now consider the Gram matrix from Eq.~\eqref{eq: Gram matrix K4},  corresponding to a scenario involving four, rather than three states, it is straightforward to define the following \emph{different} candidate matrix

\begin{equation}\label{eq: candidate matrix for imaginarity proof}
        R(\pmb{\Delta}) = \left(\begin{matrix}
            1 & r_{12} & r_{13} & r_{14}\\
            r_{12} & 1 & r_{23}e^{i\phi_{123}} & r_{24}e^{i\phi_{124}}\\
            r_{13} & r_{23}e^{-i\phi_{123}} & 1 & r_{34}e^{i\phi_{134}}\\
            r_{14} & r_{24}e^{-i\phi_{124}} & r_{34}e^{-i\phi_{134}} & 1
        \end{matrix}\right),
    \end{equation}
where now we have $$\pmb{\Delta} = (r_{12},r_{13},r_{14},r_{23},r_{24},r_{34},\Delta_{123},\Delta_{124},\Delta_{134}),$$
and $e^{i\phi_{1jk}} = \Delta_{1jk}/|\Delta_{1jk}|$ when $\Delta_{1jk} \neq 0$, and $e^{i\phi_{1jk}} = 1$ if $\Delta_{1jk} = 0$. If we define 
\begin{equation}\label{eq: W for 1234}
    \mathcal{W}_{1234} := \{(1,2),(1,3),(1,4),(2,3),(2,4),(3,4),(1,2,3),(1,2,4),(1,3,4)\},
\end{equation}
we have that $\Delta: \mathcal{W}_{1234} \to \mathbb{C}$ is the function related to the tuple described above. 

\myindent Note that, by construction, whenever $r_e \neq 0$ for all $e\in E(K_4)$ and $\pmb{\Delta}$ is quantum realizable, we have that the candidate matrix $R(\pmb \Delta)$ becomes exactly the Gram matrix described by Eq.~\eqref{eq: Gram matrix K4}. Because of that, we can proceed similarly as with~\cref{lemma: nonzero overlaps candidates determine pure state realizability} to show the following. 

\begin{lemma}[Adapted from~\cite{fernandes2024unitary}]\label{lemma: Candidate matrices real/imaginary}
    Let $R(\pmb{\Delta})$ be the candidate matrix as defined in Eq.~\eqref{eq: candidate matrix for imaginarity proof}. A tuple $\pmb{\Delta} = (r_{12},r_{13},r_{14},r_{23},r_{24},r_{34},\Delta_{123},\Delta_{124},\Delta_{134})$ where all $r_{ij}$ coordinates are non-zero is quantum realizable by pure states if, and only if, the candidate matrix $R(\pmb{\Delta})$ is positive semidefinite. 
\end{lemma}

\begin{proof}
    The proof follows the same strategy of~\cref{lemma: nonzero overlaps candidates determine pure state realizability}. If $\pmb{\Delta}\in \mathbb{C}^{\mathcal{W}_{1234}}$ is quantum realizable, $R(\pmb{\Delta}) = G_{\pmb{\psi}}^\star$ is as given by Eq.~\eqref{eq: Gram matrix K4} by construction, implying that it is positive semidefinite. On the other hand, if $R(\pmb{\Delta})$ is positive semidefinite, from the fact all $r_{ij} \neq 0$ we can use a Cholesky decomposition to find a set of vectors $\pmb{\Psi}$ such that $R(\pmb{\Delta}) = G_{\pmb{\Psi}} = U G_{\pmb{\psi}}^\star U^\dagger$ for some diagonal unitary $U$. Just as in~\cref{lemma: nonzero overlaps candidates determine pure state realizability}, this equality between matrices implies that, in fact, $R(\pmb{\Delta}) = G_{\pmb{\psi}}^\star$ from which we conclude that $\pmb{\Delta}$ is quantum realizable by the states given by some Cholesky decomposition~\footnote{Since we only assume $R(\pmb{\Delta})$ to be positive semidefinite, there is no unique Cholesky decomposition. We discuss this decomposition, and other elementary facts of matrix theory and linear algebra in~\cref{app: basic algebra}. } of $R(\pmb{\Delta})$.
\end{proof}

\begin{corollary}[Adapted from~\cite{fernandes2024unitary}]\label{corollary: real from real positive semidefinite matrix R}
    Let $R(\pmb{\Delta})$ be the candidate matrix as defined in Eq.~\eqref{eq: candidate matrix for imaginarity proof}. A tuple $\pmb{\Delta}\in \mathbb{C}^{\mathcal{W}_{1234}}$ where all $r_{ij}$ coordinates are non-zero is quantum realizable by imaginarity-free pure states, i.e. $\pmb{\Delta} \in \mathfrak{Q}(\mathcal{W}_{1234})|_{\mathrm{real}}$, if and only if $R(\pmb{\Delta})$ is real positive semidefinite.
\end{corollary}

\begin{proof}
    Clearly, if  $\pmb{\Delta} = \pmb{\Delta}(\pmb{\psi}^{\mathbb{R}})$ for some vertex labeling $\pmb{\psi}^{\mathbb{R}}: V(K_4) \to \mathcal{S}$ where $\mathcal{S}$ is an imaginarity-free set (see Def.~\ref{def: set imaginarity}) we have then that $R(\pmb{\Delta})$ is a real positive semidefinite matrix. For the converse, if $R(\pmb{\Delta})$ is real positive semidefinite, its Cholesky decomposition yields a valid tuple of real states $\pmb{\psi}^{\mathbb{R}}$ for which $\pmb{\Delta} = \pmb{\Delta}(\pmb{\psi}^{\mathbb{R}})$. In all this discussion, we have restricted to the case where $r_{ij} \neq 0$ for all $\{i,j\} \in E(K_4)$.
\end{proof}

\myindent We now show that, in fact, \emph{it is possible} to witness imaginarity using tuples of two-state overlaps. 

\begin{theorem}[Adapted from~\cite{fernandes2024unitary}]
    There exists $\pmb{r}$ that cannot be realized by imaginarity-free pure quantum states, i.e., $\mathfrak{Q}(K_4)|_{\mathrm{pure}} \setminus \mathfrak{Q}(K_4)|_{\mathrm{real}} \neq \emptyset$.
\end{theorem}

\begin{proof}
    Let $\mathcal{W}_{1234}$ given by Eq.~\eqref{eq: W for 1234}. Moreover, we let $G=K_4$ be the complete event graph of four nodes, with $V(K_4) = \{1,2,3,4\}$. Furthermore, assume that $r_{ij} \neq 0$ for all $\{i,j\} \in E(K_4)$. 
    
    \myindent From Lemma~\ref{lemma: Candidate matrices real/imaginary} and~\cref{corollary: real from real positive semidefinite matrix R} we have that  $\pmb{\Delta} \in \mathfrak{Q}(\mathcal{W}_{1234})|_{\mathrm{real}}$ iff $R(\pmb{\Delta})$ is a real \acrshort{psd} matrix. From the form of $R(\pmb{\Delta})$, this happens iff the coordinates $\Delta_{123},\Delta_{124},\Delta_{134}$ are all real-valued, happening iff  $\phi_{123},\phi_{124},\phi_{134} \in \{ 0,\pi\}$. In what follows, we momentarily denote $R(\pmb{\Delta}) = R(\pmb{r},\phi_{123},\phi_{124},\phi_{134})$.

    \myindent We can phrase our proof idea as follows: Suppose that for all pure state realizable $\pmb{r}(\pmb{\psi})$, with respect to some vertex labeling $\psi$, there exists another imaginarity-free one $\psi^{\mathbb{R}}$ such that $\pmb{r} = \pmb{r}(\pmb{\psi}) = \pmb{r}(\pmb{\psi}^{\mathbb{R}})$. This happens iff for any $\pmb{r} = \pmb{r}(\pmb{\psi})$ there exists some choice $\phi_{123},\phi_{124},\phi_{134} \in \{ 0,\pi\}$ for which $R(\pmb{r},\phi_{123},\phi_{124},\phi_{134})$ is a (real) \acrshort{psd} matrix. There are in total 8 such choices for each tuple $\pmb{r}$. We want, therefore, to find a quantum realizable tuple $\pmb{r}$ such that none of these matrices $\{R(\pmb{r},\phi_{123},\phi_{124},\phi_{134})\}_{\phi_{123},\phi_{124},\phi_{134} \in \{0,\pi\}}$ are \acrshort{psd}. 

    \myindent Take the following vertex labeling $\psi: V(K_4) \to \text{ext}(\mathcal{D}(\mathbb{C}^2))$ given by
    \begin{align*}
        \psi(1) &= \vert 0\rangle \langle 0 \vert \\
        \psi(2) &=\vert +\rangle \langle + \vert \\
        \psi(3) &= \vert -_i\rangle \langle -_i \vert \\
        \psi(4) &= \vert \pi/6,\pi/4\rangle \langle \pi/6,\pi/4 \vert 
    \end{align*}
    where $\vert \pi/6,\pi/4\rangle := \cos(\pi/6)\vert 0\rangle + e^{i\pi/4}\sin(\pi/6)\vert 1\rangle$. In this case, we have that 
    \begin{align*}
        &r_{12}(\pmb{\psi}) = r_{13}(\pmb{\psi}) = r_{23}(\pmb{\psi}) = \frac{1}{2}\\
        &r_{14}(\pmb{\psi})= \frac{3}{4},r_{24}(\pmb{\psi}) = \frac{4 + \sqrt{6}}{8},r_{34}(\pmb{\psi}) = \frac{4-\sqrt{6}}{8},
    \end{align*}
    and it is simple to numerically see that none of the aforementioned 8 matrices can be positive semidefinite since they  all have negative eigenvalues. Let us define $\lambda_{\mathrm{min}}[M] := \min \mathrm{spec}(M)$ as the smallest eigenvalue $\lambda \in \mathrm{spec}(M)$ of a matrix $M$, we then have that 
    \begin{align*}
        \lambda_{\mathrm{min}}[R_\phi(\pmb{r}(\pmb{\psi}),0,0,0)] &= −0.044984\\
        \lambda_{\mathrm{min}}[R_\phi(\pmb{r}(\pmb{\psi}),\pi,0,0)] &= −0.512315\\
        \lambda_{\mathrm{min}}[R_\phi(\pmb{r}(\pmb{\psi}),0,\pi,0)] &= −0.709002\\
        \lambda_{\mathrm{min}}[R_\phi(\pmb{r}(\pmb{\psi}),0,0,\pi)] &= −0.561292\\
        \lambda_{\mathrm{min}}[R_\phi(\pmb{r}(\pmb{\psi}),\pi,\pi,0)] &= −0.837603\\
        \lambda_{\mathrm{min}}[R_\phi(\pmb{r}(\pmb{\psi}),0,\pi,\pi)] &=−0.704281\\
        \lambda_{\mathrm{min}}[R_\phi(\pmb{r}(\pmb{\psi}),\pi,0,\pi)] &=−0.491359\\
        \lambda_{\mathrm{min}}[R_\phi(\pmb{r}(\pmb{\psi}),\pi,\pi,\pi)] &=−1.17472.
    \end{align*}
    This concludes the proof. 
\end{proof}

\myindent Therefore, we have shown that it is possible to certify imaginarity using only the values of two-state overlaps. To test the predictions above, one can take all the 8 matrices considered in the proof and use Sylvester's criterion do find a complete set of \emph{non-linear} functionals (based on the determinant of the principal minors) for each matrix. For each choice of $\phi_{123},\phi_{124},\phi_{134} \in \{0,\pi\}$ we consider the principal minors of these matrices as $M(R_\phi(\pmb{r},\phi_{123},\phi_{124},\phi_{134}))$ yielding the inequalities $$\det(M(R_\phi(\pmb{r},\phi_{123},\phi_{124},\phi_{134}))) \geq 0.$$
If, for every choice of $\phi_{123},\phi_{124},\phi_{134} \in \{0,\pi\}$, at least one of these nonlinear inequalities over the overlaps is violated, we have certified relational imaginarity. This requires the estimation of $6$ two-state overlaps. In what follows, we show that access to higher-order Bargmann invariants allows us to witness relational imaginarity from the value of a \emph{single} Bargmann invariant.  

\section{Relational coherence beyond the event graph approach}\label{sec: relational coherence higher order Bargmanns}

\myindent So far, we have focused almost exclusively on the verification of relational coherence based on quantum-realizable edge weightings. We now turn to a few basic observations about relational coherence \emph{beyond} what can be captured by two-state overlaps.

\myindent Let us fix a finite set of words $\mathcal{W} \subseteq \mathbb{N}^*$ and let $V$ be the set of positive integers  which appear in at least one of the words in $\mathcal{W}$. If $\Delta \in \mathfrak{Q}(\mathcal{W})|_{\mathrm{inc}}$, then there exists a map $\varrho: V \to \mathcal{D}(\mathcal{H})$ such that $\Delta = \Delta(\varrho)$, and the image $\varrho(V)$ is set incoherent. This implies that $[\varrho(i),\varrho(j)] = 0$ for every $i,j \in V$, and therefore we can formulate a condition analogous to that used for edge weightings $\pmb{r} \in \mathfrak{Q}(G)|_{\mathrm{inc}}$. 

\myindent In this case, the condition $\pmb{\Delta}\in \mathfrak{Q}(\mathcal{W})|_{\mathrm{inc}}$ implies that, for all words $\pmb{w} = (w_1,\ldots,w_m) \in \mathcal{W}$, 
\begin{equation}\label{eq: relational incoherence for Bargmanns}
    \Delta_{\pmb{w}}(\varrho) = \text{Tr}(\varrho(w_1)\cdots \varrho(w_m)) = \sum_{\lambda} p(\lambda|\varrho(w_1))\cdots p(\lambda|\varrho(w_m))
\end{equation}
for some orthonormal basis $\mathbb{A} = \{\vert \lambda\rangle \}_\lambda$. 

\myindent As in the case of edge weightings, it is natural to impose consistency conditions analogous to those that define the event graph polytope $\mathfrak{C}(G)$. This leads to the possibility of defining a new class of convex polytopes $\mathfrak{C}(\mathcal{W})$. A full characterization of these new polytopes is left for future work, but we highlight two simple (yet interesting) consequences of Eq.~\eqref{eq: relational incoherence for Bargmanns}.

\subsection{Equality constraints}\label{subsec: equality constraints}

\myindent We start by noticing that $\pmb{\Delta} \in \mathfrak{Q}(\mathcal{W})|_{\mathrm{inc}}$ iff for all possible permutations $\pi$ acting on a word $\pmb{w} \in \mathcal{W}$ it holds that 
\begin{equation}\label{eq: equality constraints}
    \Delta_{\pmb{w}}(\varrho) = \Delta_{\pi(\pmb{w})}(\varrho)
\end{equation}
for all possible incoherent realizations $\varrho$. This is trivial to see from Eq.~\eqref{eq: relational incoherence for Bargmanns}. Now, intriguingly, this implies that all such equality constraints can be understood as coherence witnesses. In particular, this allows one to easily witness set coherence of \emph{two}-states, for example, from the equality constraint $$\text{Tr}(\rho^2\sigma^2) = \text{Tr}(\rho\sigma\rho\sigma)$$ that must hold for all incoherent pairs of states. There are examples of coherent states---sets of real states, see~\cref{subsec: relational imaginarity}---that satisfy some of these equality constraints, showing that each such constraint is (in principle) a weaker requirement than just noncommutativity. 

\myindent Let us apply these equality constraints to an example~\citep{li2025multistateimaginaritycoherencequbit}. Recall the example we have considered in Chapter~\ref{chapter: quantum coherence} to show that $\mathcal{S}_{\mathrm{inc}}$ is non-convex. In that case, we have considered the two sets $\mathcal{S}_1,\mathcal{S}_2$ given by Eq.~\eqref{eq: set coherence nonconvex}. Let us consider the labelings for $V = \{1,2\}$, given by
\begin{equation}\label{eq: vertex label rho example}
    \varrho(1) = \vert 0\rangle \langle 0 \vert, \varrho(2) = \frac{1}{3}\vert 0\rangle \langle 0 \vert + \frac{2}{3}\vert 1\rangle \langle 1 \vert 
\end{equation}
and
\begin{equation}\label{eq: vertex label sigma example}
    \varsigma(1) = \vert +\rangle \langle + \vert, \varsigma(2) = \frac{1}{4}\vert +\rangle \langle + \vert + \frac{3}{4}\vert -\rangle \langle - \vert. 
\end{equation}
We now investigate if $\xi := \omega \varrho+(1-\omega)\varsigma$ is a set incoherent vertex labeling, where we are using
\begin{equation*}
    \xi(1) = (\omega \varrho+(1-\omega)\varsigma)(1) = \omega \varrho(1)+(1-\omega)\varsigma(1) = \left(\begin{matrix}
        \omega+\sfrac{(1-\omega)}{2} & \sfrac{(1-\omega)}{2} \\
        \sfrac{(1-\omega)}{2}&\sfrac{(1-\omega)}{2}
    \end{matrix}\right),
\end{equation*}
\begin{equation*}
    \xi(2) = (\omega \varrho+(1-\omega)\varsigma)(2) = \omega \varrho(2)+(1-\omega)\varsigma(2) = \left(\begin{matrix}
        \sfrac{\omega}{3} + \frac{1-\omega}{2} & -\sfrac{1-\omega}{4} \\
        -\sfrac{1-\omega}{4} & \sfrac{2\omega}{3}+\frac{1-\omega}{2}
    \end{matrix}\right).
\end{equation*}
To investigate this, we use the equality constraint $\Delta_{1212} = \Delta_{1122}$. In this case we have that 
\begin{align*}
    &\Delta_{1122}(\xi) = \text{Tr}(\xi(1)\xi(1)\xi(2)\xi(2)) = \frac{1}{144} (9 + 18 \omega + 94 \omega^2 - 54 \omega^3 + 13 \omega^4)\\
    &\Delta_{1212}(\xi) = \text{Tr}(\xi(1)\xi(2)\xi(1)\xi(2)) = \frac{1}{144} (9 + 18 \omega + 90 \omega^2 - 46 \omega^3 + 9 \omega^4)
\end{align*}
from which we conclude that convex combinations of set incoherent vertex labelings do not preserve the set incoherence property, since $\Delta_{1122}(\xi) \neq \Delta_{1212}(\xi)$ for all $0 < \omega< 1$. This can be seen from the fact that 
\begin{equation*}
    \Delta_{1122}(\xi)-\Delta_{1212}(\xi) = \frac{1}{36}\omega^2(1-\omega)^2.
\end{equation*}
From the above, we conclude the following proposition. 
\begin{proposition}
    Let $\mathcal{S}_{\mathrm{inc}} \subseteq 2^{\mathcal{D}(\mathcal{H})}$ be the set of all possible subsets of $\mathcal{D}(\mathcal{H})$ that are set incoherent. Then, the set of all possible vertex labelings $\varrho: V(G) \to \mathcal{S} \in \mathcal{S}_{\mathrm{inc}}$ is not convex. In other words, for $\mathcal{S}_1,\mathcal{S}_2 \in \mathcal{S}_{\mathrm{inc}}$ there are vertex labelings $\varrho: V(G) \to \mathcal{S}_1, \varsigma: V(G) \to \mathcal{S}_2$ such that $\omega \varrho+(1-\omega)\varsigma: V(G) \to \mathcal{S}$, for $\omega \in (0,1)$, where $\mathcal{S} \notin \mathcal{S}_{\mathrm{inc}}$.
\end{proposition}

\myindent This was also shown by~\cite{designolle2021set}.  Another relevance of the equality constraints given by Eq.~\eqref{eq: equality constraints} is that they witness the set imaginarity of tuples of states, something that we now explore.

\subsection{Relational imaginarity from higher-order Bargmann invariants}\label{subsec: relational imaginarity}

\myindent Recall that from~\cref{proposition: word invariants and real realizability} that we have mentioned that for words $\pmb{w} \in \mathcal{W}$ such that $\pmb{w} = \pmb{w}^*$ the quantum realizable invariants $\Delta_{\pmb{w}} \in \mathbb{R}$. In fact, it holds that for all words for which $\pmb{w} \neq \pmb{w}^*$ the failure of the equality constraint $\Delta_{\pmb{w}} = \Delta_{\pmb{w}^*}$ witnesses relational imaginarity. For example, consider the simplest case of $\mathcal{W} = \{(1,2,3)\}$. In this case, we have that $\pmb{w}^* = (1,3,2)$ and therefore the equality constraint becomes
\begin{equation*}
    \Delta_{123}(\varrho)= \Delta_{132}(\varrho) = \Delta_{123}(\varrho)^*.
\end{equation*}
For every quantum realizable $\Delta_{123} \in \mathfrak{B}_3$ it holds that
\begin{equation}
    \text{Im}[\Delta_{123}(\varrho)]= \frac{1}{2i}(\Delta_{123}(\varrho)-\Delta_{132}(\varrho)),
\end{equation}
and similarly for higher order invariants. Since we have shown in~\cref{proposition: pure and real are the same} that it is \textit{impossible} to use the values of triplets of two-state overlaps of pure states to witness relational imaginarity, for witnessing set imaginarity of triplets of states we need to estimate $\Delta_{123}(\psi)$. 

\myindent Let us show one example of the usefulness of values of higher-order invariants, in particular related to relational imaginarity, when compared to using two-state overlaps only. 

\begin{example}[Bargmann invariants can capture more relational coherence than two-state overlaps]
    Let us recall~\cref{example: realization coherent and incoherent} where we had the tuple $\pmb{r} = (\sfrac{1}{2},\sfrac{1}{2},\sfrac{1}{2})$ realizable both by a vertex labeling $\varrho^{\mathrm{cohe}}$ whose image is set coherent and given by $\{\vert 0\rangle \langle 0 \vert, \vert +\rangle \langle + \vert, \vert +_i\rangle \langle +_i \vert \}$, and by a vertex labeling $\varrho^{\mathrm{diag}}$ whose image is set incoherent and given by the singleton $\{\mathbb{1}/2\}$. In this case, $\pmb{r} = \pmb{r}(\pmb{\rho}^{\mathrm{diag}}) = \pmb{r}(\pmb{\rho}^{\mathrm{cohe}})$. 

    However, a single third-order Bargmann invariant can easily distinguish between the two realizations, due to relational imaginarity of the one related to $\varrho^{\mathrm{cohe}}$, while the other yields a real third-order invariant. Mathematically, we have that 
    \begin{equation*}
        \Delta_3(\varrho^{\mathrm{cohe}}) = \langle 0|+\rangle \langle + \vert +_i\rangle \langle +_i \vert 0\rangle \notin \mathbb{R}
    \end{equation*}
    while 
    \begin{equation*}
        \Delta_3(\varrho^{\mathrm{diag}}) = \text{Tr}(\mathbb{1})/2^3=1/4.
    \end{equation*}
\end{example}

\myindent Let us also make a remark that, among the motivations we had, we emphasized that our aim was not to substitute the usual notion of coherence with that of relational coherence. Let us then draw a simple parallel between relational imaginarity and the usual notion of basis-dependent imaginarity. 

\myindent Protocols for estimating Bargmann invariants provide a simple generic witness of basis-dependent imaginarity, connecting our formalism with the usually considered formalism initially introduced by~\cite{hickey2018quantifying}. We have that using the discussion above, there are simple witnesses of imaginarity that can be constructed for generic product states $\rho_1 \otimes \dots \otimes \rho_n$, given by $\text{Im}[C_n] := (C_n-C_n^\dagger)/2i$ where $C_n$ is the unitary cyclic operator (see Chapter~\ref{chapter: Bargmann invariants}). These Hermitian operators $\text{Im}[C_n]$ constitute operator witnesses of imaginarity (as by Def.~\ref{def: quantum resource witness}) since 
\begin{equation}
    \text{Tr}(\text{Im}[C_n] \rho_1 \otimes \dots \otimes \rho_n) = \frac{1}{2i}(\text{Tr}(C_n \rho_1 \otimes \dots \otimes \rho_n)-\text{Tr}(C_n^\dagger \rho_1 \otimes \dots \otimes \rho_n))=\text{Im}[\Delta_{12\dots n}(\varrho)]
\end{equation}
where we have used that $\text{Tr}(C_n^\dagger \rho_1 \otimes \dots \otimes \rho_n) = \text{Tr}(C_n \rho_1 \otimes \dots \otimes \rho_n)^*$.

\subsubsection{Lower bounds on the boundary of the set of all Bargmann invariants}

\myindent The possible values a Bargmann invariant $\text{Tr}(\rho_1\cdots\rho_n)$ can take is characterized by the boundary of the set $\mathfrak{B}_n$. Since the values of such invariants characterize relational coherence (in particular, their imaginary values also characterize relational imaginarity) it is important to look for complete characterizations of such sets. While these are not known in general,~\footnote{See~\cref{subsec:recent_advances_relational} where we highlight recent progress (after the submission of this thesis) that has significantly undermined this statement.} we can use families of seesaw techniques to numerically find lower bounds on the boundary $\partial \mathfrak{B}_n$. Note that since we now have complex values, because the field $\mathbb{C}$ is not ordered, we need to describe a maximization problem that returns optimal values of Bargmann invariants. To do so, we use our results from~\cref{proposition: unit disc asymptotic} and~\cref{corollary:Bm_is_convex} showing that every $\mathfrak{B}_n$ is a convex subset of the complex unit disc $\{z \in \mathbb{C} \mid |z| \leq 1\}$. 

\myindent In order to find the optimal value $\Delta_n(\varrho) := \text{Tr}(\rho_1\cdots \rho_n)$ we can then find the \emph{minimum} distance between $\Delta_n(\varrho) $ and $e^{i\theta}$ for all possible values $\theta \in [0,2\pi)$. In fact, since we know that $\Delta_n(\varrho) \in \mathfrak{B}_n \implies \Delta_n^* \in \mathfrak{B}_n$ we can even restrict to $\theta \in [0,\pi)$. In this case, we can write a seesaw optimization problem of the form 
\begin{align}\label{eq: seesaw boundary Bargmann invariants}
    &\max_{X_i \in \mathbb{C}^{d \times d}} |e^{i\theta}-\text{Tr}(\rho_1\cdots \rho_{i-1}X_i\rho_{i+1} \cdots \rho_n)|^2 \\
    &\text{subject to } X_i \geq 0\nonumber \\
    &\text{Tr}(X_i) = 1 \nonumber 
\end{align}
for all possible choices $i \in \{1,2,\dots,n\}$, and for sufficient numbers of iterations. Note that the optimization ranges over a convex set (of positive semidefinite matrices).  Using this family of \acrshort{sdp}s we have found optimal values shown in Fig.~\ref{fig: sets of Bargmanns}. For each set $\mathfrak{B}_n$ we have optimized over matrices $X_i \in \mathbb{C}^{n \times n}$. 

\section{Discussion}\label{sec: discussion Bargmann relational chapter}

\myindent This was the central Chapter of this thesis. In our view, the results herein constitute answers to Questions~\ref{question: generalize galvao and broads results} and~\ref{question: applications}. We have generalized the framework introduced by~\cite{galvao2020quantum}, provided further motivations to it, and comprehensively described and characterized novel families of coherence witnesses. 

\myindent Nevertheless, beyond having resolved some of the concrete questions we have posed in this thesis, in this Chapter we took the liberty to motivate and initiate ideas that we believe go beyond the resolution of the focused questions from Chapter~\ref{chapter: introduction}. This Chapter (and Chapter~\ref{chapter: event_graph_approach}) constitute our `bird' contributions to the body of literature. The Chapters to come (Chapter~\ref{chapter: from overlaps to noncontextuality} and Chapter~\ref{chapter: applications}) will be our `frog' contribution to it. 

\myindent We maintain the theme from the introduction, where we have planned to keep fluctuating between this `bird' and `frog' view, aiming to, as we recall, (i) bridge distinct fields of study, (ii) motivate and develop overarching tools applicable to both and (iii) investigate the novel perspectives these tools provide. In this Chapter we have focused on points (ii) and (iii). We have greatly emphasized (iii) as we believe relational coherence \emph{is} a novel perspective  provided by our results from Chapter~\ref{chapter: event_graph_approach}. In the next Chapter, we focus on (i). 

\subsection{Recent advances}\label{subsec:recent_advances_relational}

\myindent After the completion of some of the works we have mentioned here, great progress has been made in the investigation and characterization of the sets of Bargmann invariants, especially regarding the sets $\mathfrak{B}_n$. For example,~\cite{li2025bargmann} has extended the characterization of $\mathfrak{B}_n|_{\mathrm{circ}}$, the set of $n$-order Bargmann invariants realizable by sets of quantum states for which the associated Gram matrix is a circulant Gram matrix. Investigating this set, they advance---but do not resolve---conjectures~\ref{eq: difference between Bargmann sets} and~\ref{conjecture: Bn and circ Bn}. 

\myindent Attempting to resolve the problem of characterizing $\partial \mathfrak{B}_n$ such as we have done in Theorem~\ref{theorem: all third-order Bargmann invariants},~\cite{zhang2024boundaries} has provided a construction describing $\partial \mathfrak{B}^{(2)}_n$. While they do not resolve some of the main conjectures we have described in this Chapter, their work is a significant advance to the state of the art, and their tools and techniques employed shall be instrumental for future work investigating the boundary of these sets of correlations based on higher-order unitary invariants.

\myindent Shortly after the two results mentioned above, work by~\cite{xu2025bargmannnumerical} and~\cite{praptasi2025elementary} has significantly advanced this problem (of understanding the set of all possible Bargmann invariants) and completely characterized the bounds of the sets of Bargmann invariants $\mathfrak{B}_n$ for all positive integers $n$. The curves bounding $\mathfrak{B}_n$ (i.e., characterizing $\partial \mathfrak{B}_n$) are given by all points $\Delta \in \mathbb{C}$ satisfying the equation 
\begin{equation}
    \Delta^n = - \left(\frac{\sec(\sfrac{\theta}{n})}{\sec(\sfrac{\pi}{n})}\right)^ne^{i\theta},
\end{equation}
where $-\pi \leq \theta \leq \pi$. Since these are also the bounds found by~\cite{li2025bargmann} describing $\partial \mathfrak{B}_n\vert_{\rm circ}$, this shows that
\begin{equation}
    \partial \mathfrak{B}_n\vert_{\mathrm{circ}} = \partial \mathfrak{B}_n,
\end{equation}
for all $n \in \mathbb{N}$. 

\myindent These two works have resolved several of the conjectures we posed in this chapter regarding these sets of quantum unitary invariants. Namely, they have proven that~\cref{eq: difference between Bargmann sets} and ~\cref{conjecture: Bn and circ Bn} are \emph{true}. In fact, their findings indicate that 
\begin{equation}
    \mathfrak{B}_n = \mathfrak{B}_n^{(d)} = \mathfrak{B}_n\vert_{\rm pure} =  \mathfrak{B}_n\vert_{\rm circ}=\mathfrak{B}_n^{(2)}\vert_{\rm pure}=\mathfrak{B}_n^{(3)}\vert_{\rm circ}=\mathfrak{B}_n^{(d)}\vert_{\rm pure}
\end{equation}
for all $n\geq 2$ and $ d\geq 2$. Moreover, the single-qubit states realizing $\partial \mathfrak{B}_n$ are precisely the \acrshort{obg} states from~\cref{def:OBG_states}. 

\myindent Moreover,~\cite{xu2025bargmanninvariantsgaussianstates} considered a distinct restriction of $\mathfrak{B}_n$ that we have not treated previously: the restriction to \emph{Gaussian} state realization. They therefore propose studying the set
\begin{align*}
\mathfrak{B}_{n}\vert_{m\text{-}\mathrm{Gauss}} \equiv \mathfrak{G}_n^{(m)}
&:= \big\{\Delta\in\mathfrak{B}_n \;\big|\; \exists\ \{\rho_i\}_{i=1}^n\ \text{(}m\text{-mode Gaussian states)}\ \\&\hspace{5cm}\text{such that}\ \Delta=\operatorname{Tr}(\rho_1\cdots\rho_n)\big\}.
\end{align*}
In other words, $\mathfrak{G}_n^{(m)}$ denotes the set of Bargmann invariants realizable by $m$-mode Gaussian states. Here the number of modes $m$ plays a role analogous to the Hilbert-space dimension $d$ in $\mathfrak{B}_n^{(d)}$. This work formalizes several intriguing conjectures and motivates further investigation of the possible values (the range) of Bargmann invariants in the continuous-variable setting.

\subsection{Future directions}\label{subsec:future_directions_relational}

\myindent We have identified several promising directions for future research. Broadly speaking, our work introduces a new branch of coherence theory, which---if this new formalism proves fruitful---suggests an overwhelmingly vast landscape of potential developments. In what follows, however, we concentrate on specific future directions that we regard as the most significant open problems.

\myindent We have left various conjectures to be either proved, or provided with a counter-example, regarding the geometry of the sets of quantum correlations we have discussed. In our view, the most  relevant open problems regarding the geometry of these sets are the following ones: 

\begin{enumerate}
    \item Does the set equality $\mathfrak{C}(G) = \mathfrak{Q}(G)|_{\mathrm{inc}}$ holds for every event graph $G$? This can be considered the most relevant open problem in our formalism. If true, it could have an impact on the theory of \acrshort{ks} contextuality. Moreover, it would show that the convex polytopes $\mathfrak{C}(G)$ represent a complete characterization of all the possible ways that two-state overlaps can have relational coherence, an analogue of finding a `complete set of Bell inequalities' for our framework. 
    
    \item Are there some  sets of correlations defined via tuples of Bargmann invariants that are non-convex? One of the main open questions left by~\cite{designolle2021set} was on the possibility of `convexifying' the notion of set coherence. From the fact that the sets $\mathfrak{Q}(G)$ are convex, and from the fact that $\mathfrak{C}(G)$ is a convex polytope, considering relational information provided by two-state overlaps is \emph{one way} of resolving the `non-convexity' of the notion of set coherence. However, we do not believe that all such sets are necessarily convex sets. We believe, for example, that there exist event graphs $G$ and dimensions $d$ for which $\mathfrak{Q}^{(d)}(G)$ is a non-convex set.
\end{enumerate}

\myindent Some other more specific open questions that are worthy of future investigations are the following ones:

\begin{enumerate}
    \item Are there quantum realizable edge weightings that violate \emph{more than one} facet-defining inequality of $\mathfrak{C}(G)$? This aspect has parallels with Bell and \acrshort{ks} noncontextuality inequalities, since this is known to hold true for those cases. 
    \item What is the simplest event graph inequality that can be violated by stabilizer states? We have briefly mentioned the fact that some event graph inequalities are witnesses of a specific type of coherence, known as magic. As showed by~\cite{wagner2024certifying} this is not a generic feature of such inequalities, since there are event graph inequalities that \emph{can} be violated by stabilizer states. As for now, we do not know what is the \emph{simplest} such inequality. 
\end{enumerate}

\myindent Furthermore, we have focused solely on \emph{quantum realizability} and \emph{quantum coherence} in this Chapter. However, as showed by the existence of \acrshort{pr} boxes in Bell scenarios, once one has a precise notion of a scenario and of empirical correlations related to the scenario constraints, it may also be meaningful to consider \emph{post-quantum} realizations. For that to be consistent with our formalism, one needs to introduce a notion of \emph{theory-independent two-state overlaps}. It is conceivable that such a description would be helpful in analysing the connection between the event graph approach and other notions of quantum realizability where two-state overlaps play a major role. Examples of such are on the field of \emph{antidistinguishability}~\citep{leifer2020noncontextuality}, bounding $\psi$-epistemic ontological models of quantum theory~\citep{leifer2013maximally,leifer2014psi}, and the \acrshort{pbr} theorem~\citep{pusey2012onthereality}.

\chapter{From event graph inequalities to noncontextuality inequalities}\label{chapter: from overlaps to noncontextuality}

\begin{quote}
    ``\textit{It is a funny fact that in spite of many references to it, there seems to be no single result called the
Kochen--Specker theorem}.'' \\
(Andreas~\cite{doring2005kochen})
\end{quote}

\myindent Up to this point, we have provided solid theoretical evidence that event graph inequalities---i.e., facet-defining inequalities bounding the event graph polytopes $\mathfrak{C}(G)$---can be used to witness quantum coherence of a set of states and, in fact, characterize relational coherence. In this Chapter, we turn to the analysis of the relationship between coherence and contextuality. 

\myindent As mentioned back in Chapters~\ref{chapter: introduction} and~\ref{chapter: contextuality}, one can argue that the connection between coherence and contextuality has remained elusive. While some works have explored this relationship~\citep{spekkens2007evidence,catani2023whyinterference,catani2022replycommentwhyinterference}, showing that neither basis-dependent quantum coherence nor quantum entanglement are sufficient to guarantee a proof of quantum contextuality, there are a few gaps in our current understanding. To mention one, the role that \emph{set coherence} plays in the notion of generalized noncontextuality is unclear,~\footnote{As a matter of fact, a recent work by~\cite{jokinen2024nobroadcasting} has provided concrete answers solidifying the relationship between set coherence (in fact, from the related notion of \emph{no-broadcasting}) and generalized contextuality.} in contrast to the well understood role played by measurement incompatibility as shown in works by~\cite{tavakoli2020measurement} and~\cite{selby2023contextuality}. 

\myindent This Chapter has three main goals. First, we address Question~\ref{question: coherence witnesses and contextuality} from Chapter~\ref{chapter: introduction}, a central question of this thesis. We provide a way of mapping the coherence witnesses considered by~\cite{galvao2020quantum} to noncontextuality inequalities. Second, we go beyond this connection and show that, in fact, \emph{every} noncontextuality inequality from the \acrshort{csw} approach can be obtained by mapping some event graph inequality to it. Third, we also resolve Question~\ref{question: coherence witnesses and contextuality} by connecting it with the notion of generalized noncontextuality. In this manner, we are able to provide an infinite family of valid preparation  noncontextuality inequalities for the class of \acrshort{lsss} scenarios discussed in Chapter~\ref{chapter: contextuality}. To the best of our knowledge, this is the only known infinite family of valid prepare-and-measure noncontextuality inequalities. This chapter, along with the next, presents concrete applications of the formalism developed in Chapters~\ref{chapter: event_graph_approach}, and~\ref{chapter: relational coherence}. 

\myindent One key remark is in order. In noncontextuality scenarios, merely violating a given noncontextuality inequality does not, by itself, establish the failure of all noncontextual ontological models in explaining the observed statistics. A proper analysis of the scenario is always required. Consequently, it is essential to distinguish between an inequality and the scenario in which it appears. This distinction is particularly relevant to our mapping, as we identify multiple ways of relating facet-defining inequalities of event graph polytopes $\mathfrak{C}(G)$ to noncontextuality inequalities, each corresponding to a distinct scenario. The three relevant notions of noncontextuality scenarios to us, as reviewed in~\cref{chapter: contextuality}, are: (i) compatibility scenarios, (ii) the ‘scenarios’ represented by exclusivity graphs, and (iii) generalized prepare-and-measure noncontextuality scenarios.

\myindent The structure of this Chapter is as follows. In Sec.~\ref{sec: simplest connection} we describe the simplest possible mapping between facet-defining inequalities of $\mathfrak{C}(G)$ and \acrshort{ks} noncontextuality inequalities, by mapping event graphs $G$ to certain compatibility scenarios denoted by $\pmb{\Upsilon}_G$. This connection is sufficient to solve Question~\ref{question: coherence witnesses and contextuality}, relating $c_n(\pmb{r}) \leq n-2$ to the $n$-cycle \acrshort{ks} noncontextuality inequalities from Eq.~\eqref{eq: noncontextuality inequalities}, first discovered by~\cite{araujo2013all}. In Sec.~\ref{sec: relating event graph and KS} we show that this connection between event graph inequalities and \acrshort{ks} noncontextuality inequalities can be generalized. We show that the noncontextual polytope associated to a \acrshort{csw} exclusivity graph $H$ is isomorphic to a cross section of the event graph polytope $\mathfrak{C}(\nabla H)$, where $\nabla H$ is the suspension graph of $H$. Then, in Sec.~\ref{sec: relating cyclic inequalities to preparation noncontextuality}, we consider the notion of generalized noncontextuality and if we are able to resolve Question~\ref{question: coherence witnesses and contextuality} with respect to this different definition of noncontextuality, and show that in fact we can also map the $n$-cycle event graph inequalities $c_n(\pmb{r}) \leq n-2$ to preparation noncontextuality inequalities, for all integers $n \geq 3$. We conclude in Sec.~\ref{sec: discussion and further directions from event graph to noncontextuality} with a discussion and outlook. 

\section{Simplest connection and a few examples}\label{sec: simplest connection}

\myindent Let $G$ be a generic event graph. Because of the arguably simpler structure of 
 $\mathfrak{C}(G)$ compared to general noncontextuality polytopes $\mathrm{NC}(\pmb{\Upsilon})$, even if one may relate the two we intuitively expect the following:
\begin{enumerate}
    \item Not every facet-defining inequality of  $\mathfrak{C}(G)$ corresponds to some facet-defining \acrshort{ks} noncontextuality inequality in some \acrshort{ks} scenario.
    \item The converse holds as well. Not every \acrshort{ks} noncontextuality inequality corresponds to some facet-defining inequality of $\mathfrak{C}(G)$.
\end{enumerate}

\myindent Nevertheless, there is a simple argument that we can make that the faces of $\mathfrak{C}(G)$ \emph{can} also be understood as bounds on \acrshort{ks} noncontextual models. It consists of associating vertices of $G$ with \emph{measurements}, while edges identify two-measurement contexts, i.e. pairs of observables that can be jointly measured. In other words, identifying event graphs $G$ with compatibility scenarios. In this way, the weight of an edge  corresponds to the probability, with respect to a given global state, that the two incident measurements yield equal outcomes. 

\myindent A necessary and sufficient condition for the existence of a \acrshort{ks} noncontextual model whose behavior is consistent with a given edge weighting described in this manner is the existence of a global probability distribution (on outcome assignments to all measurements) whose marginals recover the correct outcome probabilities. This is the content of the Fine--Abramsky--Brandenburger (\acrshort{fab}) theorem \citep{fine1982hiddenvariablesPRL,fine1982jointJMP,abramsky2011sheaf}. See  Theorem~\ref{theorem:FAB theorem} from~\cref{chapter: contextuality}. This is therefore a link between behaviors having a \acrshort{ks} noncontextual model for a scenario, and the associated edge weighting that must be inside $\mathfrak{C}(G)$.

\myindent In summary, we formalize the discussion above via the following theorem.
\begin{theorem}\label{theorem: from G to compatibility scenarios}
    Let $G$ be any event graph, and $\mathfrak{C}(G)$ the associated event graph polytope. Define $\pmb{\Upsilon}_G := (\mathcal{M}_G, \mathcal{C}_G,\mathcal{O})$ as the compatibility scenario satisfying that $x_i \in \mathcal{M}_G$ iff $i \in V(G)$, $\gamma \in \mathcal{C}_G$,   $\gamma = \{x_i,x_j\}$ iff $\{i,j\} \in E(G)$, and, finally,  $\mathcal{O}^x = \{+1,-1\}$ for every $x \in \mathcal{M}_G$. Then, via the relation 
    \begin{equation}
        \langle x_ix_j\rangle = 2r_{ij}-1,
    \end{equation}
    every facet-defining inequality $h(\pmb{r}) \leq b$ bounding  $\mathfrak{C}(G)$ is mapped to a valid \acrshort{ks} noncontextuality inequality~\footnote{We discuss the notion of valid inequalities in Appendix~\ref{sec: convex polytopes}.} $h'(\pmb B)\leq 2b-h(\pmb{1})$ bounding  $\mathrm{NC}(\pmb{\Upsilon}_G)$. 
\end{theorem}

\begin{proof}
Let $h(\pmb{r}) \leq b$ be any facet-defining inequality from $\mathfrak{C}(G)$.  Note that, by definition of the two-point correlation functions, we have that (see Eq.~\eqref{eq: two-point correlations}) $$\langle x_ix_j\rangle = p(x_i=x_j)-p(x_i \neq x_j) = 2p(x_i=x_j)-1  \equiv 2r_{ij}- 1,$$ for every $r \in \mathfrak{C}(G)$.  For each inequality $h(\pmb{r}) \leq b$ we have that $h$ is linear and therefore we find that under the linear mapping $\pmb r \mapsto (\pmb{B}_c+\pmb{1})/2$ we have that 
\begin{equation*}
    h(\pmb{r}) \mapsto h\left(\frac{\pmb{B}_c+\pmb 1}{2}\right) = \frac{1}{2}h(\pmb{B}_c)+\frac{1}{2}h(\pmb{1}),
\end{equation*}
from which we obtain the new inequality 
\begin{align*}
    h(\pmb{B}_c) \leq 2b-h(\pmb{1}).
\end{align*}
Now, we let $\mathrm{NC}(\pmb{\Upsilon}_G) \ni B \mapsto B_c := (\langle x_ix_j \rangle )_{\{i,j\} \in E(G)}$ via the relation from Eq.~\eqref{eq: two-point correlations}. Define $h'$ to be the resulting functional, such that $h'(\pmb{B}) = h(\pmb{B}_c)$. 

\myindent It remains to show that $h'(\pmb{B}) \leq 2b-h(\pmb{1})$ bounds $\mathrm{NC}(\pmb{\Upsilon}_G)$, i.e., that $B \in \mathrm{NC}(\pmb{\Upsilon}_G)$ implies $h'(\pmb{B}) \leq 2b-h(\pmb{1})$. Suppose that there exists a behavior $ B \in \mathrm{NC}(\pmb{\Upsilon}_G)$ such that $h'(\pmb B) > 2b-h(\pmb{1})$. Then, via $\pmb{B} \mapsto \pmb{r}_B = 2\pmb{B}_c-\pmb{1}$, we would need to have that $h(\pmb{r}_B) > b$, which is absurd given that $r_B$ can be realized by jointly distributed random variables since $B$ can be as well (since it is the marginal of a global section over the contexts) due to the \acrshort{fab} theorem.  

\end{proof}
\myindent In general, this simple approach (interpreting vertices as measurements and edges as equality of outcomes in two-measurement contexts) is not sufficient to capture \acrshort{ks} contextuality in full generality~\cite[Section 2.5.3]{amaral2018graph}. An exception is, precisely, the case we have just described of dichotomic measurements~\cite[Theorem 38]{araujo2014quantum}, where equality of outcomes $x_i=x_j$ fully determines the measurement statistics, i.e. $\pmb B_c$ contains the same information as $\pmb B$. Even restricting to contextuality scenarios whose maximal contexts have size two, the facets of $\mathfrak{C}(G)$ do not correspond necessarily to facet, or even tight, \acrshort{ks} noncontextuality inequalities.  

\myindent However, there are certain examples of inequalities that \emph{are} captured by this simple construction. For example, if we take the inequality 
\begin{equation}
    c_4(\pmb{r}) = r_{12}+r_{23}+r_{34}-r_{14} \leq 2
\end{equation}
and apply $\pmb r \mapsto  (\pmb{B}_c+\pmb 1)/2$ we end up with 
\begin{align*}
 &r_{12}+r_{23}+r_{34}-r_{14} \leq 2, \mapsto \\
 &\frac{\langle x_1x_2\rangle +1}{2}+\frac{\langle x_2x_3\rangle +1}{2}+\frac{\langle x_3x_4\rangle +1}{2}-\left(\frac{\langle x_1x_4\rangle +1}{2}\right) \leq 2,\\
 &\langle x_1x_2\rangle +1+\langle x_2x_3\rangle +1+\langle x_3x_4\rangle+1 -\langle x_1x_4\rangle-1\leq 4,\\
 &\langle x_1x_2\rangle +\langle x_2x_3\rangle +\langle x_3x_4\rangle -\langle x_1x_4\rangle\leq 2,
\end{align*}
which is the \acrshort{chsh} inequality that we have introduced in Chapter~\ref{chapter: contextuality} (see Ineq.~\eqref{eq: chsh inequality}). If we let $\langle x_1x_2\rangle = 1$, we end up with the so-called original Bell inequality~\citep{larsson2014loopholes} in the minimal scenario.

\myindent With this discussion above, we are ready to solve~\cref{question: coherence witnesses and contextuality}.

\begin{corollary}
    Fix $n \in \mathbb{N}$ any, and let  $c_n(\pmb{r}) \leq n-2$ be any $n$-cycle event graph inequality given by Eq.~\eqref{eq: overlap cycle inequalities}. This inequality can be mapped to $n$-cycle \acrshort{ks} noncontextuality inequalities given by Eq.~\eqref{eq: noncontextuality inequalities}.
\end{corollary}

\begin{proof}
    Apply the same construction outlined in Theorem~\ref{theorem: from G to compatibility scenarios}, we see that for any $n\geq 3$ the mapping $G = C_n \mapsto \pmb{\Upsilon}_n$ relates the cycle event graphs to the $n$-cycle compatibility scenarios. Applying  Theorem~\ref{theorem: from G to compatibility scenarios}, transforming (any element of the class of) the $n$-cycle event graph inequalities from Eq.~\eqref{eq: overlap cycle inequalities} via $r_{ij} \mapsto \frac{\langle x_ix_j\rangle + 1}{2}$ yields an $n$-cycle \acrshort{ks} noncontextuality inequality (cf.~Eq.~\eqref{eq: noncontextuality inequalities}). 
\end{proof}

\myindent We conclude by noting that \emph{not every $n$-cycle \acrshort{ks} noncontextuality  inequality} can be reached by this method of transforming $n$-cycle event graph inequalities $c_n(\pmb{r}) \leq n-2 \to I_{\pmb{a}}^{(n)}(B)  \leq n-2$. For example, the following \acrshort{ks} noncontextuality  inequality (that is a different facet-defining inequality of $\pmb{\Upsilon}_4$) is not reached by the methods just described:
\begin{equation*}
    \langle x_1x_2\rangle -\langle x_2x_3\rangle -\langle x_3x_4\rangle -\langle x_1x_4\rangle \leq 2.
\end{equation*} 

\myindent This result addresses~\cref{question: coherence witnesses and contextuality} by concretely showing how the $n$-cycle inequalities from the event graph approach can lead to \acrshort{ks} noncontextuality inequalities. While, as a matter of fact, Theorem~\ref{theorem: from G to compatibility scenarios} provides a more general rule for constructing \acrshort{ks} noncontextuality inequalities, it is clearly limiting. We do not reach every \acrshort{ks} noncontextuality inequality of generic compatibility scenarios $\pmb{\Upsilon}$. In what follows we provide a significantly more comprehensive link between \acrshort{ks} inequalities and event graph inequalities. 

\section{Relating event graph inequalities to KS noncontextuality inequalities}\label{sec: relating event graph and KS}

\myindent In this section, we establish a formal and comprehensive connection between our framework and (Kochen--Specker) contextuality. The central result (\cref{theorem: STAB = C}) shows how our event graph formalism recovers all noncontextuality inequalities obtainable from the Cabello--Severini--Winter (\acrshort{csw}) exclusivity graph approach~\cite{cabello2014graph}.

\myindent To achieve this, we encode a contextuality setup, represented in \acrshort{csw} by an exclusivity graph $H$, by imposing exclusivity constraints on a related event graph $H_\star$. This process amounts to taking a cross-section yielding a subpolytope of the event graph polytope $\mathfrak{C}({H_\star})$. We show that the resulting facet inequalities bound noncontextual models for $H$.

\myindent In fact, we prove something \stress{stronger}. We describe an explicit isomorphism between the noncontextual polytope associated by \acrshort{csw} to the exclusivity graph $H$---the so-called \acrshort{stab} polytope---and this cross-section subpolytope of the polytope $\mathfrak{C}({H_\star})$ associated by our approach to the event graph $H_\star$. In particular, these polytopes have the same non-trivial facet-defining inequalities. These are obtainable from the inequalities that define the full (unconstrained)  polytope $\mathfrak{C}(H_\star)$ by setting some coefficients to zero. \Cref{theorem: STAB = C} thus establishes a tight correspondence between our event graph approach and a broad, well-established framework for contextuality.

\subsection{From exclusivity graphs to constrained event graphs}%

\begin{figure}[t]
    \centering
    \includegraphics[width=0.65\textwidth]{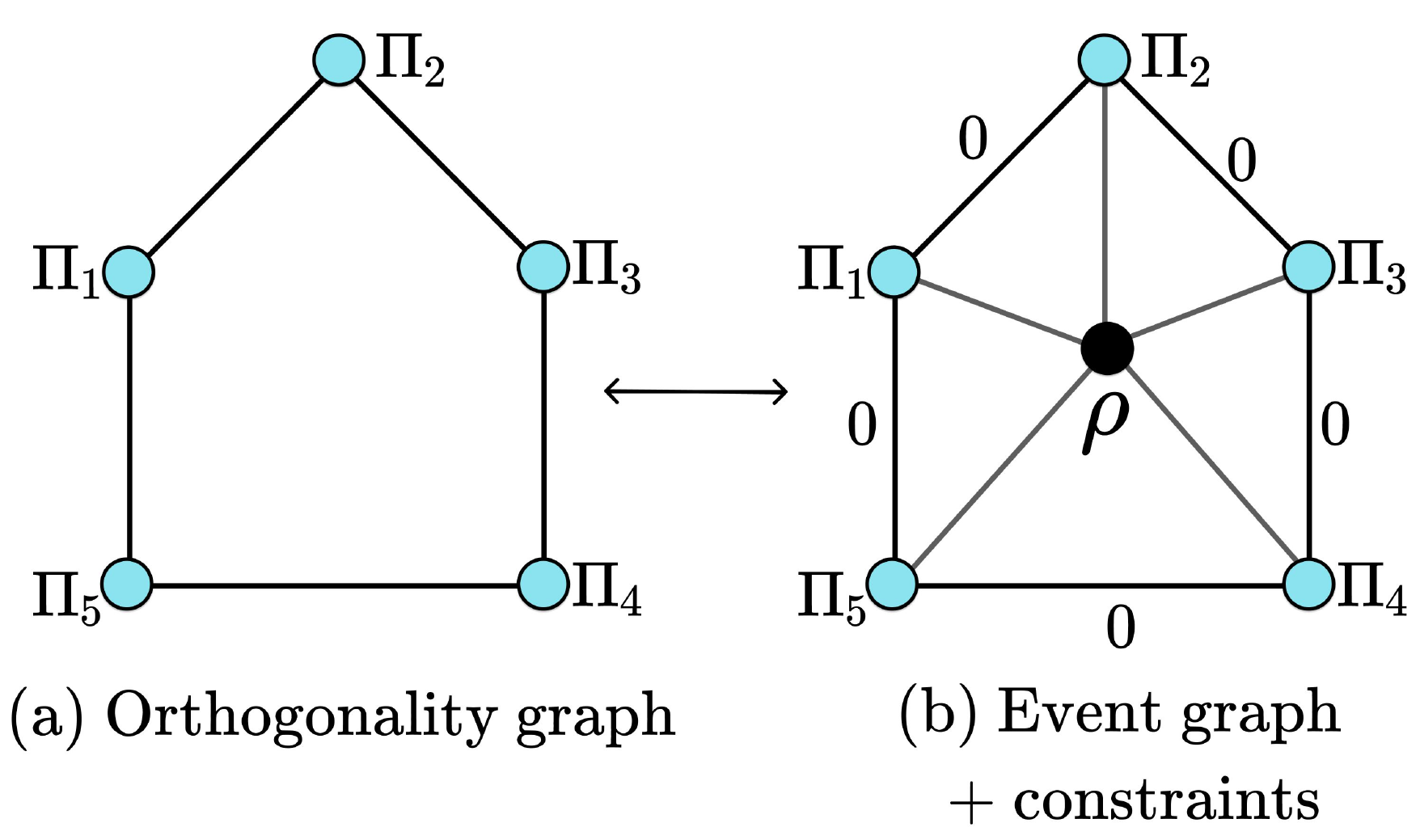}
    \caption{\textbf{Equivalence described by \cref{theorem: STAB = C} linking contextuality \emph{à la} \acrshort{csw} to event graphs.}
    The behaviors on an exclusivity graph are in bijective correspondence with edge weightings (overlap assignments) in the related event graph subject to constraints.}
    \label{fig:orthogonality_to_event}
\end{figure}

\myindent We now relate the \acrshort{csw} approach that we have presented in Chapter~\ref{chapter: contextuality}, Section~\ref{sec: KS noncontextuality}, to the event graph approach by constructing a new (event) graph $H_\star$ from any (exclusivity) graph $H$. This is obtained by adding a new vertex $\psi$ with an edge connecting it to all the vertices of $H$.  See~\cref{fig:orthogonality_to_event} for an instance of this construction for the \acrshort{kcbs} scenario and \cref{fig:ProofAppD} for a more generic description. The construction is formally described in \cref{def: G out of G prime} below.

\myindent The relevance of the new vertex $\psi$ is well known; it is usually called the `handle' (terminology introduced by~\cite{lovasz1999geometric}) and it appears in the literature on the graph approaches to \acrshort{ks} contextuality~\citep{amaral2018graph,baldijao2020classical,vandre2022quantum,ramanathan2017tightness}. Its name comes from the geometric arrangement of the vectors providing the maximal quantum violation of the \acrshort{kcbs} inequality (see Section~\ref{sec: KS noncontextuality}, Ineq.~\eqref{eq: exclusivity KCBS inequality}): the quantum state resembles the handle of an umbrella made of the vectors that describe measurement events~\citep[Example 9.1.4, Fig. 9.2, pq. 124]{lovasz1999geometric}.

\begin{definition}[Event graphs via suspension graphs of exclusivity graphs]\label{def: G out of G prime}
Let $H$ be a simple graph. Define a new graph $H_\star$ by
\begin{align*}
    V(H_\star) &:= V(H) \sqcup \enset{\psi}
    \\
    E(H_\star) &:= E(H) \cup \setdef{\enset{\psi,v}}{v \in V(H)}.
\end{align*}
Formally, the above construction of a graph $H_\star$ from a given simple graph $H$ is known as the \emph{suspension graph}~\citep[Def. 2.23, pg. 36]{amaral2018graph}, sometimes denoted as $\nabla H \equiv H_\star$. Moreover, we now make use of the cross sections $\mathfrak{C}_{G'|G}^0$ defined in Chapter~\ref{chapter: event_graph_approach} to define $\mathfrak{C}_{H|H_\star}^0 \equiv \mathfrak{C}_{H_\star}^0$ to consist of the subset of edge weightings of $\mathfrak{C}(H_\star)$ that assign value $0$ to all edges in the subgraph $H$ of the suspension graph $H_\star$, i.e.
\[
\mathfrak{C}_{H_\star}^0 := \setdef{r \in \mathfrak{C}({H_\star})}{\Forall{e\in E(H)} r_e=0} .
\]
\end{definition}

\myindent The set $\mathfrak{C}_{H_\star}^0$ is, by construction, a cross-section of the polytope $\mathfrak{C}({H_\star})$ of the event graph $H_\star$,
being its intersection with the $|V(H)|$-dimensional subspace defined by the equations $\bigwedge_{e \in E(H)}r_e=0$, as we have discussed in Chapter~\ref{chapter: event_graph_approach}.
Moreover, it is a subpolytope of $\mathfrak{C}({H_\star})$, \ie the convex hull of a subset of its vertices. These vertices are the edge $\{0,1\}$-labelings in $\text{ext}(\mathfrak{C}({H_\star}))$ that assign label $0$ to edges in $H$. If $\alpha \in \text{ext}(\mathfrak{C}_{H_\star}^0)$, and considering any $\Lambda$-realization for $\alpha$ (see Def.~\ref{def: lambda realizable edge weightings}), any two vertices adjacent in $H$ must be labelled differently.

\subsection{Recovering the \acrshort{csw} noncontextual polytope}

\myindent The edge set of the graph $H_\star$ can be partitioned into two sets: the edges already present in $H$ and the new edges of the form $\{\psi,v\}$ for $v \in V(H)$.
The latter are in one-to-one correspondence with vertices of $H$. So, there is a bijection $E(H_\star) \cong E(H) \sqcup V(H)$.

\myindent When considering the polytope $\mathfrak{C}({H_\star}) \subseteq [0,1]^{E(H_\star)}$
we adopt the convention of ordering the coordinates with the edges already in $H$ listed first, so that
\[\RR^{E(H_\star)} \cong \RR^{E(H) \sqcup V(H)} \cong \RR^{E(H)} \times \RR^{V(H)}.\]
The subpolytope $\mathfrak{C}_{H_\star}^0$ is thus written as the set of points
of $\mathfrak{C}({H_\star})$ of the form $(\pmb{0}_{H},\pmb{r})$ where 
$\pmb{0}_{H}$ is the zero vector in $\RR^{E(H)}$ (corresponding to the edges inherited from $H$)
and $\pmb{r}$ is an edge weighting of the remaining (new) edges.
In particular, the vertices of $\mathfrak{C}_{H_\star}^0$ are precisely the deterministic edge weightings that are realizable by incoherent sets of states of $H_\star$ that \emph{also} assign the label $0$ to all the edges in $H$.

\myindent We can now prove our main result of this Chapter, showing that $\mathfrak{C}^0_{H_\star}$ is indeed (isomorphic to) the polytope of \acrshort{ks} noncontextual behaviors for the exclusivity graph $H$.

\begin{figure}
    \centering
    \includegraphics[width=0.8\columnwidth]{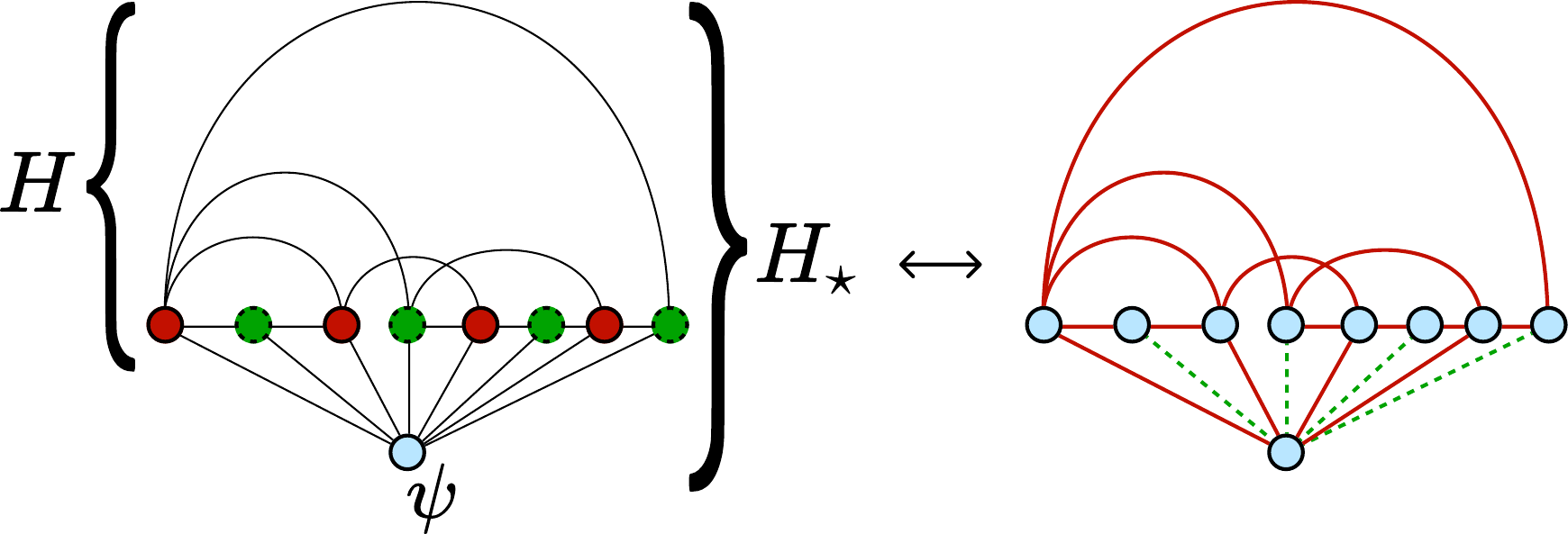}
    \caption{\textbf{Translation between vertex labelings of $H$ (that are characteristic maps of stable sets, hence vertices of $\STAB(H)$)
    and constrained edge labelings of $H_\star$ (that satisfy the conditions from Def.~\ref{def: event graph polytope}, hence vertices of  $\mathfrak{C}_{H_\star}^0)$.)}
    The left figure depicts a graph $H$, standing for a generic exclusivity graph, and its extension $H_\star$  by adjoining the handle $\psi$ and new edges $\{\psi,v\}$ for all $v \in V(H)$.
    The vertices of $H$ that are shown in green (dashed) form a stable set $S \in \mathcal{S}(H)$.
    Its characteristic map $\fdec{\chi_S}{V(H)}{\{0,1\}}$ assigns $1$ to the green (dashed) vertices and $0$ to the red (solid) vertices of $H$.
    The right figure shows how such a vertex $\{0,1\}$-labeling is translated to an edge $\{0,1\}$-labeling of $H_\star$ assigning $0$ to all the edges of $H$ (and vice-versa) as described in the proof of \cref{theorem: STAB = C}. Green (dashed) edges are labelled $1$ and red (solid) edges are labelled $0$,
    in accordance with the vertex labelings from $\chi_S$, complemented by the labels induced by exclusivity constraints, $\pmb{0}_{H}$, as described in the text.
    $S$ being stable is equivalent to the resulting edge labeling of $H_\star$ being extremal points of the associated event graph polytope.}
    \label{fig:ProofAppD}
\end{figure}

\begin{theorem}\label{theorem: STAB = C}
For any (exclusivity) graph $H$, there is an isomorphism of convex polytopes
\[\mathfrak{C}_{H_\star}^0 \; \cong \; \STAB(H)\]
between the stable set polytope (of noncontextual models) of $H$
and the subpolytope of the event graph polytope $\mathfrak{C}(H_\star)$ constrained by the exclusivity conditions.
More explicitly, this is given by the identification
\[\mathfrak{C}_{H_\star}^0 \; = \; \{\pmb{0}_H\} \times \STAB(H)\]
where $\pmb{0}_H$ is the zero vector in $\mathbb{R}^{E(H)}$.
\end{theorem}
\begin{proof}
To establish the result, we consider the vertices of these polytopes.
Per the above discussion, we have $E(H_\star) \cong E(H) \sqcup V(H)$.
Consequently, there is a bijection between
\textit{vertex} $\{0,1\}$-labelings of $H$ (equivalently, subsets of $V(H)$), on the one hand,
and \textit{edge} $\{0,1\}$-labelings of $H_\star$ that assign label $0$ to all the edges in $E(H)$, on the other.
Explicitly, to each subset of vertices $W \subseteq V(H)$ any  vertex labeling corresponds the edge-labeling of $H_\star$ defined by
\[\fdec{[\pmb{0}_H,\chi_{W}]}{E(H_\star) \cong E(H) \sqcup V(H)}{\{0,1\}},\]
as depicted in~\cref{fig:ProofAppD}. Here, $\chi_W$ is a characteristic map, as we have introduced in Chapter~\ref{chapter: contextuality}, in Eq.~\eqref{eq: characteristic map V}.

\myindent We show that this bijection restricts to a bijection between the \stress{noncontextual and transitivity preserving} assignments in both cases.
Concretely, a subset of vertices $S \subseteq V(H)$ is \stress{stable}, hence (its characteristic map $\fdec{\chi_{S}}{V(H)}{\{0,1\}}$ is) a vertex of the polytope $\STAB(H)$,
\emph{if and only if} the corresponding edge labeling $[\pmb{0}_H,\chi_{S}]$ of $H_\star$ preserves the transitivity of equality, hence corresponding to a vertex (extremal point) of the polytope $\mathfrak{C}({H_\star})$ and thus of $\mathfrak{C}_{H_\star}^0$. 

\myindent We establish the two directions of this equivalence simultaneously, recalling the characterisation of extremal points of the event graph polytope from~\cref{def: event graph polytope} and from \cref{proposition: extremal_from_realizability} in terms of $\Lambda$-realizability. Consider $H_\star$ with edge $\{0,1\}$-labeling $[\pmb{0}_H,\chi_{S}]$. The labeling fails to be in $\mathfrak{C}(H_\star)$ if and only if there is an edge with label $0$ between two vertices linked by a path consisting of edges with label $1$. Since all the edges between vertices in $H$ have label $0$, the only way to build such a path of $1$-labelled edges is via the handle $\psi$: \eg $\{u,\psi\},\, \{\psi, v\}$ where both $u$ and $v$ must belong to $S$. So, two vertices $u$ and $v$ of $H_\star$ are linked by a $1$-labelled path if and only if they both belong to $S \cup \{\psi\}$. Therefore, the labeling is in $\mathfrak{C}(H_\star)$ if and only if there is no edge with label $0$ between vertices in this set $S \cup \{\psi\}$. To further simplify this condition, note that edges between $\psi$ and a vertex from $S$ have label $1$ by construction of the second component of $[\pmb{0}_H,\chi_{S}]$, while from the first component, all edges between vertices in $H$ have label $0$. The condition of being deterministic edge weighting from $\mathrm{ext}(\mathfrak{C}(H_\star))$ is thus equivalent to there being no edges in $H$ between vertices in $S$, which is precisely to say that $S$ is stable.
\end{proof}

\subsection{Recovering all KS noncontextuality inequalities}

\myindent We established \cref{theorem: STAB = C} in terms of the vertices of the polytopes, \ie by working with their V-representations.
We now consider the relationship between their H-representations, \ie their facet-defining inequalities.\footnote{H-representation is standard terminology referring to the description of a polytope as an intersection of half-spaces, \ie in terms of (facet-defining) inequalities. The `H' in `H-representation' is not to be confused with the symbol `$H$' that we use to denote an exclusivity graph.}

\myindent Of course, there is also a bijection between the facets of $\STAB(H)$ and those of $\mathfrak{C}_{H_\star}^0$.
Given the particularly simple description of the isomorphism, whereby $\mathfrak{C}_{H_\star}^0$ is written as a product of polytopes,
we can write this correspondence explicitly.
It turns out that the facet-defining inequalities of the subpolytope $\mathfrak{C}_{H_\star}^0$
are precisely the same as the facet-defining inequalities of the stable polytope of $H$.
Moreover, these can be obtained from the inequalities defining the (unconstrained) polytope $\mathfrak{C}({H_\star})$ of the event graph $H_\star$ by setting some coefficients to zero.
We thus recover the full set of \acrshort{ks} noncontextuality inequalities (associated to some exclusivity graph $H$) from our event graph formalism.

\myindent To see this, recall that if $P$ and $Q$ are two convex polytopes with H-representations
$P = \setdef{\pmb x}{A_1 \, \pmb x \preccurlyeq \pmb{b}_1}$ 
and
$Q = \setdef{\pmb y}{A_2 \, \pmb{y} \preccurlyeq \pmb{b}_2}$
then their product has H-representation
\[
P \times Q = \setdef{(\pmb x,\pmb y)}{A_1 \, \pmb x \preccurlyeq \pmb{b}_1 \text{ and } A_2 \, \pmb y \preccurlyeq \pmb{b}_2}.
\]
Here, the notation $A \, \pmb z \preccurlyeq \pmb b$ describes a set of linear inequalities on $\pmb z$ in matrix form, with the symbol $\preccurlyeq$ standing for component-wise inequality $\leq$ between real numbers.

\myindent Applying this to
\begin{align*}
\mathfrak{C}_{H_\star}^0 &= \{\pmb{0}_{H}\} \times \STAB(H)\\
&=
\setdef{(\pmb x, \pmb y)}{\pmb x \in \{\pmb{0}_{H}\}, \pmb y \in \STAB(H)}.
\end{align*}
we obtain that the 
H-representation of $\mathfrak{C}_{H_\star}^0$ is the conjunction
of the H-representations of $\{\pmb{0}_{H}\}$ and of $\STAB(H)$.
The former consists simply of the equations $r_e = 0$ for each $e \in E(H)$,
zeroing out the first components, which corresponds to the weights of edges already in $H$.
Thus the non-trivial inequalities bounding $\mathfrak{C}_{H_\star}^0$ are thus the same as the inequalities bounding $\STAB(H)$.

\myindent Since $\mathfrak{C}_{H_\star}^0$ is obtained from  $\mathfrak{C}({H_\star})$ by intersecting with the subspace that zeroes the components corresponding to edges in $E(H)$, a complete set of inequalities for $\mathfrak{C}_{H_\star}^0$ can be obtained from the facet-defining inequalities of $\mathfrak{C}({H_\star})$
by disregarding those components, \ie setting the corresponding coefficients to zero.

\myindent This process is illustrated by the derivation of the \acrshort{kcbs} inequality. The exclusivity graph for this inequality is the $5$-cycle graph $C_5$, with neighbouring vertices representing orthogonal projectors (as we have shown in Fig.~\ref{fig: exclusivity_graph_kcbs}, Chapter~\ref{chapter: contextuality}). The suspension graph $H_\star = \nabla C_5$ is then the 6-vertex wheel graph $W_6$. In such a case, there is one specific facet-defining inequality of $\mathfrak{C}(W_6)$ that we have presented in Eq.~\eqref{eq:kcbs_event_graph_inequality} that we re-write here for convenience:
\begin{equation*}
    h_{\mathrm{KCBS}}(\pmb r):= -r_{12}-r_{23}-r_{34}-r_{45}-r_{15}+r_{16}+r_{26}+r_{36}+r_{46}+r_{56} \leq 2.
\end{equation*}
The \acrshort{kcbs} noncontextuality inequality $\sum_a\gamma_a\vert\langle \psi \vert a \rangle \vert^2 \leq \alpha(H,\gamma)$
arises as a $\mathfrak{C}_{W_6}^0$ inequality, being obtained from the inequality above from the event graph polytope $\mathfrak{C}(W_6)$ by setting to zero the coefficients relating to edges already in $H = C_5$ which in this case correspond to setting to zero $r_{12},r_{23},r_{34},r_{45},r_{15}$. In this way we obtain the inequality 
\begin{equation*}
    h_{\mathrm{KCBS}}(\pmb{0}_H,\pmb r):=+r_{16}+r_{26}+r_{36}+r_{46}+r_{56} \leq 2,
\end{equation*}
as wanted. Note that this is now the form of the \acrshort{kcbs} inequality that we have described in Eq.~\eqref{eq: exclusivity KCBS inequality}. To have a quantum violation, it suffices to assign the vertex $6$ with the handle state $\psi$ and the other 5 vertices with 5-exclusive rank-1 projectors satisfying the exclusivity conditions of the pentagon. 

\subsection{Relational coherence and KS noncontextuality are distinct notions}

\myindent The results above show that the event graph formalism is sufficiently powerful to describe all facet-defining noncontextuality inequalities. At this point, we would like to make the simple (yet important) clarification that relational coherence as captured by two-state overlaps and \acrshort{ks} noncontextuality are distinct notions, even if the relationship obtained from above may suggest a different conclusion. 

\myindent First, we note that relational coherence \emph{alone}, i.e. when we consider it as a property of generic higher order Bargmann invariants, is clearly distinct from \acrshort{ks} noncontextuality. The simplest way to see this is due to the fact that higher-order invariants that are capable of witnessing relational coherence may be described by \acrshort{ks} noncontextual models, e.g. the single-qubit stabilizer subtheory. 

\myindent But also, even if we consider relational coherence captured by two-state overlaps only, the two are also distinct. Again, there is a simple argument to make here since there exists states (and effects) having \acrshort{ks} noncontextual models that also violate $\mathfrak{C}(G)$ inequalities. To see the above, it suffices to note that sets of single-qubit states \emph{suffice} to violate these, while this statistics is always modeled by a \acrshort{ks} noncontextual model. For example, the inequality $h_{\mathrm{KCBS}}(\pmb{r}) \leq 2$ is violated by qubits when one \emph{does not} send the relevant edge-weightings to zero, and the single-qubit subtheory of quantum theory has a \acrshort{ks} noncontextual ontological model~\citep{kochen1975problem}.  

\section{Constructing preparation noncontextuality inequalities from event graphs}\label{sec: relating cyclic inequalities to preparation noncontextuality}

\myindent In this section, we map \emph{some} event graph inequalities, namely the $n$-cycle event graph inequalities, to noncontextuality inequalities bounding ontological models satisfying Spekkens's notion of preparation noncontextuality~\citep{spekkens2005contextuality}. To that end, we use the \acrshort{lsss} scenarios that have been introduced in Chapter~\ref{sec: Spekkens noncontextuality} to propose a working  \textit{definition} of preparation noncontextual edge weightings. We start, however, with a procedure similar to the one considered above where we map the information present in event graphs to this specific class of prepare-and-measure noncontextuality scenarios.  

\myindent For any given event graph $G$ with $G=(V(G),E(G))$ we construct a prepare and measure scenario as follows:

\begin{enumerate}
    \item For each vertex $v \in V(G)$ of $G$ we associate a preparation procedure $P_v$ and a binary outcome measurement $M_v$. 
    \item For each edge $e = \{v,w\} \in E(G)$ of $G$ we associate two other preparation procedures $P_{v^\perp}^e, P_{w^\perp}^e$. 
    \item To each $4$-tuple of processes $(P_v,P_{v^\perp}^e,P_w,P_{w^\perp}^e)$ for each edge $e = \{v,w\} \in E(G)$ we associate the operational equivalence $\frac{1}{2}P_v + \frac{1}{2}P_{v^\perp}^e \simeq \frac{1}{2}P_w + \frac{1}{2}P_{w^\perp}^e$.
    \item Each measurement procedure $M_v$ satisfies the operational constraints that $p(0 \vert M_v,P_v) \geq 1-\varepsilon_v$ and $p(0 \vert M_v,P_{v^\perp}^e) \leq \varepsilon_v$ for every $e \in E(G)$ for some real parameter $\varepsilon_v \geq 0$. This can be viewed as an operational account of experimental errors.
\end{enumerate}

\myindent In terms of the notation introduced earlier in Chapter~\ref{chapter: contextuality}, we can write the above construction to define what we refer generically as Lostaglio-Senno-Schmid-Spekkens (\acrshort{lsss}) scenarios $$\mathcal{PM}_{\mathrm{LSSS}} = (N_v + N_vN_e,N_v,2,\mathcal{OE}_P,\emptyset)$$ where $\mathcal{OE}_P$ correspond to the set of all operational equivalences for preparation procedures just described above in point 3, for each edge of the event graph. We also assume that the scenarios $\mathcal{PM}_{\mathrm{LSSS}}$ satisfy the constraints of $p(0\vert M_v,P_v)\geq 1-\varepsilon_v$ and $p(0\vert M_v,P_{v^\perp}^e) \leq \varepsilon_v$ since this relation is somewhat needed for the structure of noncontextual models explaining the so-called operational `confusabilities' $r_{e} \equiv p(0\vert M_v,P_w)$~\citep{schmid2018discrimination} for every edge $e=\{v,w\}$. 

\myindent If we recall Theorem~\ref{theorem: lostagio_senno} we see that, when $\varepsilon_v = 0$ for all $\varepsilon_v$, the edge weightings $r_e$ (dual to confusabilities in such graphs) relate to a notion of overlap that must be satisfied by preparation noncontextual models in such scenarios.

\begin{definition}\label{def:PNCedgeweighting}
Let $G$ be an event graph. An edge weighting $\fdec{r}{E(G)}{[0,1]}$ is said to be \stress{preparation noncontextual}
if
the edge weights are of the form in the right-hand side of \cref{equation: noncontextual overlaps}, \ie
$$r_{ij} = 1 - \|\mu_i - \mu_j\|_{_{\mathsf{TV}}},$$
for some choice of 
an (ontic) measurable space $\Lambda$ and of probability measures $\mu_i$ on $\Lambda$ for each vertex $i\in V(G)$.
\end{definition}

\myindent Above, we have considered the total variation distance described in Section~\ref{sec: Spekkens noncontextuality} and motivated by Theorem~\ref{theorem: lostagio_senno}. Using~\cref{def:PNCedgeweighting}, we can show that $n$-cycle inequalities of edge weightings bounding $\mathfrak{C}(C_n)$ can be mapped to prepare-and-measure noncontextuality inequalities in \acrshort{lsss} scenarios. To the best of our knowledge, this is the first infinite class of valid preparation noncontextuality inequalities. All other known inequalities present in the literature constitute specific instances, found using linear programming tools or providing a complete (or partial) characterization of a specific scenario. 

\subsection{Cycle inequalities witness preparation contextuality}

\myindent We now show how in the case of cycle graphs the inequalities derived from our framework serve as witnesses of preparation contextuality for operational theories satisfying the constraints defining \acrshort{lsss} scenarios. The technical result is stated in the following proposition:

\begin{proposition}\label{proposition: ideal cyclic preparation noncontextuality}
Any inequality bounding the set $\mathfrak{C}({C_n})$ cannot be violated by a preparation noncontextual edge weighting (\cref{def:PNCedgeweighting}).
\end{proposition}
\begin{proof}
For simplicity, we use addition modulo $n$ when labeling the vertices of the cycle graph $C_n$, meaning that $i = i+n$.
From the triangle inequality of the norm $\| \cdot \|_{_{\mathsf{TV}}}$ it follows that
\begin{align*}
    & \| \mu_i - \mu_{i+n-1}\|_{_{\mathsf{TV}}}  \\
    = \; & \| \mu_i \underbrace{-\mu_{i+1}+\mu_{i+1}-\dots -\mu_{i+n-2}+\mu_{i+n-2}}_{n-2\text{ zeros}}-\mu_{i+n-1}\|_{_{\mathsf{TV}}}\\
    \leq \; & \| \mu_i - \mu_{i+1}\Vert_{_{\mathsf{TV}}} + \dots + \Vert \mu_{i+n-2}-\mu_{i+n-1}\|_{_{\mathsf{TV}}} .
\end{align*}
Therefore, writing $\|\mu_{i,j}\|_{_{\mathsf{TV}}} := \|\mu_{i}-\mu_j\|_{_{\mathsf{TV}}}$ for clarity,
\[
    \Vert \mu_{i,i+n-1}\Vert_{_{\mathsf{TV}}} - \Vert \mu_{i,i+1}\Vert_{_{\mathsf{TV}}} - \dots - \Vert \mu_{i+n-2,i+n-1}\Vert_{_{\mathsf{TV}}} \leq 0 .
\]
We must now add $1$ to each term to recover the noncontextual overlaps of \cref{equation: noncontextual overlaps}.
We have $n$ terms, but since the first term has a different sign, two of these $1$s will cancel, leaving $n-2$ added to both sides of the inequality:
\begin{align*}
    -1+\Vert \mu_{i,i+n-1}\Vert_{_{\mathsf{TV}}} +1-\Vert \mu_{i,i+1}\Vert_{_{\mathsf{TV}}}  &\\
    +\;\cdots\;  + 1 - \Vert \mu_{i+n-2,i+n-1} 
    \Vert_{_{\mathsf{TV}}}  
    & \;\;\leq\;\; n-2.
\end{align*}
Recalling that $r_{ij} = 1-\Vert \mu_{i,j} \Vert_{_{\mathsf{TV}}}$, we 
recover a cycle inequality for any chosen vertex $i$:
\[
    -r_{i,i+n-1}+r_{i,i+1}+\dots+r_{i+n-2,i+n-1}\leq n-2 .
\]
\end{proof}

\myindent It is worth mentioning that~\cite{lostaglio2020contextualadvantage} have predicted the usefulness of~\cref{theorem: lostagio_senno}, i.e. this relation between preparation noncontextual models and two-state overlaps. They write:

\begin{quote}
    ``\emph{Finally, it may be possible to use the connection
between $\ell_1$ norm and confusability developed here [in Theorem~\ref{theorem: lostagio_senno}] to understand what aspects of other quantum information primitives, such as quantum teleportation, are truly nonclassical.}'' \\ 
~\citep[ pg.~7]{lostaglio2020contextualadvantage}.
\end{quote}
Indeed, our Proposition~\ref{proposition: ideal cyclic preparation noncontextuality} and our findings to be reported in Chapter~\ref{chapter: applications} consolidate their vision that this connection would be useful to uncover new generic bounds on preparation noncontextual models and contextual advantage. 

\subsection{Robust $n$-cycle inequalities}

\myindent The result from above has considered the case where $\varepsilon_v = 0$ for all $v \in V(G)$. Since we interpret these values as an operational account of experimental errors, it is interesting to look for robust descriptions of such a family of $n$-cycle noncontextuality inequalities.  There is a simple way of showing a robust version of~\cref{proposition: ideal cyclic preparation noncontextuality}.

\begin{proposition}
{Let $G=C_n$ be a cycle graph with $n\geq 3$. Then, the $n$-cycle overlap inequalities, corresponding to the non-trivial facet-defining inequalities of the convex polytope $\mathfrak{C}({C_n})$, can be mapped to robust noncontextuality inequalities of the scenario $\mathcal{PM}_{\mathrm{LSSS}}$,}
\begin{align*}
    r_{1,2}+r_{2,3}+\dots+r_{n-1,n}-r_{n,1} \leq n-2 + \varepsilon_1+\varepsilon_2+\varepsilon_3+\dots+\varepsilon_n
\end{align*}
and sign permutations of the right-hand side. 
\end{proposition}

\begin{proof}
Apply the triangle inequality satisfied by the $\ell_1$-norm, similarly to~\cref{proposition: ideal cyclic preparation noncontextuality}, while now allowing $\varepsilon_v \neq 0$ and therefore using the robust bounds provided by Theorem~\ref{theorem: lostagio_senno}. Every confusability $r_{i,i+1}$ satisfies,
\begin{equation*}
    -\varepsilon_i+1-\frac{1}{2} \Vert \mu_i - \mu_{i+1}  \Vert_1 \leq r_{i,i+1}\leq \varepsilon_i + 1 - \frac{1}{2}\Vert \mu_i - \mu_{i+1}  \Vert_1.
\end{equation*}
We conclude the argument by applying the triangle inequality to $\Vert \mu_1 - \mu_n + (\mu_2-\mu_2+\dots+\mu_{n-1}-\mu_{n-1})\Vert_1$ as before.
\end{proof}

\myindent When we treat quantum theory as an operational theory, the confusabilities $p(M_i \vert P_j)$ match the quantum realizations of two-state overlaps related to processes $P_i$ in the equivalence class defined by $\vert \phi_i \rangle$. Recall that each preparation $P_i$ has an associated measurement $M_i$---in an \acrshort{lsss} scenario---that by assumption checks whether $P_i$ was prepared. Thus, the values $\varepsilon_i$ are related to experimental errors in the sense that we allow for $p(M_i\vert P_i)=1-\varepsilon_i$ for some $\varepsilon_i>0$. Ideally, $p(M_i\vert P_i)=1$ meaning that provided we implement $M_i$, ideally we always learn with certainty that $P_i$ was implemented.  

\myindent While we have been able to map \emph{some} event graph noncontextuality inequalities to preparation noncontextuality inequalities, it is unclear if this can be generalized, meaning that there is a mapping between any facet-defining event graph inequality to \emph{some} noncontextuality inequality in some prepare and measure scenario.

\section{Discussion and further directions}\label{sec: discussion and further directions from event graph to noncontextuality}

\myindent In this Chapter we have described how the event graph polytope can be used to construct valid noncontextuality inequalities, related to the two main approaches to noncontextuality: the \acrshort{ks} approach and the Spekkens' approach, both introduced in Chapter~\ref{chapter: contextuality}.

\myindent To some extent, this Chapter has the very concrete purpose of resolving~\cref{question: coherence witnesses and contextuality}, which is a main question posed by this thesis. But beyond resolving this question, we highlight that some of our findings have independent interest for the literature of contextuality. First, we note that since it is straightforward to translate event graph inequalities to \acrshort{ks} and preparation noncontextuality inequalities, while we have focused mainly on the $n$-cycle inequalities, some of our results are broadly applicable to any of the many inequalities found in Chapter~\ref{chapter: event_graph_approach}. Moreover, as we have mentioned previously, for the case of Spekkens' notion of preparation noncontextuality our construction yields an infinite family of preparation noncontextuality inequalities, similarly to how the work by~\cite{araujo2013all} provided an infinite family of \acrshort{ks} noncontextuality inequalities, and how~\cite{collins2002bell} provided an  infinite family of Bell inequalities, which now go under the name of the \acrshort{cglmp} Bell inequalities. 

\myindent We believe that our graph-theoretic framework may be of parallel interest to research on the \acrshort{csw} approach. For instance, this approach has been recently used for applications on self-testing~\citep{bharti2019robust} and dimension witnessing~\citep{ray2021graphdimension}. We advance the use of the event graph framework for dimension witnessing in the next Chapter, but we leave open the opportunity of using the event graph approach for self-testing using insights provided by research associated with the \acrshort{csw} approach to contextuality. In our view, our approach is still in its infancy and can serve as a starting point for researchers interested in investigating the \acrshort{csw} approach, which is significantly more advanced and technically involved. It is clear that, since our inequalities recover those from the \acrshort{csw} framework, our framework can be considered equally well motivated insofar applications for the graph-theoretic scheme are considered. 

\myindent While we have been able to map \emph{some} event graph noncontextuality inequalities to preparation noncontextuality inequalities, it is unclear if this can be generalized. More precisely, it is unclear whether there is a mapping between any facet-defining event graph inequality to \emph{some} preparation noncontextuality inequality in some prepare and measure scenario. In a very concrete sense, since facet-defining inequalities of $\mathfrak{C}(G)$ can be violated by generic \textit{mixed} quantum states, it is more natural to see these as generalized noncontextuality inequalities than to \acrshort{ks} inequalities. However, we have not yet a complete such mapping. In case we could view every such inequality as a generalized noncontextuality (based on confusabilities) in some scenario this would be fairly powerful for research on Spekkens' notion of contextuality since, arguably, the facets of $\mathfrak{C}(G)$ are simpler to work with, and simpler to generalize to arbitrary families of graphs. 

\myindent It is also worth pointing out that the way we recover \acrshort{ks} noncontextuality inequalities suggests the interpretation that facet-defining inequalities of $\mathfrak{C}(\nabla H)$ for some exclusivity graph $H$ can be viewed as \emph{robust} \acrshort{ks} noncontextuality inequalities, i.e., a form of \acrshort{ks} inequalities rigorously well-defined for unsharp measurements. This is because we can view the mapping of $\mathfrak{C}(\nabla H) \mapsto \mathfrak{C}^0_{H|\nabla H}$ as a limit, hence interpreting small deviations of such a limit due to errors. In this case, we could view the different cross sections $\mathfrak{C}^\varepsilon_{H|\nabla H}$ for which $r_e = \varepsilon $ for all $e \in H$ as contributions due to the fact that the measurements were not `truly sharp', from which we have, in the example of the \acrshort{kcbs} inequality, the relation
\begin{equation}
    h_{\mathrm{KCBS}}(\pmb{\varepsilon},\pmb{r}) = r_{1,6}+r_{2,6}+r_{3,6}+r_{4,6}+r_{5,6}\leq2- 5\varepsilon.
\end{equation}
While concretely showing that this is a correct interpretation remains an open problem, we strongly believe that this is a direction worth pursuing. This would provide a better (and rigorous) understanding of robust tests of \acrshort{ks} inequalities whilst disputing claims that these are loopholes that fundamentally cannot be addressed.

\myindent To conclude, in this Chapter we have seen that there are applications of the event graph approach that stem from its connections with \acrshort{ks} noncontextuality and generalized noncontextuality inequalities. In the next Chapter we provide a concrete application that follows from this connection to the certification of quantum information advantage for a task. Moreover, we go beyond this connection showing that the event graph approach can be used for dimension witnessing, independently of its connection with noncontextuality theory. 
\chapter{Applications}\label{chapter: applications}

\begin{quote}
    ``\textit{Physics is not just about getting practical results, it is also about understanding how Nature behaves.}''\\
    (Bárbara~\cite{amaral2014phdthesis})
\end{quote}

\begin{quote}
    ``\textit{A society that permits biology to become an engineering discipline, which allows science to slip into the role of changing the living world without trying to understand it, is a danger to itself.}''\\ (Carl~\cite{woese2004newbiology})
\end{quote}

\myindent This Chapter considers applications of the framework developed in Chapters~\ref{chapter: event_graph_approach} and~\ref{chapter: relational coherence}. These selected applications are just \emph{a} motivation for our framework,  not \emph{the} main motivation for it. As the quotes above suggest, discussing applications can be insightful (see Chapter~\ref{chapter: information tasks}), but from a physics perspective their relevance lies in whether they deepen our understanding of Nature.  Practical utility alone should not be the criterion for determining what is worth investigating.

\myindent Having said that, in this Chapter we show two applications of the event graph approach: First, we show that violations of event graph inequalities can benchmark if quantum devices generate quantum coherence in Hilbert spaces of a certain minimal dimension. Second, we use the fact that our inequalities provide quantifiable bounds on the ability of preparation noncontextual models to reproduce two-state overlap statistics to prove a contextual advantage for the task of quantum interrogation.

\myindent Let us briefly comment on the relevance of benchmarking quantum devices. Usually, the task of certifying a certain property of a quantum device is described as investigating if the device in question is `truly quantum'. While this is a misleading way of phrasing (because it presumes the existence of things that are truly quantum and things that are not truly quantum) its intuition is certainly on the right track. We \emph{absolutely want} some way of experimentally certifying that a device is capable of generating quantum coherence, especially if we are interested in it due to some application. However, by `truly quantum' one often actually wants to convey the idea that the data generated by the device is \emph{truly nonclassical}, i.e., experimental results that would not have a clear classical explanation. For instance, it is of great importance to distinguish between a small quantum computer (say, having 100 noisy qubits) from a small classical simulator of a quantum computer, or to distinguish a classical random number generator from a quantum random number generator. 

\myindent Provided that a device is able to generate two-state overlap statistics, in this Chapter we apply the violation of event graph inequalities to the task of coherence certification. In fact, we certify a stringent criteria, which is that of \emph{coherence that requires} higher Hilbert space dimensions to be generated. This is a resolution of~\cref{question: applications}, which is one of the main targeted questions this thesis aims to solve. 

\myindent Let us also talk about the notion of \emph{quantum advantage} that we consider in this Chapter. We start by making a clear distinction between \emph{quantum computational advantage}~\footnote{Sometimes, rather unfortunately, termed `quantum supremacy'. } and \emph{quantum informational advantage}. While both have a similar name, they are of a significantly different nature. Since anything a classical computer can compute a quantum computer can compute as well (and vice versa), claims of quantum computational advantage are based on how \emph{efficient} a quantum computational task is, most notably in terms of time or memory required, as opposed to the performance of a classical computer. Claims of quantum computational advantage, hence, never state that `a classical computer cannot perform this calculation', but always state something of the form `a classical computer would need an unreasonably longer time to perform this calculation'. These statements are based on complexity-theoretic results, and on what is known (or believed to hold) about optimal ways of performing a certain computation.  

\myindent The concept of quantum information advantage is fundamentally different. It is not merely a matter of classical and quantum resources achieving the same performance with a difference in efficiency. Rather, one can argue that classical strategies for a given information task may be \emph{provably incapable} of achieving certain performance values that \emph{are} accessible by quantum theory. Such statements are dependent on what one understands by a `classical strategy' for implementing the information task. Once one has a working definition for what the expression `classical strategy' means (e.g. in our case, we consider that the information task can be described by some noncontextual ontological model) then usually one considers figures of merit for the performance of the information task (e.g. success rates) defined by some functional $f$. If there exists a bound on the figure of merit that can be considered, for example, something on the likes of 
\begin{equation}
    f(\pmb B) \stackrel{\text{Classical  strategies}}\leq b
\end{equation}
we have that $f(\pmb B)>b$ when $\pmb{B}$ is a quantum behavior defines, formally, a proof of quantum information advantage for the task. This is exactly what we do later for the task of quantum interrogation, and which we consider another response for~\cref{question: applications} that we have proposed as a main goal of this thesis. We consider a figure of merit $\eta$ and show that quantum theory can reach certain values of $\eta$ that  noncontextual ontological models cannot.

\myindent The structure of this Chapter is as follows. In Sec.~\ref{sec: quantum advantage for interrogation} we show how we can use the $c_3(\pmb{r}) \leq 1$ inequality to bound noncontextual models reproducing the statistical predictions of the standard quantum interrogation task. Section~\ref{sec: quantum interrogation as a PM scenario} shows the contextual advantage in an ideal case, while Sec.~\ref{appendix: robust_inter} considers a robust case that takes into account errors modeled by white noise. In Section~\ref{sec: dimension and coherence witnessing section} we show that the event graph formalism can be used to witness both coherence and dimension. We conclude in Section~\ref{sec: discussion of dimension and advantage} with a discussion and outlook. 

\section{Quantum advantage for quantum interrogation}\label{sec: quantum advantage for interrogation}

\myindent Consider the \acrshort{mzi} from Fig.~\ref{fig: Bomb_MZI},  Chapter~\ref{chapter: information tasks}, that we have used to discuss the quantum interrogation protocol 
 proposed by~\cite{elitzur1993quantum}. We reproduce the same illustration here as Fig.~\ref{fig: Bomb_MZI_PM_scenario} to ease readability. Recall that in this scenario we set $U_{\theta_1} = U_{\theta_2}^\dagger$, with $U_{\theta_1}$ describing a beam splitter (\acrshort{bs}) as in Eq.~\eqref{eq: BS}. The condition $U_{\theta_1} = U_{\theta_2}^\dagger$ is necessary  to have a dark detector. 

\subsection{Prepare and measure contextuality scenario for the quantum interrogation task}\label{sec: quantum interrogation as a PM scenario}

\myindent Let us now address operationally the content of the \acrshort{mzi} statistics arising from quantum interrogation. The  efficiency of the task---$\eta$ given by Eq.~\eqref{eq: efficiency}---arises from a specific interplay between the creation of superposition in the preparation stage, and the ability to make a coherent detection in the measurement stage. We now show that the standard bomb testing experiment satisfies the operational constraints defining a \acrshort{lsss} prepare-and-measure contextuality scenario. We start by describing the main elements of the test from the point of view of an operational probabilistic theory (\acrshort{opt}). In what follows, we consider that we have access to two beam splitters $BS_1$ and $BS_2$ that would implement (modeled quantum mechanically) the unitary operations $U_{\theta_1}$ and $U_{\theta_2}$ given by Eq.~\eqref{eq: BS}.  

\subsubsection{Preparation stage}

\begin{figure*}[t]
    \centering
    \includegraphics[width=1\linewidth]{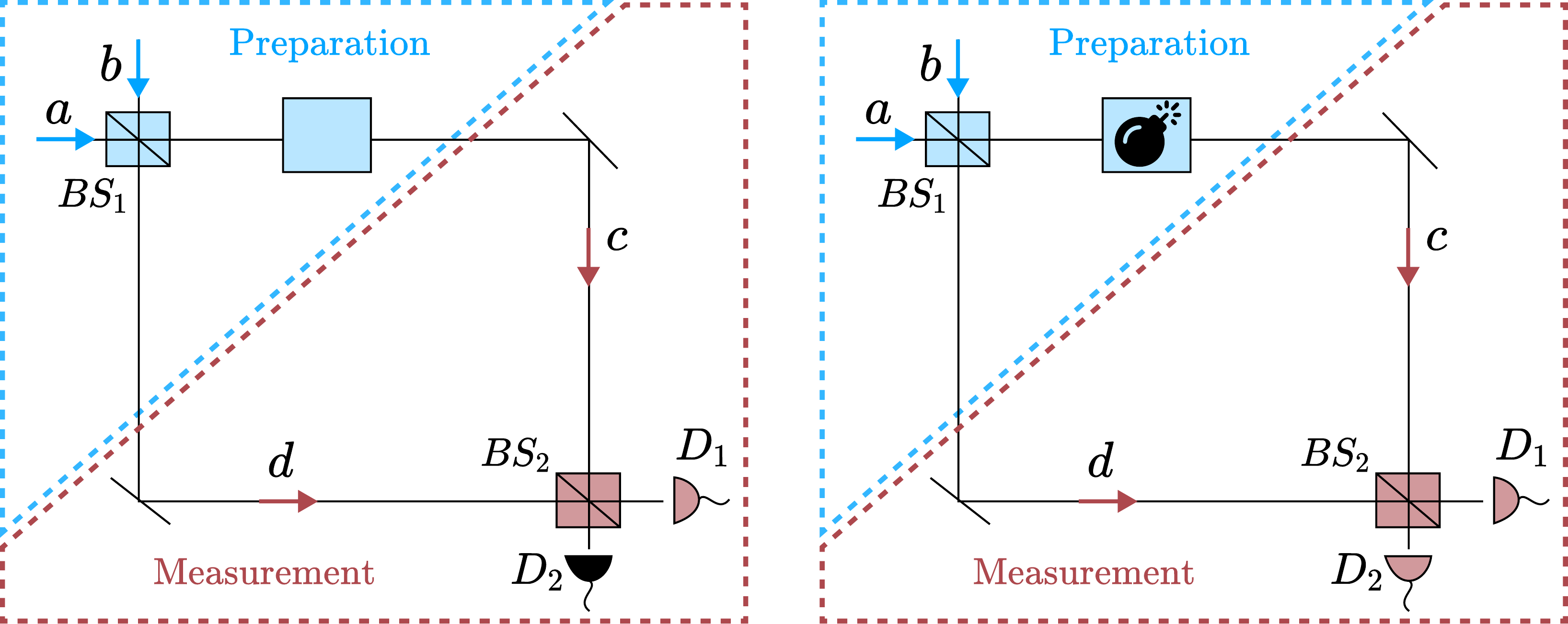}
    \caption{\textbf{The task of quantum interrogation with an \acrshort{mzi} can be viewed as defined in a prepare and measure \acrshort{lsss} scenario.} We reproduce Figure~\ref{fig: Bomb_MZI} from Chapter~\ref{chapter: contextuality} here to improve readability. When modelled quantum mechanically, we define a single-photon entering mode $a$ as the preparation of the state $\vert 0\rangle \langle 0 \vert$. Moreover, $BS_1$ and $BS_2$ are modelled quantum mechanically by a unitary $U_{\theta_1}$ and $U_{\theta_2}$, such that $U_{\theta_1} = U_{\theta_2}^\dagger$. }
    \label{fig: Bomb_MZI_PM_scenario}
\end{figure*}

\myindent We use $P_0$ to represent the preparation of a photon in mode \textit{a}, which, in quantum theory, corresponds to the preparation of the state  $\vert 0 \rangle \langle 0 \vert $. Similarly with $P_{\theta_1}$ and $P_{\theta_2}$, that we use to denote the preparations of a photon again in mode \textit{a}, followed by beam-splitters $BS_1$ or $BS_2$ that quantum mechanically implement $U_{\theta_1}$ or $U_{\theta_2}$, respectively. In Fig.~\ref{fig: Bomb_MZI_PM_scenario} we show the situation in which $BS_1$ is in the preparation stage and $BS_2$ is in the measurement stage, as this is the one used for the quantum interrogation task. We assume that we could, alternatively, have put $BS_2$ in the preparation stage and $BS_1$ in the measurement stage. In quantum theory, these preparations $P_0, P_{\theta_1}$ and $P_{\theta_2}$ just described correspond to preparing the quantum states $\vert 0\rangle \langle 0 \vert, U_{\theta_1} \vert 0 \rangle \langle 0 \vert U_{\theta_1}^\dagger$, and $U_{\theta_2}\vert 0 \rangle \langle 0 \vert U_{\theta_2}^\dagger$, respectively. 

\myindent Clearly, if we denote $P_{1}$ to represent the preparation of a photon in mode \textit{b} which, in quantum theory, corresponds to the preparation of the state $\vert 1 \rangle \langle 1 \vert$, and similarly for $P_{\theta_1^\perp}$ and $P_{\theta_2^\perp}$ substituting $\vert 0\rangle \mapsto \vert 1\rangle$ we have that the six preparation procedures $P_0,P_{\theta_1},P_{\theta_2},P_{1},P_{\theta_1^\perp},P_{\theta_2^\perp}$ satisfy the operational equivalence
\begin{equation}
    \frac{1}{2}P_0+\frac{1}{2}P_{1} \simeq \frac{1}{2}P_{\theta_1}+\frac{1}{2}P_{\theta_1^\perp} \simeq \frac{1}{2}P_{\theta_2} + \frac{1}{2}P_{\theta_2^\perp},
\end{equation}
which is the defining operational equivalence relevant to \acrshort{lsss} scenarios as described in Chapter~\ref{chapter: contextuality}. The operational equivalences from above follow from the fact that the quantum mechanical description of these procedures satisfy
\begin{equation}
    \frac{1}{2}\vert 0\rangle \langle 0 \vert +\frac{1}{2}\vert 1 \rangle \langle 1 \vert  = \frac{1}{2}U_{\theta_1}\vert 0\rangle \langle 0 \vert U_{\theta_1}^\dagger +\frac{1}{2} U_{\theta_1}\vert 1 \rangle \langle 1 \vert U_{\theta_1}^\dagger =  \frac{1}{2}U_{\theta_2}\vert 0\rangle \langle 0 \vert U_{\theta_2}^\dagger +\frac{1}{2} U_{\theta_2}\vert 1 \rangle \langle 1 \vert U_{\theta_2}^\dagger, 
\end{equation}
since all these states are equal to $\mathbb{1}/2$. 

\subsubsection{Measurement stage}

\myindent Each preparation procedure has an associated binary-outcome measurement procedure in the prepare-and-measure scenario we construct, exactly as we have described before when looking at the \acrshort{lsss} constraints. Here, in this scenario, the measurements are described by the detectors $D_1$ and $D_2$ illustrated in Fig.~\ref{fig: Bomb_MZI_PM_scenario}. We have either the possibility of letting $BS_2$ do nothing so that $M_0$ effectively just checks the path, and is modeled quantum mechanically via \begin{equation}M_0 = \{\vert 0\rangle \langle 0 \vert , \vert 1 \rangle \langle 1 \vert \}.\end{equation} Alternatively, we can let $BS_2$ perform $U_{\theta_2}$ (which is the case for the interrogation task) or $BS_1$ perform $U_{\theta_1}$ providing us with $M_{\theta_1}, M_{\theta_2}$ that are modeled quantum mechanically as \begin{equation}M_{\theta_1} = \{U_{\theta_1}\vert 0\rangle \langle 0 \vert U_{\theta_1}^\dagger , U_{\theta_1}\vert 1 \rangle \langle 1 \vert U_{\theta_1}^\dagger \}\end{equation} and \begin{equation}M_{\theta_2} = \{U_{\theta_2} \vert 0\rangle \langle 0 \vert U_{\theta_2}^\dagger , U_{\theta_2}\vert 1 \rangle \langle 1 \vert U_{\theta_2}^\dagger\},\end{equation} respectively. 

\myindent To ease notation with respect to the operational content of the scenario, we write $r_{ij} = p(0 \vert M_i, P_j)$, see Table~\ref{tab:operational_content_table}.  To conclude our mapping, showing that the \acrshort{mzi} for which we can perform the quantum interrogation task satisfies the operational requirements of an  \acrshort{lsss} scenario, we just need to note that 
\begin{equation}
    p(0|M_i,P_{i}) = 1,\quad p(0|M_i,P_{i^\perp}) = 0,
\end{equation}
because $$p(0|M_i,P_{i}) = \text{Tr}(U_i|0\rangle \langle 0 |U_i^\dagger U_i|0\rangle \langle 0 | U_i^\dagger)$$ and $$p(0|M_i,P_{i^\perp}) = \text{Tr}(U_i|1\rangle \langle 1 |U_i^\dagger U_i|0\rangle \langle 0 | U_i^\dagger) = 0,$$ where $U_i \in \{\mathbb{1},U_{\theta_1},U_{\theta_2}\}$. Clearly, this is the \emph{ideal} situation discussed in Chapters~\ref{chapter: contextuality} and~\ref{chapter: from overlaps to noncontextuality}, where $p(0|M_i,P_{i^\perp}) = \varepsilon_i = 0$. Later, in~\cref{appendix: robust_inter}, we relax this constraint, allowing for $\varepsilon_i > 0$.

\begin{table}[t]
    \centering
    \begin{tabular}{|c|c|c|c|}
    \hline
         & $P_1$ & $P_2$ & $P_3$ \\
    \hline
    $0|M_1$  &  $1$  &   $r_{\theta 0}$  &  $r_{\theta^\dagger 0}$   \\
    \hline
    $0|M_2$  &  $r_{\theta 0}$   &  $1$ &   $r_{\theta \theta^\dagger}$  \\
    \hline
    $0|M_3$  &  $r_{\theta^\dagger 0}$   &  $r_{\theta \theta^\dagger}$   &  $1$   \\
    \hline
    \end{tabular}
    \caption{\textbf{Operational symmetries over the statistics of quantum interrogation.} Notation for the prepare-and-measure statistics arising from the quantum interrogation scenario.}
    \label{tab:operational_content_table}
\end{table}

\subsubsection{Contextual advantage for quantum interrogation}

\myindent From the assumption $U_{\theta_1} = U_{\theta_2}^\dagger$, we can simply write $\theta_1 = \theta$ and $\theta_2 = \theta^\dagger$. After the first beam splitter, the presence of an object effectively corresponds to performing a measurement $M_0$, which leads to the statistics described by $r_{\theta 0} = p(0 \vert M_0,P_\theta), r_{\theta 1}=p(1\vert M_0,P_\theta)$. With reference to Fig.~\ref{fig: Bomb_MZI_PM_scenario}, in quantum theory $P_\theta$ corresponds to the preparation $\vert \psi(\theta)\rangle = U_\theta \vert 0 \rangle$, 
\begin{equation}\label{eq: preparation state bomb}
    \vert \psi(\theta)\rangle = U_\theta\vert 0\rangle =  \cos(\theta)\vert 0 \rangle + i\sin(\theta)\vert 1 \rangle. 
\end{equation}
In quantum theory, detecting the photon either in one arm or the other have the  associated probabilities $r_{\theta 0}= \vert \langle \psi(\theta) \vert 0 \rangle \vert^2$ and $ r_{\theta 1} = \vert \langle \psi (\theta) \vert 1 \rangle \vert^2$. In case of the presence of a bomb inside the device, the dark detector lights up when the photon is sent into the arm for which no explosion happens. In this event, we find the measurement statistics for the detectors to be $r_{\theta^\dagger 0} = p(\theta^\dagger\vert M_{\theta^\dagger}, P_0),r_{\theta^\dagger 1} = p({\theta^\dagger}^\perp\vert M_{\theta^\dagger}, P_0) $ whereas, in quantum theory, $r_{\theta^\dagger 1} = \vert \langle 1 \vert \psi^\dagger(\theta,0)\rangle \vert^2$, $r_{\theta^\dagger 0} = \vert \langle 0 \vert \psi^\dagger(\theta,0)\rangle \vert^2$, with $\vert \psi^\dagger(\theta,0) \rangle = U_\theta^\dagger \vert 0 \rangle $.  For the states $\vert \psi^\dagger(\theta,0)\rangle$ we have used
\begin{equation}\label{eq: detection state}
    \vert \psi^\dagger(\theta,0) \rangle = \cos(\theta)\vert 0 \rangle - i\sin(\theta)\vert 1 \rangle.
\end{equation}

\myindent Every preparation has an associated binary-outcome measurement.  Hence, the $M_\theta$ and $M_{\theta^\dagger}$ measurement procedures give $p(\theta \vert M_\theta, P_\theta) = 1$ and $p(\theta^\dagger \vert M_{\theta^\dagger}, P_{\theta^\dagger}) = 1$, and the quantity that in quantum theory is given by $r_{\theta \theta^\dagger} = \vert \langle \psi (\theta,0) \vert \psi^\dagger(\theta,0) \rangle \vert^2$ is operationally defined by the confusability $p(\theta \vert M_\theta,P_{\theta^\dagger}) = r_{\theta \theta^\dagger}$.

\myindent The efficiency of the task, given by Eq.~\eqref{eq: efficiency} $$\eta = \frac{p_{\mathrm{succ}}}{p_{\mathrm{succ}}+p_{\skull}},$$ can thus be expressed solely by the operational quantities just described. The quantity $p_{\mathrm{succ}} = r_{\theta 0}r_{\theta^\dagger 1}$ corresponds to the probability that the photon enters the \acrshort{mzi}, does not chose the path with the bomb, and then, after the second beam splitter, it makes the dark detector light up, ending up with a successful detection of the bomb without exploding it. On the contrary, $p_{\skull} = r_{\theta 1}$ corresponds to the probability that, after the first beam splitter, the photon takes the path with the bomb, which consequently explodes, failing to accomplish the task. Hence,
\begin{equation}\label{eq: operational efficiency}
    \eta = \frac{r_{\theta 0}r_{\theta^\dagger 1}}{r_{\theta 0}r_{\theta^\dagger 1}+r_{\theta 1}}.
\end{equation}

\myindent It is interesting to note that there are operationally relevant symmetries that are respected by quantum theory, and that can be imposed in the prepare-and-measure scenario. In particular, quantum theory satisfies $$r_{\theta 0} = \vert \langle \psi(\theta,0) \vert 0\rangle \vert^2 = \vert \langle \psi^\dagger(\theta,0) \vert 0 \rangle \vert^2 = r_{\theta^\dagger 0}$$ and analogously,  $r_{\theta 1} = r_{\theta^\dagger 1}$. Moreover, $$r_{\theta \theta^\dagger} = \vert \langle \psi(\theta,0)\vert \psi^\dagger(\theta,0) \rangle  \vert^2 = \vert\cos^2(\theta)-\sin^2(\theta)\vert^2= (r_{\theta 0} - r_{\theta 1})^2.$$ Therefore, we assume the following symmetries over the scenario:
\begin{align}
    &r_{\theta 0} = r_{\theta^\dagger 0},\label{eq: operational constrain 1}\\
    &r_{\theta 1} = r_{\theta^\dagger 1},\label{eq: operational constrain 2}\\
    &r_{\theta \theta^\dagger} = (r_{\theta 0} - r_{\theta 1})^2.\label{eq: operational constrain 3}
\end{align}
Under symmetries~\eqref{eq: operational constrain 1} and \eqref{eq: operational constrain 2} we can re-write the efficiency $\eta$ as a function of $r_{\theta 0}$ only,
\begin{equation}\label{eq: operational efficiency with constraints}
    \eta = \frac{p(0\vert M_0, P_\theta)}{p(0\vert M_0, P_\theta) + 1} \equiv \frac{r_{\theta 0}}{r_{\theta 0}+1},
\end{equation}
and with the symmetry \eqref{eq: operational constrain 3}, and the inequality $c_3(\pmb{r}) \leq 1$ that has been shown to be a preparation noncontextuality inequality in Chapter~\ref{chapter: event_graph_approach}, it is possible to find a clear gap bounding the efficiency achievable by any noncontextual model that attempts at explaining the quantum theory predictions,
\begin{align*}
    -r_{\theta \theta^\dagger}+r_{\theta 0} + r_{\theta^\dagger 0} \leq 1 \hspace{0.8em}&\stackrel{}{\Rightarrow}\hspace{0.8em}
    -r_{\theta \theta^\dagger} + 2r_{\theta 0} \leq 1 \hspace{0.8em}\Rightarrow \\[1.5ex]
    \frac{r_{\theta 0}}{r_{\theta 0} + 1} \leq \frac{1 + r_{\theta \theta^\dagger}}{2(r_{\theta 0}+1)} \hspace{0.8em}&\stackrel{}{\Rightarrow}\hspace{0.8em} \eta \stackrel{\mathrm{NC}}{\leq} \frac{1 + (2r_{\theta 0}-1)^2}{2(r_{\theta 0}+1)}.
\end{align*}

\begin{figure}[t]
    \centering
    \includegraphics[width=1\textwidth]{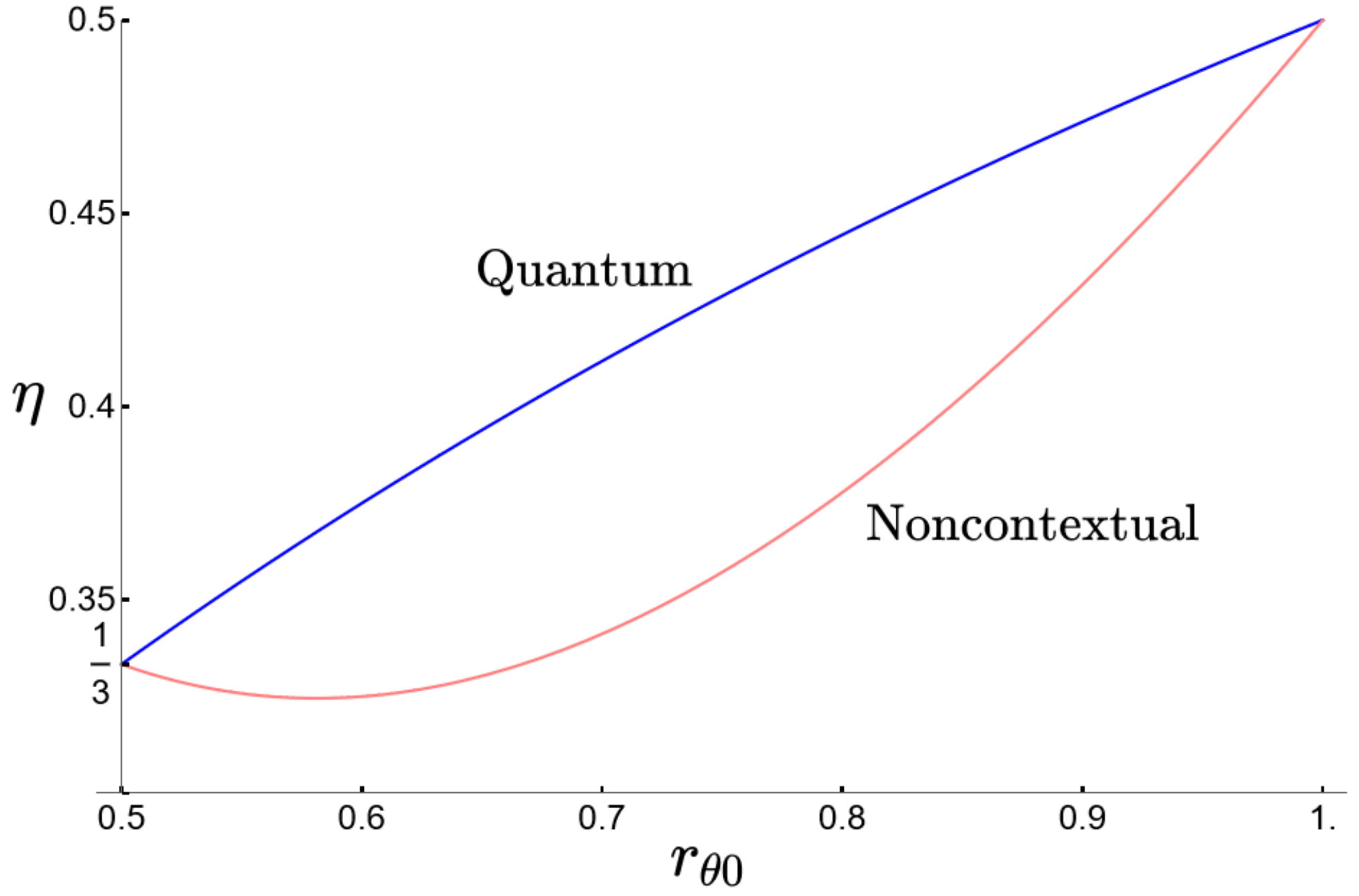}
    \caption{\textbf{Contextual advantage for quantum interrogation.} In blue, the upper curve presents the efficiency for an ideal action of asymmetric beam splitters in the MZI, achievable with standard quantum theory. In pink, the lower curve shows the optimal efficiency achievable by any noncontextual ontological model that explains the statistics of the associated prepare-and-measure scenario, while respecting two operational constraints also satisfied by quantum theory: $r_{\theta 0} = r_{\theta^\dagger 0}, r_{\theta \theta^\dagger} = (2r_{\theta 0}-1)^2$. }
    \label{fig: efficiency standard interrogation}
\end{figure}

\myindent Figure~\ref{fig: efficiency standard interrogation} shows a plot of the efficiency $\eta$, achievable by quantum theory  as described by Eq.~\eqref{eq: operational efficiency with constraints}, versus the optimal noncontextual efficiency $$\eta^{NC}_{opt} = \frac{1 + (2r_{\theta 0}-1)^2}{2(r_{\theta 0}+1)},$$ as functions of the operational quantity $r_{\theta 0}$. From this plot, it is clear that the efficiencies meet at $\sfrac{1}{3}$, corresponding to the efficiency of the original proposal of~\cite{elitzur1993quantum}, as well as that of the noncontextual toy model developed by~\cite{catani2023whyinterference}. Our results are consistent with those found by~\cite{catani2023whyinterference}, which presents a noncontextual model for the cases when $r_{\theta 0}=\sfrac{1}{2}$ and $r_{\theta 0}=1$. Moreover, our work broadly parallels the results from~\cite{catani2023aspects} and~\cite{catani2022what}, connecting contextuality theory with uncertainty relations. This connection is natural, given that wave-particle relations exemplify uncertainty relations. Our findings and those present in the aforementioned references highlight two distinct contextual aspects of quantum interference beyond the basic (noncontextual) phenomenology: one concerning the probability of success in quantum interrogation as a function of state confusability, and the other concerning the functional form of wave-particle duality relations.

\myindent We observe that asymmetric beam splitters provide contextual advantage for the task described with quantum theory, performing significantly better for some choices of $r_{\theta 0}$. The maximum gap is given by $$\max_{r_{\theta 0}}(\eta - \eta_{opt}^{NC}) \approx 0.071$$ at $r_{\theta 0} =\sqrt{3}-1 \approx 0.73$. 

\myindent One might naturally ask on what really is  the role of contextuality in this task. In the limit of highest efficiency $\eta \to \sfrac{1}{2}$ the protocol admits a noncontextual explanation. However, this limiting case is not practically useful: when $\eta \to \sfrac{1}{2}$ we have $r_{\theta0}=1$, so no quantum superposition is created inside the interferometer and the object is never probed (no photon is sent through the arm that contains it). In other words, the limit is purely mathematical and destroys the very effect the protocol is intended to exploit. Therefore, any experimentally relevant improvement of $\eta$---i.e. any improvement away from this trivial regime---must rely on contextuality. In this sense, generalized contextuality fuels the quantum advantage provided by quantum theory over noncontextual models. 

\subsection{Robust account of contextuality in quantum interrogation}\label{appendix: robust_inter}

\myindent Let us start by analysing the robustness of the effects of contextuality in the interrogation task just described due to depolarizing noise. Quantum interrogation assumes the possibility of preparing six different quantum states, and measuring over all of these states as well, where the states relevant for the quantum interrogation were described above.

\myindent Due to our results from Chapter~\ref{chapter: from overlaps to noncontextuality}, the trade-off between what can be achieved with quantum theory and any noncontextual model is described by the noncontextuality inequality
\begin{equation}\label{eq: noncontextuality inequality}
    p(0\vert M_0,P_\theta) + p(1 \vert M_0, P_{\theta^\dagger})-p(\theta \vert M_{\theta},P_{\theta^\dagger}) \leq 1 + \varepsilon_0+\varepsilon_\theta+\varepsilon_{\theta^\dagger},
\end{equation} related to the \acrshort{lsss} prepare-and-measure scenario described above. 

\myindent Recalling that, operationally, the efficiency of the task is characterized by the quantity (see Eq.~\eqref{eq: operational efficiency with constraints}) 
$$
    \eta = \frac{p(0\vert M_0, P_\theta)}{p(0\vert M_0, P_\theta) + 1},
$$
where $p(0\vert M_0, P_\theta)$ is the probability that given that the state prepared was $\vert \theta\rangle$ we have detected the photon on the path associated to state $\vert 0\rangle$. The strategy from now on is to use the theoretical description of the quantum experiment in terms of noisy quantum states affected by the channel $\mathcal{D}_\nu$. We therefore consider a depolarizing channel $\mathcal{D}_\nu$ given by
\begin{equation}\label{eq: depo channel}
    \mathcal{D}_\nu(X) := (1-\nu)X + \nu \frac{\mathbb{1}_d}{d}\text{Tr}(X),
\end{equation}
with $\mathbb{1}_d$ a $d\times d$ identity matrix, implying that $\mathbb{1}_2/2$ corresponds to the maximally mixed qubit state. The factor $\nu$ quantifies the amount of white noise.  We model measurement effects by the noisy versions $$0\vert M_0 \mapsto \mathcal{D}_\nu(\vert 0 \rangle \langle 0 \vert),\quad  1 \vert M_0 \mapsto \mathcal{D}_\nu(\vert 1\rangle \langle 1\vert )$$ and similarly for all $\theta$,  $0|M_{\theta} \mapsto \mathcal{D}_\nu(\vert \theta \rangle \langle \theta \vert)$\footnote{Technically, each measurement effect and each quantum state correspond to an equivalence class of all possible procedures that prepare/measure them. We abuse notation here, by simply saying that we map operational measurement effects into specific instances of quantum effects.}. The same  for the preparations. Therefore, we have that the optimal quantum strategy obtained under the effect of a channel $\mathcal{D}_\nu$ now reads

\begin{equation}\label{eq: robust efficiency}
    \eta = \frac{\text{Tr}(\mathcal{D}_\nu(\vert 0 \rangle \langle 0 \vert ) \mathcal{D}_\nu(\vert\theta \rangle \langle \theta \vert))}{\text{Tr}(\mathcal{D}_\nu(\vert 0 \rangle \langle 0 \vert ) \mathcal{D}_\nu(\vert\theta \rangle \langle \theta \vert)) + 1} = \frac{\text{Tr}(\rho_0\rho_\theta)}{\text{Tr}(\rho_0\rho_\theta)+1},
\end{equation}
where for simplicity we write $\rho_s \equiv \mathcal{D}_\nu(\vert s \rangle \langle s \vert) $. 

\myindent For the noncontextuality bound, we can see  $\varepsilon_0$,  $\varepsilon_\theta$ and $\varepsilon_{\theta^\dagger}$ as functions of $\nu$,
\begin{equation}
    1-\varepsilon_0 = p(0\vert M_0 , P_0) = \text{Tr}(\rho_0\rho_0) = 1+\frac{\nu^2}{2}-\nu
\end{equation}
and it is easy to see that $\varepsilon_0 = \varepsilon_\theta = \varepsilon_{\theta^\dagger}$. Therefore, $\varepsilon_i = \nu - \frac{\nu^2}{2}$ for $i \in \{0,\theta,\theta^\dagger\}$. In such a way, we have that the robust noncontextual bound provided by the prepare-and-measure noncontextuality inequality is
\begin{equation}\label{eq: noncontextual efficiency}
    \eta^{NC}(\theta,\nu) \equiv \frac{1 + \text{Tr}(\rho_\theta\rho_{\theta^\dagger}) - \text{Tr}(\rho_1\rho_{\theta^\dagger}) + 3\left(\nu - \frac{\nu^2}{2}\right)}{\text{Tr}(\rho_0 \rho_\theta) + 1}.
\end{equation}

\begin{figure}[t]
    \centering
    \includegraphics[width=\columnwidth
]{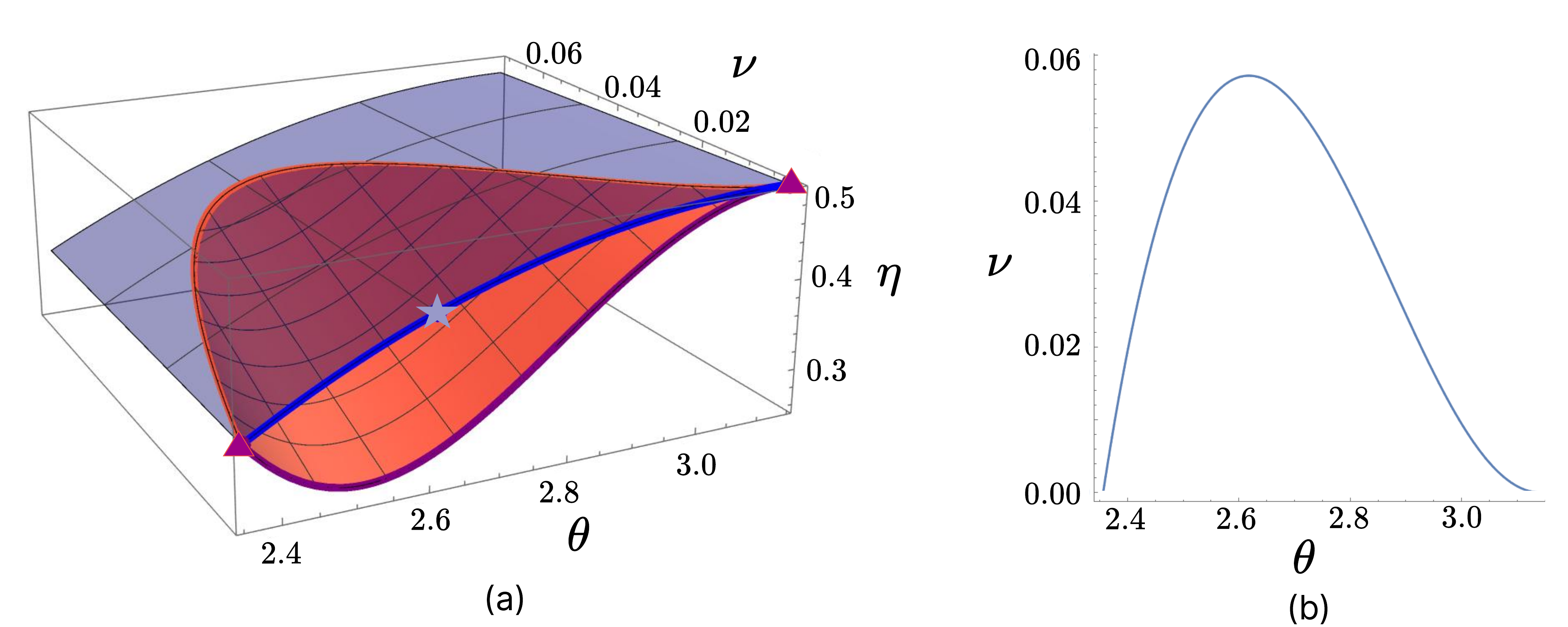}
    \caption{\textbf{Efficiency of quantum interrogation.} (a) Efficiency $\eta$ as a function of $\theta$, which encodes differences in beam splitter transmission/reflectivity. The upper (blue) curve shows quantum performance, while the lower (orange/purple) curve gives the noncontextual upper bound.  Robustness is captured by  depolarizing noise $\nu$, where  $(1-\nu)\rho +\nu \,\mathbb{1}/2$, applies to any state $\rho$ prepared in the interferometer. The purple triangles mark $\eta = \sfrac{1}{3}$ and $\eta=\sfrac{1}{2}$ achieved using the toy model from~\cite{catani2023whyinterference}. The blue star at $\theta=\sfrac{5\pi}{6}$ that can be translated into a proof of contextuality as originally presented by~\cite{spekkens2005contextuality}, with efficiency $\eta(\sfrac{5\pi}{6},0)=0.428$; contextual advantage is lost if $\nu>0.057$. (b) Curve marking, for each value of $\theta$, the noise level for which any claim of contextual advantage is lost. Visually, this curve marks the intersection of the two curved regions present in (a).}
    \label{fig:robust}
\end{figure}

\myindent Whenever $\eta > \eta^{NC}$, we observe a robust advantage provided by quantum contextuality in the interrogation task. In Fig.~\ref{fig:robust}(a) we plot both the efficiency that can be achieved with quantum theory (blue) $\eta$ and the noncontextual bound (pink) $\eta^{NC}$ as a function of the parameter $\theta$ that characterizes the transmissivity of the beam splitters in the \acrshort{mzi}, and the amount of noise $\nu$ captured by the channel from Eq.~\eqref{eq: depo channel}. The curve in which the two meet characterizes the degree of noise $\nu$ for which no advantage can be claimed. In Fig.\ref{fig:robust}(b) we plot the robustness to the noise parameter $\nu$ as a function of $\theta$ as described by the operational prepare-and-measure scenario considered. In this case, for $\nu>0.057$ we would lose the advantage in the protocol. 

\section{Dimension and coherence witnessing}\label{sec: dimension and coherence witnessing section}

\myindent In this section we show that there is a way of using the event graph framework for witnessing \emph{both} coherence and dimension by showing that there are formal gaps in the value of violations of facet-defining inequalities from $\mathfrak{C}(G)$ using different quantum realizations. In Chapter~\ref{chapter: relational coherence} we have seen how to numerically investigate the boundary of the sets of quantum correlations using \acrshort{sdp} tools, and we have already investigated different bounds for different Hilbert space dimensions. This is an important toolbox if one is interested in witnessing both coherence and dimension. Nevertheless, the task of \emph{witnessing dimension} alone, as discussed in Chapter~\ref{chapter: information tasks}, can be viewed as distinct from that of coherence witnessing. Before considering \emph{both} in a test, we discuss how one can use the event graph approach discussed in Chapter~\ref{chapter: event_graph_approach} to witness Hilbert space dimension from relational information provided by tuples of two-state overlaps $r \in \mathfrak{C}(G)$. 

\subsubsection{Witnessing dimension without witnessing relational coherence}

\myindent We start by noticing that the event graph approach (partially) extends the dimension witnessing approach considered by~\cite{galvao2020quantum}, and later experimentally tested by~\cite{giordani2021witnesses} where one is interested in witnessing Hilbert space dimension \emph{without} using relationally coherent edge weightings ${r}$. Using the terminology of this thesis,~\cite{galvao2020quantum} showed how to find bounds on relationally incoherent values ${r} \in \mathfrak{C}^{(3)}(C_3) \setminus \text{ConvHull}[\mathfrak{C}^{(2)}(C_3)]$ and ${r} \in \mathfrak{C}^{(3)}(C_3) \setminus \text{ConvHull}[\mathfrak{Q}^{(2)}(C_3)]$. For the second case, they have used knowledge from the quantum set of two-state overlaps, while for the first case they have used properties of the event graph polytope. 

\myindent Using~\cref{theorem: colorings} we can extend this same approach to every possible event graph $G$. Using~\cref{corollary: dimensionally restricted from colorability} we can see that there is a simple way to find bounds for edge weightings ${r} \in \mathfrak{C}^{(d)}(G) \setminus \text{ConvHull}[\mathfrak{C}^{(d-1)}(G)]$. To do so, we simply consider an event graph $G$ and then we solve the problem of coloring $G/\alpha$ for all  deterministic edge weightings $\alpha: E(G) \to \{0,1\} \in \mathrm{ext}(\mathfrak{C}(G))$. If we consider the convex combinations of all  deterministic edge weightings for which $G/\alpha$ is $d-1$ colorable, and take the convex hull of these, we obtain $\text{ConvHull}[\mathfrak{C}^{(d-1)}(G)]$. Formally, this is a convex polytope, and as we discuss in~\cref{sec: convex polytopes}, and as we have considered already in Chapter~\ref{chapter: event_graph_approach}, we can find facet-defining inequalities for $\text{ConvHull}[\mathfrak{C}^{(d-1)}(G)]$, from which we can then look for edge weightings $r$ that violate these inequalities. Any violation would imply witnessing $d_{\min} = d>d-1$ as a lower bound for the underlying dimension of a \emph{classical} state-space. If we consider $G=C_3$ and reference~\citep{galvao2020quantum}, the set $\mathfrak{C}^{(3)}(C_3)\setminus \mathrm{ConvHull}[\mathfrak{C}^{(2)}(C_3)]$ is provided by regions I and II from their Figure~5. The set $\mathfrak{C}^{(3)}(C_3) \setminus \text{ConvHull}[\mathfrak{Q}^{(2)}(C_3)]$ is provided by region I of the same figure. 

\myindent In our view, the more interesting test is one where we witness \emph{both} Hilbert space dimension \emph{and} some form of nonclassicality such as coherence, Bell nonlocality or contextuality.  

\subsubsection{Witnessing coherence and dimension: comparing methods}

\myindent Let us make a distinction between dimension witnessing we consider here and the dimension witnessing protocols that have presented in Chapter~\ref{chapter: information tasks}. There, we have considered dimension witnesses based on (i) the violation of Bell inequalities, (ii) the violation of \acrshort{ks} noncontextuality inequalities, and (iii) the violation of prepare and measure inequalities that assumed a restriction in the message being sent (e.g. an upper bound on the Hilbert space dimension) in order to obtain a meaningful classical-quantum divide. Each of these dimension witnessing protocols have limitations that we now overcome by using the event graph approach. 

\myindent To start commenting on these limitations, arguably, (i) cannot witness arbitrary (total) Hilbert space dimensions, but only \emph{product} Hilbert space dimensions. In order to witness $d=3$ one needs to perform a test using at least two qutrits, and assume some validity of a tensor product structure. Perhaps a more pressing limitation is the necessity of performing the specific test associated to a Bell test, which has certain stringent requirements such as entanglement generation, measurement incompatibility, and some specific circuit configuration.  Moreover, (ii) requires the assumptions related to performing a test that aimes to violate a \acrshort{ks} noncontextuality inequality (such as exclusivity conditions, the assumption of sharpness, among others), which implies that \emph{experimental deviations} from these assumptions need to be characterized and bounded, in order to appropriately perform the dimension witnessing protocol.~\footnote{The same holds to some extent for the event graph inequalities, where one needs to probe how close the statistics is from that of two-state overlaps. Experimentally, $\text{Tr}(\rho\sigma) = \text{Tr}(\sigma \rho)$ is never truly satisfied, and one needs to bound the contribution from such errors~\citep{giordani2023experimental}.} To exemplify this situation, we can consider the event graph inequality $h_{\mathrm{KCBS}}(\pmb{r}) \leq 2$ that, without the exclusivity conditions can be violated by single-qubit states (as can be seen in Table~\ref{tab: seesaw bound}) while with the exclusivity conditions it cannot be violated by qubits as it becomes equivalent to the \acrshort{kcbs} noncontextuality inequality (cf.~Chapter~\ref{chapter: from overlaps to noncontextuality}). Lastly, (iii) resolves both of the issues from (i) and (ii) just pointed out, but it \emph{does not} distinguish between sets of states defined for a sufficiently large Hilbert space that are set coherent, versus set incoherent. For (iii), if we use a certain inequality violation to witness dimension, there is always a sufficiently large Hilbert space dimension capable of violate the same inequality but using only the preparation of classical (i.e. set incoherent) quantum states (recall Fig.~\ref{fig: coherence and dimension part I}, and the discussion that follows it). 

\myindent The event graph resolves the issues mentioned for (i) and (ii), the dimension witnesses leveraging Bell and \acrshort{ks} inequalities, without the mentioned extra requirements for tests using both approaches, which are more experimentally demanding then measuring two-state overlaps. The only requirement~\footnote{That we are aware of.} for the dimension witnesses leveraging event graph inequalities are associated to assuming that the statistics is described by two-state overlaps. Because of that, we motivate that the event graph test is simpler and more broadly applicable. Moreover, it does not require an assumption on the Hilbert space dimension in order to obtain a classical-quantum divide as is required for (iii), and it is capable of distinguishing between set coherent versus set incoherent realizations, as our witnesses are actually characterized by relational coherence as discussed in Chapter~\ref{chapter: relational coherence}, which is a system agnostic concept based entirely on values of unitary-invariant correlations.

\subsection{Lower bound indicating candidates of dimension and coherence witnesses}

\myindent Consider any event graph $G$ and the set $\mathfrak{Q}(G)$ of quantum realizable edge weightings $r: E(G) \to [0,1]$. We say that a functional inequality $f(\pmb r) \leq b$ is a dimension witness if for $d' \leq d-1$, when $\pmb{r}' \in \mathfrak{Q}^{(d')}(G)$ we have $$ f(\pmb{r}') \leq b,$$
while there exists some $\pmb{r} \in \mathfrak{Q}^{(d)}(G)$ such that $f(\pmb{r})>b$. Recall Def.~\ref{def: witness} of a generic witness for comparison. According to this description, in order for $f$ to be considered a dimension witness we must have a concrete understanding of its action over the set $\mathfrak{Q}^{(d)}(G)$. More simply, it suffices to learn \emph{upper bounds} of $f$ over this set. However, as we have discussed in Chapter~\ref{chapter: relational coherence}, upper bounds are significantly harder to obtain than lower bounds, as these require the use of formal analytical tools, or the implementation of a hierarchy of semidefinite programs. 

\myindent From the criteria just described, the gaps found in Table~\ref{tab: seesaw bound}---related to the quantum realization of facet-defining inequalities with respect to different Hilbert space dimensions---\emph{cannot} be formally considered to define dimension witnesses. This happens since seesaw techniques can only find \textit{lower} bounds on optimal values of the shown functionals. Nevertheless, sufficiently many numerical experiments with seesaw semidefinite programs provide an important signal for searching dimension witnesses, and therefore we take the gaps found in Table~\ref{tab: seesaw bound} to indicate several \emph{candidates} of dimension witnesses, that can either be later shown to be true witnesses using upper bound \acrshort{sdp} relaxations or other methods. 

\myindent We also note that not every functional describing a facet-defining inequality of $\mathfrak{C}(G)$ is a dimension witness in the sense described above; for example the family of inequalities $c_n(\pmb{r}) \leq n-2$ is always maximally violated by single-qubit quantum states (as we have conjectured, and as we have shown using \acrshort{sdp} techniques up to numerical accuracy, to hold for $3 \leq n \leq 8$). Moreover, from Tab.~\ref{tab: seesaw bound} we see that $I_{(K_5,5)}(\pmb{r}) \leq 2$ and $h_{\mathrm{KCBS}}(\pmb{r}) \leq 2$ are also not a dimension witness (but only coherence witnesses). In this way, we also see that seesaw techniques are a useful and simple way to \emph{rule out} candidates of dimension witnesses based on such functionals. 

\myindent From analysing Table~\ref{tab: seesaw bound}, we have that the best candidates for dimension witnesses, identified as those having the largest gap between one dimension and another, are the facet-defining inequalities in the family denoted as $h_n(\pmb{r}) \leq 1$, defined recursively in Eq.~\eqref{eq:hn_recursively}. In what follows, we provide further conclusive evidence (but we are unable to give a formal proof for all integers $n \geq 5$) that every element of this family can indeed be viewed as a dimension witness. We proceed to show, formally, that $h_4(\pmb{r}) \leq 1$ cannot be violated by single-qubit states, and then using the \acrshort{sdp} considered in Chapter~\ref{chapter: relational coherence} we show numerically that $h_n(\pmb{r}) \leq 1$ cannot be violated by systems of dimension $d=n-2$. 

\subsection{Dimension and coherence witnessing with the \texorpdfstring{$h_n(\pmb{r}) \leq 1$}{hn<=1}  inequalities}

\myindent Using our notation $\mathfrak{Q}^{(d)}(G)$ for quantum realizations associated with $d$-dimensional Hilbert spaces (cf. Chapter~\ref{chapter: relational coherence}), we start by showing that 

\begin{equation}\label{eq: gap for the K4 graph}
    (\mathfrak{Q}^{(2)}(K_4) \setminus \mathfrak{C}(K_4)) \subsetneq (\mathfrak{Q}^{(3)}(K_4) \setminus \mathfrak{C}(K_4)).
\end{equation}

\myindent Note that the above implies that we are looking at quantum realizable edge weightings that are \emph{not} realizable by some set of  incoherent states (all with respect to the same basis). This  distinguishes our approach to  witnessing dimension to the one that considered dimension witnesses by investigating the sets $\mathfrak{C}^{(d)}(G)$, considered by~\cite{galvao2020quantum}, and~\cite{giordani2021witnesses}. 

\myindent We then show numerically that this is indeed a property that is more general, i.e., that~\cref{eq: gap for the K4 graph} holds 
\begin{equation}
    (\mathfrak{Q}^{(n-2)}(K_{n}) \setminus \mathfrak{C}(K_n)) \subsetneq (\mathfrak{Q}^{(n-1)}(K_n)\setminus \mathfrak{C}(K_n))
\end{equation}
beyond $n=4$. We conjecture it to hold for every $n$, but we only verify it numerically using semidefinite programming tools for $5 \leq n \leq 2^{12}-2$, meaning that there are relationally coherent edge weightings $r$ realizable by Hilbert spaces of dimension $n-2$ that cannot be realized by Hilbert spaces of dimension $n-1$.

\subsubsection{Relational coherence requiring qutrits}

\myindent We first show that single-qubit states cannot violate the $h_4(\pmb{r}) \leq 1$ inequality from Eq.~\eqref{eq:k4_inequalities}. 

\begin{theorem}\label{theorem: h4 inequality}
Let $\pmb{\rho} \in \mathcal{D}(\mathbb{C}^2)^4$ be any 4-tuple of quantum states. Define $r_{i,j}(\pmb{\rho}) := \mathrm{Tr}(\rho_i \rho_j)$ for $1 \leq i < j \leq 4$. Then the following inequality holds:
\[
h_4(\pmb{r}(\pmb{\rho})) := r_{1,2} + r_{1,3} + r_{1,4} - r_{2,3} - r_{2,4} - r_{3,4} \leq 1.
\]
\end{theorem}

\begin{proof}
     Note that from the results of Chapter~\ref{chapter: relational coherence}, it suffices to look at pure states. We want to maximize $h_4$ for qubit  states. The general form for this maximization procedure for any possible state is given by,
    \begin{align*}    \max_{\rho_1,\rho_2,\rho_3,\rho_4 \in \mathcal{D}(\mathcal{H})}\left(\text{Tr}(\rho_1\rho_2)+\text{Tr}(\rho_1\rho_3)+\text{Tr}(\rho_1\rho_4)-\text{Tr}(\rho_2\rho_3)-\text{Tr}(\rho_2\rho_4)-\text{Tr}(\rho_3\rho_4)\right),
    \end{align*}
where $\mathcal{D}(\mathcal{H})$ is the set of all density matrices over $\mathcal{H}\simeq \mathbb{C}^2$. In fact, if we consider only the maximization with respect to $\rho_1$ first, we see that $\rho_1$ appears only in the first 3 overlaps. This allows us to use the following relation,

\begin{equation}
    \text{Tr}(\rho_1\rho_2)+\text{Tr}(\rho_1\rho_3)+\text{Tr}(\rho_1\rho_4) = \text{Tr}\left(\rho_1\left( \sum_{i=2}^4 \rho_i \right)\right) \leq \left \Vert \sum_{i=2}^4 \rho_i \right \Vert
\end{equation}
    where in the first equality we have used linearity of the trace and for the second inequality $\Vert \cdot \Vert$ denotes the operator norm,
    \begin{equation}
        \Vert A\Vert := \sup_{v \in \mathcal{H}, v\neq 0} \frac{\Vert A v \Vert_{\mathcal{H}} }{\Vert v \Vert_{\mathcal{H}} }
    \end{equation}
    where $\Vert \cdot \Vert_{\mathcal{H}}$ is the norm that makes $\mathcal{H}$ a Hilbert space, i.e., the norm arising from the inner product.   In particular, because the sum of positive semidefinite matrices is again positive semidefinite $\sum_i \rho_i$ is positive semidefinite, and the inequality is tight meaning that there is a state $\rho_1$ for any given $\rho_2,\rho_3,\rho_4$ such that the equality holds. Therefore we end up translating our problem into  

    \begin{equation}
        \max_{\rho_2,\rho_3,\rho_4\in \mathcal{D}(\mathcal{H})}\left(\Vert \rho_2+\rho_3+\rho_4\Vert-\text{Tr}(\rho_2\rho_3)-\text{Tr}(\rho_2\rho_4)-\text{Tr}(\rho_3\rho_4))\right),
    \end{equation}
and we proceed to study the quantity $\Vert \rho_2+\rho_3+\rho_4\Vert$. Due to the invariant nature of the overlap scenarios we might choose $\rho_2,\rho_3,\rho_4$ to be the density matrices related to the pure states $\vert 0 \rangle, \vert \theta \rangle, \vert \alpha,\varphi \rangle$ defined by,
\begin{equation}
    \vert \theta \rangle = \cos\theta\vert 0\rangle + \sin\theta \vert 1 \rangle, 
\end{equation}
and
\begin{equation}
    \vert \alpha, \varphi \rangle = \cos\alpha \vert 0 \rangle + e^{i\varphi}\sin\alpha \vert 1\rangle, 
\end{equation}
with no loss of generality. This implies that we  have a relation dependent only on $3$ parameters to investigate $\theta, \alpha \in [0,\pi/2]$ and $\varphi \in [0,2\pi]$.

\myindent Using the elementary results that the operator norm $\Vert A \Vert $ is equal to the largest eigenvalue of $A$ whenever $A$ is positive semidefinite (see~\cref{theorem: spectrum radius of an operator} in the Appendix~\ref{app: basic algebra}), to calculate the operator norm we simply need to find the maximal eigenvalue of the operator sum $\vert 0 \rangle \langle 0 \vert + \vert \theta \rangle \langle \theta \vert + \vert \alpha, \varphi \rangle \langle \alpha, \varphi \vert$. This sum describes the matrix,
\begin{equation}
    \left(\begin{matrix}
    1+\cos^2\theta+\cos^2\alpha & \frac{1}{2}(\sin(2\theta)+e^{i\varphi}\sin(2\alpha))\\
    \frac{1}{2}(\sin(2\theta)+e^{-i\varphi}\sin(2\alpha)) & \sin^2\theta + \sin^2\alpha 
    \end{matrix}\right)
\end{equation}

that has the following eigenvalues,

\begin{equation}\label{eq: maximum eigenvalue}
    \lambda_{\pm} = \frac{3}{2}\pm\frac{1}{2} \sqrt{2\sin (2 \alpha ) \sin (2 \theta
   )\cos(\varphi)+4 \cos (2 \alpha )
   \cos ^2(\theta )+2  \cos (2
   \theta )+3}.
\end{equation}

The maximum eigenvalue is  given by  $\lambda_+$. After a few calculations one can find that demanding $h_4(r)>1$ corresponds to demanding that the following function must have some value larger than 0 in its image,

\begin{eqnarray*}    &g(\theta,\alpha,\varphi) =\frac{3}{2}+\frac{1}{2} \sqrt{2\sin (2 \alpha ) \sin (2 \theta
   )\cos(\varphi)+4 \cos (2 \alpha )
   \cos ^2(\theta )+2  \cos (2
   \theta )+3}\\
   &-1 - \cos^2(\theta)-\cos^2(\alpha)-\cos^2(\theta)\cos^2(\alpha) -\sin^2(\theta)\sin^2(\alpha)-\frac{1}{2}\sin(2\theta)\sin(2\alpha)\cos(\varphi).
\end{eqnarray*}
Maximizing $g(\theta,\alpha,\varphi)$, over all points $(\theta,\alpha,\varphi) \in [0,\pi/2]^2\times [0,2\pi)$ we can see that for all such points $g(\theta,\alpha,\varphi) \leq 0$ implying that $h_4(r)\leq 1$ for any pure qubit realization of $\pmb r\equiv(r_{12},r_{13},r_{14},r_{23},r_{24},r_{34})$. We conclude that pure qubit states cannot violate the $h_4(\pmb r) \leq 1$ inequality. 
\end{proof}

\myindent Theorem~\ref{theorem: h4 inequality} shows that there exists quantum coherence that is achieved by qudits that cannot be achieved with qubits. Moreover, it also says that this form of coherence is captured by relationally coherent tuples  $\pmb{r} \in \mathfrak{Q}(K_4)\setminus \mathfrak{C}(K_4)$ violating the $h_4(\pmb{r}) \leq 1$ inequality. 

\subsubsection{Set coherence requiring qudits}

\myindent The above theorem shows that the $h_4(\pmb r) \leq 1$ inequality can be used to witness both coherence and dimension. We now proceed to numerically investigate if the same property holds for the family of inequalities $h_n$ described in the main text. The results are shown in Table~\ref{tab:kn} below.

\begin{table}[ht]
    \centering
    \begin{tabular}{|*{10}{c|}}
    \cline{2-10}
    \multicolumn{1}{c|}{} & \multicolumn{9}{c|}{Dimension} \\
     \cline{2-10}
     \multicolumn{1}{c|}{} & \textbf{2} & \textbf{3} & \textbf{4} & \textbf{5} & \textbf{6} & \textbf{7} & \textbf{8} & \textbf{9} & \textbf{10} \\ 
     \hline
     $h_3$ & 1.250 & 1.250 & & & & & & &  \\ 
     $h_4$ & 1.000 & 1.333 & 1.333 & & & & & & \\ 
     $h_5$ & 0.250 & 1.000 & 1.375 & 1.375 & & & & & \\ 
    $h_6$ & -0.999 & 0.333 & 1.000 & 1.400 & 1.400 & & & & \\
    $h_7$ & -2.750 & -0.667 & 0.375 & 1.000 & 1.417 & 1.417 & & & \\
     $h_8$ & -5.000 & -2.000 & -0.500 & 0.400 & 1.000 & 1.429 & 1.417 & & \\
     $h_9$ & -7.750 & -3.667 & -1.625 & -0.400 & 0.417 & 1.000 & 1.428 & 1.429 & \\
     $h_{10}$ & -11.000 & -5.667 & -3.000 & -1.400 & -0.333 & 0.429 & 1.000 & 1.443 & 1.437\\ 
     \hline
\end{tabular}
 \caption{\textbf{Maximal values for $h_n$ inequality functionals from $K_n$ graphs.} Letting $h_n$ be the functionals, bounding incoherent models for $h_n(\pmb r)\leq 1$, we numerically investigate the maximal values of $h_n(\pmb r(\pmb{\rho}))$ for $n=3,\dots,10$ being the number of states, and $d=2,\dots,10$ being the dimension of the Hilbert space the states live in.  The violations obtained for each inequality should not decrease as we increase the dimension, some examples of this in the table result from inefficiencies of the NMaximize \textit{Mathematica} function, e.g. in violations of $h_8$ for $d=7,8$.}
    \label{tab:kn}
\end{table}

\myindent This numerical investigation shown in Table~\ref{tab:kn} was performed by maximizing the value $h_n(\pmb r(\pmb{\psi}))$ achieved for $\pmb r(\pmb{\psi}) = (\vert \langle \psi_i \vert \psi_j \rangle \vert^2)_{i,j}$ generic pure state overlaps, using maximisation functions built in the \textit{Mathematica} language, specifically, NMaximize. We also tested if $(n-2)$-dimensional states could violate the inequalities $h_n(\pmb r(\pmb \psi))\leq 1$ by randomly generating sets of quantum states from the uniform, Haar measure. For $h_6(\pmb r(\pmb{\psi}))\leq 1$ we tested $10^{10}$ sets of 6 samples of ququarts and never violated the inequality. Both sampling and numerical maximization indicate that the property holds, i.e., no set of $n$ quantum states over an $(n-2)$-dimensional Hilbert space could violate the $h_n(\pmb r)\leq 1$ inequalities.

\myindent We now recall that in Chapter~\ref{chapter: relational coherence} we have translated the problem of optimizing the values  $h_n(\pmb{r}(\pmb{\psi}))$ given some quantum realization $\pmb{r}(\pmb{\psi})$ to a semidefinite program. If one recalls Theorem~\ref{theorem: h_n_SDP}, we have shown that there is a quadratic  \acrshort{sdp} that upper bounds the optimal values of $h_n(\pmb{r}(\pmb{\psi}))$. We now proceed to numerically show the results of this optimization tool.

\begin{figure}[t]
    \centering
    \includegraphics[width=\textwidth]{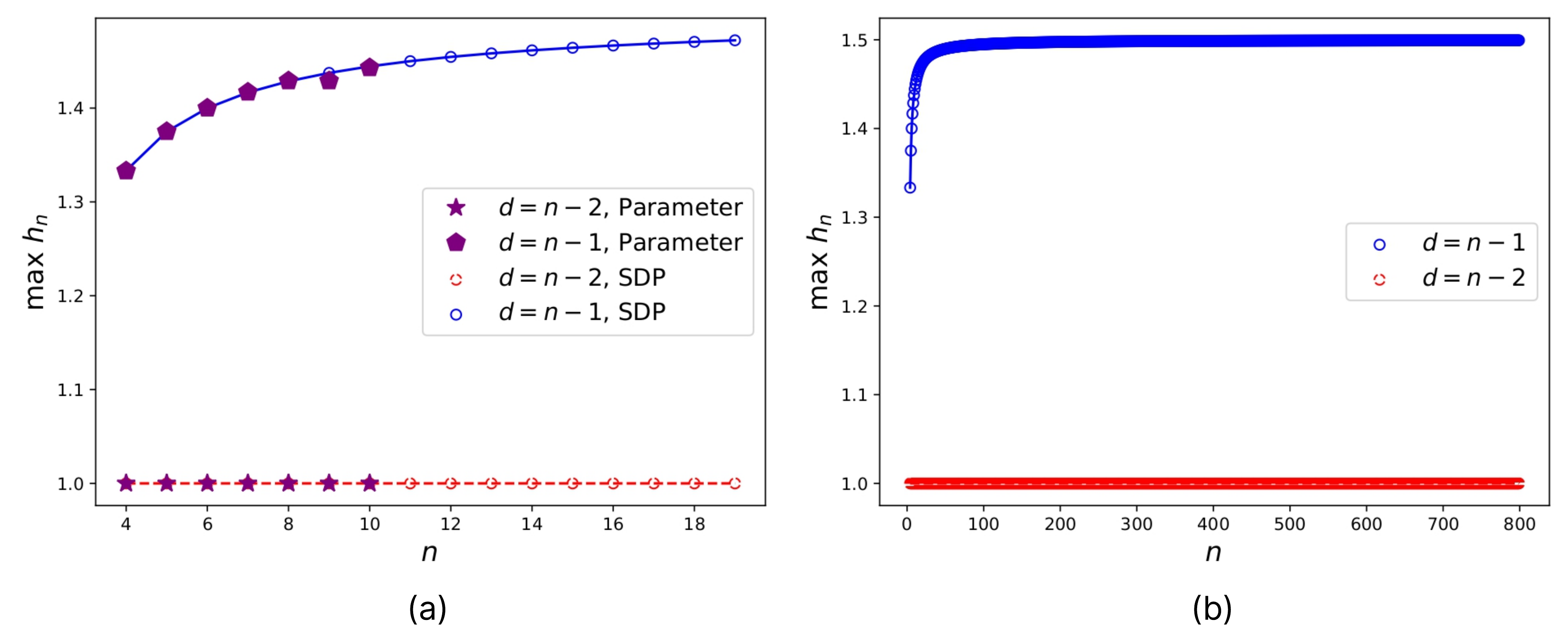}
    \caption{\textbf{Numerically testing dimension and coherence witness using semidefinite programming.} We use a quadratic \acrshort{sdp} to generate (tight) upper bounds on the maximum value of $h_n(\pmb r(\pmb{\psi}))$ for any value $n\geq 4$, for any dimension $d\leq n-1$. Plot (a) shows the result for $n$ up to 19. We compare the results obtained using \acrshort{sdp} with the results obtained by maximizing $h_n$ over all parameters of all quantum states present in a given complete graph $K_n$. Purple stars and pentagons correspond to points from Table~\ref{tab:kn}, and we use those to benchmark the \acrshort{sdp} results. Cases $n=4,5,6$ were also investigated experimentally by~\cite{giordani2023experimental}.  In Plot~(b) we perform the same numerical test for $4\leq n\leq 800$. In both plots (a) and (b) open circles correspond to maximum values of $h_n(r)$ found using the \acrshort{sdp}. Blue full lines mark violations of the $h_n(\pmb r) \leq 1$ inequality, and have $d=n-1$ while red dashed lines correspond to points that do not violate the inequality and have $\pmb r = \pmb r(\pmb{\psi})$ with $d=n-2$.}
    \label{fig: sdp}
\end{figure}

\begin{figure}[t]
    \centering
    \includegraphics[width=\textwidth]{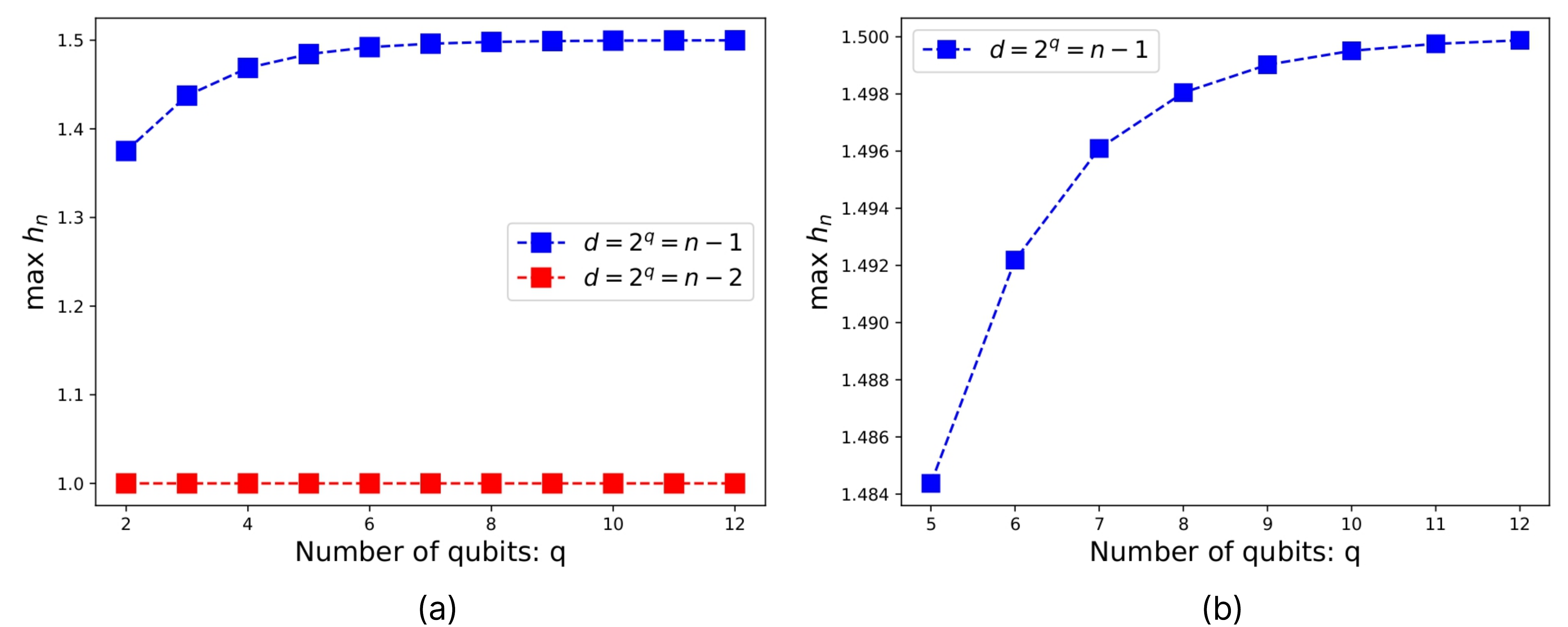}
    \caption{\textbf{Numerically testing dimension and coherence witness using semidefinite programming for multi-qubit systems}. We consider the same simulations from Fig.~\ref{fig: sdp} but using qubits, and therefore going to exponentially large Hilbert space dimensions. We see that $h_n$ remains a dimension witness for $d$ up to $2^{12}$. In Fig.~(a) we see that there is a saturation of $1.5$, and in Fig.~(b) we highlight the changes with a higher precision.}
    \label{fig: sdp qubits}
\end{figure}

\myindent The interesting aspect of transposing the problem of maximizing $h_n(\pmb r(\pmb{\psi}))$ into a quadratic \acrshort{sdp} as was done in~\cref{theorem: h_n_SDP} is that the problem becomes computationally efficient, polynomial in the memory and computational resources~\citep{tavakoli2023semidefinite}. In Figure~\ref{fig: sdp} we implement the above problem and find solutions for $n$ up to $800$. Such a gain allows us to study upper bounds of quantum violations for all integers  $4 \leq n \leq 800$. The main idea for these simulations is to provide strong numerical evidence that the family of inequalities $h_n(\pmb{r}) \leq 1$ provides a family of  dimension and coherence witnesses for all possible values of $n\geq 4$. As {this} might be interesting for when the Hilbert space dimension grows exponentially, we have also considered the case of dimension $d=2^q$ where $q$ represents number of qubits. We show numerically that $2^q-2$ dimensional states cannot violate inequalities $h_{2^q}(\pmb r)\leq 1$, up to $q=12$. The results are shown in Fig.~\ref{fig: sdp qubits}. { As our quadratic \acrshort{sdp} is well behaved, we have used interior point methods, using solvers available in $\mathtt{CVXPY}$. 

\myindent We use the Splitting Cone Solver (\acrshort{scs}) and the Python Software for Convex Optimization (\acrshort{cvxopt}), and both converged to the same values up to numerical instabilities. Results presented are those for \acrshort{scs} only. We use these solvers since they are open-source. While not devoted to quadratic \acrshort{sdp} optimization, they have converged fast for all points considered. As these are convex optimization tools applied to a quadratic optimization problem, we are not guaranteed that the results are tight; still we managed to find (local) optimal values that agreed with other methods (\textrm{NMaximize} from \textit{Mathematica}) and the experimental implementations reported by~\cite{giordani2023experimental}, providing further evidence that our dimension witnesses hold significantly beyond the regimes tested experimentally in that work. The \acrshort{sdp} code may be found in Reference~\citep{giordani2023code}.}

\section{Discussion}\label{sec: discussion of dimension and advantage}

\myindent This Chapter concretely resolves~\cref{question: applications}. We have used the formalism developed in Chapters~\ref{chapter: event_graph_approach} and~\ref{chapter: relational coherence} (that have resolved~\cref{question: generalize galvao and broads results}) to propose novel applications that had not been considered  by~\cite{galvao2020quantum}, as required by our formulation of~\cref{question: applications}. 

\myindent One resolution came from an application related to the notion of quantum information advantage, that shows relational coherence (in fact, more rigorously its relationship with preparation contextuality) to be resourceful for the task of quantum interrogation. Another resolution to~\cref{question: applications} was by considering the problem of dimension and coherence witnessing. In this case we extend the same approach considered by~\cite{galvao2020quantum}, that witnesses dimension from relationally incoherent edge-weightings $r \in \mathfrak{C}(C_3)$, to consider the more general situation for any event graph $G$ beyond $C_3$. But also, we consider the problem of witnessing \emph{both} coherence and dimension, showing that $h_4(\pmb{r}) \leq 1$ can be used to witness coherence in qutrits, while also providing significant numerical evidence that $h_n(\pmb{r}) \leq 1$ can be used to witness coherence in qudits of dimension $d=n-2$, for all integers $n \geq 3$.

\subsection{Recent advances}

\myindent To the best of our knowledge, we were the first to introduce the idea that dimension witnessing in prepare-and-measure scenarios can be used to simultaneously witness \emph{both} coherence and Hilbert space dimension. We proposed this application of the event graph in~\cite{giordani2023experimental}. Recently,~\cite{hakansson2025experimental} has further developed this idea, considering a similar scenario and to some extent their approach also has connections with contextuality. Also,~\cite{junior2025coherencedimensionalitywitnessesfractional} has implemented a test that certifies both coherence and dimensionality using the techniques we have introduced here, applied in the context of structured light.

\myindent Additionally,~\cite{hazra2024optimal} explored a similar approach where they consider the violation of generalized noncontextuality inequalities that require a certain minimal Hilbert space dimension $d_{\min}>2$. In this sense, their work also serves as a proposal for probing nonclassical features that necessitate a certain Hilbert space dimension. From our perspective, testing the contextuality of a device provides the strongest possible benchmark for a quantum system, as this phenomenon is particularly sensitive to specific types of errors (as we have seen in Sec.~\ref{appendix: robust_inter}). As such tests are also demanding, certifying coherence from the violation of event graph inequalities can be viewed as a complementary certification task.  

\myindent We believe that in the near future, stringent tests capable of benchmarking both Hilbert space dimension and quantum coherence, quantum entanglement, or quantum contextuality could become a standard benchmarking requirement for characterizing and validating certain quantum devices, such as quantum random number generators or integrated photonic interferometers.

\subsection{Future work}

\myindent A precise understanding of potential loopholes in experimental tests violating event graph inequalities remains unexplored. In particular, the impact of experimental errors on such violations, in the case of standard noise channels such as depolarization or dephasing channels, warrants further investigation. A key question is to quantify the extent to which deviations in observed statistics $p(\rho|M_\rho,P_\sigma) = \text{Tr}({\rho}\sigma)$ such as $$|p(\rho|M_\rho,P_\sigma) - p(\sigma|M_\sigma,P_\rho) |<\varepsilon_{\mathrm{cyc}},$$ depending on $\varepsilon_{\mathrm{cyc}} \neq 0$, influence the validity of our results. This source of error can be interpreted as a deviation from the assumption that the prepare and measure statistics is given by two-state overlaps. Future work should aim to determine the robustness of our findings under realistic noise models and explore refinements of our results (such as we have done in Sec.~\ref{appendix: robust_inter}) to account for such discrepancies.

\myindent Another open question is whether there exists an example of relational imaginarity (cf. Chapter~\ref{chapter: relational coherence}) that inherently requires a Hilbert space dimension $d \geq 3$. From our current results, all known instances of relational imaginarity do not \emph{necessarily} require Hilbert space dimensions greater than 2, in the sense that every relationally coherent edge weighting can be attained with sets of single-qubit quantum states. Investigating higher-dimensional cases could lead to interesting benchmarks that consider not only coherence, but other types of relational coherence as well.

\myindent It is important to note that the ability to simultaneously witness coherence and dimension from relationally coherent edge weightings is not specific to the event graphs $G=K_n$, for some $n$.  Rather, these cases represent instances where we have been able to establish rigorous statements regarding the separation between the two sets of quantum-realizable edge weightings. A natural question for future research is whether more effective dimension and coherence witnesses can be identified.  Investigating alternative graph structures may yield stronger or more experimentally robust inequalities for distinguishing these sets.

\chapter{Conclusions and perspectives}

\begin{quote}
    ``\emph{This has been an overly concise discussion of an ambitious research program (and one that may ultimately fail).}''\\(Sean~\cite{carroll2022reality})
\end{quote}

\myindent Reflecting on the introductory Chapter~\ref{chapter: introduction}, we can say that this thesis has taken us both `high into the sky as birds', exploring broad conceptual landscapes, and `back to the ground as frogs', delving into intricate technical details of formal mathematical results. We have engaged with highly technical contributions, such as the detailed characterization of facet-defining event graph inequalities and the precise description of the boundaries of novel sets of quantum correlations, defined through tuples of Bargmann invariants. At the same time, we have introduced and motivated an entirely new branch of coherence theory, driven by the investigation of such sets of correlations beyond those traditionally considered in the literature.

\myindent From a broader perspective, we now revisit the overarching goals (i), (ii), and (iii) outlined in Chapter~\ref{chapter: introduction} and discuss how they have been addressed through our contributions.

\myindent \textbf{(i) Bridging distinct fields of study:} One of our key objectives was to establish connections between coherence and contextuality (cf.~\cref{question: coherence witnesses and contextuality}). We have argued extensively, particularly in Chapter~\ref{chapter: from overlaps to noncontextuality}, that our work achieves this by applying the formalism developed in Chapter~\ref{chapter: event_graph_approach}. While our contributions in this direction may be considered modest, we believe they provide meaningful insights into the role of relational information---captured by unitary invariants and by a holistic understanding of experimental scenarios rather than isolated nonclassical features of specific processes. Furthermore, we have examined how set coherence, the operational description of noncommutativity of observables in quantum state preparation, plays a role in the violation of noncontextuality inequalities. Concretely, these contributions are our solutions to Question~\ref{question: coherence witnesses and contextuality}.

\myindent \textbf{(ii) Developing overarching tools applicable to multiple fields:} In our view, this constitutes the main contribution of this thesis, even though it was not the initial primary goal of this PhD project. The framework developed in Chapter~\ref{chapter: event_graph_approach}, the notion of relational coherence, and the study of quantum correlation sets described by tuples of Bargmann invariants in Chapter~\ref{chapter: relational coherence} exemplify the creation of tools designed for a specific application but with potential relevance far beyond their original motivation. We have provided both conceptual justification and technical development for these tools in the aforementioned chapters. These are our solutions to Question~\ref{question: generalize galvao and broads results}. 

\myindent \textbf{(iii) Exploring novel perspectives and applications:} Another central goal of this thesis was to investigate how the tools we developed could offer new perspectives and solutions to existing problems across multiple subfields. As discussed in Chapter~\ref{chapter: applications}, we demonstrated that our framework has concrete applications beyond its initial scope. Specifically, our results have implications for research in prepare-and-measure scenarios, quantum advantages over classical information processing, quantum contextuality, and dimension witnessing. We have shown that the theoretical tools introduced in this thesis can serve as competitive alternatives to existing approaches in these domains. We have therefore argued that these contributions are our solutions to Question~\ref{question: applications}.

\section{Limitations}

\myindent Our event graph approach is inherently limited by the ability to obtain and characterize novel and interesting facet-defining inequalities. Numerically, a complete characterization of event graph polytopes, as we have done in this thesis, is only feasible for event graphs that are sufficiently sparse or have a small number of vertices. For dense event graphs with a large number of vertices, our numerical tools can, at best, provide partial characterizations of facets and vertices. Therefore, while many results can still be obtained for simple event graphs with symmetry (such as cycle graphs, wheel graphs, and complete graphs), the long-term applicability of our formalism depends on advances in numerical methods for vertex and facet enumeration, convex optimization, graph theory, and invariant theory. Fortunately, these are some of the most actively researched areas in mathematics and computer science.

\myindent Conceptually, the notion of relational coherence is highly abstract and broadly applicable. While this generality is an interesting feature, it also presents a challenge: it is difficult to pinpoint a single operational task where its relevance is as self-evident as, for instance, the role of quantum entanglement in quantum teleportation. Some may view this broad applicability as a strength, while others may see it as a limitation. While we do not consider this abstract nature to be a drawback, and have provided several motivations for it in Chapter~\ref{chapter: relational coherence}, we acknowledge that our arguments may not be universally compelling.

\myindent Furthermore, one could argue that generalized contextuality offers a more well-motivated and broadly applicable notion of nonclassicality. We agree with this perspective and do not claim that our framework provides a universally valid or comprehensive characterization of nonclassicality. Instead, we view it as a more modest approach, suited to specific contexts. 

\myindent A major limitation of our framework is its focus on quantum states and quantum effects (which are dual to some quantum state). As we have already mentioned, we believe that committing to a specific ``portion'' of a quantum experiment to describe its nonclassicality or resourcefulness is inherently restrictive. A focus solely on states provides only a partial understanding of nonclassical phenomena. A more complete perspective would involve studying nonclassicality dynamically (by considering quantum channels) and structurally (by analyzing causal restrictions, such as those in Bell tests). We recognize this as a significant limitation of our approach, and we do not currently see a way to resolve it---nor are we certain that it should be resolved. %We believe, however, that this is an important consideration for future research.

\section{Future directions}

\myindent At the end of each Chapter in Part II, we have highlighted various future directions related to the specific topics discussed. Because of that, we take this opportunity to discuss more speculative directions that, in our view, could significantly advance our research agenda and increase its consideration in the literature. While we have provided several promising paths, we here outline what one might call `high-risk' directions, as they are more speculative but would lead to more impactful contributions.

\myindent We start with an application for Bell network scenarios. The sets of quantum correlations in such scenarios are famously known to be \emph{non-convex}. This makes progress difficult, since few (if any) simple theoretical criteria exist for understanding such sets of quantum correlations. It would be interesting to show that, for the specific subset of Network scenarios where sources generate states $\rho_1,\dots,\rho_n \in \mathcal{D}(\mathcal{H})$ all defined with respect to the same Hilbert space dimension, one can directly infer gaps in the sets of quantum correlations from the values of the associated Bargmann invariants $\Delta_n(\rho_1,\dots,\rho_n)$. For example, it could be that $\Delta_n(\rho_1,\dots,\rho_n) \in \mathbb{C}$ is a simple criterion for guaranteeing a gap between real quantum theory versus complex-valued quantum theory, shown to exist by~\cite{renou2021quantum}. Clearly, more broadly speaking, it could be that such values provide simple criteria for understanding other aspects of such scenarios. Given the highly non-trivial nature of such non-convex sets of correlations, and their importance for quantum communication, obtaining such simple criteria would be of great theoretical relevance.

\myindent We also comment on an application not considered here but often viewed as a main application of Bargmann invariants: the characterization of multi-photon indistinguishability. Since, as we have discussed in Chapter~\ref{chapter: Bargmann invariants}, Bargmann invariants are crucial for correctly describing the output statistics of multi-photon interferometers, and are also relevant for characterizing bosonic indistinguishability (e.g. the \acrshort{hom} effect), one main line of inquiry might be that of \emph{developing an operational resource theory} of multiphoton indistinguishability \emph{based solely on relational information} provided by two-state overlaps and higher order Bargmann invariants. We envision that such a resource theory could provide formal methods for the quantification and characterization of various interesting phenomena powered by bosonic indistinguishability. Due to its theoretical, foundational and industrial relevance, we believe that this is one of the most exciting future research directions to take, but also one of the most challenging ones. Some preliminary work in this direction was initiated by~\cite{annoni2025incoherentbehavior}. In this regard, we believe our contributions from Sec.~\ref{chapter: relational coherence}, especially those regarding geometric aspects of the sets of Bargmann invariants, shall be of great relevance.

\myindent From a more foundational perspective, we could highlight that Bargmann invariants are also connected to quasiprobability representations of quantum theory, a research direction that we have pursued in~\cite{wagner2024quantumcircuits} and~\cite{schmid2024kirkwood}. Such representations have been related to advantages in quantum metrology~\citep{arvidssonShukur2020quantum}. While generalized contextuality has been thoroughly investigated, and its relevance to quantum advantage is a vivid research topic, research connecting contextuality to advantages in quantum metrology and quantum sensing has been (somewhat surprisingly) lacking. We believe that via the connection between Bargmann invariants and contextuality that we have outlined in Chapter~\ref{chapter: from overlaps to noncontextuality}, and the relevance of Bargmann invariants to (multi)parameter estimation in some specific scenarios (such as those considered by~\cite{arvidssonShukur2020quantum} or~\cite{carollo2019quantumness}) proofs of quantum contextual advantage for metrology tasks may be one step away. A straightforward way of obtaining such a proof might be to consider the re-writing of quantum Fisher information using Bargmann invariants as was done for example by~\cite{jenne2022unbounded}, use our results from Chapters~\ref{chapter: event_graph_approach} and~\ref{chapter: relational coherence} to investigate which optimal values this quantum Fisher information can achieve when restricted to satisfy noncontextual bounds provided by, for example, the preparation $n$-cycle noncontextuality inequalities.

\myindent One of the best aspects of our formalism is that it is profoundly tailored to be experimentally appealing and simple. To this extent, it provides a formal novel toolbox for experimentally investigating quantum resources. While witnessing coherence is practically an experimentally simple task in various platforms, there are others in which witnessing coherence may be considered \emph{the} most important open question. For example, certifying the generation of quantum coherent states of matter in biological systems such as photosynthetic complexes would represent a groundbreaking achievement for the field of quantum biology. Similarly, unambiguously verifying quantum superposition of macroscopic systems could be a first step towards tests capable of verifying the quantum nature of the gravitational interaction. Because our framework has a solid foundational background, focusing on specific instances where physical systems are modeled by the relevant dynamics (describing the phenomena above) could lead to new experimental routes for certifying coherence in such extreme, non-trivial, and highly relevant scenarios.

\renewcommand{\baselinestretch}{1}
{\footnotesize
\bibliographystyle{plainnat}
\bibliography{dissertation.bib}

@misc{OEISBell,
    author = {{OEIS Foundation Inc.}},
    title = {The {B}ell or exponential numbers},
    howpublished = {Entry A000110 in The On-line Encyclopedia of Integer Sequences, \url{https://oeis.org/A000110}},
    year = {2022}
    }

@book{boole1854investigation,
title={An investigation of the laws of thought: On which are founded the mathematical theories of logic and probabilities},
  author={Boole, George},
  year={1854},
  publisher={Macmillan},
  address={London}
}

@article{hilbert1890ueber,
  title        = {{Ueber die Theorie der algebraischen Formen}},
  author       = {Hilbert, David},
  journal      = {Mathematische Annalen},
  volume       = {36},
  number       = {4},
  pages        = {473--534},
  year         = {1890},
  month        = {Dec},
  publisher    = {Springer Science and Business Media LLC},
  issn         = {1432-1807},
  doi          = {10.1007/bf01208503},
  url          = {http://dx.doi.org/10.1007/BF01208503}
}

@inbook{minkowski1989allgemeine, 
    title={Allgemeine {L}ehrsätze über die konvexen {P}olyeder}, 
    ISBN={9783709195369}, 
    ISSN={0233-0962}, 
    url={http://dx.doi.org/10.1007/978-3-7091-9536-9_5}, 
    DOI={10.1007/978-3-7091-9536-9_5}, 
    booktitle={Ausgewählte Arbeiten zur Zahlentheorie und zur Geometrie}, 
    publisher={Springer Vienna, 1989}, 
    author={Minkowski, Hermann}, 
    year={1897}, 
    pages={121–139} 
}

@book{loria1902spezielle,
  title={Spezielle algebraische und transscendente ebene Kurven: Theorie und Geschichte},
  author={Loria, Gino and Sch{\"u}tte, Fritz},
  volume={5},
  year={1902},
  publisher={BG Teubner}
}

@book{wieleitner1908spezielle,
  title={Spezielle ebene kurven},
  author={Wieleitner, Heinrich},
  volume={56},
  year={1908},
  publisher={GJ G{\"o}schen}
}

@book{minkowski1911convexhull,
  author    = {Hermann Minkowski and Hermann Weyl and Andreas Speiser and David Hilbert},
  title     = {Gesammelte {A}bhandlungen von {H}ermann {M}inkowski},
  year      = {1911},
  publisher = {B. G. Teubner},
  address   = {Leipzig},
  pages     = {157--161},
}

@article{steinitz1916bedingt,
  author    = {Ernst Steinitz},
  title     = {Bedingt konvergente {R}eihen und konvexe {S}ysteme {VI}, {VII}},
  journal   = {Journal f{\"u}r die reine und angewandte Mathematik},
  volume    = {146},
  pages     = {1--52},
  year      = {1916},
  doi       = {10.1515/crll.1916.146.1},
  s2cid     = {122897233},
}

@article{noether1918invariante,
  title={{Invariante Variationsprobleme}},
  author={Noether, Emmy},
  journal={Nachrichten von der Gesellschaft der Wissenschaften zu G{\"o}ttingen, Mathematisch-Physikalische Klasse},
  year={1918},
  pages={235--257},
  url={https://eudml.org/doc/59024}
}

@article{deBroglie1927lamecanique,
  author    = {de Broglie, Louis},
  title     = {{La mécanique ondulatoire et la structure atomique de la matière et du rayonnement}},
  journal   = {Journal de Physique et le Radium},
  volume    = {8},
  number    = {5},
  pages     = {225--241},
  year      = {1927},
  publisher = {EDP Sciences},
  issn      = {0368-3842},
  doi       = {10.1051/jphysrad:0192700805022500},
  url       = {http://dx.doi.org/10.1051/jphysrad:0192700805022500}
}

@book{wigner1931gruppen,
  title={Gruppentheorie},
  author={Wigner, E. P.},
  publisher={Frederick Vieweg und Sohn},
  address={Braunschweig, Germany},
  year={1931},
  pages={251--254}
}

@book{vonNeumann1932mathematische,
  author       = {von Neumann, John},
  title        = {{Mathematische Grundlagen der Quantenmechanik}},
  series       = {Die Grundlehren der mathematischen Wissenschaften},
  volume       = {38},
  edition      = {2},
  publisher    = {Springer-Verlag Berlin Heidelberg},
  year         = {1932},
  isbn         = {978-3-642-61409-5},
  doi          = {10.1007/978-3-642-61409-5},
  url          = {https://doi.org/10.1007/978-3-642-61409-5},
  pages        = {IX, 262},
  note         = {Originally published as Volume 38 of Die Grundlehren der Mathematischen Wissenschaften, in 1932},
  address      = {Berlin, Heidelberg}
}

@book{kolmogorov1933grundbegriffe,
  author       = {Kolmogoroff, A.},
  title        = {{Grundbegriffe der Wahrscheinlichkeitsrechnung}},
  series       = {Ergebnisse der Mathematik und ihrer Grenzgebiete. 1. Folge},
  publisher    = {Springer Berlin, Heidelberg},
  year         = {1933},
  address      = {Berlin},
  isbn         = {978-3-642-49596-0},
  doi          = {10.1007/978-3-642-49888-6},
  url          = {https://doi.org/10.1007/978-3-642-49888-6},
  edition      = {1},
  pages        = {62},
  note         = {Springer-Verlag Berlin Heidelberg 1933. eBook published: 02 July 2013}
}

@article{kirkwood1933quantum,
  title = {{Quantum Statistics of Almost Classical Assemblies}},
  author = {Kirkwood, John G.},
  journal = {Phys. Rev.},
  volume = {44},
  issue = {1},
  pages = {31--37},
  numpages = {0},
  year = {1933},
  month = {Jul},
  publisher = {American Physical Society},
  doi = {10.1103/PhysRev.44.31},
  url = {https://link.aps.org/doi/10.1103/PhysRev.44.31}
}

@article{weyl1934elementare, 
    title={Elementare {T}heorie der konvexen {P}olyeder}, 
    volume={7}, 
    ISSN={1420-8946}, 
    url={http://dx.doi.org/10.1007/BF01292722}, 
    DOI={10.1007/bf01292722}, 
    number={1}, 
    journal={Commentarii Mathematici Helvetici}, 
    publisher={European Mathematical Society - EMS - Publishing House GmbH}, 
    author={Weyl, H.}, 
    year={1934}, 
    month=dec, 
    pages={290–306} 
}

@article{bell1934exponential,
  title={Exponential numbers},
  author={Bell, Eric T.},
  journal={The American Mathematical Monthly},
  volume={41},
  number={7},
  pages={411--419},
  year={1934},
  publisher={Taylor {\&} Francis},
  doi = {10.1080/00029890.1934.11987615},
  url = {https://www.tandfonline.com/doi/abs/10.1080/00029890.1934.11987615}
}

@article{hermann1935dienaturphilosophischen,
  author    = {Hermann, Grete},
  title     = {{Die naturphilosophischen Grundlagen der Quantenmechanik}},
  journal   = {Die Naturwissenschaften},
  volume    = {23},
  number    = {42},
  pages     = {718--721},
  year      = {1935},
  month     = {Oct},
  publisher = {Springer Science and Business Media LLC},
  issn      = {1432-1904},
  doi       = {10.1007/bf01491142},
  url       = {http://dx.doi.org/10.1007/bf01491142}
}

@article{schrodinger1935gegenwartige,
  title        = {{Die gegenwärtige Situation in der Quantenmechanik}},
  author       = {Schrödinger, E.},
  journal      = {Die Naturwissenschaften},
  volume       = {23},
  number       = {48},
  year         = {1935},
  month        = {Nov},
  pages        = {807–812},
  publisher    = {Springer Science and Business Media LLC},
  doi          = {10.1007/bf01491891},
  url          = {http://dx.doi.org/10.1007/BF01491891},
  issn         = {1432-1904}
}

@article{einstein1935can,
  title = {{Can Quantum-Mechanical Description of Physical Reality Be Considered Complete?}},
  author = {Einstein, A. and Podolsky, B. and Rosen, N.},
  journal = {Phys. Rev.},
  volume = {47},
  issue = {10},
  pages = {777--780},
  numpages = {0},
  year = {1935},
  month = {May},
  publisher = {American Physical Society},
  doi = {10.1103/PhysRev.47.777},
  url = {https://link.aps.org/doi/10.1103/PhysRev.47.777}
}

@article{bohr1935can,
  title = {{Can Quantum-Mechanical Description of Physical Reality be Considered Complete?}},
  author = {Bohr, Niels},
  journal = {Phys. Rev.},
  volume = {48},
  issue = {8},
  pages = {696--702},
  numpages = {0},
  year = {1935},
  month = {Oct},
  publisher = {American Physical Society},
  doi = {10.1103/PhysRev.48.696},
  url = {https://link.aps.org/doi/10.1103/PhysRev.48.696}
}

@article{weaver1942Tschirnhausen,
  title        = {{On the Cubic of Tschirnhausen}},
  author       = {Weaver, J. H.},
  journal      = {National Mathematics Magazine},
  volume       = {16},
  number       = {8},
  pages        = {371},
  year         = {1942},
  month        = {May},
  publisher    = {JSTOR},
  issn         = {1539-5588},
  doi          = {10.2307/3028893},
  url          = {http://dx.doi.org/10.2307/3028893}
}

@article{dirac1945analogy,
  title = {On the Analogy Between Classical and Quantum Mechanics},
  author = {Dirac, Paul A. M.},
  journal = {Rev. Mod. Phys.},
  volume = {17},
  issue = {2-3},
  pages = {195--199},
  numpages = {0},
  year = {1945},
  month = {Apr},
  publisher = {American Physical Society},
  doi = {10.1103/RevModPhys.17.195},
  url = {https://link.aps.org/doi/10.1103/RevModPhys.17.195}
}

@article{bohm1952suggested_one,
  title = {{A Suggested Interpretation of the Quantum Theory in Terms of "Hidden" Variables. I}},
  author = {Bohm, David},
  journal = {Phys. Rev.},
  volume = {85},
  issue = {2},
  pages = {166--179},
  numpages = {0},
  year = {1952},
  month = {Jan},
  publisher = {American Physical Society},
  doi = {10.1103/PhysRev.85.166},
  url = {https://link.aps.org/doi/10.1103/PhysRev.85.166}
}

@article{bohm1952suggested_two,
  title = {{A Suggested Interpretation of the Quantum Theory in Terms of "Hidden" Variables. II}},
  author = {Bohm, David},
  journal = {Phys. Rev.},
  volume = {85},
  issue = {2},
  pages = {180--193},
  numpages = {0},
  year = {1952},
  month = {Jan},
  publisher = {American Physical Society},
  doi = {10.1103/PhysRev.85.180},
  url = {https://link.aps.org/doi/10.1103/PhysRev.85.180}
}

@book{kolmogorov1956foundations,
  author       = {Kolmogorov, A. N.},
  title        = {{Foundations of the Theory of Probability}},
  publisher    = {Chelsea Publishing Company},
  year         = {1956},
  address      = {New York},
  note         = {Translated from the German by N. Morrison. Original title: \textit{Grundbegriffe der Wahrscheinlichkeitrechnung}, Berlin 1933}
}

@article{fano1957description,
  title = {{Description of States in Quantum Mechanics by Density Matrix and Operator Techniques}},
  author = {Fano, U.},
  journal = {Rev. Mod. Phys.},
  volume = {29},
  issue = {1},
  pages = {74--93},
  numpages = {0},
  year = {1957},
  month = {Jan},
  publisher = {American Physical Society},
  doi = {10.1103/RevModPhys.29.74},
  url = {https://link.aps.org/doi/10.1103/RevModPhys.29.74}
}

@article{stueckelberg1960quantum,
  title={Quantum theory in real {H}ilbert space},
  author={Stueckelberg, Ernst CG},
  journal={Helv. Phys. Acta},
  volume={33},
  number={727},
  pages={458},
  year={1960},
  doi={https://doi.org/10.5169/seals-113093}
}

@article{specker1960dielogik,
  author    = {Specker, Ernst},
  title     = {{Die Logik nicht gleichzeitig entscheidbarer Aussagen}},
  journal   = {Dialectica},
  volume    = {14},
  number    = {2--3},
  pages     = {239--246},
  year      = {1960},
  month     = {Sep},
  publisher = {Verein philosophie.ch},
  issn      = {1746-8361},
  doi       = {10.1111/j.1746-8361.1960.tb00422.x},
  url       = {http://dx.doi.org/10.1111/j.1746-8361.1960.tb00422.x}
}

@article{stueckelberg1961quantum,
  title={Quantum theory in real {H}ilbert space {II} ({A}ddenda and {E}rrats)},
  author={Stueckelberg von Breidenbach, Ernst Carl Gerlach and Guenin, Marcel},
  journal={Helv. Phys. Acta},
  volume={34},
  number={6/7},
  pages={621--628},
  year={1961}
}

@article{halperin1962onthegrammatrix,
    title={On the {G}ram {M}atrix},
    volume={5},
    ISSN={1496-4287},
    url={http://dx.doi.org/10.4153/CMB-1962-027-1},
    DOI={10.4153/cmb-1962-027-1},
    number={3},
    journal={Canadian Mathematical Bulletin},
    publisher={Canadian Mathematical Society},
    author={Halperin, Israel},
    year={1962},
    month=sep,
    pages={265–280}
}

@article{vorobyev1962consistent,
  author    = {Vorob’ev, N. N.},
  title     = {{Consistent Families of Measures and Their Extensions}},
  journal   = {Theory of Probability \& Its Applications},
  volume    = {7},
  number    = {2},
  pages     = {147--163},
  year      = {1962},
  month     = {Jan},
  publisher = {Society for Industrial \& Applied Mathematics (SIAM)},
  doi       = {10.1137/1107014},
  url       = {http://dx.doi.org/10.1137/1107014},
  issn      = {1095-7219}
}

@article{pearcy1962complete,
  title        = {{A Complete Set of Unitary Invariants for \(3 \times 3\) Complex Matrices}},
  author       = {Pearcy, Carl},
  journal      = {Transactions of the American Mathematical Society},
  volume       = {104},
  number       = {3},
  pages        = {425},
  year         = {1962},
  month        = {Sep},
  publisher    = {JSTOR},
  issn         = {0002-9947},
  doi          = {10.2307/1993786},
  url          = {http://dx.doi.org/10.2307/1993786}
}

@article{uhlhorn1963representation,
title = {Representation of symmetry transformations in quantum mechanics},
author = {Uhlhorn, U},
doi = {},
url = {https://www.osti.gov/biblio/4695249}, journal = {Arkiv Fysik},
number = {},
volume = {23},
year = {1963},
month = {1},
pages={307–340}
}

@article{sudarshan1963equivalence,
  title = {{Equivalence of Semiclassical and Quantum Mechanical Descriptions of Statistical Light Beams}},
  author = {Sudarshan, E. C. G.},
  journal = {Phys. Rev. Lett.},
  volume = {10},
  issue = {7},
  pages = {277--279},
  numpages = {0},
  year = {1963},
  month = {Apr},
  publisher = {American Physical Society},
  doi = {10.1103/PhysRevLett.10.277},
  url = {https://link.aps.org/doi/10.1103/PhysRevLett.10.277}
}

@article{glauber1963coherent,
  title = {Coherent and Incoherent States of the Radiation Field},
  author = {Glauber, Roy J.},
  journal = {Phys. Rev.},
  volume = {131},
  issue = {6},
  pages = {2766--2788},
  numpages = {0},
  year = {1963},
  month = {Sep},
  publisher = {American Physical Society},
  doi = {10.1103/PhysRev.131.2766},
  url = {https://link.aps.org/doi/10.1103/PhysRev.131.2766}
}

@article{bell1964ontheEPRparadox,
  author = {Bell, John S.},
  title = {On the {E}instein {P}odolsky {R}osen paradox},
  journal = {Physics Physique Fizika},
  volume = {1},
  number = {3},
  pages = {195--200},
  month = {Nov},
  year = {1964},
  publisher = {American Physical Society},
  doi = {10.1103/PhysicsPhysiqueFizika.1.195},
  URL = {https://journals.aps.org/ppf/abstract/10.1103/PhysicsPhysiqueFizika.1.195}
}

@article{bargmann1964note,
  title={{Note on Wigner’s Theorem on Symmetry Operations}},
  author={Bargmann, Valentine},
  journal={Journal of Mathematical Physics},
  volume={5},
  number={7},
  year={1964},
  month= {Jul},
  pages={862--868},
  publisher={AIP Publishing},
  doi={10.1063/1.1704188},
  url={http://dx.doi.org/10.1063/1.1704188},
  issn={1089-7658}
}

@article{bell1966ontheproblem,
  title = {{On the Problem of Hidden Variables in Quantum Mechanics}},
  author = {Bell, John S.},
  journal = {Rev. Mod. Phys.},
  volume = {38},
  issue = {3},
  pages = {447--452},
  numpages = {0},
  year = {1966},
  month = {Jul},
  publisher = {American Physical Society},
  doi = {10.1103/RevModPhys.38.447},
  url = {https://link.aps.org/doi/10.1103/RevModPhys.38.447}
}

@article{grunbaum1969convex,
  title={Convex polytopes},
  author={Gr{\"u}nbaum, Branko and Shephard, Geoffrey C},
  journal={Bulletin of the London Mathematical Society},
  volume={1},
  number={3},
  pages={257--300},
  year={1969},
  publisher={Citeseer}
}

@article{clauser1969proposed,
  title = {{Proposed Experiment to Test Local Hidden-Variable Theories}},
  author = {Clauser, John F. and Horne, Michael A. and Shimony, Abner and Holt, Richard A.},
  journal = {Phys. Rev. Lett.},
  volume = {23},
  issue = {15},
  pages = {880--884},
  numpages = {0},
  year = {1969},
  month = {Oct},
  publisher = {American Physical Society},
  doi = {10.1103/PhysRevLett.23.880},
  url = {https://link.aps.org/doi/10.1103/PhysRevLett.23.880}
}

@article{dieudonne1970invariant,
  title={Invariant theory, old and new},
  author={Dieudonné, Jean A. and Carrell, James B.},
  journal={Advances in Mathematics},
  volume={4},
  number={1},
  year={1970},
  month={feb},
  pages={1--80},
  publisher={Elsevier BV},
  doi={10.1016/0001-8708(70)90015-0},
  url={http://dx.doi.org/10.1016/0001-8708(70)90015-0},
  issn={0001-8708}
}

@book{lawrence1972catalog,
  title={A catalog of special plane curves},
  author={Lawrence, J Dennis},
  year={1972},
  publisher={Courier Corporation}
}

@article{kochen1975problem,
	author = {Kochen, Simon and Specker, Ernst},
	title = {The problem of hidden variables in quantum mechanics},
	journal = {Journal of Mathematics and Mechanics},
    volume = {17},
    number = {1},
    pages = {59--87},
    month = {July},
    year = {1967},
    publisher = {Indiana University Mathematics Department},
    doi={https://doi.org/10.1007/978-3-0348-9259-9_21},
    url = {https://link.springer.com/chapter/10.1007/978-3-0348-9259-9_21}
}

@article{lewis1979counterfactuals,
  title={David K. Lewis. \textit{Counterfactuals}. Harvard University Press, Cambridge, Mass., 1973, x + 150 pp.},
  volume={44},
  number={2},
  journal={Journal of Symbolic Logic},
  pages={278--281},
  publisher={Cambridge University Press (CUP)},
  author={Parry, William T.},
  year={1979},
  month={jun},
  DOI={10.2307/2273738},
  url={http://dx.doi.org/10.2307/2273738},
  ISSN={1943-5886}
}

@book{takesaki1979theory,
  title        = {{Theory of Operator Algebras I}},
  author       = {Takesaki, Masamichi},
  edition      = {1},
  publisher    = {Springer-Verlag New York Inc.},
  address      = {New York, NY},
  year         = {1979},
  doi          = {10.1007/978-1-4612-6188-9},
  isbn         = {978-1-4612-6190-2},
  eisbn        = {978-1-4612-6188-9},
  pages        = {VIII, 418}
}

@article{rehder1980projections,
  title={When do projections commute?},
  author={Rehder, W},
  journal={Zeitschrift f{\"u}r Naturforschung A},
  volume={35},
  number={4},
  pages={437--441},
  year={1980},
  publisher={Verlag der Zeitschrift f{\"u}r Naturforschung}
}

@article{suppes1981when,
  title        = {{When are probabilistic explanations possible?}},
  author       = {Suppes, Patrick and Zanotti, Marino},
  journal      = {Synthese},
  volume       = {48},
  pages        = {191--199},
  year         = {1981},
  month        = {Aug},
  doi          = {10.1007/BF01063886},
  url          = {https://doi.org/10.1007/BF01063886},
  issn         = {0039-7857}
}

@article{fine1982hiddenvariablesPRL,
  title = {{Hidden Variables, Joint Probability, and the Bell Inequalities}},
  author = {Fine, Arthur},
  journal = {Phys. Rev. Lett.},
  volume = {48},
  issue = {5},
  pages = {291--295},
  numpages = {0},
  year = {1982},
  month = {Feb},
  publisher = {American Physical Society},
  doi = {10.1103/PhysRevLett.48.291},
  url = {https://link.aps.org/doi/10.1103/PhysRevLett.48.291}
}

@article{fine1982jointJMP,
  title={Joint distributions, quantum correlations, and commuting observables},
  author={Fine, Arthur},
  journal={Journal of Mathematical Physics},
  volume={23},
  number={7},
  pages={1306--1310},
  month = {Jun},
  year={1982},
  publisher={American Institute of Physics},
  doi = {10.1063/1.525514},
  url={https://doi.org/10.1063/1.525514}
}

@article{bell1982impossible,
  title = {On the impossible pilot wave},
  author = {Bell, John S.},
  journal = {Foundations of Physics},
  volume = {12},
  number = {10},
  pages = {989–999},
  year = {1982},
  month = {Oct},
  publisher = {Springer Science and Business Media LLC},
  doi = {10.1007/bf01889272},
  url = {http://dx.doi.org/10.1007/BF01889272},
  issn = {1572-9516}
}

@article{page1983evolution,
  title = {Evolution without evolution: Dynamics described by stationary observables},
  author = {Page, Don N. and Wootters, William K.},
  journal = {Phys. Rev. D},
  volume = {27},
  issue = {12},
  pages = {2885--2892},
  numpages = {0},
  year = {1983},
  month = {Jun},
  publisher = {American Physical Society},
  doi = {10.1103/PhysRevD.27.2885},
  url = {https://link.aps.org/doi/10.1103/PhysRevD.27.2885}
}

@article{berry1984quantal,
  title={Quantal phase factors accompanying adiabatic changes},
  author={Berry, Michael Victor},
  journal={Proceedings of the Royal Society of London. A. Mathematical and Physical Sciences},
  volume={392},
  number={1802},
  pages={45--57},
  year={1984},
  publisher={The Royal Society London},
  doi = {https://doi.org/10.1098/rspa.1984.0023},
  url={https://royalsocietypublishing.org/doi/10.1098/rspa.1984.0023}
}

@article{deutsch1985universal,
  title     = {{Quantum theory, the Church–Turing principle and the universal quantum computer}},
  author    = {Deutsch, David},
  journal   = {Proceedings of the Royal Society of London. A. Mathematical and Physical Sciences},
  volume    = {400},
  number    = {1818},
  pages     = {97--117},
  year      = {1985},
  month     = {Jul},
  issn      = {0080-4630},
  publisher = {The Royal Society},
  url       = {http://dx.doi.org/10.1098/rspa.1985.0070},
  doi       = {10.1098/rspa.1985.0070}
}

@book{biggs1986graph,
  title={Graph Theory, 1736-1936},
  author={Biggs, Norman and Lloyd, E Keith and Wilson, Robin J},
  year={1986},
  publisher={Oxford University Press}
}

@article{tsirelson1987quantumanalogues, title={Quantum analogues of the Bell inequalities. The case of two spatially separated domains}, volume={36}, ISSN={1573-8795}, url={http://dx.doi.org/10.1007/BF01663472}, DOI={10.1007/bf01663472}, number={4}, journal={Journal of Soviet Mathematics}, publisher={Springer Science and Business Media LLC}, author={Tsirel’son, B. S.}, year={1987}, month=feb, pages={557–570} }

@article{hong1987measurement,
  title = {Measurement of subpicosecond time intervals between two photons by interference},
  author = {Hong, Chung Ki and Ou, Zheyu Jeff and Mandel, Leonard},
  journal = {Phys. Rev. Lett.},
  volume = {59},
  issue = {18},
  pages = {2044--2046},
  numpages = {0},
  year = {1987},
  month = {Nov},
  publisher = {American Physical Society},
  doi = {10.1103/PhysRevLett.59.2044},
  url = {https://link.aps.org/doi/10.1103/PhysRevLett.59.2044}
}

@article{landau1988empirical, title={Empirical two-point correlation functions}, volume={18}, ISSN={1572-9516}, url={http://dx.doi.org/10.1007/BF00732549}, DOI={10.1007/bf00732549}, number={4}, journal={Foundations of Physics}, publisher={Springer Science and Business Media LLC}, author={Landau, Lawrence J.}, year={1988}, month=apr, pages={449–460} }

@article{werner1989quantum,
  title = {{Quantum states with Einstein-Podolsky-Rosen correlations admitting a hidden-variable model}},
  author = {Werner, Reinhard F.},
  journal = {Phys. Rev. A},
  volume = {40},
  issue = {8},
  pages = {4277--4281},
  numpages = {0},
  year = {1989},
  month = {Oct},
  publisher = {American Physical Society},
  doi = {10.1103/PhysRevA.40.4277},
  url = {https://link.aps.org/doi/10.1103/PhysRevA.40.4277}
}

@incollection{jaynes1990complexity,
  title = {Complexity, Entropy, and the Physics of Information},
  booktitle = {Complexity, Entropy And The Physics Of Information},
  editor = {W. H. Zurek},
  author = {Jaynes, E. T.},
  publisher = {Addison / CRC Press},
  year = {1990},
  pages = {381},
  note = {Reprinted in 2018 by CRC Press, ISBN={9780429502880}},
  url = {http://dx.doi.org/10.1201/9780429502880},
  doi = {10.1201/9780429502880}
}

@article{deutsch1992quantumcomputation,
  title     = {{Quantum computation}},
  author    = {Deutsch, David},
  journal   = {Physics World},
  volume    = {5},
  number    = {6},
  pages     = {57--61},
  year      = {1992},
  month     = {Jun},
  issn      = {2058-7058},
  publisher = {IOP Publishing},
  url       = {http://dx.doi.org/10.1088/2058-7058/5/6/38},
  doi       = {10.1088/2058-7058/5/6/38}
}

@article{donkin1992invariants,
  title={Invariants of several matrices},
  author={Donkin, Stephen},
  journal={Inventiones mathematicae},
  volume={110},
  pages={389--401},
  year={1992},
  publisher={Springer-Verlag}
}

@article{simon1993Bargmann,
  title = {{Bargmann invariant and the geometry of the G\"uoy effect}},
  author = {Simon, R. and Mukunda, N.},
  journal = {Phys. Rev. Lett.},
  volume = {70},
  issue = {7},
  pages = {880--883},
  numpages = {0},
  year = {1993},
  month = {Feb},
  publisher = {American Physical Society},
  doi = {10.1103/PhysRevLett.70.880},
  url = {https://link.aps.org/doi/10.1103/PhysRevLett.70.880}
}

@article{mermin1993hidden,
      title = {{Hidden variables and the two theorems of John Bell}},
      author = {Mermin, N. David},
      journal = {Rev. Mod. Phys.},
      volume = {65},
      issue = {3},
      pages = {803--815},
      numpages = {0},
      year = {1993},
      month = {Jul},
      publisher = {American Physical Society},
      doi = {10.1103/RevModPhys.65.803},
      url = {https://link.aps.org/doi/10.1103/RevModPhys.65.803}
}

@article{elitzur1993quantum, 
    title={Quantum mechanical interaction-free measurements}, 
    volume={23}, 
    ISSN={1572-9516}, 
    url={http://dx.doi.org/10.1007/BF00736012}, 
    DOI={10.1007/bf00736012}, 
    number={7}, 
    journal={Foundations of Physics}, 
    publisher={Springer Science and Business Media LLC}, 
    author={Elitzur, Avshalom C. and Vaidman, Lev}, 
    year={1993}, 
    month= {Jul}, 
    pages={987–997} 
}

@article{popescu1994nonlocality,
  author    = {Popescu, Sandu and Rohrlich, Daniel},
  title     = {{Quantum Nonlocality as an Axiom}},
  journal   = {Foundations of Physics},
  volume    = {24},
  number    = {3},
  pages     = {379–385},
  year      = {1994},
  month     = {Mar},
  ISSN      = {1572-9516},
  publisher = {Springer Science and Business Media LLC},
  DOI       = {10.1007/bf02058098},
  url       = {http://dx.doi.org/10.1007/BF02058098}
}

@book{palmer1994banach,
  title        = {{Banach Algebras and the General Theory of *-Algebras, Vol. I}},
  author       = {Palmer, Theodore W.},
  series       = {Encyclopedia of Mathematics and its Applications},
  volume       = {49},
  publisher    = {Cambridge University Press},
  address      = {Cambridge},
  year         = {1994},
  note         = {MR 95c:46002}
}

@article{reck1994experimental,
  title = {Experimental realization of any discrete unitary operator},
  author = {Reck, Michael and Zeilinger, Anton and Bernstein, Herbert J. and Bertani, Philip},
  journal = {Phys. Rev. Lett.},
  volume = {73},
  issue = {1},
  pages = {58--61},
  numpages = {0},
  year = {1994},
  month = {Jul},
  publisher = {American Physical Society},
  doi = {10.1103/PhysRevLett.73.58},
  url = {https://link.aps.org/doi/10.1103/PhysRevLett.73.58}
}

@article{pittman1995optical,
  title = {Optical imaging by means of two-photon quantum entanglement},
  author = {Pittman, T. B. and Shih, Y. H. and Strekalov, D. V. and Sergienko, A. V.},
  journal = {Phys. Rev. A},
  volume = {52},
  issue = {5},
  pages = {R3429--R3432},
  numpages = {0},
  year = {1995},
  month = {Nov},
  publisher = {American Physical Society},
  doi = {10.1103/PhysRevA.52.R3429},
  url = {https://link.aps.org/doi/10.1103/PhysRevA.52.R3429}
}

@article{kwiat1995interactionfree,
  title = {{Interaction-Free Measurement}},
  author = {Kwiat, Paul and Weinfurter, Harald and Herzog, Thomas and Zeilinger, Anton and Kasevich, Mark A.},
  journal = {Phys. Rev. Lett.},
  volume = {74},
  issue = {24},
  pages = {4763--4766},
  numpages = {0},
  year = {1995},
  month = {Jun},
  publisher = {American Physical Society},
  doi = {10.1103/PhysRevLett.74.4763},
  url = {https://link.aps.org/doi/10.1103/PhysRevLett.74.4763}
}

@article{kwiat1996quantumseeing,  
  title={{Quantum Seeing in the Dark}},  
  author={Kwiat, Paul and Weinfurter, Harald and Zeilinger, Anton},  
  journal={{Scientific American}},  
  volume={275},  
  number={5},  
  pages={72--78},  
  year={1996},  
  url={http://www.jstor.org/stable/24993449}  
}

@article{rovelli1996relational,
  title        = {{Relational Quantum Mechanics}},
  author       = {Rovelli, Carlo},
  journal      = {International Journal of Theoretical Physics},
  volume       = {35},
  number       = {8},
  pages        = {1637--1678},
  year         = {1996},
  month        = {Aug},
  doi          = {10.1007/BF02302261},
  url          = {https://doi.org/10.1007/BF02302261}
}

@article{horodecki1996necessary,
  title        = {{On the Necessary and Sufficient Conditions for Separability of Mixed Quantum States}},
  author       = {Horodecki, Michał and Horodecki, Paweł and Horodecki, Ryszard},
  journal      = {Physics Letters A},
  volume       = {223},
  number       = {1},
  pages        = {1--8},
  year         = {1996},
  publisher    = {Elsevier BV}
}

@misc{vaidman1996interaction,
      title={{Interaction-Free Measurements}}, 
      author={Lev Vaidman},
      year={1996},
      eprint={quant-ph/9610033},
      archivePrefix={arXiv},
      howpublished={arXiv:quant-ph/9610033 [quant-ph]},
      primaryClass={quant-ph},
      url={https://arxiv.org/abs/quant-ph/9610033}, 
}

@book{rockafellar1997convex,
  author    = {R. Tyrrell Rockafellar},
  title     = {Convex Analysis},
  edition   = {Reprint, Revised},
  series    = {Princeton Mathematical Series},
  volume    = {57},
  year      = {1997},
  publisher = {Princeton University Press},
  address   = {Princeton, NJ},
  isbn      = {0691015864},
  pages     = {451},
  note      = {Princeton Landmarks in Mathematics and Physics},
  doi       = {https://doi.org/10.1515/9781400873173}
}

@article{mcMullen1970upper_bound_theorem,
  author       = {McMullen, P.},
  title        = {{The Maximum Numbers of Faces of a Convex Polytope}},
  journal      = {Mathematika},
  volume       = {17},
  number       = {2},
  pages        = {179--184},
  year         = {1970},
  month        = dec,
  doi          = {10.1112/S0025579300002850},
  url          = {http://dx.doi.org/10.1112/S0025579300002850},
  issn         = {2041-7942},
  publisher    = {Wiley}
}

@article{molmer1997optical,
  title = {Optical coherence: A convenient fiction},
  author = {M\o{}lmer, Klaus},
  journal = {Phys. Rev. A},
  volume = {55},
  issue = {4},
  pages = {3195--3203},
  numpages = {0},
  year = {1997},
  month = {Apr},
  publisher = {American Physical Society},
  doi = {10.1103/PhysRevA.55.3195},
  url = {https://link.aps.org/doi/10.1103/PhysRevA.55.3195}
}

@article{white1998interactionfree,
  title = {{``Interaction-free'' imaging}},
  author = {White, Andrew G. and Mitchell, Jay R. and Nairz, Olaf and Kwiat, Paul G.},
  journal = {Phys. Rev. A},
  volume = {58},
  issue = {1},
  pages = {605--613},
  numpages = {0},
  year = {1998},
  month = {Jul},
  publisher = {American Physical Society},
  doi = {10.1103/PhysRevA.58.605},
  url = {https://link.aps.org/doi/10.1103/PhysRevA.58.605}
}

@article{barnes1998common,
 ISSN = {00029939, 10886826},
 URL = {http://www.jstor.org/stable/118614},
 author = {Bruce A. Barnes},
 journal = {Proceedings of the American Mathematical Society},
 number = {4},
 pages = {1055--1061},
 publisher = {American Mathematical Society},
 title = {{Common Operator Properties of the Linear Operators RS and SR}},
 urldate = {2025-01-08},
 volume = {126},
 year = {1998}
}

@misc{ambainis1998densequantumcodinglower,
      title={Dense Quantum Coding and a Lower Bound for 1-way Quantum Automata}, 
      author={Andris Ambainis and Ashwin Nayak and Amnon Ta-Shma and Umesh Vazirani},
      year={1998},
      eprint={quant-ph/9804043},
      archivePrefix={arXiv},
      primaryClass={quant-ph},
      url={https://arxiv.org/abs/quant-ph/9804043}, 
      doi={https://doi.org/10.48550/arXiv.quant-ph/9804043}
}

@article{kwiat1999high,
  title = {{High-Efficiency Quantum Interrogation Measurements via the Quantum Zeno Effect}},
  author = {Kwiat, P. G. and White, A. G. and Mitchell, J. R. and Nairz, O. and Weihs, G. and Weinfurter, H. and Zeilinger, A.},
  journal = {Phys. Rev. Lett.},
  volume = {83},
  issue = {23},
  pages = {4725--4728},
  numpages = {0},
  year = {1999},
  month = {Dec},
  publisher = {American Physical Society},
  doi = {10.1103/PhysRevLett.83.4725},
  url = {https://link.aps.org/doi/10.1103/PhysRevLett.83.4725}
}

@article{rudolph2000better,
  title = {{Better Schemes for Quantum Interrogation in Lossy Experiments}},
  author = {Rudolph, T.},
  journal = {Phys. Rev. Lett.},
  volume = {85},
  issue = {14},
  pages = {2925--2928},
  numpages = {0},
  year = {2000},
  month = {Oct},
  publisher = {American Physical Society},
  doi = {10.1103/PhysRevLett.85.2925},
  url = {https://link.aps.org/doi/10.1103/PhysRevLett.85.2925}
}

@article{jozsa2000distinguishability,
  title = {Distinguishability of states and von {N}eumann entropy},
  author = {Jozsa, Richard and Schlienz, J\"urgen},
  journal = {Phys. Rev. A},
  volume = {62},
  issue = {1},
  pages = {012301},
  numpages = {11},
  year = {2000},
  month = {Jun},
  publisher = {American Physical Society},
  doi = {10.1103/PhysRevA.62.012301},
  url = {https://link.aps.org/doi/10.1103/PhysRevA.62.012301}
}

@book{loudon2000quantum,
    title = {{The Quantum Theory of Light}},
    author = {Loudon, Rodney},
    publisher = {Oxford University Press},
    year = {2000},
    edition = {3rd},
    series = {Oxford Science Publications},
    address = {Oxford, UK},
    pages = {448},
    ISBN = {978-0198501770}
}

@article{zetie2000does,
    title = {{How does a Mach-Zehnder interferometer work?}},
    author = {Zetie, K. P. and Adams, S. F. and Tocknell, R. M.},
    journal = {Physics Education},
    volume = {35},
    number = {1},
    pages = {46},
    year = {2000},
    publisher = {IOP Publishing Ltd},
    doi = {10.1088/0031-9120/35/1/308},
    url = {https://iopscience.iop.org/article/10.1088/0031-9120/35/1/308}
}

@article{terhal2000bell,
  title        = {{Bell Inequalities and the Separability Criterion}},
  author       = {Terhal, Barbara M.},
  journal      = {Physics Letters A},
  volume       = {271},
  number       = {5–6},
  pages        = {319–326},
  year         = {2000},
  month        = {Jul},
  publisher    = {Elsevier BV},
  doi          = {10.1016/S0375-9601(00)00401-1},
  url          = {http://dx.doi.org/10.1016/S0375-9601(00)00401-1},
  issn         = {0375-9601}
}

@article{horodecki2001separability,
  title        = {{Separability of n-Particle Mixed States: Necessary and Sufficient Conditions in Terms of Linear Maps}},
  author       = {Horodecki, Michał and Horodecki, Paweł and Horodecki, Ryszard},
  journal      = {Physics Letters A},
  volume       = {283},
  number       = {1–2},
  pages        = {1–7},
  year         = {2001},
  month        = {May},
  publisher    = {Elsevier BV},
  doi          = {10.1016/S0375-9601(01)00142-6},
  url          = {http://dx.doi.org/10.1016/S0375-9601(01)00142-6},
  issn         = {0375-9601}
}

@article{eggeling2001separability,
  title = {Separability properties of tripartite states with $U\otimes U\otimes U$ symmetry},
  author = {Eggeling, T. and Werner, R. F.},
  journal = {Phys. Rev. A},
  volume = {63},
  issue = {4},
  pages = {042111},
  numpages = {15},
  year = {2001},
  month = {Mar},
  publisher = {American Physical Society},
  doi = {10.1103/PhysRevA.63.042111},
  url = {https://link.aps.org/doi/10.1103/PhysRevA.63.042111}
}

@article{mukunda2001Bargmann,
  title = {Bargmann invariants and off-diagonal geometric phases for multilevel quantum systems: A unitary-group approach},
  author = {Mukunda, N. and Arvind and Chaturvedi, S. and Simon, R.},
  journal = {Phys. Rev. A},
  volume = {65},
  issue = {1},
  pages = {012102},
  numpages = {10},
  year = {2001},
  month = {Dec},
  publisher = {American Physical Society},
  doi = {10.1103/PhysRevA.65.012102},
  url = {https://link.aps.org/doi/10.1103/PhysRevA.65.012102}
}

@article{mitchison2001counterfactual,
  title={Counterfactual computation},
  author={Mitchison, Graeme and Jozsa, Richard},
  journal={Proceedings of the Royal Society of London. Series A: Mathematical, Physical and Engineering Sciences},
  volume={457},
  number={2009},
  pages={1175--1193},
  year={2001},
  month={may},
  publisher={The Royal Society},
  DOI={10.1098/rspa.2000.0714},
  url={http://dx.doi.org/10.1098/rspa.2000.0714},
  ISSN={1471-2946}
}

@book{west2001introduction,
  title     = {Introduction to Graph Theory},
  author    = {West, Douglas B.},
  year      = {2001},
  edition   = {2nd},
  publisher = {Prentice Hall},
  address   = {Upper Saddle River, NJ},
  isbn      = {978-0130144003}
}

@article{brukner2001conceptual,
      title = {{Conceptual inadequacy of the Shannon information in quantum measurements}},
      author = {Brukner, \ifmmode \check{C}\else \v{C}\fi{}aslav and Zeilinger, Anton},
      journal = {Phys. Rev. A},
      volume = {63},
      issue = {2},
      pages = {022113},
      numpages = {10},
      year = {2001},
      month = {Jan},
      publisher = {American Physical Society},
      doi = {10.1103/PhysRevA.63.022113},
      url = {https://link.aps.org/doi/10.1103/PhysRevA.63.022113}
}

@article{buhrman2001quantumfingerprinting,
  title = {{Quantum Fingerprinting}},
  author = {Buhrman, Harry and Cleve, Richard and Watrous, John and de Wolf, Ronald},
  journal = {Phys. Rev. Lett.},
  volume = {87},
  issue = {16},
  pages = {167902},
  numpages = {4},
  year = {2001},
  month = {Sep},
  publisher = {American Physical Society},
  doi = {10.1103/PhysRevLett.87.167902},
  url = {https://link.aps.org/doi/10.1103/PhysRevLett.87.167902}
}

@article{ambainis2002dense,
    title = {Dense quantum coding and quantum finite automata},
    author = {Ambainis, Andris and Nayak, Ashwin and Ta-Shma, Amnon and Vazirani, Umesh},
    journal = {Journal of the ACM},
    volume = {49},
    number = {4},
    pages = {496--511},
    year = {2002},
    month = {Jul},
    publisher = {Association for Computing Machinery (ACM)},
    ISSN = {1557-735X},
    url = {http://dx.doi.org/10.1145/581771.581773},
    DOI = {10.1145/581771.581773}
}

@article{caves2002conditions,
  title = {Conditions for compatibility of quantum-state assignments},
  author = {Caves, Carlton M. and Fuchs, Christopher A. and Schack, R\"udiger},
  journal = {Phys. Rev. A},
  volume = {66},
  issue = {6},
  pages = {062111},
  numpages = {11},
  year = {2002},
  month = {Dec},
  publisher = {American Physical Society},
  doi = {10.1103/PhysRevA.66.062111},
  url = {https://link.aps.org/doi/10.1103/PhysRevA.66.062111}
}

@article{molnar2002orthogonality,
  title        = {{Orthogonality Preserving Transformations on Indefinite Inner Product Spaces: Generalization of Uhlhorn’s Version of Wigner’s Theorem}},
  author       = {Molnár, Lajos},
  journal      = {Journal of Functional Analysis},
  volume       = {194},
  number       = {2},
  pages        = {248--262},
  year         = {2002},
  month        = {Oct},
  publisher    = {Elsevier BV},
  issn         = {0022-1236},
  doi          = {10.1006/jfan.2002.3970},
  url          = {http://dx.doi.org/10.1006/jfan.2002.3970}
}

@phdthesis{galvao2002foundations,
    title = {Foundations of quantum theory and quantum information applications},
    author = {Ernesto F. Galvão},
    year = {2002},
    school = {University of Oxford},
    type = {{Ph.D. Thesis}},
    eprint = {quant-ph/0212124v1},
    archivePrefix = {arXiv},
    primaryClass = {quant-ph},
    note = {Available at \url{https://arxiv.org/abs/quant-ph/0212124}}
}

@article{herzog2002minimum,
  title = {Minimum-error discrimination between subsets of linearly dependent quantum states},
  author = {Herzog, Ulrike and Bergou, J\'anos A.},
  journal = {Phys. Rev. A},
  volume = {65},
  issue = {5},
  pages = {050305},
  numpages = {4},
  year = {2002},
  month = {May},
  publisher = {American Physical Society},
  doi = {10.1103/PhysRevA.65.050305},
  url = {https://link.aps.org/doi/10.1103/PhysRevA.65.050305}
}

@book{nielsen2002quantum,
	Author = {Michael Nielsen and Isaac Chuang},
	Isbn = {0521635039},
	Publisher = {Cambridge University Press},
	Title = {Quantum Computation and Quantum Information},
	Year = {2000}
}

@misc{rudolph2002rebit,
      title={A 2 rebit gate universal for quantum computing}, 
      author={Terry Rudolph and Lov Grover},
      year={2002},
      howpublished={arXiv:quant-ph/0210187 [quant-ph]},
      doi={https://doi.org/10.48550/arXiv.quant-ph/0210187},
      url={https://arxiv.org/abs/quant-ph/0210187}
}

@article{collins2002bell,
  title={Bell inequalities for arbitrarily high-dimensional systems},
  author={Collins, Daniel and Gisin, Nicolas and Linden, Noah and Massar, Serge and Popescu, Sandu},
  journal={Phys. Rev. Lett.},
  volume={88},
  number={4},
  pages={040404},
  month = {Jan},
  year={2002},
  publisher = {American Physical Society},
  doi = {10.1103/PhysRevLett.88.040404},
  url = {https://journals.aps.org/prl/abstract/10.1103/PhysRevLett.88.040404}
}

@misc{klyachko2002coherent,
         title = {Coherent states, entanglement, and geometric invariant theory},
         author = {Alexander Klyachko },
         year = {2002},
         eprint = {quant-ph/0206012v1},
         archivePrefix = {arXiv},
         primaryClass ={quant-ph},
         url = {https://arxiv.org/abs/quant-ph/0206012},
         doi = {
https://doi.org/10.48550/arXiv.quant-ph/0206012}
        }

@misc{aharonov2003simple,
      title={A {S}imple {P}roof that {T}offoli and {H}adamard are {Q}uantum {U}niversal}, 
      author={Dorit Aharonov},
      year={2003},
      howpublished={arXiv:quant-ph/0301040 [quant-ph]},
      doi={https://doi.org/10.48550/arXiv.quant-ph/0301040},
      URL = {https://arxiv.org/abs/quant-ph/0301040}
}

@article{fujii2003exchange,
  title={{Exchange gate on the qudit space and Fock space}},
  author={Fujii, Kazuyuki},
  journal={Journal of Optics B: Quantum and Semiclassical Optics},
  volume={5},
  number={6},
  year={2003},
  month={oct},
  pages={S613--S618},
  publisher={IOP Publishing},
  doi={10.1088/1464-4266/5/6/011},
  url={http://dx.doi.org/10.1088/1464-4266/5/6/011},
  issn={1741-3575}
}

@article{semrl2003generalized,
  title        = {{Generalized Symmetry Transformations on Quaternionic Indefinite Inner Product Spaces: An Extension of Quaternionic Version of Wigner’s Theorem}},
  author       = {Šemrl, Peter},
  journal      = {Communications in Mathematical Physics},
  volume       = {242},
  number       = {3},
  pages        = {579--584},
  year         = {2003},
  month        = {Oct},
  publisher    = {Springer Science and Business Media LLC},
  issn         = {0010-3616},
  doi          = {10.1007/s00220-003-0957-7},
  url          = {http://dx.doi.org/10.1007/s00220-003-0957-7}
}

@article{zurek2003decoherence,
  title = {Decoherence, einselection, and the quantum origins of the classical},
  author = {Zurek, Wojciech Hubert},
  journal = {Rev. Mod. Phys.},
  volume = {75},
  issue = {3},
  pages = {715--775},
  numpages = {0},
  year = {2003},
  month = {May},
  publisher = {American Physical Society},
  doi = {10.1103/RevModPhys.75.715},
  url = {https://link.aps.org/doi/10.1103/RevModPhys.75.715}
}

@article{mukunda2003Bargmann,
  title = {Bargmann invariants, null phase curves, and a theory of the geometric phase},
  author = {Mukunda, N. and Arvind and Ercolessi, E. and Marmo, G. and Morandi, G. and Simon, R.},
  journal = {Phys. Rev. A},
  volume = {67},
  issue = {4},
  pages = {042114},
  numpages = {16},
  year = {2003},
  month = {Apr},
  publisher = {American Physical Society},
  doi = {10.1103/PhysRevA.67.042114},
  url = {https://link.aps.org/doi/10.1103/PhysRevA.67.042114}
}

@article{mukunda2003Wigner,
  title={{Wigner rotations, Bargmann invariants and geometric phases}},
  author={Mukunda, N. and Aravind, P. K. and Simon, R.},
  journal={Journal of Physics A: Mathematical and General},
  volume={36},
  number={9},
  year={2003},
  month={feb},
  pages={2347--2370},
  publisher={IOP Publishing},
  doi={10.1088/0305-4470/36/9/312},
  url={http://dx.doi.org/10.1088/0305-4470/36/9/312},
  issn={0305-4470}
}

@article{gurvits2004classical,
  title        = {{Classical Complexity and Quantum Entanglement}},
  author       = {Gurvits, Leonid},
  journal      = {Journal of Computer and System Sciences},
  volume       = {69},
  number       = {3},
  pages        = {448–484},
  year         = {2004},
  month        = {Nov},
  publisher    = {Elsevier BV},
  doi          = {10.1016/j.jcss.2004.06.003},
  url          = {http://dx.doi.org/10.1016/j.jcss.2004.06.003},
  issn         = {0022-0000}
}

@article{fukuda2004frequently,
  title={Frequently asked questions in polyhedral computation},
  author={Fukuda, Komei and others},
  journal={ETH, Zurich, Switzerland},
  volume={85},
  pages={10--35},
  year={2004},
  link={https://citeseerx.ist.psu.edu/document?repid=rep1&type=pdf&doi=496ba304b69c49620aca90f0faea6b5c3b21b601}
}

@article{woese2004newbiology, title={{A New Biology for a New Century}}, volume={68}, ISSN={1098-5557}, url={http://dx.doi.org/10.1128/mmbr.68.2.173-186.2004}, DOI={10.1128/mmbr.68.2.173-186.2004}, number={2}, journal={Microbiology and Molecular Biology Reviews}, publisher={American Society for Microbiology}, author={Woese, Carl R.}, year={2004}, month=jun, pages={173–186} }

@misc{smolin2004continuous,
      title={The Continuous Variable Quantum Teleportation Controversy}, 
      author={John A. Smolin},
      year={2004},
      eprint={quant-ph/0407009},
      archivePrefix={arXiv},
      primaryClass={quant-ph},
      url={https://arxiv.org/abs/quant-ph/0407009}, 
      doi={https://doi.org/10.48550/arXiv.quant-ph/0407009},
      howpublished={arXiv:quant-ph/0407009 [quant-ph] }
}

@article{chefles2004physicaltransformations,
  title       = {{On the Existence of Physical Transformations Between Sets of Quantum States}},
  author      = {Chefles, Anthony and Jozsa, Richard and Winter, Andreas},
  journal     = {International Journal of Quantum Information},
  volume      = {02},
  number      = {01},
  pages       = {11–21},
  year        = {2004},
  month       = {Mar},
  publisher   = {World Scientific Pub Co Pte Lt},
  doi         = {10.1142/s0219749904000031},
  url         = {http://dx.doi.org/10.1142/S0219749904000031},
  ISSN        = {1793-6918}
}

@misc{ahnefeld2025coherenceresourcephaseestimation,
      title={Coherence as a resource for phase estimation}, 
      author={Felix Ahnefeld and Thomas Theurer and Martin B. Plenio},
      year={2025},
      eprint={2505.18544},
      archivePrefix={arXiv},
      primaryClass={quant-ph},
      url={https://arxiv.org/abs/2505.18544}, 
}

@misc{masanes2005extremalquantumcorrelationsn,
      title={Extremal quantum correlations for N parties with two dichotomic observables per site}, 
      author={Ll. Masanes},
      year={2005},
      eprint={quant-ph/0512100},
      archivePrefix={arXiv},
      primaryClass={quant-ph},
      url={https://arxiv.org/abs/quant-ph/0512100}, 
}

@article{schlosshauer2005decoherence,
  title = {Decoherence, the measurement problem, and interpretations of quantum mechanics},
  author = {Schlosshauer, Maximilian},
  journal = {Rev. Mod. Phys.},
  volume = {76},
  issue = {4},
  pages = {1267--1305},
  numpages = {0},
  year = {2005},
  month = {Feb},
  publisher = {American Physical Society},
  doi = {10.1103/RevModPhys.76.1267},
  url = {https://link.aps.org/doi/10.1103/RevModPhys.76.1267}
}

@book{alexandrov2005convex,
  title        = {Convex Polyhedra},
  author       = {A.D. Alexandrov},
  series       = {Springer Monographs in Mathematics},
  publisher    = {Springer Berlin, Heidelberg},
  year         = {2005},
  doi          = {10.1007/b137434},
  isbn         = {978-3-540-23158-5},
  eisbn        = {978-3-540-26340-1},
  url          = {https://doi.org/10.1007/b137434},
  edition      = {1},
  pages        = {XII, 542},
  address      = {Berlin, Heidelberg},
  month        = feb,
  note         = {Original Russian edition published by Gosudarstv. Izdat. Tekhn. Teor. Lit., Moscow-Leningrad, 1950},
  isbnsoft     = {978-3-642-06215-5},
  numillustrations = {165 b/w illustrations},
  copyright    = {Springer-Verlag Berlin Heidelberg 2005},
  eissn        = {2196-9922},
  issn         = {1439-7382},
  seriesvolume = {R0}
}

@article{doring2005kochen,  
  title={{Kochen–Specker Theorem for von Neumann Algebras}},  
  author={Döring, Andreas},  
  journal={{International Journal of Theoretical Physics}},  
  volume={44},  
  number={2},  
  pages={139--160},  
  year={2005},  
  month={feb},  
  publisher={{Springer Science and Business Media LLC}},  
  issn={1572-9575},  
  doi={10.1007/s10773-005-1490-6},  
  url={http://dx.doi.org/10.1007/s10773-005-1490-6}  
}

@article{bravyi2005universal,
  title = {Universal quantum computation with ideal Clifford gates and noisy ancillas},
  author = {Bravyi, Sergey and Kitaev, Alexei},
  journal = {Phys. Rev. A},
  volume = {71},
  issue = {2},
  pages = {022316},
  numpages = {14},
  year = {2005},
  month = {Feb},
  publisher = {American Physical Society},
  doi = {10.1103/PhysRevA.71.022316},
  url = {https://link.aps.org/doi/10.1103/PhysRevA.71.022316}
}

@article{pironio2005lifting,
  title     = {Lifting {B}ell inequalities},
  author    = {Pironio, Stefano},
  journal   = {Journal of Mathematical Physics},
  volume    = {46},
  number    = {6},
  pages     = {062112},
  year      = {2005},
  month     = {Jun},
  publisher = {AIP Publishing},
  doi       = {10.1063/1.1928727},
  url       = {http://dx.doi.org/10.1063/1.1928727},
  issn      = {1089-7658}
}

@article{spekkens2005contextuality,
  title = {Contextuality for preparations, transformations, and unsharp measurements},
  author = {Spekkens, R. W.},
  journal = {Phys. Rev. A},
  volume = {71},
  issue = {5},
  pages = {052108},
  numpages = {17},
  year = {2005},
  month = {May},
  publisher = {American Physical Society},
  doi = {10.1103/PhysRevA.71.052108},
  url = {https://link.aps.org/doi/10.1103/PhysRevA.71.052108}
}

@article{landsman2006Whenchampionsmeed,
  author    = {Landsman, N. P.},
  title     = {{When champions meet: Rethinking the Bohr–Einstein debate}},
  journal   = {Studies in History and Philosophy of Science Part B: Studies in History and Philosophy of Modern Physics},
  volume    = {37},
  number    = {1},
  pages     = {212–242},
  year      = {2006},
  month     = {Mar},
  issn      = {1355-2198},
  doi       = {10.1016/j.shpsb.2005.10.002},
  url       = {http://dx.doi.org/10.1016/j.shpsb.2005.10.002},
  publisher = {Elsevier BV}
}

@article{bartlett2006dialogue,
  title        = {{Dialogue Concerning Two Views on Quantum Coherence: Factist and Fictionist}},
  author       = {Bartlett, Stephen D. and Rudolph, Terry and Spekkens, Robert W.},
  journal      = {International Journal of Quantum Information},
  volume       = {4},
  number       = {1},
  pages        = {17--43},
  year         = {2006},
  month        = {Feb},
  publisher    = {World Scientific Pub Co Pte Ltd},
  issn         = {1793-6918},
  doi          = {10.1142/S0219749906001591},
  url          = {https://doi.org/10.1142/S0219749906001591}
}

@article{acin2006grothendieck,
  title = {Grothendieck's constant and local models for noisy entangled quantum states},
  author = {Ac\'{\i}n, Antonio and Gisin, Nicolas and Toner, Benjamin},
  journal = {Phys. Rev. A},
  volume = {73},
  issue = {6},
  pages = {062105},
  numpages = {5},
  year = {2006},
  month = {Jun},
  publisher = {American Physical Society},
  doi = {10.1103/PhysRevA.73.062105},
  url = {https://link.aps.org/doi/10.1103/PhysRevA.73.062105}
}

@article{hosten2006counterfactual,
  title={Counterfactual quantum computation through quantum interrogation},
  author={Hosten, Onur and Rakher, Matthew T. and Barreiro, Julio T. and Peters, Nicholas A. and Kwiat, Paul G.},
  journal={Nature},
  volume={439},
  number={7079},
  pages={949--952},
  year={2006},
  month={feb},
  publisher={Springer Science and Business Media LLC},
  DOI={10.1038/nature04523},
  url={http://dx.doi.org/10.1038/nature04523},
  ISSN={1476-4687}
}

@book{peres2006quantumtheory,
  editor    = {Peres, Asher},
  title     = {{Quantum Theory: Concepts and Methods}},
  series    = {Fundamental Theories of Physics},
  publisher = {Springer Dordrecht},
  year      = {2006},
  edition   = {1},
  pages     = {XIV, 450},
  isbn      = {978-0-306-47120-9},
  doi       = {10.1007/0-306-47120-5},
  url       = {https://doi.org/10.1007/0-306-47120-5},
  copyright = {Springer Science+Business Media B.V. 2002},
  topics    = {Quantum Physics, Philosophy of Science, Popular Science, general}
}

@article{spekkens2007evidence,
  title = {Evidence for the epistemic view of quantum states: A toy theory},
  author = {Spekkens, Robert W.},
  journal = {Phys. Rev. A},
  volume = {75},
  issue = {3},
  pages = {032110},
  numpages = {30},
  year = {2007},
  month = {Mar},
  publisher = {American Physical Society},
  doi = {10.1103/PhysRevA.75.032110},
  url = {https://link.aps.org/doi/10.1103/PhysRevA.75.032110}
}

@article{hosten2008observation,
	doi = {10.1126/science.1152697},
	url = {https://doi.org/10.1126/science.1152697},
	year = 2008,
	month = {Feb},
	publisher = {American Association for the Advancement of Science ({AAAS})},
	volume = {319},
	number = {5864},
	pages = {787--790},
	author = {Onur Hosten and Paul Kwiat},
	title = {Observation of the {S}pin {H}all {E}ffect of {L}ight via {W}eak {M}easurements},
	journal = {Science}
}

@article{cabello2008experimentally,
  title = {{Experimentally Testable State-Independent Quantum Contextuality}},
  author = {Cabello, Ad\'an},
  journal = {Phys. Rev. Lett.},
  volume = {101},
  issue = {21},
  pages = {210401},
  numpages = {4},
  year = {2008},
  month = {Nov},
  publisher = {American Physical Society},
  doi = {10.1103/PhysRevLett.101.210401},
  url = {https://link.aps.org/doi/10.1103/PhysRevLett.101.210401}
}

@article{brunner2008testing,
  title = {{Testing the Dimension of Hilbert Spaces}},
  author = {Brunner, Nicolas and Pironio, Stefano and Acin, Antonio and Gisin, Nicolas and M\'ethot, Andr\'e Allan and Scarani, Valerio},
  journal = {Phys. Rev. Lett.},
  volume = {100},
  issue = {21},
  pages = {210503},
  numpages = {4},
  year = {2008},
  month = {May},
  publisher = {American Physical Society},
  doi = {10.1103/PhysRevLett.100.210503},
  url = {https://link.aps.org/doi/10.1103/PhysRevLett.100.210503}
}

@article{wehner2008lowerbound,
  title = {Lower bound on the dimension of a quantum system given measured data},
  author = {Wehner, Stephanie and Christandl, Matthias and Doherty, Andrew C.},
  journal = {Phys. Rev. A},
  volume = {78},
  issue = {6},
  pages = {062112},
  numpages = {8},
  year = {2008},
  month = {Dec},
  publisher = {American Physical Society},
  doi = {10.1103/PhysRevA.78.062112},
  url = {https://link.aps.org/doi/10.1103/PhysRevA.78.062112}
}

@article{perez2008unbounded,
    title = {{Unbounded Violation of Tripartite Bell Inequalities}},
    author = {P{\'e}rez-Garc{\'i}a, D. and Wolf, M. M. and Palazuelos, C. and Villanueva, I. and Junge, M.},
    journal = {Communications in Mathematical Physics},
    volume = {279},
    number = {2},
    pages = {455--486},
    year = {2008},
    month = {Feb},
    publisher = {Springer Science and Business Media LLC},
    ISSN = {1432-0916},
    url = {http://dx.doi.org/10.1007/s00220-008-0418-4},
    DOI = {10.1007/s00220-008-0418-4}
}

@article{pitowsky2008newbell,
    title = {{New Bell inequalities for the singlet state: Going beyond the Grothendieck bound}},
    author = {Pitowsky, Itamar},
    journal = {Journal of Mathematical Physics},
    volume = {49},
    number = {1},
    year = {2008},
    month = {Jan},
    publisher = {AIP Publishing},
    ISSN = {1089-7658},
    url = {http://dx.doi.org/10.1063/1.2826227},
    DOI = {10.1063/1.2826227}
}

@article{pal2008efficiency,
  title = {{Efficiency of higher-dimensional Hilbert spaces for the violation of Bell inequalities}},
  author = {P\'al, K\'aroly F. and V\'ertesi, Tam\'as},
  journal = {Phys. Rev. A},
  volume = {77},
  issue = {4},
  pages = {042105},
  numpages = {9},
  year = {2008},
  month = {Apr},
  publisher = {American Physical Society},
  doi = {10.1103/PhysRevA.77.042105},
  url = {https://link.aps.org/doi/10.1103/PhysRevA.77.042105}
}

@article{vertesi2008generalized,
  title = {{Generalized Clauser-Horne-Shimony-Holt inequalities maximally violated by higher-dimensional systems}},
  author = {V\'ertesi, T. and P\'al, K. F.},
  journal = {Phys. Rev. A},
  volume = {77},
  issue = {4},
  pages = {042106},
  numpages = {8},
  year = {2008},
  month = {Apr},
  publisher = {American Physical Society},
  doi = {10.1103/PhysRevA.77.042106},
  url = {https://link.aps.org/doi/10.1103/PhysRevA.77.042106}
}

@article{klyachko2008simple,
  title = {{Simple Test for Hidden Variables in Spin-1 Systems}},
  author = {Klyachko, Alexander A. and Can, M. Ali and Binicio\ifmmode \breve{g}\else \u{g}\fi{}lu, Sinem and Shumovsky, Alexander S.},
  journal = {Phys. Rev. Lett.},
  volume = {101},
  issue = {2},
  pages = {020403},
  numpages = {4},
  year = {2008},
  month = {Jul},
  publisher = {American Physical Society},
  doi = {10.1103/PhysRevLett.101.020403},
  url = {https://link.aps.org/doi/10.1103/PhysRevLett.101.020403}
}

@article{McKague2009simulating,
	doi = {10.1103/physrevlett.102.020505},
	url = {https://doi.org/10.1103/physrevlett.102.020505},
	year = 2009,
	month = {Jan},
	publisher = {American Physical Society ({APS})},
	volume = {102},
	number = {2},
	author = {Matthew McKague and Michele Mosca and Nicolas Gisin},
          pages = {020505},
	title = {Simulating {Q}uantum {S}ystems {U}sing {R}eal {H}ilbert {S}paces},
	journal = {Phys. Rev. Lett.}
}

@article{kirchmair2009state,
  title        = {State-independent experimental test of quantum contextuality},
  author       = {Kirchmair, G. and Zähringer, F. and Gerritsma, R. and others},
  journal      = {Nature},
  volume       = {460},
  pages        = {494--497},
  year         = {2009},
  doi          = {10.1038/nature08172},
  url          = {https://doi.org/10.1038/nature08172},
  month     = {May}
}

@article{dyson2009birds,
  title={Birds and frogs},
  author={Dyson, Freeman},
  journal={Notices of the AMS},
  volume={56},
  number={2},
  pages={212--223},
  year={2009},
  url={https://www.ams.org/notices/200902/rtx090200212p.pdf?utm_referrer=https%3A%2F%2Fdzen.ru%2Fmedia%2Fid%2F63ebe0724a49c867e356a1a2%2F654f57784115eb087fae8b51}
}

@article{spekkens2009preparation,
  title = {{Preparation Contextuality Powers Parity-Oblivious Multiplexing}},
  author = {Spekkens, Robert W. and Buzacott, D. H. and Keehn, A. J. and Toner, Ben and Pryde, G. J.},
  journal = {Phys. Rev. Lett.},
  volume = {102},
  issue = {1},
  pages = {010401},
  numpages = {4},
  year = {2009},
  month = {Jan},
  publisher = {American Physical Society},
  doi = {10.1103/PhysRevLett.102.010401},
  url = {https://link.aps.org/doi/10.1103/PhysRevLett.102.010401}
}

@book{arora2009computational,
  title     = {{Computational Complexity: A Modern Approach}},
  author    = {Arora, Sanjeev and Barak, Boaz},
  publisher = {Cambridge University Press},
  year      = {2009},
  month     = {Apr},
  isbn      = {9780511804090},
  url       = {http://dx.doi.org/10.1017/CBO9780511804090},
  doi       = {10.1017/cbo9780511804090}
}

@book{bacciagaluppi2009quantum,
  author    = {Bacciagaluppi, Guido and Valentini, Antony},
  title     = {{Quantum Theory at the Crossroads: Reconsidering the 1927 Solvay Conference}},
  publisher = {Cambridge University Press},
  year      = {2009},
  month     = {Oct},
  isbn      = {9781107698314},
  doi       = {10.1017/cbo9781139194983},
  url       = {http://dx.doi.org/10.1017/CBO9781139194983}
}

@article{dixon2009ultrasensitive,
  title = {{Ultrasensitive Beam Deflection Measurement via Interferometric Weak Value Amplification}},
  author = {Dixon, P. Ben and Starling, David J. and Jordan, Andrew N. and Howell, John C.},
  journal = {Phys. Rev. Lett.},
  volume = {102},
  issue = {17},
  pages = {173601},
  numpages = {4},
  year = {2009},
  month = {Apr},
  publisher = {American Physical Society},
  doi = {10.1103/PhysRevLett.102.173601},
  url = {https://link.aps.org/doi/10.1103/PhysRevLett.102.173601}
}

@article{hardy2010why,
         title = {{Why Physics Needs Quantum Foundations}},
         author = {Lucien Hardy and Robert Spekkens },
         year = {2010},
         eprint = {1003.5008v1},
         archivePrefix = {arXiv},
         primaryClass ={quant-ph},
        doi = {https://doi.org/10.48550/arXiv.1003.5008
},
        note = {Available at: \url{https://arxiv.org/abs/1003.5008}}
        }

@article{harrigan2010einstein,
  author    = {Harrigan, Nicholas and Spekkens, Robert W.},
  title     = {{Einstein, Incompleteness, and the Epistemic View of Quantum States}},
  journal   = {Foundations of Physics},
  volume    = {40},
  number    = {2},
  pages     = {125--157},
  year      = {2010},
  month     = {Jan},
  publisher = {Springer Science and Business Media LLC},
  issn      = {1572-9516},
  doi       = {10.1007/s10701-009-9347-0},
  url       = {http://dx.doi.org/10.1007/s10701-009-9347-0}
}

@article{brunner2010measuringsmall,
	doi = {10.1103/physrevlett.105.010405},
	url = {https://doi.org/10.1103/physrevlett.105.010405},
	year = 2010,
	month = {Jul},
	publisher = {American Physical Society ({APS})},
	volume = {105},
        pages = {010405},
	number = {1},
	author = {Nicolas Brunner and Christoph Simon},
	title = {Measuring {S}mall {L}ongitudinal {P}hase {S}hifts: {W}eak {M}easurements or {S}tandard {I}nterferometry?},
	journal = {Phys. Rev. Lett.}
}

@misc{hermens2010quantummechanicsrealismintuitionism,
      title={{Quantum Mechanics: From Realism to Intuitionism}}, 
      author={Ronnie Hermens},
      year={2010},
      eprint={1002.1410},
      archivePrefix={arXiv},
      primaryClass={quant-ph},
      url={https://arxiv.org/abs/1002.1410}, 
}

@article{gallego2010device,
  title = {Device-Independent Tests of Classical and Quantum Dimensions},
  author = {Gallego, Rodrigo and Brunner, Nicolas and Hadley, Christopher and Ac\'{\i}n, Antonio},
  journal = {Phys. Rev. Lett.},
  volume = {105},
  issue = {23},
  pages = {230501},
  numpages = {4},
  year = {2010},
  month = {Nov},
  publisher = {American Physical Society},
  doi = {10.1103/PhysRevLett.105.230501},
  url = {https://link.aps.org/doi/10.1103/PhysRevLett.105.230501}
}

@article{badziag2010pentagrams,
  title        = {Pentagrams and Paradoxes},
  author       = {Badziag, Piotr and Bengtsson, Ingemar and Cabello, Ad\'an and Granström, Helena and Larsson, Jan-Åke},
  journal      = {Foundations of Physics},
  volume       = {41},
  number       = {3},
  pages        = {414--423},
  year         = {2010},
  month        = {Mar},
  publisher    = {Springer Science and Business Media LLC},
  issn         = {1572-9516},
  doi          = {10.1007/s10701-010-9433-3},
  url          = {http://dx.doi.org/10.1007/s10701-010-9433-3}
}

@article{bollen2010direct,
  title = {{Direct measurement of the Kirkwood-Rihaczek distribution for the spatial properties of a coherent light beam}},
  author = {Bollen, Viktor and Sua, Yong Meng and Lee, Kim Fook},
  journal = {Phys. Rev. A},
  volume = {81},
  issue = {6},
  pages = {063826},
  numpages = {11},
  year = {2010},
  month = {Jun},
  publisher = {American Physical Society},
  doi = {10.1103/PhysRevA.81.063826},
  url = {https://link.aps.org/doi/10.1103/PhysRevA.81.063826}
}

@article{wootters2010entanglement,
	doi = {10.1007/s10701-010-9488-1},
	url = {https://doi.org/10.1007/s10701-010-9488-1},
	year = 2010,
	month = {Jul},
	publisher = {Springer Science and Business Media {LLC}},
	volume = {42},
	number = {1},
	pages = {19--28},
	author = {William K. Wootters},
	title = {Entanglement {S}haring in {R}eal-{V}ector-{S}pace {Q}uantum {T}heory},
	journal = {Foundations of Physics}
}

@inproceedings{aaronson2011bosonsampling,
  title       = {{The computational complexity of linear optics}},
  author      = {Aaronson, Scott and Arkhipov, Alex},
  booktitle   = {Proceedings of the Forty-Third Annual ACM Symposium on Theory of Computing},
  series      = {STOC'11},
  pages       = {333--342},
  year        = {2011},
  month       = {Jun},
  publisher   = {ACM},
  url         = {http://dx.doi.org/10.1145/1993636.1993682},
  doi         = {10.1145/1993636.1993682}
}

@book{zhang2011matrix,
  author    = {Fuzhen Zhang},
  title     = {{Matrix Theory: Basic Results and Techniques}},
  publisher = {Springer Science \& Business Media},
  address   = {New York},
  year      = {2011}
}

@article{briet2011generalized,
    title = {{A Generalized Grothendieck Inequality and Nonlocal Correlations that Require High Entanglement}},
    author = {Briët, Jop and Buhrman, Harry and Toner, Ben},
    journal = {Communications in Mathematical Physics},
    volume = {305},
    number = {3},
    pages = {827--843},
    year = {2011},
    month = {Jun},
    publisher = {Springer Science and Business Media LLC},
    ISSN = {1432-0916},
    url = {http://dx.doi.org/10.1007/s00220-011-1280-3},
    DOI = {10.1007/s00220-011-1280-3}
}

@article{hermens2011theproblem,
  author    = {Hermens, Ronnie},
  title     = {The problem of contextuality and the impossibility of experimental metaphysics thereof},
  journal   = {Studies in History and Philosophy of Science Part B: Studies in History and Philosophy of Modern Physics},
  volume    = {42},
  number    = {4},
  pages     = {214--225},
  year      = {2011},
  month     = {Nov},
  issn      = {1355-2198},
  doi       = {10.1016/j.shpsb.2011.06.001},
  url       = {http://dx.doi.org/10.1016/j.shpsb.2011.06.001},
  publisher = {Elsevier BV}
}

@article{hofmann2011uncertainty,
  title = {Uncertainty limits for quantum metrology obtained from the statistics of weak measurements},
  author = {Hofmann, Holger F.},
  journal = {Phys. Rev. A},
  volume = {83},
  issue = {2},
  pages = {022106},
  numpages = {5},
  year = {2011},
  month = {Feb},
  publisher = {American Physical Society},
  doi = {10.1103/PhysRevA.83.022106},
  url = {https://link.aps.org/doi/10.1103/PhysRevA.83.022106}
}

@article{abramsky2011sheaf,
  author    = {Abramsky, Samson and Brandenburger, Adam},
  title     = {{The Sheaf-Theoretic Structure of Non-Locality and Contextuality}},
  journal   = {New Journal of Physics},
  volume    = {13},
  number    = {11},
  pages     = {113036},
  year      = {2011},
  month     = {Nov},
  ISSN      = {1367-2630},
  publisher = {IOP Publishing},
  DOI       = {10.1088/1367-2630/13/11/113036},
  url       = {http://dx.doi.org/10.1088/1367-2630/13/11/113036}
}

@article{hardy2011limited,
	doi = {10.1007/s10701-011-9616-6},
	url = {https://doi.org/10.1007/s10701-011-9616-6},
	year = 2011,
	month = {Dec},
	publisher = {Springer Science and Business Media {LLC}},
	volume = {42},
	number = {3},
	pages = {454--473},
	author = {Lucien Hardy and William K. Wootters},
	title = {Limited {H}olism and {R}eal-{V}ector-{S}pace {Q}uantum {T}heory},
	journal = {Foundations of Physics}
}

@article{kedem2012usingtechnical,
	doi = {10.1103/physreva.85.060102},
	url = {https://doi.org/10.1103/physreva.85.060102},
	year = 2012,
	month = {Jun},
	publisher = {American Physical Society ({APS})},
	volume = {85},
	number = {6},
        pages = {060102},
	author = {Yaron Kedem},
	title = {Using technical noise to increase the signal-to-noise ratio of measurements via imaginary weak values},
	journal = {Phys. Rev. A}
}

@book{Ziegler2012,
  title={Lectures on polytopes},
  author={Ziegler, G{\"u}nter M},
  volume={152},
  year={2012},
  publisher={Springer Science \& Business Media}
}

@article{johansson2012qutip,  
  title={{QuTiP: An open-source Python framework for the dynamics of open quantum systems}},  
  author={Johansson, J. R. and Nation, P. D. and Nori, Franco},  
  journal={Computer Physics Communications},  
  volume={183},  
  number={8},  
  pages={1760--1772},  
  year={2012},  
  month={aug},  
  publisher={Elsevier BV},  
  doi={10.1016/j.cpc.2012.02.021},  
  url={http://dx.doi.org/10.1016/j.cpc.2012.02.021},  
  issn={0010-4655}  
}

@article{hendrych2012experimental,
    title = {Experimental estimation of the dimension of classical and quantum systems},
    author = {Hendrych, Martin and Gallego, Rodrigo and Mi{\v{c}}uda, Michal and Brunner, Nicolas and Ac{\'i}n, Antonio and Torres, Juan P.},
    journal = {Nature Physics},
    volume = {8},
    number = {8},
    pages = {588--591},
    year = {2012},
    month = {Jun},
    publisher = {Springer Science and Business Media LLC},
    ISSN = {1745-2481},
    url = {http://dx.doi.org/10.1038/nphys2334},
    DOI = {10.1038/nphys2334}
}

@book{brondsted2012introduction,
  title={An introduction to convex polytopes},
  author={Brondsted, Arne},
  volume={90},
  year={2012},
  publisher={Springer Science \& Business Media}
}

@article{pusey2012onthereality,
  author    = {Pusey, Matthew F. and Barrett, Jonathan and Rudolph, Terry},
  title     = {{On the Reality of the Quantum State}},
  journal   = {Nature Physics},
  volume    = {8},
  number    = {6},
  pages     = {475–478},
  year      = {2012},
  month     = {May},
  issn      = {1745-2481},
  publisher = {Springer Science and Business Media LLC},
  doi       = {10.1038/nphys2309},
  url       = {http://dx.doi.org/10.1038/nphys2309}
}

@article{colbeck2012freerandomness,
  author    = {Colbeck, Roger and Renner, Renato},
  title     = {{Free Randomness Can Be Amplified}},
  journal   = {Nature Physics},
  volume    = {8},
  number    = {6},
  pages     = {450–453},
  year      = {2012},
  month     = {May},
  issn      = {1745-2481},
  publisher = {Springer Science and Business Media LLC},
  doi       = {10.1038/nphys2300},
  url       = {http://dx.doi.org/10.1038/nphys2300}
}

@article{modi2012discord,
      title = {The classical-quantum boundary for correlations: Discord and related measures},
      author = {Modi, Kavan and Brodutch, Aharon and Cable, Hugo and Paterek, Tomasz and Vedral, Vlatko},
      journal = {Rev. Mod. Phys.},
      volume = {84},
      issue = {4},
      pages = {1655--1707},
      numpages = {0},
      year = {2012},
      month = {Nov},
      publisher = {American Physical Society},
      doi = {10.1103/RevModPhys.84.1655},
      url = {https://link.aps.org/doi/10.1103/RevModPhys.84.1655}
}

@article{hofmann2012complex,
  title        = {{Complex Joint Probabilities as Expressions of Reversible Transformations in Quantum Mechanics}},
  author       = {Hofmann, Holger F.},
  journal      = {New Journal of Physics},
  volume       = {14},
  number       = {4},
  year         = {2012},
  month        = {Apr},
  pages        = {043031},
  publisher    = {IOP Publishing},
  doi          = {10.1088/1367-2630/14/4/043031},
  url          = {http://dx.doi.org/10.1088/1367-2630/14/4/043031},
  issn         = {1367-2630}
}

@article{flammia2012quantum,
  title={Quantum tomography via compressed sensing: error bounds, sample complexity and efficient estimators},
  author={Flammia, Steven T. and Gross, David and Liu, Yi-Kai and Eisert, Jens},
  fjournal={New Journal of Physics},
  journal={New J. Phys.},
  volume={14},
  number={9},
  pages={095022},
  year={2012},
  doi={10.1088/1367-2630/14/9/095022},
  publisher={IOP Publishing},
  url = {https://iopscience.iop.org/article/10.1088/1367-2630/14/9/095022}
}

@article{araujo2013all,
  title = {All noncontextuality inequalities for the $n$-cycle scenario},
  author = {Ara\'ujo, Mateus and Quintino, Marco T\'ulio and Budroni, Costantino and Cunha, Marcelo Terra and Cabello, Ad\'an},
  journal = {Phys. Rev. A},
  volume = {88},
  issue = {2},
  pages = {022118},
  numpages = {7},
  year = {2013},
  month = {Aug},
  publisher = {American Physical Society},
  doi = {10.1103/PhysRevA.88.022118},
  url = {https://link.aps.org/doi/10.1103/PhysRevA.88.022118}
}

@article{garcia2013swap,
  title = {{swap test and Hong-Ou-Mandel effect are equivalent}},
  author = {Garcia-Escartin, Juan Carlos and Chamorro-Posada, Pedro},
  journal = {Phys. Rev. A},
  volume = {87},
  issue = {5},
  pages = {052330},
  numpages = {10},
  year = {2013},
  month = {May},
  publisher = {American Physical Society},
  doi = {10.1103/PhysRevA.87.052330},
  url = {https://link.aps.org/doi/10.1103/PhysRevA.87.052330}
}

@article{salih2013protocol,
  title = {{Protocol for Direct Counterfactual Quantum Communication}},
  author = {Salih, Hatim and Li, Zheng-Hong and Al-Amri, M. and Zubairy, M. Suhail},
  journal = {Phys. Rev. Lett.},
  volume = {110},
  issue = {17},
  pages = {170502},
  numpages = {5},
  year = {2013},
  month = {Apr},
  publisher = {American Physical Society},
  doi = {10.1103/PhysRevLett.110.170502},
  url = {https://link.aps.org/doi/10.1103/PhysRevLett.110.170502}
}

@book{matousek2013lectures,
  title        = {{Lectures on Discrete Geometry}},
  editor       = {Jiří Matoušek},
  series       = {Graduate Texts in Mathematics},
  publisher    = {Springer New York, NY},
  year         = {2002},
  volume       = {212},
  edition      = {1},
  pages        = {XVI, 486},
  doi          = {10.1007/978-1-4613-0039-7},
  isbn         = {978-0-387-95373-1},
  isbnsoft     = {978-0-387-95374-8},
  eisbn        = {978-1-4613-0039-7},
  url          = {https://doi.org/10.1007/978-1-4613-0039-7},
  copyright    = {Springer-Verlag New York 2002},
  topics       = {Geometry, Convex and Discrete Geometry},
  seriesissn   = {0072-5285},
  serieseissn  = {2197-5612},
  ebookpackages = {Springer Science Book Archive \& Business Media}
}

@article{kaibel2003isomorphism,
  title        = {{On the Complexity of Polytope Isomorphism Problems}},
  author       = {Kaibel, Volker and Schwartz, Alexander},
  journal      = {Graphs and Combinatorics},
  volume       = {19},
  number       = {2},
  pages        = {215–230},
  year         = {2003},
  month        = jun,
  doi          = {10.1007/s00373-002-0503-y},
  url          = {http://dx.doi.org/10.1007/s00373-002-0503-y},
  issn         = {1435-5914},
  publisher    = {Springer Science and Business Media LLC}
}

@article{li2013relationship,
  title = {Relationship between semi- and fully-device-independent protocols},
  author = {Li, Hong-Wei and Mironowicz, Piotr and Paw\l{}owski, Marcin and Yin, Zhen-Qiang and Wu, Yu-Chun and Wang, Shuang and Chen, Wei and Hu, Hong-Gang and Guo, Guang-Can and Han, Zheng-Fu},
  journal = {Phys. Rev. A},
  volume = {87},
  issue = {2},
  pages = {020302},
  numpages = {4},
  year = {2013},
  month = {Feb},
  publisher = {American Physical Society},
  doi = {10.1103/PhysRevA.87.020302},
  url = {https://link.aps.org/doi/10.1103/PhysRevA.87.020302}
}

@article{moroder2013device,
  title = {{Device-Independent Entanglement Quantification and Related Applications}},
  author = {Moroder, Tobias and Bancal, Jean-Daniel and Liang, Yeong-Cherng and Hofmann, Martin and G\"uhne, Otfried},
  journal = {Phys. Rev. Lett.},
  volume = {111},
  issue = {3},
  pages = {030501},
  numpages = {5},
  year = {2013},
  month = {Jul},
  publisher = {American Physical Society},
  doi = {10.1103/PhysRevLett.111.030501},
  url = {https://link.aps.org/doi/10.1103/PhysRevLett.111.030501}
}

@article{leifer2013maximally,
  title = {{Maximally Epistemic Interpretations of the Quantum State and Contextuality}},
  author = {Leifer, Matthew S. and Maroney, Owen J. E.},
  journal = {Phys. Rev. Lett.},
  volume = {110},
  issue = {12},
  pages = {120401},
  numpages = {5},
  year = {2013},
  month = {Mar},
  publisher = {American Physical Society},
  doi = {10.1103/PhysRevLett.110.120401},
  url = {https://link.aps.org/doi/10.1103/PhysRevLett.110.120401}
}

@article{amselem2013comment,
  title = {{Comment on ``State-Independent Experimental Test of Quantum Contextuality in an Indivisible System''}},
  author = {Amselem, Elias and Bourennane, Mohamed  and Budroni, Costantino and Cabello, Ad\'an and G\"uhne, Otfried and Kleinmann, Matthias and Larsson, Jan-Åke and Wie\ifmmode \acute{s}\else \'{s}\fi{}niak, Marcin },
  journal = {Phys. Rev. Lett.},
  volume = {110},
  issue = {7},
  pages = {078901},
  numpages = {1},
  year = {2013},
  month = {Feb},
  publisher = {American Physical Society},
  doi = {10.1103/PhysRevLett.110.078901},
  url = {https://link.aps.org/doi/10.1103/PhysRevLett.110.078901}
}

@book{lidar2013quantumerrorcorrection,
  title       = {{Quantum Error Correction}},
  editor      = {Lidar, Daniel A. and Brun, Todd A.},
  publisher   = {Cambridge University Press},
  year        = {2013},
  month       = {Sep},
  url         = {https://doi.org/10.1017/CBO9781139034807},
  doi         = {10.1017/CBO9781139034807},
  isbn        = {9781139034807},
  note        = {Online publication date: September 2013},
  address     = {University of Southern California}
}

@book{wilde2013quantum,
  title     = {Quantum Information Theory},
  author    = {Wilde, Mark M.},
  publisher = {Cambridge University Press},
  year      = {2013},
  month     = {Apr},
  isbn      = {9781139525343},
  doi       = {10.1017/CBO9781139525343},
  url       = {http://dx.doi.org/10.1017/CBO9781139525343}
}

@article{antoniya2013real,
  title = {Real-vector-space quantum theory with a universal quantum bit},
  author = {Aleksandrova, Antoniya and Borish, Victoria and Wootters, William K.},
  journal = {Phys. Rev. A},
  volume = {87},
  issue = {5},
  pages = {052106},
  numpages = {23},
  year = {2013},
  month = {May},
  publisher = {American Physical Society},
  doi = {10.1103/PhysRevA.87.052106},
  url = {https://link.aps.org/doi/10.1103/PhysRevA.87.052106}
}

@article{eltschka2013negativity,
  title = {{Negativity as an Estimator of Entanglement Dimension}},
  author = {Eltschka, Christopher and Siewert, Jens},
  journal = {Phys. Rev. Lett.},
  volume = {111},
  issue = {10},
  pages = {100503},
  numpages = {4},
  year = {2013},
  month = {Sep},
  publisher = {American Physical Society},
  doi = {10.1103/PhysRevLett.111.100503},
  url = {https://link.aps.org/doi/10.1103/PhysRevLett.111.100503}
}

@article{brunner2013dimension,
  title = {{Dimension Witnesses and Quantum State Discrimination}},
  author = {Brunner, Nicolas and Navascu\'es, Miguel and V\'ertesi, Tam\'as},
  journal = {Phys. Rev. Lett.},
  volume = {110},
  issue = {15},
  pages = {150501},
  numpages = {4},
  year = {2013},
  month = {Apr},
  publisher = {American Physical Society},
  doi = {10.1103/PhysRevLett.110.150501},
  url = {https://link.aps.org/doi/10.1103/PhysRevLett.110.150501}
}

@article{spekkens2014statusofoutcome, title={The Status of Determinism in Proofs of the Impossibility of a Noncontextual Model of Quantum Theory}, volume={44}, ISSN={1572-9516}, url={http://dx.doi.org/10.1007/s10701-014-9833-x}, DOI={10.1007/s10701-014-9833-x}, number={11}, journal={Foundations of Physics}, publisher={Springer Science and Business Media LLC}, author={Spekkens, Robert W.}, year={2014}, month=sep, pages={1125–1155} }

@mastersthesis{araujo2014quantum,
  author  = {Ara{\'u}jo, Mateus},
  title   = {Quantum realism and quantum surrealism},
  school  = {Universidade Federal de Minas Gerais (UFMG)},
  month   = {Jun},
  year    = {2012},
  note    = {Available at \href{https://arxiv.org/abs/1208.6283}{arXiv:1208.6283 [quant-ph]}},
}

@article{allahverdyan2014nonequilibrium,
  title = {Nonequilibrium quantum fluctuations of work},
  author = {Allahverdyan, A. E.},
  journal = {Phys. Rev. E},
  volume = {90},
  issue = {3},
  pages = {032137},
  numpages = {9},
  year = {2014},
  month = {Sep},
  publisher = {American Physical Society},
  doi = {10.1103/PhysRevE.90.032137},
  url = {https://link.aps.org/doi/10.1103/PhysRevE.90.032137}
}

@inbook{chudnovsky2014cliques, title={Cliques and stable sets in undirected graphs}, ISBN={9788876425257}, url={http://dx.doi.org/10.1007/978-88-7642-525-7_2}, DOI={10.1007/978-88-7642-525-7_2}, booktitle={Geometry, Structure and Randomness in Combinatorics}, publisher={Scuola Normale Superiore}, author={Chudnovsky, Maria}, year={2014}, pages={19–25} }

@article{leifer2014psi,
  title = {$\ensuremath{\psi}$-Epistemic Models are Exponentially Bad at Explaining the Distinguishability of Quantum States},
  author = {Leifer, Matthew S.},
  journal = {Phys. Rev. Lett.},
  volume = {112},
  issue = {16},
  pages = {160404},
  numpages = {4},
  year = {2014},
  month = {Apr},
  publisher = {American Physical Society},
  doi = {10.1103/PhysRevLett.112.160404},
  url = {https://link.aps.org/doi/10.1103/PhysRevLett.112.160404}
}

@article{larsson2014loopholes,
  title        = {{Loopholes in Bell inequality tests of local realism}},
  author       = {Larsson, Jan-Åke},
  journal      = {Journal of Physics A: Mathematical and Theoretical},
  volume       = {47},
  number       = {42},
  pages        = {424003},
  year         = {2014},
  month        = {Oct},
  doi          = {10.1088/1751-8113/47/42/424003},
  url          = {https://doi.org/10.1088/1751-8113/47/42/424003},
  publisher    = {IOP Publishing Ltd}
}

@article{chiribella2014dilation, title={Dilation of states and processes in operational-probabilistic theories}, volume={172}, ISSN={2075-2180}, url={http://dx.doi.org/10.4204/EPTCS.172.1}, DOI={10.4204/eptcs.172.1}, journal={Electronic Proceedings in Theoretical Computer Science}, publisher={Open Publishing Association}, author={Chiribella, Giulio}, year={2014}, month=dec, pages={1–14} }

@article{dressel2014colloquium,
  title = {Colloquium: Understanding quantum weak values: Basics and applications},
  author = {Dressel, Justin and Malik, Mehul and Miatto, Filippo M. and Jordan, Andrew N. and Boyd, Robert W.},
  journal = {Rev. Mod. Phys.},
  volume = {86},
  issue = {1},
  pages = {307--316},
  numpages = {10},
  year = {2014},
  month = {Mar},
  publisher = {American Physical Society},
  doi = {10.1103/RevModPhys.86.307},
  url = {https://link.aps.org/doi/10.1103/RevModPhys.86.307}
}

@phdthesis{amaral2014phdthesis,
    title = {The {E}xclusivity {P}rinciple and the {S}et o {Q}uantum {C}orrelations},
    author = {Bárbara Amaral},
    year = {2014},
    school = {Universidade Federal de Minas Gerais},
    type = {{Ph.D. Thesis}},
    eprint = {1502.03235},
    archivePrefix = {arXiv},
    primaryClass = {quant-ph},
    note = {Available at \url{https://arxiv.org/pdf/1502.03235}}
}

@article{mironowicz2014properties,
  title = {Properties of dimension witnesses and their semidefinite programming relaxations},
  author = {Mironowicz, Piotr and Li, Hong-Wei and Paw\l{}owski, Marcin},
  journal = {Phys. Rev. A},
  volume = {90},
  issue = {2},
  pages = {022322},
  numpages = {13},
  year = {2014},
  month = {Aug},
  publisher = {American Physical Society},
  doi = {10.1103/PhysRevA.90.022322},
  url = {https://link.aps.org/doi/10.1103/PhysRevA.90.022322}
}

@article{guhne2014bounding,
  title = {Bounding the quantum dimension with contextuality},
  author = {G\"uhne, Otfried and Budroni, Costantino and Cabello, Ad\'an and Kleinmann, Matthias and Larsson, Jan-\AA{}ke},
  journal = {Phys. Rev. A},
  volume = {89},
  issue = {6},
  pages = {062107},
  numpages = {11},
  year = {2014},
  month = {Jun},
  publisher = {American Physical Society},
  doi = {10.1103/PhysRevA.89.062107},
  url = {https://link.aps.org/doi/10.1103/PhysRevA.89.062107}
}

@article{bowles2014certifying,
  title = {Certifying the Dimension of Classical and Quantum Systems in a Prepare-and-Measure Scenario with Independent Devices},
  author = {Bowles, Joseph and Quintino, Marco T\'ulio and Brunner, Nicolas},
  journal = {Phys. Rev. Lett.},
  volume = {112},
  issue = {14},
  pages = {140407},
  numpages = {5},
  year = {2014},
  month = {Apr},
  publisher = {American Physical Society},
  doi = {10.1103/PhysRevLett.112.140407},
  url = {https://link.aps.org/doi/10.1103/PhysRevLett.112.140407}
}

@article{brunner2014bell,
  title = {Bell nonlocality},
  author = {Brunner, Nicolas and Cavalcanti, Daniel and Pironio, Stefano and Scarani, Valerio and Wehner, Stephanie},
  journal = {Rev. Mod. Phys.},
  volume = {86},
  issue = {2},
  pages = {419--478},
  numpages = {60},
  year = {2014},
  month = {Apr},
  publisher = {American Physical Society},
  doi = {10.1103/RevModPhys.86.419},
  url = {https://link.aps.org/doi/10.1103/RevModPhys.86.419}
}

@article{leifer2014isthe,
  author    = {Leifer, Matthew S.},
  title     = {{Is the Quantum State Real? An Extended Review of $\psi$-ontology Theorems}},
  journal   = {Quanta},
  volume    = {3},
  number    = {1},
  pages     = {67},
  year      = {2014},
  month     = {Nov},
  publisher = {Quanta},
  issn      = {1314-7374},
  doi       = {10.12743/quanta.v3i1.22},
  url       = {http://dx.doi.org/10.12743/quanta.v3i1.22}
}

@article{cabello2014graph,
  title = {{Graph-Theoretic Approach to Quantum Correlations}},
  author = {Cabello, Ad\'an and Severini, Simone and Winter, Andreas},
  journal = {Phys. Rev. Lett.},
  volume = {112},
  issue = {4},
  pages = {040401},
  numpages = {5},
  year = {2014},
  month = {Jan},
  publisher = {American Physical Society},
  doi = {10.1103/PhysRevLett.112.040401},
  url = {https://link.aps.org/doi/10.1103/PhysRevLett.112.040401}
}

@article{pironio2014allCHSHpolytopes,
  title     = {{All Clauser–Horne–Shimony–Holt polytopes}},
  author    = {Pironio, Stefano},
  journal   = {Journal of Physics A: Mathematical and Theoretical},
  volume    = {47},
  number    = {42},
  pages     = {424020},
  year      = {2014},
  month     = {Oct},
  publisher = {IOP Publishing},
  doi       = {10.1088/1751-8113/47/42/424020},
  url       = {http://dx.doi.org/10.1088/1751-8113/47/42/424020},
  issn      = {1751-8121}
}

@article{navascues2014characterization,
  title = {{Characterization of Quantum Correlations with Local Dimension Constraints and Its Device-Independent Applications}},
  author = {Navascu\'es, Miguel and de la Torre, Gonzalo and V\'ertesi, Tam\'as},
  journal = {Phys. Rev. X},
  volume = {4},
  issue = {1},
  pages = {011011},
  numpages = {13},
  year = {2014},
  month = {Jan},
  publisher = {American Physical Society},
  doi = {10.1103/PhysRevX.4.011011},
  url = {https://link.aps.org/doi/10.1103/PhysRevX.4.011011}
}

@article{levi2014quantitative,
  title     = {A Quantitative Theory of Coherent Delocalization},
  author    = {Levi, Federico and Mintert, Florian},
  journal   = {New Journal of Physics},
  volume    = {16},
  number    = {3},
  pages     = {033007},
  year      = {2014},
  month     = {Mar},
  publisher = {IOP Publishing},
  doi       = {10.1088/1367-2630/16/3/033007},
  url       = {http://dx.doi.org/10.1088/1367-2630/16/3/033007},
  issn      = {1367-2630}
}

@article{baumgratz2014quantifying,
  title = {Quantifying {C}oherence},
  author = {Baumgratz, Tillmann and Cramer, Marcus and Plenio, Martin B.},
  journal = {Phys. Rev. Lett.},
  volume = {113},
  issue = {14},
  pages = {140401},
  numpages = {5},
  year = {2014},
  month = {Sep},
  publisher = {American Physical Society},
  doi = {10.1103/PhysRevLett.113.140401},
  URL = {https://journals.aps.org/prl/abstract/10.1103/PhysRevLett.113.140401}
}

@article{lemos2014quantum,
    title = {Quantum imaging with undetected photons},
    author = {Lemos, Gabriela Barreto and Borish, Victoria and Cole, Garrett D. and Ramelow, Sven and Lapkiewicz, Radek and Zeilinger, Anton},
    journal = {Nature},
    volume = {512},
    number = {7515},
    pages = {409--412},
    year = {2014},
    month = {aug},
    publisher = {Springer Science and Business Media LLC},
    doi = {10.1038/nature13586},
    url = {http://dx.doi.org/10.1038/nature13586},
    issn = {1476-4687}
}

@article{lahiri2015theory,
  title = {Theory of quantum imaging with undetected photons},
  author = {Lahiri, Mayukh and Lapkiewicz, Radek and Lemos, Gabriela Barreto and Zeilinger, Anton},
  journal = {Phys. Rev. A},
  volume = {92},
  issue = {1},
  pages = {013832},
  numpages = {8},
  year = {2015},
  month = {Jul},
  publisher = {American Physical Society},
  doi = {10.1103/PhysRevA.92.013832},
  url = {https://link.aps.org/doi/10.1103/PhysRevA.92.013832}
}

@article{terhal2015quantumerrorcorrection,
  title       = {{Quantum error correction for quantum memories}},
  author      = {Terhal, Barbara M.},
  journal     = {Rev. Mod. Phys.},
  volume      = {87},
  issue       = {2},
  pages       = {307--346},
  numpages    = {40},
  year        = {2015},
  month       = {Apr},
  publisher   = {American Physical Society},
  doi         = {10.1103/RevModPhys.87.307},
  url         = {https://link.aps.org/doi/10.1103/RevModPhys.87.307}
}

@article{shchesnovich2015partial,
  title = {Partial indistinguishability theory for multiphoton experiments in multiport devices},
  author = {Shchesnovich, V. S.},
  journal = {Phys. Rev. A},
  volume = {91},
  issue = {1},
  pages = {013844},
  numpages = {16},
  year = {2015},
  month = {Jan},
  publisher = {American Physical Society},
  doi = {10.1103/PhysRevA.91.013844},
  url = {https://link.aps.org/doi/10.1103/PhysRevA.91.013844}
}

@article{navascues2015bounding,
  title = {{Bounding the Set of Finite Dimensional Quantum Correlations}},
  author = {Navascu\'es, Miguel and V\'ertesi, Tam\'as},
  journal = {Phys. Rev. Lett.},
  volume = {115},
  issue = {2},
  pages = {020501},
  numpages = {5},
  year = {2015},
  month = {Jul},
  publisher = {American Physical Society},
  doi = {10.1103/PhysRevLett.115.020501},
  url = {https://link.aps.org/doi/10.1103/PhysRevLett.115.020501}
}

@article{kujala2015necessary,
  title = {Necessary and Sufficient Conditions for an Extended Noncontextuality in a Broad Class of Quantum Mechanical Systems},
  author = {Kujala, Janne V. and Dzhafarov, Ehtibar N. and Larsson, Jan-\AA{}ke},
  journal = {Phys. Rev. Lett.},
  volume = {115},
  issue = {15},
  pages = {150401},
  numpages = {5},
  year = {2015},
  month = {Oct},
  publisher = {American Physical Society},
  doi = {10.1103/PhysRevLett.115.150401},
  url = {https://link.aps.org/doi/10.1103/PhysRevLett.115.150401}
}

@article{navascues2015characterizing,
  title = {Characterizing finite-dimensional quantum behavior},
  author = {Navascu\'es, Miguel and Feix, Adrien and Ara\'ujo, Mateus and V\'ertesi, Tam\'as},
  journal = {Phys. Rev. A},
  volume = {92},
  issue = {4},
  pages = {042117},
  numpages = {15},
  year = {2015},
  month = {Oct},
  publisher = {American Physical Society},
  doi = {10.1103/PhysRevA.92.042117},
  url = {https://link.aps.org/doi/10.1103/PhysRevA.92.042117}
}

@incollection{wootters2015optimal,
	doi = {10.1007/978-94-017-7303-4_2},
	url = {https://doi.org/10.1007/978-94-017-7303-4_2},
	year = 2015,
	month = {Dec},
	publisher = {Springer Netherlands},
	pages = {21--43},
	author = {William K. Wootters},
	title = {Optimal {I}nformation {T}ransfer and {R}eal-{V}ector-{S}pace {Q}uantum {T}heory},
	booktitle = {Fundamental Theories of Physics}
}

@inproceedings{odonnel2015quantumspectrum,
  title={Quantum spectrum testing},
  author={O'Donnell, Ryan and Wright, John},
  booktitle={Proceedings of the forty-seventh annual ACM symposium on Theory of computing},
  pages={529--538},
  doi={10.1145/2746539.2746582},
  year={2015},
  publisher={ACM},
  url = {https://doi.org/10.1145/2746539.2746582}
}

@article{donohue2015identifying,
  title = {Identifying nonconvexity in the sets of limited-dimension quantum correlations},
  author = {Donohue, John Matthew and Wolfe, Elie},
  journal = {Phys. Rev. A},
  volume = {92},
  issue = {6},
  pages = {062120},
  numpages = {11},
  year = {2015},
  month = {Dec},
  publisher = {American Physical Society},
  doi = {10.1103/PhysRevA.92.062120},
  url = {https://link.aps.org/doi/10.1103/PhysRevA.92.062120}
}

@article{yao2015quantum,
  title = {Quantum coherence in multipartite systems},
  author = {Yao, Yao and Xiao, Xing and Ge, Li and Sun, C. P.},
  journal = {Phys. Rev. A},
  volume = {92},
  issue = {2},
  pages = {022112},
  numpages = {7},
  year = {2015},
  month = {Aug},
  publisher = {American Physical Society},
  doi = {10.1103/PhysRevA.92.022112},
  url = {https://link.aps.org/doi/10.1103/PhysRevA.92.022112}
}

@misc{deronde2016unscrambling,
           title = {{Unscrambling the Quantum Omelette of Epistemic and Ontic Contextuality: Classical Contexts and Quantum Reality}},
          author = {Christian de Ronde},
            year = {2016},
             url = {https://philsci-archive.pitt.edu/12198/}
}

@article{chien2016characterization,
  title        = {{A Characterization of Projective Unitary Equivalence of Finite Frames and Applications}},
  author       = {Chien, Tuan-Yow and Waldron, Shayne},
  journal      = {SIAM Journal on Discrete Mathematics},
  volume       = {30},
  number       = {2},
  pages        = {976--994},
  year         = {2016},
  month        = {Jan},
  publisher    = {Society for Industrial \& Applied Mathematics (SIAM)},
  issn         = {1095-7146},
  doi          = {10.1137/15m1042140},
  url          = {http://dx.doi.org/10.1137/15m1042140}
}

@article{clements2016optimal,
    title = {Optimal design for universal multiport interferometers},
    author = {Clements, William R. and Humphreys, Peter C. and Metcalf, Benjamin J. and Kolthammer, W. Steven and Walmsley, Ian A.},
    journal = {Optica},
    volume = {3},
    number = {12},
    pages = {1460},
    year = {2016},
    month = {Dec},
    publisher = {Optica Publishing Group},
    doi = {10.1364/OPTICA.3.001460},
    url = {http://dx.doi.org/10.1364/OPTICA.3.001460}
}

@article{sikora2016minimum,
  title = {{Minimum Dimension of a Hilbert Space Needed to Generate a Quantum Correlation}},
  author = {Sikora, Jamie and Varvitsiotis, Antonios and Wei, Zhaohui},
  journal = {Phys. Rev. Lett.},
  volume = {117},
  issue = {6},
  pages = {060401},
  numpages = {5},
  year = {2016},
  month = {Aug},
  publisher = {American Physical Society},
  doi = {10.1103/PhysRevLett.117.060401},
  url = {https://link.aps.org/doi/10.1103/PhysRevLett.117.060401}
}

@article{sun2016experimental,
  title = {Experimental realization of dimension witnesses based on quantum state discrimination},
  author = {Sun, Yong-Nan and Liu, Zhao-Di and Sun, Jun and Chen, Geng and Xu, Xiao-Ye and Wu, Yu-Chun and Tang, Jian-Shun and Han, Yong-Jian and Li, Chuan-Feng and Guo, Guang-Can},
  journal = {Phys. Rev. A},
  volume = {94},
  issue = {5},
  pages = {052313},
  numpages = {8},
  year = {2016},
  month = {Nov},
  publisher = {American Physical Society},
  doi = {10.1103/PhysRevA.94.052313},
  url = {https://link.aps.org/doi/10.1103/PhysRevA.94.052313}
}

@article{lo2016experimental,
    title = {{Experimental Violation of Bell Inequalities for Multi-Dimensional Systems}},
    author = {Lo, Hsin-Pin and Li, Che-Ming and Yabushita, Atsushi and Chen, Yueh-Nan and Luo, Chih-Wei and Kobayashi, Takayoshi},
    journal = {Scientific Reports},
    volume = {6},
    number = {1},
    year = {2016},
    month = {Feb},
    publisher = {Springer Science and Business Media LLC},
    ISSN = {2045-2322},
    url = {http://dx.doi.org/10.1038/srep22088},
    DOI = {10.1038/srep22088}
}

@phdthesis{kunjwal2016contextualitykochenspeckertheorem,
    title={{Contextuality beyond the Kochen-Specker theorem}}, 
    author={Ravi Kunjwal},
    year={2016},
    school = {Institute of Mathematical Sciences, Chennai},
    type = {{Ph.D. Thesis}},
    eprint={1612.07250},
    archivePrefix={arXiv},
    primaryClass={quant-ph},
    note = {Available at \url{https://arxiv.org/abs/1612.07250}}
}

@article{heinosaari2016invitation,
  author    = {Heinosaari, Teiko and Miyadera, Takayuki and Ziman, Mário},
  title     = {{An Invitation to Quantum Incompatibility}},
  journal   = {Journal of Physics A: Mathematical and Theoretical},
  volume    = {49},
  number    = {12},
  pages     = {123001},
  year      = {2016},
  month     = {Feb},
  issn      = {1751-8121},
  doi       = {10.1088/1751-8113/49/12/123001},
  url       = {http://dx.doi.org/10.1088/1751-8113/49/12/123001},
  publisher = {IOP Publishing}
}

@article{yao2016maximal,
  title = {Maximal coherence in a generic basis},
  author = {Yao, Yao and Dong, G. H. and Ge, Li and Li, Mo and Sun, C. P.},
  journal = {Phys. Rev. A},
  volume = {94},
  issue = {6},
  pages = {062339},
  numpages = {9},
  year = {2016},
  month = {Dec},
  publisher = {American Physical Society},
  doi = {10.1103/PhysRevA.94.062339},
  url = {https://link.aps.org/doi/10.1103/PhysRevA.94.062339}
}

@article{coecke2016mathematical,
  title       = {{A mathematical theory of resources}},
  author      = {Coecke, Bob and Fritz, Tobias and Spekkens, Robert W.},
  journal     = {Information and Computation},
  volume      = {250},
  pages       = {59–86},
  year        = {2016},
  month       = {Oct},
  publisher   = {Elsevier BV},
  doi         = {10.1016/j.ic.2016.02.008},
  url         = {http://dx.doi.org/10.1016/j.ic.2016.02.008},
  ISSN        = {0890-5401}
}

@article{yu2016total,
  title        = {{Total Quantum Coherence and Its Applications}},
  author       = {Yu, Chang-shui and Yang, Si-ren and Guo, Bao-qing},
  journal      = {Quantum Information Processing},
  volume       = {15},
  number       = {9},
  year         = {2016},
  month        = {Jun},
  pages        = {3773–3784},
  publisher    = {Springer Science and Business Media LLC},
  doi          = {10.1007/s11128-016-1376-y},
  url          = {http://dx.doi.org/10.1007/s11128-016-1376-y},
  issn         = {1573-1332}
}

@article{killoran2016converting,
  title = {Converting Nonclassicality into Entanglement},
  author = {Killoran, Nathan and Steinhoff, Frank E. S. and Plenio, Martin B.},
  journal = {Phys. Rev. Lett.},
  volume = {116},
  issue = {8},
  pages = {080402},
  numpages = {6},
  year = {2016},
  month = {Feb},
  publisher = {American Physical Society},
  doi = {10.1103/PhysRevLett.116.080402},
  url = {https://link.aps.org/doi/10.1103/PhysRevLett.116.080402}
}

@phdthesis{wright2016learn,
  title={How to learn a quantum state},
  author={Wright, John},
  year={2016},
  school={Carnegie Mellon University},
  url={http://reports-archive.adm.cs.cmu.edu/anon/2016/abstracts/16-108.html}
}

@inproceedings{odonnell2016efficient,
	doi = {10.1145/2897518.2897544},
	url = {https://doi.org/10.1145/2897518.2897544},
	year = 2016,
	month = {Jun},
	publisher = {{ACM}},
    pages={899--912},
	author = {Ryan O'Donnell and John Wright},
	title = {Efficient quantum tomography},
	booktitle = {Proceedings of the forty-eighth annual {ACM} symposium on Theory of Computing}
}

@article{marvian2016unspeakable,
  title     = {How to quantify coherence: Distinguishing speakable and unspeakable notions},
  author    = {Marvian, Iman and Spekkens, Robert W.},
  journal   = {Phys. Rev. A},
  volume    = {94},
  issue     = {5},
  pages     = {052324},
  year      = {2016},
  month     = {Nov},
  publisher = {American Physical Society},
  doi       = {10.1103/PhysRevA.94.052324},
  url       = {https://link.aps.org/doi/10.1103/PhysRevA.94.052324}
}

@article{yao2016frobenius,
  title        = {{Frobenius-Norm-Based Measures of Quantum Coherence and Asymmetry}},
  author       = {Yao, Yao and Dong, G. H. and Xiao, Xing and Sun, C. P.},
  journal      = {Scientific Reports},
  volume       = {6},
  number       = {1},
  year         = {2016},
  month        = {Aug},
  publisher    = {Springer Science and Business Media LLC},
  doi          = {10.1038/srep32010},
  url          = {http://dx.doi.org/10.1038/srep32010},
  issn         = {2045-2322}
}

@incollection{henk2017basic,
  title={Basic properties of convex polytopes},
  author={Henk, Martin and Richter-Gebert, J{\"u}rgen and Ziegler, G{\"u}nter M},
  booktitle={Handbook of discrete and computational geometry, Chapter 15},
  pages={383--413},
  year={2017},
  publisher={Chapman and Hall/CRC},
  url={https://www.csun.edu/~ctoth/Handbook/chap15.pdf}
}

@misc{ying2025quantumtheoryneedscomplex,
      title={On whether quantum theory needs complex numbers: the foil theories perspective}, 
      author={Yìlè Yīng and Maria Ciudad Alañón and Daniel Centeno and Jacopo Surace and Marina Maciel Ansanelli and Ruizhi Liu and David Schmid and Robert W. Spekkens},
      year={2025},
      eprint={2506.08091},
      archivePrefix={arXiv},
      primaryClass={quant-ph},
      url={https://arxiv.org/abs/2506.08091}, 
}

@article{liu2017resourcedestroying,
  title = {Resource Destroying Maps},
  author = {Liu, Zi-Wen and Hu, Xueyuan and Lloyd, Seth},
  journal = {Phys. Rev. Lett.},
  volume = {118},
  issue = {6},
  pages = {060502},
  numpages = {6},
  year = {2017},
  month = {Feb},
  publisher = {American Physical Society},
  doi = {10.1103/PhysRevLett.118.060502},
  url = {https://link.aps.org/doi/10.1103/PhysRevLett.118.060502}
}

@article{gour2017quantumresource,
  title = {Quantum resource theories in the single-shot regime},
  author = {Gour, Gilad},
  journal = {Phys. Rev. A},
  volume = {95},
  issue = {6},
  pages = {062314},
  numpages = {11},
  year = {2017},
  month = {Jun},
  publisher = {American Physical Society},
  doi = {10.1103/PhysRevA.95.062314},
  url = {https://link.aps.org/doi/10.1103/PhysRevA.95.062314}
}

@misc{wang2017intrinsic,
         title = {Intrinsic basis-independent quantum coherence measure},
         author = {Wei-Chen Wang and Mao-Fa Fang and Min Yu },
         year = {2017},
         eprint = {1701.05110v5},
         archivePrefix = {arXiv},
         primaryClass ={quant-ph},
         howpublished = {arXiv:1701.05110v5 [quant-ph]},
         url={https://arxiv.org/abs/1701.05110}
        }

@article{ramanathan2017tightness,
  title = {Tightness of correlation inequalities with no quantum violation},
  author = {Ramanathan, Ravishankar and Quintino, Marco T\'ulio and Sainz, Ana Bel\'en and Murta, Gl\'aucia and Augusiak, Remigiusz},
  journal = {Phys. Rev. A},
  volume = {95},
  issue = {1},
  pages = {012139},
  numpages = {15},
  year = {2017},
  month = {Jan},
  publisher = {American Physical Society},
  doi = {10.1103/PhysRevA.95.012139},
  url = {https://link.aps.org/doi/10.1103/PhysRevA.95.012139}
}

@article{vanHimbeeck2017semidevice,
  doi = {10.22331/q-2017-11-18-33},
  url = {https://doi.org/10.22331/q-2017-11-18-33},
  title = {Semi-device-independent framework based on natural physical assumptions},
  author = {Van Himbeeck, Thomas and Woodhead, Erik and Cerf, Nicolas J. and Garc{\'{i}}a-Patr{\'{o}}n, Ra{\'{u}}l and Pironio, Stefano},
  journal = {{Quantum}},
  issn = {2521-327X},
  publisher = {{Verein zur F{\"{o}}rderung des Open Access Publizierens in den Quantenwissenschaften}},
  volume = {1},
  pages = {33},
  month = {Nov},
  year = {2017}
}

@article{cong2017witnessing,
  title = {{Witnessing Irreducible Dimension}},
  author = {Cong, Wan and Cai, Yu and Bancal, Jean-Daniel and Scarani, Valerio},
  journal = {Phys. Rev. Lett.},
  volume = {119},
  issue = {8},
  pages = {080401},
  numpages = {5},
  year = {2017},
  month = {Aug},
  publisher = {American Physical Society},
  doi = {10.1103/PhysRevLett.119.080401},
  url = {https://link.aps.org/doi/10.1103/PhysRevLett.119.080401}
}

@article{vicente2017shared,
  title = {Shared randomness and device-independent dimension witnessing},
  author = {de Vicente, Julio I.},
  journal = {Phys. Rev. A},
  volume = {95},
  issue = {1},
  pages = {012340},
  numpages = {11},
  year = {2017},
  month = {Jan},
  publisher = {American Physical Society},
  doi = {10.1103/PhysRevA.95.012340},
  url = {https://link.aps.org/doi/10.1103/PhysRevA.95.012340}
}

@article{wehner2006tsirelson,
  title = {{Tsirelson bounds for generalized Clauser-Horne-Shimony-Holt inequalities}},
  author = {Wehner, Stephanie},
  journal = {Phys. Rev. A},
  volume = {73},
  issue = {2},
  pages = {022110},
  numpages = {5},
  year = {2006},
  month = {Feb},
  publisher = {American Physical Society},
  doi = {10.1103/PhysRevA.73.022110},
  url = {https://link.aps.org/doi/10.1103/PhysRevA.73.022110}
}

@misc{wagner2025structuretheoremcomplexvaluedquasiprobability,
      title={A structure theorem for complex-valued quasiprobability representations of physical theories}, 
      author={Rafael Wagner and Roberto D. Baldijão and Matthias Salzger and Yìlè Yīng and David Schmid and John H. Selby},
      year={2025},
      eprint={2509.10949},
      archivePrefix={arXiv},
      primaryClass={quant-ph},
      url={https://arxiv.org/abs/2509.10949},
      doi={
https://doi.org/10.48550/arXiv.2509.10949
}
}

@article{weilenmann2025partial,
  title = {Partial Independence Suffices to Rule Out Real Quantum Theory Experimentally},
  author = {Weilenmann, Mirjam and Gisin, Nicolas and Sekatski, Pavel},
  journal = {Phys. Rev. Lett.},
  volume = {135},
  issue = {18},
  pages = {180201},
  numpages = {7},
  year = {2025},
  month = {Oct},
  publisher = {American Physical Society},
  doi = {10.1103/3fv7-p8cs},
  url = {https://link.aps.org/doi/10.1103/3fv7-p8cs}
}

@article{faehrmann2025intheshadowofhadamard,
  title = {{In the Shadow of the Hadamard Test: Using the Garbage State for Good and Further Modifications}},
  author = {Faehrmann, Paul K. and Eisert, Jens and Kueng, Richard},
  journal = {Phys. Rev. Lett.},
  volume = {135},
  issue = {15},
  pages = {150603},
  numpages = {8},
  year = {2025},
  month = {Oct},
  publisher = {American Physical Society},
  doi = {10.1103/cqjw-kl8s},
  url = {https://link.aps.org/doi/10.1103/cqjw-kl8s}
}

@misc{li2025multistateimaginaritycoherencequbit,
      title={Multi-state imaginarity and coherence in qubit systems}, 
      author={Mao-Sheng Li and Rafael Wagner and Lin Zhang},
      year={2025},
      eprint={2507.14878},
      archivePrefix={arXiv},
      primaryClass={quant-ph},
      url={https://arxiv.org/abs/2507.14878},
      doi={https://doi.org/10.48550/arXiv.2507.14878}
}

@misc{azado2025measuringunitaryinvariantsquantum,
      title={Measuring unitary invariants with the quantum switch}, 
      author={Pedro C. Azado and Rafael Wagner and Rui Soares Barbosa and Ernesto F. Galvão},
      year={2025},
      eprint={2508.02345},
      archivePrefix={arXiv},
      primaryClass={quant-ph},
      url={https://arxiv.org/abs/2508.02345},
      doi={https://doi.org/10.48550/arXiv.2508.02345}
}

@article{hirsch2017betterlocalhidden,
  doi = {10.22331/q-2017-04-25-3},
  url = {https://doi.org/10.22331/q-2017-04-25-3},
  title = {Better local hidden variable models for two-qubit {W}erner states and an upper bound on the {G}rothendieck constant {$K_G(3)$}},
  author = {Hirsch, Flavien and Quintino, Marco T{\'{u}}lio and V{\'{e}}rtesi, Tam{\'{a}}s and Navascu{\'{e}}s, Miguel and Brunner, Nicolas},
  journal = {{Quantum}},
  issn = {2521-327X},
  publisher = {{Verein zur F{\"{o}}rderung des Open Access Publizierens in den Quantenwissenschaften}},
  volume = {1},
  pages = {3},
  month = {Apr},
  year = {2017}
}

@article{diviansky2017qutrit,
  title = {Qutrit witness from the {G}rothendieck constant of order four},
  author = {Divi\'anszky, P\'eter and Bene, Erika and V\'ertesi, Tam\'as},
  journal = {Phys. Rev. A},
  volume = {96},
  issue = {1},
  pages = {012113},
  numpages = {11},
  year = {2017},
  month = {Jul},
  publisher = {American Physical Society},
  doi = {10.1103/PhysRevA.96.012113},
  url = {https://link.aps.org/doi/10.1103/PhysRevA.96.012113}
}

@misc{cabello2017whattolearn,
            doi = {10.48660/17070034},
            url = {https://pirsa.org/17070034},
            author = {Cabello, Ad\'an},
            language = {en},
            title = {{What do we learn about quantum theory from Kochen-Specker quantum contextuality?}},
            publisher = {Perimeter Institute},
            year = {2017},
            month = {Jul},
            note         = {Accessed: 08/03/2025}
}

@book{landsman2017foundations,
  author    = {Landsman, Klaas},
  title     = {{Foundations of Quantum Theory: From Classical Concepts to Operator Algebras}},
  series    = {Fundamental Theories of Physics},
  volume    = {188},
  year      = {2017},
  publisher = {Springer},
  address   = {Cham, Switzerland},
  isbn      = {978-3-319-51776-6},
  isbn_e    = {978-3-319-51777-3},
  doi       = {10.1007/978-3-319-51777-3},
  url       = {https://library.oapen.org/bitstream/handle/20.500.12657/27966/1/1002033.pdf}
}

@article{menssen2017distinguishability,
      title = {Distinguishability and Many-Particle Interference},
      author = {Menssen, Adrian J. and Jones, Alex E. and Metcalf, Benjamin J. and Tichy, Malte C. and Barz, Stefanie and Kolthammer, W. Steven and Walmsley, Ian A.},
      journal = {Phys. Rev. Lett.},
      volume = {118},
      issue = {15},
      pages = {153603},
      numpages = {6},
      year = {2017},
      month = {Apr},
      publisher = {American Physical Society},
      doi = {10.1103/PhysRevLett.118.153603},
      url = {https://link.aps.org/doi/10.1103/PhysRevLett.118.153603}
}

@article{chin2017generalizedcoherence,
  title       = {{Generalized coherence concurrence and path distinguishability}},
  author      = {Chin, Seungbeom},
  journal     = {Journal of Physics A: Mathematical and Theoretical},
  volume      = {50},
  number      = {47},
  pages       = {475302},
  year        = {2017},
  month       = {Nov},
  publisher   = {IOP Publishing},
  doi         = {10.1088/1751-8121/aa908d},
  url         = {http://dx.doi.org/10.1088/1751-8121/aa908d},
  ISSN        = {1751-8121}
}

@article{hu2017maximal,
  title = {Maximum coherence in the optimal basis},
  author = {Hu, Ming-Liang and Shen, Shu-Qian and Fan, Heng},
  journal = {Phys. Rev. A},
  volume = {96},
  issue = {5},
  pages = {052309},
  numpages = {7},
  year = {2017},
  month = {Nov},
  publisher = {American Physical Society},
  doi = {10.1103/PhysRevA.96.052309},
  url = {https://link.aps.org/doi/10.1103/PhysRevA.96.052309}
}

@article{chin2017coherencerankmixedstates,
  title = {Coherence number as a discrete quantum resource},
  author = {Chin, Seungbeom},
  journal = {Phys. Rev. A},
  volume = {96},
  issue = {4},
  pages = {042336},
  numpages = {12},
  year = {2017},
  month = {Oct},
  publisher = {American Physical Society},
  doi = {10.1103/PhysRevA.96.042336},
  url = {https://link.aps.org/doi/10.1103/PhysRevA.96.042336}
}

@article{theurer2017resource,
  title = {Resource Theory of Superposition},
  author = {Theurer, T. and Killoran, N. and Egloff, D. and Plenio, M. B.},
  journal = {Phys. Rev. Lett.},
  volume = {119},
  issue = {23},
  pages = {230401},
  numpages = {6},
  year = {2017},
  month = {Dec},
  publisher = {American Physical Society},
  doi = {10.1103/PhysRevLett.119.230401},
  url = {https://link.aps.org/doi/10.1103/PhysRevLett.119.230401}
}

@article{streltsov2017colloquium,
  title     = {Colloquium: Quantum coherence as a resource},
  author    = {Streltsov, Alexander and Adesso, Gerardo and Plenio, Martin B.},
  journal   = {Reviews of Modern Physics},
  volume    = {89},
  number    = {4},
  pages     = {041003},
  year      = {2017},
  month     = {Oct},
  publisher = {American Physical Society},
  doi       = {10.1103/RevModPhys.89.041003},
  url       = {https://link.aps.org/doi/10.1103/RevModPhys.89.041003}
}

@book{dariano2017quantum,
  title={Quantum theory from first principles: An informational approach},
  author={D'Ariano, Giacomo Mauro and Chiribella, Giulio and Perinotti, Paolo},
  year={2017},
  publisher={Cambridge University Press},
  doi = {10.1017/9781107338340},
  url = {https://www.cambridge.org/core/books/quantum-theory-from-first-principles/4B583F61C12E168F55FBCC2664ADB750}
}

@article{cardoso2019classical,
  title = {Classical imaging with undetected light},
  author = {Cardoso, A. C. and Berruezo, L. P. and \'Avila, D. F. and Lemos, G. B. and Pimenta, W. M. and Monken, C. H. and Saldanha, P. L. and P\'adua, S.},
  journal = {Phys. Rev. A},
  volume = {97},
  issue = {3},
  pages = {033827},
  numpages = {5},
  year = {2018},
  month = {Mar},
  publisher = {American Physical Society},
  doi = {10.1103/PhysRevA.97.033827},
  url = {https://link.aps.org/doi/10.1103/PhysRevA.97.033827}
}

@book{asimov_trilogia_fundacao1951,
  author    = {Isaac Asimov},
  title     = {Trilogia da Fundação},
  publisher = {Aleph},
  year      = {2019},  
  isbn      = {978-8576571971},
  language  = {Brazilian Portuguese},
  pages     = {600},  
  notes     = {First pubblished in 1951},
  edition   = {Gebundene Ausgabe},
}

@article{carroll2021mereology,
  title = {Quantum mereology: Factorizing Hilbert space into subsystems with quasiclassical dynamics},
  author = {Carroll, Sean M. and Singh, Ashmeet},
  journal = {Phys. Rev. A},
  volume = {103},
  issue = {2},
  pages = {022213},
  numpages = {23},
  year = {2021},
  month = {Feb},
  publisher = {American Physical Society},
  doi = {10.1103/PhysRevA.103.022213},
  url = {https://link.aps.org/doi/10.1103/PhysRevA.103.022213}
}

@article{amaral2018necessary, 
    title={Necessary conditions for extended noncontextuality in general sets of random variables}, 
    volume={59}, 
    ISSN={1089-7658}, 
    url={http://dx.doi.org/10.1063/1.5024885}, 
    DOI={10.1063/1.5024885}, 
    number={7}, 
    journal={Journal of Mathematical Physics}, publisher={AIP Publishing}, 
    author={Amaral, Bárbara and Duarte, Cristhiano and Oliveira, Roberto I.}, 
    year={2018}, 
    month= {Jul},
    pages={072202}
}

@article{goh2018geometry,
  title = {Geometry of the set of quantum correlations},
  author = {Goh, Koon Tong and Kaniewski, Jedrzej and Wolfe, Elie and V\'ertesi, Tam\'as and Wu, Xingyao and Cai, Yu and Liang, Yeong-Cherng and Scarani, Valerio},
  journal = {Phys. Rev. A},
  volume = {97},
  issue = {2},
  pages = {022104},
  numpages = {20},
  year = {2018},
  month = {Feb},
  publisher = {American Physical Society},
  doi = {10.1103/PhysRevA.97.022104},
  url = {https://link.aps.org/doi/10.1103/PhysRevA.97.022104}
}

@article{halpern2018quasiprobability,
	doi = {10.1103/physreva.97.042105},
	url = {https://doi.org/10.1103/physreva.97.042105},
	year = 2018,
	month = {Apr},
	publisher = {American Physical Society ({APS})},
	volume = {97},
	pages={042105},
	number = {4},
	author = {Nicole Yunger Halpern and Brian Swingle and Justin Dressel},
	title = {Quasiprobability behind the out-of-time-ordered correlator},
	fjournal = {Phys. Rev. A},
    journal = {Phys. Rev. A}
}

@book{harary2018graph,
  title={Graph theory (on Demand Printing of 02787)},
  author={Harary, Frank},
  year={2018},
  publisher={CRC Press}
}

@article{aguilar2018certifying,
  title = {{Certifying an Irreducible 1024-Dimensional Photonic State Using Refined Dimension Witnesses}},
  author = {Aguilar, Edgar A. and Farkas, Máté and Mart\'{\i}nez, Daniel and Alvarado, Mat\'{\i}as and Cari\~ne, Jaime and Xavier, Guilherme B. and Barra, Johanna F. and Ca\~nas, Gustavo and Paw\l{}owski, Marcin and Lima, Gustavo},
  journal = {Phys. Rev. Lett.},
  volume = {120},
  issue = {23},
  pages = {230503},
  numpages = {6},
  year = {2018},
  month = {Jun},
  publisher = {American Physical Society},
  doi = {10.1103/PhysRevLett.120.230503},
  url = {https://link.aps.org/doi/10.1103/PhysRevLett.120.230503}
}

@article{bong2018strong,
  title = {{Strong Unitary and Overlap Uncertainty Relations: Theory and Experiment}},
  author = {Bong, Kok-Wei and Tischler, Nora and Patel, Raj B. and Wollmann, Sabine and Pryde, Geoff J. and Hall, Michael J. W.},
  journal = {Phys. Rev. Lett.},
  volume = {120},
  issue = {23},
  pages = {230402},
  numpages = {5},
  year = {2018},
  month = {Jun},
  publisher = {American Physical Society},
  doi = {10.1103/PhysRevLett.120.230402},
  url = {https://link.aps.org/doi/10.1103/PhysRevLett.120.230402}
}

@article{pusey2018simplest,
  title = {Robust preparation noncontextuality inequalities in the simplest scenario},
  author = {Pusey, Matthew F.},
  journal = {Phys. Rev. A},
  volume = {98},
  issue = {2},
  pages = {022112},
  numpages = {8},
  year = {2018},
  month = {Aug},
  publisher = {American Physical Society},
  doi = {10.1103/PhysRevA.98.022112},
  url = {https://link.aps.org/doi/10.1103/PhysRevA.98.022112}
}

@article{schmid2018all,
      title = {All the noncontextuality inequalities for arbitrary prepare-and-measure experiments with respect to any fixed set of operational equivalences},
      author = {Schmid, David and Spekkens, Robert W. and Wolfe, Elie},
      journal = {Phys. Rev. A},
      volume = {97},
      issue = {6},
      pages = {062103},
      numpages = {12},
      year = {2018},
      month = {Jun},
      publisher = {American Physical Society},
      doi = {10.1103/PhysRevA.97.062103},
      url = {https://link.aps.org/doi/10.1103/PhysRevA.97.062103}
}

@article{frauchiger2018quantum,
  author    = {Frauchiger, Daniela and Renner, Renato},
  title     = {{Quantum Theory Cannot Consistently Describe the Use of Itself}},
  journal   = {Nature Communications},
  volume    = {9},
  number    = {1},
  year      = {2018},
  month     = {Sep},
  issn      = {2041-1723},
  pages     = {article no. 3711}, 
  publisher = {Springer Science and Business Media LLC},
  doi       = {10.1038/s41467-018-05739-8},
  url       = {http://dx.doi.org/10.1038/s41467-018-05739-8}
}

@article{amaral2018noncontextual,
  title = {{Noncontextual Wirings}},
  author = {Amaral, Bárbara and Cabello, Ad\'an and Cunha, Marcelo Terra and Aolita, Leandro},
  journal = {Phys. Rev. Lett.},
  volume = {120},
  issue = {13},
  pages = {130403},
  numpages = {6},
  year = {2018},
  month = {Mar},
  publisher = {American Physical Society},
  doi = {10.1103/PhysRevLett.120.130403},
  url = {https://link.aps.org/doi/10.1103/PhysRevLett.120.130403}
}

@article{ringbauer2018certification,
  title = {Certification and Quantification of Multilevel Quantum Coherence},
  author = {Ringbauer, Martin and Bromley, Thomas R. and Cianciaruso, Marco and Lami, Ludovico and Lau, W. Y. Sarah and Adesso, Gerardo and White, Andrew G. and Fedrizzi, Alessandro and Piani, Marco},
  journal = {Phys. Rev. X},
  volume = {8},
  issue = {4},
  pages = {041007},
  numpages = {12},
  year = {2018},
  month = {Oct},
  publisher = {American Physical Society},
  doi = {10.1103/PhysRevX.8.041007},
  url = {https://link.aps.org/doi/10.1103/PhysRevX.8.041007}
}

@article{carollo2018uhlmann,
	doi = {10.1038/s41598-018-27362-9},
	url = {https://doi.org/10.1038/s41598-018-27362-9},
	year = 2018,
	month = {Jun},
	publisher = {Springer Science and Business Media {LLC}},
	volume = {8},
        pages={9852},
	number = {1},
	author = {Angelo Carollo and Bernardo Spagnolo and Davide Valenti},
	title = {Uhlmann curvature in dissipative phase transitions},
	journal = {Scientific Reports}
}

@article{streltsov2018maximal,
  title        = {{Maximal Coherence and the Resource Theory of Purity}},
  author       = {Streltsov, Alexander and Kampermann, Hermann and Wölk, Sabine and Gessner, Manuel and Bruß, Dagmar},
  journal      = {New Journal of Physics},
  volume       = {20},
  number       = {5},
  year         = {2018},
  month        = {May},
  pages        = {053058},
  publisher    = {IOP Publishing},
  doi          = {10.1088/1367-2630/aac484},
  url          = {http://dx.doi.org/10.1088/1367-2630/aac484},
  issn         = {1367-2630}
}

@article{shchesnovich2018collective,
  title = {Collective phases of identical particles interfering on linear multiports},
  author = {Shchesnovich, V. S. and Bezerra, M. E. O.},
  journal = {Phys. Rev. A},
  volume = {98},
  issue = {3},
  pages = {033805},
  numpages = {12},
  year = {2018},
  month = {Sep},
  publisher = {American Physical Society},
  doi = {10.1103/PhysRevA.98.033805},
  url = {https://link.aps.org/doi/10.1103/PhysRevA.98.033805}
}

@article{hickey2018quantifying,
  title        = {{Quantifying the Imaginarity of Quantum Mechanics}},
  author       = {Hickey, Alexander and Gour, Gilad},
  journal      = {Journal of Physics A: Mathematical and Theoretical},
  volume       = {51},
  number       = {41},
  year         = {2018},
  month        = {Sep},
  pages        = {414009},
  publisher    = {IOP Publishing},
  doi          = {10.1088/1751-8121/aabe9c},
  url          = {http://dx.doi.org/10.1088/1751-8121/aabe9c},
  issn         = {1751-8121}
}

@book{amaral2018graph,
  title={{On Graph Approaches to Contextuality and their Role in Quantum Theory}},
  author={Amaral, Bárbara and Cunha, Marcelo Terra},
  series = {SpringerBriefs in Mathematics},
  month = {July},
  year={2018},
  publisher={Springer Cham},
  doi = {10.1007/978-3-319-93827-1},
  URL = {https://link.springer.com/book/10.1007/978-3-319-93827-1}
}

@article{schmid2018discrimination,
  title     = {Contextual advantage for state discrimination},
  author    = {Schmid, David and Spekkens, Robert W.},
  journal   = {Phys. Rev. X},
  volume    = {8},
  number    = {1},
  pages     = {011015},
  year      = {2018},
  month     = {Feb},
  publisher = {American Physical Society},
  doi       = {10.1103/PhysRevX.8.011015},
  url       = {https://link.aps.org/doi/10.1103/PhysRevX.8.011015}
}

@article{amaral2019characterizing,
  title = {Characterizing and quantifying extended contextuality},
  author = {Amaral, Bárbara and Duarte, Cristhiano},
  journal = {Phys. Rev. A},
  volume = {100},
  issue = {6},
  pages = {062103},
  numpages = {12},
  year = {2019},
  month = {Dec},
  publisher = {American Physical Society},
  doi = {10.1103/PhysRevA.100.062103},
  url = {https://link.aps.org/doi/10.1103/PhysRevA.100.062103}
}

@incollection{wigderson2019mathematics,
  title={Mathematics and computation},
  author={Wigderson, Avi},
  booktitle={Mathematics and Computation},
  year={2019},
  publisher={Princeton University Press},
  address={Princeton, NJ},
  url={https://press.princeton.edu/books/hardcover/9780691189130/mathematics-and-computation}
}

@article{gonzalez2019otoc,
  title = {{Out-of-Time-Ordered-Correlator Quasiprobabilities Robustly Witness Scrambling}},
  author = {Gonz\'alez Alonso, Jos\'e Ra\'ul and Yunger Halpern, Nicole and Dressel, Justin},
  journal = {Phys. Rev. Lett.},
  volume = {122},
  issue = {4},
  pages = {040404},
  numpages = {7},
  year = {2019},
  month = {Feb},
  publisher = {American Physical Society},
  doi = {10.1103/PhysRevLett.122.040404},
  url = {https://link.aps.org/doi/10.1103/PhysRevLett.122.040404}
}

@article{zhang2019interactionfree,
    title = {Interaction-free ghost-imaging of structured objects},
    author = {Zhang, Yingwen and Sit, Alicia and Bouchard, Fr\'{e}d\'{e}ric and Larocque, Hugo and Grenapin, Florence and Cohen, Eliahu and Elitzur, Avshalom C. and Harden, James L. and Boyd, Robert W. and Karimi, Ebrahim},
    journal = {Optics Express},
    volume = {27},
    number = {3},
    pages = {2212},
    year = {2019},
    month = {Jan},
    publisher = {Optica Publishing Group},
    doi = {10.1364/OE.27.002212},
    url = {http://dx.doi.org/10.1364/OE.27.002212},
    issn = {1094-4087}
}

@article{shi2019semideviceindependent,
  title = {Semi-device-independent characterization of quantum measurements under a minimum overlap assumption},
  author = {Shi, Weixu and Cai, Yu and Brask, Jonatan Bohr and Zbinden, Hugo and Brunner, Nicolas},
  journal = {Phys. Rev. A},
  volume = {100},
  issue = {4},
  pages = {042108},
  numpages = {9},
  year = {2019},
  month = {Oct},
  publisher = {American Physical Society},
  doi = {10.1103/PhysRevA.100.042108},
  url = {https://link.aps.org/doi/10.1103/PhysRevA.100.042108}
}

@article{pusey2019contextuality,
         title = {Contextuality without access to a tomographically complete set},
         author = {Matthew F. Pusey and Lídia del Rio and Bettina Meyer },
         year = {2019},
         eprint = {1904.08699v1},
         archivePrefix = {arXiv},
         primaryClass ={quant-ph},
         url={https://doi.org/10.48550/arXiv.1904.08699}
        }

@misc{spekkens2019ontological,
         title = {The ontological identity of empirical indiscernibles: Leibniz's methodological principle and its significance in the work of Einstein},
         author = {Robert W. Spekkens },
         year = {2019},
         eprint = {1909.04628v1},
         archivePrefix = {arXiv},
         primaryClass ={physics.hist-ph},
        howpublished={arXiv:1909.04628v1 [physics.hist-ph]},
        ur={https://arxiv.org/abs/1909.04628}
        }

@article{xu2019necessary,
  title = {Necessary and sufficient condition for contextuality from incompatibility},
  author = {Xu, Zhen-Peng and Cabello, Ad\'an},
  journal = {Phys. Rev. A},
  volume = {99},
  issue = {2},
  pages = {020103},
  numpages = {7},
  year = {2019},
  month = {Feb},
  publisher = {American Physical Society},
  doi = {10.1103/PhysRevA.99.020103},
  url = {https://link.aps.org/doi/10.1103/PhysRevA.99.020103}
}

@mastersthesis{wagner2020resourcetheory,
  author       = {Wagner, Rafael},
  title        = {{Resource Theory for Generalized Contextuality}},
  school       = {Instituto de Física, University of São Paulo},
  year         = {2020},
  address      = {São Paulo},
  month        = {Nov},
  doi          = {10.11606/D.43.2020.tde-17112020-134143},
  url          = {https://doi.org/10.11606/D.43.2020.tde-17112020-134143}
}

@article{akhilesh2020geometric,
  title={{Geometric phases for finite-dimensional systems—The roles of Bargmann invariants, null phase curves, and the Schwinger–Majorana SU(2) framework}},
  author={Akhilesh, K. S. and Arvind and Chaturvedi, S. and Mallesh, K. S. and Mukunda, N.},
  journal={Journal of Mathematical Physics},
  volume={61},
  number={7},
  year={2020},
  month= {Jul},
  publisher={AIP Publishing},
  doi={10.1063/1.5124865},
  url={http://dx.doi.org/10.1063/1.5124865},
  issn={1089-7658}
}

@article{tavakoli2020measurement,
  title = {Measurement incompatibility and steering are necessary and sufficient for operational contextuality},
  author = {Tavakoli, Armin and Uola, Roope},
  journal = {Phys. Rev. Res.},
  volume = {2},
  issue = {1},
  pages = {013011},
  numpages = {7},
  year = {2020},
  month = {Jan},
  publisher = {American Physical Society},
  doi = {10.1103/PhysRevResearch.2.013011},
  url = {https://link.aps.org/doi/10.1103/PhysRevResearch.2.013011}
}

@article{jones2020multiparticle,
  title = {{Multiparticle Interference of Pairwise Distinguishable Photons}},
  author = {Jones, Alex E. and Menssen, Adrian J. and Chrzanowski, Helen M. and Wolterink, Tom A. W. and Shchesnovich, Valery S. and Walmsley, Ian A.},
  journal = {Phys. Rev. Lett.},
  volume = {125},
  issue = {12},
  pages = {123603},
  numpages = {6},
  year = {2020},
  month = {Sep},
  publisher = {American Physical Society},
  doi = {10.1103/PhysRevLett.125.123603},
  url = {https://link.aps.org/doi/10.1103/PhysRevLett.125.123603}
}

@article{arvidssonShukur2020quantum,
	doi = {10.1038/s41467-020-17559-w},
	url = {https://doi.org/10.1038/s41467-020-17559-w},
	year = 2020,
	month = {Jul},
	publisher = {Springer Science and Business Media {LLC}},
	volume = {11},
	number = {1},
	pages={3775},
	author = {David R. M. Arvidsson-Shukur and Nicole Yunger Halpern and Hugo V. Lepage and Aleksander A. Lasek and Crispin H. W. Barnes and Seth Lloyd},
	title = {Quantum advantage in postselected metrology},
	fjournal = {Nature Communications},
	journal = {Nat. Comm.}
}

@article{poderini2020criteria,
  title = {Criteria for nonclassicality in the prepare-and-measure scenario},
  author = {Poderini, Davide and Brito, Samura\'{\i} and Nery, Ranieri and Sciarrino, Fabio and Chaves, Rafael},
  journal = {Phys. Rev. Res.},
  volume = {2},
  issue = {4},
  pages = {043106},
  numpages = {13},
  year = {2020},
  month = {Oct},
  publisher = {American Physical Society},
  doi = {10.1103/PhysRevResearch.2.043106},
  url = {https://link.aps.org/doi/10.1103/PhysRevResearch.2.043106}
}

@misc{schmid2021unscrambling,
      title={Unscrambling the omelette of causation and inference: The framework of causal-inferential theories}, 
      author={David Schmid and John H. Selby and Robert W. Spekkens},
      year={2021},
      eprint={2009.03297},
      archivePrefix={arXiv},
      primaryClass={quant-ph},
      url={https://arxiv.org/abs/2009.03297}, 
      doi = {https://doi.org/10.48550/arXiv.2009.03297}
}

@article{brod2021Bosons,
  title        = {{Bosons vs. Fermions – A computational complexity perspective}},
  author       = {Brod, Daniel Jost},
  journal      = {Revista Brasileira de Ensino de Física},
  volume       = {43},
  number       = {suppl 1},
  year         = {2021},
  publisher    = {FapUNIFESP (SciELO)},
  doi          = {10.1590/1806-9126-rbef-2020-0403},
  url          = {http://dx.doi.org/10.1590/1806-9126-RBEF-2020-0403},
  issn         = {1806-1117}
}

@article{miklin2021universalscheme,
  doi = {10.22331/q-2021-04-06-424},
  url = {https://doi.org/10.22331/q-2021-04-06-424},
  title = {A universal scheme for robust self-testing in the prepare-and-measure scenario},
  author = {Miklin, Nikolai and Oszmaniec, Micha{\l{}}},
  journal = {{Quantum}},
  issn = {2521-327X},
  publisher = {{Verein zur F{\"{o}}rderung des Open Access Publizierens in den Quantenwissenschaften}},
  volume = {5},
  pages = {424},
  month = {Apr},
  year = {2021}
}

@article{foulds2021controlledSWAP,
  title={{The controlled SWAP test for determining quantum entanglement}},
  author={Foulds, Steph and Kendon, Viv and Spiller, Tim},
  journal={Quantum Science and Technology},
  volume={6},
  number={3},
  year={2021},
  month={apr},
  pages={035002},
  publisher={IOP Publishing},
  doi={10.1088/2058-9565/abe458},
  url={http://dx.doi.org/10.1088/2058-9565/abe458},
  issn={2058-9565}
}

@article{leifer2020noncontextuality,
  title = {Noncontextuality inequalities from antidistinguishability},
  author = {Leifer, Matthew and Duarte, Cristhiano},
  journal = {Phys. Rev. A},
  volume = {101},
  issue = {6},
  pages = {062113},
  numpages = {11},
  year = {2020},
  month = {Jun},
  publisher = {American Physical Society},
  doi = {10.1103/PhysRevA.101.062113},
  url = {https://link.aps.org/doi/10.1103/PhysRevA.101.062113}
}

@misc{tezzin2020contextualitybydefault,
      title={Contextuality-by-default for behaviours in compatibility scenarios}, 
      author={Alisson Tezzin and Rafael Wagner and Bárbara Amaral},
      year={2020},
      doi={https://doi.org/10.48550/arXiv.2008.02273},
      howpublished={arXiv preprint: 2008.02273 [quant-ph]},
      url={https://arxiv.org/abs/2008.02273}
}

@article{chaturvedi2020quantum,
  doi = {10.22331/q-2020-10-21-345},
  url = {https://doi.org/10.22331/q-2020-10-21-345},
  title = {Quantum prescriptions are more ontologically distinct than they are operationally distinguishable},
  author = {Chaturvedi, Anubhav and Saha, Debashis},
  journal = {{Quantum}},
  issn = {2521-327X},
  publisher = {{Verein zur F{\"{o}}rderung des Open Access Publizierens in den Quantenwissenschaften}},
  volume = {4},
  pages = {345},
  month = {Oct},
  year = {2020}
}

@article{bong2020strong,
  author    = {Bong, Kok-Wei and Utreras-Alarcón, Aníbal and Ghafari, Farzad and Liang, Yeong-Cherng and Tischler, Nora and Cavalcanti, Eric G. and Pryde, Geoff J. and Wiseman, Howard M.},
  title     = {{A Strong No-Go Theorem on the Wigner’s Friend Paradox}},
  journal   = {Nature Physics},
  volume    = {16},
  number    = {12},
  year      = {2020},
  month     = {Aug},
  pages     = {1199–1205},
  issn      = {1745-2481},
  publisher = {Springer Science and Business Media LLC},
  doi       = {10.1038/s41567-020-0990-x},
  url       = {http://dx.doi.org/10.1038/s41567-020-0990-x}
}

@article{wolfe2020quantifyingbell,
  doi = {10.22331/q-2020-06-08-280},
  url = {https://doi.org/10.22331/q-2020-06-08-280},
  title = {Quantifying {B}ell: the {R}esource {T}heory of {N}onclassicality of {C}ommon-{C}ause {B}oxes},
  author = {Wolfe, Elie and Schmid, David and Sainz, Ana Bel{\'{e}}n and Kunjwal, Ravi and Spekkens, Robert W.},
  journal = {{Quantum}},
  issn = {2521-327X},
  publisher = {{Verein zur F{\"{o}}rderung des Open Access Publizierens in den Quantenwissenschaften}},
  volume = {4},
  pages = {280},
  month = {Jun},
  year = {2020}
}

@article{uola2019quantifying,
  title = {{Quantifying Quantum Resources with Conic Programming}},
  author = {Uola, Roope and Kraft, Tristan and Shang, Jiangwei and Yu, Xiao-Dong and G\"uhne, Otfried},
  journal = {Phys. Rev. Lett.},
  volume = {122},
  issue = {13},
  pages = {130404},
  numpages = {6},
  year = {2019},
  month = {Apr},
  publisher = {American Physical Society},
  doi = {10.1103/PhysRevLett.122.130404},
  url = {https://link.aps.org/doi/10.1103/PhysRevLett.122.130404}
}

@article{bharti2019robust,
  title = {{Robust Self-Testing of Quantum Systems via Noncontextuality Inequalities}},
  author = {Bharti, Kishor and Ray, Maharshi and Varvitsiotis, Antonios and Warsi, Naqueeb Ahmad and Cabello, Ad\'an and Kwek, Leong-Chuan},
  journal = {Phys. Rev. Lett.},
  volume = {122},
  issue = {25},
  pages = {250403},
  numpages = {6},
  year = {2019},
  month = {Jun},
  publisher = {American Physical Society},
  doi = {10.1103/PhysRevLett.122.250403},
  url = {https://link.aps.org/doi/10.1103/PhysRevLett.122.250403}
}

@inproceedings{abramsky2019comonadic,
  title       = {{A comonadic view of simulation and quantum resources}},
  author      = {Abramsky, Samson and Barbosa, Rui Soares and Karvonen, Martti and Mansfield, Shane},
  booktitle   = {2019 34th Annual ACM/IEEE Symposium on Logic in Computer Science (LICS)},
  pages       = {1–12},
  year        = {2019},
  month       = {Jun},
  publisher   = {IEEE},
  doi         = {10.1109/lics.2019.8785677},
  url         = {http://dx.doi.org/10.1109/LICS.2019.8785677}
}

@article{amaral2019resource,
  author    = {Amaral, Bárbara},
  title     = {{Resource Theory of Contextuality}},
  journal   = {Philosophical Transactions of the Royal Society A: Mathematical, Physical and Engineering Sciences},
  volume    = {377},
  number    = {2157},
  pages     = {20190010},
  year      = {2019},
  month     = {Sep},
  publisher = {The Royal Society},
  ISSN      = {1471-2962},
  DOI       = {10.1098/rsta.2019.0010},
  url       = {http://dx.doi.org/10.1098/rsta.2019.0010}
}
\printindex
}
\appendix
\renewcommand\chaptername{Appendix}

\part{Appendices}

\chapter{Mathematical preliminaries}\label{app: basic algebra}

\begin{quote}
    ``\emph{P. O senhor pode provar que essa matemática é válida?}\newline 
    \emph{R. Apenas para outro matemático.}'' \\
    (Isaac~\cite{asimov_trilogia_fundacao1951}, Fundação, Livro I)
\end{quote}

\myindent This appendix presents a concise overview of the foundational mathematical results and constructions from operator algebra and matrix analysis that support this thesis. While Part I introduced the conceptual framework and Part II detailed our original contributions, the material collected here and in the following Appendices consists of well-established definitions, theorems, and technical tools that are frequently used but not proved or elaborated upon in the main text.

\myindent For the sake of brevity, we restrict our focus to essential results that were omitted from the main exposition---either because they are not original contributions of this thesis, or because their technical nature lies outside the scope of Parts I and II. This separation is also intended to aid clarity and facilitate the evaluation of the results presented herein. When appropriate, we provide references to standard literature for further details and proofs.

\myindent It should be emphasized that this appendix is not intended to serve as a self-contained \emph{introduction} to the relevant mathematical background. Rather, it functions as a supporting compendium that complements the main text.

\myindent The structure of this Appendix is as follows: Section~\ref{sec: elementary constructions_alg} reviews fundamental mathematical concepts including set operations, relations, functions, and equivalence relations. Section~\ref{sec: Hilbert spaces and algebras of operators} introduces Hilbert spaces, operator algebras, and key results about operator spectra. Section~\ref{sec: convex cone of positive operators} covers convex sets, positive operators, and matrix algebras, including important decompositions like the Cholesky decomposition. Finally, Section~\ref{sec: groups and group actions} discusses group theory concepts with emphasis on unitary groups and their actions.

\section{Elementary constructions}\label{sec: elementary constructions_alg}

\myindent In most parts of this thesis we consider compositions of sets such as Cartesian products, unions, and intersections as usual. All of these concepts are covered in elementary treatments of (naive) set theory. A key distinction for our purposes, emphasized in the main text, is that \textit{tuples} carry both a notion of order and a notion of repetition, whereas \textit{sets} do not. To capture repetition without order, one can introduce the notion of a \emph{multiset}, in which each element is assigned a natural number indicating its multiplicity—i.e., the number of times it appears in the set

\begin{definition}[Multisets]
    Let $X$ be a set. A \emph{multiset} $M$ over $X$ is a function
    \[
        M : X \to \mathbb{N}_0
    \]
    where $\mathbb{N}_0 = \{0, 1, 2, \ldots\}$ denotes the set of non-negative integers. For each $x \in X$, the value $M(x)$ is called the \emph{multiplicity} of $x$ in $M$. The \emph{support} of the multiset is $\mathrm{supp}(M) := \{x \in X \mid M(x) > 0\}$.
\end{definition}

\myindent We will use multisets later to define multigraphs. While they do not explicitly appear in this thesis, since event graphs are simple graphs, we believe that future work developing event graph theory could extend the tools described to consider multigraphs. 

\myindent In~\cref{chapter: event_graph_approach} we made use of a special type of union between two sets called the \emph{disjoint union}. 

\begin{definition}[Disjoint union]
    Let $A$ and $B$ be two sets, their \emph{disjoint union} $A \sqcup B$ is given by the set 
    \begin{equation}
        A \sqcup B := (A \times \{0\}) \cup (B \times \{1\}).
    \end{equation}
\end{definition}
Intuitively, while the union `forgets' the parent sets (e.g. $\{0,1,3\} \cup \{0,1,2\} = \{0,1,2,3\}$) the disjoint union does not (e.g. $\{1\} \sqcup \{2,3\} = \{(1,0),(2,1),(3,1)\}$). 

\myindent Many of the constructions we have considered in this thesis rely on  \emph{equivalence relations}, which partition sets into disjoint classes of `equivalent' elements. For example, in Part I we have encountered the notion of an operational equivalence between procedures (see for instance~\cref{def: preparation operational equivalences}), and in Part II we have used the notion of equivalence relation between deterministic edge-weightings (cf.~\cref{def: equivalence relation in graphs from paths}). Relations are defined as follows.

\begin{definition}[Relation]
    Given two sets $X$ and $Y$ a \emph{relation} $R$ on $X \times Y$ is a subset $R \subseteq X \times Y$. Its \emph{domain} denoted as $\mathrm{dom}(R)$ is the set of all elements $x \in X$ such that $(x,y) \in R$ for some $y \in Y$. Its \emph{image} denoted as $\mathrm{im}(R)$ is the set of all $y \in Y$ such that $(x,y) \in R$ for some $x \in X$. 
\end{definition}

\myindent A relation $R$ on set $X$ is a subset $R \subseteq X \times X$. By definition, for any  relation $R$ on $X \times Y$ we have that $\mathrm{dom}(R) \subseteq X$ while $\mathrm{im}(R) \subseteq Y$. As discussed in the beginning of Part II, if $(x,y) \in R$ we write $xRy$ or use the notation $x \sim y$. Whenever $(x,y) \notin R$ we write $x \not\sim y$ instead. Both notations $R$ and $\sim$ are used interchangeably. Equivalence relations are relations satisfying a few additional axioms.

\begin{definition}[Equivalence relation]
    A relation $R$ on a set $X$ is an \emph{equivalence relation} if it is:
    \begin{enumerate}
        \item \emph{Reflexive}:  $(x,x) \in R$ for all $x \in X$.
        \item \textit{Symmetric}: $(x,y) \in R$ then $(y,x) \in R$. 
        \item \textit{Transitive}: $(x,y),(y,z) \in R$ then $(x,z) \in R$.
    \end{enumerate}
\end{definition}

\myindent Relations on a set $X$ that are reflexive and transitive (but not necessarily symmetric) are called \emph{preorders} or \emph{quasiorders}.
A \emph{partial equivalence relation} is usually taken to be symmetric and transitive (but not necessarily reflexive on the whole set). For example, the relation $\leq$ on $\mathbb{R}$ is reflexive and transitive but not symmetric, hence an example of a preorder that is not an equivalence relation. The relation $=$ on the same set is an example of an equivalence relation. An equivalence relation $R$  provides a partition of the set $X$ into disjoint subsets called \emph{equivalence classes}. 

\begin{definition}[Equivalence class]
    Let $R$ be an equivalence relation on a set $X$. An \emph{equivalence class} $[x]$ associated to an object $x \in X$ is a subset of $X$ defined by all the elements $y \in X$ such that $(x,y) \in R$. In other words, for every $x \in X$ we define its associated equivalence class as
    \begin{equation}
        [x] := \{y \in X \mid (x,y) \in R\}.
    \end{equation}
\end{definition}

\myindent From above, if $\sim$ is an equivalence relation we have that $x \sim y$ iff $[x] = [y]$. Moreover, if $x \not\sim y$ then $[x] \neq [y]$. Any given element of $X$ is in one and only one equivalence class. In fact, $X$ is the disjoint union of all the  equivalence classes $[x]$.  The set of equivalence classes is called the \emph{quotient}.

\begin{definition}[Quotient set]
    Given an equivalence relation $\sim$ on a set $X$ the \emph{quotient set} $$X/\sim \,\,\,\,:= \{[x] \mid x \in X\}$$ is the set of all its equivalence classes. 
\end{definition}

\myindent We have encountered an example of a quotient set in~\cref{def: possibly loopy graph}. In that case, we had that the set of vertices of a graph was given by equivalence classes. An even simpler example is the case for $X= \mathbb{Z}$ the set of integers and the relation $\sim$ given by congruence modulo two. This makes $\mathbb{Z}/\sim \,\,= \{[0],[1]\}$. Not every relation is a \emph{function}. 

\begin{definition}[Functions]
    A \emph{function} $f$ on a set $X \times Y$ is a relation such that $\mathrm{dom}(f) = X$ and $(x,y),(x,y') \in f$ if and only if $y=y'$. In this case, we write the relation as $f:X \to Y$, and the image of the relation $f$ is denoted as $\mathrm{im}(f) = f(X)$.
\end{definition}

\myindent For example, Chapter~\ref{chapter: event_graph_approach} defines a function $\mathfrak{C}$ from the set of all event graphs to $0/1$-polytopes, $G \mapsto \mathfrak{C}(G) \subseteq [0,1]^{|V(G)|}$.  Given two sets $X$ and $Y$ we denote the set of all functions from $X$ to $Y$ as $Y^X$. For example, the set $\mathbb{R}^\Lambda$ denotes the set of all functions $f:\Lambda \to \mathbb{R}$. Suppose that the domain is a finite discrete set $X = \{x_1,x_2,\dots,x_n\}$. In this case, each  function $f: X \to Y$ is uniquely defined by  $\pmb{f} = (f(x_i))_{i=1}^n$ yielding a bijective equivalence between elements in $Y^X$ and those  in $\underbrace{Y \times \dots \times Y}_{n\text{ times}} = Y^n$. 

\myindent Injective functions $f:X \to Y$ are those for which $x = y$ implies that $f(x) = f(y)$ (equivalently, $x \neq y$ implies that $f(x) \neq f(y)$). Surjective ones are those for which $Y=f(X)$. Injective functions are also sometimes referred to as `one-to-one' functions. Moreover, when one says that $f$ is a function from $X$ \emph{onto} $Y$, `onto' is implicitly indicating that the function $f$ is surjective. 

\myindent As another remark on notation, functions $f: X \to Y$ are also often denoted as $X \stackrel{f}{\to}Y$ and its action on the elements of the corresponding sets is usually denoted with another arrow $x \mapsto f(x)$. 

\myindent The cardinality of a set $|X|$ is defined via a notion of a function. We say that $|X| = n$ if there exists a bijective function $g: X \to \{1,\dots,n\} \equiv [n]$. Note that the notation for the set $\{1,2,\dots,n\} \equiv [n]$---which we employ in this thesis---is the \emph{same} as the notation for an equivalence class yet there exists no equivalence relation for which the set described can be an equivalence class, since it would not partition $\mathbb{N}$ into disjoint sets. 

\myindent Note that the definition of a function $f$ on $X \times Y$ forces us to consider it as a relation for which the domain $\mathrm{dom}(f)$ is equal to $X$. If we are interested in understanding the action of $f$ over a subset $A \subseteq X$ we use the notion of \emph{restriction}.  

\begin{definition}[Restriction of a function]
    Given any function $f: X \to Y$ and $A \subseteq X$, the \emph{restriction of $f$ over $A$} is defined by the function $f|_A: A \to Y$ where $f|_A(x)=f(x)$ for every $x \in A$.    
\end{definition}

\myindent If we provide further structure to the sets $X,Y$ the functions often receive different names depending on the context o their use. For examples akin to quantum foundations and quantum information, functions between spaces (vector spaces or normed spaces) are called \emph{operators}, and functions from a space to $\mathbb{R}$ or $\mathbb{C}$ are called \emph{functionals}. We now proceed to describe a few of the relevant structures that we can add to sets that are present in this thesis.

\section{Hilbert spaces and algebras of operators}\label{sec: Hilbert spaces and algebras of operators}

\myindent A \emph{vector space} \( V \) over a field \( \mathbb{K} \) (typically \( \mathbb{R} \) or \( \mathbb{C} \)) is a set equipped with two relations: vector addition \( + : V \times V \to V \), and scalar multiplication \( \cdot : \mathbb{K} \times V \to V \), 
satisfying the usual axioms (associativity and commutativity of addition, existence of additive identity and inverses, compatibility of scalar multiplication with field multiplication, and distributivity of scalar and vector addition). 

\myindent In~\cref{def:kernel_loops} we have used the notion of the \emph{kernel} of a function to define event graph polytopes. The kernel of a function acting on a vector space is defined as follows.

\begin{definition}[Kernel]
    Let $X,Y$ be vector spaces and $0$ be the zero element in $Y$. For any function $f: X \to Y$, the \emph{kernel} of $f$, $\mathrm{Ker}(f) \subseteq X$, is defined by the set  $$\mathrm{Ker}(f) := \{x \in X \mid f(x) = 0\}.$$ 
\end{definition}

\myindent If one consider $f$ to be a function between groups (defined later in this appendix), the kernel is defined with respect to the identity element of that group. In this thesis, we have mostly used the notion of kernel to apply to edge weightings $r: E \to [0,1]$, where our choice of $0$ is precisely that of $0 \in \mathbb{R}$.

\myindent Quantum mechanics is structured around the notion of \emph{normed} vector spaces, which are vector spaces equipped~\footnote{A set $X$ \emph{equipped with} a relation $R$ is sometimes denoted as $(X,R)$ to indicate that it is the \emph{pair} that defines the space. In this sense, vector spaces are often denoted as $(V,+,.)$, algebras as $(\mathcal{A},+,.,\circ)$, and so on. If it is clear from the context, we simply write $V$ instead of $(V,+,.)$ and $\mathcal{A}$ instead of $(\mathcal{A},+,.,\circ)$.} with a norm.

\begin{definition}[Normed space]
A \emph{normed space} is a vector space \( V \) over a field \( \mathbb{K} \) equipped with a function \( \|\cdot\| : V \to \mathbb{R} \), called a \emph{norm}, such that for all \( x, y \in V \) and all \( \lambda \in \mathbb{K} \):
\begin{enumerate}
    \item \( \|x\| \geq 0 \), and \( \|x\| = 0 \) if and only if \( x = 0 \),
    \item \( \|\lambda x\| = |\lambda| \cdot \|x\| \),
    \item \( \|x + y\| \leq \|x\| + \|y\| \).
\end{enumerate}
\end{definition}

\myindent Norms induce a canonical metric given by the distance via $d(\cdot, \cdot) := \Vert \cdot - \cdot \Vert$. A normed vector space is  \emph{complete} if every $\Vert \cdot \Vert$-Cauchy sequence $(v_n)_{n \in \mathbb{N}}$ of elements in the vector space (defined with respect to the induced metric $d$) converges to some $v \in V$. Complete normed spaces are  \emph{Banach spaces}.  The central subclass of Banach spaces for quantum mechanics are the  \emph{Hilbert spaces}, which are Banach spaces with a norm that has an associated inner-product (sometimes called \emph{scalar} product). 

\begin{definition}[Inner-product]
    An inner-product $\langle \cdot, \cdot\rangle$ on a vector space $V$ relative to a field $\mathbb{C}$ is a function $\langle \cdot, \cdot \rangle : V \times V \to \mathbb{C}$ satisfying for all $x,y,z \in V$ and $\alpha,\beta \in \mathbb{C}$:
    \begin{enumerate}
        \item $\langle x,x\rangle \geq 0$, with equality iff $x=0$.
        \item $\langle x,y\rangle = \langle y,x\rangle^*,$
        \item $\langle x,\alpha y + \beta z\rangle = \alpha\langle x,y\rangle + \beta\langle x,z\rangle$.
    \end{enumerate}
\end{definition}

\myindent In this thesis we have used (as usual in quantum information science) the so-called bra-ket notation for inner-products. In this case, we have that $\langle \cdot ,\cdot \rangle \equiv \langle \cdot  | \cdot \rangle. $ 

\myindent Vector spaces equipped with inner-products are also normed spaces since one can define the canonical norm induced by any inner-product $\Vert \cdot \Vert_{\langle \cdot ,\cdot \rangle}: V \to \mathbb{R}$ using $x \mapsto \Vert x \Vert_{\langle \cdot,\cdot \rangle} := \sqrt{\langle x,x\rangle}$.

\begin{theorem}[Jordan and von Neumann]
    Let $V$ be any normed vector space over $\mathbb{C}$. If the norm satisfies the parallelogram identity, i.e., 
    \begin{equation}
        \Vert x-y\Vert^2 + \Vert x+y\Vert^2 = 2\Vert x \Vert ^2 + 2\Vert y\Vert^2,
    \end{equation}
    for every $x,y\in V$ then there is a unique inner-product $\langle \cdot, \cdot \rangle : V \times V \to \mathbb{C}$ such that $\Vert \cdot \Vert = \sqrt{\langle \cdot, \cdot\rangle }$. 
\end{theorem}

\myindent The parallelogram identity implies the existence of an inner product via the polarization identity. In the complex case as above the polarization identity uses complex conjugation. For real vectors spaces, the recovery comes from the polarization identity formula:
\begin{equation}
    \langle x,y\rangle = \frac{1}{4}\sum_{n=1}^2(-1)^n \Vert x+(-1)^n \, y\Vert ^2.
\end{equation}
For complex vector spaces, the recovery comes from the polarization identity formula:
\begin{equation}
    \langle x,y\rangle = \frac{1}{4}\sum_{n=1}^4(-i)^n \Vert x+i^n \, y\Vert ^2.
\end{equation}

\begin{definition}[Hilbert space (as a normed space)]
A \emph{Hilbert space} is a Banach space with a norm that satisfies the parallellogram identity. 
\end{definition}

\myindent Equivalently, due to the Jordan von Neumann theorem, we can define Hilbert spaces using the inner-product directly. 

\begin{definition}[Hilbert space (as an inner-product space)]
A \emph{Hilbert space} is a complete inner product space, i.e., a vector space \( \mathcal{H} \) over \( \mathbb{C} \) or \( \mathbb{R} \) equipped with an inner product \( \langle \cdot, \cdot \rangle \) such that the norm \( \|x\| := \sqrt{\langle x, x \rangle} \) turns \( \mathcal{H} \) into a Banach space.
\end{definition}

\myindent In finite dimensions, every normed vector space is automatically a Banach space, and every inner product induces a norm, making every finite-dimensional inner product space a Hilbert space. 

\myindent Vector states in quantum mechanics $\vert \psi \rangle $ are elements of Hilbert spaces, while most of the constructions we have considered in this thesis---such as density matrices or \acrshort{povm} elements---are elements of spaces of linear functions $A:\mathcal{H} \to \mathcal{H}$. These are called \emph{linear operators}, and the set of all such functions forms a vector space under pointwise addition and scalar multiplication, commonly denoted $\mathcal{L}(\mathcal{H})$.

\myindent If we equip $\mathcal{L}(\mathcal{H})$ with a norm, and consider the subset of all linear operators having a finite norm, we end up with the normed space $\mathcal{B}(\mathcal{H})$ of bounded linear operators. To every $A \in \mathcal{B}(\mathcal{H})$ we can associate a unique operator $A^*$ called the adjoint of $A$.

\begin{definition}[Adjoint]
    Let $A \in \mathcal{B}(\mathcal{H})$ be a bounded linear operator on a Hilbert space $\mathcal{H}$. The \emph{adjoint} of $A$, denoted by $A^*$, is the unique operator in $\mathcal{B}(\mathcal{H})$ such that for all $|\psi\rangle, |\phi\rangle \in \mathcal{H}$,
    \begin{equation}
        \langle \psi | A \phi \rangle = \langle A^* \psi | \phi \rangle.
    \end{equation}
\end{definition}

\myindent If we consider the case of linear operators acting between different Hilbert spaces $A: \mathcal{H}_1 \to \mathcal{H}_2$, whose space we denote as $\mathcal{L}(\mathcal{H}_1,\mathcal{H}_2)$, there are special structure preserving maps (i.e., maps that preserve properties of the space) between Hilbert spaces.

\begin{definition}[Isometries and unitaries between Hilbert spaces]
Let $\mathcal{H}_1$ and $\mathcal{H}_2$ be (real or complex) Hilbert spaces, and let $U : \mathcal{H}_1 \to \mathcal{H}_2 $ in $ \mathcal{L}(\mathcal{H}_1,\mathcal{H}_2)$ be a linear map.

\begin{itemize}
    \item $U$ is called an \emph{isometry} if it preserves inner products, i.e., for all $x, y \in \mathcal{H}_1$,
    \[
    \langle Ux, Uy \rangle = \langle x, y \rangle.
    \]
    Equivalently, $U^* U = \mathrm{id}_{\mathcal{H}_1}$.

    \item $U$ is called \emph{unitary} if it is an isometry and surjective. Equivalently, $U$ is a bijective linear map satisfying
    \[
    U^* U = \mathrm{id}_{\mathcal{H}_1} \quad \text{and} \quad UU^* = \mathrm{id}_{\mathcal{H}_2}.
    \]
\end{itemize}
\end{definition}

\myindent In finite-dimensions, for every given norm the set of linear operators and the set of bounded operators with respect to that norm coincide. Therefore, the distinction in notation $\mathcal{L}(\mathcal{H})$ and $\mathcal{B}(\mathcal{H})$ is merely a distinction where we consider one as a normed space, while the other as merely a vector space. The space $\mathcal{L}(\mathcal{H})$, when equipped with the operator norm $$A \mapsto \Vert A\Vert := \sup_{\vert \psi \rangle \in \mathcal{H},\vert \psi \rangle \neq 0} \frac{\Vert A \vert \psi \rangle \Vert }{\Vert \vert \psi \rangle \Vert},$$ has a subspace $\mathcal{B}(\mathcal{H})$ that is a Banach space (but not a Hilbert space). If we equip $\mathcal{L}(\mathcal{H})$ with the Hilbert--Schmidt norm $$A \mapsto \Vert A \Vert_{\mathrm{HS}} :=  \sqrt{\text{Tr}(A^* A)},$$ the subspace of all elements in $A$ having finite norm in this case is denoted as $\mathrm{HS}(\mathcal H)$---called Hilbert--Schmidt operators---and is a Hilbert space (and therefore a Banach space)
$$\mathrm{HS}(\mathcal H):=\{A\in\mathcal B(\mathcal H)\mid \mathrm{Tr}(A^*A)<\infty\}$$ when equipped with the Hilbert--Schmidt inner product. In the case of finite-dimensional Hilbert spaces every bounded linear operator is also a Hilbert--Schmidt operator, which implies that as sets $\mathcal{B}(\mathcal{H})$ coincides with $\mathrm{HS}(\mathcal H)$.

\myindent The Hilbert--Schmidt norm described above arises from the Hilbert--Schmidt inner-product $\langle A,B\rangle_{\mathrm{HS}} = \text{Tr}(A^*B)$ where $A^*$ is the \emph{adjoint} operator defined via the inner product of $\mathcal{H}$. The adjoint is an example of an involution (see Def.~\ref{def: involution} below).

\myindent The normed space of bounded operators acting on a Hilbert space $\mathcal{B}(\mathcal{H})$ has an additional structure that is not common to Hilbert spaces $\mathcal{H}$ which is that of a \emph{product} between elements, defined by the standard composition of two functions $A\circ B(\vert\psi\rangle ) \equiv AB(\vert\psi\rangle ) = A(B(\vert \psi \rangle))$. In this case, $\mathcal{B}(\mathcal{H})$ is moreover an \emph{algebra}. 

\begin{definition}[Algebra]
    A vector space $\mathcal{A}$ over the field  $\mathbb{C}$ is an \emph{algebra} if is endowed with a binary bilinear operation $\circ: \mathcal{A}\times \mathcal{A} \to \mathcal{A}$. 
\end{definition}

\myindent For any two elements $A,B \in \mathcal{A}$ we denote the new element $\circ (A,B) = A \circ B \equiv AB$ as usually. The algebra is said to be commutative or Abelian if $AB = BA$, for all pairs of elements $A,B$. Similarly, the algebra is said to be associative if $(AB)C = A(BC)$ for all three elements $A, \, B$, and $C$ in the algebra. An algebra $\mathcal{A}$ is \emph{unital} if there exists an element $\mathbb{1}$ (called `unit') such that for all $A$ we have that $A\mathbb{1} = \mathbb{1}A = A$. If it exists, the unit is necessarily unique,~\footnote{Since $\mathbb{1}=\mathbb{1}\mathbb{1}' = \mathbb{1}'$.} and when it does not exist there exists a standard procedure to transform the algebra into a unital one.

\begin{definition}[Involutions]\label{def: involution}
An \emph{involution} on an algebra \( \mathcal{A} \) is a map \( * : \mathcal{A} \to \mathcal{A} \) such that for all \( a, b \in \mathcal{A} \) and \( \lambda \in \mathbb{K} \):
\begin{itemize}
    \item \( (A^*)^* = A \),
    \item \( (A + B)^* = A^* + B^* \),
    \item \( (\lambda A)^* = \overline{\lambda} A^* \),
    \item \( (AB)^* = B^* A^* \).
\end{itemize}
An algebra equipped with such a structure is called a $*$-algebra. 
\end{definition}

\myindent Elements $A \in \mathcal{A}$ for which $A=A^*$ are called selfadjoint. In the bra-ket notation, when considering the algebra $\mathcal{B}(\mathcal{H})$ and the involution defined via the inner product of $\mathcal{H}$, a generic operator acting on a ket is denoted as $A\vert \phi \rangle \equiv \vert A\phi \rangle$, and therefore its associated inner product with another ket as $\langle \phi | A\phi \rangle$. Whenever $A$ is selfadjoint (and only in such a case) we write $\langle \phi|A \psi \rangle = \langle \phi | A | \psi \rangle = \langle A\phi |\psi \rangle$. The normed spaces of bounded operators $\mathcal{B}(\mathcal{H})$ equipped with the operator norm are special types of algebras known as Banach algebras.~\footnote{In fact, these are the basic examples of $C^*$- and $W^*$-algebras, which are the elementary constructions in any formal treatment of algebraic quantum field theory.}

\begin{definition}[Banach algebras and $C^*$ algebras]
A \emph{Banach algebra} is an associative algebra \( \mathcal{A} \) over \( \mathbb{K} \), which is also a Banach space with norm \( \|\cdot\| \) satisfying:
\[
\|AB\| \leq \|A\| \cdot \|B\| \quad \text{for all } A, B \in \mathcal{A}.
\]
A $C^*$ algebra is a Banach $*$-algebra for which $\Vert A^* A\Vert = \Vert A \Vert^2$, for every element $A \in \mathcal{A}$. 
\end{definition}

\myindent Therefore, algebraic treatments of the linear operators acting on Hilbert spaces consider the space of bounded linear operators equipped with the operator norm, because in this case $\mathcal{B}(\mathcal{H})$ is a $C^*$-algebra. If one emphasizes the treatment of $\mathcal{B}(\mathcal{H})$ as a Hilbert space then one equips this space with the Hilbert--Schmidt norm, losing however the property of being a Banach algebra since there are elements $A,B$ for which $\Vert AB \Vert_{\mathrm{HS}} \not\leq \Vert A\Vert_{\mathrm{HS}} \cdot \Vert B \Vert_{\mathrm{HS}}$. 

\myindent A special class of elements from $\mathcal{B}(\mathcal{H})$ are the projectors. We say that $P \in \mathcal{B}(\mathcal{H})$ is a projector if $P=P^*=P^2$. For the study of multivariate traces, the following relation is relevant. 

\begin{proposition}[\cite{rehder1980projections}]
    For two projections $P$ and $Q$ in a complex Hilbert space $\mathcal{H}$, $[P,Q] = 0$ iff $PQP = QPQ$. 
\end{proposition}

\myindent Since states separate generic selfadjoint operators (i.e. two selfadjoint elements $A \neq B$ iff there exists a state $\rho$ such that $\text{Tr}(\rho A) \neq \text{Tr}(\rho B)$) we have that two projectors commute $[P,Q] =0$ iff 
\begin{equation}
    \text{Tr}(\rho PQP) = \text{Tr}(\rho QPQ)
\end{equation}
for every state $\rho \in \mathcal{D}(\mathcal{H})$. This was recently pointed out by~\cite{tezzin2025ontologicalmodelsadequatelyrepresent}.

\myindent If we consider the spectrum $\mathrm{spec}(A)$ of an operator $A \in \mathcal{B}(\mathcal{H})$ of a complex Banach space given by $$\mathrm{spec}(A) :=  \mathbb{C} \setminus \{a \in \mathbb{C} \mid (A-a\,\mathrm{id}_{\mathcal{H}}) \text{ has an inverse in }\mathcal{B}(\mathcal{H})\},$$ where $\mathrm{id}_{\mathcal{H}}: \mathcal{H} \to \mathcal{H}$ is the identity operator $\mathrm{id}_{\mathcal{H}}\vert \psi \rangle = \vert \psi \rangle $ we have that it holds the following theorem.~\footnote{The spectrum of an operator $\mathcal{B}(\mathcal{H})$ is never empty.}

\begin{theorem}
    Let $\mathcal{B}(\mathcal{H})$ be any associative Banach $*$-algebra of operators acting on a Hilbert space $\mathcal{H}$. Then, for all $A \in \mathcal{B}(\mathcal{H})$ the spectrum  of the adjoint is equal to the complex-valued spectrum, i.e. $$\mathrm{spec}(A^*) = \mathrm{spec}(A)^* := \{\lambda^* \mid \lambda \in \mathrm{spec}(A) \}.$$
\end{theorem}

\myindent We remark that the above theorem is valid for any associative $*$-algebra. It shows that $A$ is a selfadjoint operator (in any such algebras) iff its spectrum is entirely contained in the real section of $\mathbb{C}$. In an Banach algebra, there are non-trivial things that one can show about the spectrum of the \emph{product} of operators. 

\begin{lemma}[Jacobson's lemma]\label{lemma: Jacobson's lemma}
    Let $\mathcal{B}$ be a Banach algebra and let $A, B \in \mathcal{B}$ any. Then, denoting $\mathrm{spec}(A)$ as the spectrum of an operator $A$ we have that
    \begin{equation}
        \mathrm{spec}(AB)\setminus \{0\} = \mathrm{spec}(BA) \setminus \{0\}.
    \end{equation}
\end{lemma}

\myindent Note that in an Abelian algebra this statement is trivial, and can even be made stronger since in this case $\mathrm{spec}(AB) = \mathrm{spec}(BA)$.  This lemma is useful in quantum theory when analyzing products of operators, especially in composite systems or Gram matrix formulations. The proof of the above lemma can be found in several references~\citep{barnes1998common,palmer1994banach,takesaki1979theory}. 

\myindent Properties of the spectrum of an operator are connect to norm topological properties.

\begin{theorem}\label{theorem: spectrum radius of an operator}
    Let $\mathcal{H}$ be any (real or complex) Hilbert space, and $\Vert \cdot \Vert$ denote the operator norm. For every selfadjoint element $A \in \mathcal{B}(\mathcal{H})$,
    \begin{equation}
        \Vert A \Vert = \max_{\lambda \in \mathrm{spec}(A)}|\lambda|.
    \end{equation}
\end{theorem}

\myindent We have used this result in~\cref{chapter: applications} for proving one of our main results of that section. Selfadjoint operators, i.e. those satisfying $A=A^*$, have real spectrum but possibly \emph{negative}. Quantum states described as density matrices, as well as  \acrshort{povm} elements, are described as \emph{positive elements} of the algebras $\mathcal{B}(\mathcal{H})$. 

\section{Convex cone of positive operators}\label{sec: convex cone of positive operators}

\myindent We start by reviewing elementary facts from convex theory. 

\begin{definition}[Convex sets]
   Let $V$ be a real vector space. A subset $S\subseteq V$ is convex if for every pair of elements $x,y \in S$ and every real scalar $\alpha \in [0,1]$, their convex combination is also an element of $S$, i.e.  $\alpha x + (1-\alpha)y \in S$. 
\end{definition}

\myindent Examples of convex sets are the empty set and any singleton subset of a vector space. There is a relation between convex sets and convex functions provided by the associated notions of epigraph and hypograph. 

\begin{definition}[Convex and concave functions]
Let \( S \subseteq \mathbb{R}^n \) be a convex set. A function \( f: S \to \mathbb{R} \) is called \emph{convex} if, for all \( x, y \in S \) and all \( \lambda \in [0,1] \), we have
\[
f(\lambda x + (1 - \lambda) y) \leq \lambda f(x) + (1 - \lambda) f(y).
\]
A function $f$ is \emph{concave} if $-f$ is a convex function. 
\end{definition}

\begin{proposition}
    Let $\mathrm{epi}(f):=\{(x,t)\in S \times \mathbb{R} \mid  t \geq f(x)\}$ and $\mathrm{hypo}(f):=\{(x,t)\in S \times \mathbb R \mid  t \leq f(x)\}$ denote, respectively, the epigraph and the hypograph of a function $f: S \to \mathbb{R}$. Suppose that $S$ is a convex set. Then, 
    \begin{enumerate}
    \item $f$ is a convex function iff its epigraph $\mathrm{epi}(f)$ is a convex set.
    \item $f$ is a concave function iff its hypograph $\mathrm{hypo}(f)$ is a convex set.
    \end{enumerate}
\end{proposition}

\myindent Assuming that $S \neq \emptyset$ we can also define the convex hull of $S$, denoted as $\mathrm{ConvHull}(S)$, as the smallest convex subset of $V$ containing $S$, i.e.
   \begin{equation}
       \mathrm{ConvHull}(S) := \bigcap_{\begin{array}{c}C \supseteq S\\C\text{ is convex}\end{array}}C .
   \end{equation} 

\myindent This is a well-defined notion, i.e. the convex hull necessarily yields a convex set since the following proposition holds.

\begin{proposition}
    Let $V$ be a real vector space and $(S_i)_{i \in I} \subseteq V$ be any family of convex subsets. The intersection $\bigcap_{i \in I}S_i$ is also a convex subset of $V$.
\end{proposition}

\myindent The convex hull of any nonempty set $M$ is equal to the following set:

\begin{equation}
    \mathrm{ConvHull}(S) = \left \{ \alpha_1x_1+\dots+\alpha_nx_n \mid x_1,\dots,x_n \in S, \alpha_1,\dots,\alpha_n \geq 0, \sum_{i=1}^n \alpha_i = 1, n \in \mathbb{N} \right\}.
\end{equation}

\myindent In general, it is \emph{not} sufficient to just take all possible convex combinations $\alpha x + (1-\alpha)y$. For $n$-dimensional vector spaces $V$, Caratheodory's convexity theorem shows that it suffices to take all convex combinations of $n+1$ elements in the subset $S \subseteq V$ in order to obtain $\mathrm{ConvHull}(S)$. 

\begin{theorem}[Caratheodory's convexity theorem]
    Let $V$ be any $n$-dimensional real vector space, every vector $x$ in the convex hull of a subset $S \subseteq V$ can be written as a convex combination of at most $n + 1$ vectors $x_1,\dots,x_{n+1}$ in  $S$.
\end{theorem}

\myindent In order to state the Krein--Milman theorem that we used in~\cref{chapter: event_graph_approach} we  introduce the notion of extremal points, which will also be important in~\cref{sec: convex polytopes}.

\begin{definition}[Extreme subsets and points]
    Let $V$ be any real vector space and $S \subseteq V$. We say that $S'$ is an extreme subset of $S$ if for every $x,y \in S$ and $\alpha \in (0,1)$ such that $\alpha x + (1-\alpha)y \in S'$, it implies that $x,y \in S'$. We say that $x$ is an extreme point if $\{x\}$ is an extreme subset of $S$. The set of all extreme points of a set $S$ is denoted as $\mathrm{ext}(S)$.
\end{definition}

\myindent In essence, an extreme point is one that cannot be written as a convex combination of any other point. In principle, generic subsets $S$ of a vector space, it is possible that $\text{ext}(S) = \emptyset$. Let us assume that the vector space in question is a metric space $(V,d)$. These are the situations we consider because we either study $\mathbb{R}^d$---when considering the event graph polytopes---or the normed spaces $\mathcal{H}$ and $\mathcal{B}(\mathcal{H})$. Both are examples of metric spaces, the former with respect to the standard Euclidean metric of $\mathbb{R}^d$, and the letter with respect to the  metric induced by the norm. In this case, we can even further note that there is a notion of `closure' of a set, that is provided by the metric $d$, and we denote it by $\overline{X}$. Moreover, we also have a notion of `compact' sets since the metric induces a topology. For normed spaces with the metric induced by the norm, this topology is called the norm topology. The Krein--Milman theorem guarantees that for all the spaces we have considered in this thesis, compact convex subsets are given by the closure of the convex hull of its extremal points. 

\begin{theorem}[Krein--Milman theorem]
    Let $(V,d)$ be any metric vector space over $\mathbb{R}$ and $K\subseteq V$ any nonempty compact subset. 
    \begin{itemize}
        \item If $(V,d)$ is a locally convex space, $\mathrm{ext}(K) \neq \emptyset$ and $K \subseteq \overline{\mathrm{ConvHull}(\mathrm{ext}(K))}.$
        \item If, moreover,  $K$ is convex, then $K = \overline{\mathrm{ConvHull}(\mathrm{ext}(K))}.$
    \end{itemize}
\end{theorem}

\myindent This theorem shows that states---viewed either as normalized positive linear functionals acting on some algebra, or as is usual in quantum information as density matrices---are given by the convex combination of pure states. Moreover, states are pure iff they are extremal, and we can write $\mathcal{P}(\mathcal{H}_1) = \mathrm{ext}(\mathcal{D}(\mathcal{H}))$ where $\mathcal{P}(\mathcal{H}_1) = \{\vert \psi \rangle \langle \psi \vert \mid \vert \psi \rangle \in \mathcal{H}, \Vert \vert \psi \rangle \Vert = 1\}$. 

\myindent For the case of subsets $S \subseteq \mathbb{R}^d$, where $\mathbb{R}^d$ is viewed as a normed vector space with respect to the usual Euclidean norm it holds the Heine--Borel theorem for compact sets.

\begin{theorem}[Heine--Borel theorem]\label{theorem: HeineBorel}
    Let $S \subseteq \mathbb{R}^d$. Then $S$ is compact iff $S$ is closed and bounded. 
\end{theorem}

\myindent ~\cref{theorem: HeineBorel} is so important that when considering generic metric spaces one says that this space has a `Heine--Borel property' if compact becomes equivalent to closed and bounded. In this regard, finite-dimensional normed spaces are particularly relevant as they always have this property. 

\begin{theorem}[Bolzano--Weierstrass  theorem]\label{theorem: BolzanoWeierstrass}
    Let $V$ be any (real or complex) finite dimensional normed space. Then any subset $S \subseteq V$ is compact iff it is closed and bounded. 
\end{theorem}

\myindent Having described convexity in great depth, we now proceed to describe what is known as a convex pointed cone. The most relevant example for us will be that of positive linear bounded operators acting on Hilbert spaces.

\begin{definition}[Convex pointed cone]
    Let $V$ be a vector space and $S \subseteq V$. We say that $S$ is a convex pointed cone if it satisfies:
    \begin{itemize}
        \item If $x \in S, \alpha \in \mathbb{R}_{\geq 0}$ then $\alpha x \in S$.
        \item If $x,y \in S$ then $x+y \in S$. 
        \item $S \cap -S = \{0\}$.
    \end{itemize}
\end{definition}

\myindent Given any $\mathcal{B}(\mathcal{H})$ the set of positive operators $A \in \mathcal{B}(\mathcal{H})^+$ defined as 
\begin{equation}
    \mathcal{B}(\mathcal{H})^+ = \{A \in \mathcal{B}(\mathcal{H}) \mid \langle \psi| A\psi\rangle \geq 0, \forall \ket \psi \in \mathcal{H}\},
\end{equation}
there are a few elementary crucial aspects of elements in $\mathcal{B}(\mathcal{H})^+$. To start,  $\mathcal{B}(\mathcal{H})^+$ is an example of a (closed) convex pointed cone. Moreover, $A \in \mathcal{B}(\mathcal{H})^+$ iff $\mathrm{spec}(A) \subseteq [0,\infty)$, from which we conclude that the operator norm of positive elements is equal to  $\Vert A \Vert = \max_{\lambda \in \mathrm{spec}(A)} \lambda$. Moreover, there is the following proposition that characterizes positive elements as the product of other operators: 

\begin{proposition}\label{prop: positive elements}
    Let $A \in \mathcal{B}(\mathcal{H})$. The following are assertions are equivalent:
    \begin{enumerate}
        \item $A$ is a positive element, i.e. $\mathrm{spec}(A) \subseteq [0,\infty)$.
        \item $A=B^2$ for some $B \in \mathcal{B}(\mathcal{H})$ selfadjoint.
        \item $A=C^*C$ for some $C \in \mathcal{B}(\mathcal{H})$.
    \end{enumerate}
\end{proposition}

\myindent Pointed convex cones $S \subseteq \mathcal{B}(\mathcal{H})$ induce a partial order on the set of bounded operators acting on a Hilbert space. Given any such cones they induce the relation $$A \leq B :\iff  B-A \in S.$$ It is not difficult to show that this induces a partial order on the set $\mathcal{B}(\mathcal{H})$. In general, whenever we write $A \leq B$, unless specified the choice of convex cone is precisely $\mathcal{B}(\mathcal{H})^+.$ 

\subsection{Matrix algebras}\label{sec: matrix algebras of operators}

\myindent Every finite-dimensional normed space of bounded linear operators acting on a $d$-dimensional Hilbert space $\mathcal{H}$ is isometrically isomorphic (up to a choice of basis) to a normed space of matrices over $\mathbb{C}$, (or $\mathbb{R}$ if the Hilbert space is real). Shortly, $\mathcal{B}(\mathcal{H}) \simeq \mathrm{Mat}_d(\mathbb{C}) \simeq \mathbb{C}^{d \times d}$, where $\matdCC$ is our notation for the set of $d \times d$ complex matrices. Given a choice of orthonormal basis for $\mathcal{H}$ every element $A \in \mathcal{B}(\mathcal{H})$ can be associated to a matrix $(A)_{ij} \in \matdCC$. Norms on $\mathcal{B}(\mathcal{H})$ translate to matrix norms on $\matdCC$. Because of that, in this section we simply focus on the case where $\mathcal{H} = \mathbb{C}^d$ for some $d$ which implies that $\mathcal{B}(\mathbb{C}^d) \simeq \matdCC$. 

\myindent Fixing $\mathbb{C}^d$ as our Hilbert space of interest with respect to the usual inner product $\langle \psi|\phi \rangle = \sum_k \psi_k^*\phi_k$ from textbook quantum information theory, $\matdCC$ is the vector space of $d \times d $ matrices with respect to standard matrix summation and scalar multiplication. The involution in this case becomes Hermitian conjugate $\dagger: \matdCC \to \matdCC$, $A^\dagger := (A^T)^*$. Selfadjoint operators in this case are Hermitian matrices $A = A^\dagger$. If we consider the operator norm as the sup norm,  $\matdCC$ becomes a $C^*$-algebra. In finite dimensions, the spectrum of a matrix coincides with its set of eigenvalues, and a matrix $A$ is Hermitian iff $\mathrm{spec}(A) \subseteq \mathbb{R}$. 

\myindent The convex pointed cone $\mathcal{B}(\mathbb{C}^d)^+$ subset of  $\mathcal{B}(\mathcal{H})$ is dual to the set of all positive semidefinite matrices (\acrshort{psd}), that we denote as $\matdCC^+$. From what we have seen above, we learn that: (i) The operator norm of a \acrshort{psd} matrix is equal to its largest eigenvalue; (ii) $A$ is \acrshort{psd} iff all its eigenvalues are real and nonnegative; (iii) $A$ is \acrshort{psd} iff there exists another Hermitian matrix $B$---called the square root of $A$---such that $A =  B^2$; (iv) $A$ is \acrshort{psd} iff there exists $C$ such that $A=C^\dagger C$. The choices of $B$ and $C$ are not unique and different choices yield different ways of described $A$, which are called \emph{decompositions of $A$}. One important fact that follows from this last remark is the following theorem:

\begin{corollary}[Positive semidefinite matrices are always Gram matrices]
    A matrix $A \in \matdCC^+$ iff there exists some $d$-tuple of vectors $\pmb{v}=(v_i)_{i=1}^d$ such that $A = G_{\pmb{v}}$, i.e., $A$ is a Gram matrix.
\end{corollary}

\begin{proof}
Since \( A \in \matdCC^+ \), there exists a matrix \( C \in \matdCC \) such that \( A = C^\dagger C \). Let \( \pmb{v} = (v_i)_{i=1}^d \) be the tuple of column vectors of \( C \). Then
\[
A_{ij} = (C^\dagger C)_{ij} = \langle v_i| v_j \rangle,
\]
so \( A \) is the Gram matrix of \( \pmb{v} \).
\end{proof}

\myindent One important decomposition is the \emph{Cholesky decomposition}. This decomposition uses the notion of a lower triangular matrix. 

\begin{definition}[Lower triangular matrix]
    A square matrix \( L \in \matdCC\) is called \emph{lower triangular} if all its entries above the main diagonal are zero, that is,
    \[
        (L)_{ij} = 0 \quad \text{for all } i < j.
    \]
    Equivalently, \( L \) has the form
    \[
        L = 
        \begin{pmatrix}
        \ell_{11} & 0         & \cdots & 0 \\
        \ell_{21} & \ell_{22} & \cdots & 0 \\
        \vdots    & \vdots    & \ddots & \vdots \\
        \ell_{d1} & \ell_{d2} & \cdots & \ell_{dd}
        \end{pmatrix}.
    \]
\end{definition}

\begin{theorem}[Cholesky decomposition]\label{theorem: Cholesky decomposition}
    A matrix $A \in \matdCC^+$ iff there exists a lower triangular matrix $L$ with nonnegative diagonal entries such that $A=LL^\dagger$. If $A$ is positive \emph{definite} the matrix $L$ is unique and diagonal entries are strictly positive.
\end{theorem}

Note that \emph{not every} \acrshort{psd} matrix (and hence not every Gram matrix) has a unique Cholesky decomposition.

\myindent There is an important result that relates the operator norm of a Gram matrix of vectors and the operator norm of the sum of rank-1 projectors.

\begin{theorem}
    Let $\mathcal{H}$ be a $d$-dimensional Hilbert space and $\pmb{\Psi} \in \mathcal{H}^n$. Then, we have that 
    \begin{equation}
        \left\Vert \sum_{k=1}^n \vert \psi_k\rangle \langle \psi_k \vert  \right\Vert = \Vert G_{\pmb{\Psi}}\Vert 
    \end{equation}
    where $\Vert \cdot \Vert$ denotes the operator norm. 
\end{theorem}

\begin{proof}
    Let $Y$ be a $d \times n$ matrix whose columns are given by the vectors $\vert \psi_k \rangle$. In terms of its components, $Y$ is  defined via $Y_{ij} := \langle i|\psi_j\rangle$. In this case, we have that $Y^\dagger$ is the $n \times d$ matrix   $$(Y^\dagger)_{ij} = Y_{ji}^* = \langle j \vert \psi_i\rangle^* = \langle \psi_i \vert j \rangle. $$ With that we find
    \begin{equation*}
        (YY^\dagger)_{ij} = \sum_{k=1}^n Y_{ik}(Y^\dagger)_{kj} =\sum_k \langle i\vert \psi_k \rangle \langle \psi_k \vert j \rangle = \left\langle i \left \vert \sum_k \vert \psi_k \rangle \langle \psi_k \vert \right\vert  j \right\rangle  \Rightarrow YY^\dagger = \sum_{k=1}^n \vert \psi_k \rangle \langle \psi_k \vert
    \end{equation*}
    and  
    \begin{equation*}
        (Y^\dagger Y)_{ij} = \sum_{\ell=1}^d (Y^\dagger)_{i\ell}(Y)_{\ell j} = \sum_{\ell=1}^d \langle \psi_i|\ell\rangle \langle \ell | \psi_j\rangle = \langle \psi_i|\psi_j\rangle = (G_{\pmb{\Psi}})_{ij} \implies Y^\dagger Y = G_{\pmb{\Psi}}.
    \end{equation*}
    From Jacobson's~\cref{lemma: Jacobson's lemma}, and from the fact that the operator norm can be defined as an optimization over the spectrum of an operator, we conclude the proof.
\end{proof}

\myindent Note that the result above is valid for vectors in $\mathcal{H}$ and not only the quantum vector states which are those in $\mathcal{H}_1 = \{\vert \psi \rangle \in \mathcal{H} \mid \Vert \vert \psi \rangle \Vert = 1\}$.

\myindent One specific subset of \( d \times d \) complex matrices of relevance to us is the set of \emph{circulant matrices}.

\begin{definition}[Circulant matrix]
    A matrix \( A \in \matdCC \) is said to be \emph{circulant} if there exists a vector \( \pmb{c} = (c_0, c_1, \dots, c_{d-1}) \in \mathbb{C}^d \) such that each row of \( A \) is a cyclic right-shift of the previous one. That is,
    \[
        A = \begin{pmatrix}
        c_0 & c_1 & c_2 & \cdots & c_{d-1} \\
        c_{d-1} & c_0 & c_1 & \cdots & c_{d-2} \\
        c_{d-2} & c_{d-1} & c_0 & \cdots & c_{d-3} \\
        \vdots & \vdots & \vdots & \ddots & \vdots \\
        c_1 & c_2 & c_3 & \cdots & c_0
        \end{pmatrix}.
    \]
    We may also write \( A = \mathrm{circ}(\pmb{c}) \), and refer to \( \pmb{c} \) as the \emph{generating vector} of the circulant matrix.
\end{definition}

\myindent Circulant matrices have been considered in Chapter~\ref{chapter: relational coherence} as they appeared as the relevant types of Gram matrices useful for the characterization of the sets of all $n$-order Bargmann invariants $\mathfrak{B}_n$. There are various intriguing properties that the set of circulant matrices satisfy. For example, the set of \( d \times d \) circulant matrices forms a \emph{commutative subalgebra} of \( \matdCC \). This happens because this set is closed under addition, scalar multiplication, and matrix multiplication, and that every circulant matrix can be simultaneously diagonalized by the discrete Fourier transform $F$. This implies that for any two $d \times d$ circulant matrices $A = \mathrm{circ}(\pmb{c})$ and $B = \mathrm{circ}(\pmb{d})$ we have that $A=F^\dagger D_A F, B = F^\dagger D_B F$, with $D_A,D_B$ two diagonal matrices, implying that  
\begin{equation}
    AB = F^\dagger D_AF F^\dagger D_B F = F^\dagger D_A D_BF = F^\dagger D_B D_AF = F^\dagger D_BF F^\dagger D_AF = BA,
\end{equation}
where we have used that diagonal matrices commute. We state the claim of diagonalization of circulant matrices as a theorem.

\begin{theorem}[Diagonalization of circulant matrices via the discrete Fourier transform]\label{theorem: DFT and circulant matirces}
    Let \( A = \mathrm{circ}(c_0, c_1, \dots, c_{d-1}) = \mathrm{circ}{(\pmb{c})} \) be a circulant matrix. Define the discrete Fourier transform matrix \( F \in \mathbb{C}^{d \times d} \) with entries
    \[
        F_{jk} = \frac{1}{\sqrt{d}} \omega^{jk}, \quad \text{for } j, k \in \{0, 1, \dots, d-1\},
    \]
    where \( \omega = e^{2\pi i / d} \) is a primitive \( d \)-th root of unity.  Then \( A \) is diagonalized by \( F \), i.e.,
    \[
        A = F^\dagger \, \mathrm{diag}(\lambda_0, \lambda_1, \dots, \lambda_{d-1}) \, F,
    \]
    where the eigenvalues \( \lambda_k \) are given by the discrete Fourier transform of the generating vector \( \pmb{c} \):
    \[
        \lambda_k = \sum_{j=0}^{d-1} c_j \, \omega^{jk}, \quad \text{for } k = 0, 1, \dots, d-1.
    \]
\end{theorem}

\myindent From the theorem above we see that the interesting aspect of such matrices is that for any $d$, their eigenvalues and their eigenvectors are easily characterized. 

\begin{theorem}[Adapted from reference~\citep{zhang2011matrix}]\label{theorem: zhangs theorem}
    Let $A, B \in \matdCC^+$. Then,
    \begin{enumerate}
        \item[(i)] $\mathrm{Tr}(AB) \leq \mathrm{Tr}(A)\mathrm{Tr}(B)$.
        \item[(ii)] Denoting $\Vert A \Vert = \alpha, \Vert B \Vert = \beta$ the largest eigenvalues of $A$ and $B,$ respectively, then \begin{equation}-\frac{\alpha \beta}{4} \mathbb{1} \leq AB + BA \leq 2 \alpha \beta \mathbb{1.}\end{equation}
    \end{enumerate}
\end{theorem}

\begin{proof}
    For part (i), using the Cauchy--Schwarz inequality for the Hilbert--Schmidt inner product we end up with 
    \begin{equation}
        \text{Tr}(AB) = \vert \text{Tr}(AB)\vert \leq \Vert A \Vert_{\mathrm{HS}} \Vert B \Vert_{\mathrm{HS}} = \sqrt{\text{Tr}(A^2)}\sqrt{\text{Tr}(B^2)} \leq \text{Tr}(A)\text{Tr}(B).
    \end{equation}
    Above we have used that $A,B$ are \acrshort{psd} (hence Hermitian), that the product of \acrshort{psd} matrices has nonnegative trace, and that for \acrshort{psd} matrices it holds that $0 \leq \sqrt{\text{Tr}(A^2)} \leq \text{Tr}(A).$  For part (ii), the result holds trivially true if $A$ and $B$ are the zero matrix. Therefore, assume that $A,B \neq \mathbb{0} \in \matdCC$. We start by re-scaling $A$ and $B$ to $A'= A/\Vert A \Vert, B'=B/\Vert B \Vert $ so we can prove instead that 
    \begin{equation}
        -\frac{\mathbb 1}{4} \leq A'B'+B'A' \leq 2\mathbb 1.
    \end{equation}
    For the lower bound, we use 
    \begin{align*}
        &\mathbb 0 \leq (A'+B'-\frac{\mathbb 1}{2})^2 = (A'+B')^2-(A'+B')+\frac{\mathbb 1}{4}\\&=A'^2+B'^2+A'B'+B'A'-A'-B'+\frac{\mathbb 1}{4}\\
        &\leq A'B'+B'A'+\frac{\mathbb 1}{4},
    \end{align*}
    where we have used that $\mathbb 0 \leq A',B' \leq \mathbb{1}$, and the fact that this implies that for both $A',B'$ it holds that $(A')^2\leq A'$. For the upper bound, 
    \begin{equation*}
        \mathbb 0 \leq (A'-B')^2 = A'^2+B'^2-A'B'-B'A' \leq 2\mathbb 1-A'B'-B'A'. 
    \end{equation*}
\end{proof}

This result was used by~\cite{allahverdyan2014nonequilibrium} to show that the smallest possible real part of a Bargmann invariant is equal to $-\sfrac{1}{8}$. This can be seen by noticing that 
\begin{equation}
    \text{Re}[\text{Tr}(\psi_1\psi_2\psi_3)] = \text{Tr}\left( \psi_1 \frac{\psi_2\psi_3 + \psi_3\psi_2}{2} \right )
\end{equation}
and since $\psi_2,\psi_3$ are rank-1 projectors, applying~\cref{theorem: zhangs theorem} one has that $$\text{Re}[\text{Tr}(\psi_1\psi_2\psi_3)] \geq -\frac{1}{8} \text{Tr}(\psi_1 \mathbb 1) = -\frac{1}{8}.$$

\myindent One relevant equivalent description of \acrshort{psd} matrices is the so-called \emph{Sylvester's criterion}. In order to state the criterion as a theorem, we need to define the notion of a principal minor of a Hermitean matrix $M$.

\begin{definition}[Principal minors]
    Let $M \in \matdCC$, and let $I = \{i_1, i_2, \dots, i_k\} \subseteq \{1, \dots, d\}$ be an index set with $1 \leq i_1 < i_2 < \cdots < i_k \leq d$. The \emph{principal minor} of $M$ corresponding to the index set $I$ is the determinant of the $k \times k$ submatrix of $M$ formed by selecting the rows and columns indexed by $I$. That is,
    \[
        \det(M[I, I]) := \det\left( (M_{i_p i_q})_{p,q=1}^k \right),
    \]
    where $M[I, I]$ denotes the submatrix of $M$ with rows and columns from the same index set $I$.
\end{definition}

\begin{example}
    Consider the following Hermitian (in this case, symmetric) matrix
    \[
    M = \begin{bmatrix}
        1 & 2 & 3 \\
        2 & 4 & 5 \\
        3 & 5 & 6
    \end{bmatrix}.
    \]
    \begin{enumerate}
        \item The principal minor corresponding to the index set $I = \{1\}$ is just the $(1,1)$ entry of $M$:
        \[
        \det(M[\{1\}, \{1\}]) = \det([1]) = 1.
        \]
        
        \item The principal minor corresponding to the index set $I = \{1, 3\}$ is the determinant of the $2 \times 2$ submatrix using rows and columns $1$ and $3$:
        \[
        M[\{1,3\}, \{1,3\}] = \begin{bmatrix}
            1 & 3 \\
            3 & 6
        \end{bmatrix},
        \]
        where 
        \[
        \det(M[\{1,3\}, \{1,3\}]) = 1\cdot 6 - 3\cdot 3 = -3.
        \]
    \end{enumerate}
    As we will see shortly, Sylvester's theorem will guarantee that $M$ is not \acrshort{psd}, since the determinant above is negative (indeed, it has one eigenvalue which is approximately $-0.5$).
\end{example}

\myindent Notably, checking all principal minors is not efficient, as there $O(2^d)$ principal minors for a $d \times d$ matrix. Nevertheless, one can show the following result:

\begin{theorem}[Sylvester]\label{theorem: Sylvesters criteria}
    A Hermitian matrix $M$ is positive definite iff all \emph{leading} principal minors are strictly positive. Moreover, a Hermitian matrix $M \in \matdCC$ is positive-\emph{semi}definite if and only if \emph{all} principal minors of $M$ are nonnegative.
\end{theorem}

\subsection{Groups and group actions}\label{sec: groups and group actions}

\myindent Another structure relevant to us is that of a \emph{group}. For example, we have considered the group of unitary operations acting on a Hilbert space $U:\mathcal{H} \to \mathcal{H}$ in~\ref{chapter: Bargmann invariants}. A group is defined as follows. 

\begin{definition}[Groups]
A \emph{group} \( G \) is a set equipped with a binary operation \( \circ_G: G \times G \to G \) that satisfies the following axioms:
\begin{enumerate}
    \item \emph{Associativity:} For all \( a, b, c \in G \), we have
    \[
        (a \circ_G b) \circ_G c = a \circ_G (b \circ_G c).
    \]
    \item \emph{Identity:} There exists an element \( e \in G \) such that for all \( g \in G \),
    \[
        g \circ_G e = g = e \circ_G g.
    \]
    \item \emph{Inverses:} For every \( g \in G \), there exists an element \( g^{-1} \in G \) such that
    \[
        g \circ_G g^{-1} = e = g^{-1} \circ_G g.
    \]
\end{enumerate}
\end{definition}

\myindent For simplicity, we write $h\circ_G g \equiv hg$. For the purposes of this thesis, one of the most relevant groups to consider is the \emph{unitary group}, which is the group of unitary maps acting on Hilbert spaces, and with the binary group operation given by composition of maps. 

\begin{example}[Unitary group]
    The \emph{unitary group} \( U(n) \) is the group of complex unitary matrices \( U \in \mathcal{B}(\mathbb{C}^n) \). The group operation is matrix multiplication. The inverse \( U^\dagger \) denotes the conjugate transpose of \( U \), and \( \mathbb 1 \) is the identity matrix, as well as the identity element of the group.

    \myindent A special case is the group \( U(1) \), consisting of all complex numbers of unit modulus:
    \[
        U(1) = \{ z \in \mathbb{C} : |z| = 1 \},
    \]
    with the group operation given by complex multiplication. The identity element is \( 1 \in \mathbb{C} \), and the inverse of \( z \in U(1) \) is its complex conjugate \( \overline{z} \).
\end{example}

\myindent The main operations considered on tuples of bounded operators are described by group actions of the unitary group $U(n)$ on tuples of bounded operators (or on tuples of vector states).  

\begin{definition}[Group actions]\label{def: group action}
    A group action of a group $G$ on a set $X$ is defined by a map $\mathrm{act}:G \times X \to X$ satisfying, for every element $x \in X$:
    \begin{enumerate}
        \item[(a)] $\mathrm{act}(e,x)=x$, where $e$ is the identity element of $G$, and
        \item[(b)] $\mathrm{act}(g,\mathrm{act}(h,x)) = \mathrm{act}(gh,x)$ for all $g,h \in G$.
    \end{enumerate}
\end{definition}

\myindent If we take, for example, the operation $M_{\pmb{\theta}}$ is formally a group action of the group $U(1)^n$ which is the direct $n$-product of the $U(1)$ group, on the set of tuples $\pmb{\Psi} \in \mathcal{H}^n$. For example, since $\pmb{1}=(1,\dots,1)$ is the identity element of $U(1)^n$ we have that $M_{\pmb{0}}(\pmb{\Psi}) = \pmb{\Psi}$ for every $\pmb{\Psi}$. Also, for any $\pmb{\theta},\pmb{\theta}'$ we have that $M_{\pmb{\theta}}(M_{\pmb{\theta}'}(\pmb{\Psi})) = M_{\pmb{\theta}+\pmb{\theta}'}(\pmb{\Psi})$, for every $\pmb{\Psi}$. These two properties show that gauge-transformations are group actions. Similarly, one can see that unitary-transformations and \acrshort{pu}-transformations are also group actions. 

\myindent Every group action of a group $G$ on a set $X$ determines an equivalence relation on $X$. 

\begin{proposition}
    Suppose that $G$ acts on $X$ via the group action act. Let $ \sim_{\mathrm{act}} \in X \times X$ be a relation defined as:
    \begin{equation}
        x \sim_{\mathrm{act}} y \iff \exists g \in G: x=\mathrm{act}(g,y).
    \end{equation}
    Then, $\sim_{\mathrm{act}}$ is an equivalence relation. 
\end{proposition}

\begin{proof}
    Since $\mathrm{act}(e,x)=x$ we see that for every $x \in X$ it holds that $x \sim_{\mathrm{act}}x$. Suppose that x $\sim_{\mathrm{act}} y$. Then,   $x=\mathrm{act}(g,y)$ and we have that $$y=\mathrm{act}(g^{-1}g,y) = \mathrm{act}(g^{-1},\mathrm{act}(g,y)) = \mathrm{act}(g^{-1},x),$$ since $G$ is group and $\mathrm{act}$ is a group action.    
\end{proof}

\myindent Therefore, the relations defined on tuples such as those in Def.~\ref{def: equivalences_of_tuples} can all be viewed as relations defined with respect to certain group actions of the unitary group on sets of tuples. This implies that these are all equivalence relations. 

\chapter{Graph theory}\label{sec: graph theory}

\begin{quote}
    ``\emph{A tournament is a hero if and only if it is a celebrity.}''\\
    (Maria~\cite{chudnovsky2014cliques})
\end{quote}

\myindent Graph theory is a branch of mathematics where terms like \emph{imperfect}, \emph{perfect}, \emph{simple}, \emph{hero}, \emph{tournament}, \emph{celebrity}, and even \emph{galaxy} take on precise meanings. In this appendix, we collect definitions and background from graph theory that support the mathematical framework developed throughout the  thesis. These notions appear primarily in Chapter~\ref{chapter: event_graph_approach}, where the event graph formalism is introduced, and in Chapters~\ref{chapter: Bargmann invariants} and~\ref{chapter: relational coherence}, where we explore frame graphs and their applications to unitary equivalence.

We also touch upon technical aspects relevant to the event graph construction and the graph-theoretic approach to \acrshort{ks} contextuality. The definitions provided here are standard and can be found in many graph theory textbooks; they are included for the reader's convenience and to make the thesis self-contained. For a good introduction to graph theory we refer to the book by~\cite{west2001introduction}. As we have already pointed out in~\ref{chapter: contextuality}, the main introductory reference to the graph theory applied to contextuality is the book by~\cite{amaral2018graph}. 

\myindent The structure of this Appendix is as follows. Section~\ref{sec: elementary constructions} introduces fundamental graph-theoretic concepts including graph types (directed, undirected, simple, and multigraphs), walks and paths, connectivity, and graph isomorphism. Section~\ref{sec: concepts applied to event graphs} focuses on concepts specific to event graphs, including vertex and edge labelings, chromatic number, equality labelings, $\Lambda$-realizability, and event graph gluing operations. Section~\ref{sec: concepts applied to contextuality} covers graph-theoretic tools for analyzing contextuality, including stable sets, independence number, and suspension graphs, with applications to compatibility and exclusivity graphs reviewed in the main text.

\section{Elementary constructions}\label{sec: elementary constructions}

\myindent Graphs are one of the most basic and useful concepts in mathematics. We can even start without a definition, and just by using pencil and paper, drawing dots and lines connecting them. The mathematical theory one constructs by taking this serious is far from elementary, and has profound counter-intuitive consequences~\footnote{Perhaps something similar happens with the theory of \textit{knots}.}. Mathematically, graphs are defined as ordered pairs of two sets.

\begin{definition}[Graphs]\label{def: directed and undirected graphs}
    A \emph{graph} is a mathematical structure consisting of a set of vertices and a set of edges connecting them. Graphs can be:
    \begin{enumerate}
        \item \emph{Directed}: A directed graph is an ordered pair $(V, E)$ where $V$ is a set of vertices, and $E \subseteq V \times V$ is a set of ordered pairs of vertices, called \emph{directed edges}. An edge $(v,w) \in E$ `points' from vertex $v$ to vertex $w$.
        \item \emph{Undirected}: An undirected graph is an ordered pair $(V, E)$ where $V$ is a set of vertices, and $E \subseteq \{\{v,w\} \mid v,w \in V\}$ is a set of unordered pairs of vertices. An edge $\{v,w\} \in E$ connects $v$ and $w$ with no direction.
    \end{enumerate}
\end{definition}

\myindent For a given graph $G$, we may also denote its set of vertices by $V(G)$ and its set of edges by $E(G)$ when we wish to refer to them explicitly. Let $e \in E$ be an edge of an undirected graph, if we have two distinct vertices $v,v' \in e$ we say that $v$ and $v'$ are adjacent (or neighbours). 

\myindent Directed graphs are also called \emph{digraphs} and directed edges are also called \textit{arcs} or \textit{arrows}. In the language of relations introduced before, a directed graph is a pair of a set $V$ and a relation $R$ on $V$. Note that in our definition of a graph, for both directed or undirected graphs, it is allowed for the graph to have \emph{loops}. 

\begin{definition}[Loops]\label{def: loops in a graph}
    A \emph{loop} is an edge that connects a vertex to itself. In a directed graph, a loop is an edge of the form $(v,v)$; in an undirected graph, a loop is $\{v,v\}=\{v\}$.
\end{definition}

\myindent Moreover, one can also have a situation in which many edges are `repeated' in a graph. This is fairly common as some natural operations on graphs may map a graph (or a collection of graphs) to something that doesn't fit the definition we gave for a graph, but a more generic construction called a \emph{multigraph.}~\footnote{Different authors have different names for multigraphs. What we define here as a multigraph is also known as a pseudograph~\citep{harary2018graph}.}

\begin{definition}[Multigraphs]\label{def: multigraph}
    A \emph{multigraph} is an ordered pair $G = (V, E)$ where $V$ is a finite set of vertices and $E$ is a multiset over the set of unordered pairs of vertices, i.e., it is equipped with a function
        \[
            \sharp : \left\{ \{v,w\} \mid v, w \in V \right\} \to \mathbb{N}_0.
        \]
    
    That is, for each pair $\{v,w\}$, the value $\sharp(\{v,w\})$ indicates the number of (parallel) edges between $v$ and $w$. 
\end{definition}

\myindent The class of graphs that was central to this work is that of \textit{simple graphs}.

\begin{definition}[Simple graphs]\label{def: simple graphs}
    A simple graph $G$ is an ordered pair $(V,E)$ of two sets. Elements of $V$ are called \emph{vertices} or \emph{nodes} of the graph, while  $E$ is a set of pairs $e \equiv \{v,w\}$, with $v,w \in V$, and which  $e \in E$ elements are called \emph{edges} of the graph.
\end{definition}

\myindent Event graphs are simple graphs. Graphs described as above are called simple since edges are undirected (i.e., $\{v,w\} = \{w,v\}$), nodes have no loops (i.e., $\{v,v\} \notin E$), and two nodes can pertain to at most \textit{one} edge (i.e. $\forall e_1,e_2 \in E$ such that  $v,w \in e_1 \cap e_2$ with $v \neq w$, then $e_1 = e_2$). A graph is said to be \emph{finite} if the number of vertices is finite, i.e. $|V|<\infty$, and is said to be \emph{complete} if $E = \{\{v,w\}\,|\, \forall v,w \in V\}$, i.e., $E$ has every possible pair of elements from $V$. The complete graphs having $n=\vert V\vert$ nodes are denoted as $K_n$. In the following example we define various other graphs that have been encountered in~\cref{chapter: event_graph_approach}:

\begin{example}[Simple graphs of relevance]
We now list several examples of simple graphs. Let \( G = (V, E) \) be a finite simple graph. We say that \( G \) is a:
\begin{enumerate}
    \item \emph{Complete graph} on \( n \) vertices, written \( G = K_n \), if \( |V| = n \), and every pair of distinct vertices \( v_i, v_j \in V \) is connected by an edge; that is, \( \{v_i, v_j\} \in E \) for all \( i \ne j \).

    \item \emph{Cycle graph} on \( n \geq 3 \) vertices, written \( G = C_n \), if \( V = \{v_1, \dots, v_n\} \) and 
    \[
    E = \big\{ \{v_i, v_{i+1}\} \mid 1 \leq i \leq n \big\}, 
    \]
    where the indices are taken modulo \( n \) (i.e., \( v_{n+1} := v_1 \)).

    \item \emph{Bipartite graph} on \( n + m \) vertices if its vertex set can be partitioned into two disjoint sets \( V_1 \) and \( V_2 \), with \( |V_1| = n \) and \( |V_2| = m \), such that every edge connects a vertex in \( V_1 \) to one in \( V_2 \), and no edge connects vertices within the same part; that is, 
    \[
    E \subseteq \big\{ \{v, w\} \mid v \in V_1,\ w \in V_2 \big\}.
    \]

    \item \emph{Complete bipartite graph} on \( n + m \) vertices, written \( G = K_{n,m} \), if it is a bipartite graph with 
    \[
    E = \big\{ \{v, w\} \mid v \in V_1,\ w \in V_2 \big\}.
    \]

    \item \emph{Wheel graph} on \( n+1 \) vertices, written \( G = W_n \), if \( V = \{v_1, \dots, v_n, \star\} \), and
    \[
    E = E_1 \cup E_2,
    \]
    where
    \[
    E_1 = \big\{ \{v_i, v_{i+1}\} \mid 1 \leq i \leq n \big\}, \quad \text{with } v_{n+1} := v_1,
    \]
    and
    \[
    E_2 = \big\{ \{\star, v_i\} \mid 1 \leq i \leq n \big\}.
    \]
    That is, a cycle of \( n \) vertices plus one central vertex connected to all others.
\end{enumerate}
\end{example}

\myindent It is possible to depict simple graphs as we do in, e.g.,~\cref{fig: 3-cycle frame graph}, where each element from $V$ is depicted as a node (circle) while each element $\{v,w\}$ of $E$ is depicted as a line drawn between nodes $v$ and $w$. From now on, whenever we say `graph' we mean `simple graph' unless stated otherwise. 

\begin{definition}[Walks and paths]\label{def: walks and paths of a graph}
    A \emph{vertex-based walk} in a graph is a finite sequence of vertices $v_0, v_1, \dots, v_n$ such that $\{v_i, v_{i+1}\} \in E$ for all $i$. A \emph{vertex-based path} is a walk in which all vertices are distinct. A \emph{closed path} is a walk that starts and ends at the same vertex, and in which all intermediate vertices are distinct. The \emph{length} of a path is given by the number of edges $\{v_i,v_{i+1}\}$ in $E$.

    \myindent An \emph{edge-based walk} is a finite sequence of edges $e_1, e_2, \dots, e_n \in E$ such that $e_i \cap e_{i+1} \neq \emptyset$ for all $i$. If the sequence contains no repeated edges, it is called an \emph{edge-based path}. The walk is said to go from a vertex $v$ to a vertex $w$ if $v \in e_1$ and $w \in e_n$. An \emph{edge-based closed path} is an edge-based walk with $e_1 \cap e_n \neq \emptyset$ and no repeated edges.
\end{definition}

\myindent These two notions of paths are essentially equivalent for simple graphs. For a given vertex-based walk $v_0,\dots,v_n$ the sequence $\{v_i,v_{i+1}\}$ defines an edge-based walk, and the converse also holds. The distinction is merely on which sequence one wants to focus on: since we have considered edge-weighted graphs, it is more natural to consider the edge-based definition. 

\myindent For any graph $G$, the pair $(v,v)$ is a valid closed path. We call this the \emph{path of length $0$}. Hence, a cycle graph $C_n$ is described by a path of length $n$. To any path $(v_0,\dots,v_n)$ (or $(e_0,\dots,e_n)$) the backwards path is given by the word involution, i.e., $(v_n,\dots,v_0)$ (or $(e_n, \dots, e_0)$). 

\myindent A graph is said to be \textit{connected} if there exists a way to move along any two vertices of $V$ using the edges of the graph.

\begin{definition}[Connected graphs]
    A graph $G = (V, E)$ is said to be \emph{connected} if for every pair of vertices $v, w \in V$, there exists a sequence of vertices (forming a \emph{path}) that begins at $v$ and ends at $w$. In an undirected graph, this means there exists a finite sequence of vertices $(v_0, v_1, \dots, v_k)$ such that $v_0 = v$, $v_k = w$, and $\{v_i, v_{i+1}\} \in E$ for all $i = 0, \dots, k-1$.
\end{definition}

\myindent Graphs that fail to be connected are said to be, naturally, \emph{disconnected}. We let the set of all possible simple connected finite graphs to be denoted as $\mathcal{G}$. These are the most relevant instances of \emph{event graphs}. 

\begin{definition}[Subgraphs]
We say that $G' = (V',E')$ is a subgraph of $G = (V,E)$ if $V' \subseteq V$ and $E' \subseteq E$, and denote the set of all possible subgraphs of $G$ as $\text{sub}(G)$.
\end{definition}

\myindent An important class of subgraphs for us is that of the \textit{spanning tree}. 

\begin{definition}[Tree and spanning tree]
    A \emph{tree} is a connected graph with no cycles. A \emph{spanning tree} of a connected graph $G = (V,E)$ is a subgraph $\tau = (V,E')$, where $E' \subseteq E$ that is a tree.
\end{definition}

\begin{definition}[Graph with a cut vertex]
    A graph $G = (V,E)$ has a \emph{cut vertex} $v \in V$ if the subgraph induced by $V \setminus \{v\}$ is disconnected.
\end{definition}

\myindent Two graphs $G_1,G_2$ may have different sets of vertices (and therefore also different sets of edges) but be equivalent, in the intuitive sense of having the same connections (i.e. the same graphical representation). To make this notion formal, we introduce the notion of a graph isomorphism.

\begin{definition}[Graph isomorphism]
    Let $G_1 = (V_1, E_1)$ and $G_2 = (V_2, E_2)$ be graphs. We say that $G_1$ and $G_2$ are \emph{isomorphic}, and write
    \[
        G_1 \simeq G_2,
    \]
    if there exists a bijective function $f : V_1 \to V_2$ such that for all $v, w \in V_1$,
    \[
        \{v, w\} \in E_1 \iff \{f(v), f(w)\} \in E_2.
    \]
    The function $f$ is called a \emph{graph isomorphism} between $G_1$ and $G_2$.
\end{definition}

\section{Concepts applied to event graphs}\label{sec: concepts applied to event graphs}

\myindent The event graph approach is structured around the idea of bounding specific types of realizations of edge weightings $r: E(G) \to [0,1]$ of a graph $G$. These are a specific kind of edge labelings. 

\begin{definition}[labelings and colouring]\label{def: vertex labeling and colouring}
Let $G$ be an event graph. A \textit{vertex labeling} by a set $\Lambda$,
or a \textit{vertex $\Lambda$-labeling} for short, is a function $\lambda: V(G) \to \Lambda$ assigning to each vertex a label from $\Lambda$. It is called a \textit{colouring} if $\{v,w\} \in E(G)$ implies $\lambda(v) \neq \lambda(w)$. The graph $G$ is said to be $k$-colourable for $k \in \mathbb{N}$ when it admits a colouring by a set of size $k$.
Similarly, an \textit{edge labeling}
by a set $\Lambda$, or an \textit{edge $\Lambda$-labeling} for short, is a function $\alpha:E(G)\to \Lambda$ assigning a label from $\Lambda$ to each edge. When $\Lambda=[0,1]$, we call this an \textit{edge weighting}. When $\Lambda = \{0,1\}$ we call this a deterministic edge weighting or an edge $\{0,1\}$-labeling.
\end{definition}

\begin{definition}[Chromatic number]
    The \textit{chromatic number} of a graph $G$, written $\chi(G)$, is the smallest $k \in \mathbb{N}$ such that $G$ is $k$-colorable. 
\end{definition}

\myindent In~\cref{chapter: event_graph_approach} we consider different notions of realizations in terms of jointly distributed random variables. Here, we can recall that there is a relation between the idea of having the edge weightings to be realizable by joint distributions with the idea of edge and vertex labelings. To see this, we introduce the definition of an \textit{equality labeling}. Our goal here will be to re-structure~\cref{def: lambda realizable edge weightings} using the notion of an \emph{equality labeling}.

\begin{definition}[Equality labeling]
Let $G$ be an event graph. Given any vertex labeling
$\lambda:V(G) \to \Lambda$,
its \textit{equality labeling}
$\epsilon_\lambda$ is
the edge $\{0,1\}$-labeling 
given by:
\begin{align*}
\epsilon_\lambda& \,(\{v,w\}) \,:=\,
\delta_{\lambda(v),\lambda(w)} = \begin{cases} 1 & \text{if $\lambda(v)=\lambda(w)$,} \\ 0 & \text{if $\lambda(v) \neq \lambda(w)$.}\end{cases}
\end{align*}
\end{definition}

\myindent Using this description, we are interested in characterizing the edge $\{0,1\}$-labelings that arise as equality labelings of vertex labelings. These will characterize the vertices of the event graph polytopes. In other words, we simply want to find a formal relationship between edge labelings and vertex labelings, where the latter can be interpreted as a certain type of labeling of random variables taking values on the alphabet $\Lambda$. This leads us to the notion of $\Lambda$ realizability via equality labelings (which is equivalent to the one given in~\cref{def: lambda realizable edge weightings}, now describing formally the primitive of equality labelings).

\begin{definition}[$\Lambda$ realizability]
An edge $\{0,1\}$-labeling $\alpha$ is said to be
\textit{$\Lambda$-realizable}
if it is the equality labeling of some vertex $\Lambda$-labeling,
i.e. if $\alpha = \epsilon_\lambda$ for some $\lambda: V(G) \to \Lambda$.
If $\Lambda$ has size $k \in \mathbb{N}$, we say that $\alpha$ is $k$-realizable.
\end{definition}

\myindent Finally, here we recall the formal definition of \emph{gluing} to event graphs that we have considered in~\cref{chapter: event_graph_approach}. 

\begin{definition}[Gluing]
Given graphs $G_1$ and $G_2$, and tuples of vertices
\begin{align*}
\pmb{v}_1&=(v_1^1, \ldots, v_1^k) \in V(G_1)^k,\\
\pmb{v}_2&=(v_2^1, \ldots, v_2^k) \in V(G_2)^k,
\end{align*}
the \emph{gluing of $G_1$ and $G_2$ along $\pmb{v}_1$ and $\pmb{v}_2$}, written $G_1 +_{\pmb{v}_1=\pmb{v}_2} G_2$, is the graph obtained by taking the disjoint union $G_1+G_2$ and identifying the vertices $v_1^j$ and $v_2^j$ for $j = 1,\ldots, k$. Explicitly: its vertices are
\[
    V(G_1 +_{\pmb{v}_1=\pmb{v}_2} G_2) \;:=\; O_1 \sqcup O_2 \sqcup N ,
\]   
where  $O_i :=  V(G_i) \setminus \{v_i^1, \ldots, v_i^k\}$  is the set of vertices of $G_i$ not being identified and
$N = \{v^1, \ldots v^k\}$ is a set of `new' vertices (i.e. not in either $G_i$); its edges are
\[
E(G_1 +_{\pmb{v}_1=\pmb{v}_2} G_2) \;:=\; E_1 \cup E_2 ,
\]
where $E_i$ is equal to $E(G_i)$ but with occurrences of $v_i^j$ replaced by the new $v^j$.
\end{definition}

\section{Concepts applied to \acrshort{ks} contextuality}\label{sec: concepts applied to contextuality}

\myindent In contextuality theory, compatibility and exclusivity structures are encoded using graphs. The translation between these structures must be treated with care, particularly when constructing the correct contextuality scenario from a given compatibility graph. This section summarizes important definitions relevant to the study of graph-theoretic approaches to \acrshort{ks} contextuality.

\myindent We start by defining the set of all stable sets $\mathcal{S}(H)$ of a graph $H$. 

\begin{definition}\label{def: stable set}
Let $H$ be a graph. A subset $S \subseteq V(H)$ of vertices is called a \emph{stable set} if no two vertices of $S$ are adjacent in $H$, i.e. for all $v, w \in S$, $\{v,w\} \not\in E(H)$. Write $\mathcal{S}(H)$ for the set of stable sets of $H$.
\end{definition}

\myindent The size of the largest stable set is called the independence number. 

\begin{definition}[Independence number]\label{def_app:independence_number}
    Let $\mathcal{S}(G)$ denote the set of all stable sets $S$ of the graph $G$. The \emph{independence number} $\alpha(G)$ of a graph $G = (V,E)$ is the size of the largest stable set $S \in \mathcal{S}(G)$ of $G$, i.e.
    \begin{equation}
        \alpha(G) = \max_{S \in \mathcal{S}(G)}\vert S \vert.
    \end{equation}
    The \emph{weighted} independence number of a vertex-weighted graph $(V,\vartheta)$, with vertex-weighting $\vartheta: V(G) \to \mathbb{R}_{\geq 0}$ is 
    \begin{equation}
        \alpha(G,\vartheta) = \max_{S \in \mathcal{S}(G)}\sum_{v\in S}\vartheta(v).
    \end{equation}
\end{definition}

\myindent The independence number of a graph is used to bound noncontextual realizability~\citep{cabello2014graph}. In~\ref{chapter: from overlaps to noncontextuality} we have made use of the following graph mapping referred to as the \emph{suspension}.

\begin{definition}[Suspension graph]
    The \emph{suspension} $\nabla G$ of a graph $G = (V,E)$ is the graph obtained by adding a new vertex $v_0$ to $G$, and connecting it to every vertex in $V$, i.e., $\nabla G = (V \cup \{v_0\}, E \cup \{\{v_0,v\} \mid v \in V\})$.
\end{definition}

\myindent For instance, the wheel graphs $W_n$ are always the suspension graph of a $n$-cycle graph $C_n$, i.e., $W_n = \nabla C_n$. The added vertex to this graph is called the \emph{handle}. 
\chapter{Convex polytopes}\label{sec: convex polytopes}

\begin{quote}
    ``\emph{Before the beginning of this century, three events can be picked out as being of
 the utmost importance for the theory of convex polytopes. The first was the publication of Euclid's Elements which, as Sir D'Arcy Thompson once remarked, was
 intended as a treatise on the five regular (Platonic) 3-polytopes, and not as an introduction to elementary geometry. The second was the discovery in the eighteenth
 century  of the celebrated Euler's Theorem connecting the numbers of
 vertices, edges and polygonal faces of a convex polytope in $E^3$. Not only is this a
 result of great generality, but it initiated the combinatorial theory of polytopes.
 The third event occurred about a century later with the discovery of polytopes in
 $d \geq 4 $ dimensions. This has been attributed to the Swiss mathematician Ludwig
 Schlafli; it happened at a time when very few mathematicians (Cayley, Grassmann,
 Mobius) realised that geometry in more than three dimensions was possible.}''\newline 
 ~\citep{grunbaum1969convex}
\end{quote}

\myindent A complementary statement to that of~\cite{grunbaum1969convex}  might be that the application of convex polytope theory in quantum foundations - and subsequently in quantum information and computation - could be considered a significant new development of great importance to the field of convex polytopes.  Convex polytopes are a fundamental construction to this thesis, and we have already provided strong arguments in favor of their relevance to quantum information in~\cref{chapter: contextuality} and in~\cref{chapter: event_graph_approach}. In this appendix, we present the basics of convex polytope theory which can be found in standard books on the topic~\citep{Ziegler2012,brondsted2012introduction,matousek2013lectures}.  

\myindent The structure of this Appendix is as follows: Section~\ref{sec: V representation} introduces V-representations of polytopes through convex hulls of finite sets of points and discusses extremal points and minimal representations. Section~\ref{sec: H representation} presents the dual H-representation via intersections of half-spaces defined by linear inequalities. Section~\ref{sec: Main theorem for polytopes} establishes the fundamental duality between these representations through the Minkowski-Weyl theorem. Section~\ref{sec: Properties of polytopes} examines key properties including compactness, different types of inequalities (valid, tight, face and facet-defining), and the structure of faces. Section~\ref{sec: relevant classes of polytopes} covers important classes of polytopes including simplexes, simplicial polytopes, cyclic polytopes, and 0/1-polytopes. Finally, Section~\ref{sec: comparing polytopes} discusses methods for comparing polytopes through subpolytopes, projections, isomorphisms, and various composition operations.

\section{V-representation}\label{sec: V representation}

\myindent The elementary mathematical structures for investigating convex polytopes are convex and affine sets, as well as linear mappings between such sets~\citep{mordukhovich2022convex}. A \emph{convex polyhedron} is a set in $\mathbb{R}^d$ given by a finite intersection of (bounded or unbounded) closed half-spaces~\citep{alexandrov2005convex}. A convex polytope is a \emph{bounded} convex polyhedron. We say that a subset $A$ of any vector space is convex (affine) if it is closed under convex (affine) combinations: For any $\{a_i\}_{i=1}^m \subseteq A$ we have that $\sum_i \lambda_i a_i \in A$ is an affine combination of points if $\sum_i \lambda_i = 1$ and $\{\lambda_i\}_i \in \mathbb{R}$. A convex combination further requires $\lambda_i\ge 0$ for all $i$. 

\myindent Convex polytopes $\mathfrak{P} \subseteq \mathbb{R}^d$ are bounded convex sets that can be equivalently defined in a number of ways. Two fundamental ways of defining $\mathfrak{P}$ are: 1) as the convex hull of a set of points in $\mathbb{R}^d$, 2) as the intersection of a finite set of half-spaces, that will be defined shortly. These two descriptions of a convex polytope $\mathfrak{P}$ are \emph{dual} and define what are known as the V- and H-representations of $\mathfrak{P}$, respectively. Intuitively, convex polytopes are the generalization of a convex polygon to higher dimensions.

\myindent Any finite set of points $K = \{x_1,\dots,x_n\}\in\mathbb{R}^d$, defines a convex polytope $\mathfrak{P}$ via the convex hull
\begin{equation*}
 \mathrm{ConvHull}(K) = \left\{ \sum_{j=1}^n \lambda_jx_j\ :\ \lambda_j\geq 0,\ \sum_j \lambda_j = 1\right\},
\end{equation*}
where $\mathfrak{P} := \mathrm{ConvHull}(K)$.  Each such set $K$ defines what is called as a \emph{V-representation of $\mathfrak{P}$}. As an example, if we recall~\cref{chapter: event_graph_approach}, for the case of the event graph polytopes $\mathfrak{C}(G)$ associated to simple graphs $G$ we have $\mathfrak{C}(G) = \mathrm{ConvHull}(\mathcal{V}_G)$ with $\mathcal{V}_G$ described in~\cref{def: event graph polytope}.  This description of convex polytopes (as described) is not unique, as two finite sets of points $K \neq K'$ can have the \emph{same} convex hull. 

\myindent If we define $\text{ext}(\mathfrak{P})$ as the set of points that are not convex combinations of other points, from the fact that $\mathfrak{P}$ is a convex compact set we can use the Krein-Milman theorem (which for $\mathbb{R}^d$ was first shown by~\cite{minkowski1911convexhull} and~\cite{steinitz1916bedingt}) to conclude that  
\begin{equation}\label{eq: minimal V representation}
\mathfrak{P} := \text{ConvHull}(\text{ext}(\mathfrak{P})).
\end{equation}
In words, any convex polytope can be fully characterized by the set of extremal points, which provides a minimal representation of $\mathfrak{P}$ that is called \emph{the minimal} V-representation of $\mathfrak{P}$. The minimal V-representation is unique. The points $\text{ext}(\mathfrak{P})$ are called the \emph{vertices} of $\mathfrak{P}$. 

\myindent We say that a vertex is \emph{degenerate} if it is contained in more than $\dim(\mathfrak{P})$ facets. More generally, we say that an entire polyhedron is degenerate if it has \emph{at least one} degenerate vertex.

\section{H-representation}\label{sec: H representation}

\myindent We can also define convex polytopes via the intersection of specific convex sets known as \emph{half-spaces}, specified by \emph{hyper-planes}. These are sets that can be specified by finite families of convex-linear functionals. Let $h:\mathbb{R}^d \to \mathbb{R}$, we say that $h$ is linear if $$h(a\pmb x+b \pmb y) = ah(\pmb x)+bh(\pmb y).$$
If it only holds that $h(a\pmb x+(1-a)\pmb{y}) \leq ah(\pmb x)+(1-a)h(\pmb y)$ for any $a \in [0,1]$ we say $h$ is convex, and if $-h$ is convex we say that $h$ is instead concave. We say that any mapping $h$ is \emph{convex-linear} if the image $h(\pmb{z})$ is a mixture of $h(\pmb{x}),h(\pmb{y})$ whenever the preimage $\pmb{z}$ is \emph{the same} mixture of $\pmb{x}$ and $\pmb{y}$~\citep{wolfe2020quantifyingbell}. Every linear map is also convex-linear (and affine) but the converse is not necessarily true. This can be seen from the fact that linear functionals must satisfy that $h(0) = 0$, i.e., they must preserve the vector space identity, while convex-linear maps \emph{do not} need to satisfy this property.  

\myindent Convex linear functionals $h$ define half-spaces via inequalities. Each pair of functional $h$ and real scalar $z$ defines a predicate $h(\pmb x) \leq z$ on variables $\pmb x$ that we call an \emph{inequality}. The set of points $H_{h,z}^{\bullet} := \{\pmb x \in \mathbb{R}^d \mid h(\pmb x) = z\}$ is the (affine) \emph{hyperplane} associated to $h$ and $z$. This hyperplane separates $\mathbb{R}^d$ into two open \emph{half-spaces} defined similarly via the conditions $h(\pmb x)<z$ and $h(\pmb x)>z$. We are interested in the \emph{closed} half-spaces defined by $h$ and $z$ given by 
\begin{equation}
    H_{h,z} := \{\pmb x \in \mathbb{R}^d \mid h(\pmb x) \leq z\}.
\end{equation}
Note that $H_{h,z}^{\bullet} = H_{h,z} \cap H_{-h,-z}$. 

\myindent Without loss of generality, $h$ can be taken as a linear functional when considering half-spaces as for every convex-linear functional $h$ and scalar $z$ there exists a linear functional $h'$ and scalar $z'$ such that $H_{h,z} = H_{h',z'}$. 

\myindent If we take then a finite collection ${(h_i,z_i)}_{i=1}^m$ of pairs of linear functionals and scalars we can define the set $\mathfrak{P}\subseteq \mathbb{R}^d$ via the intersection
\begin{equation}
    \mathfrak{P} := \bigcap_{i=1}^m H_{h_i,z_i}
\end{equation}
of all such closed halfspaces. Any such description is known as an \emph{H-representation of the convex polytope $\mathfrak{P}$}. Again, this representation is not unique in general. Also, the fact that it is a \emph{finite} intersection is crucial as infinite intersections of closed halfspaces are not necessarily polytopes. 

\myindent We can also represent this intersection in a compact notation using matrices $A: \mathbb{R}^d \to \mathbb{R}^m$ such that $(A\pmb x)_i \equiv h_i(\pmb x)$ implying that there exists $A$ for which
\begin{equation}
    \mathfrak{P} \equiv \mathfrak{P}(A,z) := \{\pmb{x} \in \mathbb{R}^d \mid A\pmb{x} \preceq \pmb{z}\}
\end{equation}
where $\pmb{z} = (z_1,z_2,\dots,z_m) \in \mathbb{R}^m$. The set $\mathfrak{P}$ is convex from the convexity of each halfspace $H_{h_i,z_i}$, and the fact that intersections preserve convexity, but it is not evident that $\mathfrak{P}$ is truly a convex polytope via our definition in terms of some V-representation. The fact that this is indeed the case, and that V- and H- are dual representations of the same object, is a fundamental theorem for convex polytope theory.

\section{Main theorem for convex polytopes}\label{sec: Main theorem for polytopes}

\myindent The two representations seem nonequivalent at first, but the fundamental theorem for convex polytopes, also known as the Minkowski--Weyl theorem~\citep{minkowski1989allgemeine,weyl1934elementare,fukuda2004frequently}, guarantees a duality between the two.

\begin{theorem}[Main theorem for convex polytopes]
A subset $\mathfrak{P}\in\mathbb{R}^d$ is defined as the convex hull of a finite set of points if, and only if, it is a bounded intersection of finitely many closed half-spaces. 
\end{theorem}

\begin{figure}[t]
    \centering
    \includegraphics[width=0.45\textwidth]{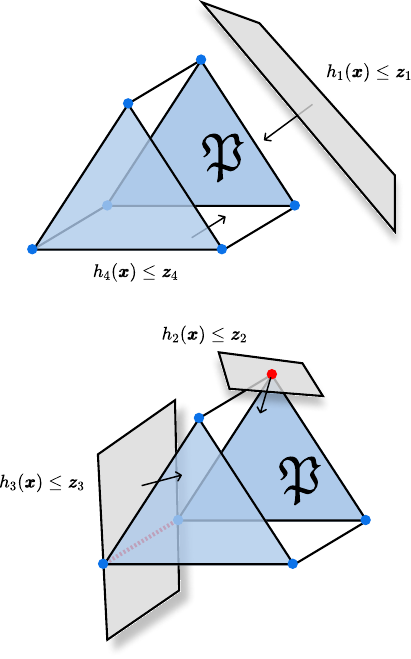}
    \caption{\textbf{Valid, tight, face and facet-defining inequalities for a convex polytope $\mathfrak{P}$}. A convex polytope $\mathfrak{P}$ is shown. Each hyper-plane showed together with a direction vector defines a specific choice of closed half-space. $h_1(\pmb{x}) \leq z_1$ defines a valid inequality for $P$, that is not tight neither face or facet-defining. $h_2(\pmb{x}) \leq z_2$ defines a tight valid inequality that is face-defining for the highlighted zero face of $\mathfrak{P}$. $h_3(\pmb{x}) \leq z_3$ is a valid tight face-defining inequality for the $1$-face highlighted. The inequality $h_4(\pmb{x}) \leq z_4$ is a facet-defining inequality for $\mathfrak{P}$. The minimal H-representation of $\mathfrak{P}$ uses $5$ facet-defining inequalities and the minimal V-representation of $\mathfrak{P}$ uses $6$ extremal points. }
    \label{fig:inequality_types}
\end{figure}

\section{Properties of convex polytopes}\label{sec: Properties of polytopes}

\myindent One particularly simple aspect of convex polytopes is that they are compact sets.

\begin{theorem}\label{theorem: polytopes are compact}
    Let $\mathfrak{P} \subseteq \mathbb{R}^d$ be a convex polytope. Then, $\mathfrak{P}$ is compact. 
\end{theorem}

\begin{proof}
    There exists some finite set of points $K$ such that $\mathfrak{P} = \mathrm{ConvHull}(K)$. Since $K$ is finite let $K = \{\alpha_1, \alpha_2, \dots, \alpha_{n}\}$. Define the $n$-simplex  $\mathfrak S^n :=\{(a_1,\dots,a_n)\in\mathbb{R}^n: a_i\ge0,\ \sum_i a_i=1\}$.  Let also $\Phi: \mathfrak{S}^{n} \to \mathbb{R}^d$ be defined such that
    \begin{equation*}
        \Phi(a_1,a_2,\dots,a_n) = \sum_{i=1}^n a_i \alpha_i.
    \end{equation*}
    Note that $\Phi$ is continuous with respect to the usual Euclidean norm topology on $\mathbb{R}^n$ and $\Phi(\mathfrak{S}^{n}) = \mathfrak{P}$. Since the simplex $\mathfrak{S}^{n}$ is a compact set and the image of a continuous function on a compact set is again compact, we conclude that $\mathfrak{P}$ is compact.
\end{proof}

\myindent In particular, we then have that $\mathfrak{C}(G)$ for every event graph $G$ is also compact. While H-representations are not unique, as it is the case for V-representations of $\mathfrak P$ via $\text{ext}(\mathfrak P)$ we can define a notion of a \emph{minimal H-representation} that is given by a finite complete set of facet-defining inequalities. 

\myindent Let $\mathfrak{P}$ be a convex polytope. We say that a linear inequality $h(\pmb{x}) \leq z$ is \emph{valid} for $\mathfrak P$ if every point in $\mathfrak P$ satisfies it, i.e., if
$$\mathfrak P \cap \{\pmb{x} \in \mathbb{R}^d \mid h(\pmb{x}) \leq z\} = \mathfrak{P}.$$
We say that it is \emph{tight} if it is valid and there exists at least one element $\pmb x \in \mathfrak P$ such that $h(\pmb x) = z$. It is also commonly used the term \emph{supporting} to denote hyperplanes $H^{\bullet}$ such that ${H}^{\bullet} \cap S \neq \emptyset$ for any convex closed bounded set $S$, \emph{and} if $S$ is in one of the hyperspaces defined by the hyperplane $H$. To define the notion of face and facet-defining inequalities we need to introduce subsets $\mathfrak F \subseteq \mathfrak P$ that we call faces. Given some valid inequality $h(\pmb x) \leq z$ for $\mathfrak P$, a face $\mathfrak F_{h,z}$ of $\mathfrak P$ is a convex subset defined as
\begin{equation}\label{eq:faces}
\mathfrak F_{h,z} := \{\pmb x\in \mathfrak P: h(\pmb x)=z\}.
\end{equation}
Every face $\mathfrak F_{h,z}$ is a convex polytope, and because of that has its own faces. Every face of a face of $\mathfrak P$ is again a face of $\mathfrak P$, albeit of a lower dimension. If $h(\pmb x)\leq z$ is valid but not tight then $\mathfrak F_{h,z} = \emptyset$. We say that $\mathfrak F_{h,z}$ is \emph{proper} when it is neither empty nor equal to the polytope, meaning that, $\mathfrak F_{h,z} \neq \mathfrak P$. 

\myindent Any subset $S \subseteq \mathbb{R}^d$ has an associated dimension described by the dimension of the \emph{affine span} (also known as the affine hull) of $S$. Similarly to the convex hull, the affine span of a set $S$ is defined by the set of all possible affine combinations of elements of $S$, and we denote it as $\text{Aff}(S)$. The dimension of $\text{Aff}(S)$ is determined by the number of affinely independent points in $S$ that span $\text{Aff}(S)$. If there are $k+1$ affinely independent points that span $\text{Aff}(S)$ then we say that the dimension of $S$ is equal to $k$ and write $\dim(S) = k$. In general, for any $S \subseteq \mathbb{R}^d$ we have that $\dim(S) \leq d$. We will refer to the \emph{dimension of a polytope $\mathfrak P$} as the dimension of its affine hull as describes, i.e., 
\begin{equation}\label{eq: polytope_dimension}
\text{dim}(\mathfrak P) := \text{dim}[\text{Aff}(\mathfrak P)].
\end{equation}
A face $\mathfrak F_{h,z}$ is said to be a $k$-face if $\dim(\mathfrak F_{h,z}) = k$. The minimal proper $0$-faces of $\mathfrak P$ are the extremal points of $\mathfrak P$ and the maximal proper faces of $\mathfrak P$ of dimension $\dim(\mathfrak P)-1$ are called \emph{facets}. Faces of dimension $\dim(\mathfrak P)-2$ are called \emph{ridges}. We say that a valid inequality $h(\pmb x) \leq z$ is \emph{facet-defining} if, and only if, $\mathfrak F_{h,z}$ is a facet of $\mathfrak P$. Let $\{\mathfrak F_i\}_{i=1}^m$ be all the facets of a convex polytope $\mathfrak P$, associated to facet-defining inequalities $\{h_i(\pmb x)\leq z_i\}_{i=1}^m$. If we write in this case, 
\begin{equation*}
    \mathfrak P = \bigcap_{i=1}^m \mathfrak \{\pmb x \in \mathbb{R}^d \mid h_i(\pmb x)\leq z_i\},
\end{equation*}
this is the so-called \emph{minimal H-representation} of the polytope $\mathfrak P$. (Note: each facet is $\mathfrak F_i=\{\pmb x\in\mathfrak P: h_i(\pmb x)= z_i\}$, but the intersection in the H-representation is of the half-spaces, \emph{not} of the facet sets.) This representation is unique for any convex polytope. 

\myindent Another relevant notion when investigating generic convex sets is that of an \emph{exposed point}~\citep{matousek2013lectures}. Given any nonempty convex set $S \subseteq \mathbb{R}^d$ we say that a point $\pmb s \in S$ is \emph{exposed} if there exists some hyperplane $H^\bullet$ such that $H^\bullet \cap S = \{\pmb s\}$ where $\{\pmb s\}$ is the singleton set having just one element, and such that $S$ is entirely contained in \emph{one and only one} of the closed half-spaces associated to the hyperplane  $H^\bullet$. Since extremal points of convex polytopes are always exposed, the exposed points of any convex polytope are the vertices, and only the vertices. The converse does not hold for more general sets, as not every extremal point of a compact convex set is exposed~\cite[Fig.~1, pg.~2]{goh2018geometry}. Along the same line we can generalize the idea of an exposed point to \emph{exposed faces}~\citep{rockafellar1997convex}. We say that a face $\mathfrak F_{h,z} = \{\pmb s \in S \mid h(\pmb x) = z\}$ associated to a face-defining inequality functional $h$ is an exposed face of $S$ if $z = \max_{\pmb s \in S}h(\pmb s)$. 

\myindent Given any inequality $h(\pmb x) \leq z$, checking if it is facet-defining for a convex polytope $\mathfrak P$ amounts to checking that the following holds: (i) There exists a set of points $\{\pmb f_i\}_i \subseteq \mathfrak P$ such that $h(\pmb f_i) = z$ for every $i$, (ii) the set $\{\pmb f_i\}_i$ is affinely independent; (iii) The affine dimension of $\{\pmb f_i\}_i$ given by $\dim[\text{Aff}(\{\pmb f_i\}_i)]$ is equal to $\dim(\mathfrak P)-1$. These imply that $\text{Aff}(\{\pmb f_i\}_i) \cap \mathfrak P = \mathfrak F_{h,z}$ is a facet of the polytope $\mathfrak P$.

\section{Relevant classes of polytopes}\label{sec: relevant classes of polytopes}

\myindent One standard family of convex polytopes is given by the \emph{ canonical $d$-simplexes}, where we take $K = \{\pmb e_i\}_{i=1}^d$ to be the canonical basis of $\mathbb{R}^d$. In general, $d$-simplexes are convex hulls of any $(d+1)$ affinely independent points.  Some polytopes have a simple characterization of faces. For example, we say that a convex polytope is \emph{simplicial} if every proper face is a simplex.  A specific type of simplicial polytopes, known as \emph{cyclic polytopes} are relevant as they are provably the polytopes that have the largest number possible of facets for a given number of vertices that any convex polytope can have. This fact results from the \textit{McMullen's upper bound theorem}~\citep{mcMullen1970upper_bound_theorem}. A cyclic polytope is constructed as follows: Take some family of strictly increasing real numbers $t_1 < t_2 < \dots < t_n$ and define the vectors $\pmb x(t_i) = (t_i, t_i^2, \dots, t_i^d) \in \mathbb{R}^d,$ for some value $d$ equal to the dimension of the ambient vector space $\mathbb{R}^d$. A cyclic polytope $\mathfrak{P}_d(n)$ is defined as the convex hull of the sequence of vectors $\{\pmb x(t_i)\}_i$, i.e., $$\mathfrak{P}_d(n):= \text{conv}(\{\pmb x(t_1),\pmb x(t_2),\ldots,\pmb x(t_n)\}).$$ Every cyclic polytope is what is known as a \emph{neighbourly polytope}, which is one where every subset of $\lfloor{d/2}\rfloor$ or less is the vertex set of a face of the polytope (in other words, the convex combination of any such subset yields some face of the polytope).

\myindent The hypercube is the convex polytope defined by $[0,1]^d$ and the class of $0/1$ polytopes constitutes the class of all possible vertex induced subpolytopes of the hypercube. One such example is the canonical $d$-simplex. Most convex polytopes considered in quantum information and foundations are 0/1-polytopes (when written using behavior functions) because they are convex combinations of deterministic behaviors. Local polytopes, prepare-and-measure classical polytopes, Kochen--Specker noncontextuality polytopes are examples. Another example of such polytopes are Birkhoff polytopes.

\section{Comparing polytopes}\label{sec: comparing polytopes}

\myindent In this section we will give an overview of some simple mappings acting on convex polytopes, and how to compare two (or more) polytopes. 

\myindent Given two polytopes $\mathfrak P$ and $\mathfrak P'$ we say that $\mathfrak P'$ is a \emph{subpolytope} of $\mathfrak P$ if as \emph{sets} $\mathfrak P' \subseteq \mathfrak P$. We say moreover that $\mathfrak P'$ is a \emph{vertex induced subpolytope} of $\mathfrak P$ if $\text{ext}(\mathfrak P') \subseteq \text{ext}(\mathfrak P)$. It is a proper subpolytope if $\mathfrak P' \neq \mathfrak P$ and $\mathfrak P' \neq \emptyset$. For relevant examples: every vertex induced subpolytope of a $d$-simplex is again a simplex. Moreover, the local polytope of a Bell scenario is always a vertex induced subpolytope of the non-signaling polytope. 

\myindent For any given polytope $\mathfrak P \subseteq \mathbb{R}^d$ we define a projection $\pi$ as the mapping $\pi: \mathbb{R}^d \to \mathbb{R}^k$ that `forgets' the $k$ coordinates of points $\pmb x \in \mathbb{R}^d$. If we write $\pmb x=(x_1,x_2,\dots,x_d)$ and let $i \in [d] \equiv \{1,2,\dots,d\}$ then we define the \emph{canonical projection} $ \pi_i: \mathbb{R}^d \to \mathbb{R}^{d-1}$ such that the image $ \pi_i(\mathfrak P)$ is given by
\begin{align}
    \pi_i(\mathfrak P) &= \bigr\{(x_1,\dots,x_{i-1},x_{i+1},\dots,x_d) \in \mathbb{R}^{d-1} \mid \nonumber \\ 
    &(x_1,\dots,x_{i-1},x_i,x_{i+1},\dots,x_d) \in \mathfrak{P} \bigr\}
\end{align}
Every other projection $\pi:\mathbb{R}^d \to \mathbb{R}^k$ is defined as a composition $\pi_{i_1}\circ\pi_{i_2}\circ\dots \circ\pi_{i_{d-k}}$, for some sequence $\{i_j\}_j \subseteq [d]$. If $\mathfrak P$ is a convex polytope then $\pi(\mathfrak P)$ is also a convex polytope, for every $\pi$.

\myindent Polytopes can be isomorphic in two ways: geometrical and combinatorial. We say that two convex polytopes $\mathfrak P_1\subseteq \mathbb{R}^{d_1}$ and $\mathfrak P_2\subseteq \mathbb{R}^{d_2}$ are \emph{geometrically isomorphic} if they are \emph{affinely equivalent}~\citep{henk2017basic}, which implies that there exists an affine mapping $\phi: \mathbb{R}^{d_1} \to \mathbb{R}^{d_2}$ from $\mathfrak P_1$ to $\mathfrak P_2$ such that when restricted to the affine spans $\phi|_{\text{Aff}}: \text{Aff}(\mathfrak P_1) \to \text{Aff}(\mathfrak P_2)$ is bijective. We say that two convex polytopes are \emph{combinatorially isomorphic}~\citep{kaibel2003isomorphism} if their face lattices are isomorphic. The face lattice~\citep{matousek2013lectures} is the lattice defined by all faces $\mathfrak F$ of a convex polytope $\mathfrak P$, with partial order defined by set inclusion. Deciding if two polytopes are isomorphic is believed to be NP-complete in general due to its connection to the graph isomorphism problem, but some computationally tractable situations are known~\citep{kaibel2003isomorphism}. 

\myindent It is possible that two polytopes are combinatorially isomorphic, but are \emph{not} geometrically isomorphic. Every $d$-simplex is geometrically isomorphic to the canonical $d$-simplex we have introduced earlier. 

\myindent Given two convex polytopes $\mathfrak P_1$ and $\mathfrak P_2$ we have that the intersection $\mathfrak P_1 \cap \mathfrak P_2$, the Cartesian product $\mathfrak P_1 \times \mathfrak P_2$, the Minkowski sum $\mathfrak P_1+_M \mathfrak P_2$, the affine image $\phi(\mathfrak P_1)$ of an affine map $\phi$, the join $\mathfrak P_1 * \mathfrak P_2$, and the (sub)direct sum $\mathfrak P_1 \oplus \mathfrak P_2$ are again all convex polytopes.

\myindent Given two polytopes $\mathfrak P_1$ and $\mathfrak P_2$ and H-representations $\{H_i^{(1)}\}_{i=1}^{m_1}$, $\{H_i^{(2)}\}_{i=1}^{m_2}$ that are completely characterized by $A_s\in \mathbb{R}^{d_s}\times \mathbb{R}^{m_s},\pmb z_s \in \mathbb{R}^{m_s}$ with $s=1,2$ and $d_1=d_2=d$ we can find the intersection polytope $\mathfrak P_1 \cap \mathfrak P_2$ by 
\begin{equation}
    \mathfrak P_1 \cap \mathfrak P_2 := \{\pmb x \in \mathbb{R}^d \mid A_1\pmb x\preceq \pmb z_1, A_2\pmb x\preceq \pmb z_2\},
\end{equation}
which implies that finding $\mathfrak P_1 \cap \mathfrak P_2$ is easy once we have an H-representation of both $\mathfrak P_1$ and $\mathfrak P_2$.

\clearpage
\ifodd\thepage\else
\null
\clearpage
\fi

\null
\thispagestyle{empty}
\pagecolor{PANTONE1807C}
\afterpage{\nopagecolor}
\newpage

\begin{backcover}
\thispagestyle{empty}
\pagecolor{PANTONECoolGray7C}
\afterpage{\nopagecolor}
\color{white}
{~\vfill
\noindent
I am grateful to the Quantum Portugal Initiative and FCT—Fundação para a Ciência e a Tecnologia (Portugal) for funding my research through the PhD Grant SFRH/BD/151199/2021.
\vfill ~}
\end{backcover}

\end{document}